\documentclass[aps,prb,twocolumn]{revtex4-1}

\usepackage{hyperref}
\usepackage[utf8]{inputenc}
\usepackage{amsmath}
\usepackage{bm}
\usepackage{graphicx}
\usepackage{verbatim}
\usepackage{color}

\newcommand{\etal}{\textit{et al.}}
\newcommand{\eg}{\textit{e.g.}}
\newcommand{\ie}{\textit{i.e.}}
\newcommand{\cf}{\textit{cf.}}
\newcommand{\angstrom}{\textup{\AA}}

\begin{document}

\title{First-principles theory of field-effect doping in transition-metal dichalcogenides: Structural properties, electronic structure, Hall coefficient, and electrical conductivity}
\date{\today}
\author{Thomas Brumme}
\email{Thomas.Brumme@impmc.upmc.fr}
\author{Matteo Calandra}
\author{Francesco Mauri}
\affiliation{CNRS, UMR 7590}
\affiliation{Sorbonne Universit\'{e}s, UPMC Univ Paris 06, IMPMC - Institut de Min\'{e}ralogie, de Physique des Mat\'{e}riaux, et de Cosmochimie, 4 place Jussieu, F-75005, Paris, France}

\begin{abstract}
We investigate how field-effect doping affects the structural properties, the electronic structure and
the Hall coefficient of few-layers transition metal dichalcogenides by using density-functional theory.
We consider mono-, bi-, and trilayers of the H polytype of MoS$_2$, MoSe$_2$, MoTe$_2$, WS$_2$, and WSe$_2$ and provide a full
database of electronic structures and Hall coefficients for hole and electron doping. We find that, for
both electron and hole doping, the electronic structure depends on the number of layers and cannot be described by a rigid band shift.
Furthermore, it is important to relax the structure under the asymmetric electric field.
Interestingly, while the width of the conducting channel depends on the doping, the number of occupied bands at each
given $\mathbf{k}$ point is almost uncorrelated with the thickness of the doping-charge distribution.
Finally, we calculate within the constant-relaxation-time approximation the electrical conductivity
and the inverse Hall coefficient. We demonstrate that in some cases the charge determined by Hall-effect measurements
can deviate from the real charge by up to 50\%. For hole-doped MoTe$_2$ the Hall charge has even the wrong polarity at
low temperature. We provide the mapping between the doping charge and the Hall coefficient.
In the appendix we present
more than 250 band structures for all doping levels of the transition-metal dichalcogenides considered
within this work.
\end{abstract}


\maketitle

\section{Introduction}
Since the rise of graphene\cite{geim2007} and the discovery of topological insulators\cite{kane2010}
a lot of interesting physics has been found in systems with reduced dimensions.
Other two-dimensional (2D) material, such as monolayers or few-layer systems (nanolayers) of
transition-metal dichalcogenides\cite{wang2012,fiori2014,ganatra2014,jariwala2014,xu2014}
(TMDs) are gaining importance because of their intrinsic band gap.
TMDs are MX$_2$-type compounds where M is a transition metal (\eg, M = Mo, W)
and X represents a chalcogen (S, Se, Te). These materials form layered structures in which
the different X--M--X layers are held together by weak van der Waals forces.
Thus, similar to graphene, one can easily extract single or few layers from the bulk compound
using the mechanical-exfoliation or other experimental techniques. 

Doping these nanolayers with field-effect transistors (FETs) is particularly
appealing\cite{konenkamp1988,radisavljevic2011,ye2012,das2013_1,das2013_2,najmaei2013,radisavljevic2013,wu2013,yuan2013,chuang2014,lu2014,ovchinnikov2014,roesner2014,yuan2014}
as it allows for the exploration the semiconducting, metallic, superconducting, and charge-density-wave
regimes in reduced dimensionality.
Furthermore, the TMDs are promising materials to realize valleytronics, \ie, the usage of the
valley index of carriers to process information\cite{wu2013,mak2014,xu2014}.

Despite these challenging experimental perspectives, the understanding
of structural, electronic, and transport
properties at high electric field in FET configuration is still limited,
particularly in the physically relevant case of multilayer samples.
Previous theoretical works \cite{ge2013, roesner2014}
analyzed the high doping limit of $20\:\mathrm{nm}$ thick
MoS$_2$ flakes\cite{ye2012}, relevant for ionic-liquid based FETs,
by assuming that only the topmost layer is doped uniformly.
However, it is unclear to what extent the doping of thick flakes can be
modeled in this approximation as the thickness
of the conductive channel is not experimentally accessible.

A second crucial issue is the determination of the doping charge. Usually the charge is
determined via Hall-effect measurements. However, the interpretation of Hall experiments
assumes a 2D electron-gas model, most likely valid only in the low doping regime. In
TMDs, due to the multivalley electronic structure, this assumption is highly questionable.

In this paper we solve these issues and provide a thorough study of structural,
electronic and transport properties and of their changes under field-effect doping for TMDs.
We use our recently developed first-principles theoretical approach to model doping in
field-effect devices\cite{brumme2014}. The method allows for calculation of the electronic
structure as well as complete structural relaxation in field-effect configuration using
density-functional theory (DFT). We apply our approach to the H polytype of MoS$_2$,
MoSe$_2$, MoTe$_2$, WS$_2$, and WSe$_2$.

The paper is organized as follows: In section \ref{sec:DFT} we summarize the parameters and methods used within the paper
and the relaxed geometries of the bulk TMDs. Then we will first show the results for the undoped mono-, bi-, and
trilayer systems (section \ref{sec:uncharged}).
After a brief discussion on the quantum capacitance (\ref{sec:quancap}), we will
investigate the changes of the geometry and the electronic structure of the
different TMDs in sections \ref{sec:phases} and \ref{sec:dope}, respectively.
Finally, we will focus on how the doping-charge concentration can be determined experimentally
by Hall-effect measurements. We will show in section \ref{sec:hall} that the results of such a measurement cannot be
interpreted within a 2D electron-gas model but that the specific band structure of the TMDs and its
changes in a field-effect setup need to be taken into account.
In section \ref{sec:cond_dos} we will furthermore investigate the density of states (DOS) at the Fermi energy
and the electrical conductivity as function of doping.
In the end, we will summarize our results and draw some final conclusion in section \ref{sec:conclusion}.
In the appendix we provide a full database of electronic structures for all doping levels considered in this work (in total more than 250 calculations, Figs.~28--63)
and Hall coefficients (Figs.~64--66) of mono- , bi-, and trilayer dichalcogenides as a function of doping.

\section{Computational details}
\label{sec:DFT}
All calculations were performed within the framework of DFT using the {\sc Quantum ESPRESSO} package\cite{quantumespresso}
which uses a plane-wave basis set to describe the valence-electron wave function and charge density.
We employed full-relativistic, projector augmented wave potentials\cite{blochl1994}.
While the local-density approximation is known to underestimate the lattice parameters, generalized-gradient
approximations for the exchange-correlation energy overestimate the out-of-plane lattice constant (see, \eg, Ref.~\citenum{roldan2014} and references therein).
In our FET setup a correct description of the interlayer distance is however very important. Accordingly,
we choose the Perdew-Burke-Ernzerhof functional\cite{pbe96} (PBE) for the exchange-correlation energy
and furthermore included dispersion corrections\cite{grimme2006} (D2). This also
leads to the best agreement with both the experimental in-plane and out-of-plane lattice parameters (\cf, Tab.~\ref{tab:lattice}).
A comparison between PBE and LDA for MoS$_2$ can be found in the appendix (Fig.~27).

Using the experimental lattice parameters of the H polytype of the bulk structures\cite{podberezskaya2001} as starting geometry,
we minimized the total energy as a function of the lattice parameters until it changed by less then $2\:\mathrm{meV}$.
For the molybdenum-containing dichalcogenides we used a cutoff of $50\:\mathrm{Ry}$ and $410\:\mathrm{Ry}$ ($1\:\mathrm{Ry}\approx13.6\:\mathrm{eV}$)
for the wave functions and the charge density, respectively, while for the tungsten dichalcogenides we chose $55\:\mathrm{Ry}$/$410\:\mathrm{Ry}$.
The Brillouin zone (BZ) integration has been performed with a Monkhorst-Pack grid\cite{monkhorst1976} of $16\times16\times4$ $\mathbf{k}$ points
and using a Gaussian broadening of $0.002\:\mathrm{Ry}\approx27\:\mathrm{meV}$. The convergence with respect to
the number of $\mathbf{k}$ points as well as the wave-function and charge-density cutoff has been checked.
The self-consistent solution of the Kohn-Sham equations was obtained when the total energy changed by less
than $10^{-9}\:\mathrm{Ry}$ and the maximum force on all atoms was less than $5\cdot10^{-4}\:\mathrm{Ry}\:a_0^{-1}$
($a_0\approx0.529177\:\angstrom$ is the Bohr radius).
The lattice parameters thus determined are given in Tab.~\ref{tab:lattice} and agree within 2\% with the experimental values.
\begin{table}
 \caption{\label{tab:lattice}Comparison of the calculated and experimental\cite{podberezskaya2001} lattice parameters.}
 \begin{ruledtabular}
 \begin{tabular}{lcccc}
          &    a calc.    &    a exp.     &    c calc.    &    c exp.\\
\hline
  MoS$_2$ & $3.197\:\angstrom$ & $3.160\:\angstrom$ & $12.38\:\angstrom$ & $12.29\:\angstrom$ \\
  MoSe$_2$& $3.328\:\angstrom$ & $3.289\:\angstrom$ & $13.07\:\angstrom$ & $12.93\:\angstrom$ \\
  MoTe$_2$& $3.536\:\angstrom$ & $3.518\:\angstrom$ & $14.00\:\angstrom$ & $13.97\:\angstrom$ \\
  WS$_2$  & $3.190\:\angstrom$ & $3.153\:\angstrom$ & $12.15\:\angstrom$ & $12.32\:\angstrom$ \\
  WSe$_2$ & $3.341\:\angstrom$ & $3.282\:\angstrom$ & $12.87\:\angstrom$ & $12.96\:\angstrom$ 
 \end{tabular}
 \end{ruledtabular}
\end{table}

The final geometry of the bulk system was used as the starting geometry for the relaxation of the layered-2D systems in an FET setup. To achieve this, the size of
the unit cell was fixed in plane and the perpendicular size was increased such that the vacuum region between the repeated images was at least $23\:\angstrom$.
The layers were stacked as in the H polytype of the bulk compound.
The BZ integration has been performed with a Monkhorst-Pack grid of $64\times64\times1$ $\mathbf{k}$ points
for the charged systems and $16\times16\times1$ $\mathbf{k}$ points for the neutral ones. In order to correctly determine the Fermi energy
in the charged systems, we performed a non-self-consistent calculation on a denser $\mathbf{k}$-point grid of at least $90\times90\times1$
points starting from the converged charge density. All other parameters were the same as in the calculations for the bulk systems.
For the total-energy calculations of the 1T and 1T' polytype of MoS$_2$ and WSe$_2$ under FET doping, we doubled the unit-cell size along one in-plane direction and
correspondingly halved the number of $\mathbf{k}$ points along this direction.

Several methods have been developed to study electrostatics in periodic boundary condition\cite{otani2006,brumme2014,andreussiPRB2014,andreussiJCP2014}
with different experimental geometries.
We used our recently developed method\cite{brumme2014} as it is tailored for the
FET setup and allows for structural optimization.
The dipole for the dipole correction\cite{brumme2014,bengtsson1999} was placed at $z_\mathrm{dip}=d_\mathrm{dip}/2$ with $d_\mathrm{dip}=0.01\:L$
and $L$ being the unit-cell size in the direction perpendicular to the 2D plane -- $L$ changed for the different calculations and was between
$34\:\angstrom$ and $48\:\angstrom$. The charged plane modeling the gate electrode\cite{brumme2014} was placed close to the dipole at
$z_\mathrm{mono}=0.011\:L$.
A potential barrier with a height of $V_0=2.5\:\mathrm{Ry}$ and a width of $d_b=0.1\:L$ was used
in order to prevent the ions from moving too close to the gate electrode.
The final results were found to be independent of the separation of the dipole planes, as well as the barrier height and width as long as it is high
or thick enough to ensure that the electron density at the position of the dipole and the gate electrode is zero,
$\rho^e\left(z_\mathrm{mono}\right)=\rho^e\left(z_\mathrm{dip}\right)=0$. As we will often give the doping-charge concentration per unit cell $\mathrm{n}$ (\ie,
in charge per unit cell, $e$/unit cell, with the elementary charge $e\approx1.602\times10^{-19}\:\mathrm{C}$), we summarized in Tab.~\ref{tab:conversion} the conversion
to charge-carrier concentration per area $n$ (in $\mathrm{cm}^{-2}$) for the different dichalcogenides and two typical doping-charge concentrations of $\mathrm{n}=0.01\:e$/unit cell and $\mathrm{n}=0.15\:e$/unit cell.
Throughout the paper we will use $n<0$ and $n>0$ for electron and hole doping, respectively. Typical charge-carrier concentrations
which can be achieved in experiments using either solid-state dielectrics such as SiO$_2$ or ionic-liquid based FETs can be found in Tab.~\ref{tab:experimental}.
Note that in principle those maximum concentrations could be possible for all TMDs\cite{bulletin_aps} even if we did not find references for, \eg,
ionic-liquid-based field-effect doping of MoSe$_2$.
\begin{table}
 \caption{\label{tab:conversion}Conversion for the doping-charge concentration $\mathrm{n}$ (in $e$/unit cell)
 to charge-carrier concentration $n$ per area $\mathrm{cm}^{-2}$ for the different dichalcogenides and two typical
 doping-charge concentrations of $\mathrm{n}=0.01\:e$/unit cell and $\mathrm{n}=0.15\:e$/unit cell.}
 \begin{ruledtabular}
 \begin{tabular}{lcc}
          & $\mathrm{n}=0.01\:e$/unit cell & $\mathrm{n}=0.15\:e$/unit cell\\
\hline
  MoS$_2$ & $n\approx0.1127\cdot10^{14}\:\mathrm{cm}^{-2}$& $n\approx1.6911\cdot10^{14}\:\mathrm{cm}^{-2}$\\
  MoSe$_2$& $n\approx0.1042\cdot10^{14}\:\mathrm{cm}^{-2}$& $n\approx1.5636\cdot10^{14}\:\mathrm{cm}^{-2}$\\
  MoTe$_2$& $n\approx0.0924\cdot10^{14}\:\mathrm{cm}^{-2}$& $n\approx1.3856\cdot10^{14}\:\mathrm{cm}^{-2}$\\
  WS$_2$  & $n\approx0.1135\cdot10^{14}\:\mathrm{cm}^{-2}$& $n\approx1.7027\cdot10^{14}\:\mathrm{cm}^{-2}$\\
  WSe$_2$ & $n\approx0.1035\cdot10^{14}\:\mathrm{cm}^{-2}$& $n\approx1.5521\cdot10^{14}\:\mathrm{cm}^{-2}$
 \end{tabular}
 \end{ruledtabular}
\end{table}
\begin{table}
 \caption{\label{tab:experimental}Maximum experimental charge-carrier concentration $n$ per area $\mathrm{cm}^{-2}$
 for the different dichalcogenides using either a solid-state or an ionic-liquid based FET. References are given after
 the charge-carrier concentration.}
 \begin{ruledtabular}
 \begin{tabular}{lcll}
          & Polarity & Solid-state FET & Ionic-liquid FET\\
\hline
  MoS$_2$ & n-type   & $\approx-3.6\cdot10^{13}$, [\!\!\citenum{radisavljevic2013}]& $\approx-9.8\cdot10^{14}$, [\!\!\citenum{lee2014}]\\
          & p-type   & ---                                                         & --- \\
  MoSe$_2$& n-type   & $\approx-3.9\cdot10^{12}$, [\!\!\citenum{pradhan2014_1}]    & --- \\
          & p-type   & $\approx+2.0\cdot10^{12}$, [\!\!\citenum{pradhan2014_1}]    & --- \\
  MoTe$_2$& n-type   & $\approx-1.3\cdot10^{13}$, [\!\!\citenum{pradhan2014_2}]    & $\approx-1.1\cdot10^{13}$, [\!\!\citenum{lezama2014}]\\
          & p-type   & ---                                                         & $\approx+1.8\cdot10^{13}$, [\!\!\citenum{lezama2014}]\\
  WS$_2$  & n-type   & $\approx-1.0\cdot10^{14}$, [\!\!\citenum{ovchinnikov2014}]  & $\approx-4.0\cdot10^{14}$, [\!\!\citenum{jo2015}]\\
          & p-type   & ---                                                         & $\approx+3.5\cdot10^{13}$, [\!\!\citenum{braga2012}]\\
  WSe$_2$ & n-type   & ---                                                         & $\approx-1.4\cdot10^{14}$, [\!\!\citenum{yuan2013}]\\
          & p-type   & $\approx+9.0\cdot10^{12}$, [\!\!\citenum{pradhan2015}]      & $\approx+1.9\cdot10^{15}$, [\!\!\citenum{yuan2013}]
 \end{tabular}
 \end{ruledtabular}
\end{table}

In order to calculate the Hall tensor $R_{ijk}(T;E_F)$, we used the BoltzTraP code\cite{madsen2006} to determine the conductivity tensors
in Eqs.~(\ref{eq:sigmaab_Tmu}) and (\ref{eq:sigmaabg_Tmu}) (see section \ref{sec:hall} for more details).
We fitted the band structure for each doping of the different TMDs by using 55-times more plane waves than bands and used afterwards
the in-plane components of the energy-projected tensors to calculate the Hall coefficient
$R_{xyz}(T;E_F)$ for temperature $T$ and chemical potential $E_F$. $R_{xyz}(T;E_F)$ is the only relevant Hall coefficient for
our 2D systems assuming that the magnetic field is applied perpendicular to the layers. We checked the convergence by
calculating $R_{xyz}(T;E_F)$ with increasing number of $\mathbf{k}$ points and found that the results for the $64\times64\times1$
grid and the dense grid of the non-self-consistent calculation are the same. 

\section{Results}

\subsection{Electronic structure of TMDs}
\label{sec:uncharged}

In the following we will first briefly summarize the results for the undoped TMDs before investigating the
changes under field-effect doping.
We focus on the changes in the valence-band maximum and the conduction-band minimum with changing
transition metal or chalcogen and compare them with other results found in literature\cite{roldan2014}.

\begin{figure}
 \includegraphics[width=0.46\textwidth,clip=]{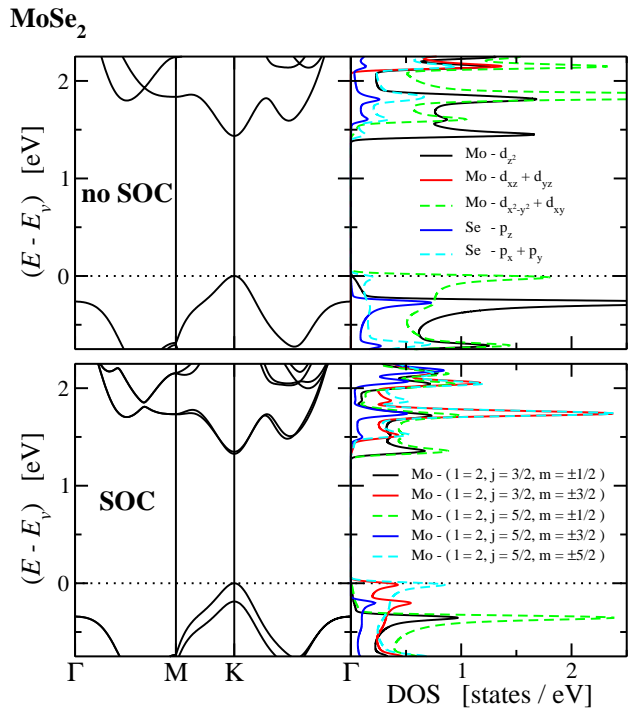}
 \caption{\label{fig:MoSe2_overview}(color online) Band structure and density of states projected onto atomic orbitals for monolayer MoSe$_2$
          without (upper panel) and with (lower panel) including spin-orbit coupling. The energy is given relative to the valence-band
          maximum $E_v$.
          As apparent in the case without SOC, mainly
          in-plane states contribute to the valence-band maximum near K (Mo $d_{x^2-y^2}$ and $d_{xy}$, Se $p_{x/y}$), while the maximum
          near $\Gamma$ is formed by out-of-plane states (Mo $d_{z^2}$, Se $p_{z}$). This also holds in the SOC case where the valence-band
          states with mainly in-plane character can be found near K ($j=5/2, m_j=\pm\,5/2,\pm\,3/2$ and $j=3/2, m_j=\pm\,3/2$) while those
          with more out-of-plane character can be found near $\Gamma$ ($m_j=\pm\,1/2$ for both $j=5/2$ and $j=3/2$, see Ref.~\citenum{corso2005}
          for the states in terms of spherical harmonics). On the other hand, the conduction-band minimum near K has mainly contributions
          from out-of-plane Mo and in-plane Se states (without SOC: Mo $d_{z^2}$ and Se $p_{x/y}$, with SOC: Mo $m_j=\pm\,1/2$ for both $j=5/2$
          and $j=3/2$).}
\end{figure}

Figure \ref{fig:MoSe2_overview} shows the band structure and the projected density of states (pDOS) for monolayer MoSe$_2$ with and without
including spin-orbit coupling (SOC). Monolayer molybdenum diselenide is (as most TMDs) a direct-band-gap semiconductor with a DFT gap of about $1.329\:\mathrm{eV}$
at the K point. Our calculated gap is smaller by $83\:\mathrm{meV}$ than the one in Ref.~\citenum{roldan2014} which can be attributed to our slightly
larger in-plane lattice parameter as the size of the band gap decreases with increasing lattice constant\cite{peelaers2012,yun2012,shi2013,guzman2014}.
This is due to the fact that the valence-band maximum at K is formed by in-plane states of both the transition metal and the chalcogen\cite{guzman2014}.
On the other hand, the conduction-band minimum at K is mainly formed by out-of-plane Mo states ($d_{z^2}$ without SOC and $m_j=\pm\,1/2$ for both $j=5/2$ and $j=3/2$ including SOC)
and in-plane states of the chalcogen.
The valence-band maximum near $\Gamma$ has basically only out-of-plane states of Mo and Se as can be seen in Fig.~\ref{fig:MoSe2_overview}.
This will become very important for hole doping of the nanolayers in an FET setup -- depending on which valley is doped (K or $\Gamma$) one can expect different doping-charge
distributions. Energetically very close to the conduction-band minimum at K is a minimum half-way between K and $\Gamma$. The corresponding point in
$\mathbf{k}$ space is called Q in literature (sometimes $\Lambda$ or $\Lambda_\mathrm{min}$ as it is a minimum along the $\Lambda$ line, from $\Gamma$ to K)
even if it is not a high symmetry point of the BZ. This is also why this minimum does not lie exactly at the
same point for the different TMDs and its position can even change if the number of layers is increased. The states close to the Q point have a stronger
in-plane character and can thus also lead to a different doping-charge distribution if the doping occurs mainly at this point in the BZ.
The same results for the character of the different valleys were also obtained in Refs.~\citenum{cappelluti2013,shi2013,guzman2014}.

The different character of the states in the different valleys is even more important for the other TMDs.
From sulfur to tellurium the difference between the minimum at K and Q decreases: for MoS$_2$ the minimum at K is lower by $279\:\mathrm{meV}$
while it is only $154\:\mathrm{meV}$ and $72\:\mathrm{meV}$ lower for MoSe$_2$ and MoTe$_2$, respectively. The change in the case of the tungsten
dichalcogenides is much lower which however might be due to the stronger spin-orbit splitting of the bands near Q compared the splitting at K.
The band structures of all undoped TMDs are summarized in the appendix, Figs.~19--23.

\begin{figure}
 \includegraphics[width=0.46\textwidth,clip=]{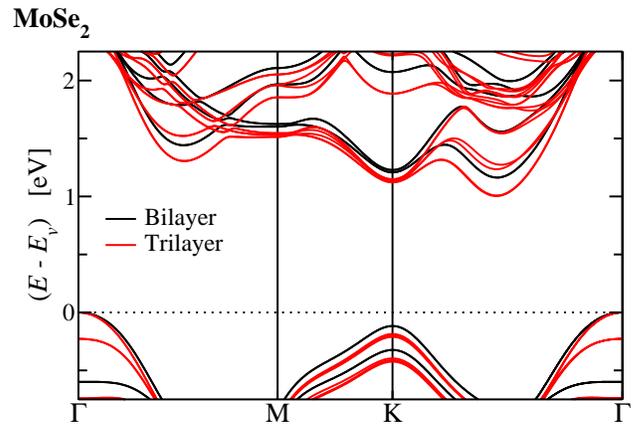}
 \caption{\label{fig:MoSe2_bitri}Band structure for bi- and trilayer MoSe$_2$. The energy is given relative to the valence-band
          maximum $E_v$. Molybdenum diselenide changes (as all TMDs
          investigated in this paper) from a direct-band-gap semiconductor to an indirect one when the number of layers is increased.}
\end{figure}
Increasing the number of layers in TMDs leads to a well-known change from a direct-band-gap semiconductor to an indirect
one\cite{mak2010,splendiani2010,cappelluti2013,roldan2014,ruppert2014,zhangyi2014} as shown in Fig.~\ref{fig:MoSe2_bitri}.
The change of the direct band gap at K with increasing the number of layers is much smaller than the changes at $\Gamma$ or Q.
This is due to the small hybridization between different layers at K as those states have only in-plane chalcogen character. On the other hand,
both valleys at $\Gamma$ and Q have contributions from Se $p_z$ states. Accordingly,
the maximum (minimum) at $\Gamma$ (Q) shift up (down) in energy which eventually leads to an indirect band gap between $\Gamma$ and Q (see also Ref.~\citenum{cappelluti2013} for an in-depth analyses).
The transition between the indirect gap $\Gamma$ $\rightarrow$ K and $\Gamma$ $\rightarrow$ Q happens at different number of layers and occurs either in the bilayer case (MoSe$_2$),
in the trilayer case (WS$_2$ and WSe$_2$), or in the bulk limit (MoS$_2$, MoTe$_2$).
Depending on the level of theory that was used, one can also find very different results in literature when this transition occurs.
Most calculations were done for MoS$_2$ for which the transition either already occurs for the bilayer\cite{ellis2011,liu2012,yun2012} or with larger number of layers\cite{cheiwchanchamnangij2012,zhang2014,zunger2015}.
In fact, Ramasubramaniam \etal{} have shown in Ref.~\citenum{ramasubramaniam2011} that using the experimental bulk distance between the layers in the bilayer case also leads to an indirect band gap
$\Gamma$ $\rightarrow$ Q while relaxation using PBE+D2 leads to an indirect gap $\Gamma$ $\rightarrow$ K. Also for other TMDs one can find different results\cite{bhattacharyya2012,kumar2012,yun2012,debbichi2014,zunger2015}.
For a comprehensive review of the theoretical papers see also Ref.~\citenum{roldan2014} and references therein.
Molybdenum ditelluride is especially peculiar, since the calculations show that its valence-band maximum is located at K even in the trilayer
case. Furthermore, we find that the difference between the maximum at $\Gamma$ and K is only $26\:\mathrm{meV}$ in bulk MoTe$_2$ which is in agreement
with the experimental results in Ref.~\citenum{boker2001}.

In section \ref{sec:dope} we will see that, \eg, the varying difference between the conduction-band minimum at K and Q for the different TMDs will also
lead to a different thickness of the conductive channel for electron doping in an FET setup, while nearly all TMDs will behave similarly under
hole doping. However, before investigating the changes under field-effect doping we want to focus on another problem that can make it difficult
to dope a 2D system -- the quantum capacitance.

\subsection{Quantum capacitance}
\label{sec:quancap}

A prominent example in which the quantum capacitance hinders the doping via field-effect setup is graphene. Due
to its linear dispersion relation at the K points, the charge that can be induced in an FET setup is much smaller
than the corresponding charge at the gate electrode\cite{fang2007,xia2009}. Thus, doping concentrations exceeding
$10^{13}\:\mathrm{cm}^{-2}$ are hardly achievable using common dieletrics such as SiO$_2$ or HfO$_2$.
Similarly, the quantum capacitance in nanolayers of TMDs could also reduce the amount of induced charge.
In the following we want to show that TMDs are however quite different from graphene and that the quantum
capacitance is not relevant in their case as soon as the Fermi energy is within the conduction or valence band.
In fact, experimentally doping concentrations in the order of $10^{14}\:\mathrm{cm}^{-2}$ are possible using ionic-liquid based
FETs\cite{ye2012,taniguchi2012,zhang2012,yuan2013,jo2015} (\cf, Tab.~\ref{tab:experimental}).

\begin{figure}
 \includegraphics[width=0.46\textwidth,clip=]{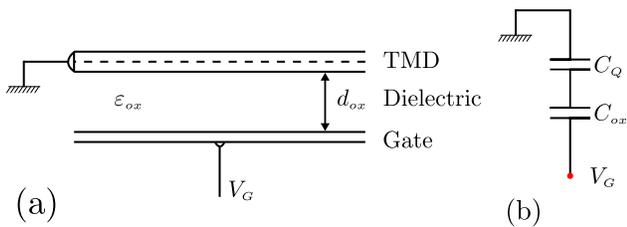}
 \caption{\label{fig:circuit}
          (a) Schematic illustration of an FET setup in which the 2D metallic system is separated from the gate electrode by a dielectric
          with dielectric constant $\varepsilon_{ox}$ of thickness $d_{ox}$.
          (b) Equivalent circuit for the overall capacitance seen at the gate electrode.}
\end{figure}
The term ``quantum capacitance'' was first used by Serge Luryi\cite{luryi1988} in order to develop an equivalent
circuit model to describe the incomplete screening of an electric field by a 2D electron gas. When a 2D metallic
system is contacted by a gate electrode (separated by a dielectric as shown in Fig.~\ref{fig:circuit}(a)), the
electric field generated by the charges on the dielectric surface leads to a shift of the Fermi level of the 2D metal.
This effect results in a modified capacitance with respect to the geometrical capacitance $C_{ox}$ per unit surface
\begin{align}
\label{eq:geometricalcap}
 C_{ox} &= \varepsilon_{ox}\varepsilon_0\:d_{ox}^{-1},
\end{align}
obtained for a capacitor having a dielectric constant $\varepsilon_{ox}$ and thickness $d_{ox}$.

As shown in Fig.~\ref{fig:circuit}(b), the quantum capacitance $C_Q$ is in series with that of the dielectric, namely
\begin{align}
\label{eq:fullcap}
 \frac{1}{C} &= \frac{1}{C_{ox}} + \frac{1}{C_Q}.
\end{align}
Under the simplified assumption that both $C_{ox}$ and $C_Q$ are independent of the applied gate voltage $V_G$, the charge
induced in the 2D system $n_Q$ can be written as:
\begin{align}
\label{eq:charge}
 n_Q &= \left\{
 \begin{array}{cl}
   0                    &\forall\:V_G\in(V_v,V_c)\\
   C\left(V_v-V_G\right)  &\forall\:V_G<V_v\\
   C\left(V_c-V_G\right)  &\forall\:V_G>V_c
 \end{array}
 \right.
\end{align}
where $V_v$ and $V_c$ are the onset potentials to fill the valence-band minimum or conduction-band maximum, respectively.
Thus, when $C_Q\gg C_{ox}$, we have
\begin{align}
\label{eq:capacitance}
 C &= C_{ox}\:\left(1+\frac{C_{ox}}{C_Q}\right)^{-1} \approx C_{ox}.
\end{align}
and we regain the classical result, \ie, the charge that can be induced in the 2D system depends only on the applied gate voltage
and the capacitance between gate and sample.

As shown in Refs.~\citenum{luryi1988,john2004}, the quantum capacitance in this case and for gate voltages larger (smaller) than
the onset potential of the conduction band (valence band) is given by
\begin{align}
\label{eq:quancap}
 C_Q &= \frac{g_sg_vmq^2}{\pi \hbar^2},
\end{align}
where $g_s$ and $g_v$, $m$, and $q$ are the spin and valley degeneracies, the effective mass, and the charge, respectively.
Depending on (i) the dielectric thickness $d_{ox}$, (ii) the effective mass of the 2D metallic system, and (iii) the number of valleys $\nu=g_s\,g_v$,
the quantum capacitance $C_Q$ can be relevant or not.

\begin{table}
 \caption{\label{tab:quancap}Geometrical capacitances for parallel-plate capacitors with $100\:\mathrm{nm}$ SiO$_2$
          (relative permittivity $\varepsilon_r=3.9$), $10\:\mathrm{nm}$ HfO$_2$ ($\varepsilon_r\approx20$),
          and an $1\:\mathrm{nm}$ thick ionic-liquid (IL) FET ($\varepsilon_r\approx15$), quantum capacitance for
          electron doping of the K valley ($m\approx0.5m_0$) and hole doping of the $\Gamma$ valley ($m\approx m_0$) of MoS$_2$ ($\nu=4$),
          and the overall capacitance of a setup as shown in Fig.~\ref{fig:circuit}(b). All capacitances are given in units
          of $\mathrm{\mu F\:cm^{-2}}$.}
 \begin{ruledtabular}
 \begin{tabular}{lcc}
              & $C_Q^{\mathrm{K}}=133.9$ & $C_Q^{\mathrm{\Gamma}}=267.7$ \\
\hline
 $C_\mathrm{SiO_2}=0.035$ & $C=0.035$ & $C=0.035$\\
 $C_\mathrm{HfO_2}=1.771$ & $C=1.748$ & $C=1.759$\\
 $C_\mathrm{IL}=13.28$    & $C=12.08$ & $C=12.65$
 \end{tabular}
 \end{ruledtabular}
\end{table}
Table \ref{tab:quancap} shows the total capacitance, geometrical capacitance for typical gate dielectrics, and the quantum capacitance for electron doping
of the K valley or hole doping of the $\Gamma$ valley of MoS$_2$ (the effective masses of the different valleys were taken from Ref.~\citenum{yun2012}).
As the geometrical capacitance of a parallel-plate capacitor increases with decreasing thickness of the dielectric, an ionic-liquid (IL) FET
with an $1\:\mathrm{nm}$ thick electric double layer (inner Helmholtz plane) shows the largest deviation of the total capacitance $C$ from 
the geometrical capacitance $C_\mathrm{IL}$ -- even for a thin dielectric layer of $10\:\mathrm{nm}$ HfO$_2$, $C\approx0.986\:C_\mathrm{HfO_2}$.
Yet, for TMDs the number of valleys increases with increasing doping (valleys at K and Q or K and $\Gamma$ for electron or hole doping, respectively),
which also leads to a considerably larger DOS at the Fermi energy than for a single quadratic band. Thus, the quantum capacitance $C_Q$ is further increased\cite{john2004}
which in turn leads to $C\approx C_{ox}$. Similar results have been found by Nan Ma and Debdeep Jena in Ref.~\citenum{ma2015} who also
provide a detailed description of the low-doping regime $n<10^{13}\:\mathrm{cm}^{-2}$

\subsection{Structural changes under field-effect doping}
\label{sec:phases}
The doping via FET setup has only a minor influence on the structure of the TMDs -- the changes are much smaller than
for, \eg, ZrNCl\cite{brumme2014}. This is due to the weak polarity of the bond between the transition-metal and the chalcogen.
The largest change can be found for the layer thickness ($z$ component of the chalcogen--chalcogen vector) of the layer closest
to the charged plane representing the gate electrode -- the layer thickness increases by $\approx0.06\:\angstrom$ for a large
electron doping of $\mathrm{n}=-0.3\:e$/unit cell ($n\approx-\,3.16\cdot10^{14}\:\mathrm{cm}^{-2}$) and decreases by
$\approx0.02\:\angstrom$ for a large hole doping of $\mathrm{n}=+0.3\:e$/unit cell.
This change is mainly due to the increase/decrease of the chalcogen--transition-metal bond length of those being closest to the gate -- we
find $\approx+0.04\:\angstrom$ for $\mathrm{n}=-0.3\:e$/unit cell and $\approx-0.02\:\angstrom$ for $\mathrm{n}=+0.3\:e$/unit cell.
Accordingly, there is also a small change in the angle between the first chalcogen, the transition-metal, and the second chalcogen
of up to $+0.9^{\circ}$ ($-0.4^{\circ}$) for large electron (hole) doping. 
Note that even if the structural changes seem to be small it is still important to relax the system in the FET setup. Otherwise,
the band structure can be quite different for doping $|\mathrm{n}|>0.15\:e$/unit cell as exemplified in Fig.~24 in the appendix.

\begin{figure}
 \begin{tabular}[t]{lcr}
  \begin{tabular}[b]{@{}c@{}}
      1H-MoS$_2$\\ \includegraphics[width=0.08\textwidth]{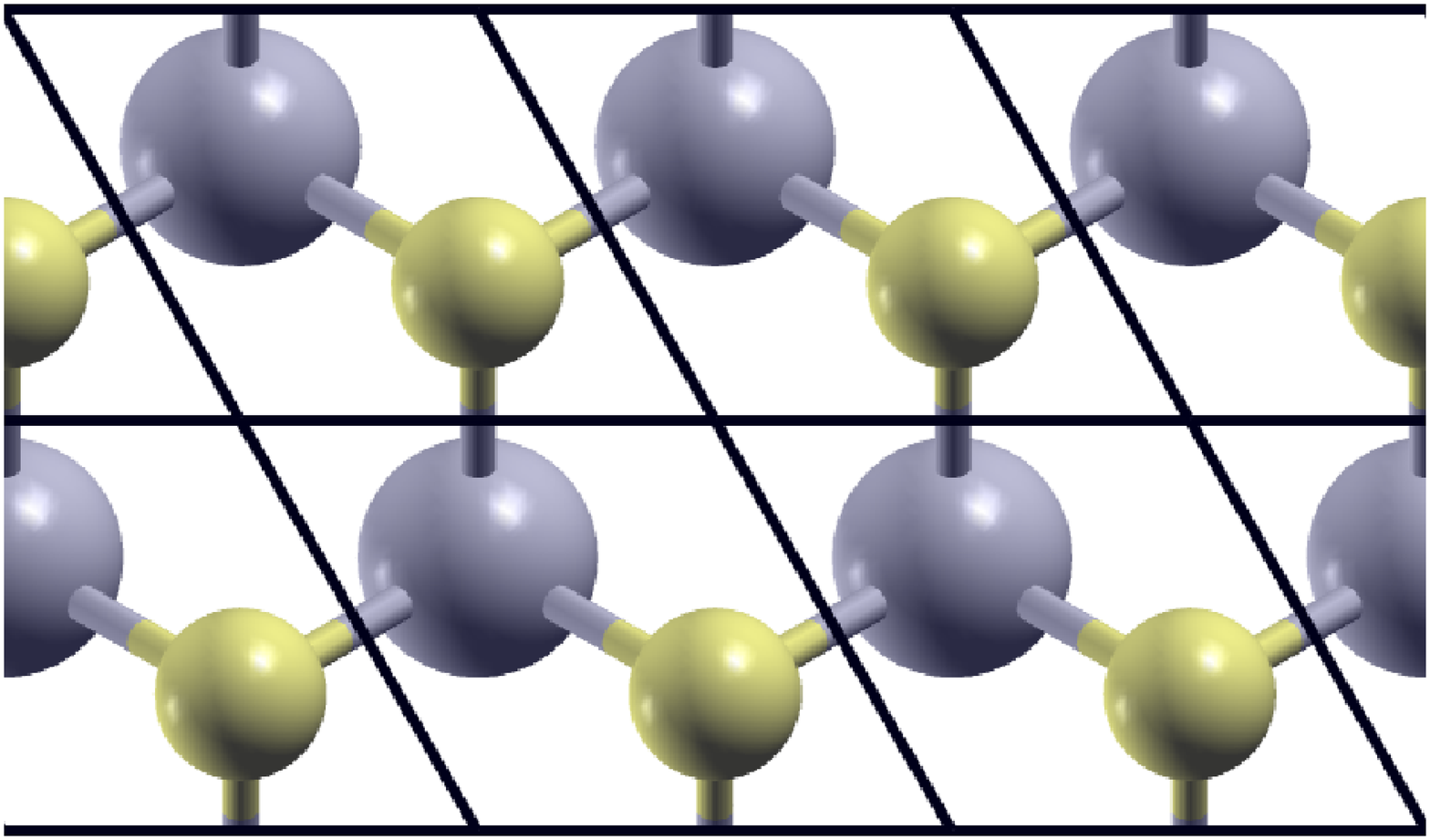}
   \\ 1T-MoS$_2$\\ \includegraphics[width=0.08\textwidth]{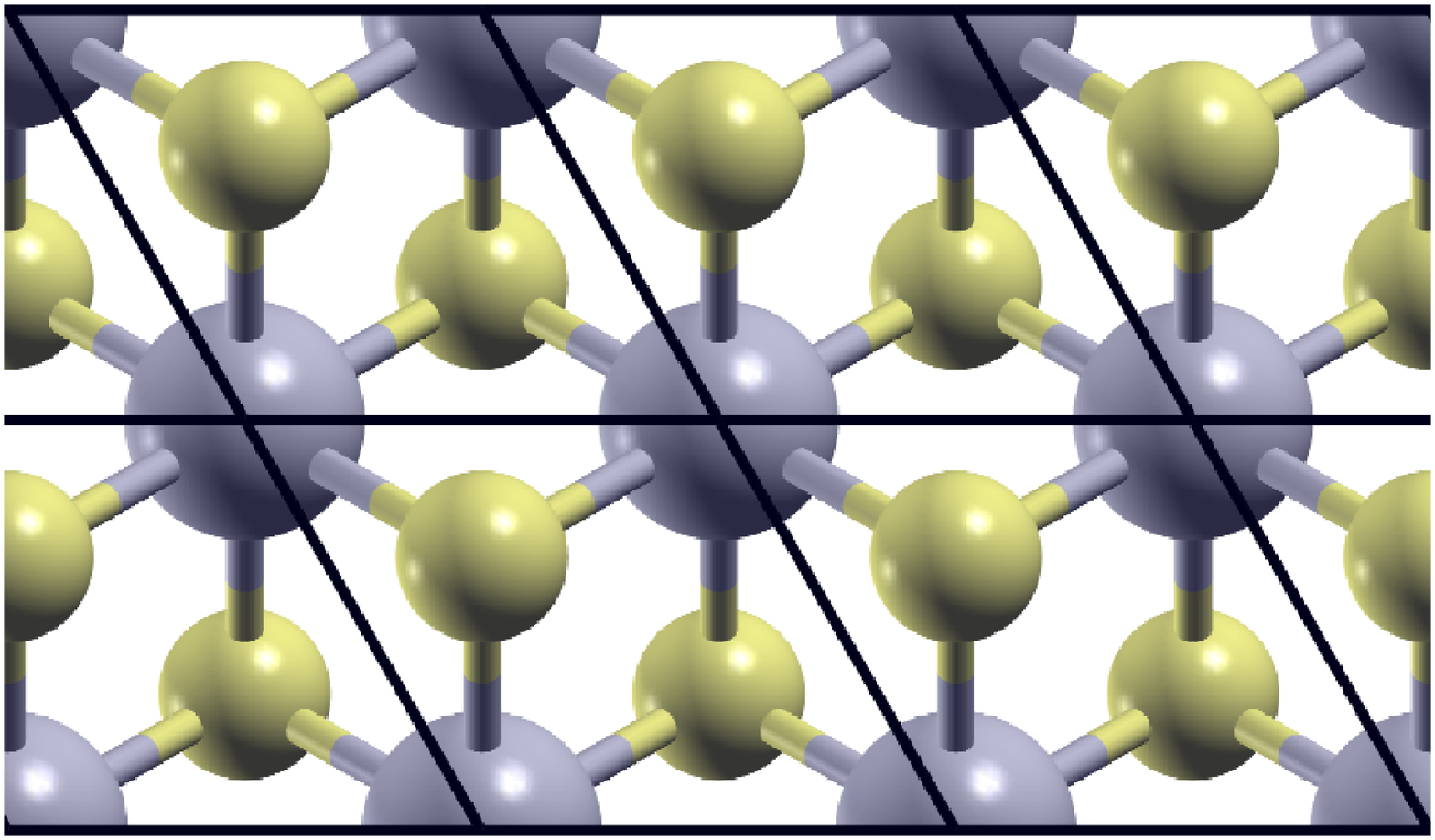}
   \\1T'-MoS$_2$\\ \includegraphics[width=0.08\textwidth]{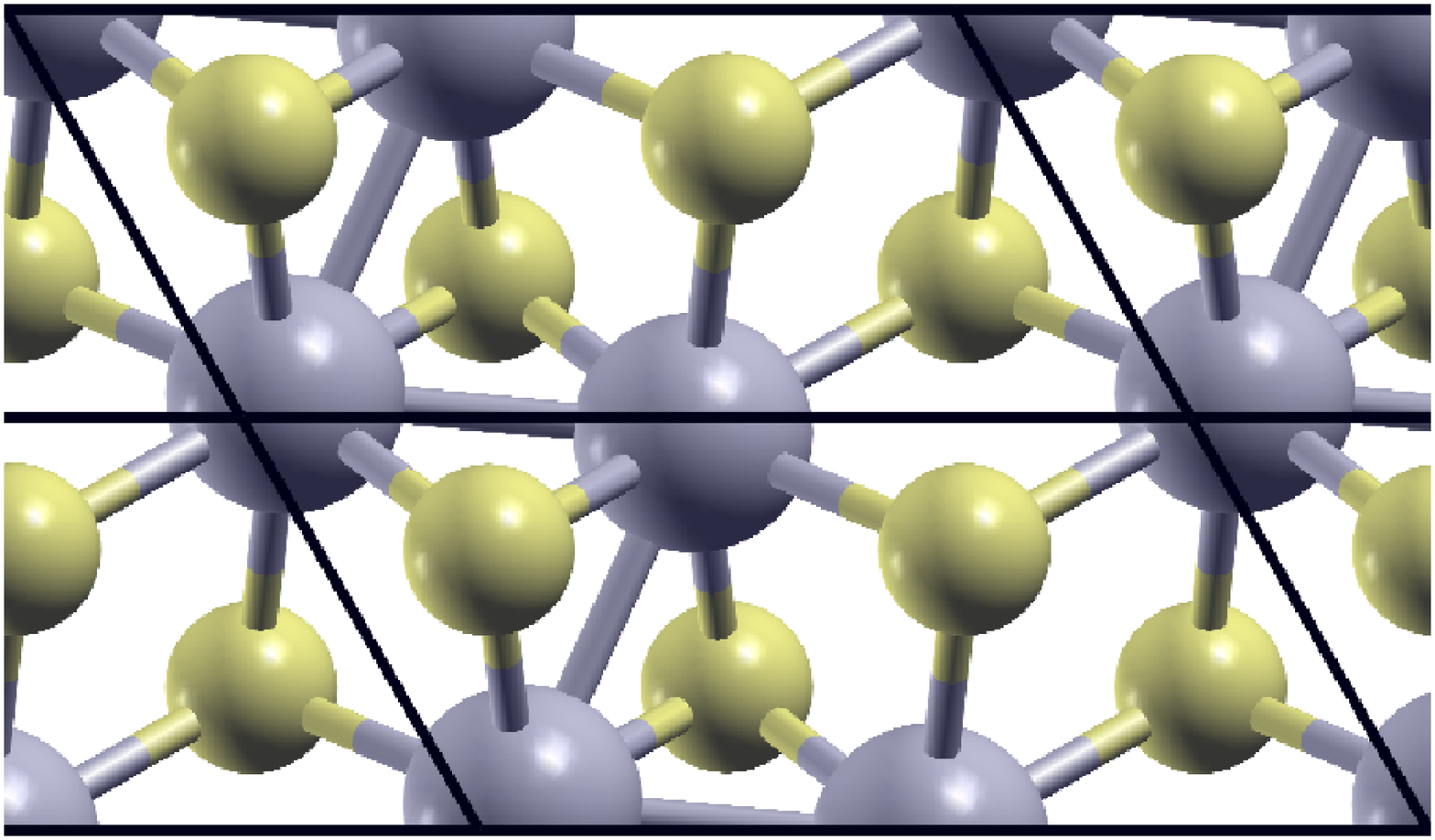} \\ \\
  \end{tabular}
  &
  \hspace{0.3cm}\includegraphics[width=0.37\textwidth]{figure4d.eps}
 \end{tabular}
 \caption{\label{fig:MoS2_phases}(left) Structure of 3 different structural phases of monolayer MoS$_2$. The H polytype is the one found in the bulk
          compound where the coordination of the molybdenum (gray) is trigonal prismatic. The T polytype with octahedral coordination can change to
          the T' polytype in which the molybdenum atoms form zig-zag chains. The sulfur atoms are shown in yellow.
          (right) With increasing electron doping the difference between the total energy of 1T/1T' structure ($E^\mathrm{tot}_{\text{x-MoS}_2}$)
          and the total energy of the 1H polytype ($E^\mathrm{tot}_{\text{1H-MoS}_2}$) decreases. For a doping larger than $\mathrm{n}=-0.44\:e$/unit cell
          ($n\approx-4.96\cdot10^{14}\:\mathrm{cm}^{-2}$) 1T'-MoS$_2$ is the lowest-energy structure.}
\end{figure}
Even if the internal structure changes only slightly under FET doping, electron doping could induce a phase transition where the structure
of the full nanolayer system is altered.
It is well known that lithium\cite{somoano1973,enyashin2012} or potassium\cite{wypych1992} intercalated MoS$_2$ can undergo a phase transition
in which the Mo coordination changes from a direct-band-gap, semiconducting, trigonal-prismatic structure (labeled 2H, ``2'' as there are two layers in the unit cell) to a metallic,
octahedral one (1T).
It has been found experimentally\cite{sandoval1991,eda2012} and shown theoretically \cite{calandra2013,kan2014}, that also monolayer MoS$_2$ can undergo this phase transition.
In the calculations, for a high electron doping by H (Ref.~\citenum{calandra2013}) or Li (Ref.~\citenum{kan2014}) adsorption of $\mathrm{n}\approx-0.35\:e$/unit cell or $\mathrm{n}=-0.44\:e$/unit cell, respectively,
the octahedral phases such as the 1T or 1T' phases become more stable than the 1H phase. In the 1T' phase the molybdenum forms zig-zag chains like tungsten in
WTe$_2$\cite{tang1990,mar1992,crossley1994,augustin2000,watson2002,ali2014}.
Such a transition was also found experimentally by rhenium doping of WS$_2$ nanotubes\cite{enyashin2011} and monolayer MoS$_2$\cite{lin2014}, and by transfer of hot electrons generated in
gold nanoparticles to monolayer MoS$_2$\cite{kang2014}.
In order to determine if the FET setup can lead to such a phase transition for electron doping, we compare in Fig.~\ref{fig:MoS2_phases} the
total energy for the monolayer structures of 1T-MoS$_2$ and 1T'-MoS$_2$ ($E^\mathrm{tot}_{\text{x-MoS}_2}$) with the total energy of the 1H polytype
($E^\mathrm{tot}_{\text{1H-MoS}_2}$). We find that the 1T' structure becomes more stable for electron doping
larger than $\mathrm{n}=-0.44\:e$/unit cell in close agreement with the results of Refs.~\citenum{calandra2013,kan2014}.
Thus, it seems that the interaction between the H/Li atoms and the MoS$_2$ layer has only a minor influence on the phase stability as the transition occurs
in our FET setup at the same doping. We also calculated the energy difference between 1T'-WSe$_2$ and 1H-WSe$_2$, $(E^\mathrm{tot}_{\text{1T'-WSe}_2}-E^\mathrm{tot}_{\text{1H-WSe}_2})$,
for a few electron-doping concentrations and found that for a concentration of $\mathrm{n}=-0.35\:e$/unit cell the 1T' polytype becomes more stable by $63\:\mathrm{meV}$.
In the following, we will thus only consider doping of the H polytype with electron concentrations $\mathrm{n}\geq-0.35\:e$/unit cell as it is the most stable structure
found in nature and is often used to prepare the samples by the mechanical-cleavage method.

\subsection{Band structure in FET setup}
\label{sec:dope}

\begin{figure*}[thp]
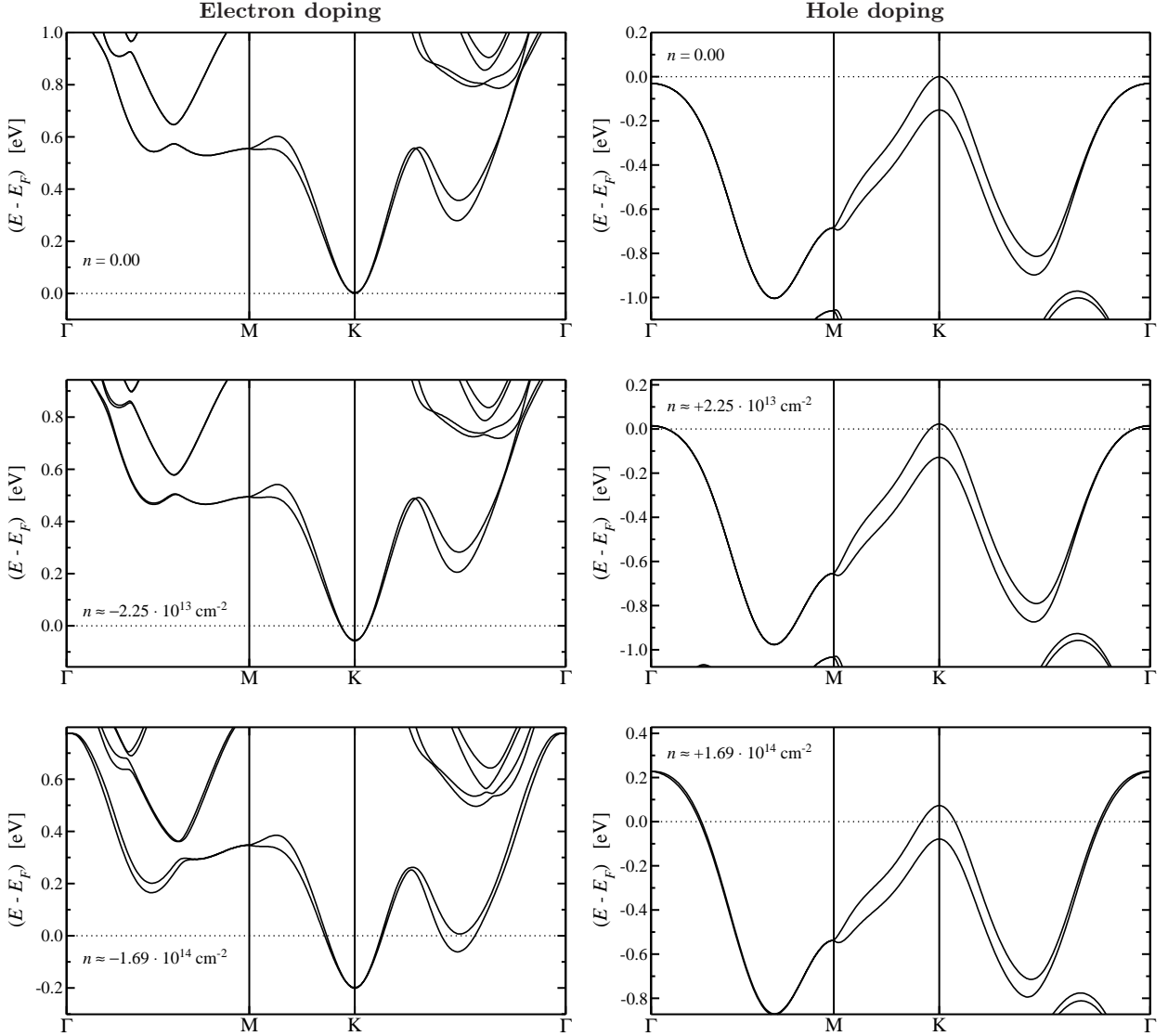

 \begin{tabular}{ccc}
 \multicolumn{3}{c}{{\Large\bf\underline{Monolayer MoS$_2$}}}\\
 {\bf Electron doping}& &{\bf Hole doping}\\
 \includegraphics[scale=0.28,clip=]{figure5a.eps}& &\includegraphics[scale=0.28,clip=]{figure5b.eps}\\
 & &\\
 \includegraphics[scale=0.28,clip=]{figure5c.eps}& &\includegraphics[scale=0.28,clip=]{figure5d.eps}\\
 & &\\
 \includegraphics[scale=0.28,clip=]{figure5e.eps}& &\includegraphics[scale=0.28,clip=]{figure5f.eps}\\
 \end{tabular}
 \caption{\label{fig:MoS2singlebands}Band structure for different FET induced doping of monolayer MoS$_2$.
          The figures in the left column are for electron doping while the right column shows the hole doping case.
          For monolayer MoS$_2$ mainly the valleys at K are filled and only for a high doping of
          $n\approx\pm1.69\cdot10^{14}\:\mathrm{cm}^{-2}$ ($\mathrm{n}=\pm\,0.15\:e$/unit cell) a small amount of charge is in the
          maximum at $\Gamma$ (hole doping) or in the minimum at Q (electron doping).
          The band structures for more doping levels and all the other TMDs can be found in the appendix (Figs.~28--63).}
\end{figure*}
\begin{figure*}[thp]
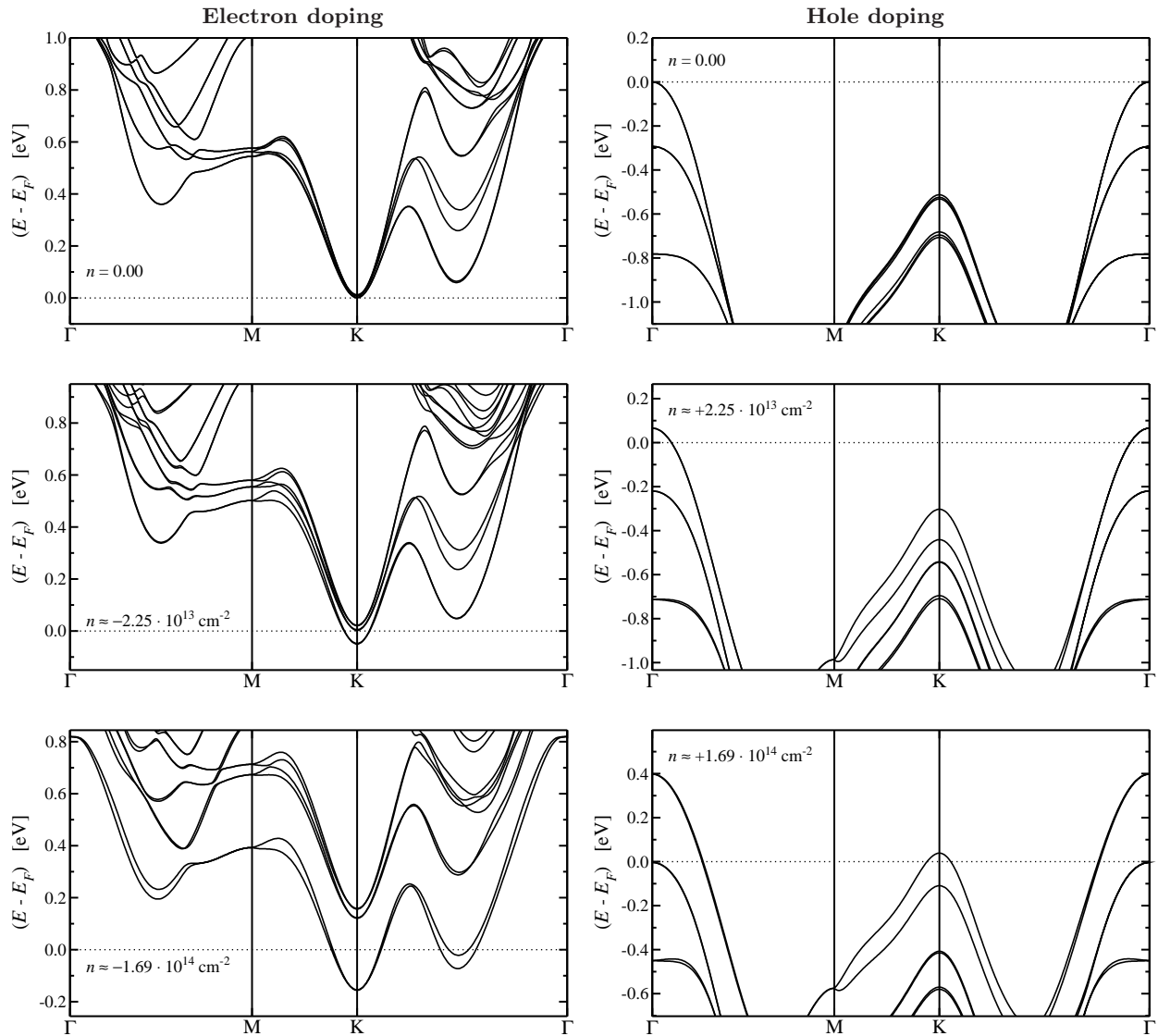

 \begin{tabular}{ccc}
 \multicolumn{3}{c}{{\Large\bf\underline{Trilayer MoS$_2$}}}\\
 {\bf Electron doping}& &{\bf Hole doping}\\
 \includegraphics[scale=0.28,clip=]{figure6a.eps}& &\includegraphics[scale=0.28,clip=]{figure6b.eps}\\
 & &\\
 \includegraphics[scale=0.28,clip=]{figure6c.eps}& &\includegraphics[scale=0.28,clip=]{figure6d.eps}\\
 & &\\
 \includegraphics[scale=0.28,clip=]{figure6e.eps}& &\includegraphics[scale=0.28,clip=]{figure6f.eps}\\
 \end{tabular}
 \caption{\label{fig:MoS2triplebands}Band structure for different FET induced doping of trilayer MoS$_2$.
          The figures in the left column are for electron doping while the right column shows the hole doping case.
          In contrast to the monolayer case, the doping at $\Gamma$/Q (hole/electron doping) is more important in trilayer MoS$_2$.
          The band structures for more doping levels and all the other TMDs can be found in the appendix (Figs.~28--63).}
\end{figure*}
In the following section, we want to investigate the influence of field-effect doping on the electronic properties of the TMDs.
The doping via FET setup changes the band structure considerably as exemplified in Figs.~\ref{fig:MoS2singlebands} and \ref{fig:MoS2triplebands}, which
show the band structures for different electron and hole doping levels for mono- and trilayer MoS$_2$, respectively.
In the appendix we also demonstrate that it is important to correctly model the FET setup by comparing the band structures
of mono- and trilayer MoSe$_2$ calculated with a compensating jellium background to those calculated with our method (Figs.~25, 26).
Furthermore, we also provide the band structures for more doping levels and all the other TMDs in the appendix (in total more than 250 calculations, Figs.~28--63).
We summarized the evolution of the band structure with increasing doping in the left panel of Figs.~\ref{fig:MoS2_bandpos_charge}--\ref{fig:WSe2_bandpos_charge}
which show the position of the different band extrema with respect to the Fermi energy.
Additionally, the right panel in those figures shows the relative amount of doping charge per valley
given by
\begin{align}
 n_\alpha&=\frac{e}{N_\alpha}\sum_{\mathbf{k}\in\Omega_\alpha}
                             \sum_{i}^{\varepsilon_1\leq\varepsilon_{i,\mathbf{k}}\leq\varepsilon_2}\left|\psi_{i,\mathbf{k}}\right|^2.
\end{align}
Here $\alpha=\{\Gamma,\mathrm{K},\mathrm{Q}\}$, $\Omega_\alpha$ defines the subset of $\mathbf{k}$ points which are closer to, \eg, $\alpha=\Gamma$ than to any
$\alpha\neq\Gamma$, $N_\alpha$ is the total number of those $\mathbf{k}$ points, and $\varepsilon_{i,\mathbf{k}}$ is the eigenenergy for band $i$ at $\mathbf{k}$.
The interval $[\varepsilon_1,\varepsilon_2]$ is always chosen such that the probability density is integrated between an energy within the former band gap
$E_m$ and the Fermi energy $E_F$ of the doped system, \ie, $[E_F,E_m]$ and $[E_m,E_F]$
for hole and electron doping, respectively.

\subsubsection*{Electron doping}
\begin{figure*}[thp]
 \includegraphics[height=0.41\textheight,clip=]{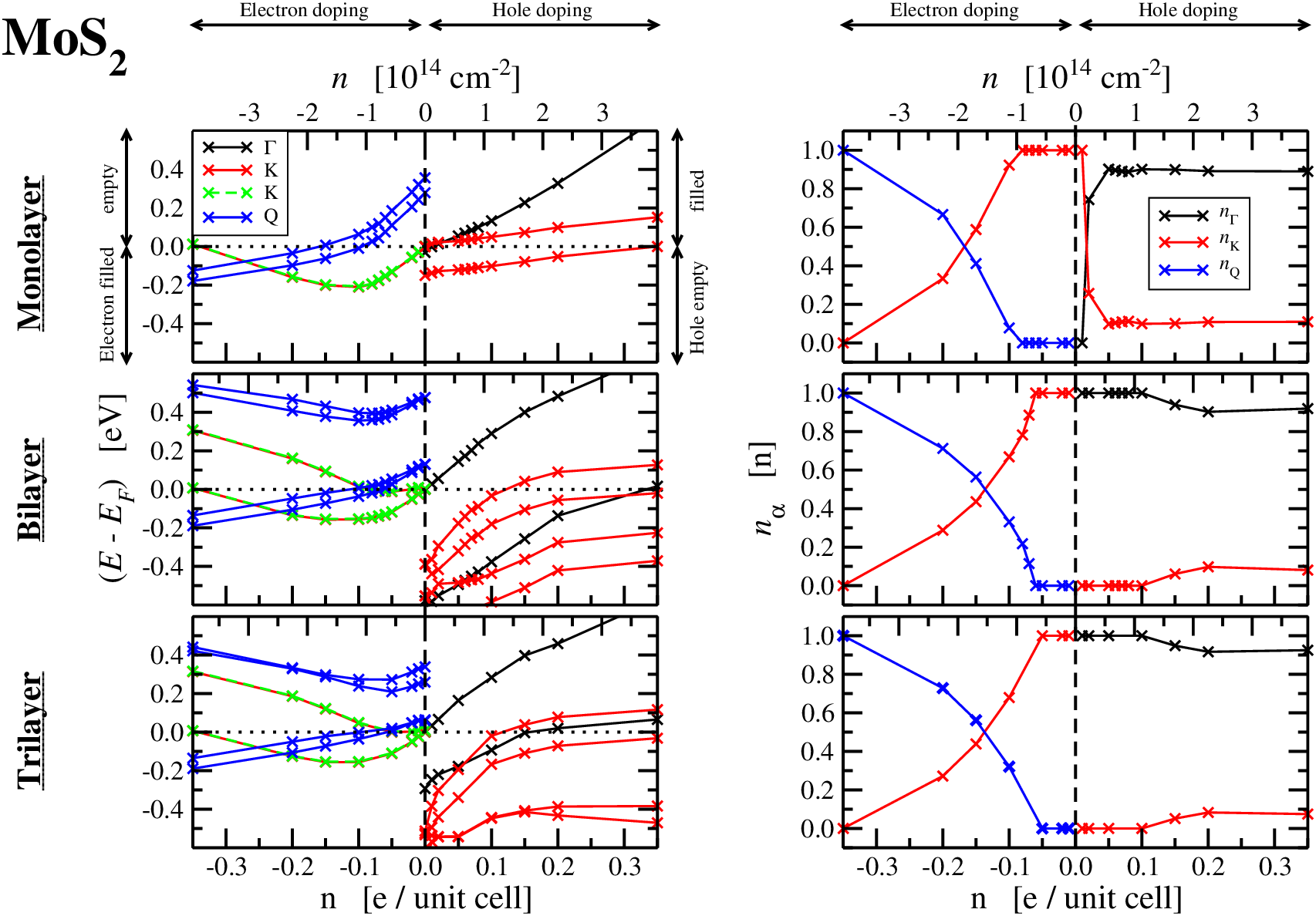}
 \caption{\label{fig:MoS2_bandpos_charge}Position of the different band minima/maxima $E_i$ (left panel) with respect to the Fermi level and 
          relative amount of doping charge per valley $n_\alpha$ (right panel) as a function of doping for
          mono-, bi-, and trilayer MoS$_2$. ``Q'' labels the conduction-band minimum half-way between K and $\Gamma$.
          In each graph, the scale and the units for the lower $x$ axis are given in the lowest
          graph, while those of the upper $x$ axis are given in the uppermost graph. Two different line styles
          for the two spin-orbit-split conduction-band minima at K were used in order to enhance the readability. Lines are guides for the eye.}
\end{figure*}

For n-type doping of monolayer MoS$_2$ (as for all monolayer TMDs) the doping charge first occupies the extrema at K. For small doping (as long as only
one valley is doped) the bands are rigidly shifted.
However, as soon as a second valley is close to the Fermi energy, the doping cannot
be described by a rigid shift of the bands anymore. For electron doping, the down shift of the bands at K slows down and, as the valley
at Q starts to get occupied, is eventually reversed into an up shift. Finally, for high electron doping, the K valley is unoccupied
again and the doping charge is solely localized around Q. Comparing our results of the changes in the conduction band for
electron doping of monolayer MoS$_2$ (Figs.~\ref{fig:MoS2singlebands}, \ref{fig:MoS2_bandpos_charge}) with literature shows that it is important
to correctly model the system -- while the authors of Ref.~\citenum{ge2014} find an up shift of the Q valley with increasing electron
doping, we see a down shift. The opposite shift in Ref.~\citenum{ge2014} might be due to the free-electron states at $\Gamma$ (\ie, the
states in the vacuum between the repeated images, \cf, also Ref.~\citenum{topsakal2012}) which approach the Fermi energy with increasing doping.
Also the authors of Refs.~\citenum{topsakal2012,ge2013,roesner2014} find a down shift of the Q valley further supporting our results
even if in those works the asymmetric electric field in an FET has not been taken into account.
The amount of doped electrons needed to have the charge completely localized at Q 
depends on the TMD (\ie, the initial energy difference between the minimum at K and Q) and is larger than $n=-2.2\cdot10^{14}\:\mathrm{cm}^{-2}$
($\mathrm{n}\approx-0.2\:e$/unit cell): the transition occurs for MoS$_2$ at $n\approx-3.83\cdot10^{14}\:\mathrm{cm}^{-2}$, for MoSe$_2$ at $n\approx-2.5\cdot10^{14}\:\mathrm{cm}^{-2}$,
for MoTe$_2$ at $n\approx-2.22\cdot10^{14}\:\mathrm{cm}^{-2}$, for WS$_2$ at $n\approx-3.41\cdot10^{14}\:\mathrm{cm}^{-2}$, and for WSe$_2$ at $n\approx-3.31\cdot10^{14}\:\mathrm{cm}^{-2}$.
Please note, that using LDA could slightly change these results as it leads to a smaller unit cell.
The compressive in-plain strain would then reduce the difference between the conduction-band minima at K and Q\cite{peelaers2012,yun2012,guzman2014}
and thus decrease the doping-charge concentration needed to solely dope the valley at Q (\cf, Fig.~27).

\begin{figure*}[p]
 \includegraphics[height=0.41\textheight,clip=]{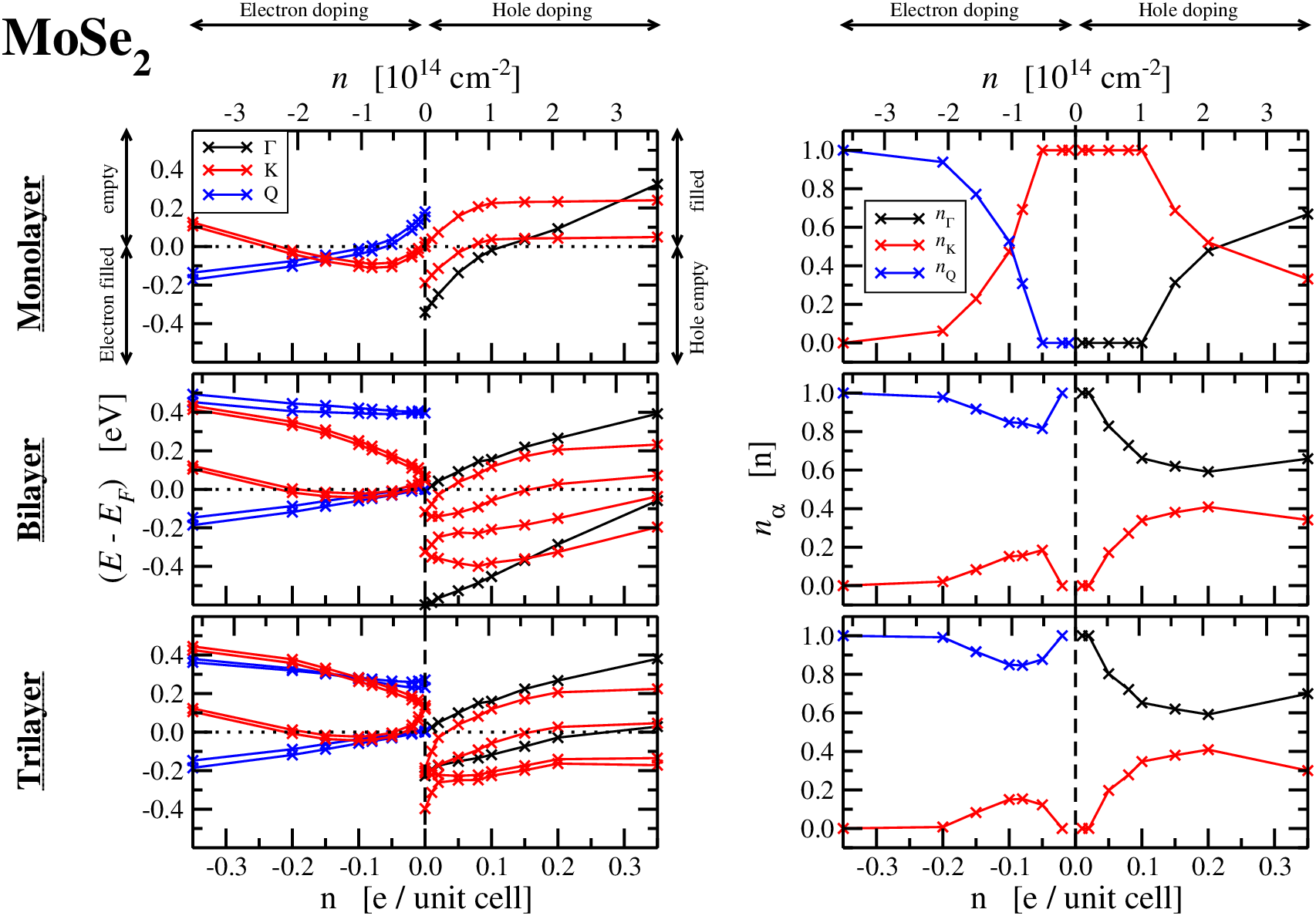}
 \caption{\label{fig:MoSe2_bandpos_charge}Position of the different band minima/maxima $E_i$ (left panel) with respect to the Fermi level and 
          relative amount of doping charge per valley $n_\alpha$ (right panel) as a function of doping for
          mono-, bi-, and trilayer MoSe$_2$. ``Q'' labels the conduction-band minimum half-way between K and $\Gamma$.
          In each graph, the scale and the units for the lower $x$ axis are given in the lowest
          graph, while those of the upper $x$ axis are given in the uppermost graph. Lines are guides for the eye.}
\end{figure*}

\begin{figure*}[p]
 \includegraphics[height=0.41\textheight,clip=]{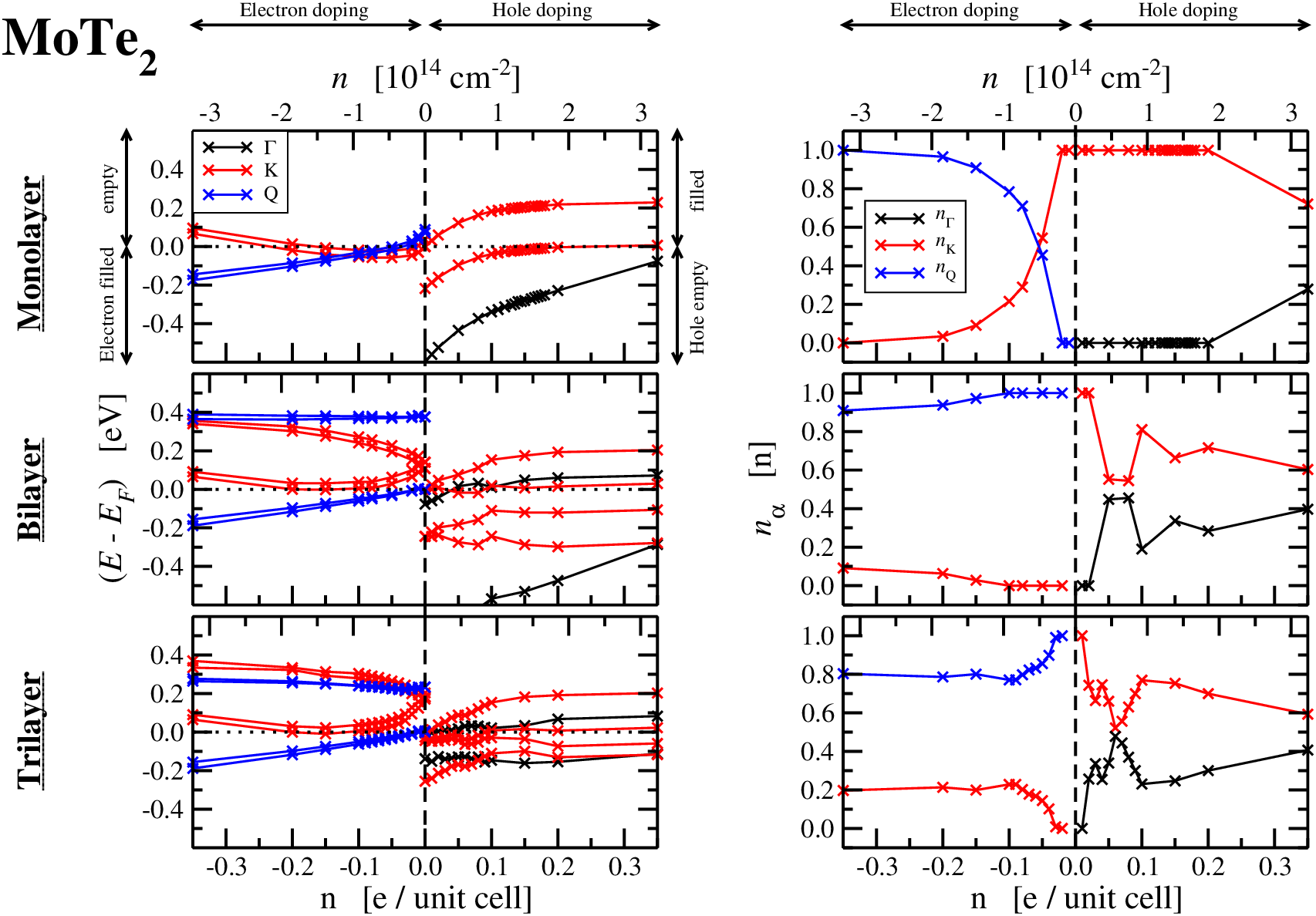}
 \caption{\label{fig:MoTe2_bandpos_charge}Position of the different band minima/maxima $E_i$ (left panel) with respect to the Fermi level and 
          relative amount of doping charge per valley $n_\alpha$ (right panel) as a function of doping for
          mono-, bi-, and trilayer MoTe$_2$. ``Q'' labels the conduction-band minimum half-way between K and $\Gamma$.
          In each graph, the scale and the units for the lower $x$ axis are given in the lowest
          graph, while those of the upper $x$ axis are given in the uppermost graph. Lines are guides for the eye.}
\end{figure*}

\begin{figure*}[p]
 \includegraphics[height=0.41\textheight,clip=]{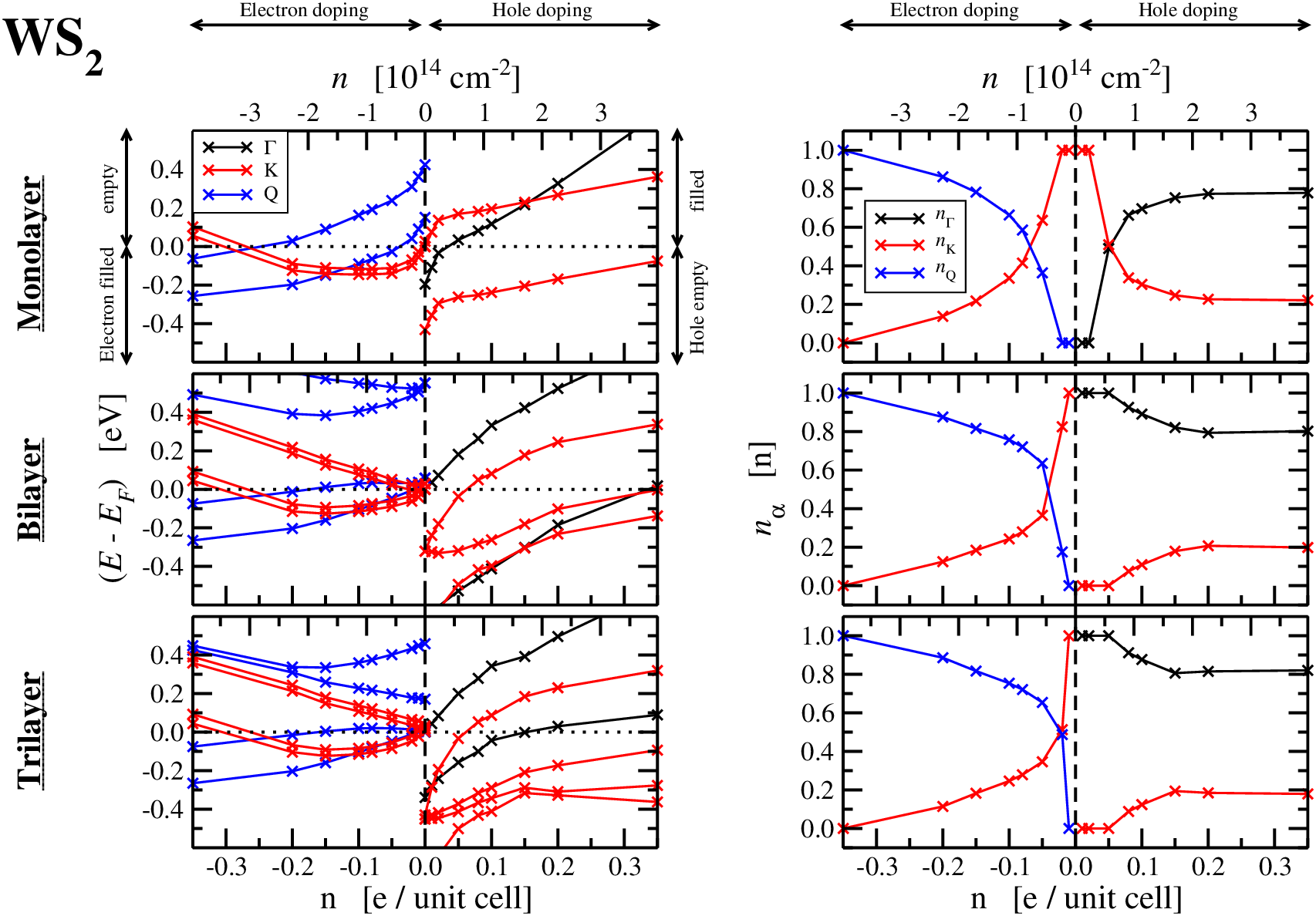}
 \caption{\label{fig:WS2_bandpos_charge}Position of the different band minima/maxima $E_i$ (left panel) with respect to the Fermi level and 
          relative amount of doping charge per valley $n_\alpha$ (right panel) as a function of doping for
          mono-, bi-, and trilayer WS$_2$. ``Q'' labels the conduction-band minimum half-way between K and $\Gamma$.
          In each graph, the scale and the units for the lower $x$ axis are given in the lowest
          graph, while those of the upper $x$ axis are given in the uppermost graph. Lines are guides for the eye.}
\end{figure*}

\begin{figure*}[htp]
 \includegraphics[height=0.41\textheight,clip=]{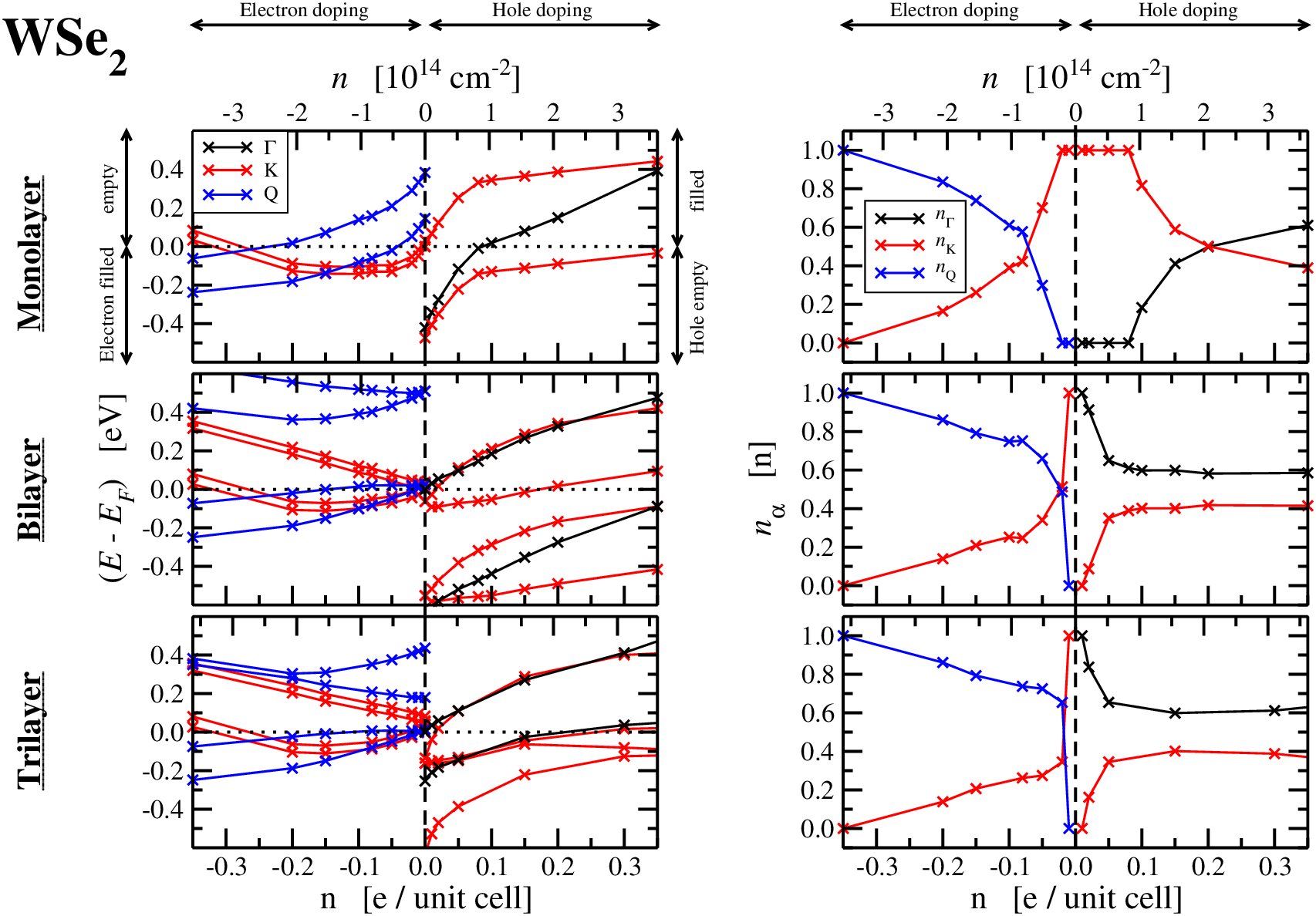}
 \caption{\label{fig:WSe2_bandpos_charge}Position of the different band minima/maxima $E_i$ (left panel) with respect to the Fermi level and 
          relative amount of doping charge per valley $n_\alpha$ (right panel) as a function of doping for
          mono-, bi-, and trilayer WSe$_2$. ``Q'' labels the conduction-band minimum half-way between K and $\Gamma$.
          In each graph, the scale and the units for the lower $x$ axis are given in the lowest
          graph, while those of the upper $x$ axis are given in the uppermost graph. Lines are guides for the eye.}
\end{figure*}

In multilayer MoS$_2$, WS$_2$, and WSe$_2$ first the valley at K is doped and both valleys at K and Q are occupied until $n\approx-3.3\cdot10^{14}\:\mathrm{cm}^{-2}$ while
in bi- and trilayer MoSe$_2$ and MoTe$_2$ the order is reversed: electrons first occupy the valley at Q and the doping at K is always smaller.
This is due to the minimum at Q being lower in energy than the
one at K in the undoped system (see Figs.~\ref{fig:MoSe2_bandpos_charge}, \ref{fig:MoTe2_bandpos_charge}).
For a doping of $n\lesssim-2.1\cdot10^{14}\:\mathrm{cm}^{-2}$ for MoSe$_2$ the K valley is even unoccupied.
Yet, one can expect that for small electron doping ($|n|<10^{13}\:\mathrm{cm}^{-2}$) of thick samples (more than 3 layers) of MoS$_2$, WS$_2$, and WSe$_2$ the electrons will
also first occupy the Q valley as this valley is lowered in energy with increasing number of layers.

\subsubsection*{Hole doping}
For p-type doping of the monolayer TMDs, the doping charge first occupies the extrema at K.
However, in contrast to the electron-doping case, in the high-hole-doping limit ($\mathrm{n}>+0.2\:e$/unit cell, $n\gtrsim+2.1\cdot10^{14}\:\mathrm{cm}^{-2}$)
of the monolayer TMDs both valleys at $\Gamma$ and K are occupied -- the relative amount of doping charge in the $\Gamma$ valley is even higher than
that in K. The transition when the $\Gamma$ valley is more occupied than the K valley again depends on the TMD, \ie, on the initial
energy difference between the valence-band maxima. In the case of MoS$_2$ for example even a small doping is enough to dope more the
$\Gamma$ valley. The only exception from this picture is MoTe$_2$ for which K is always more occupied even for $\mathrm{n}=+0.35\:e$/unit cell
($n\approx+3.23\cdot10^{14}\:\mathrm{cm}^{-2}$, \cf, Fig.~\ref{fig:MoTe2_bandpos_charge}).

\begin{figure}
 \centering
 \includegraphics[width=0.46\textwidth,clip=]{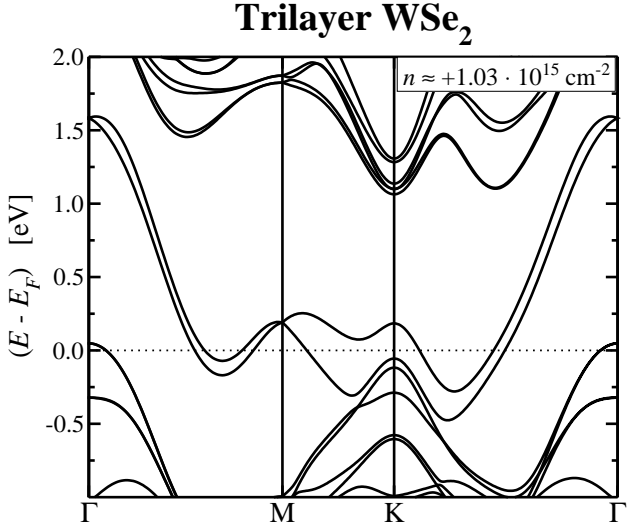}
 \caption{\label{fig:WSe2_bands_high_hole}Band structure of trilayer WSe$_2$ for a high hole doping of $n\approx+1.03\cdot10^{15}\:\mathrm{cm}^{-2}$
          ($\mathrm{n}=+1\:e$/unit cell).}
\end{figure}

Hole doping of the bilayer and trilayer TMDs is again very similar: in all investigated compounds (except MoTe$_2$)
first the conduction-band maximum at $\Gamma$ is occupied, while for higher doping also K starts to get filled. Most interestingly,
in some cases the second band at K is never occupied. For trilayer WS$_2$ and WSe$_2$ it is even pushed down in energy, effectively
increasing the splitting of the spin-orbit-split bands.
For WSe$_2$ it is possible to achieve even higher hole-doping concentrations than shown in Fig.~\ref{fig:WSe2_bandpos_charge}
by using an IL-FET\cite{yuan2013}. We thus calculated for trilayer WSe$_2$ also some higher doping cases
(up to $\mathrm{n}=+1\:e$/unit cell, or $n\approx+1\cdot10^{15}\:\mathrm{cm}^{-2}$) in order to understand
what might happen for such a high doping.
Our calculated band structure is very different from the one in Ref.~\citenum{yuan2013} which was calculated without proper
treatment of the FET setup.
For a doping of one hole per unit cell the second band at K is pushed down below the Fermi energy and also the first band at K is lowered. 
The band structure in Fig.~\ref{fig:WSe2_bands_high_hole} shows that the former band gap is closed and the first two bands are nearly unoccupied. 
However, since these are bands localized on the first layer close to the
gate, we believe that for such a high doping the ions of the ionic liquid might start to interact with WSe$_2$.
In our simplified model without inclusion of the full dielectric it is difficult to prove this statement
and we will thus leave this interesting problem for future investigations and will concentrate here on lower doping values.

\subsubsection*{Conductive channel}
Another important property in order to understand different experiments on FET doping on TMDs is the doping-charge layer thickness, \ie,
the size and shape of the conductive channel created and influenced by the gate voltage. 
To visualize the conductive channel we calculated the planar-averaged doping-charge distribution along $z$,
\begin{align}
 \rho^\mathrm{dop}_{||}\left(z\right)&=\frac{e}{\Omega_{2D}\,N_k}\int dA\sum_{i,\mathbf{k}}^{\varepsilon_1\leq\varepsilon_{i,\mathbf{k}}\leq\varepsilon_2}\left|\psi_{i,\mathbf{k}}\left(\mathbf{r}\right)\right|^2
\end{align}
Here $\Omega_{2D}$ is the unit cell area and $N_k$ is the total number of $\mathbf{k}$ points.
The interval $[\varepsilon_1,\varepsilon_2]$ is defined as above, \ie, $[E_F,E_m]$ and $[E_m,E_F]$
for hole and electron doping, respectively.
This property not only reveals the thickness of the conductive channel but also the relative distribution among the different layers.

\begin{figure}
 \includegraphics[width=0.46\textwidth,clip=]{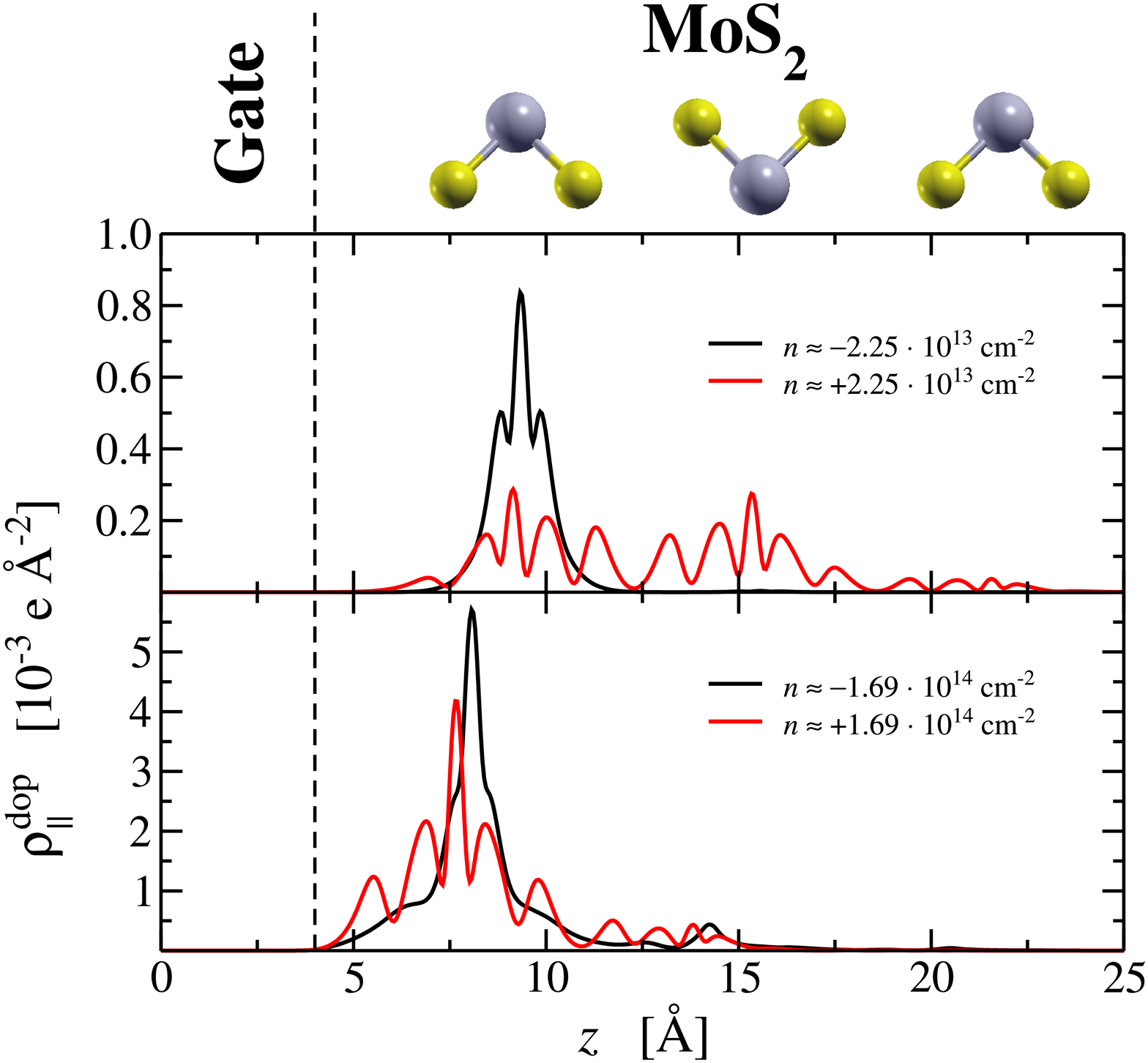}
 \caption{\label{fig:MoS2_idlos}Planar-averaged doping-charge distribution along $z$ for trilayer MoS$_2$ for a doping of $n\approx\pm2.25\cdot10^{13}\:\mathrm{cm}^{-2}$
          ($\mathrm{n}=\pm\,0.02\:e$/unit cell, upper panel) and $n\approx\pm1.69\cdot10^{14}\:\mathrm{cm}^{-2}$
          ($\mathrm{n}=\pm\,0.15\:e$/unit cell, lower panel). The dashed line within the graph indicates the end of the barrier potential, while
          the sketch above shows the position of the atoms (gray -- Mo, yellow -- S).}
\end{figure}
In our calculations for low hole doping the charge is nearly evenly distributed between the first two layers with
only small contributions at the third layer as can be seen in the upper panel of Fig.~\ref{fig:MoS2_idlos}.
Increasing the doping (\ie, increasing the gate voltage, lower panel of Fig.~\ref{fig:MoS2_idlos}) the charge is more and more localized on the layer
closest to the gate with a negligible amount of holes on the third layer.
In the case of electron doping of multilayer MoS$_2$, the charge is localized on the first layer as only the K valley is doped (small contribution
of the out-of-plane states of sulfur). Only for higher doping when the minimum at Q is occupied a small amount of charge can be found
on the second layer (\cf, lower panel of Fig.~\ref{fig:MoS2_idlos} and right panel of Fig.~\ref{fig:MoS2_bandpos_charge}).
One can also see that the asymmetry in the doping-charge distribution is more pronounced for hole doping and that the system in this case moves closer to the dielectric.
Furthermore, for low hole doping the doping charge on the second layer is even slightly larger than the one on the first.

\begin{figure}
 \includegraphics[scale=0.39,clip=]{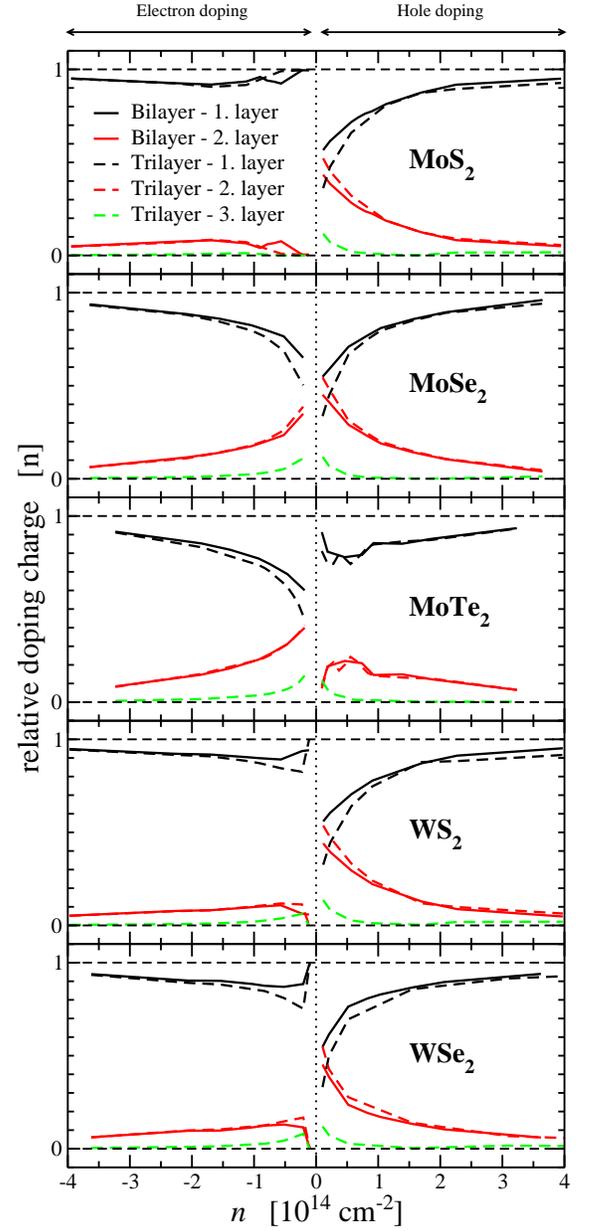}
 \caption{\label{fig:charge}Relative doping charge per layer for bi- and trilayer of the different dichalcogenides with increasing FET doping.}
\end{figure}

Figure \ref{fig:charge} summarizes the doping-charge distribution for all bi- and trilayer TMDs.
In the hole-doping case all TMDs behave similarly except MoTe$_2$: for low doping the holes are delocalized over the
first two layers with only small contributions in the third layer and, thus, the bi- and trilayer systems are nearly the same. Increasing the doping leads
to stronger localization of the charge within the first layer and an effective narrowing of the conductive channel. One can also easily understand why
MoTe$_2$ behaves differently: in all multilayer TMDs first the valley at $\Gamma$ is doped, since it is the valence-band maximum, while in MoTe$_2$ the K valley is the
maximum. As we have seen in section \ref{sec:uncharged} the states close to $\Gamma$ have large out-of-plane contributions of both the transition metal and
the chalcogen atom, while the valley at K is composed only of in-plane states. Increasing the hole doping however also leads for MoTe$_2$ to a small doping
of the $\Gamma$ valley. Thus, the amount of charge within the second layer increases slightly in the beginning. The small kink for trilayer MoTe$_2$
close to $n=+0.46\cdot10^{14}\:\mathrm{cm}^{-2}$ is due to both bands at $\Gamma$ and K being close to the Fermi energy. In this low doping limit (per valley)
the calculation would require an infinite number of $\mathbf{k}$
points to fully converge the results. Further increase of the doping leads as for all TMDs to a larger
screening of the electric field and therefore to the stronger localization within the first layer as can also be seen in the lower panel Fig.~\ref{fig:MoS2_idlos}.

For electron doping we can divide the different TMDs into two different classes: (i) those in which the conductive channel for low doping
($n\approx-10^{13}\:\mathrm{cm}^{-2}$) has a thickness
of just one layer (MoS$_2$, WS$_2$, WSe$_2$) and (ii) those with a three-layer-thick channel (MoSe$_2$, MoTe$_2$).
Using the results of Figs.~\ref{fig:MoS2_bandpos_charge}--\ref{fig:WSe2_bandpos_charge}
one can see that in the TMDs of class (i) initially the K valley is doped while in (ii) the Q valley is occupied.
Since the chalcogen states close to the conduction-band minimum at K have mainly in-plane character (in contrast to the transition-metal states which
have $d_{z^2}$ character), the hybridization between the layers is small
and the electrons are more localized within the first layer. The chalcogen states close to Q on the other hand have a large out-of-plane
contribution which leads to a stronger hybridization between the layers. With increasing doping the electric field of the gate is more and more
screened and the size of the conductive channel reduces to one layer. Furthermore, one can also understand why the tungsten dichalcogenides have a steep
increase of the channel thickness in the beginning while this is not the case for MoS$_2$: first, the difference between the conduction-band minimum
at K and Q is much smaller in multilayer WS$_2$/WSe$_2$ than in MoS$_2$ and second, a small electron doping can also results in an effective separation
of the single (doped) layer from the multilayer system. The difference between the conduction-band minimum at K and Q for monolayer MoS$_2$ is however
twice as large as in the tungsten systems ($\approx300\:\mathrm{meV}$ for MoS$_2$ and $\approx150\:\mathrm{meV}$ for WS$_2$/WSe$_2$). Thus, the valley
at the Q point is doped much earlier in WS$_2$/WSe$_2$ than in MoS$_2$.

\subsubsection*{Number of occupied bands}
\begin{figure}
 \includegraphics[scale=0.39,clip=]{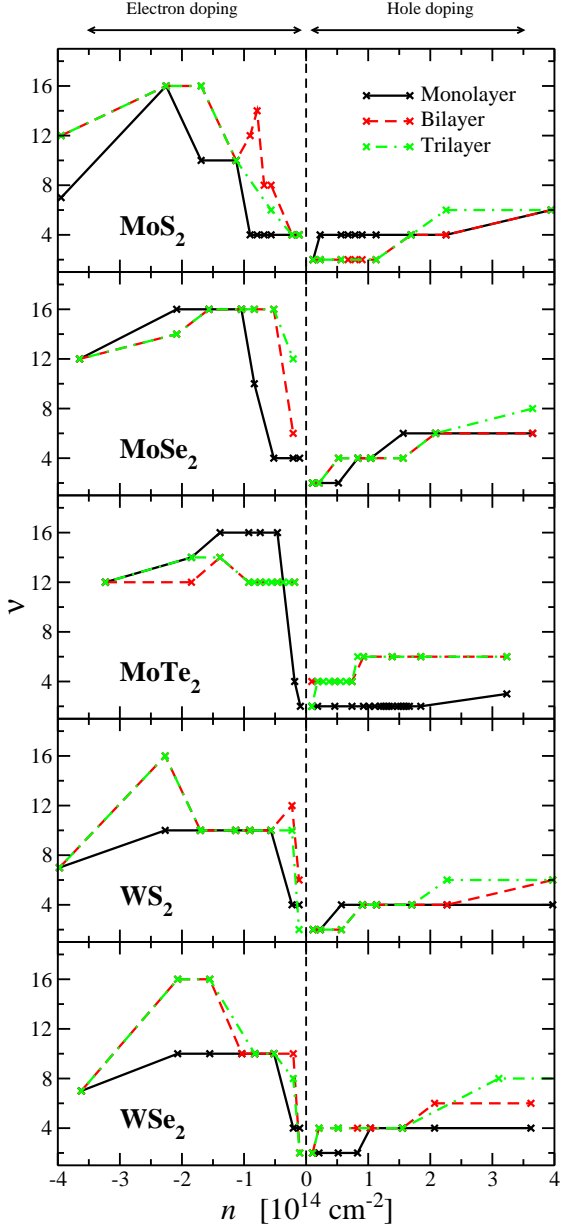}
 \caption{\label{fig:numbands}Total spin-valley degeneracy $\nu$ as function of doping for the mono-,
          bi-, and trilayer system of all investigated TMDs. The total spin-valley degeneracy has been
          calculated by summing the spin and valley degeneracies of all doped valleys.
          Lines are guides for the eye.}
\end{figure}
It is important to note, that the thickness of the doping-charge distribution, the number of occupied bands at a given $\mathbf{k}$ point,
and the number of TMD layers (\ie, the system size) are uncorrelated.
Indeed, as can be seen in Fig.~\ref{fig:numbands}, the total
spin-valley degeneracy $\nu$ can be quite similar for different number of layers,
whereas the doping charge is localized on one or two layers as seen in the previous section.
Here the total spin-valley degeneracy has been calculated by counting the number of valleys within the interval $[\varepsilon_1,\varepsilon_2]$
as defined above:
\begin{align}
 \nu&=\sum_\alpha\nu^\alpha=\sum g_s^\alpha\,g_v^\alpha.
\end{align}
Here $g_s^\alpha$ and $g_v^\alpha$ are the spin and valley degeneracies of the valley at $\mathbf{k}$ point $\alpha=\{\Gamma,\mathrm{K},\mathrm{Q}\}$.
For electron doping $\nu$ is much higher than in the hole doping case as the valley degeneracy for the conduction-band
minimum close to Q is $g_v=6$ (spin degeneracy $g_s=1$). Thus, as soon as the valley at Q is doped, $\nu$ increases drastically
by 6 or 12 depending on whether only one or both spin-orbit-split bands are filled. For some TMDs a high electron doping of $n<-3.5\cdot10^{14}\:\mathrm{cm}^{-2}$
leads to a large lowering of the minimum at Q -- so much that it is actually not a valley anymore but a ring around $\Gamma$. This
can be seen by the minimum between $\Gamma$ and M which appears in the band structure (\cf, Figs.~\ref{fig:MoS2singlebands}, \ref{fig:MoS2triplebands} and
band structures in the appendix). In this case, we do not count it as six independent valleys but as one.

In contrast, for high hole doping ($n\gtrsim+2\cdot10^{14}\:\mathrm{cm}^{-2}$) of multilayer TMDs the number of occupied
valleys is either $\nu=6$ or $\nu=8$ as only bands at K ($g_v=2$, $g_s=1$) and $\Gamma$ ($g_v=1$, $g_s=2$) are doped.
In the monolayer systems often two valleys are doped -- the valence-band maximum at K and either
the second spin-orbit-split band at K or the valence-band maximum at $\Gamma$.
Accordingly, the difference between the valence-band maximum at K and at $\Gamma$ determines
the doping-charge concentration needed in order to dope two valleys at different points in the BZ.
The spin-degenerate maximum at $\Gamma$ is occupied for monolayer
MoS$_2$, MoSe$_2$, WS$_2$, and WSe$_2$ for hole doping larger
than $+1\cdot10^{13}\:\mathrm{cm}^{-2}$, $+5\cdot10^{13}\:\mathrm{cm}^{-2}$, $+2\cdot10^{13}\:\mathrm{cm}^{-2}$, and $+8\cdot10^{13}\:\mathrm{cm}^{-2}$, respectively.
Again monolayer MoTe$_2$ is exceptional due to the large difference of $\approx600\:\mathrm{meV}$ between the maximum at K
and $\Gamma$. Just for a very high hole doping of $n\approx+3.23\cdot10^{14}\:\mathrm{cm}^{-2}$ ($\mathrm{n}=+0.35\:e$/unit cell)
the second band at K is occupied and a ``ring'' around and close to $\Gamma$ appears (which is again counted as 1).

Up to now we have focused on the changes in the electronic structure of the different TMDs with increasing doping and we saw that for, \eg, high
electron doping the charge is mainly localized around the Q point. We now want to investigate how the amount of doping charge and thus the
number of occupied bands and the thickness of the conductive channel is determined experimentally.

\subsection{Hall-effect measurements}
\label{sec:hall}

In order to determine the doping charge in a sample, one commonly performs a Hall experiment as the inverse Hall coefficient is
directly proportional to the charge-carrier density $n$ in the case of parabolic, isotropic bands.
This, however, is already a crude approximation which, as we will see below, can lead to large differences between
the charge-carrier density thus calculated and the real density within the sample.

We closely follow the work of Madsen and Singh\cite{madsen2006} and sketch the calculation of the Hall coefficient (or more specifically, the Hall tensor $R_{ijk}$)
within Boltzmann transport theory. In the presence of an electric field $\mathbf{E}$ and a magnetic field $\mathbf{B}$, the electric current $\mathbf{j}$ can be written as
\begin{align}
\label{eq:boltzmann}
 j_\alpha &= \sigma_{\alpha\beta}E_\beta + \sigma_{\alpha\beta\gamma}E_\beta\:B_\gamma+\cdots,
\end{align}
with the conductivity tensors $\sigma_{\alpha\beta}$ and $\sigma_{\alpha\beta\gamma}$, and the indices denoting
the spatial dimensions. Here and henceforth, we will always adopt Einstein's sum convention, according to which
whenever an index occurs twice in a single-term expression, the summation is carried out over all possible values
of this index.
The Hall tensor is defined as
\begin{align}
\label{eq:hallcoefficients}
 R_{ijk}\left(T;E_F\right) &= \frac{\mathbf{E}^{ind}_j}{\mathbf{j}^{appl}_i\:\mathbf{B}^{appl}_k} = \left(\sigma^{-1}\right)_{\alpha j}\,\sigma_{\alpha\beta k}\,\left(\sigma^{-1}\right)_{i\beta},
\end{align}
where $\mathbf{E}^{ind}_j$ is the electric field along direction $j$ which is induced by the applied magnetic field $\mathbf{B}^{appl}_k$ and
current $\mathbf{j}^{appl}_i$ along direction $k$ and $i$, respectively.

Within the relaxation-time approximation the conductivity tensors $\sigma_{\alpha\beta}$ and $\sigma_{\alpha\beta\gamma}$
for temperature $T$ and chemical potential $E_F$ are given in the 2D case by\cite{madsen2006}
\begin{align}
\label{eq:sigmaab_Tmu}
 \sigma_{\alpha\beta}\left(T;E_F\right) &=
    \frac{q^2}{\left(2\pi\right)^2}\int\:\tau_{i,\mathbf{k}}\:v_\alpha^{i,\mathbf{k}}\:v_\beta^{i,\mathbf{k}}\notag\\
    &\quad\times\left[-\frac{\partial f_{E_F}\left(T;\varepsilon_{i,\mathbf{k}}\right)}{\partial\varepsilon}\right]\:d^2\mathbf{k},\\
\label{eq:sigmaabg_Tmu}
 \sigma_{\alpha\beta\gamma}\left(T;E_F\right) &=
    \frac{q^3}{\left(2\pi\right)^2}\int\:\tau^2_{i,\mathbf{k}}\:\epsilon_{\gamma u \nu}\:
    v_\alpha^{i,\mathbf{k}}\:v_\nu^{i,\mathbf{k}}\:\left(M_{\beta u}^{i,\mathbf{k}}\right)^{-1}\notag\\
    &\quad\times\left[-\frac{\partial f_{E_F}\left(T;\varepsilon_{i,\mathbf{k}}\right)}{\partial\varepsilon}\right]\:d^2\mathbf{k}.
\end{align}
Here, $q=\pm e$ is the charge of the charge carriers in band $\varepsilon_{i,\mathbf{k}}$ with momentum $\mathbf{k}$,
$f_{E_F}(T;\varepsilon)$ is the Fermi function $f_{E_F}(T;\varepsilon)=(\exp[(\varepsilon-E_F)/(k_BT)]+1)^{-1}$,
and $\epsilon_{\gamma u \nu}$ is the Levi-Civita symbol. The relaxation time $\tau_{i,\mathbf{k}}$ in principle is dependent on both the band
index $i$ and the $\mathbf{k}$ vector direction.
Furthermore, $v_\alpha^{i,\mathbf{k}}$ is the group velocity
\begin{align}
\label{eq:vgroup}
 v_\alpha^{i,\mathbf{k}} &= \frac{1}{\hbar}\:\frac{\partial\varepsilon_{i,\mathbf{k}}}{\partial k_\alpha}
\end{align}
and $\left(M_{\beta u}^{i,\mathbf{k}}\right)^{-1}$ the inverse mass tensor
\begin{align}
\label{eq:masstensor}
 \left(M_{\beta u}^{i,\mathbf{k}}\right)^{-1} &= \frac{1}{\hbar^2}\:\frac{\partial^2\varepsilon_{i,\mathbf{k}}}{\partial k_\beta\partial k_u}.
\end{align}

To show that the inverse Hall coefficient $R_{xyz}$ is proportional to $n$, we start by assuming bands with quadratic dispersion.
The dispersion relation for a quadratic, isotropic band in 2D is given by
\begin{align}
\label{eq:quadband}
 \varepsilon_{i,k} &= \frac{\hbar^2 k^2}{2m_i},
\end{align}
with $k^2=k^2_x+k^2_y$. The group velocity is then $v_\alpha^{i,k_\alpha}=\hbar k_\alpha/m_i$ while the mass tensor in Eq.~(\ref{eq:masstensor})
is for each band a diagonal matrix with $m_i^{-1}$ on the diagonal.
In the zero temperature limit and assuming an $i$- and $\mathbf{k}$-independent relaxation time $\tau_{i,\mathbf{k}}=\tau(E_F)$ the conductivity distributions
in Eqs.~(\ref{eq:sigmaab_Tmu}) and (\ref{eq:sigmaabg_Tmu}) are given by
\begin{align}
\label{eq:sigmaaaquad_0Kmu}
 \sigma_{\alpha\alpha}\left(0;E_F\right) &= \sum_i\frac{q^2\:\tau}{m_i}\:n_i,\\
\label{eq:sigmaabgquad_0Kmu}
 \sigma_{\alpha\beta\gamma}\left(0;E_F\right) &= -\sum_i\frac{q^3\:\tau^2}{m_i^2}\:\epsilon_{\alpha\beta\gamma}\:n_i,
\end{align}
where $n_i$ is the charge-carrier density in band $i$.
Finally, assuming that the magnetic field is applied perpendicular to the 2D system along $z$, we get for the Hall coefficient
\begin{align}
\label{eq:hallcoefficients_0K}
 R_{xyz}\left(0;E_F\right) &= \frac{\sum_i m^{-2}_i\:n_i}{q\:\left(\sum_i m^{-1}_i\:n_i\right)^2}.
\end{align}
Thus, only for valley-independent effective mass $m_i=m$ the inverse Hall coefficient is directly proportional to the doping-charge concentration $\mathrm{n}=n\:q$.
The results in Eqs.~(\ref{eq:sigmaaaquad_0Kmu}), (\ref{eq:sigmaabgquad_0Kmu}), and (\ref{eq:hallcoefficients_0K}) also hold for a 3D system,
however with the important difference that the conductivity in two dimensions is independent of the mass of the charge carriers, since the
density $n$ is proportional to $m$. Furthermore, the Hall coefficient is inversely proportional to the mass $m$ in the 2D case.
\begin{figure}
 \includegraphics[scale=0.39,clip=]{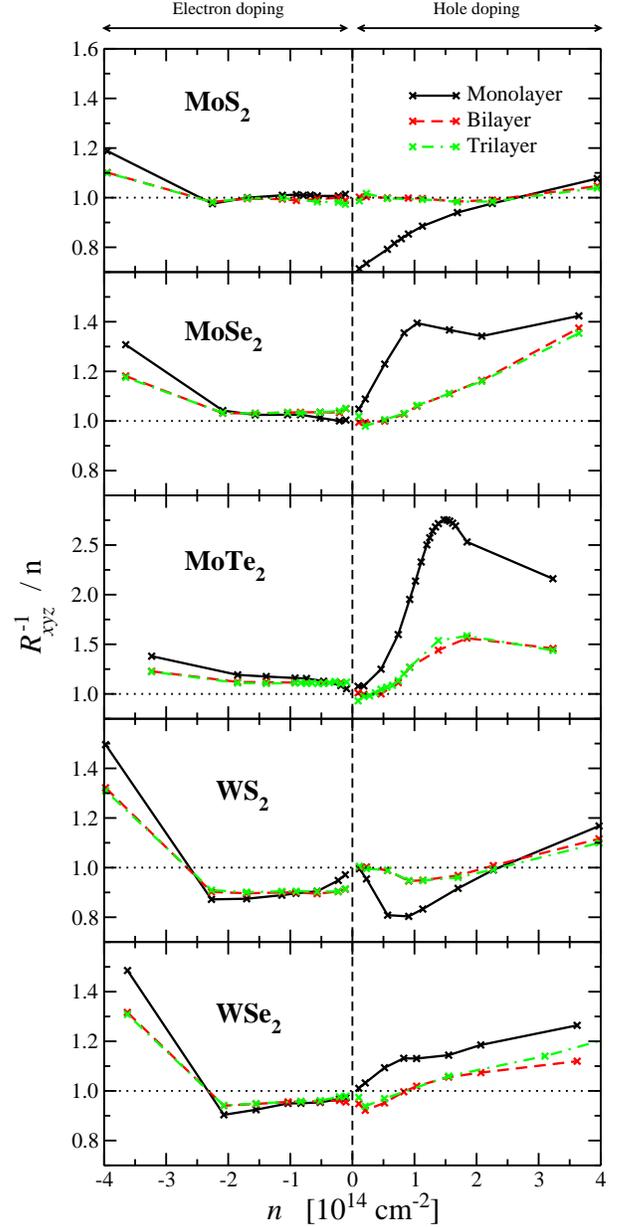}
 \caption{\label{fig:hallcoefficient}Ratio of the inverse Hall coefficient $R^{-1}_{xyz}$ to the doping charge $\mathrm{n}$
          as a function of doping for the mono- (black, solid), bilayer (red, dashed), and trilayer (green, dash-dotted)
          of all investigated TMDs for a temperature of $T=300\:\mathrm{K}$.
          Note also the different range of the ordinate in the case of MoTe$_2$. }
\end{figure}
\begin{figure}
 \includegraphics[width=0.46\textwidth,clip=]{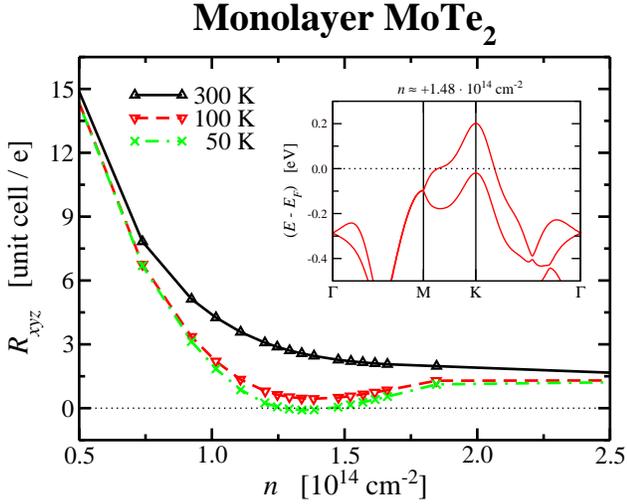}
 \caption{\label{fig:MoTe2_hall}Hall coefficient $R_{xyz}$ as a function of doping for monolayer MoTe$_2$ and temperatures $T=300\:\mathrm{K}$, $T=100\:\mathrm{K}$, and $T=50\:\mathrm{K}$.
          The inset shows the band structure for a critical doping of $n\approx1.48\cdot10^{14}\:\mathrm{cm}^{-2}$ ($\mathrm{n}=+0.16\:e$/unit cell). The specific band structure of MoTe$_2$ with the nearly linear dispersion along K $\rightarrow$ $\Gamma$
          and the changing sign of the effective mass when increasing the distance to K leads to the large difference between $R^{-1}_{xyz}$ and $\mathrm{n}$.}
\end{figure}

In the constant-relaxation-time approximation ($\tau_{i,\mathbf{k}}=\tau(E_F)$) and for hexagonal symmetry (such as in the TMDs),
the Hall tensor in Eq.~(\ref{eq:hallcoefficients}) has only two independent coefficients\cite{lovett1999} (in-plane and out-of-plane component).
However, it is important to remember that this simple equation for the Hall coefficient $R_{xyz}$, Eq.~(\ref{eq:hallcoefficients_0K}), is only valid as long as there are only bands
with isotropic, quadratic dispersion and if $\tau_{i,\mathbf{k}}=\tau(E_F)$.
For small doping this might be a good approximation but in the high doping case this approximation can break down --
especially, as soon as a minimum/maximum with non-quadratic dispersion starts to get filled.

Figure \ref{fig:hallcoefficient} shows the ratio of the inverse Hall coefficient $R^{-1}_{xyz}(T;E_F)$ (calculated with the BoltzTraP code\cite{madsen2006}
using Eqs.~(\ref{eq:hallcoefficients})--(\ref{eq:sigmaabg_Tmu}) and assuming $\tau_{i,\mathbf{k}}=\tau(E_F)$) to the doping-charge concentration $\mathrm{n}$,
as a function of doping for the mono-, bi-, and trilayer systems for $T=300\:\mathrm{K}$.
The comparison between $T=0\:\mathrm{K}$ and $T=300\:\mathrm{K}$ which can be found in the appendix, Figs.~64--66,
shows that the temperature has only a minor influence on the inverse Hall coefficient except for MoTe$_2$ (which we will also discuss separately).

For most of the studied systems, the inverse Hall coefficient $R^{-1}_{xyz}(300\:\mathrm{K};E_F)$ shows a strong deviation from the doping charge $\mathrm{n}$
for large doping -- the doping-charge concentration calculated using the inverse Hall coefficient can be $1.5$ times larger than the real concentration.
In fact, $R^{-1}_{xyz}(300\:\mathrm{K};E_F)\approx\mathrm{n}$ is true only in two cases (\cf, Eq.~(\ref{eq:hallcoefficients_0K})):
either (i) if mainly one valley with parabolic dispersion is doped or 
(ii) if the doping charge is split between several valleys with quadratic bands which however have similar effective masses.

Case (i) holds only for small hole doping ($n\lesssim+5\cdot10^{13}\:\mathrm{cm}^{-2}$) of all multilayer TMDs
as mainly the $\Gamma$ valley is doped. The deviation of the inverse Hall coefficient from the doping
charge $\mathrm{n}$ increases with increasing doping of the valence-band maximum at K.
Accordingly, $R^{-1}_{xyz}(300\:\mathrm{K};E_F)\approx\mathrm{n}$ for a larger range of hole
doping of multilayer MoS$_2$ than in the other TMDs as the doping at K is negligible.
For high hole doping ($n\gtrsim+2\cdot10^{14}\:\mathrm{cm}^{-2}$) the deviation is due to the
non-parabolicity of the bands near K. If however the charge is split between $\Gamma$ and K
the difference can be explained by the much larger effective mass of the $\Gamma$ valley
(\ie, in variance with case (ii)). Assuming $\mathrm{n}_\Gamma\approx\mathrm{n}_\text{K}=\mathrm{n}/2$,
the ratio of the inverse Hall coefficient to the doping charge
(\cf, Eq.~(\ref{eq:hallcoefficients_0K})) simplifies to
\begin{align}
 \frac{R_{xyz}\left(0;E_F\right)^{-1}}{\mathrm{n}} &= \frac{1}{2}+\frac{m_\Gamma\,m_\text{K}}{m^2_\Gamma+m^2_\text{K}},
\end{align}
which is always smaller than $1$ for $m_\Gamma\neq m_\text{K}$. This also explains why the ratio
approaches $\approx0.7$ for small hole doping of monolayer MoS$_2$. At $T=300\:\mathrm{K}$
$\mathrm{n}$ is split between both maxima which have however very different masses\cite{yun2012} -- $m_\Gamma=3.524\,m_0$
and $m_\text{K}=0.637\,m_0$.

The agreement between $R^{-1}_{xyz}(300\:\mathrm{K};E_F)$ and $\mathrm{n}$ is much better
for electron doping up to $n\approx-2\cdot10^{14}\:\mathrm{cm}^{-2}$ even if both conduction-band minima at K and Q
are occupied. This is due to the similar effective masses of those two valleys. As in the hole-doping case 
the difference between $R^{-1}_{xyz}(300\:\mathrm{K};E_F)$ and $\mathrm{n}$ increase for larger doping
($n\lesssim-2\cdot10^{14}\:\mathrm{cm}^{-2}$) which is due to the increasing non-parabolicity of the bands.

The only case where the model of a 2D electron gas with constant relaxation time gives reasonable results
for the doping-charge concentration within a large range of both electron and hole concentrations is multilayer MoS$_2$.
Since the agreement between $R^{-1}_{xyz}(T;E_F)$ and $\mathrm{n}$ can be much better in other systems as exemplified
for CoSb$_3$ in Ref.~\citenum{madsen2006}, the deviations
shown in Fig.~\ref{fig:hallcoefficient} point out problems if the specific band structure is not taken into account.
Once again, MoTe$_2$ is particularly interesting because the inverse Hall coefficient $R^{-1}_{xyz}(300\:\mathrm{K};E_F)$
is nearly three times bigger than $\mathrm{n}$ for a hole doping of $n\approx+1.4\cdot10^{14}\:\mathrm{cm}^{-2}$.

In order to understand the origin of this behavior of $R^{-1}_{xyz}/\mathrm{n}$ for hole doping of monolayer MoTe$_2$, we plot $R_{xyz}$ as a function of
temperature in Fig.~\ref{fig:MoTe2_hall}. As the temperature is reduced, the Hall coefficient $R_{xyz}(T;E_F)$ decreases and
for $T\leq50\:\mathrm{K}$ it even changes the sign. The result for $T=50\:\mathrm{K}$ in Fig.~\ref{fig:MoTe2_hall} indicates that in the range of
$+1.25\cdot10^{14}\:\mathrm{cm}^{-2}\leq n\leq+1.5\cdot10^{14}\:\mathrm{cm}^{-2}$ the Hall coefficient $R_{xyz}$ changes twice the sign. The band structure for a
doping of $n\approx+1.48\cdot10^{14}\:\mathrm{cm}^{-2}$ (inset in Fig.~\ref{fig:MoTe2_hall})
shows that the valence band has at least two inflection points close to the K point which cross the Fermi energy with increasing doping.
Accordingly, the effective mass changes the sign and thus also the conductivity tensor $\sigma_{\alpha\beta\gamma}$ and $R_{ijk}$.
Furthermore, the nearly linear dispersion along K $\rightarrow$ $\Gamma$ leads to $m$ $\rightarrow$ $0$ and thus to the large difference between $R^{-1}_{xyz}$ and $\mathrm{n}$.
For bi- and trilayer MoTe$_2$ the deviations are smaller and at higher doping values. This is due to the finite contribution of the doping charge at the $\Gamma$ point which
leads to a smaller doping around K.

\subsection{Conductivity and DOS at the Fermi energy}
\label{sec:cond_dos}

\begin{figure*}
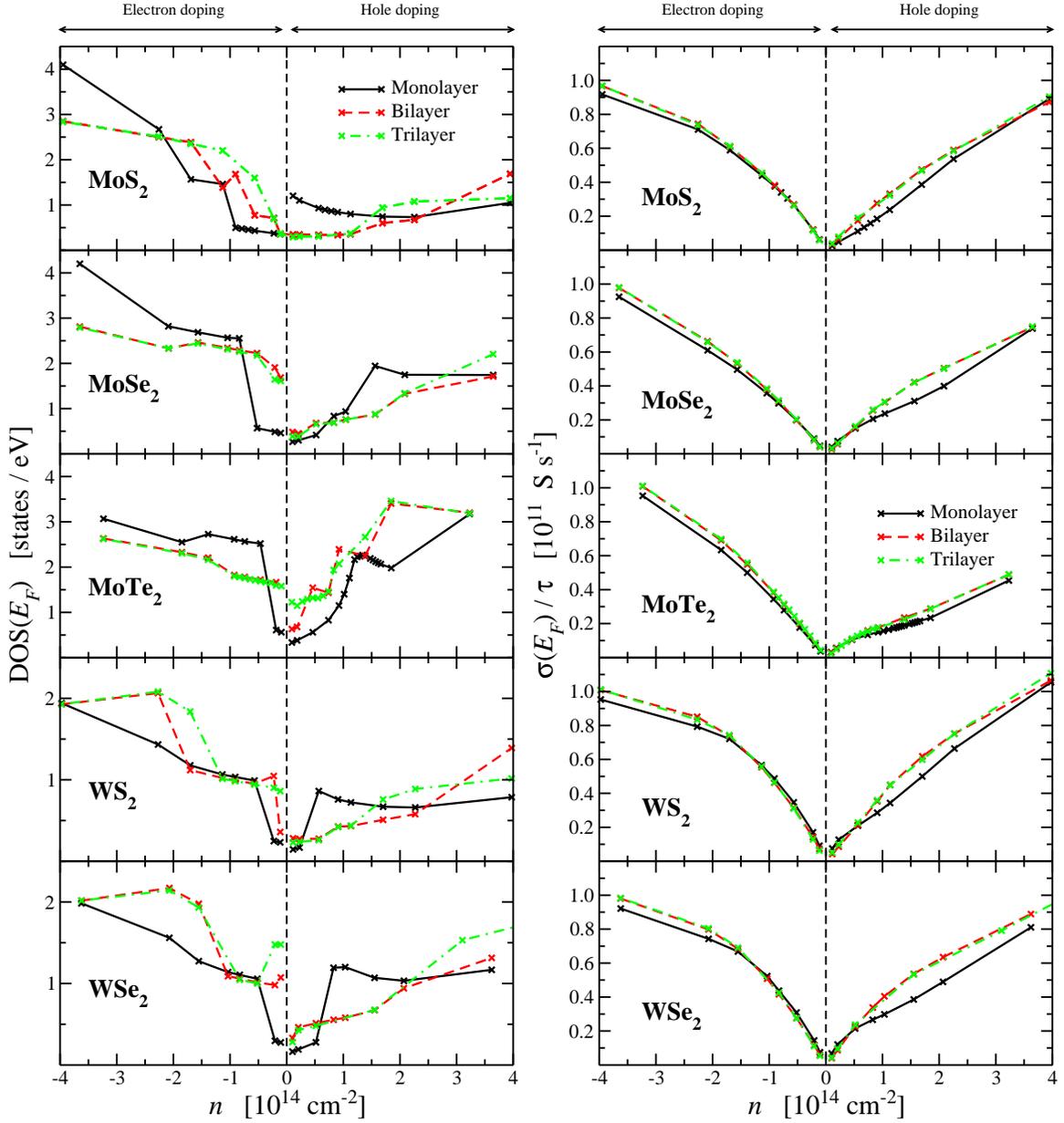

 \includegraphics[scale=0.39,clip=]{figure18a.eps}
 \includegraphics[scale=0.39,clip=]{figure18b.eps}
 \caption{\label{fig:dos_cond_ef}DOS at the Fermi energy $E_F$ (left panel, $T=0\:\mathrm{K}$)
          and in-plane conductivity $\sigma_{xx}/\tau$ (right panel, $T=300\:\mathrm{K}$) as a function of doping
          for the mono- (black, solid), bilayer (red, dashed), and trilayer (green, dash-dotted) of all investigated TMDs.
          Note also the different range of the ordinate for DOS($E_F$) in the case of WS$_2$ and WSe$_2$.
          Lines are guides for the eye.}
\end{figure*}
In the end, we want to briefly analyze the DOS at the Fermi energy $E_F$ and the in-plane conductivity $\sigma_{xx}/\tau$
in Fig.~\ref{fig:dos_cond_ef}. Both were calculated using the fitted band structure of BoltzTraP.
The DOS at the Fermi energy $E_F$ in the left-hand panel shows, that the doping charge cannot always be described
with quadratic, isotropic bands in 2D (as also shown above by the behavior of $R^{-1}_{xyz}$).
In this case, $\mathrm{DOS}(E_F)$ would be constant and would have steps as soon as another band crosses $E_F$.
In fact, $\mathrm{DOS}(E_F)$ has steps and those can be related to crossing bands, but for high doping it can deviate
from a simple 2D electron gas.

For hole doping of monolayer MoS$_2$, MoSe$_2$, WS$_2$,
and WSe$_2$ the DOS at the Fermi energy is nearly constant as soon as the $\Gamma$ valley is doped, \ie, for doping larger than
$n\approx+0.1\cdot10^{14}\:\mathrm{cm}^{-2}$, $n\approx+1.5\cdot10^{14}\:\mathrm{cm}^{-2}$, $n\approx+0.5\cdot10^{14}\:\mathrm{cm}^{-2}$, and
$n\approx+0.9\cdot10^{14}\:\mathrm{cm}^{-2}$, respectively. Also for hole-doping smaller than $n\leq+1.1\cdot10^{14}\:\mathrm{cm}^{-2}$
of multilayer MoS$_2$ the DOS is constant as only the $\Gamma$ valley is doped in this regime.
The non-constant behavior for hole doping of the other multilayer systems increases with increasing
doping of the K valley: it is more pronounced for MoSe$_2$ and WSe$_2$ than for MoS$_2$ and WS$_2$
(\cf, Figs.~\ref{fig:MoS2_bandpos_charge}--\ref{fig:WSe2_bandpos_charge}).
In the case of MoTe$_2$, where mainly the valence bands at K are doped, a description with 2D, quadratic, isotropic bands
completely fails.

For n-type doping of all TMDs the DOS at the Fermi energy shows a quasi 2D behavior.
It has steps when the conduction-band minimum at K or Q enters the bias window
and is nearly constant in between. However, for larger electron doping the non-constant behavior increases.
This is due to the stronger deviation of the spin-orbit-split conduction band at Q from a quadratic
dispersion.

The in-plane conductivity $\sigma_{xx}/\tau$ is another measure for the deviation from quadratic, isotropic bands in 2D.
As can be seen in Eq.~(\ref{eq:sigmaaaquad_0Kmu}), for a perfect 2D electron gas the conductivity would be an increasing linear
function of the doping-charge concentration $\mathrm{n}$.
Most interestingly, the right-hand panel of Fig.~\ref{fig:dos_cond_ef} shows that $\sigma_{xx}/\tau$
weakly differs from one TMD to the other and from the monolayer to the multilayer case. Only for the tungsten
dichalcogenides the conductivity has a small nonlinear component for n-type doping. This might be due to the
stronger SOC of tungsten which leads also to larger deviation of the Q valley from a quadratic dispersion.

\section{Conclusions}
\label{sec:conclusion}

In this work, we have calculated from \textit{ab initio} the structural, electronic, and transport properties for \mbox{mono-,}
bi-, and trilayer TMDs in field-effect configuration. We have first investigated the structural changes of the TMDs under
field-effect doping. We found that the internal structure is only slightly changed
but that it is nevertheless important to fully relax the system. We also showed that high electron doping
can induce a phase transition from the 1H to the 1T' structure. In accordance with literature this transition can
however only occur for electron doping larger than $\mathrm{n}\leq-0.35\:e$/unit cell. Therefore, we concentrated
on smaller doping of the H polytype as it is the most stable structure found in nature.

The band structure and thus also the transport properties can be changed considerably under field-effect doping.
We have shown that most TMDs behave similarly under hole doping while for electron doping they can be divided
into two different classes: one in which the conductive channel has a width of approximately two layers for small doping (MoSe$_2$
and MoTe$_2$) and one in which the charge is localized within the first layer. This can be attributed to the relative position
of the conduction-band minimum at K and Q in the multilayer TMDs. In the former class Q is lower than K
and the doping charge first occupies the states in the Q valley. Since these states have a large $p_z$ contribution
of chalcogen states, the hybridization between the layers is larger and the electrons are more delocalized. Additionally,
for the tungsten dichalcogenides the difference between K and Q is smaller than for MoS$_2$ and thus, as the electron
doping is increased, the charge rapidly starts to occupy states at Q. This leads to an increase of the width of the
conductive channel to approximately two layers. For high electron doping only the Q valley is occupied in all investigated TMDs
(also in the monolayer systems) and the width of the conductive channel is reduced to one layer.

Under hole doping most TMDs behave similarly: in the monolayer case first the valence-band maximum at K is occupied
while in the multilayer case it is the $\Gamma$ valley. This also leads to the delocalization of the doping charge
over more layers as the states at $\Gamma$ have large chalcogen $p_z$ character. For large hole doping also in the
monolayer case the doping at $\Gamma$ is larger than those at K. Only exception from this picture is MoTe$_2$ in which
the valence-band maximum for the undoped compound is always at K even in the multilayer case. Accordingly, the doping
charge within the K valley is always larger than those at $\Gamma$.

However, even if the thickness of the doping-charge distribution is approximately two layers for multilayer TMDs,
the number of doped valleys can be comparable to the monolayer case. The main difference is between n-type and
p-type doping: as the valley degeneracy for the conduction-band minimum at Q is $g_v=6$, the total number of
doped valleys can be as large as $\nu=16$ while for hole doping the maximum is $\nu=8$.

In the next part, we have seen that a Hall-effect measurement can often not directly be used in order to determine the
charge-carrier concentration under the assumption of quadratic bands -- the charge
thus determined can be up to $1.5$ times larger than the real doping charge within the sample. 
For MoTe$_2$ the Hall coefficient $R_{xyz}$ even changes the sign due to the changing curvature of the valence
band. Thus, an interpretation based on parabolic bands would lead to an incorrect sign of the charge of the
carriers.
Even if this can only be seen at low temperatures, the Hall-effect measurement still largely overestimates the
doping charge concentration if one does not take into account the specific band structure. Only in the case of
multilayer MoS$_2$ the inverse Hall coefficient is directly proportional to the doping-charge concentration for
a large range of electron and hole doping.

In this work, we have shown that it is important to correctly model an FET setup. The changes in the electronic and
transport properties cannot be described with both the rigid doping and the uniform-background doping approach.
We provide not only a full database of electronic structure of mono- , bi-, and trilayer dichalcogenides as a
function of doping, but also a mapping between the doping charge and the Hall coefficient.

\begin{acknowledgments}
The authors acknowledge financial support of the Graphene Flagship and of the French National ANR funds
within the \textit{Investissements d'Avenir programme} under reference ANR-13-IS10-0003-01.
Computer facilities were provided by PRACE, CINES, CCRT and IDRIS.
\end{acknowledgments}

\bibliography{dichalcogenides}

\clearpage
\onecolumngrid
\appendix
\section{Band structure of the undoped TMDs}
\begin{figure*}[hbp]
 \centering
 \includegraphics[width=0.43\textwidth,clip=]{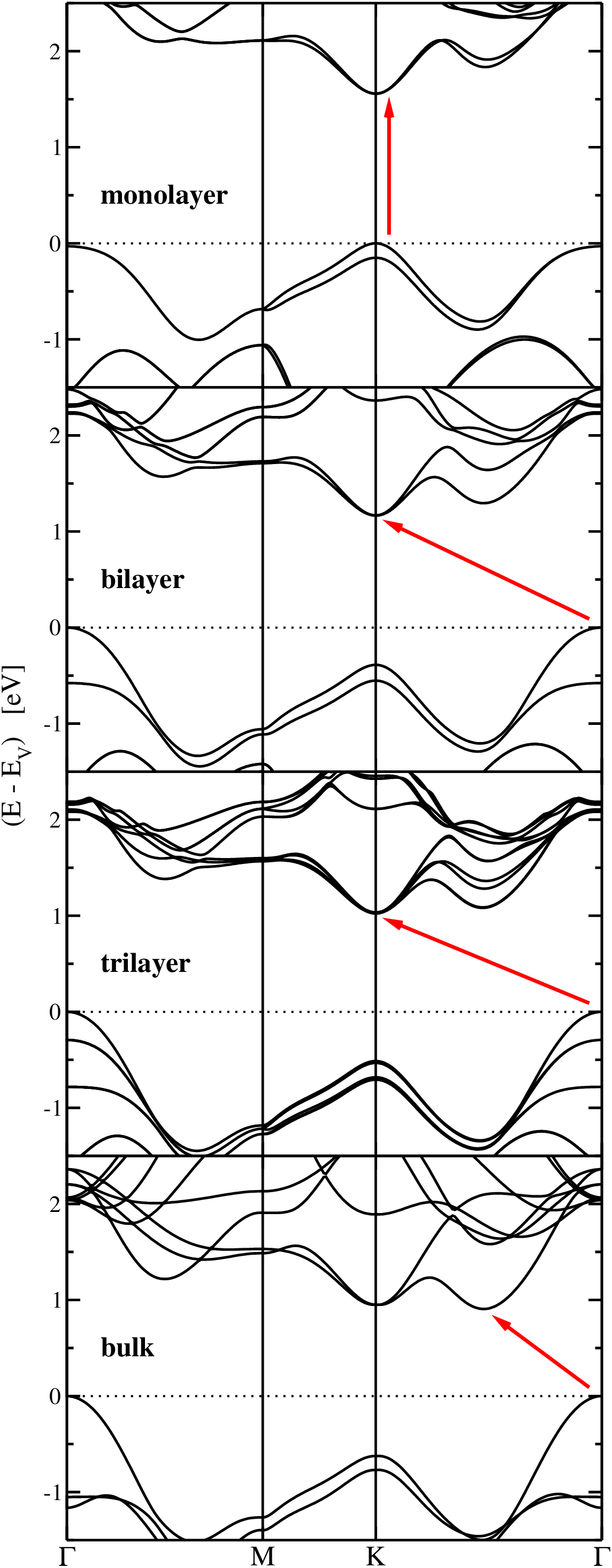}
 \caption{Band structure of mono-, bi-, trilayer, and bulk MoS$_2$. The arrow indicates the lowest-energy transition.}
\end{figure*}
\begin{figure*}[hbp]
 \centering
 \includegraphics[width=0.43\textwidth,clip=]{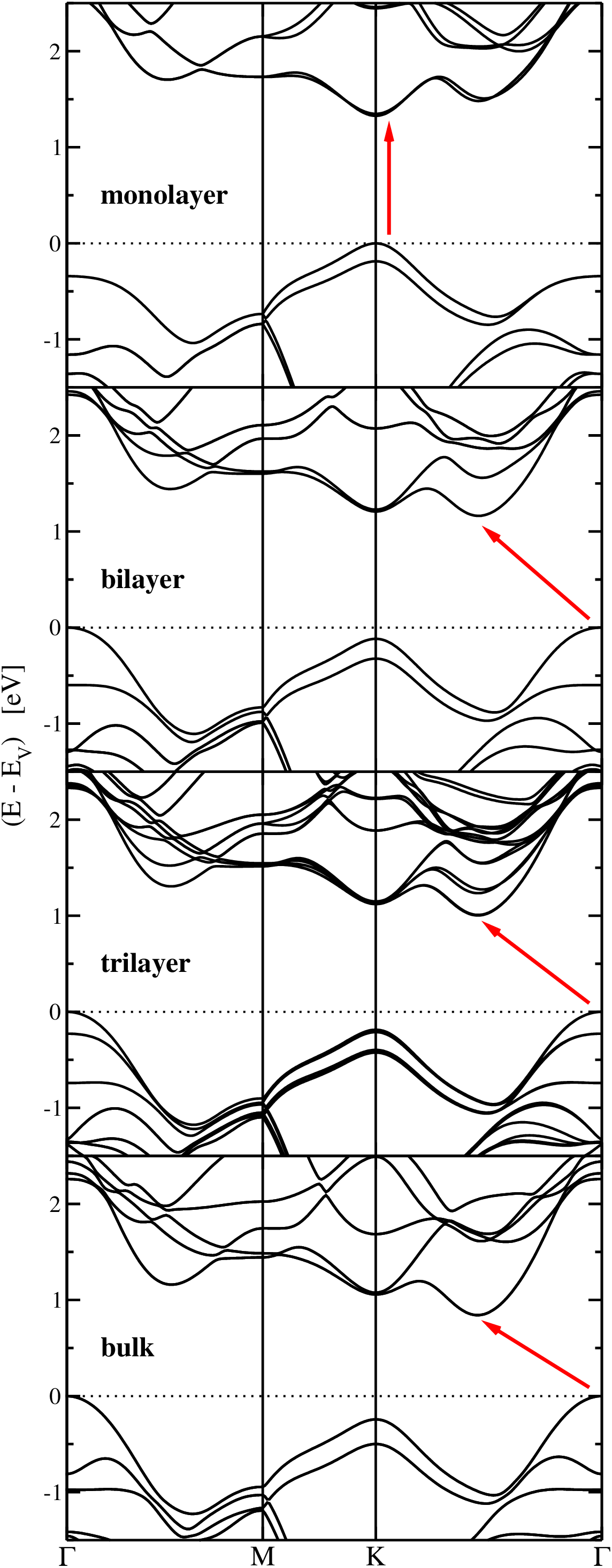}
 \caption{Band structure of mono-, bi-, trilayer, and bulk MoSe$_2$. The arrow indicates the lowest-energy transition.}
\end{figure*}
\begin{figure*}[hbp]
 \centering
 \includegraphics[width=0.43\textwidth,clip=]{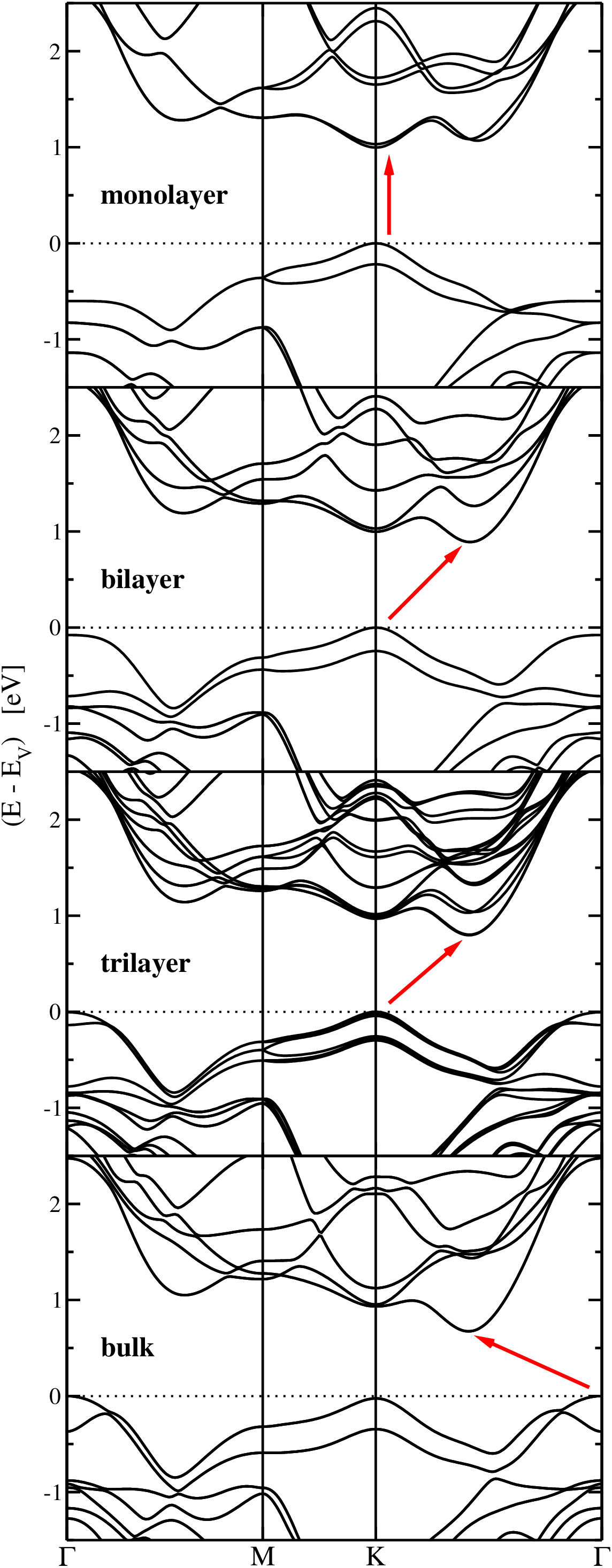}
 \caption{Band structure of mono-, bi-, trilayer, and bulk MoTe$_2$. The arrow indicates the lowest-energy transition.}
\end{figure*}
\begin{figure*}[hbp]
 \centering
 \includegraphics[width=0.43\textwidth,clip=]{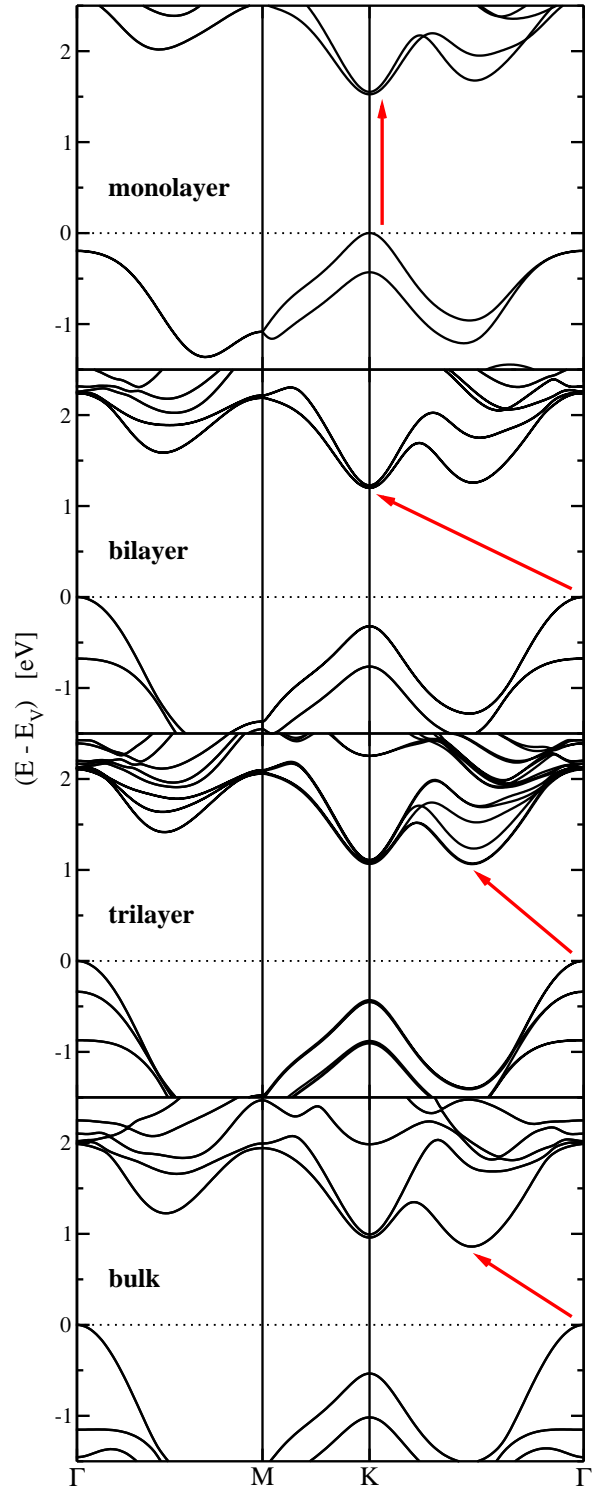}
 \caption{Band structure of mono-, bi-, trilayer, and bulk WS$_2$. The arrow indicates the lowest-energy transition.}
\end{figure*}
\begin{figure*}[hbp]
 \centering
 \includegraphics[width=0.43\textwidth,clip=]{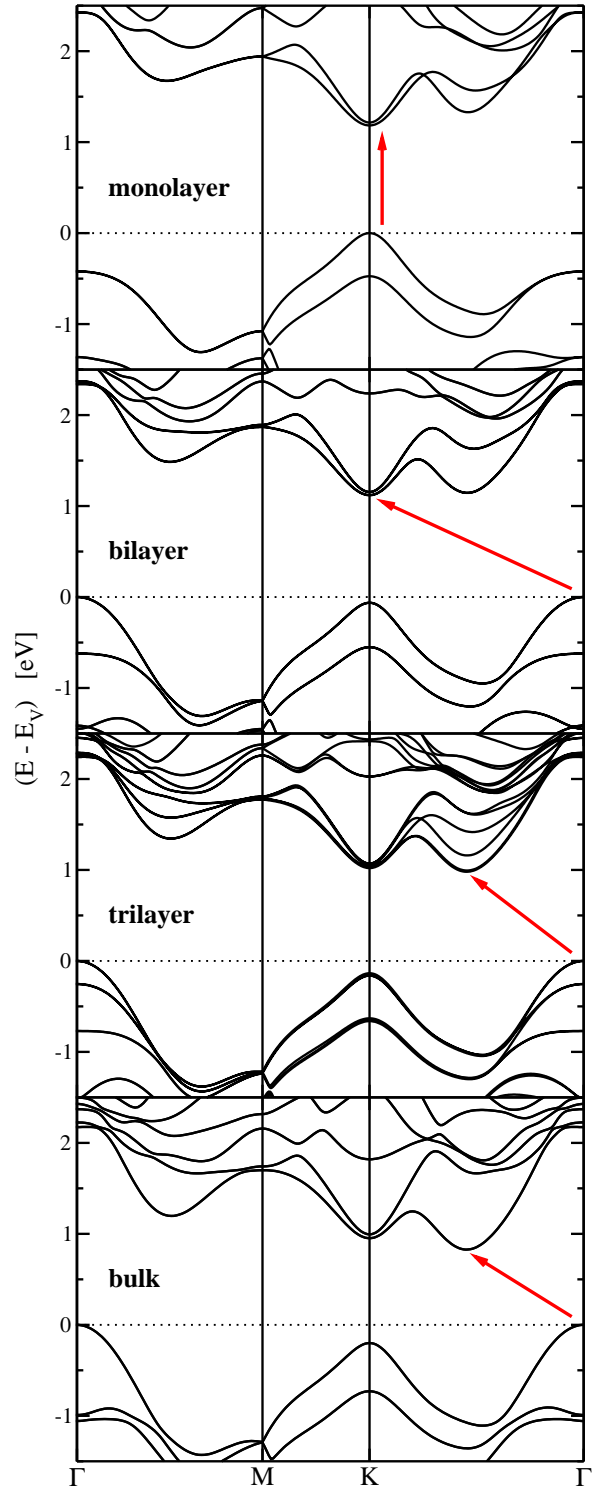}
 \caption{Band structure of mono-, bi-, trilayer, and bulk WSe$_2$. The arrow indicates the lowest-energy transition.}
\end{figure*}

\clearpage
\section{Influence of structural relaxation in FET setup}
\begin{figure*}[hbp]
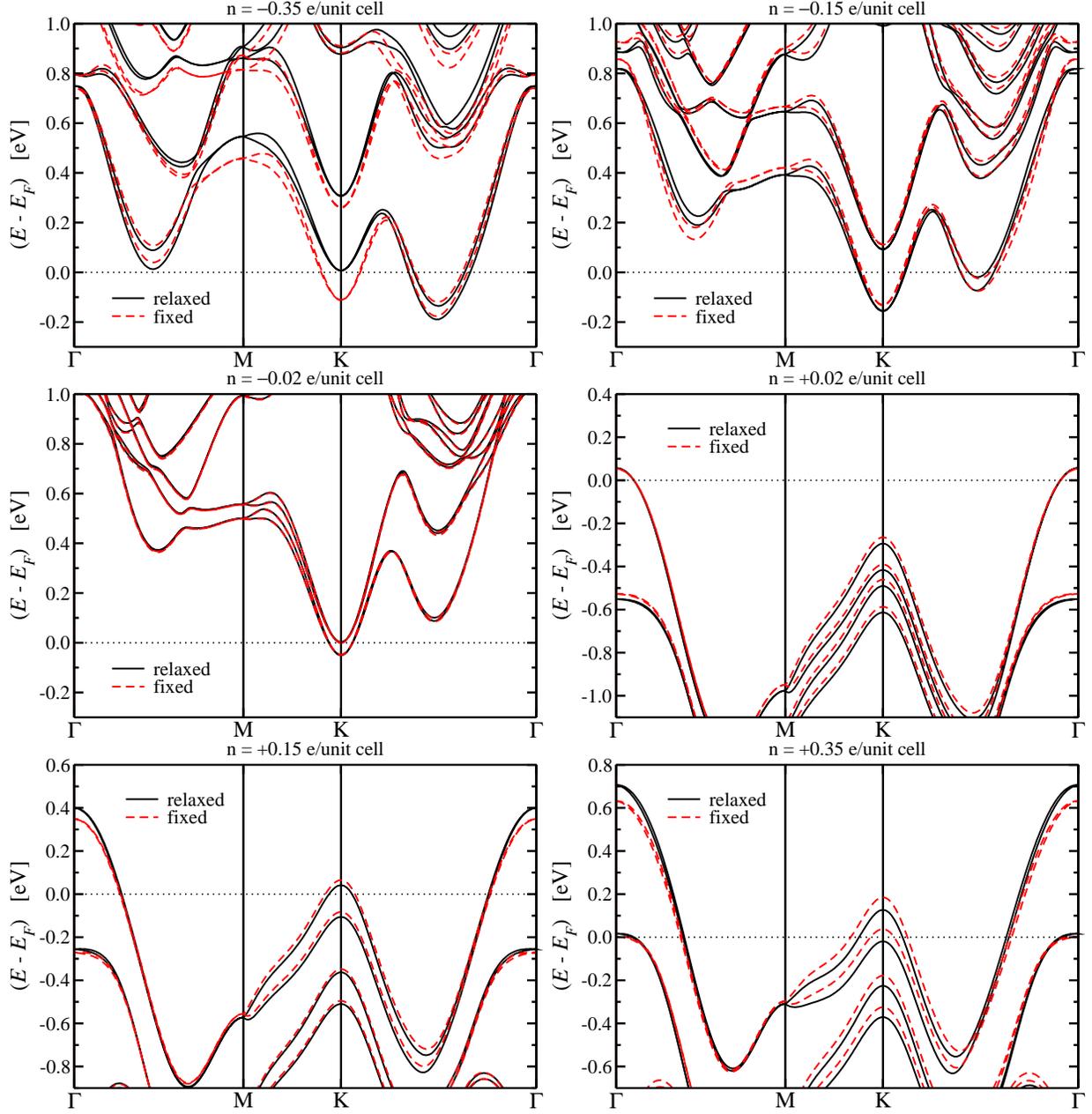

 \centering
 \includegraphics[width=0.45\textwidth,clip=]{compare_norelax_-0.35.eps}
 \includegraphics[width=0.45\textwidth,clip=]{compare_norelax_-0.15.eps}
 \includegraphics[width=0.45\textwidth,clip=]{compare_norelax_-0.02.eps}
 \includegraphics[width=0.45\textwidth,clip=]{compare_norelax_+0.02.eps}
 \includegraphics[width=0.45\textwidth,clip=]{compare_norelax_+0.15.eps}
 \includegraphics[width=0.45\textwidth,clip=]{compare_norelax_+0.35.eps}
 \caption{Comparison of the band structures of doped bilayer MoS$_2$ with and without relaxation.
          The doping is indicated in the labels. For the fixed structures we used the relaxed
          geometry of undoped bilayer MoS$_2$ and placed the system at approximately the same
          distance to the barrier potential as in the case of the fully relaxed system.}
\end{figure*}

\clearpage
\section{Influence of the asymmetric electric field in an FET}
\begin{figure*}[hbp]
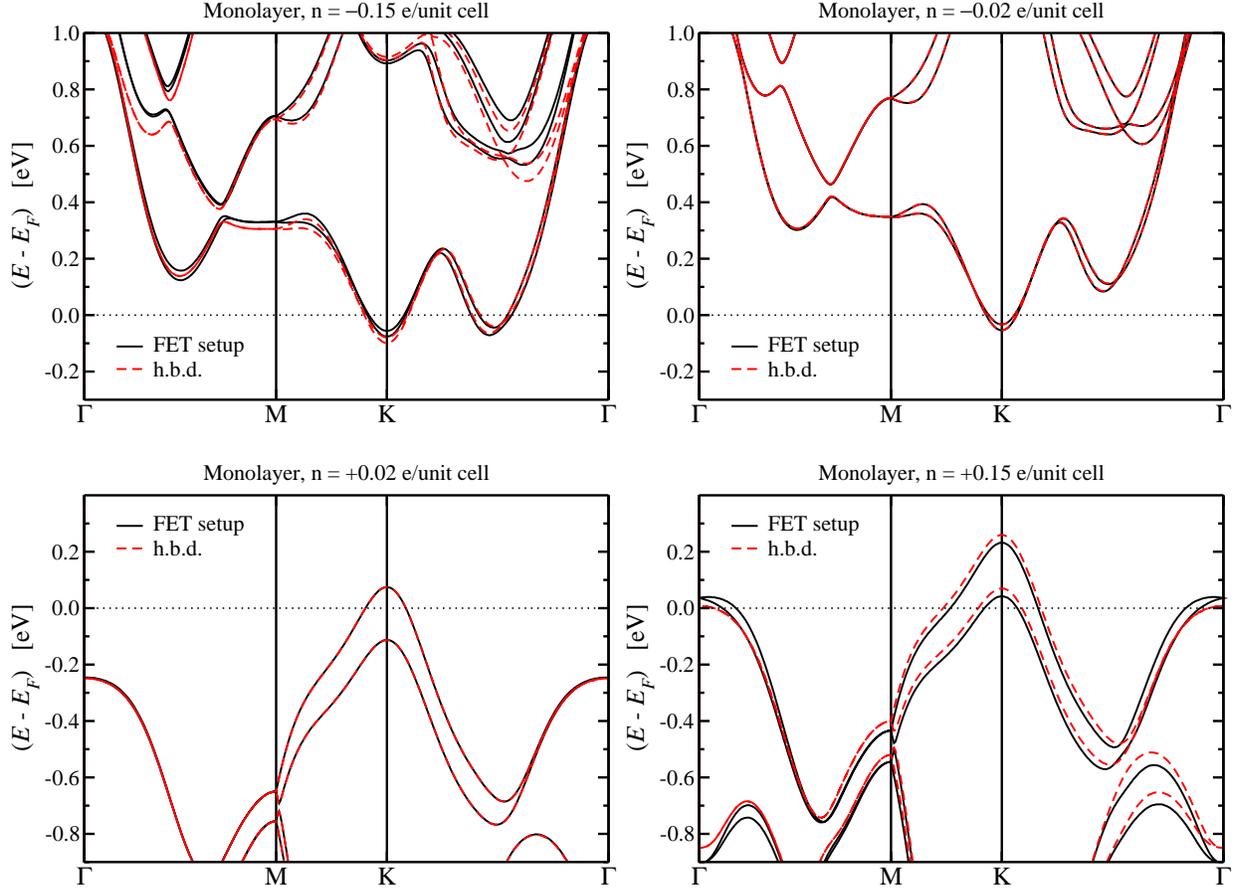

 \centering
 \includegraphics[width=0.45\textwidth,clip=]{compare_hbd_sgl_-0.15.eps}
 \includegraphics[width=0.45\textwidth,clip=]{compare_hbd_sgl_-0.02.eps}\vspace*{0.5cm}
 \includegraphics[width=0.45\textwidth,clip=]{compare_hbd_sgl_+0.02.eps}
 \includegraphics[width=0.45\textwidth,clip=]{compare_hbd_sgl_+0.15.eps}
 \caption{Comparison of the band structures of doped monolayer MoSe$_2$ calculated with a compensating
          jellium background (homogeneous-background doping, ``h.b.d.'') and the proper FET setup.
          The doping is indicated in the labels. All systems have been fully relaxed.
          The difference is negligible for a small doping of $\mathrm{n}=\pm0.02\:e$/unit cell, but for a higher
          doping of $\mathrm{n}=\pm0.15\:e$/unit cell one can see clear differences not only in the relative amount
          of doping in the different valleys but also in the curvature of the bands, \ie, their effective masses.
          Furthermore, it is difficult to model the system properly with a compensating jellium background for electron doping
          larger than $\mathrm{n}=-0.15\:e$/unit cell due to the interlayer states crossing the Fermi energy.}
\end{figure*}
\begin{figure*}[hbp]
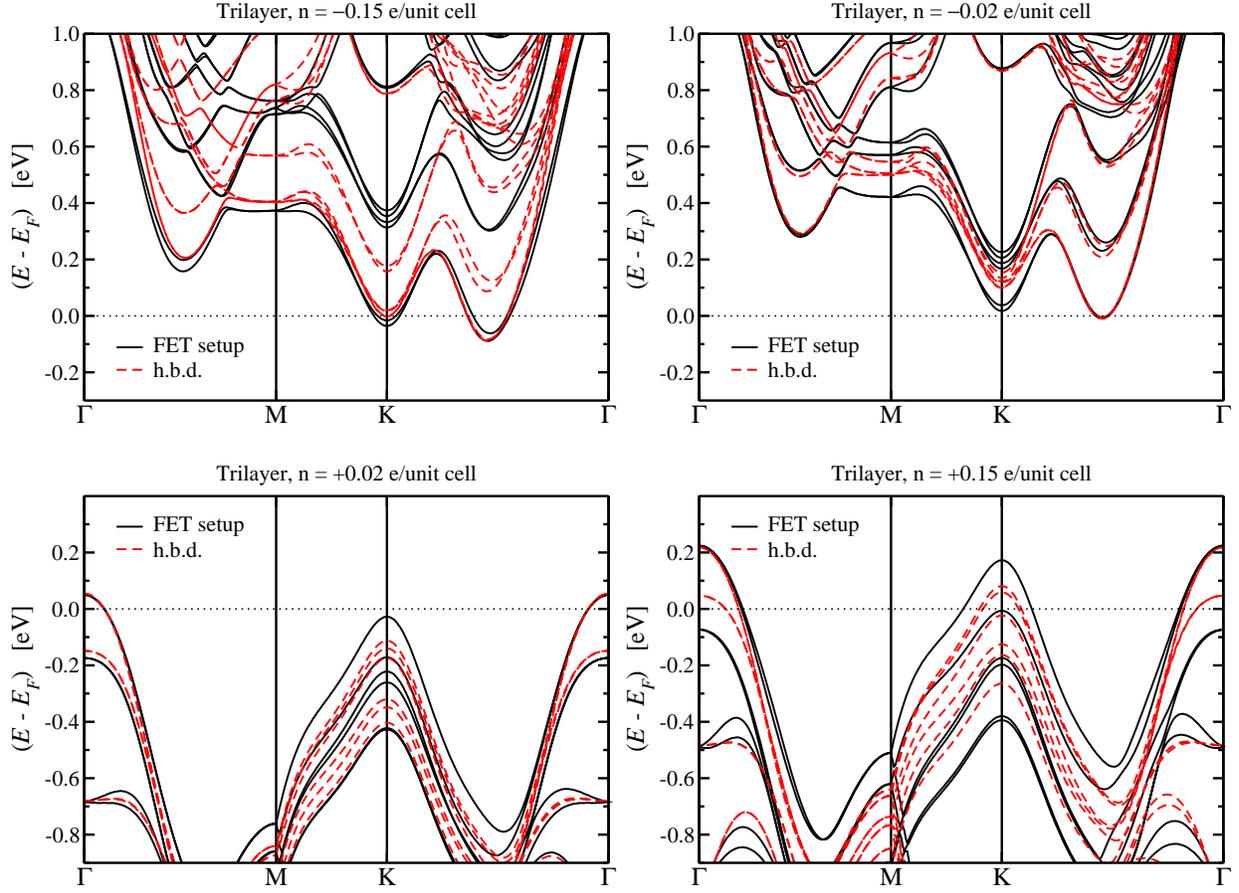

 \centering
 \includegraphics[width=0.45\textwidth,clip=]{compare_hbd_tri_-0.15.eps}
 \includegraphics[width=0.45\textwidth,clip=]{compare_hbd_tri_-0.02.eps}\vspace*{0.5cm}
 \includegraphics[width=0.45\textwidth,clip=]{compare_hbd_tri_+0.02.eps}
 \includegraphics[width=0.45\textwidth,clip=]{compare_hbd_tri_+0.15.eps}
 \caption{Comparison of the band structures of doped trilayer MoSe$_2$ calculated with a compensating
          jellium background (homogeneous background doping, ``h.b.d.'') and the proper FET setup.
          The doping is indicated in the labels. All systems have been fully relaxed.
          The difference is again small for a low doping of $\mathrm{n}=\pm0.02\:e$/unit cell.
          However, in contrast to the monolayer case, in the FET setup other states are close to the Fermi energy which
          will change the Hall coefficient for finite temperature.
          For a higher doping of $\mathrm{n}=\pm0.15\:e$/unit cell one can see clear differences not only in the relative amount
          of doping in the different valleys but also in the curvature of the bands, \ie, their effective masses.
          The differences are also larger for high hole doping as shown in the lower right panel.}
\end{figure*}

\clearpage
\section{Influence of the exchange-correlation approximation}
\begin{figure*}[hbp]
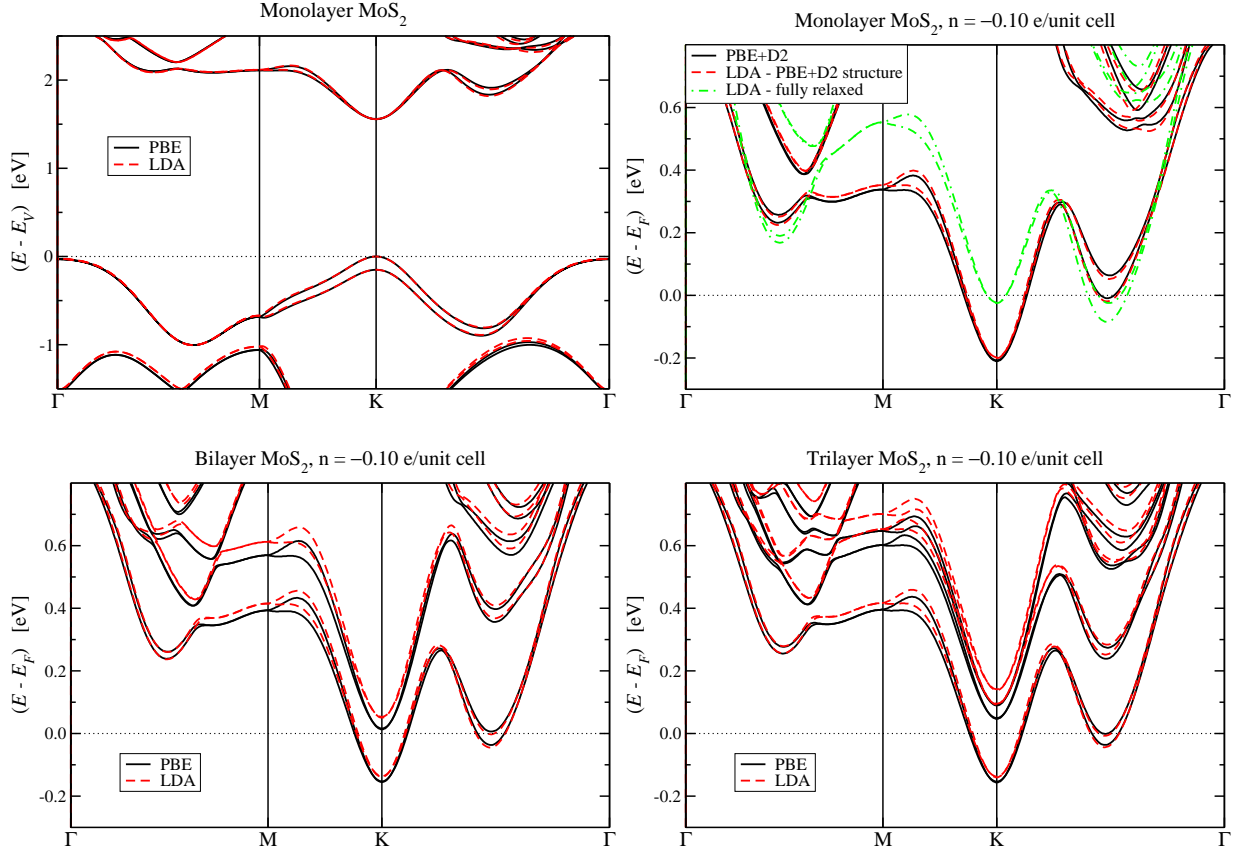

 \centering
 \includegraphics[width=0.45\textwidth,clip=]{bands_sgl_LDA.eps}
 \includegraphics[width=0.45\textwidth,clip=]{bands_sgl_-0.10_LDA.eps}\vspace*{0.5cm}
 \includegraphics[width=0.45\textwidth,clip=]{bands_dbl_-0.10_LDA.eps}
 \includegraphics[width=0.45\textwidth,clip=]{bands_tri_-0.10_LDA.eps}
 \caption{Comparison of the band structures of (doped) MoS$_2$ calculated using either
          PBE+D2 or LDA for the exchange-correlation energy. In the LDA case we used the final,
          fully-relaxed geometry of the PBE+D2 calculations except for the doped monolayer
          for which we also show the results for the LDA-relaxed unit cell (``LDA - fully relaxed'').
          One can see only minor differences in the relative occupation of the conduction-band
          minima at K or Q if the geometry of PBE+D2 is used. This is due to the different interlayer spacing in the case of
          PBE+D2 or LDA. Thus, as we used the geometry of the PBE+D2 calculations, LDA leads to minor
          changes in the bands which are doped. That this is only a result of the different
          interlayer spacing is further confirmed by the fact that the difference in the case of
          the monolayer is negligible. In the fully relaxed LDA case the differences are much
          larger which is due to the compressive strain as the LDA lattice parameter is 2.2\% smaller
          than the one calculated using PBE+D2.}
\end{figure*}

\clearpage
\section{Band structure of the doped TMDs}
\subsection{Molybdenum disulfide}
\begin{figure*}[hbp]
 \centering
 \includegraphics[width=0.31\textwidth,clip=,angle=-90]{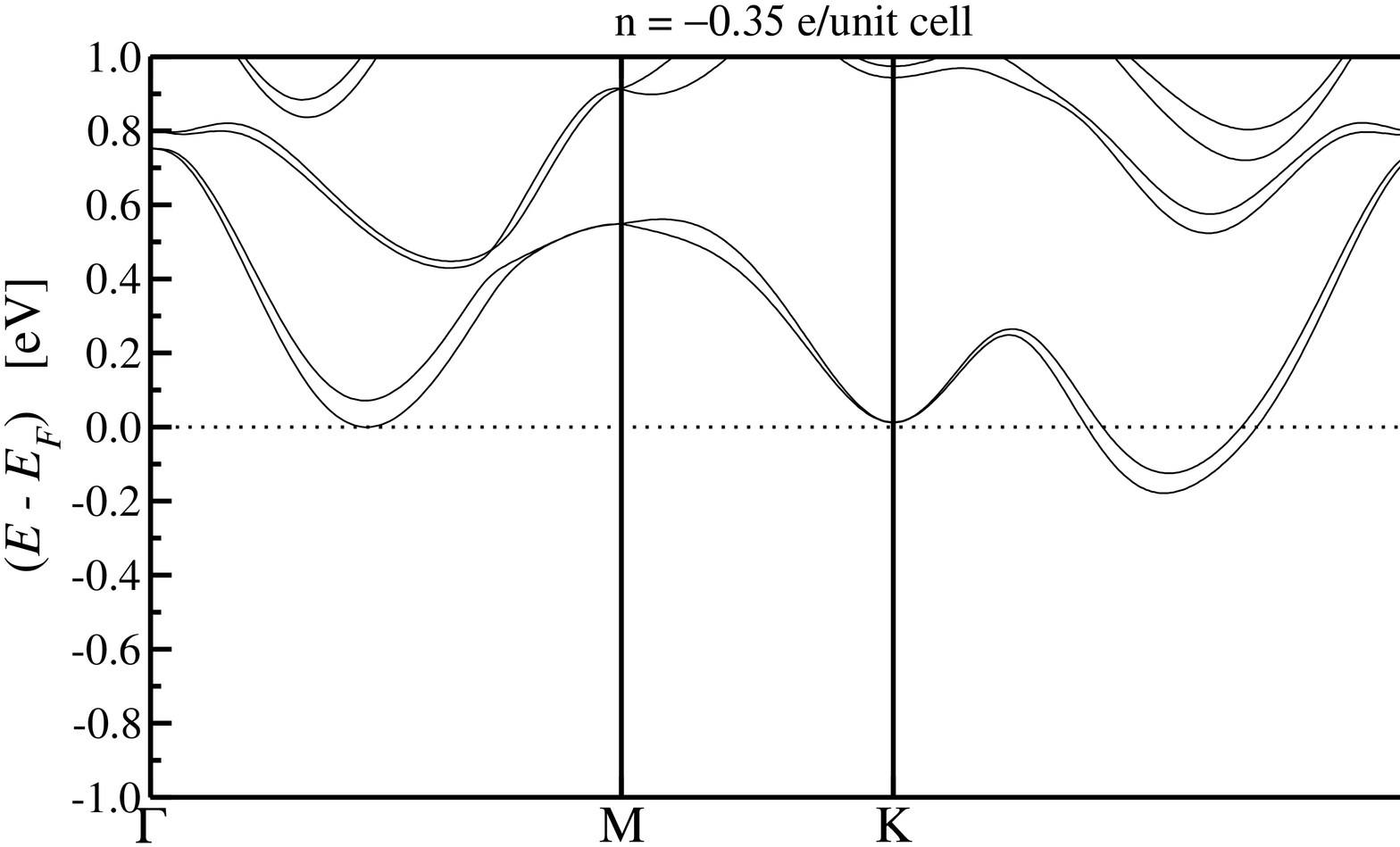}
 \includegraphics[width=0.31\textwidth,clip=,angle=-90]{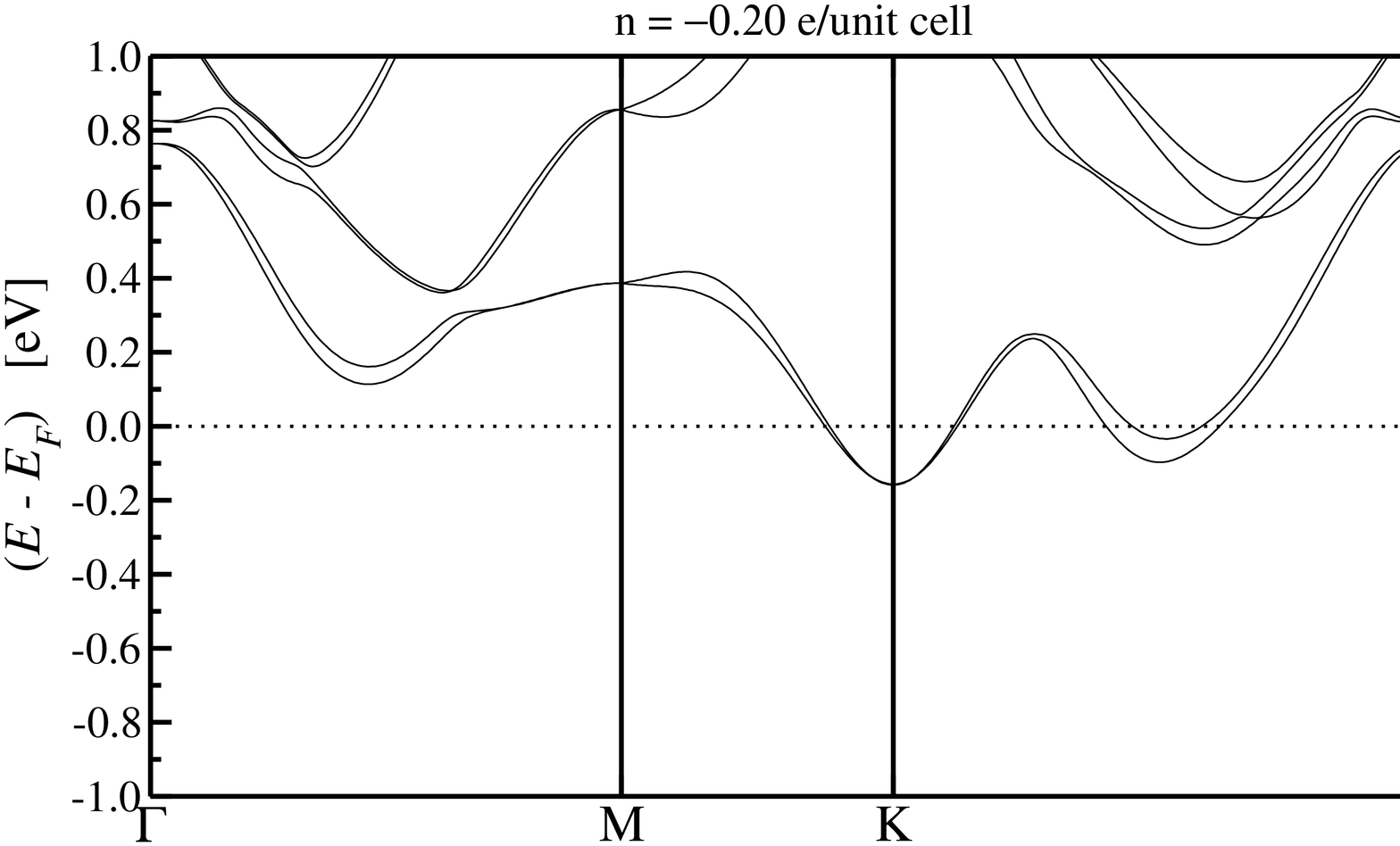}
 \includegraphics[width=0.31\textwidth,clip=,angle=-90]{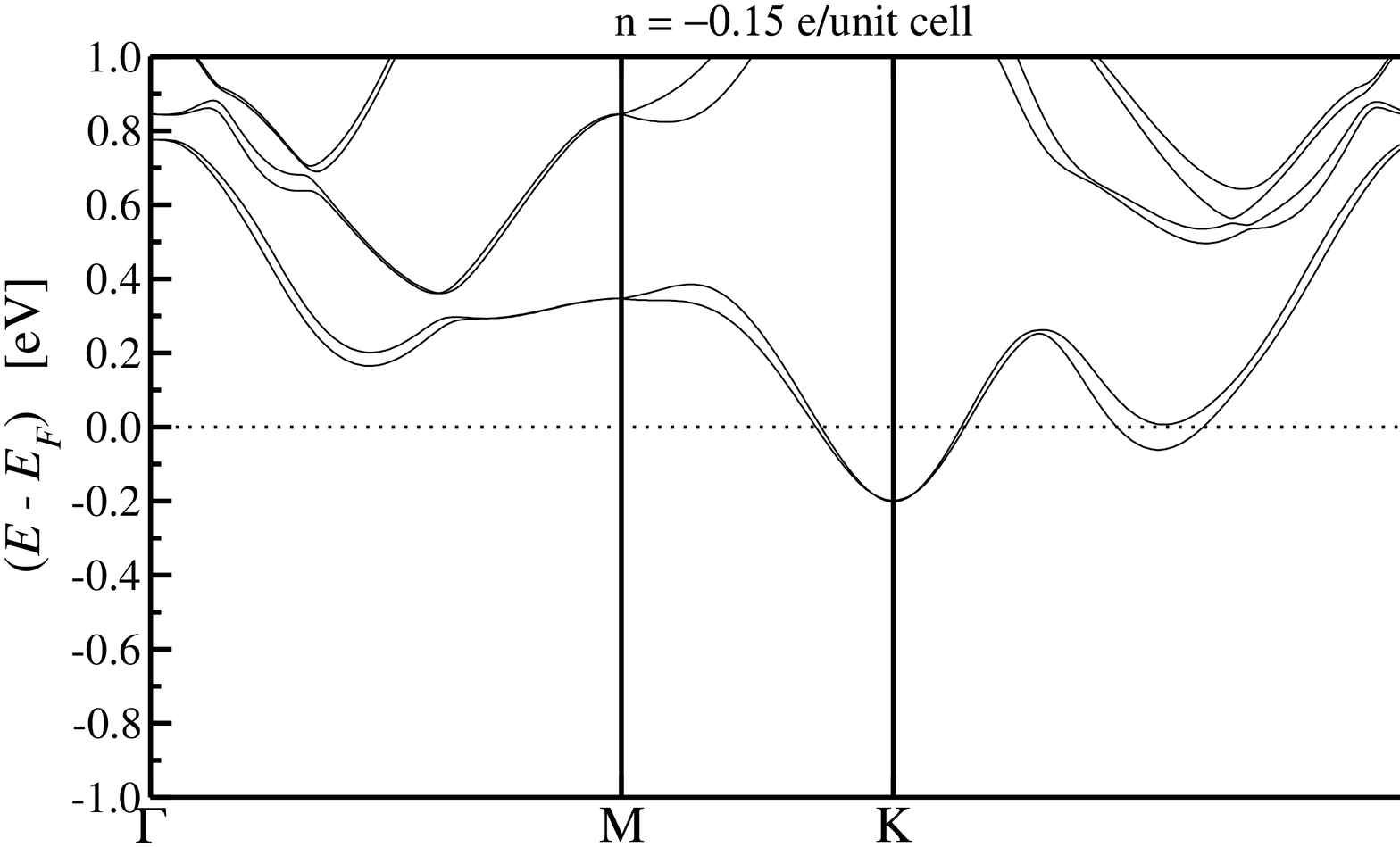}
 \includegraphics[width=0.31\textwidth,clip=,angle=-90]{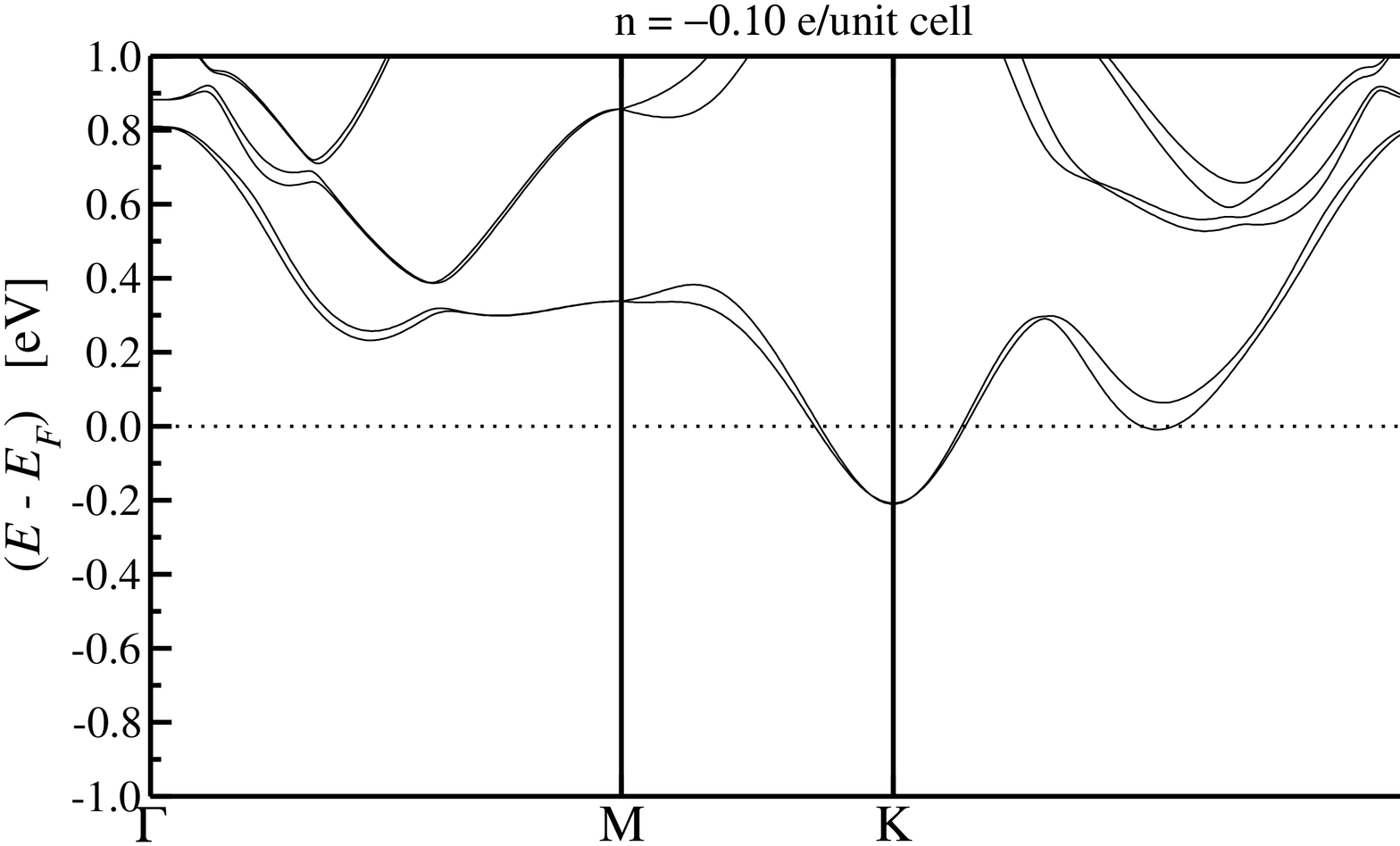}
 \includegraphics[width=0.31\textwidth,clip=,angle=-90]{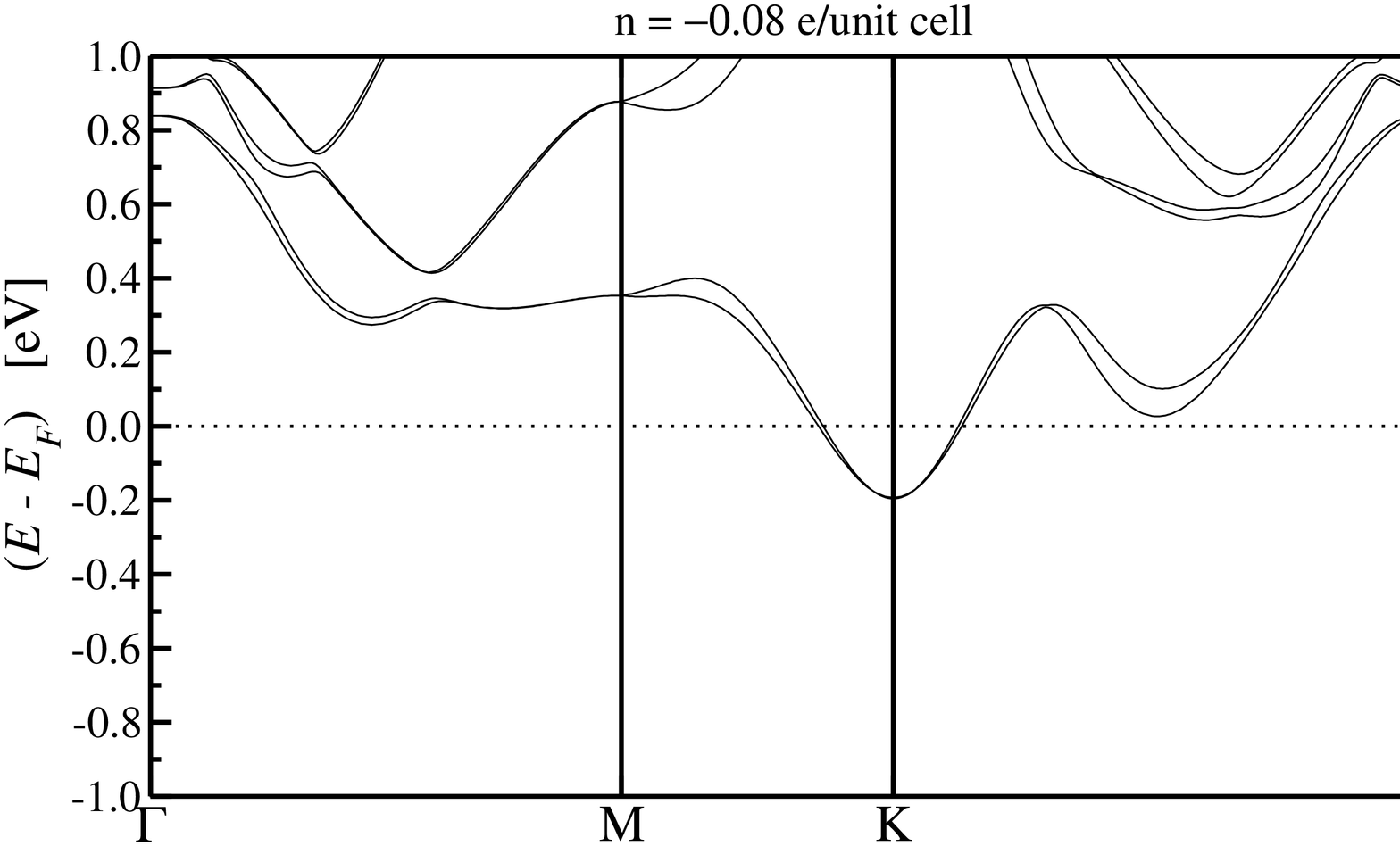}
 \includegraphics[width=0.31\textwidth,clip=,angle=-90]{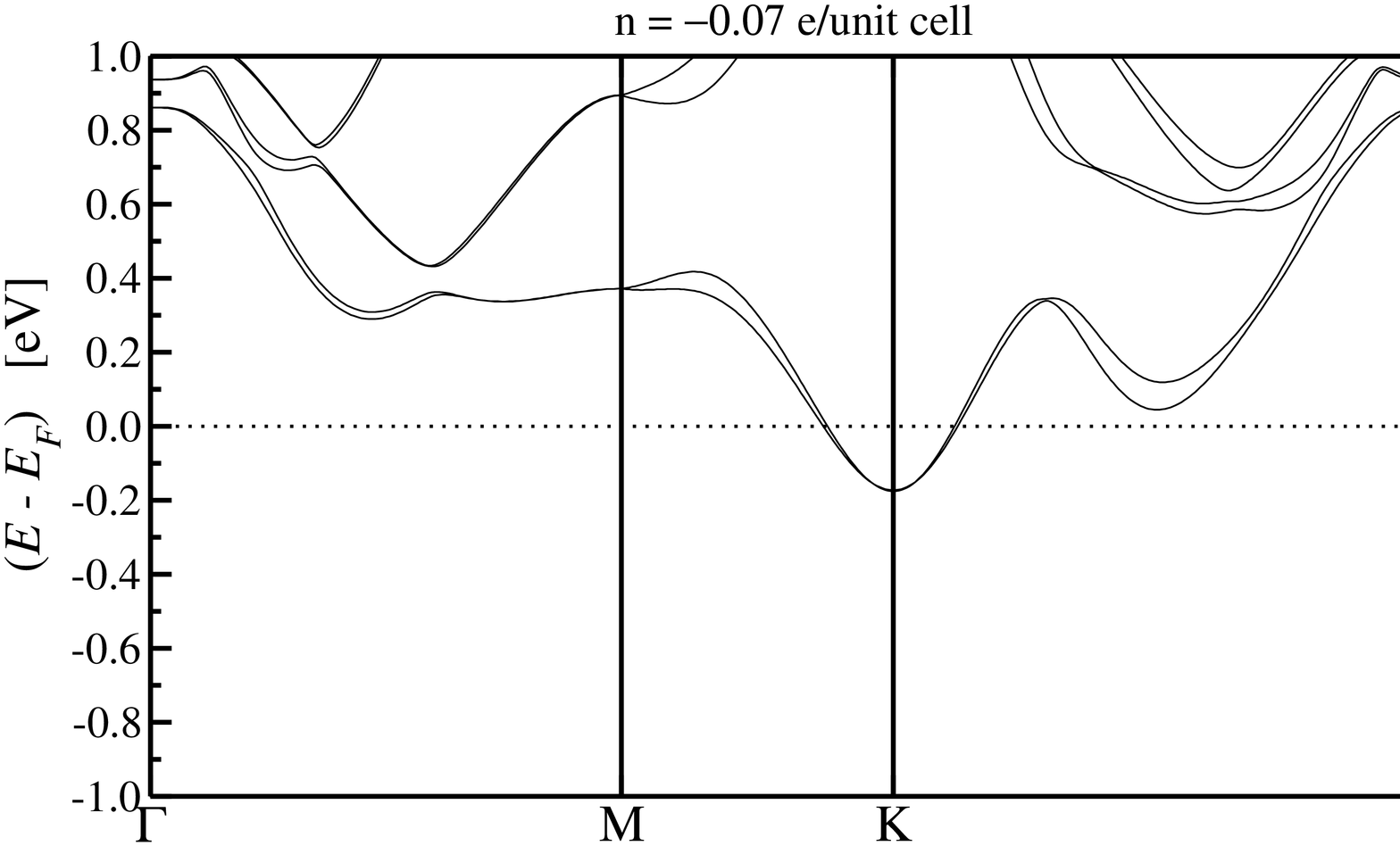}
 \includegraphics[width=0.31\textwidth,clip=,angle=-90]{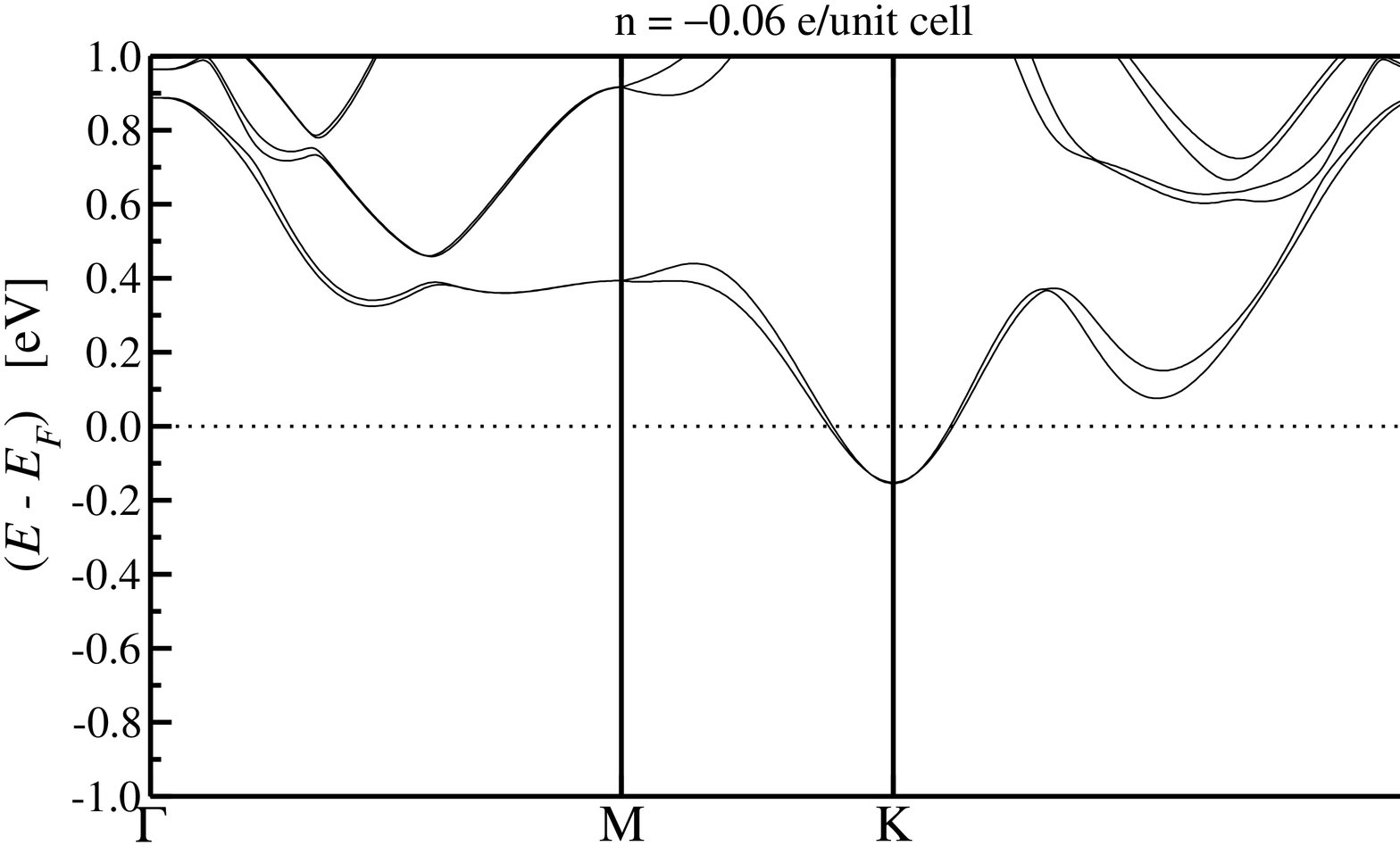}
 \includegraphics[width=0.31\textwidth,clip=,angle=-90]{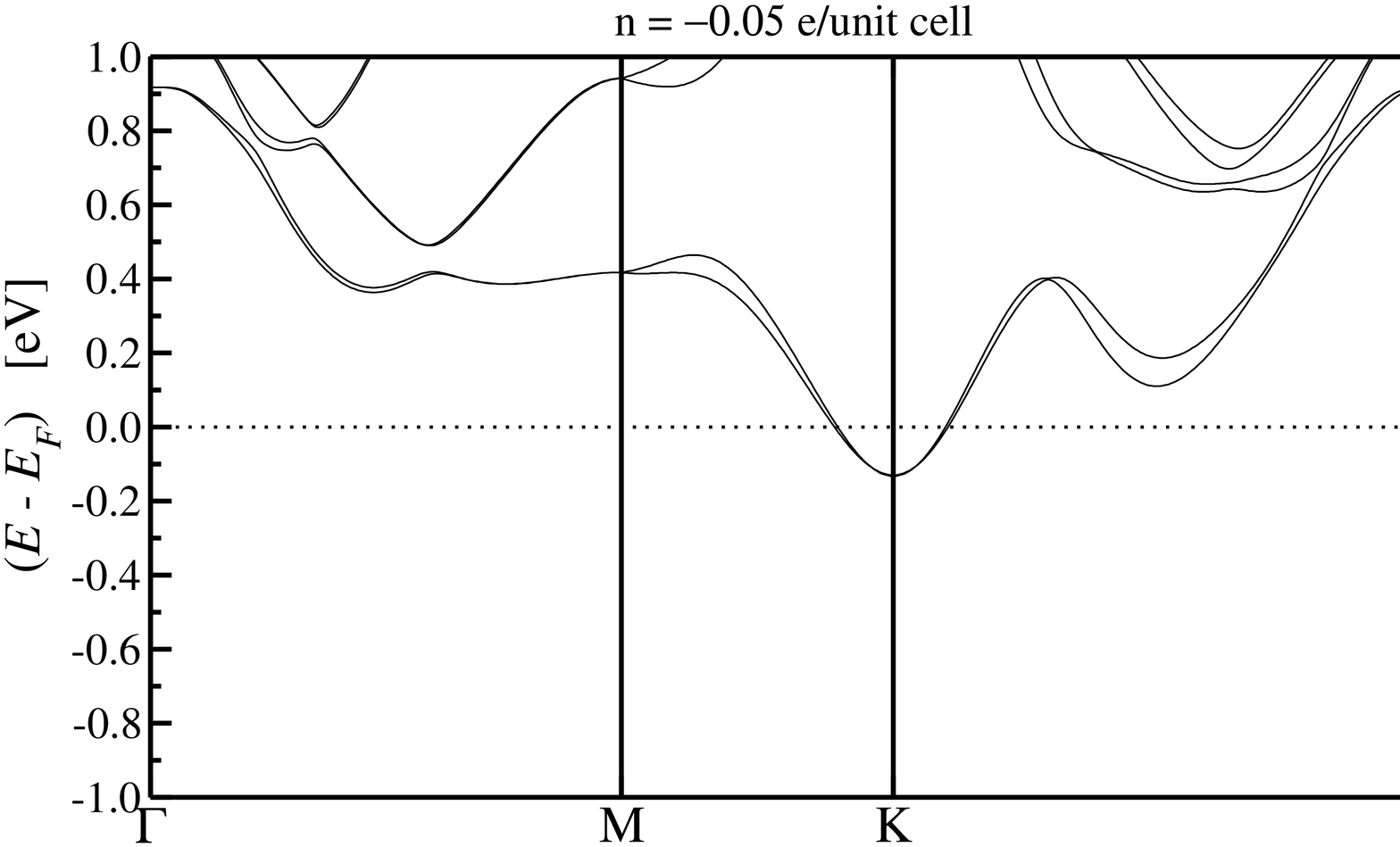}
 \caption{Band structure of monolayer MoS$_2$ for different doping as indicated in the labels.}
\end{figure*}
\begin{figure*}[hbp]
 \centering
 \includegraphics[width=0.31\textwidth,clip=,angle=-90]{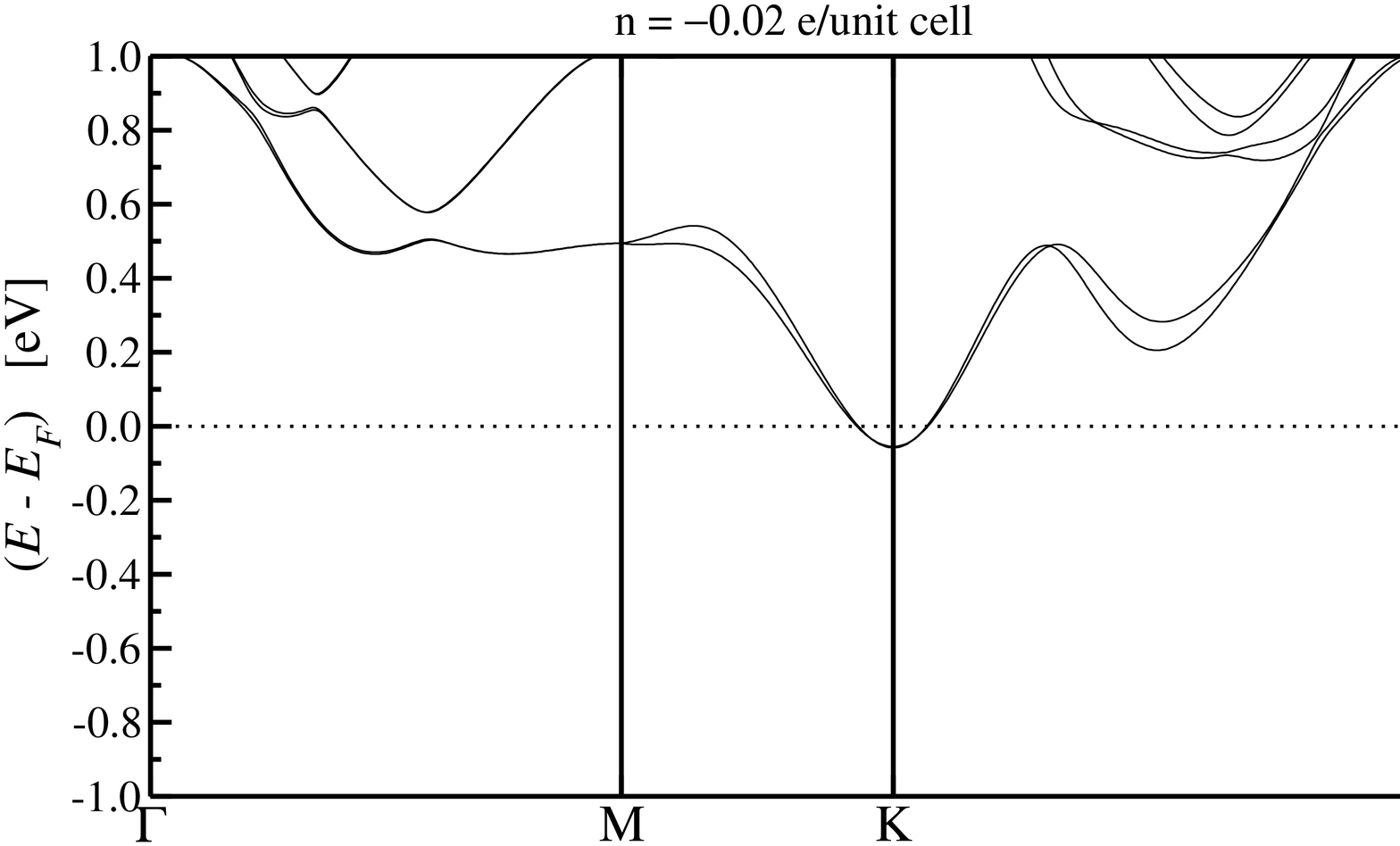}
 \includegraphics[width=0.31\textwidth,clip=,angle=-90]{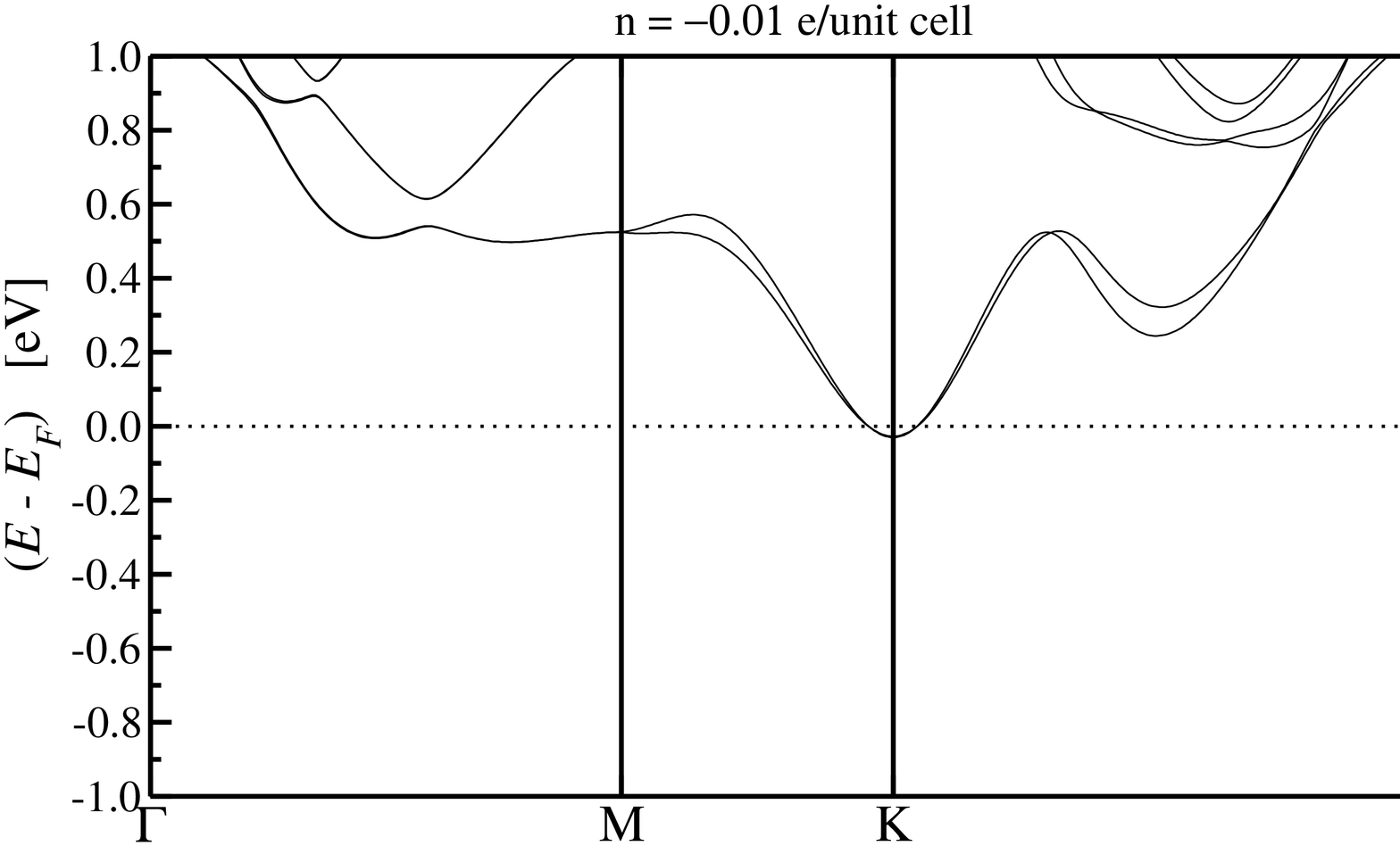}
 \includegraphics[width=0.31\textwidth,clip=,angle=-90]{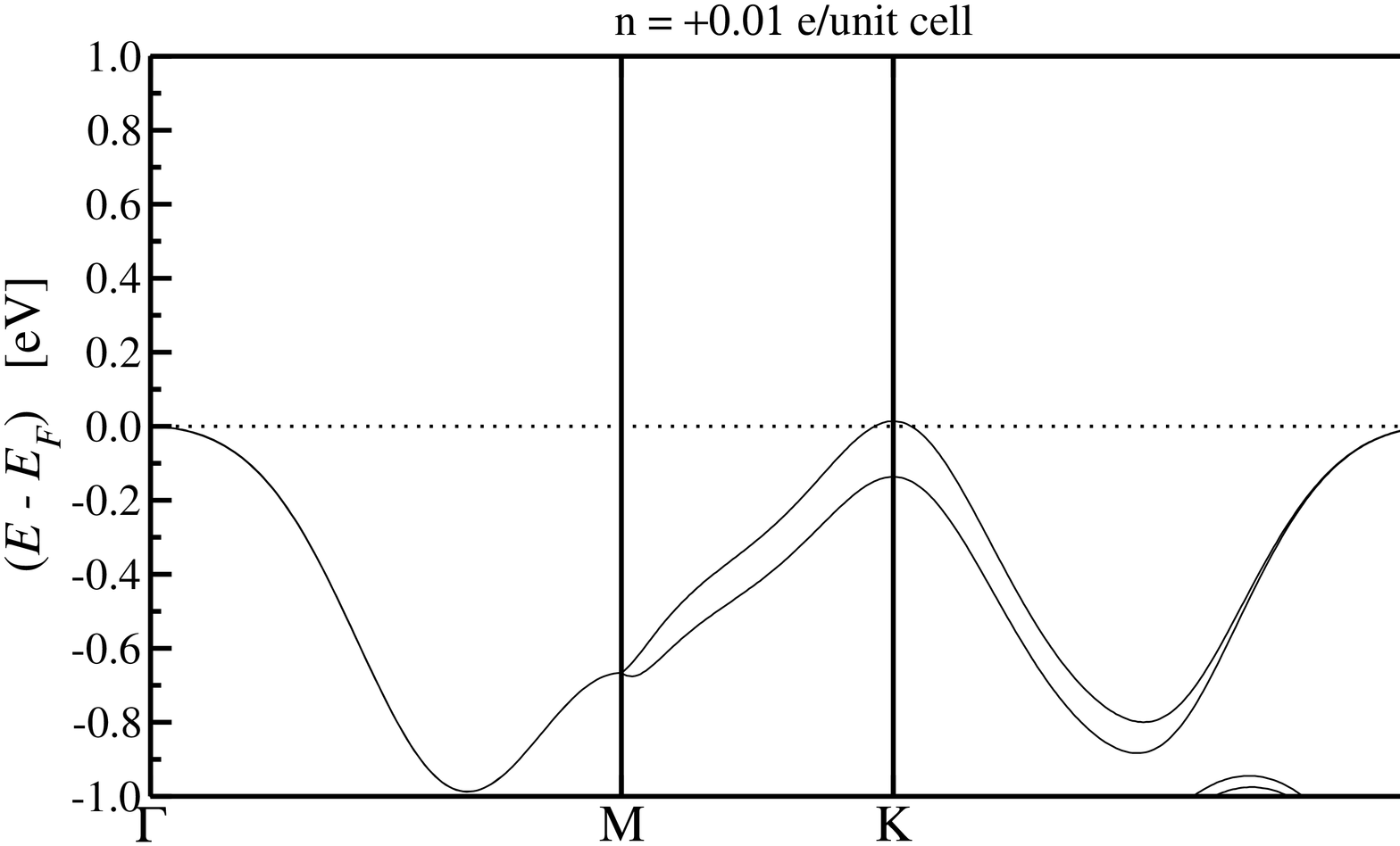}
 \includegraphics[width=0.31\textwidth,clip=,angle=-90]{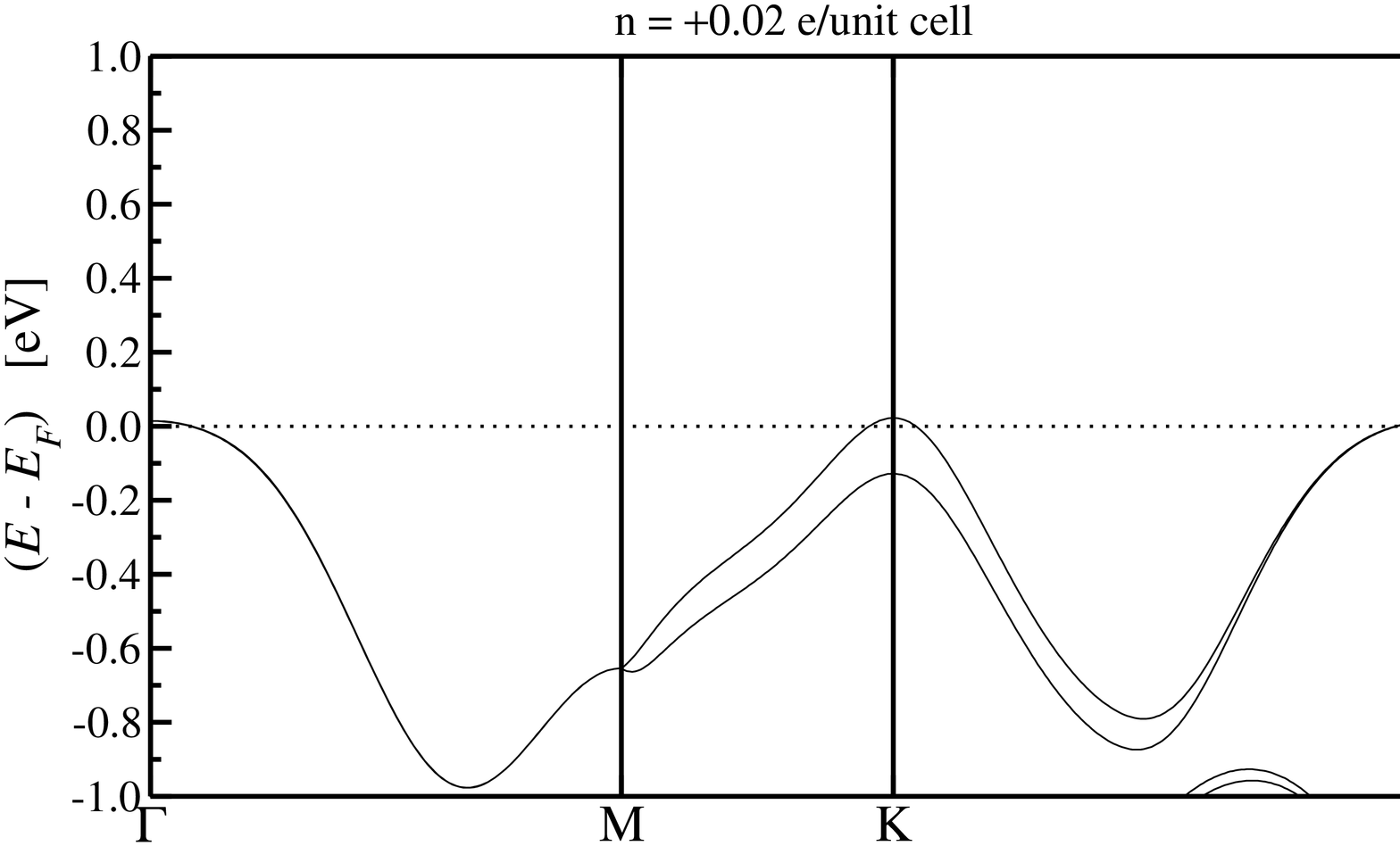}
 \includegraphics[width=0.31\textwidth,clip=,angle=-90]{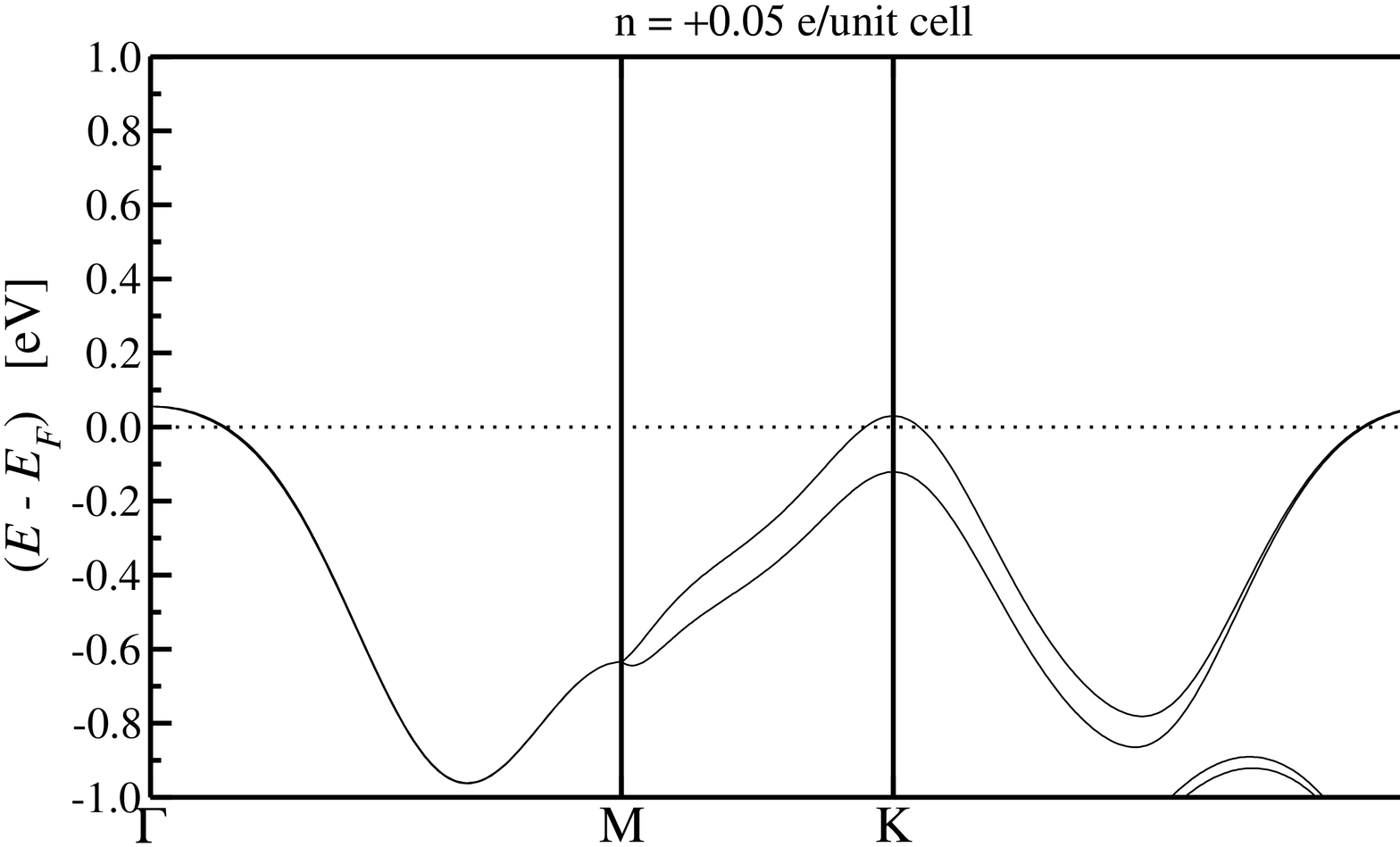}
 \includegraphics[width=0.31\textwidth,clip=,angle=-90]{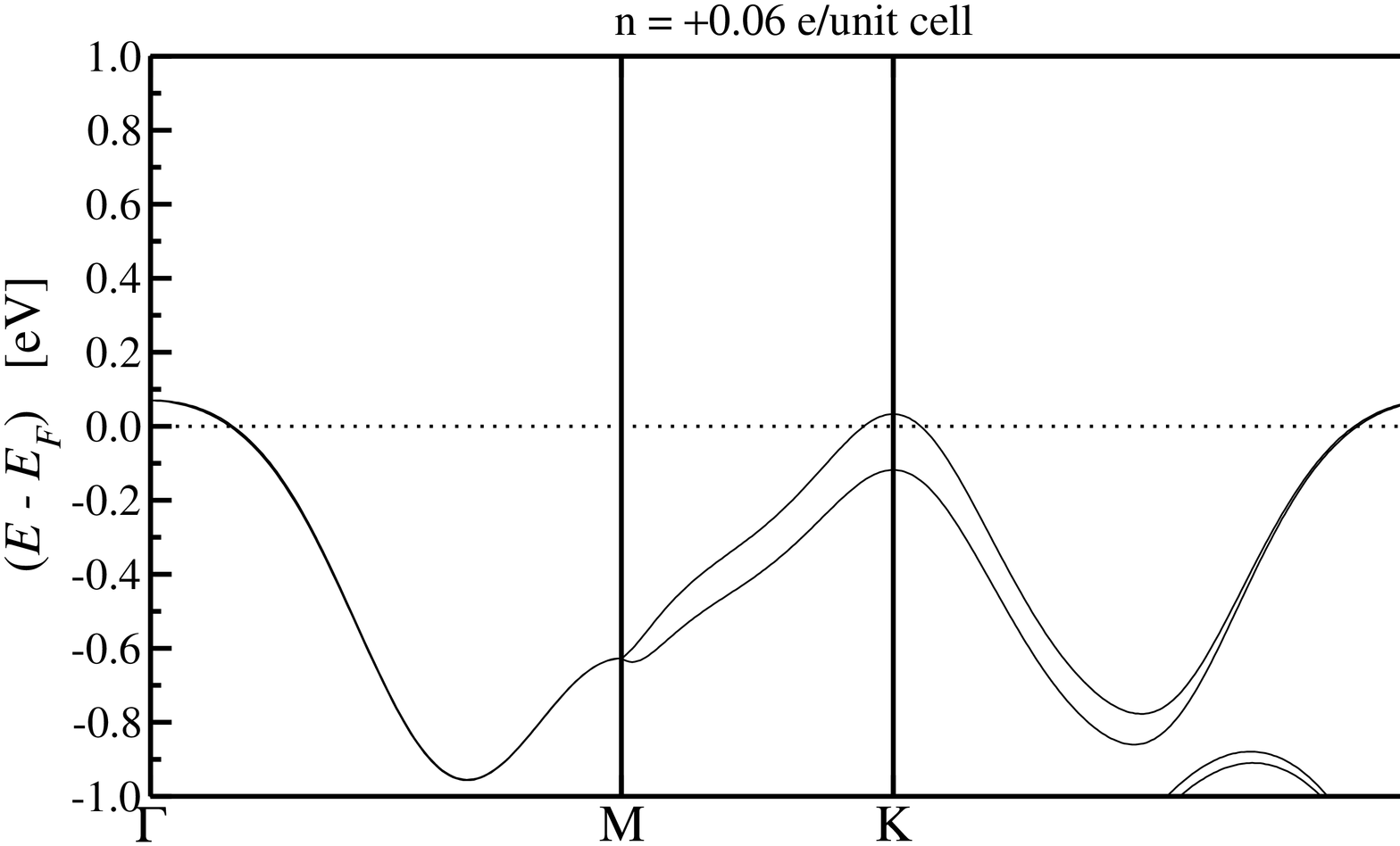}
 \includegraphics[width=0.31\textwidth,clip=,angle=-90]{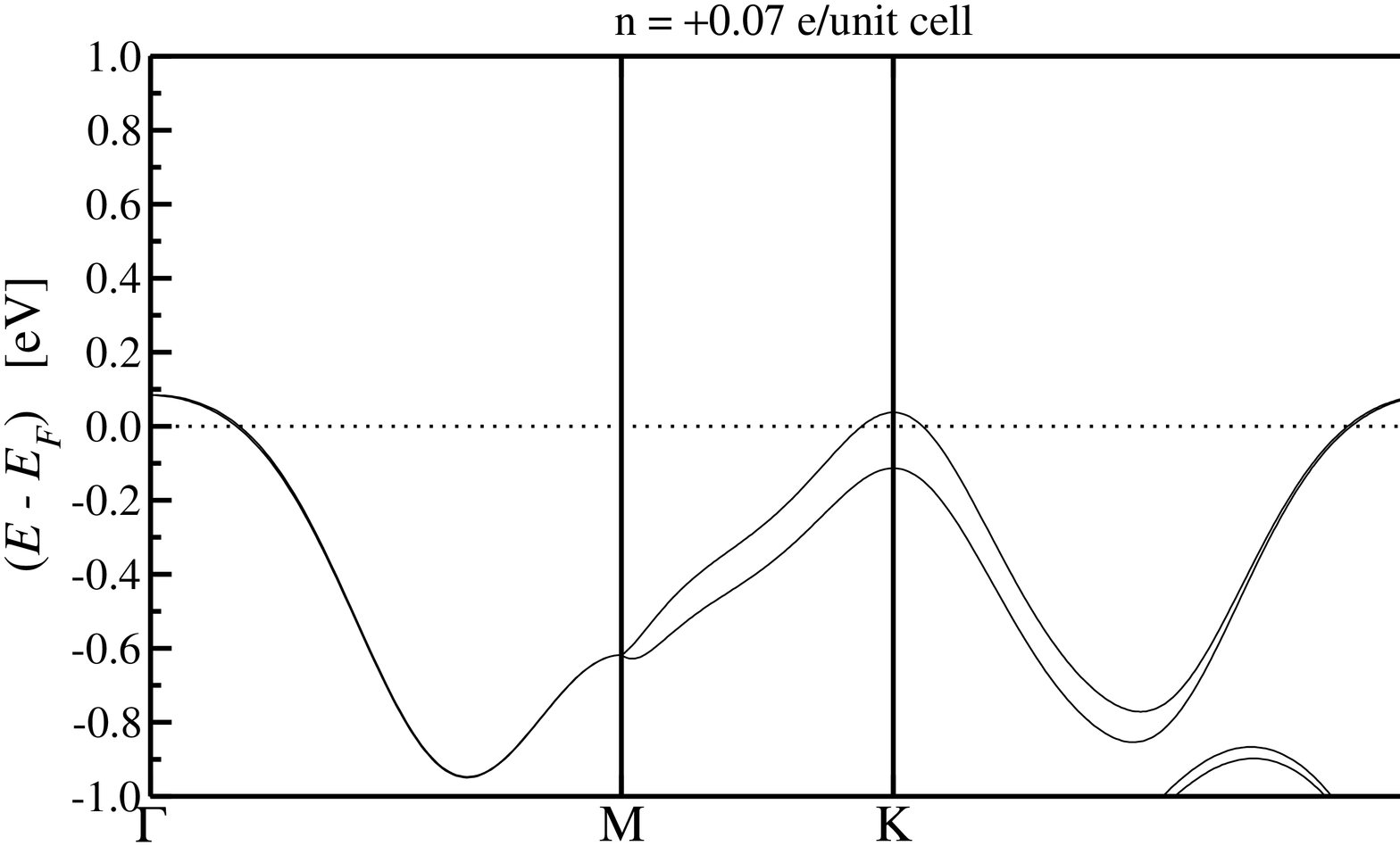}
 \includegraphics[width=0.31\textwidth,clip=,angle=-90]{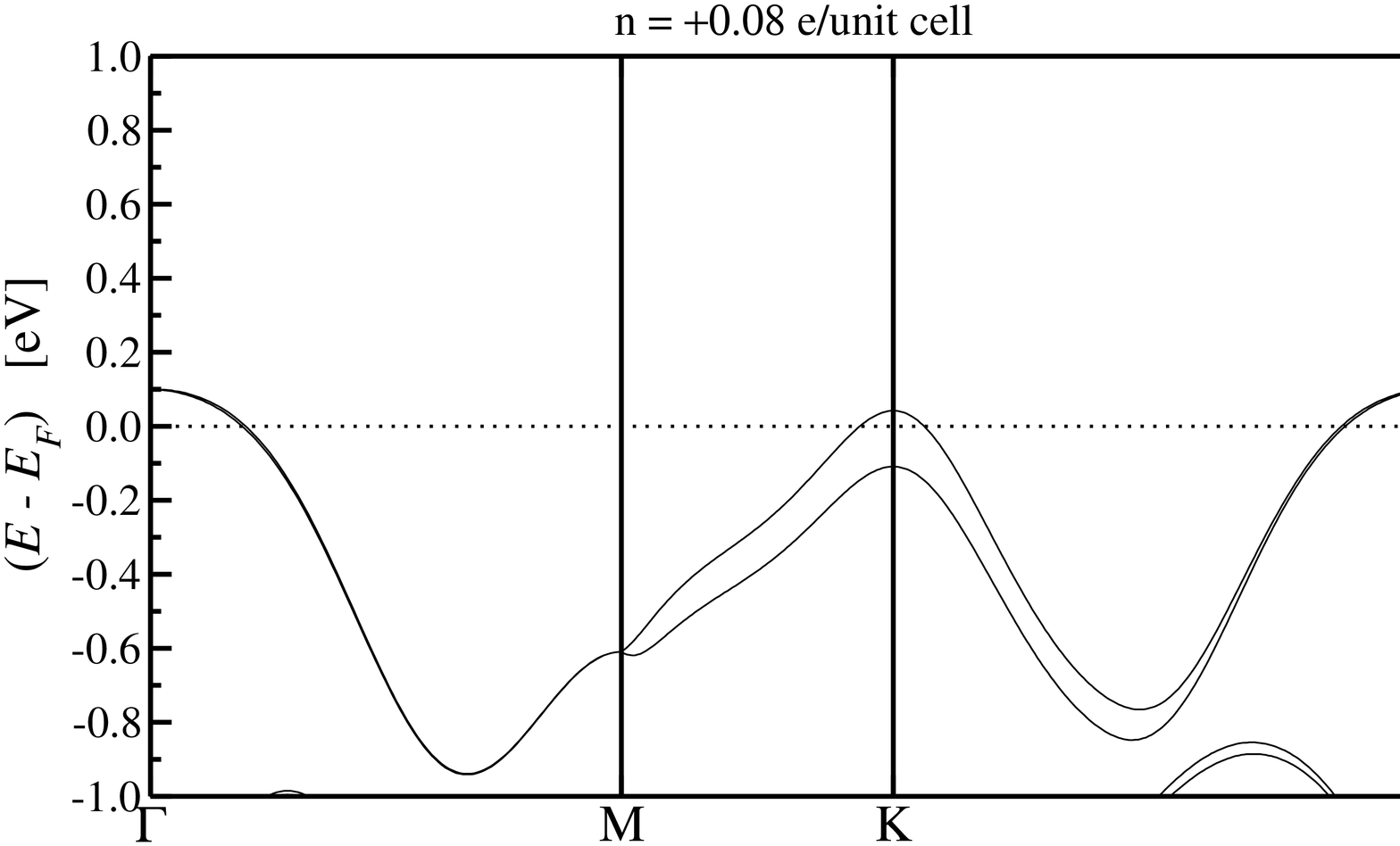}
 \caption{Band structure of monolayer MoS$_2$ for different doping as indicated in the labels.}
\end{figure*}
\begin{figure*}[hbp]
 \centering
 \includegraphics[width=0.31\textwidth,clip=,angle=-90]{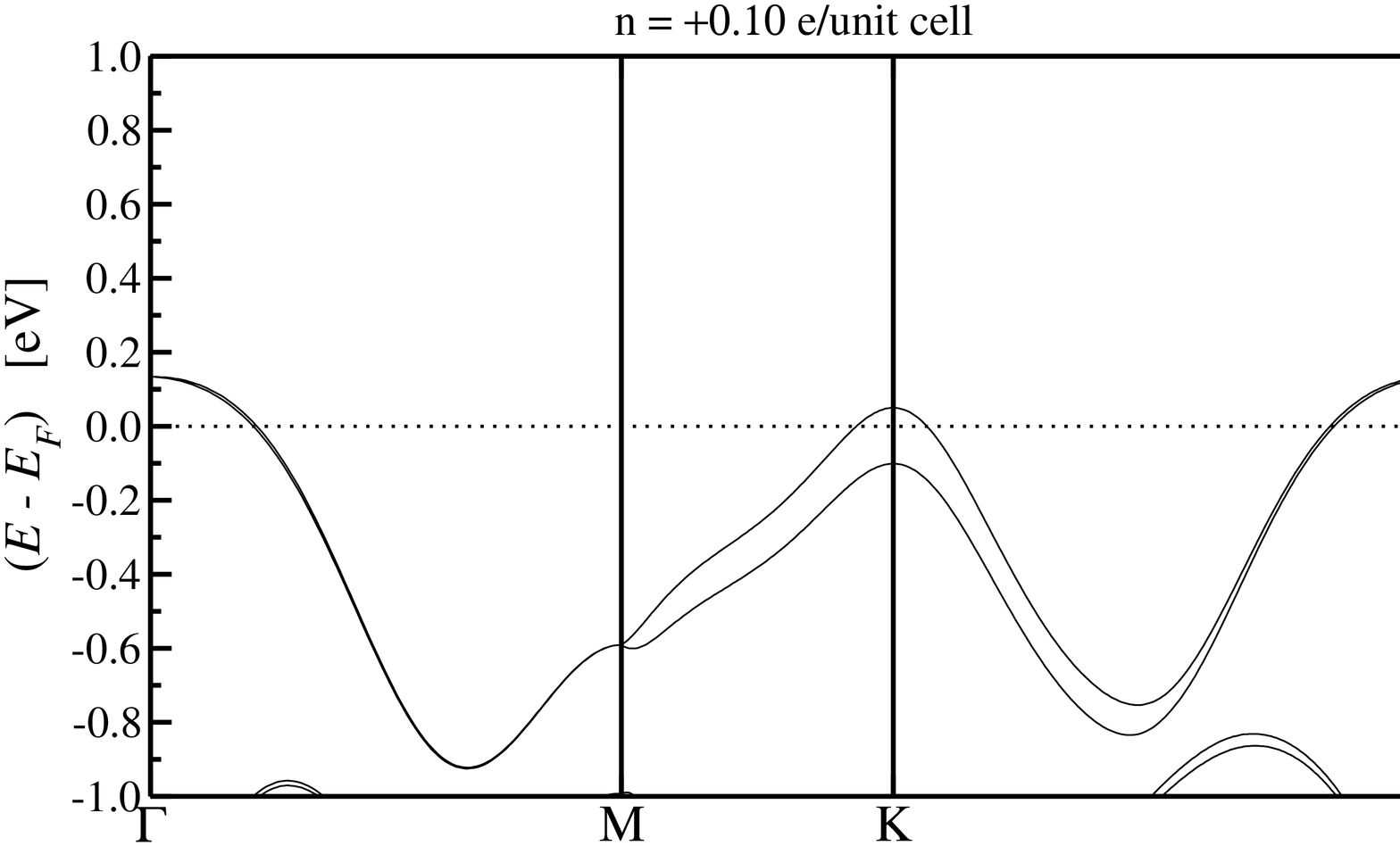}
 \includegraphics[width=0.31\textwidth,clip=,angle=-90]{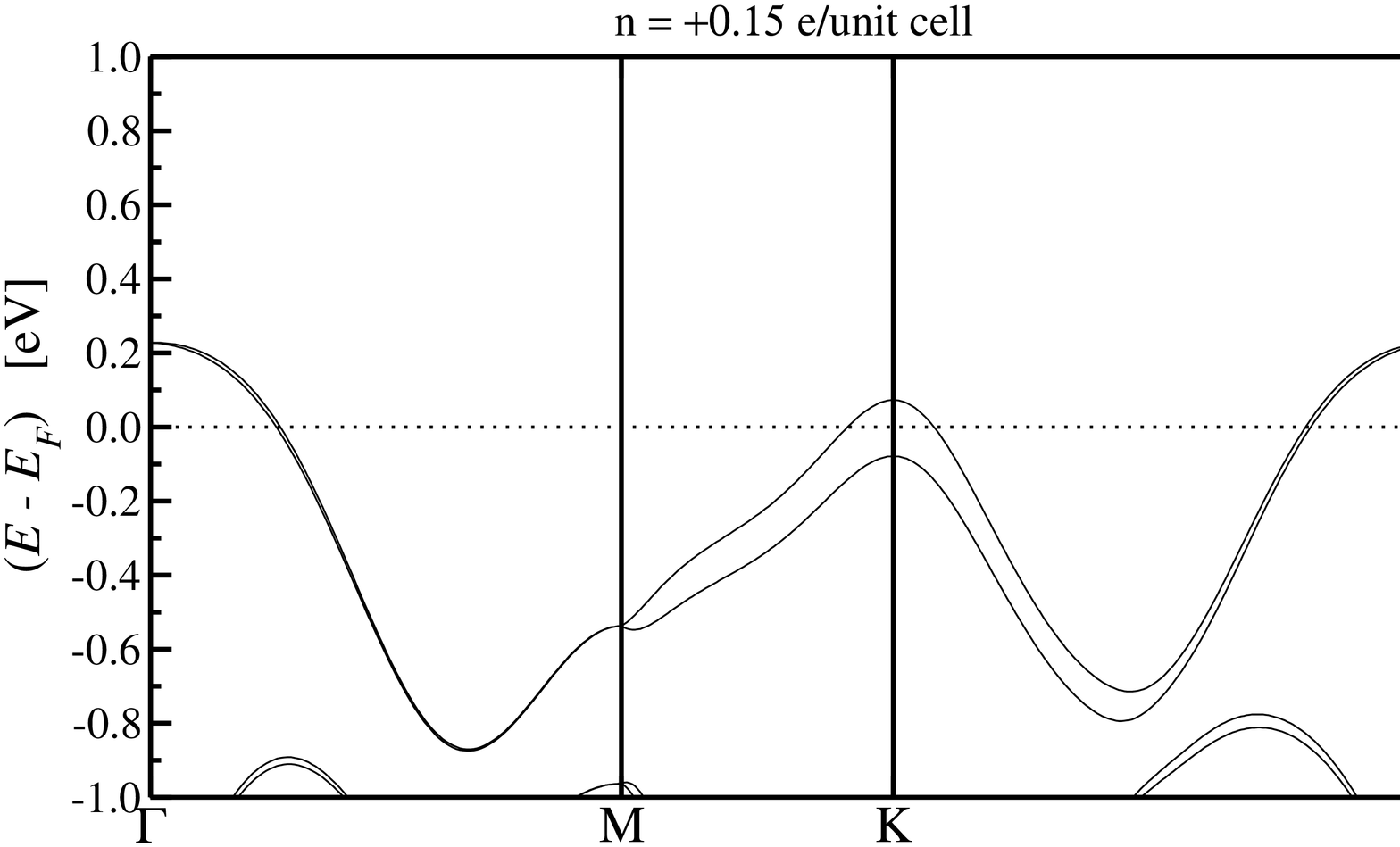}
 \includegraphics[width=0.31\textwidth,clip=,angle=-90]{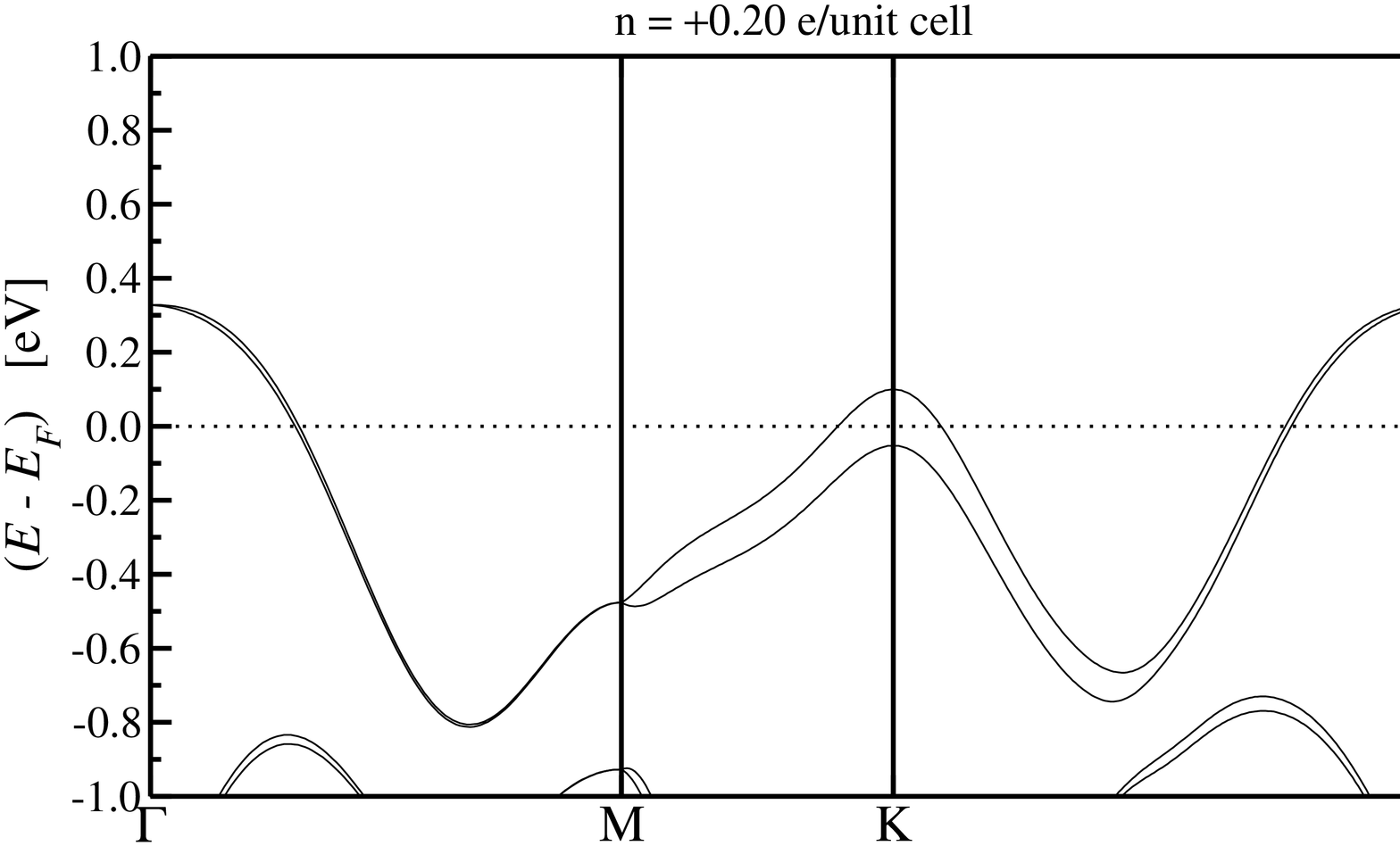}
 \includegraphics[width=0.31\textwidth,clip=,angle=-90]{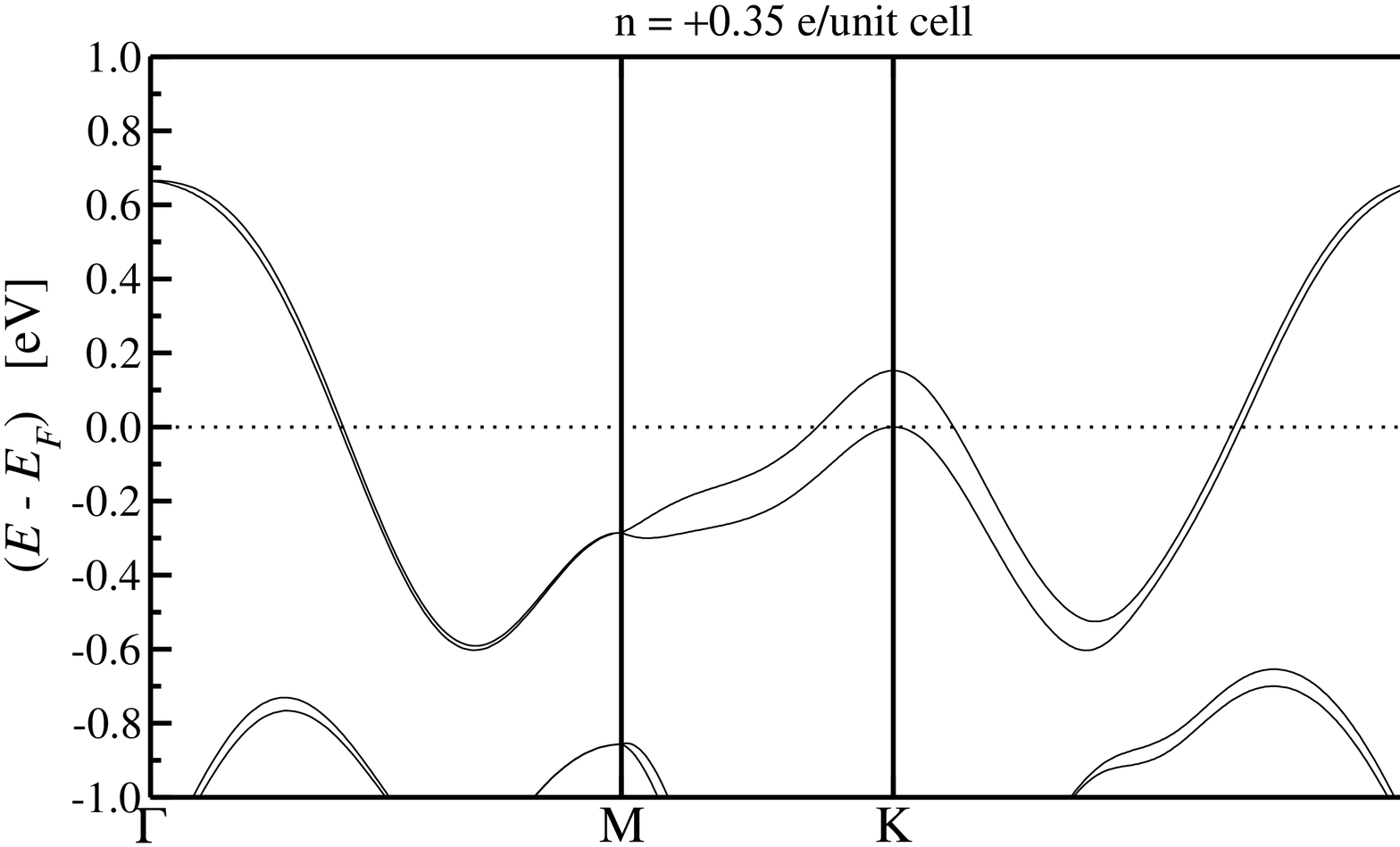}
 \caption{Band structure of monolayer MoS$_2$ for different doping as indicated in the labels.}
\end{figure*}
\begin{figure*}[hbp]
 \centering
 \includegraphics[width=0.31\textwidth,clip=,angle=-90]{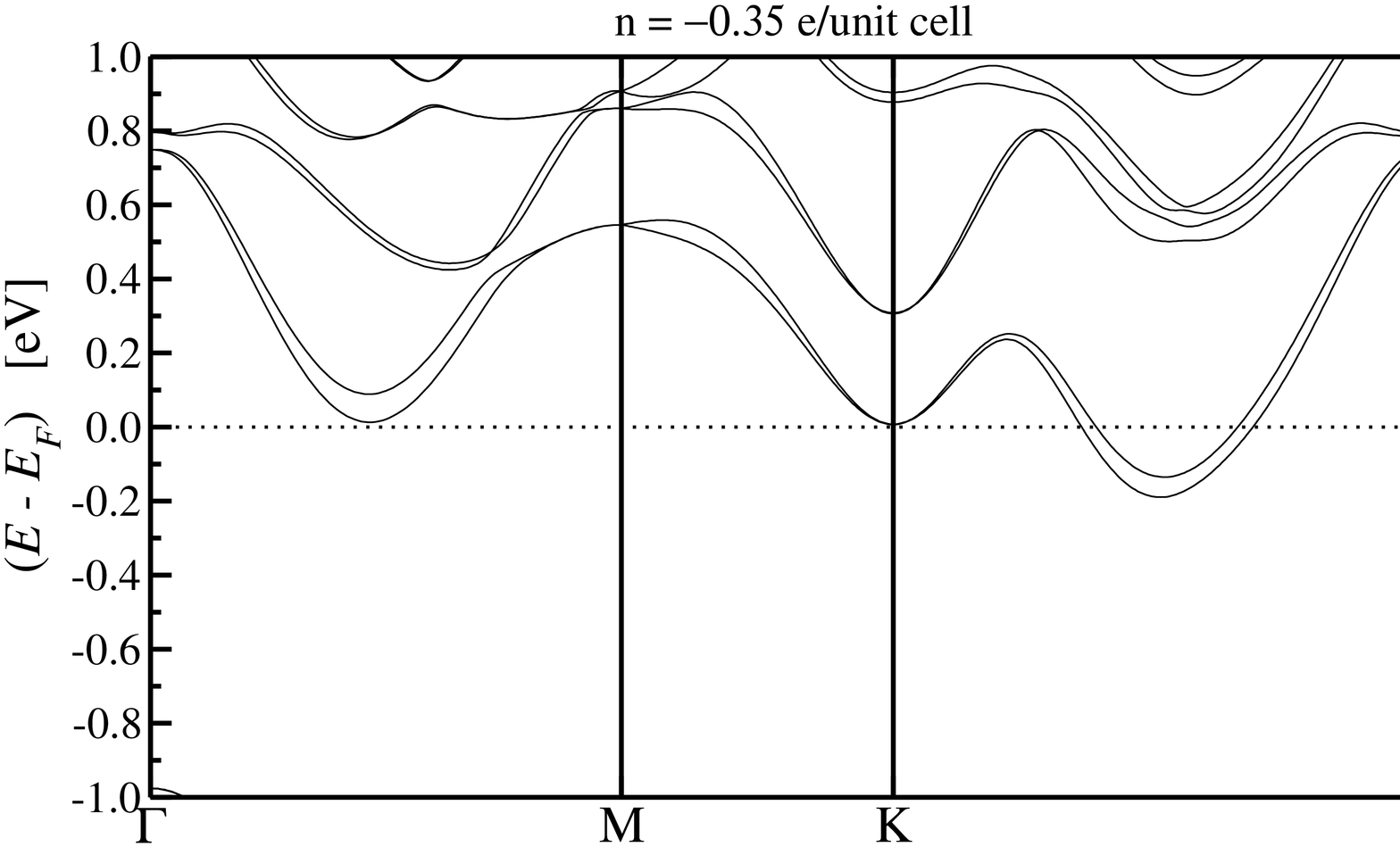}
 \includegraphics[width=0.31\textwidth,clip=,angle=-90]{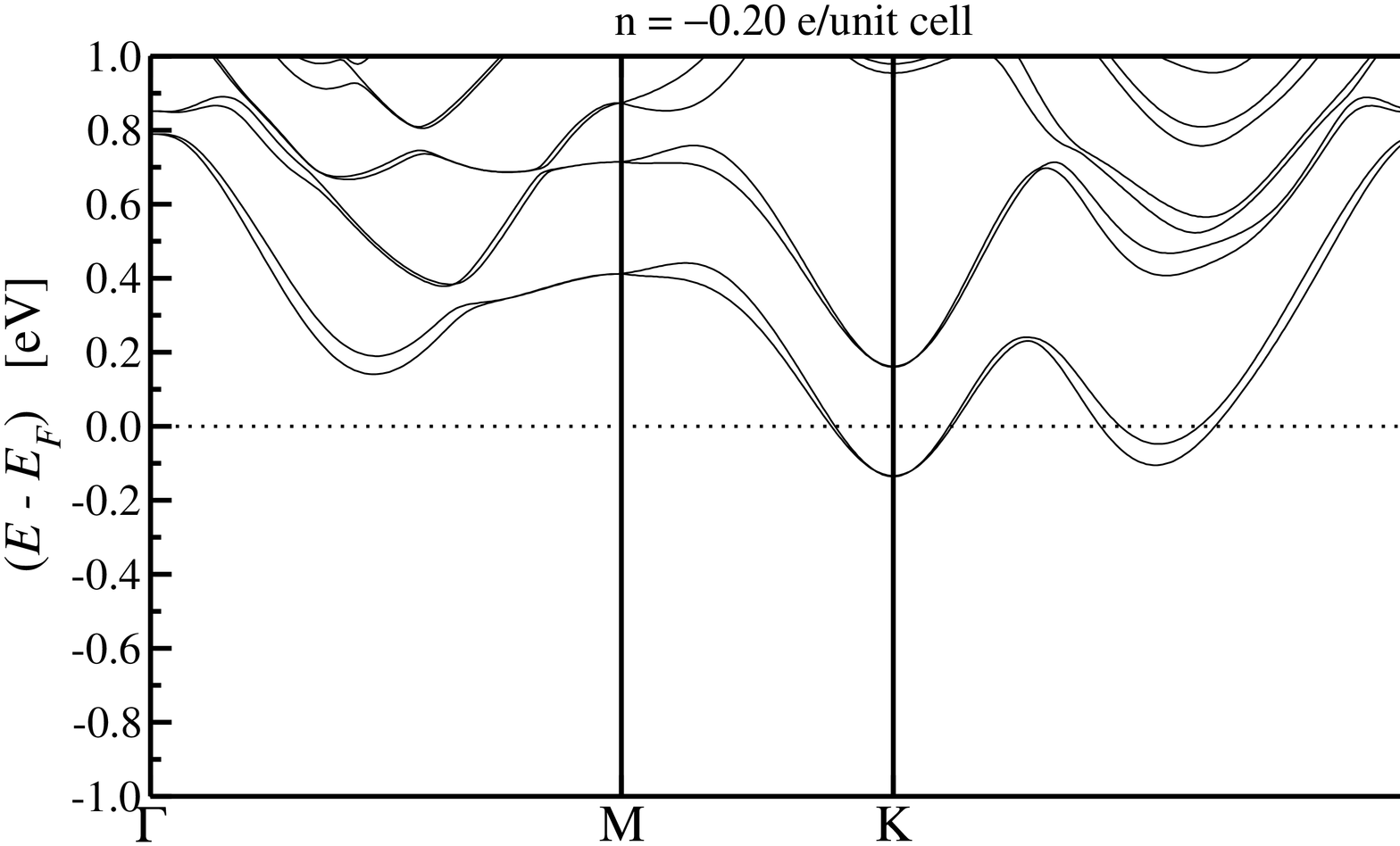}
 \includegraphics[width=0.31\textwidth,clip=,angle=-90]{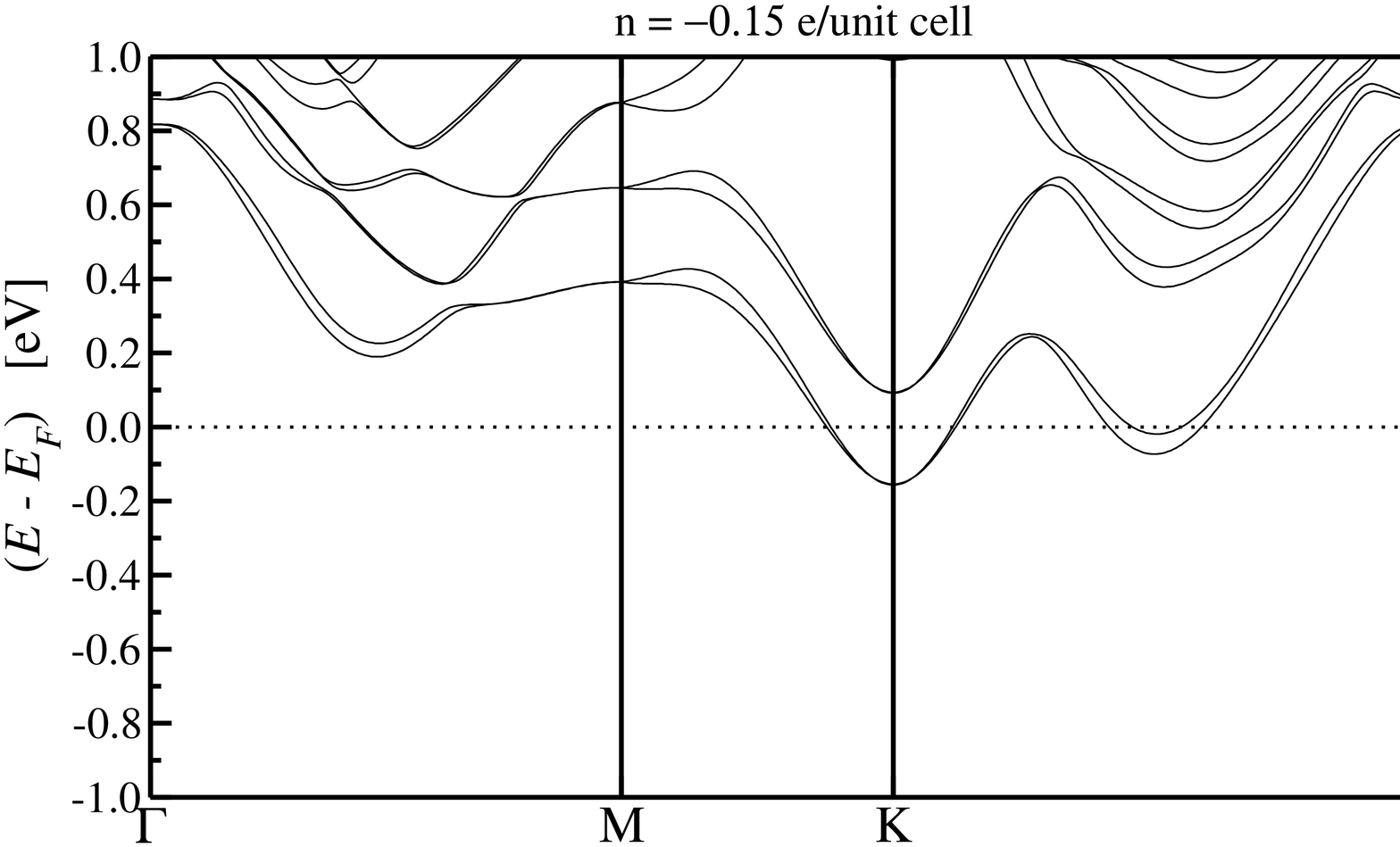}
 \includegraphics[width=0.31\textwidth,clip=,angle=-90]{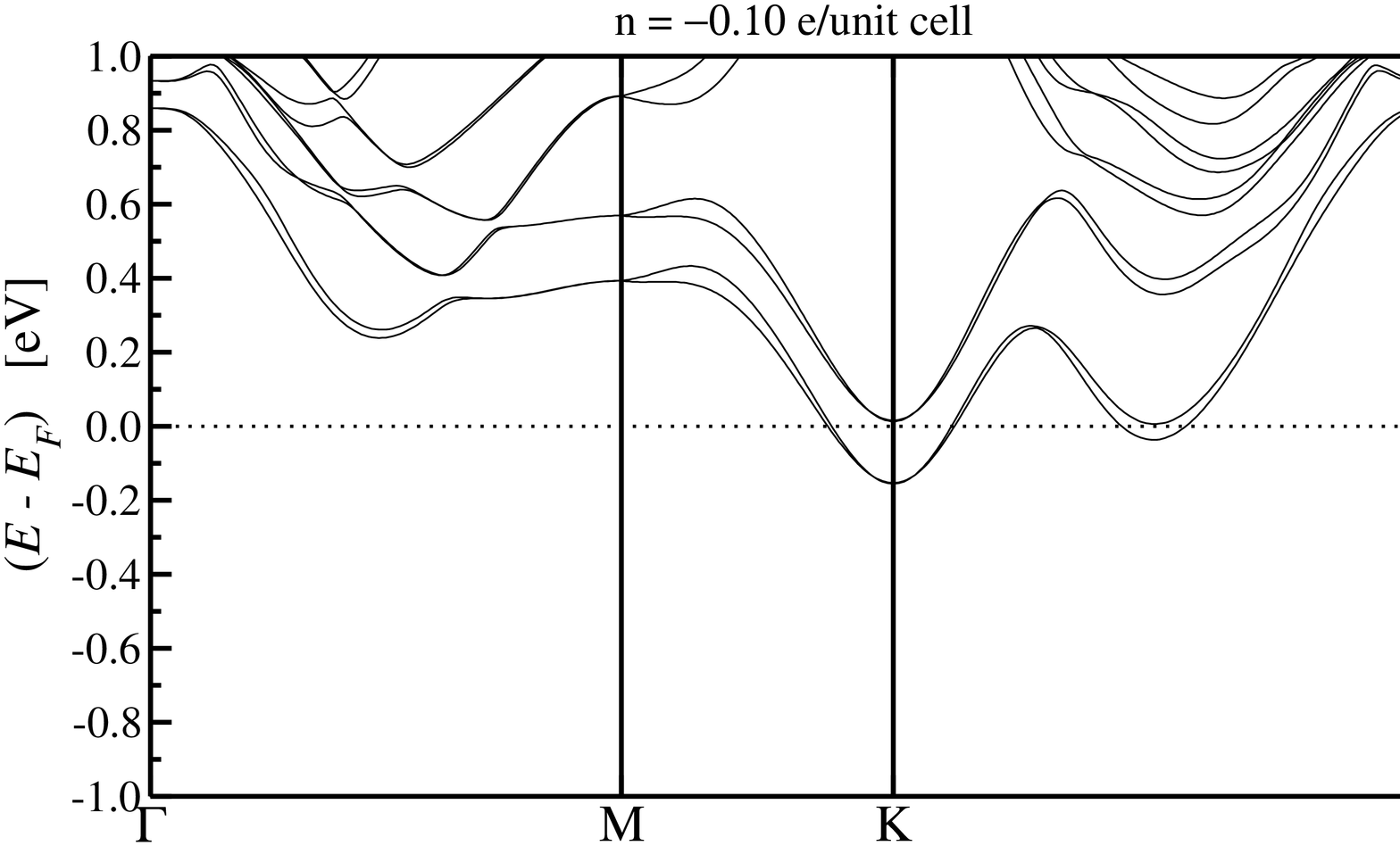}
 \includegraphics[width=0.31\textwidth,clip=,angle=-90]{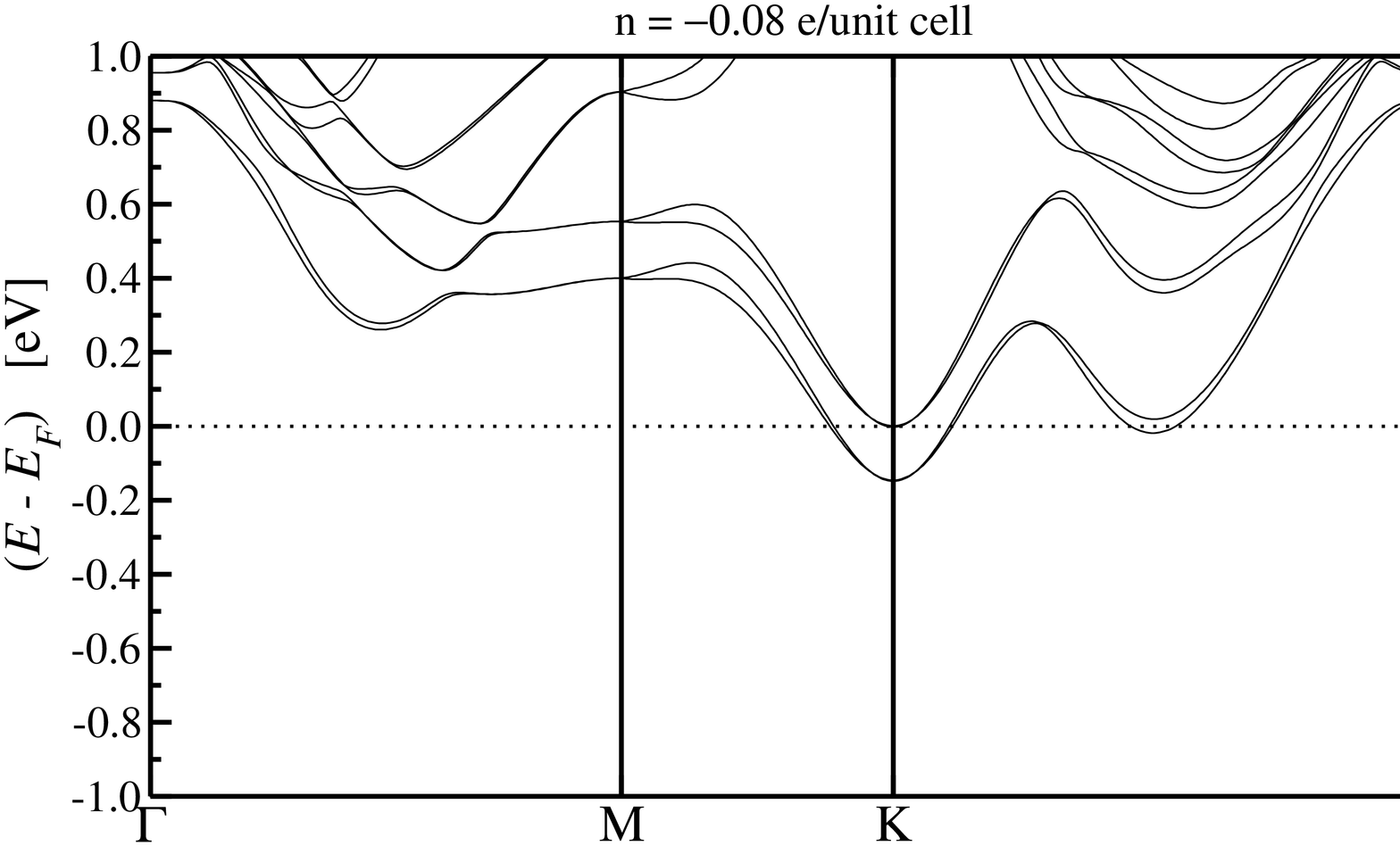}
 \includegraphics[width=0.31\textwidth,clip=,angle=-90]{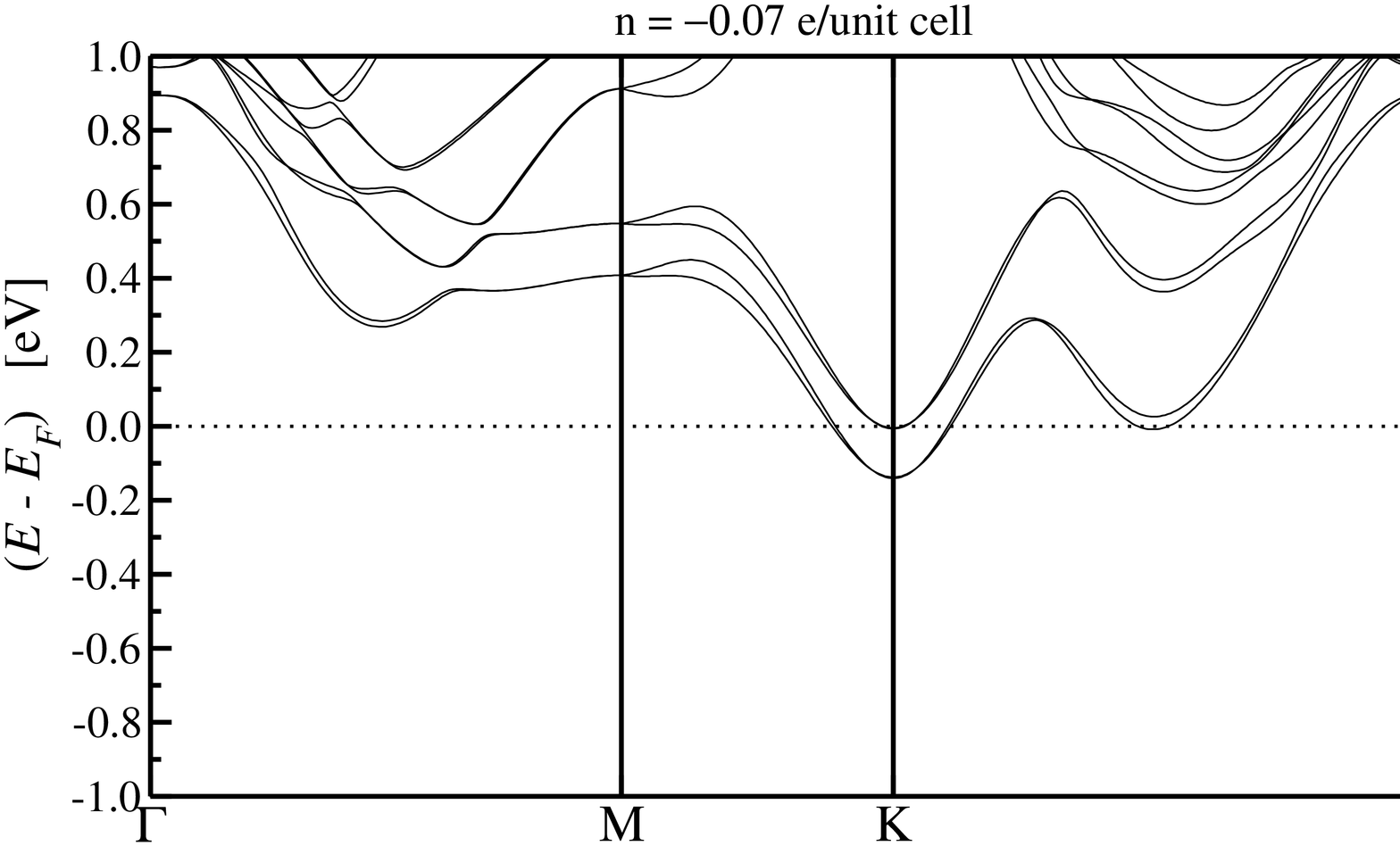}
 \includegraphics[width=0.31\textwidth,clip=,angle=-90]{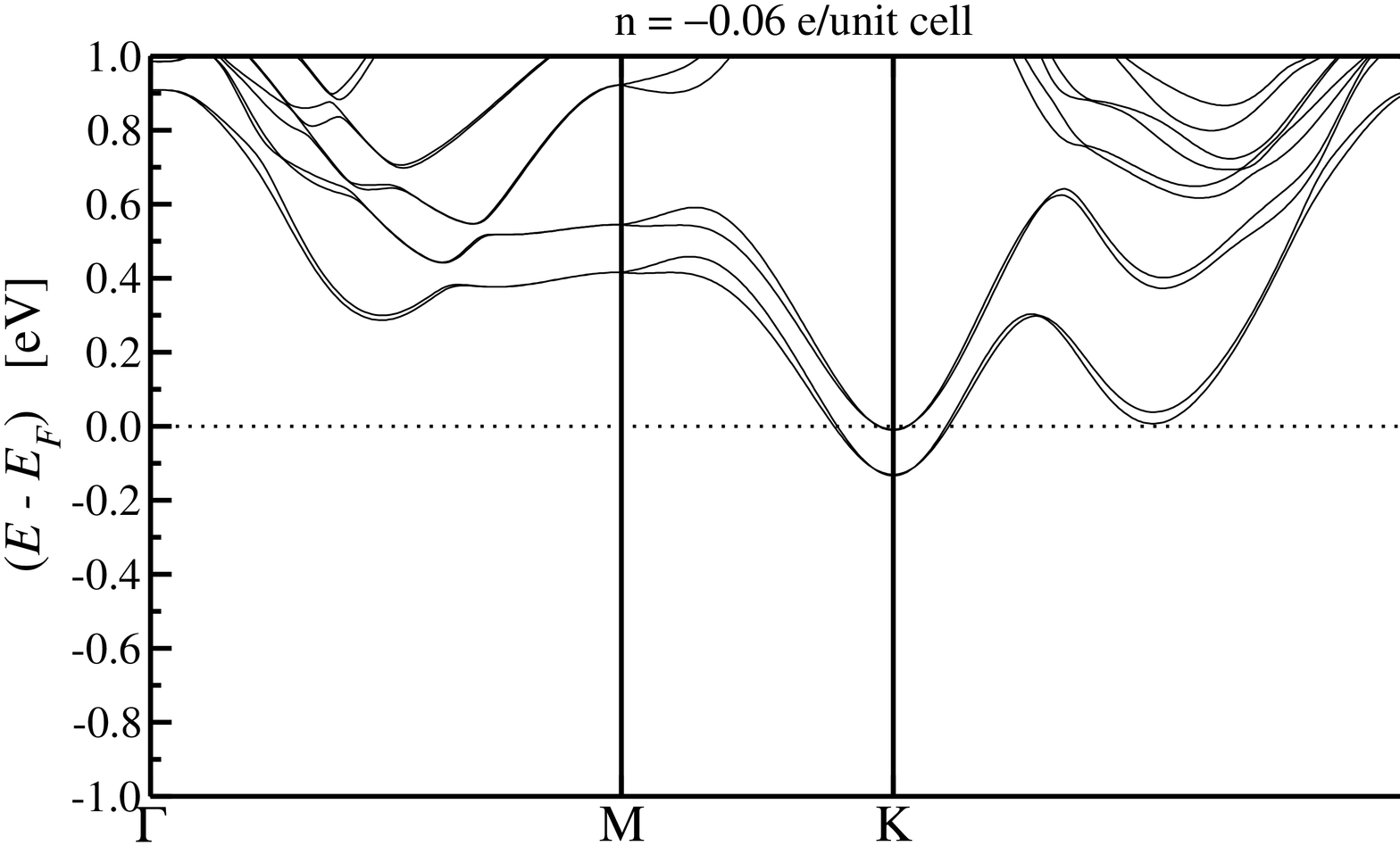}
 \includegraphics[width=0.31\textwidth,clip=,angle=-90]{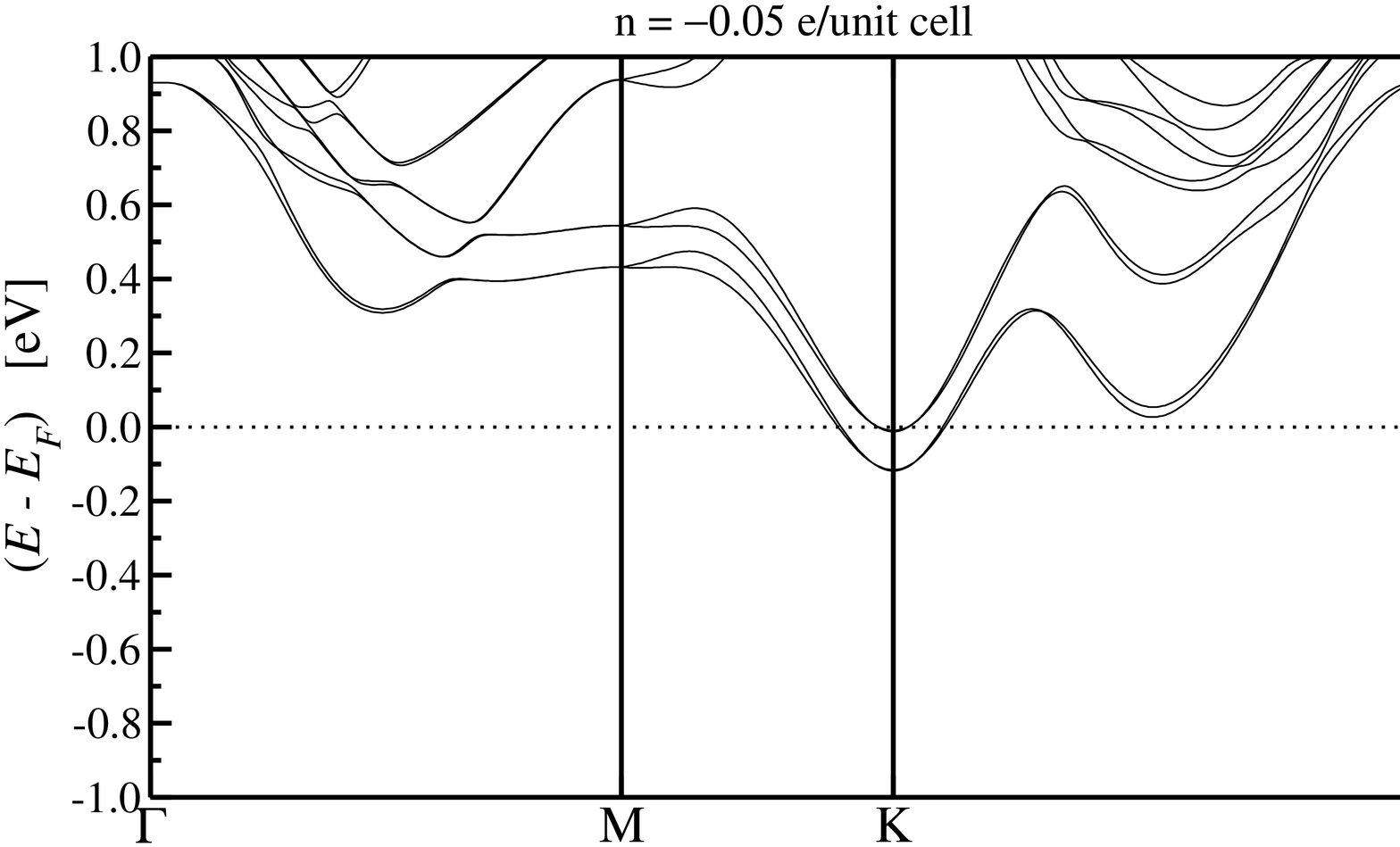}
 \caption{Band structure of bilayer MoS$_2$ for different doping as indicated in the labels.}
\end{figure*}
\begin{figure*}[hbp]
 \centering
 \includegraphics[width=0.31\textwidth,clip=,angle=-90]{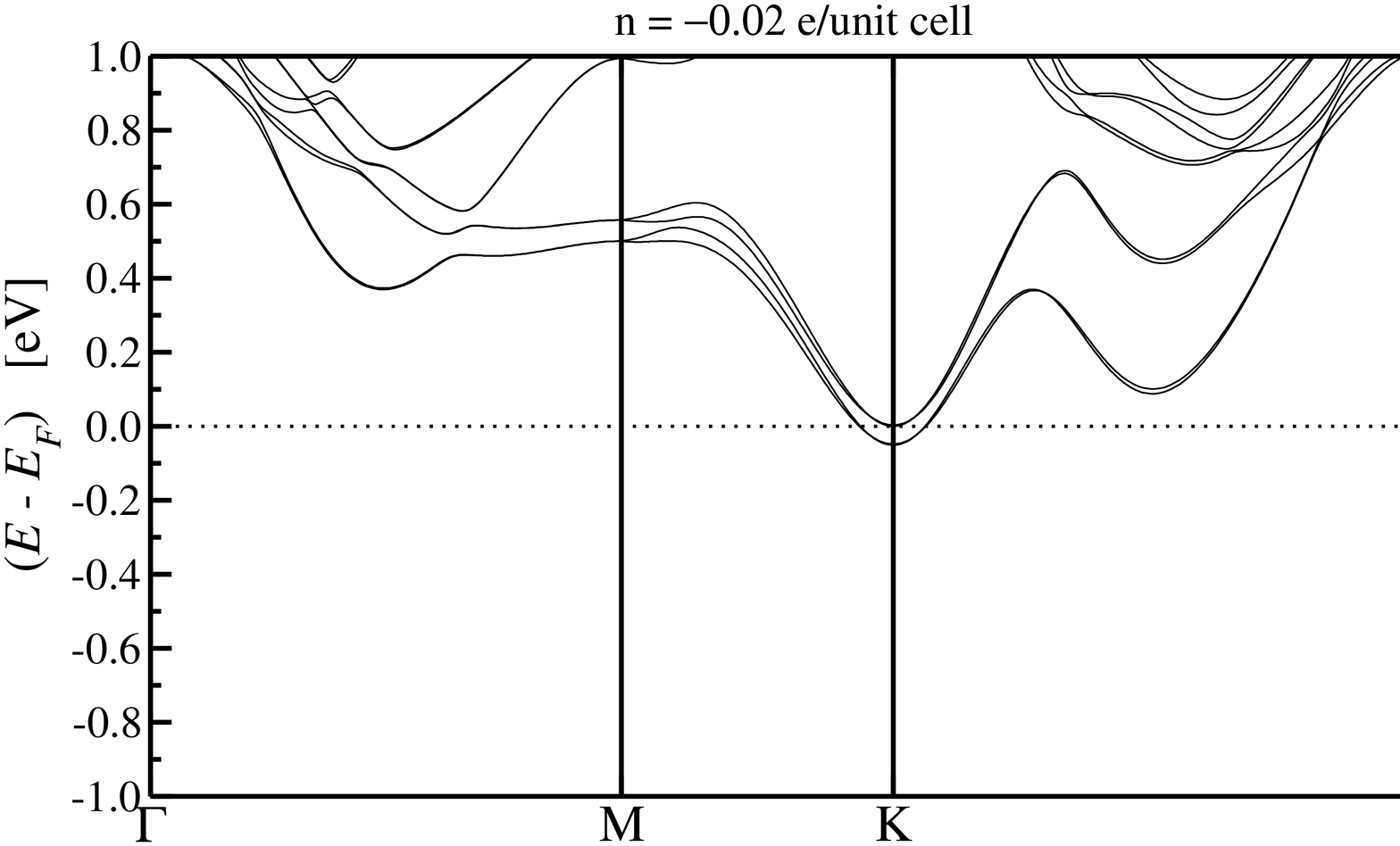}
 \includegraphics[width=0.31\textwidth,clip=,angle=-90]{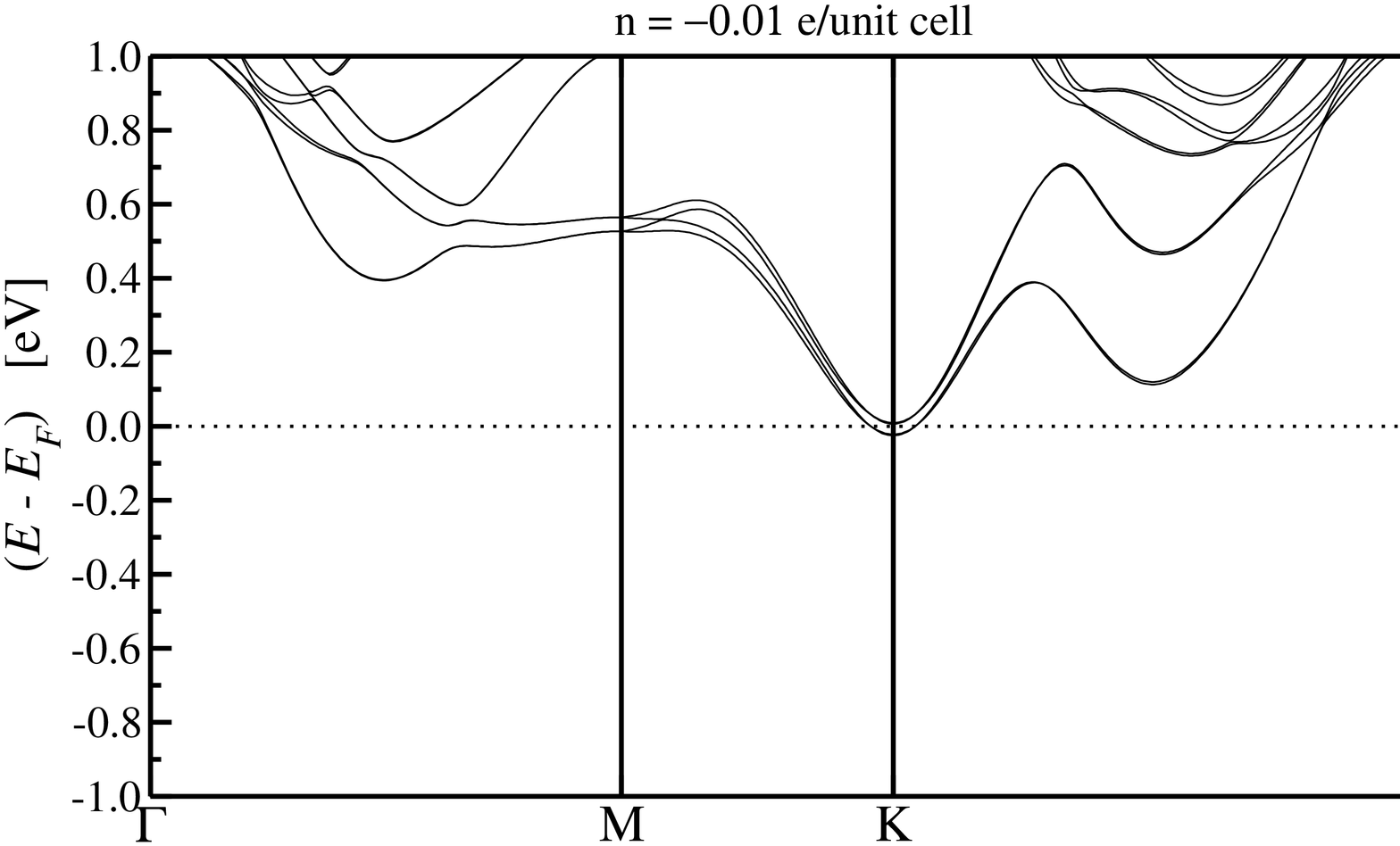}
 \includegraphics[width=0.31\textwidth,clip=,angle=-90]{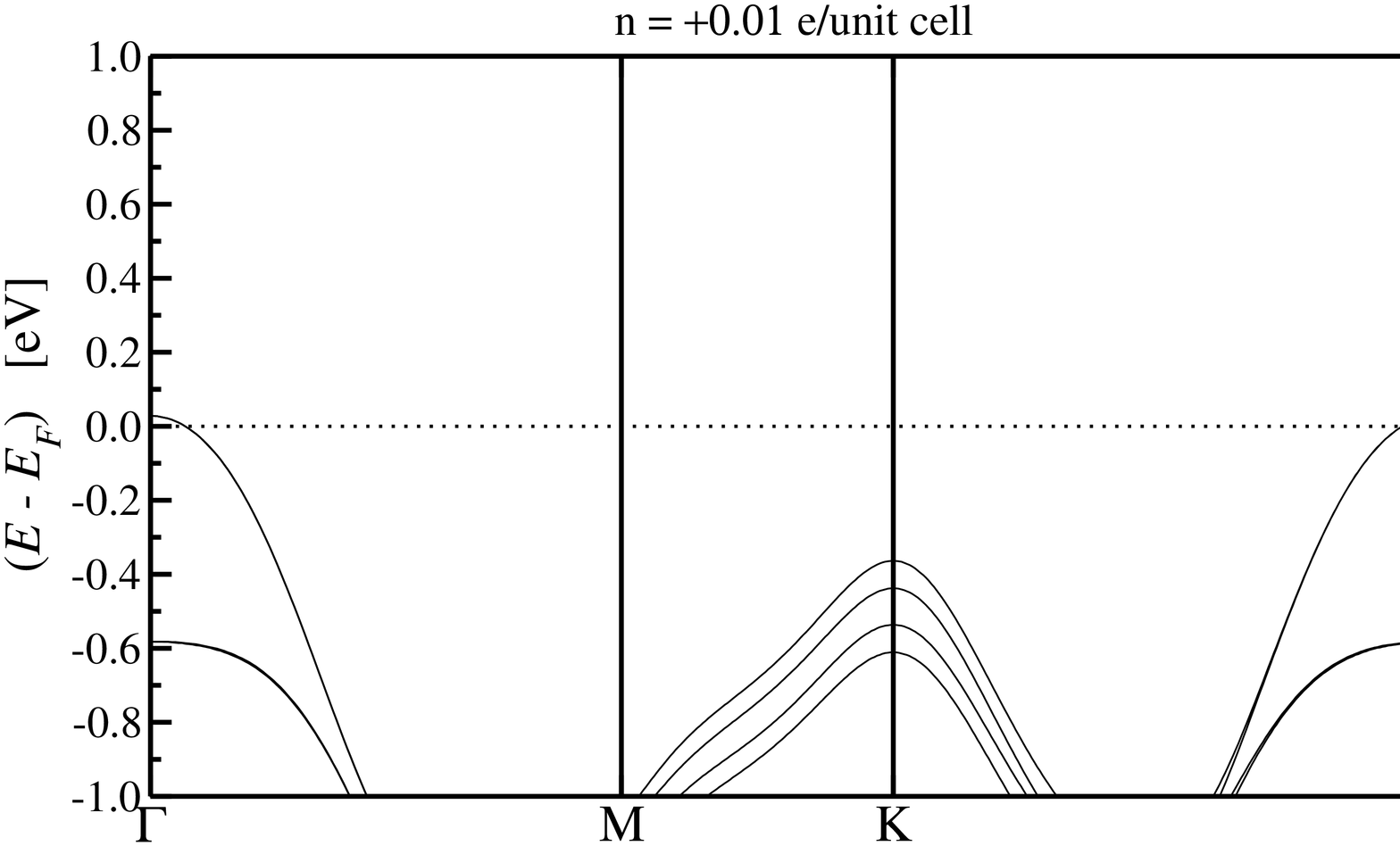}
 \includegraphics[width=0.31\textwidth,clip=,angle=-90]{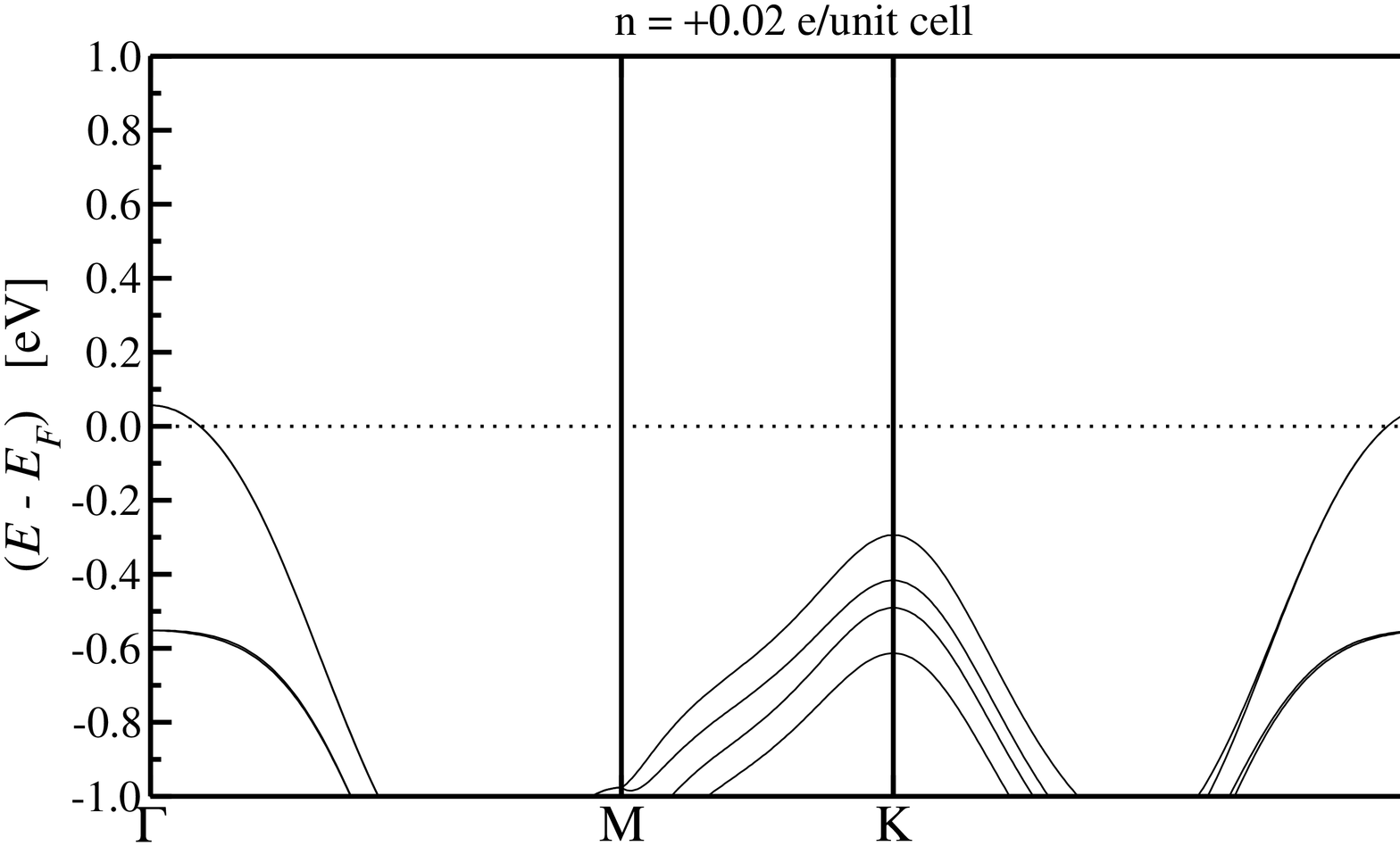}
 \includegraphics[width=0.31\textwidth,clip=,angle=-90]{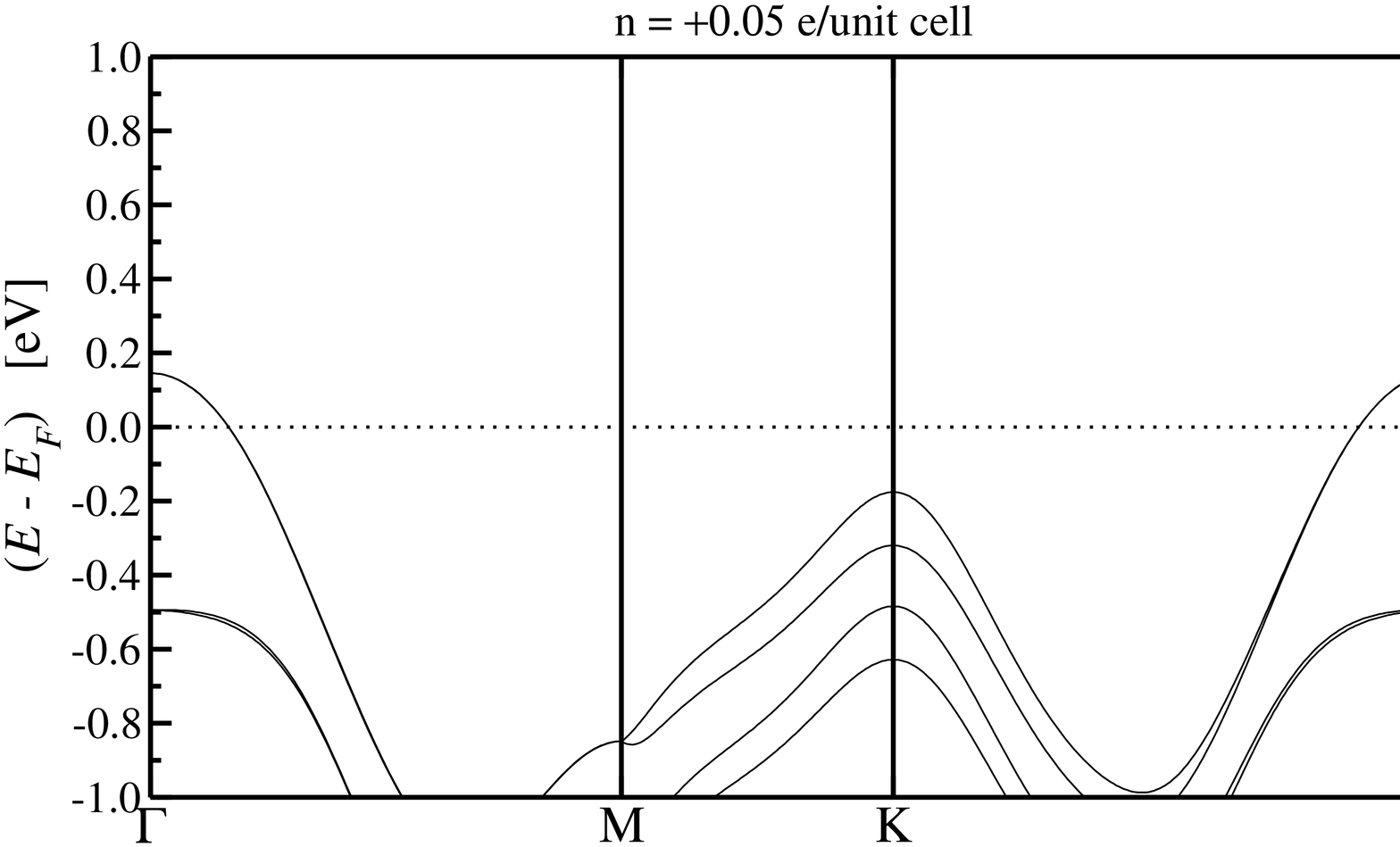}
 \includegraphics[width=0.31\textwidth,clip=,angle=-90]{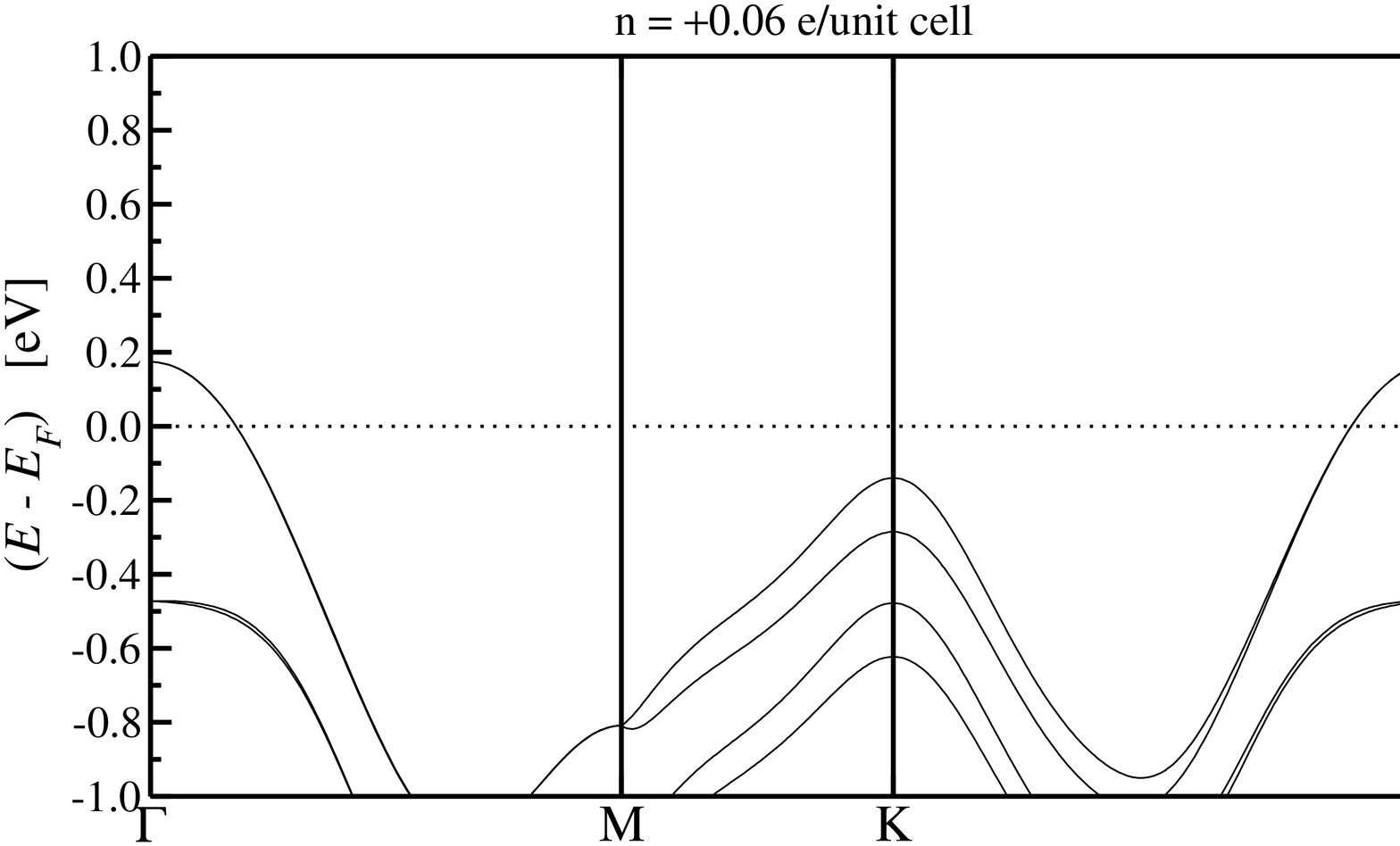}
 \includegraphics[width=0.31\textwidth,clip=,angle=-90]{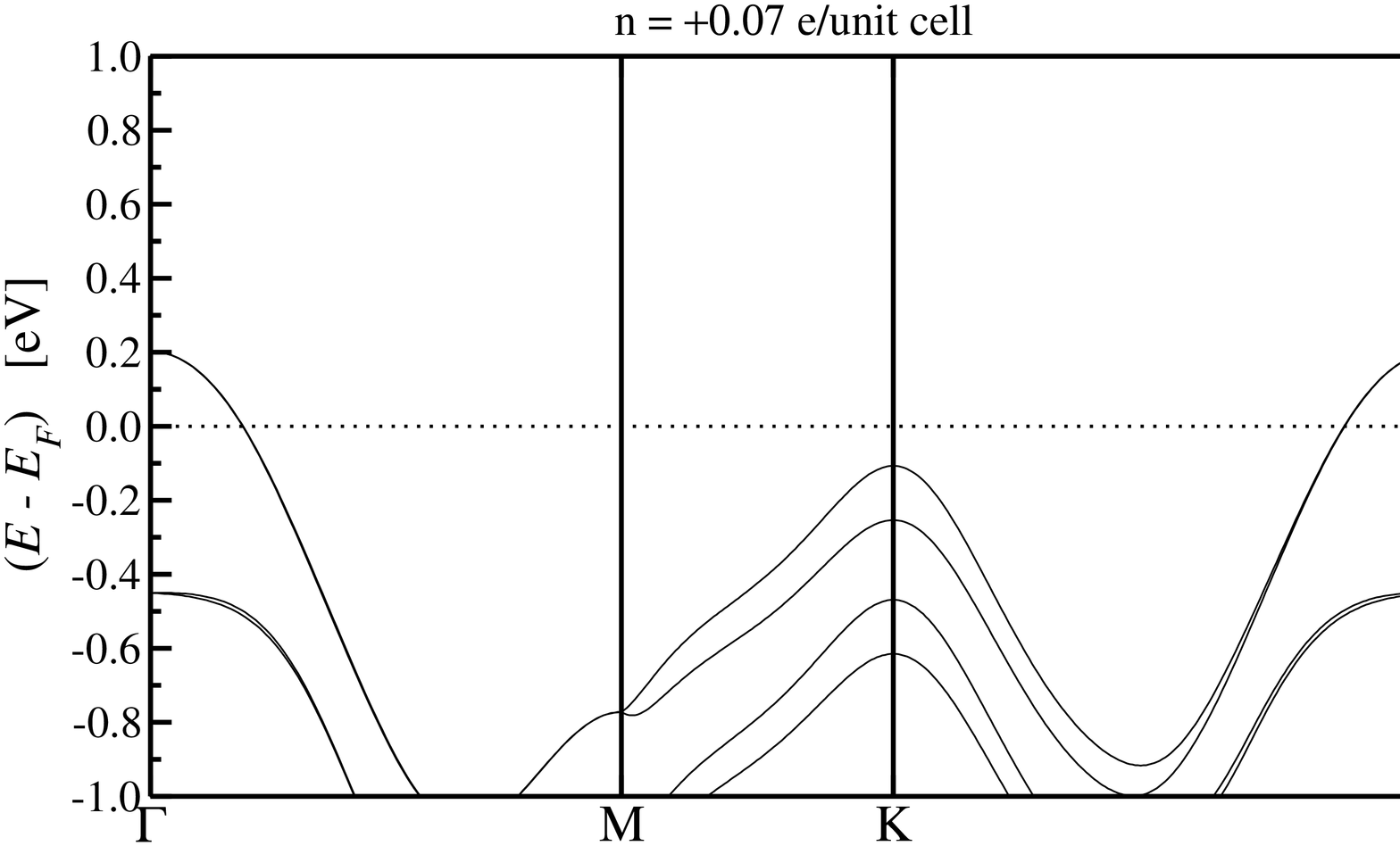}
 \includegraphics[width=0.31\textwidth,clip=,angle=-90]{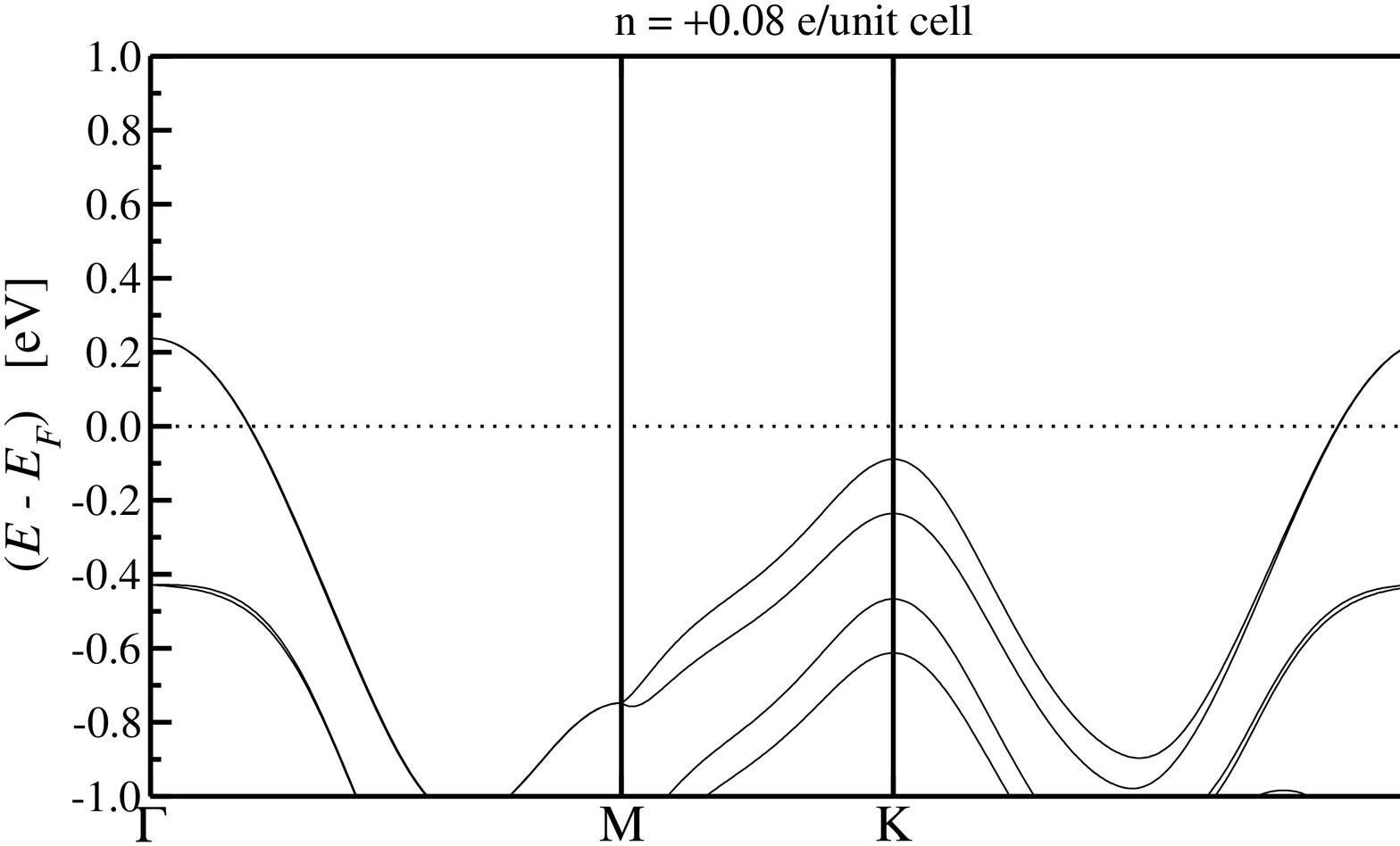}
 \caption{Band structure of bilayer MoS$_2$ for different doping as indicated in the labels.}
\end{figure*}
\begin{figure*}[hbp]
 \centering
 \includegraphics[width=0.31\textwidth,clip=,angle=-90]{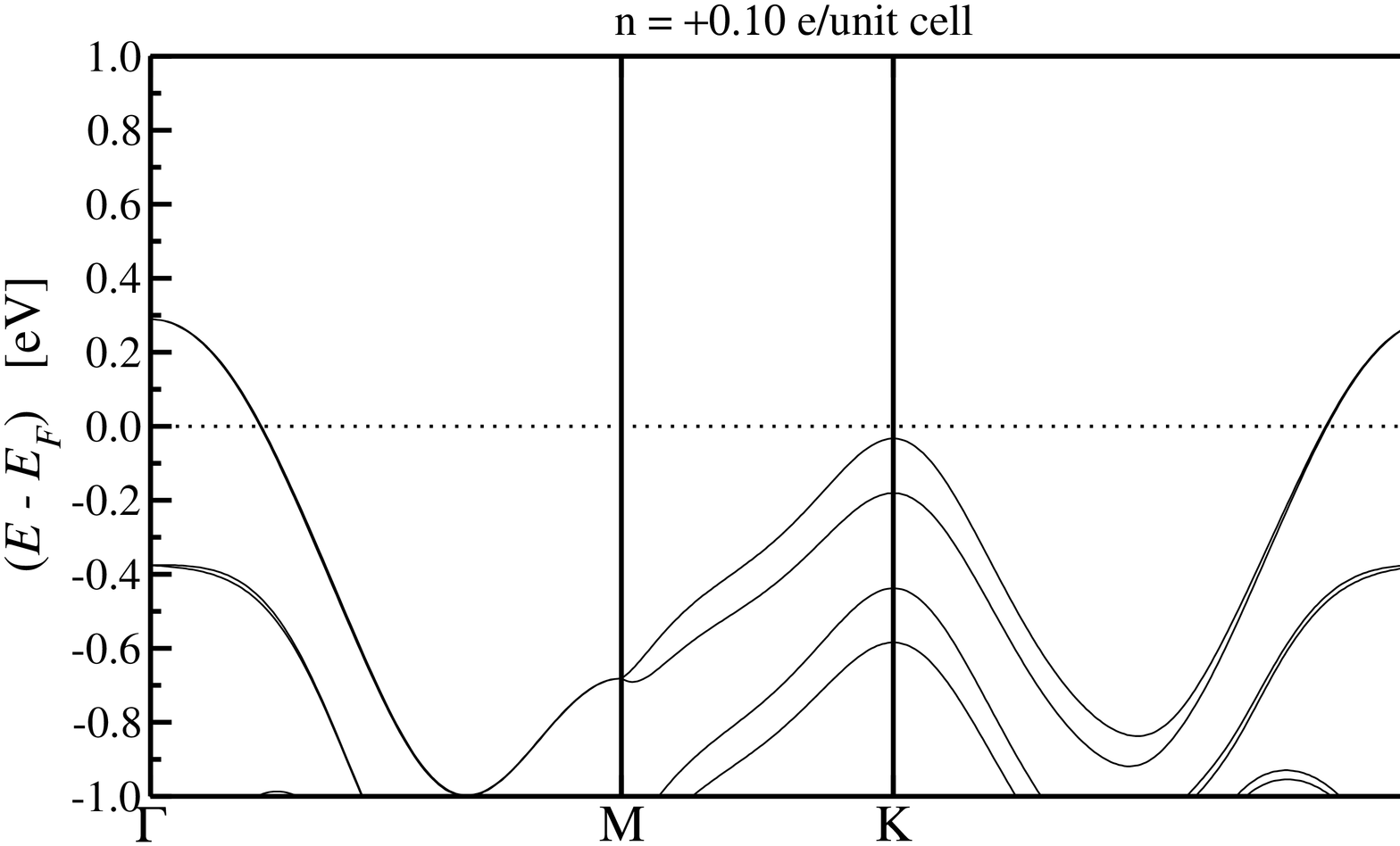}
 \includegraphics[width=0.31\textwidth,clip=,angle=-90]{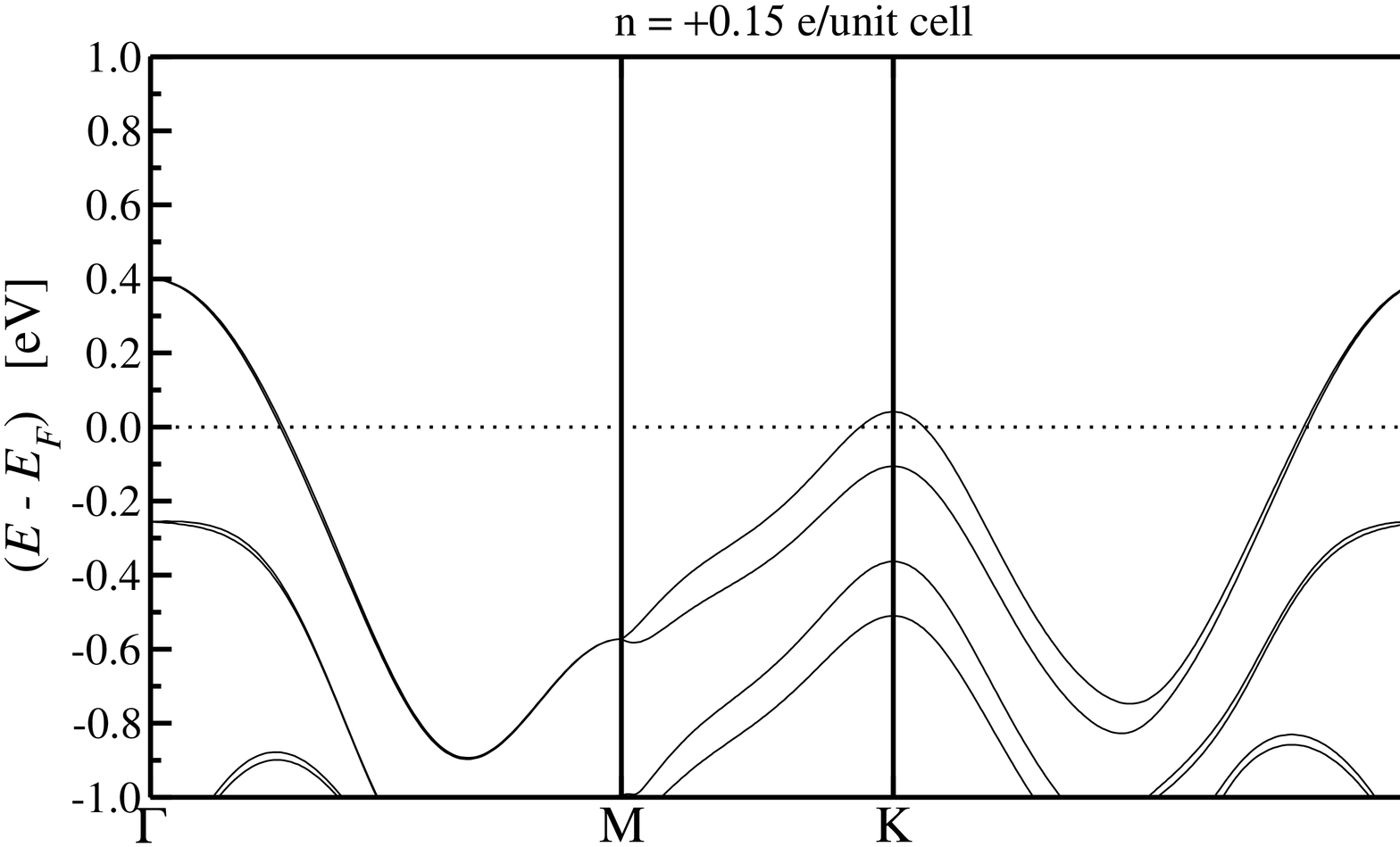}
 \includegraphics[width=0.31\textwidth,clip=,angle=-90]{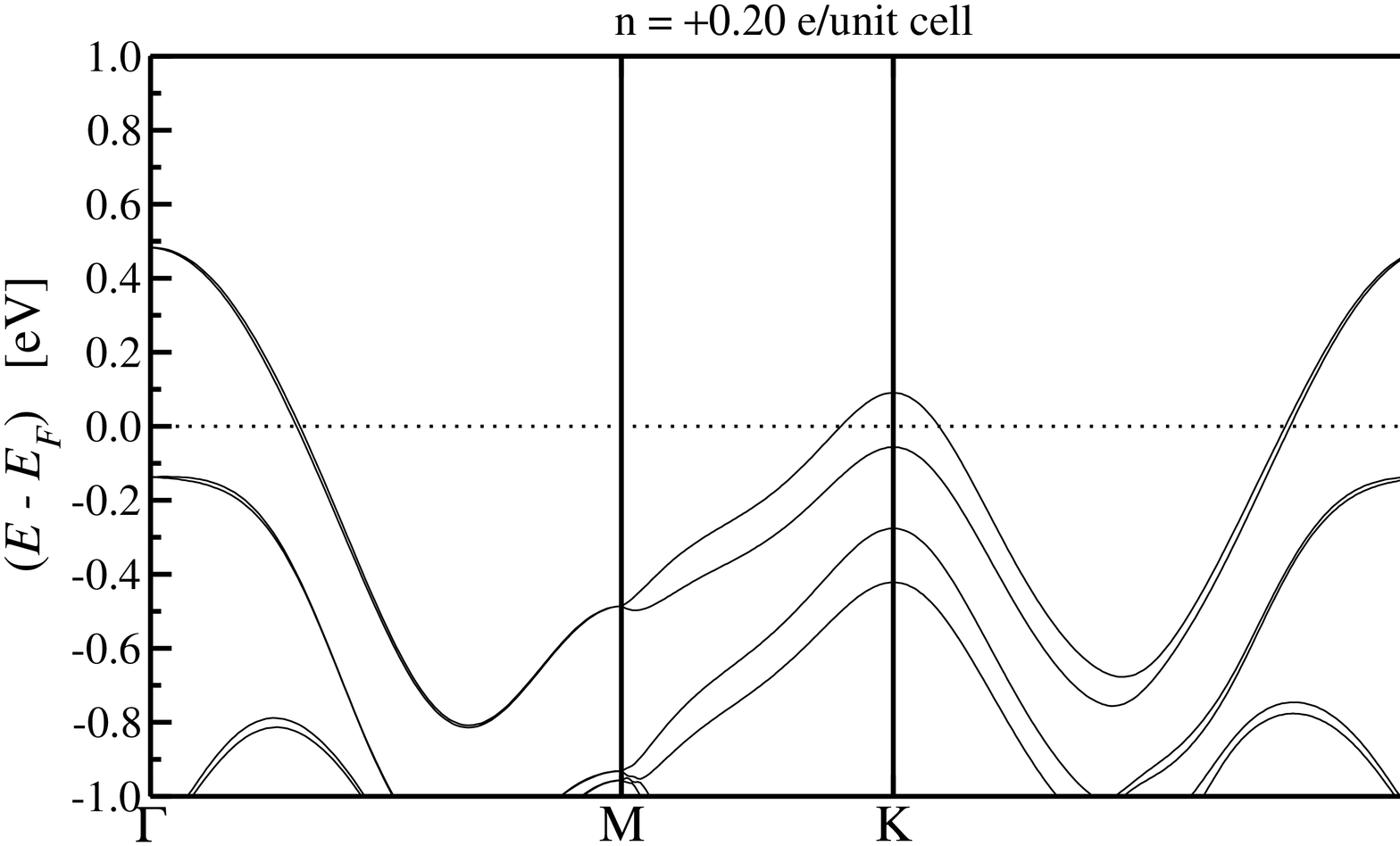}
 \includegraphics[width=0.31\textwidth,clip=,angle=-90]{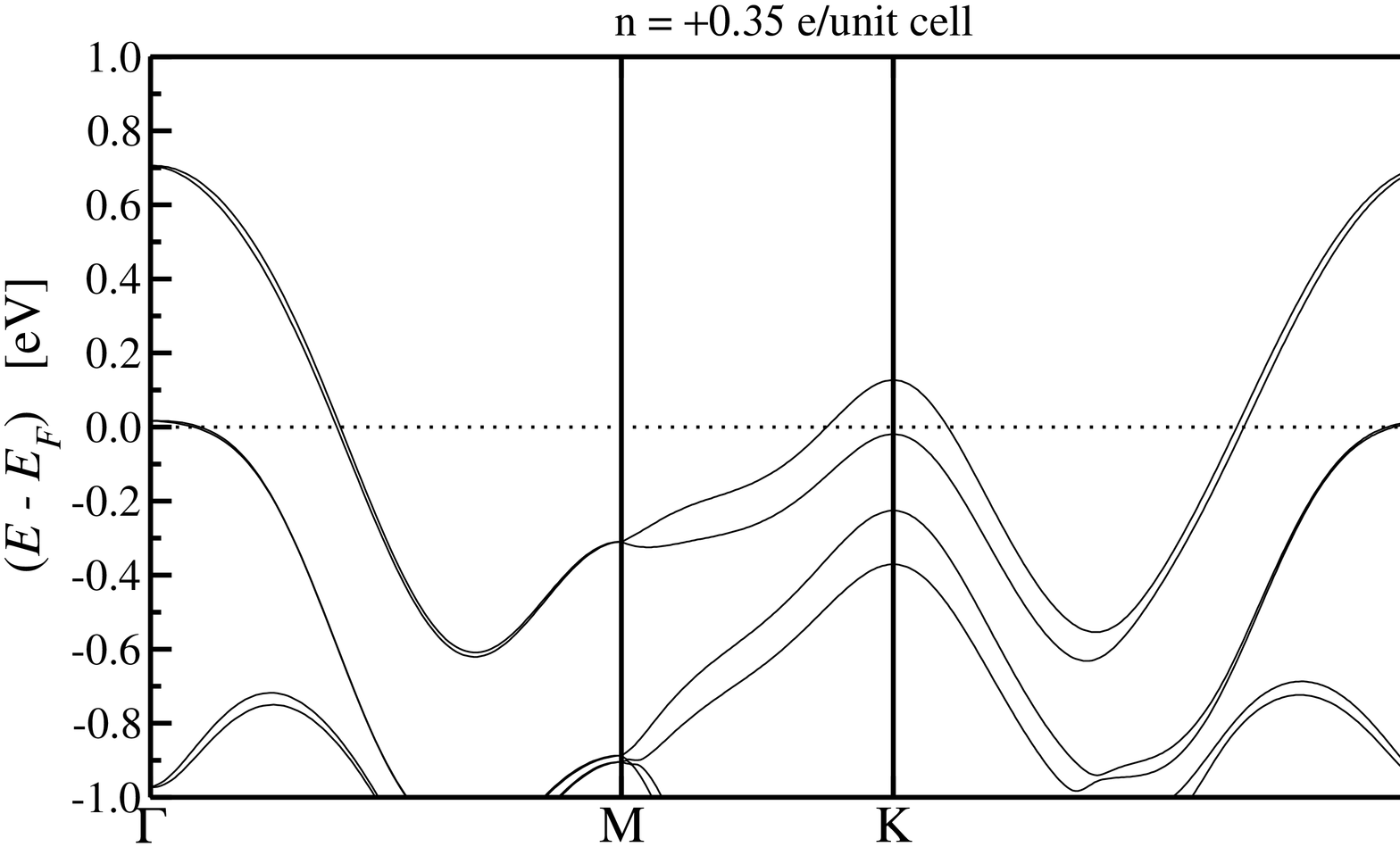}
 \caption{Band structure of bilayer MoS$_2$ for different doping as indicated in the labels.}
\end{figure*}
\begin{figure*}[hbp]
 \centering
 \includegraphics[width=0.31\textwidth,clip=,angle=-90]{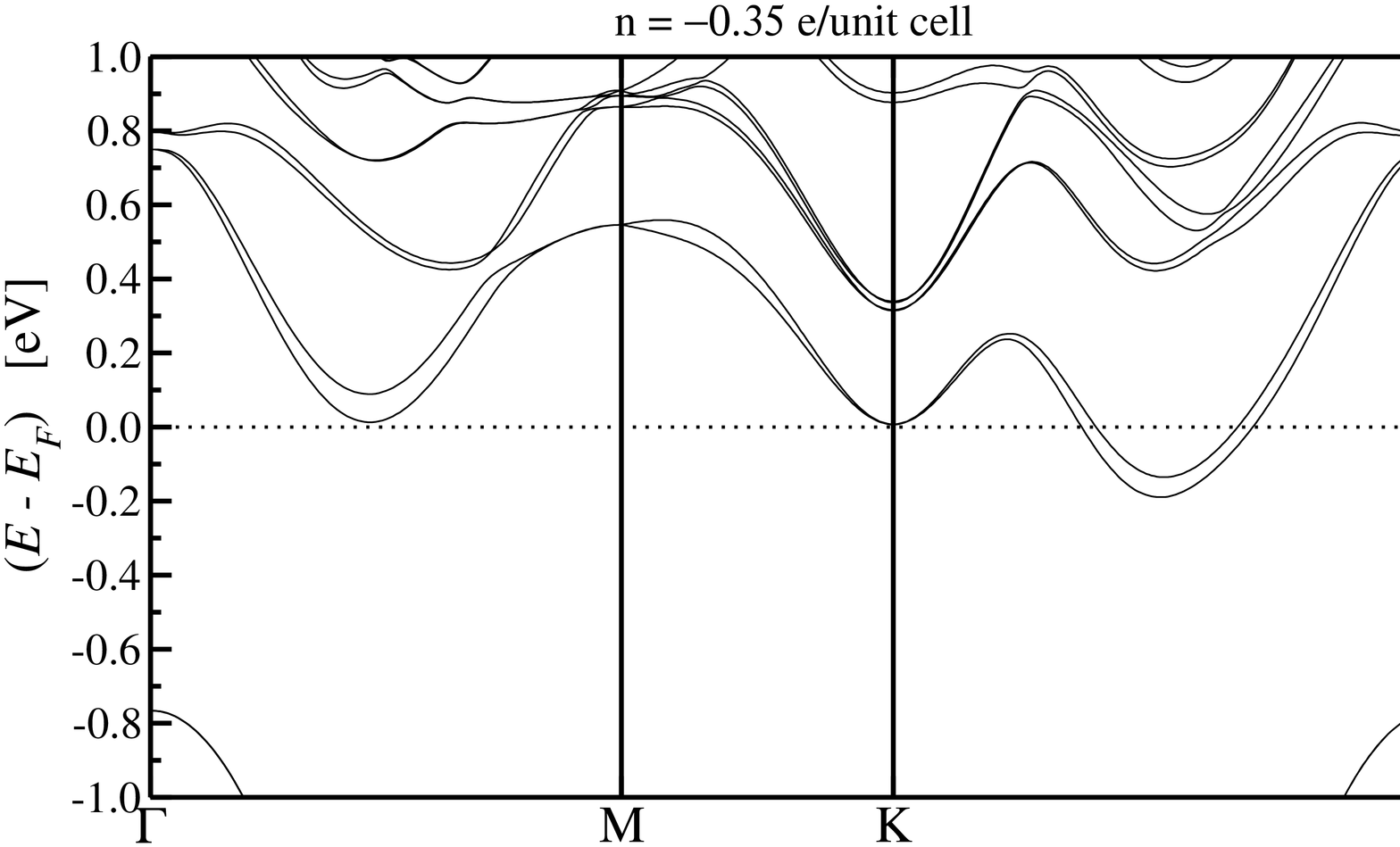}
 \includegraphics[width=0.31\textwidth,clip=,angle=-90]{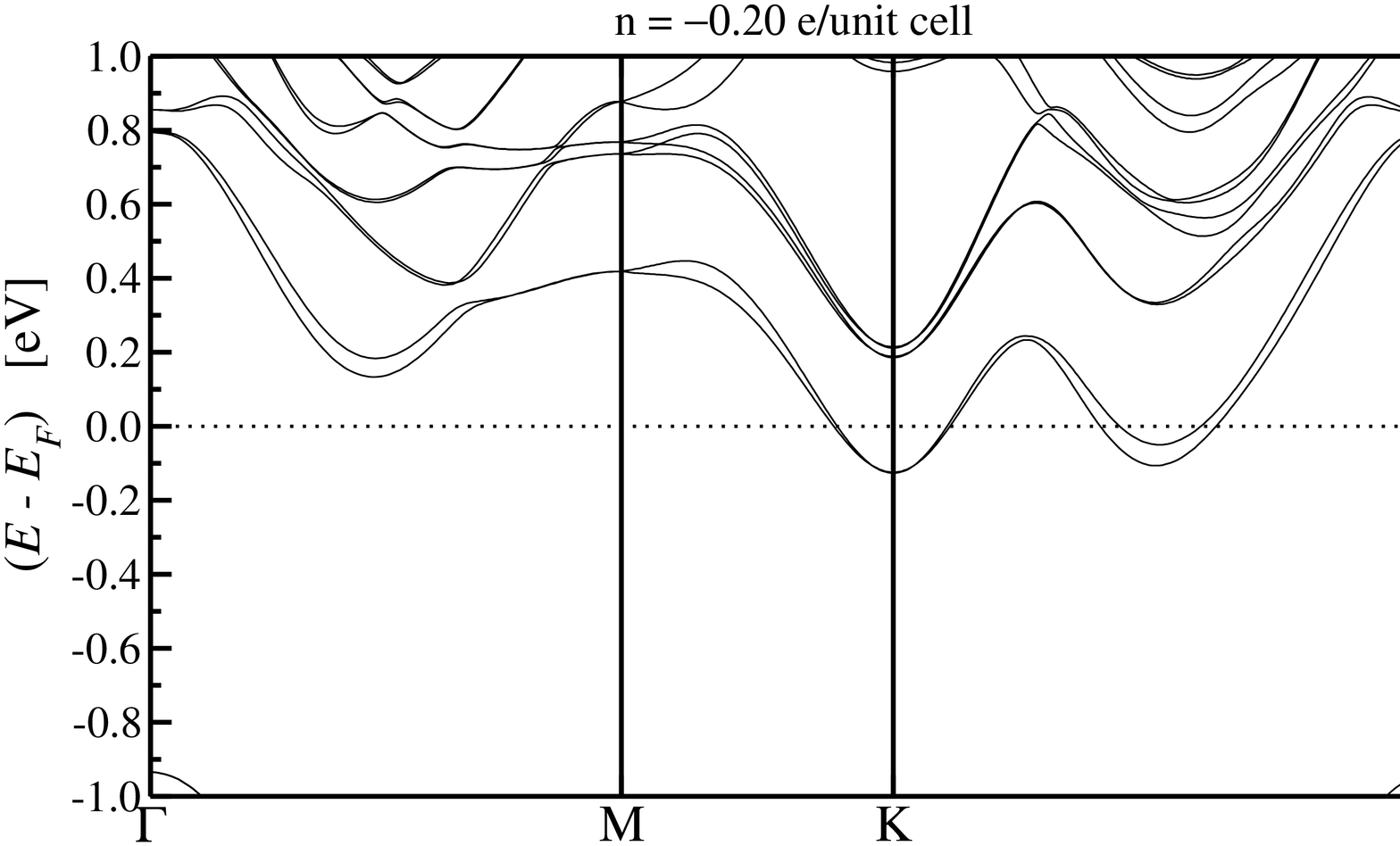}
 \includegraphics[width=0.31\textwidth,clip=,angle=-90]{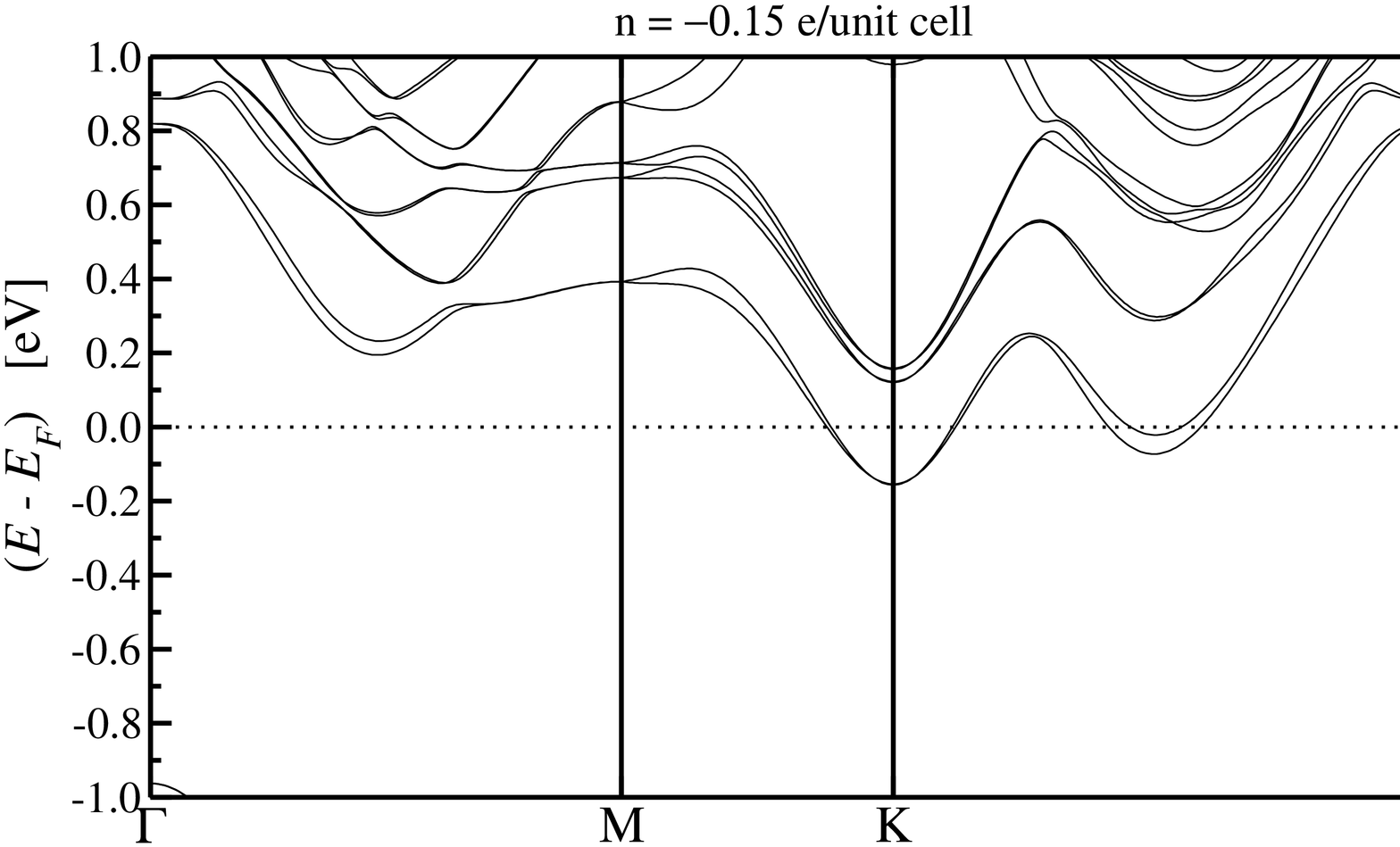}
 \includegraphics[width=0.31\textwidth,clip=,angle=-90]{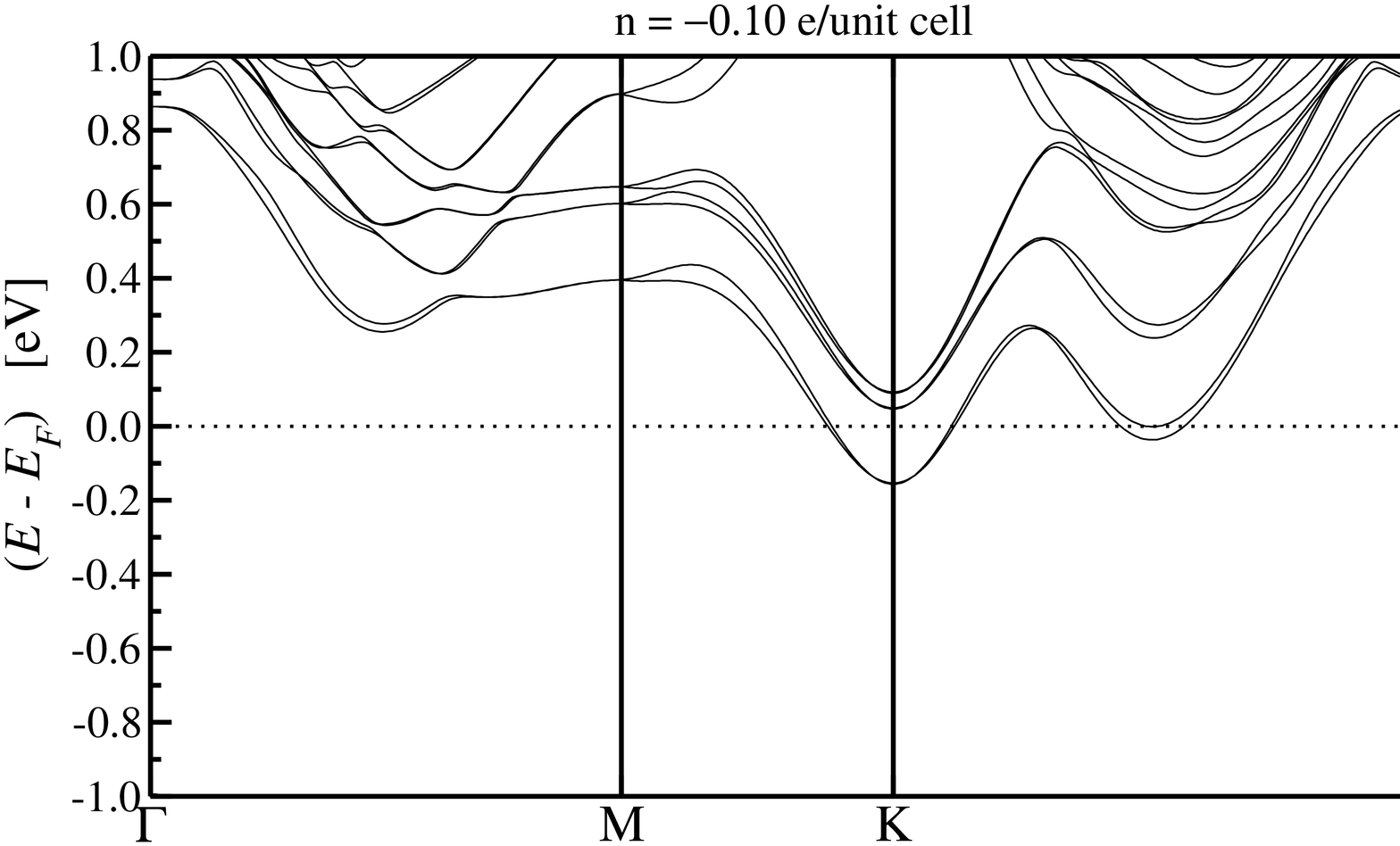}
 \includegraphics[width=0.31\textwidth,clip=,angle=-90]{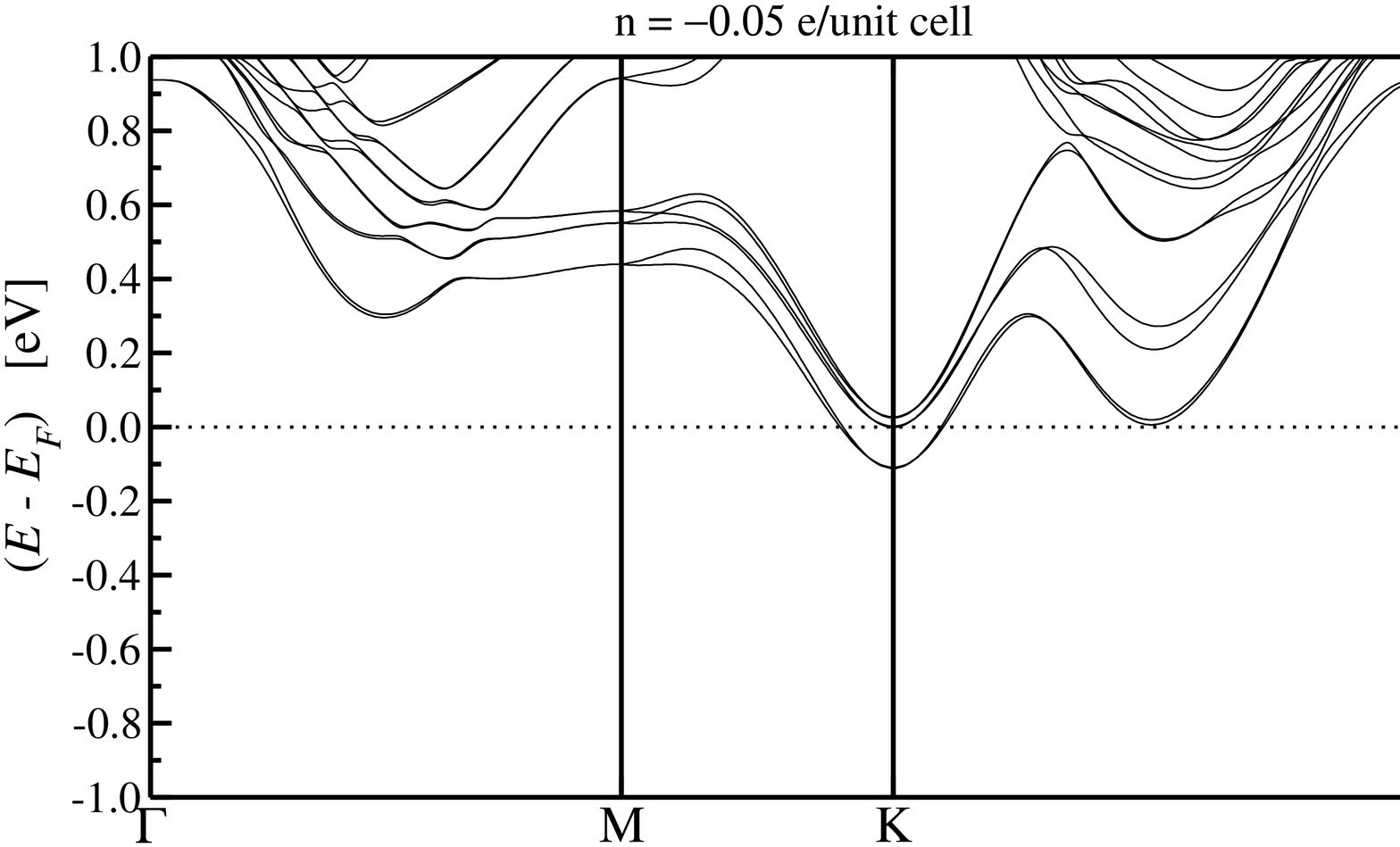}
 \includegraphics[width=0.31\textwidth,clip=,angle=-90]{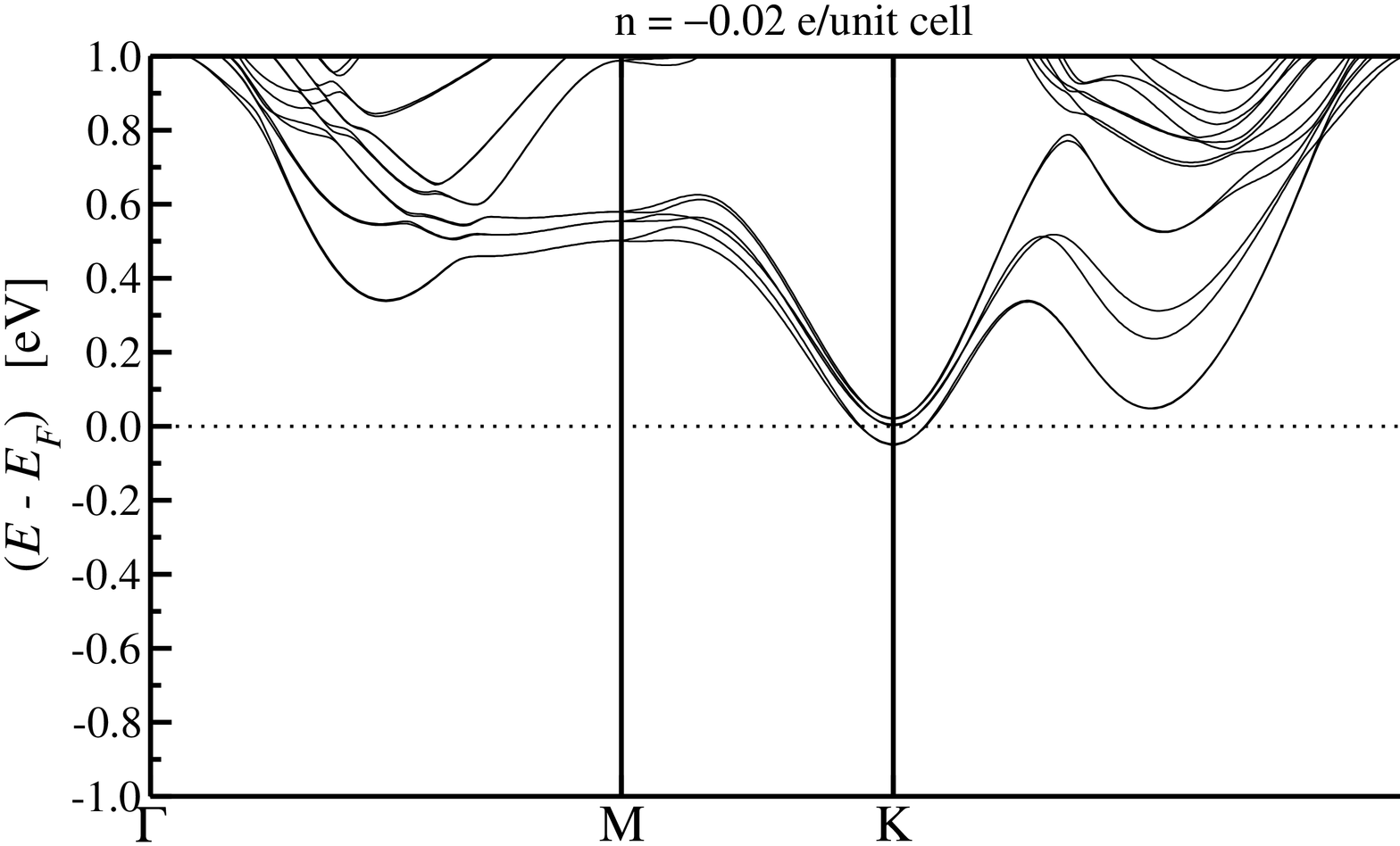}
 \includegraphics[width=0.31\textwidth,clip=,angle=-90]{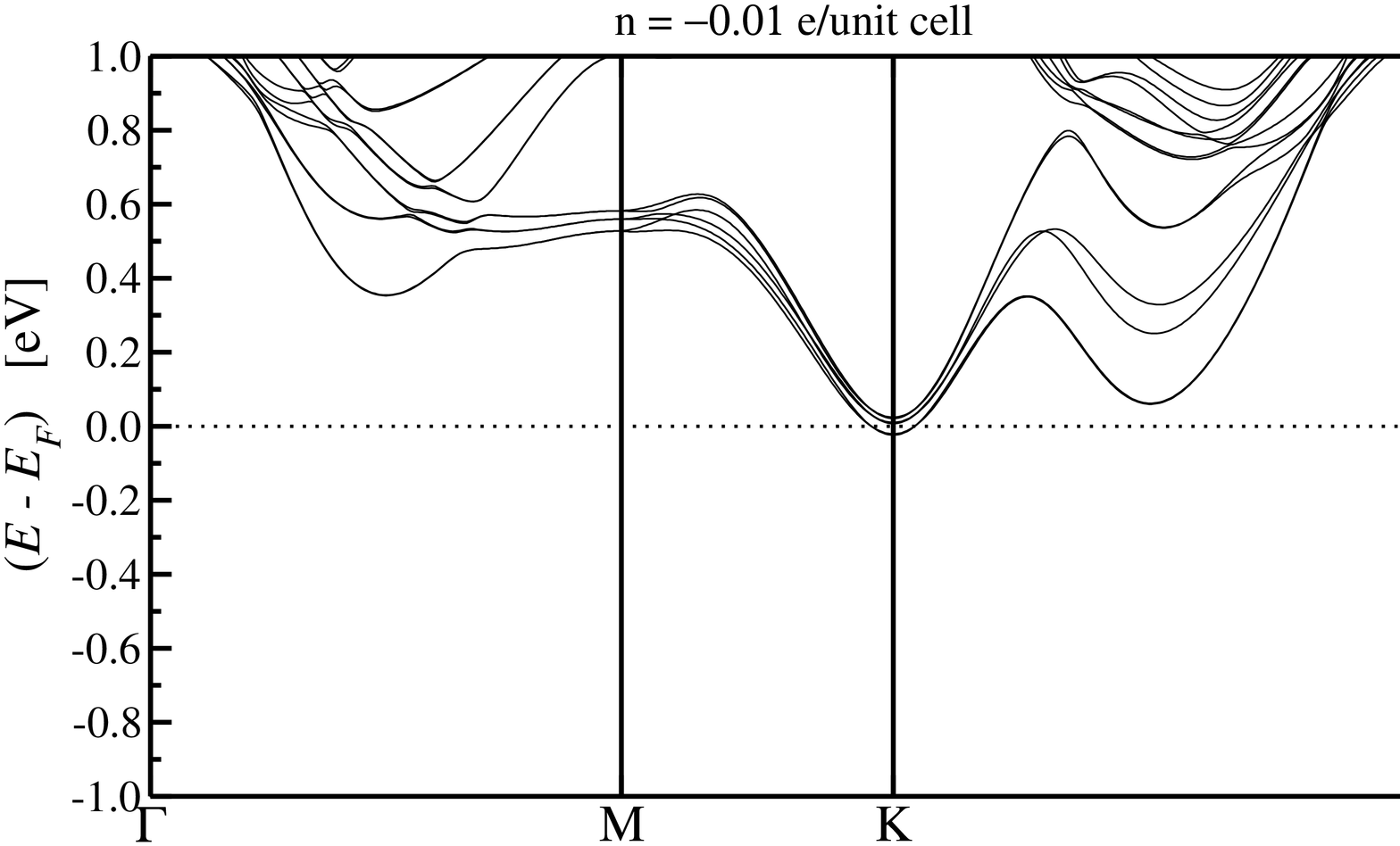}
 \includegraphics[width=0.31\textwidth,clip=,angle=-90]{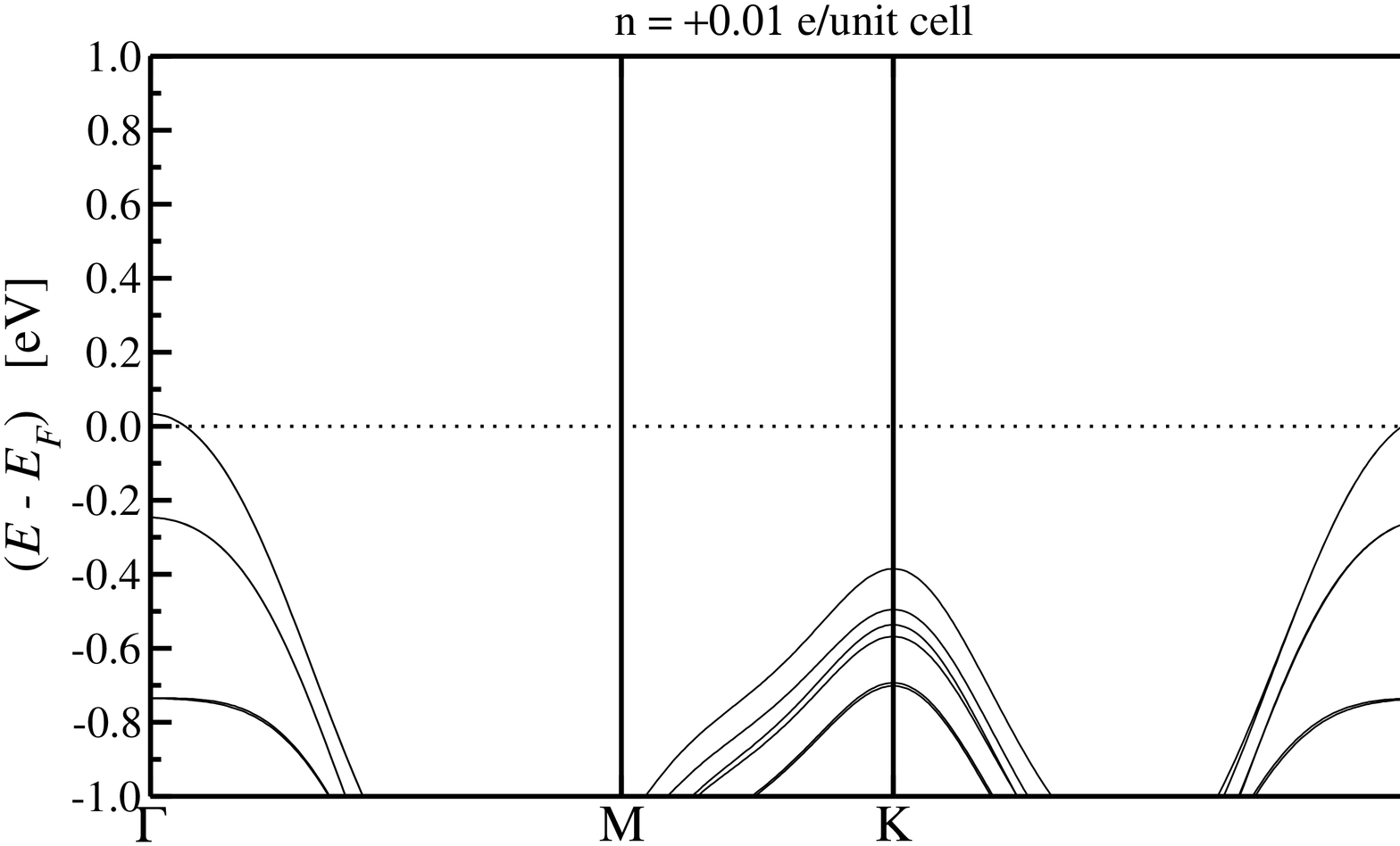}
 \caption{Band structure of trilayer MoS$_2$ for different doping as indicated in the labels.}
\end{figure*}
\begin{figure*}[hbp]
 \centering
 \includegraphics[width=0.31\textwidth,clip=,angle=-90]{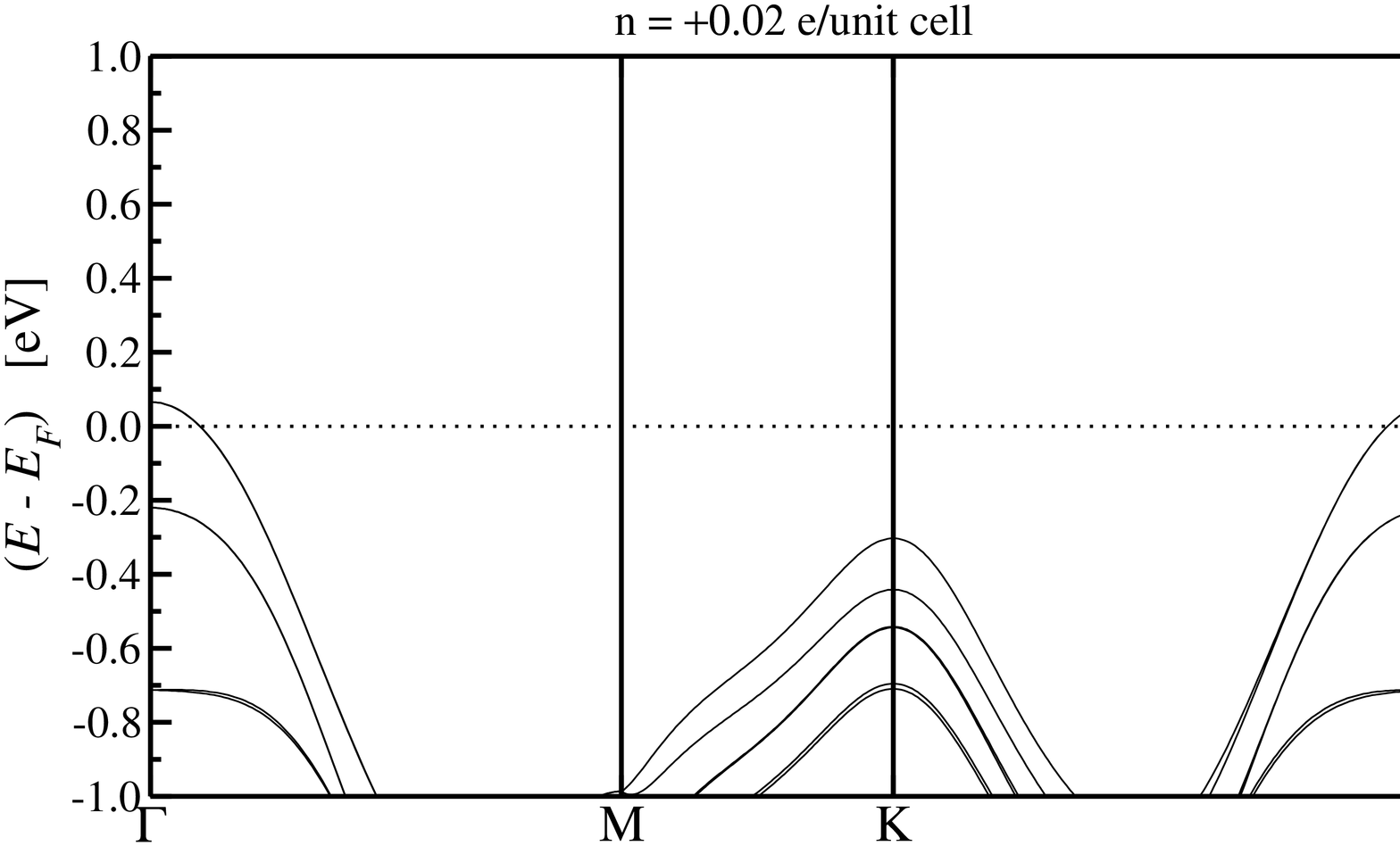}
 \includegraphics[width=0.31\textwidth,clip=,angle=-90]{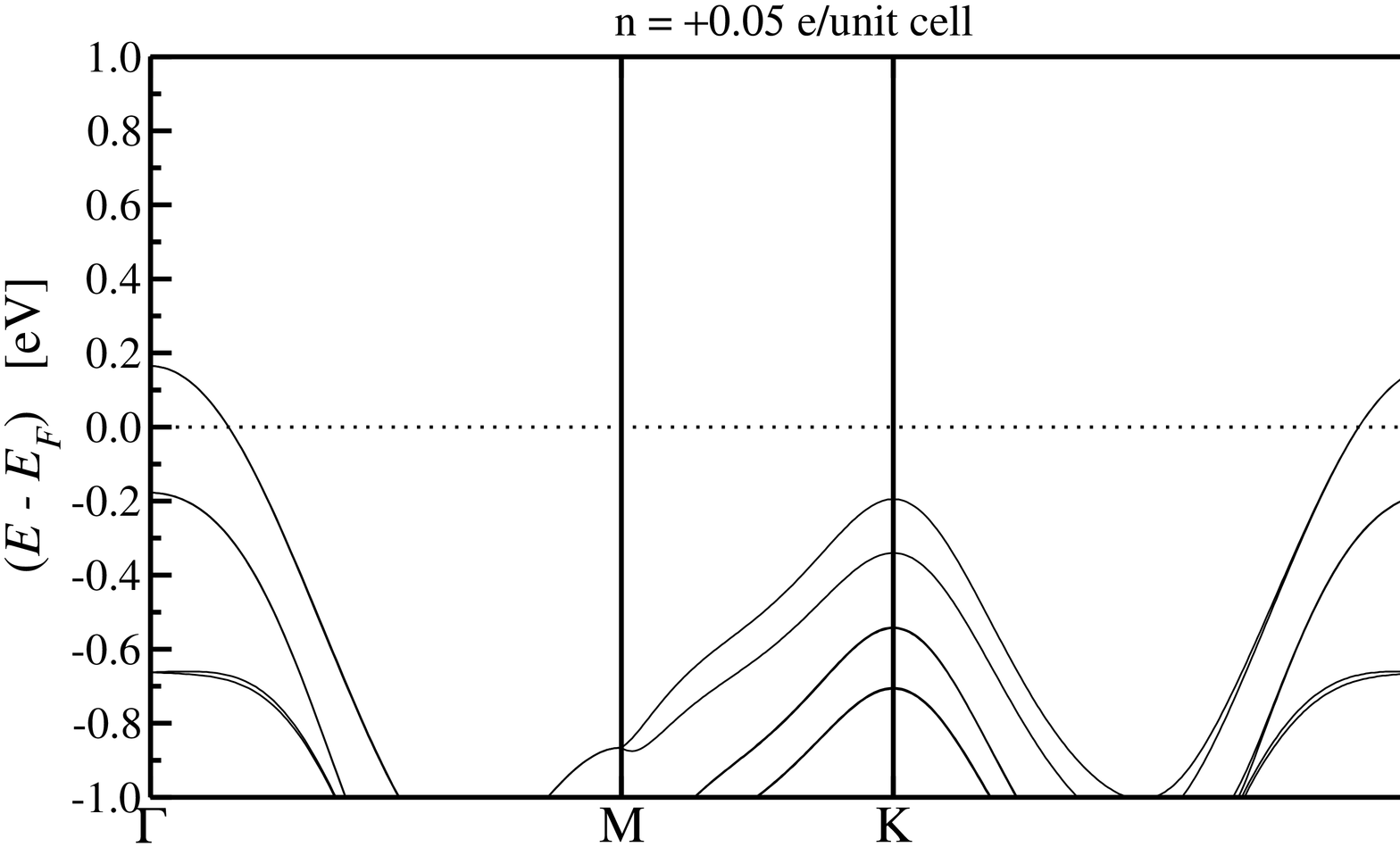}
 \includegraphics[width=0.31\textwidth,clip=,angle=-90]{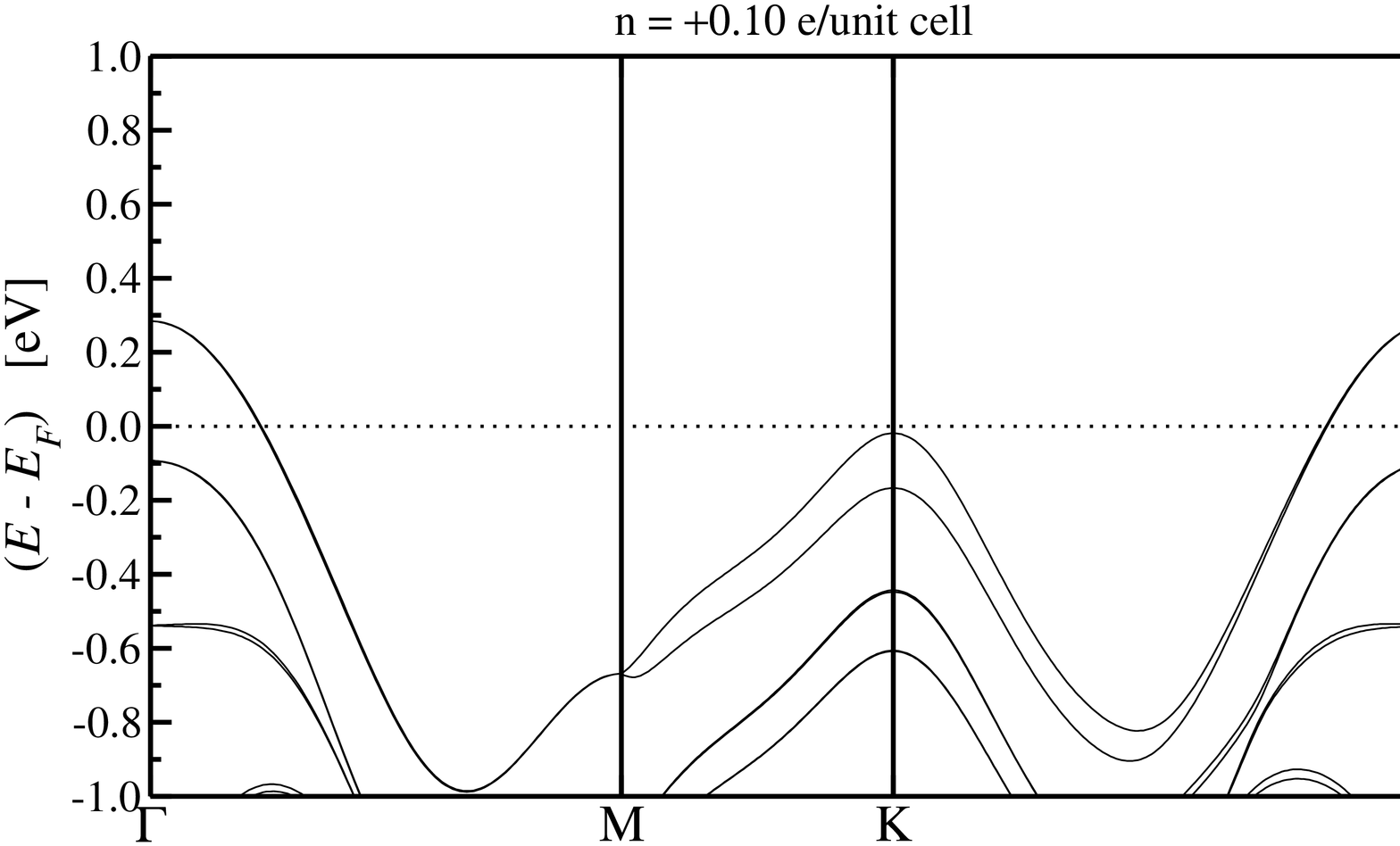}
 \includegraphics[width=0.31\textwidth,clip=,angle=-90]{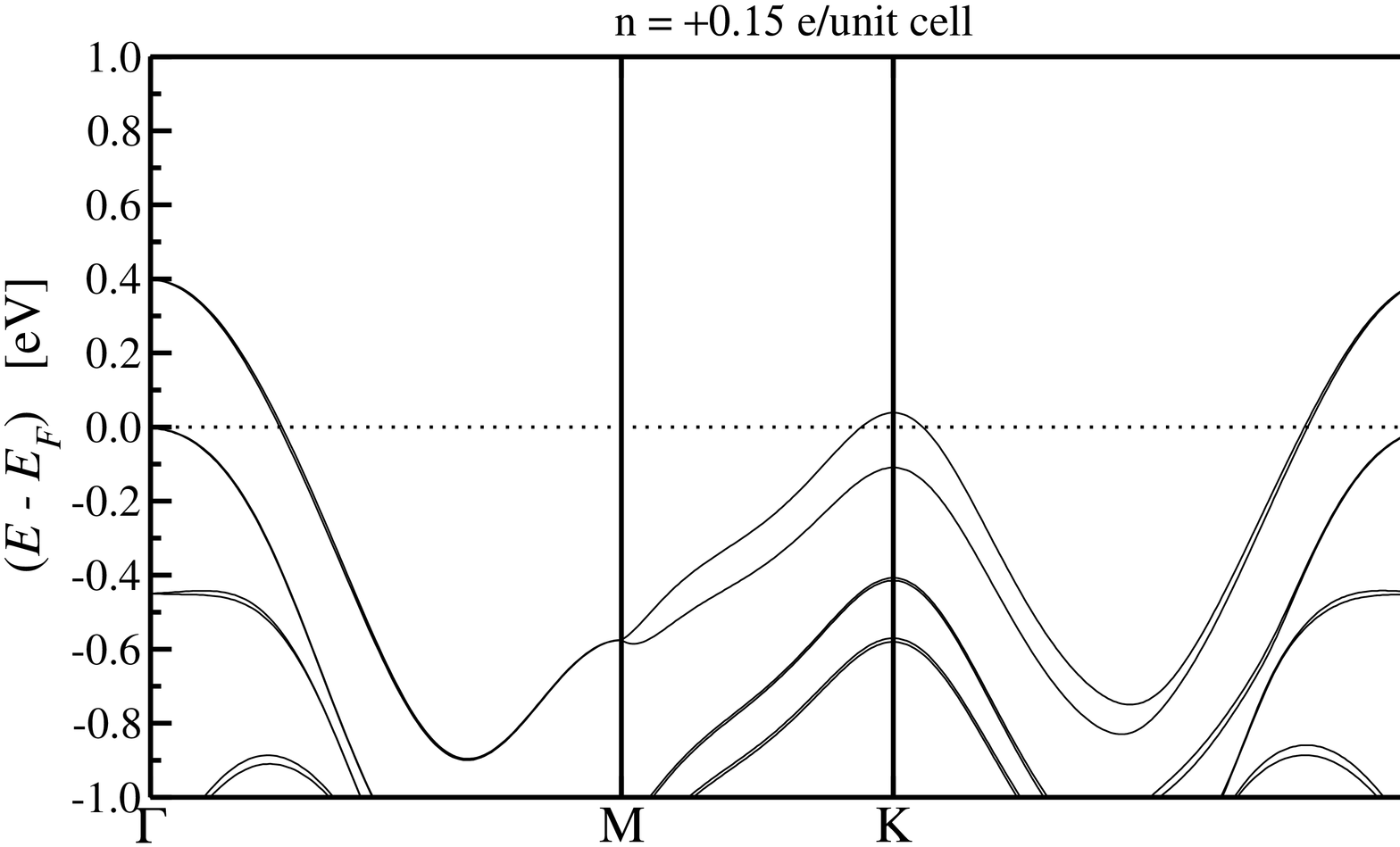}
 \includegraphics[width=0.31\textwidth,clip=,angle=-90]{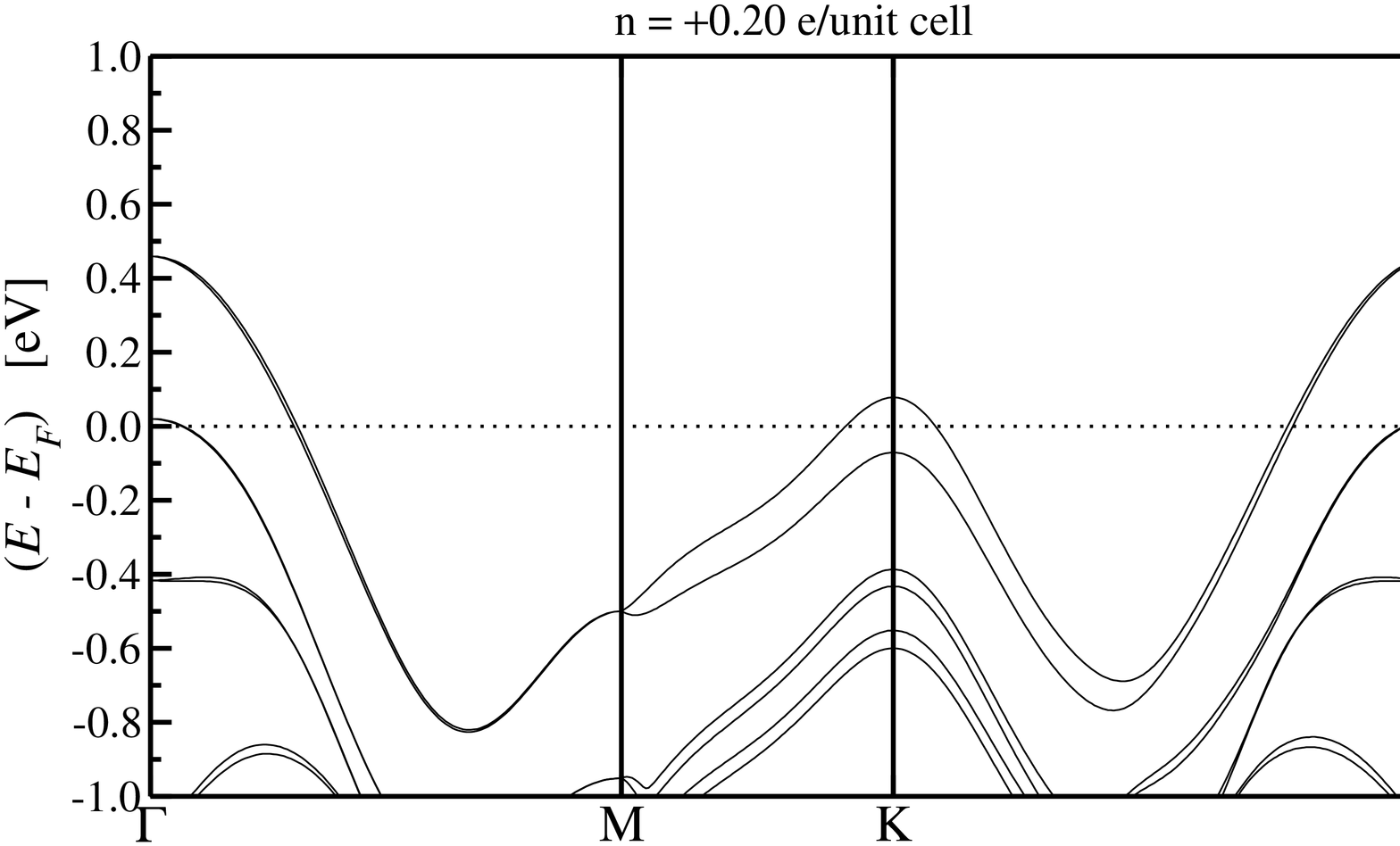}
 \includegraphics[width=0.31\textwidth,clip=,angle=-90]{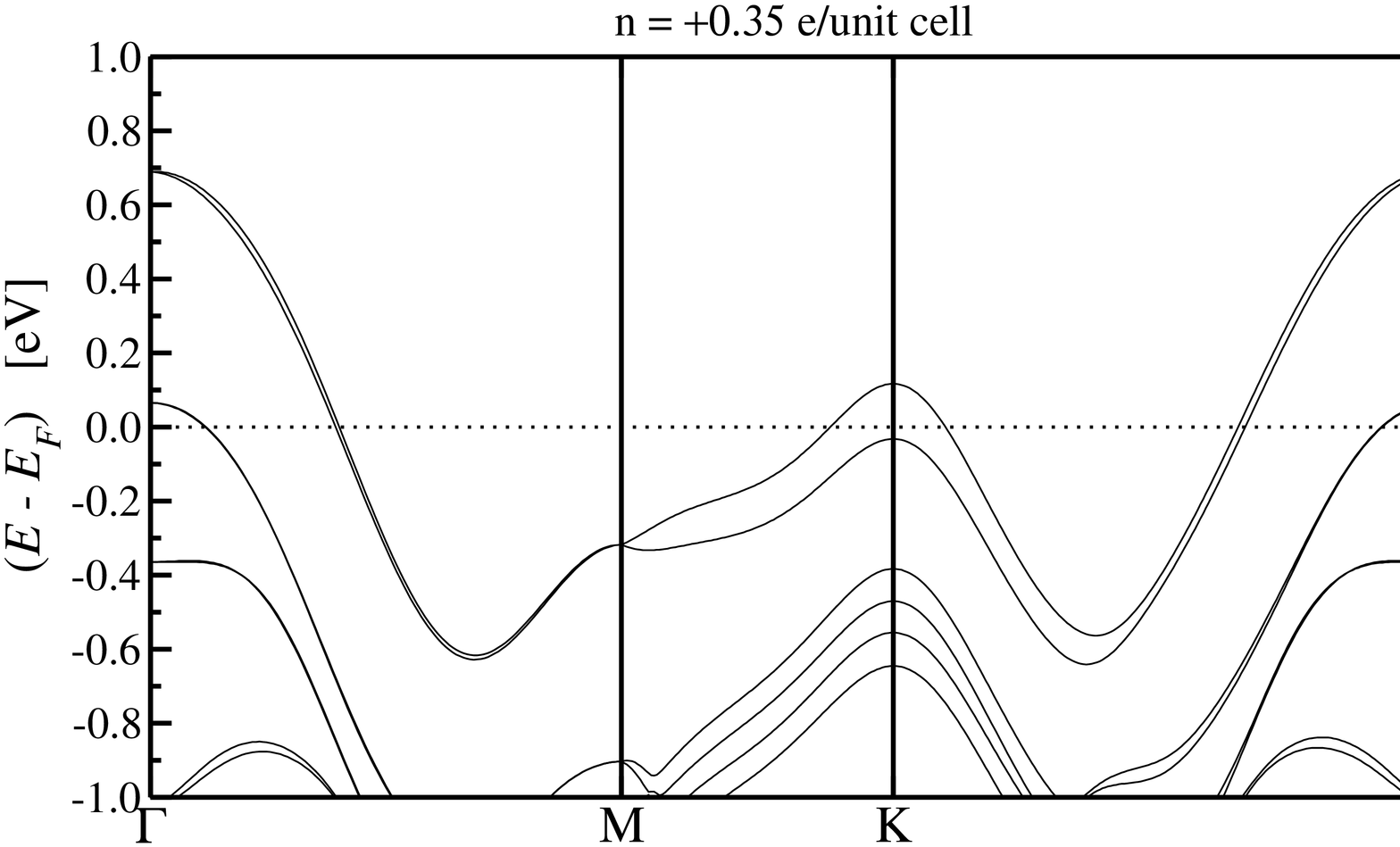}
 \caption{Band structure of trilayer MoS$_2$ for different doping as indicated in the labels.}
\end{figure*}

\clearpage
\subsection{Molybdenum diselenide}
\begin{figure*}[hbp]
 \centering
 \includegraphics[width=0.31\textwidth,clip=,angle=-90]{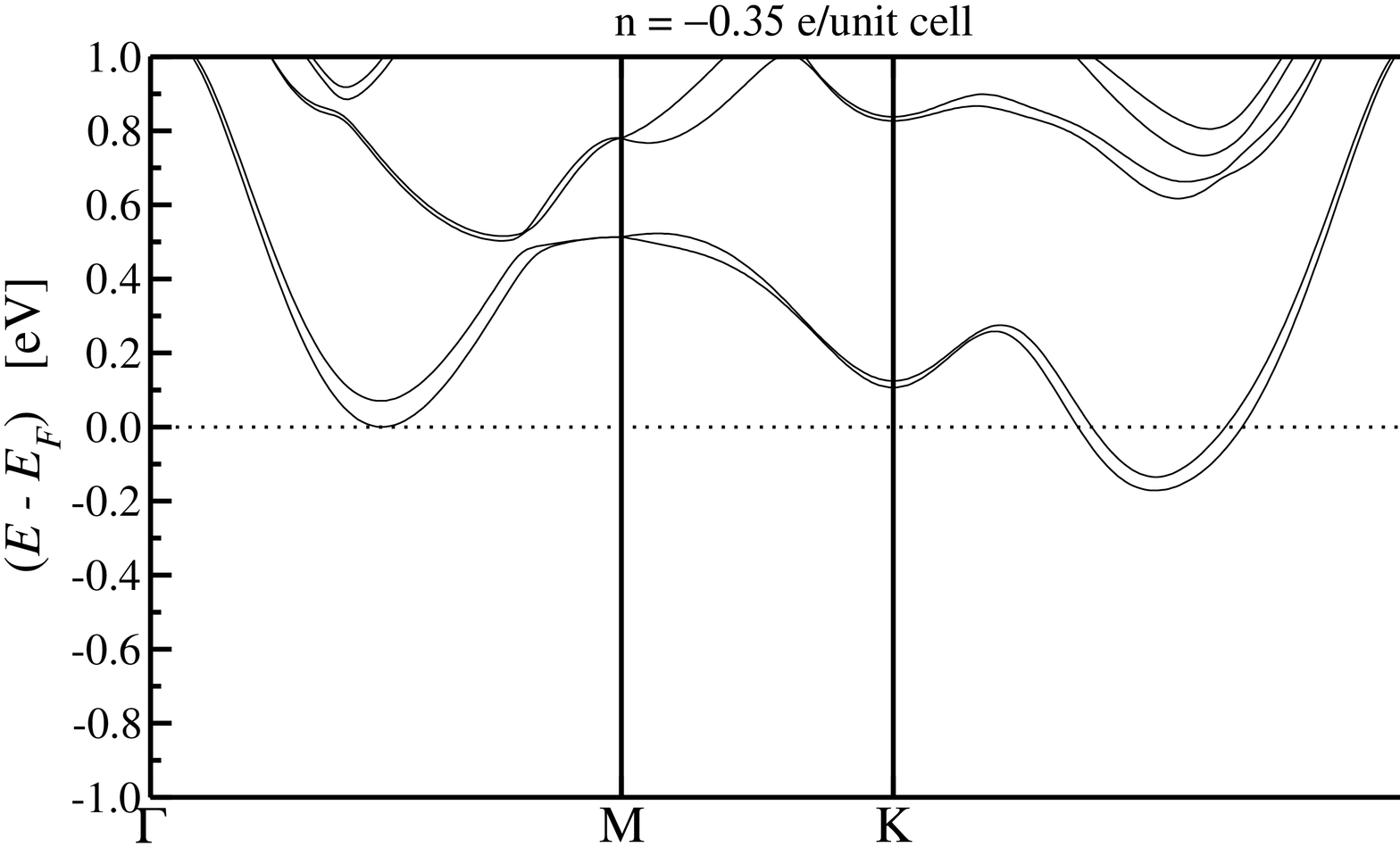}
 \includegraphics[width=0.31\textwidth,clip=,angle=-90]{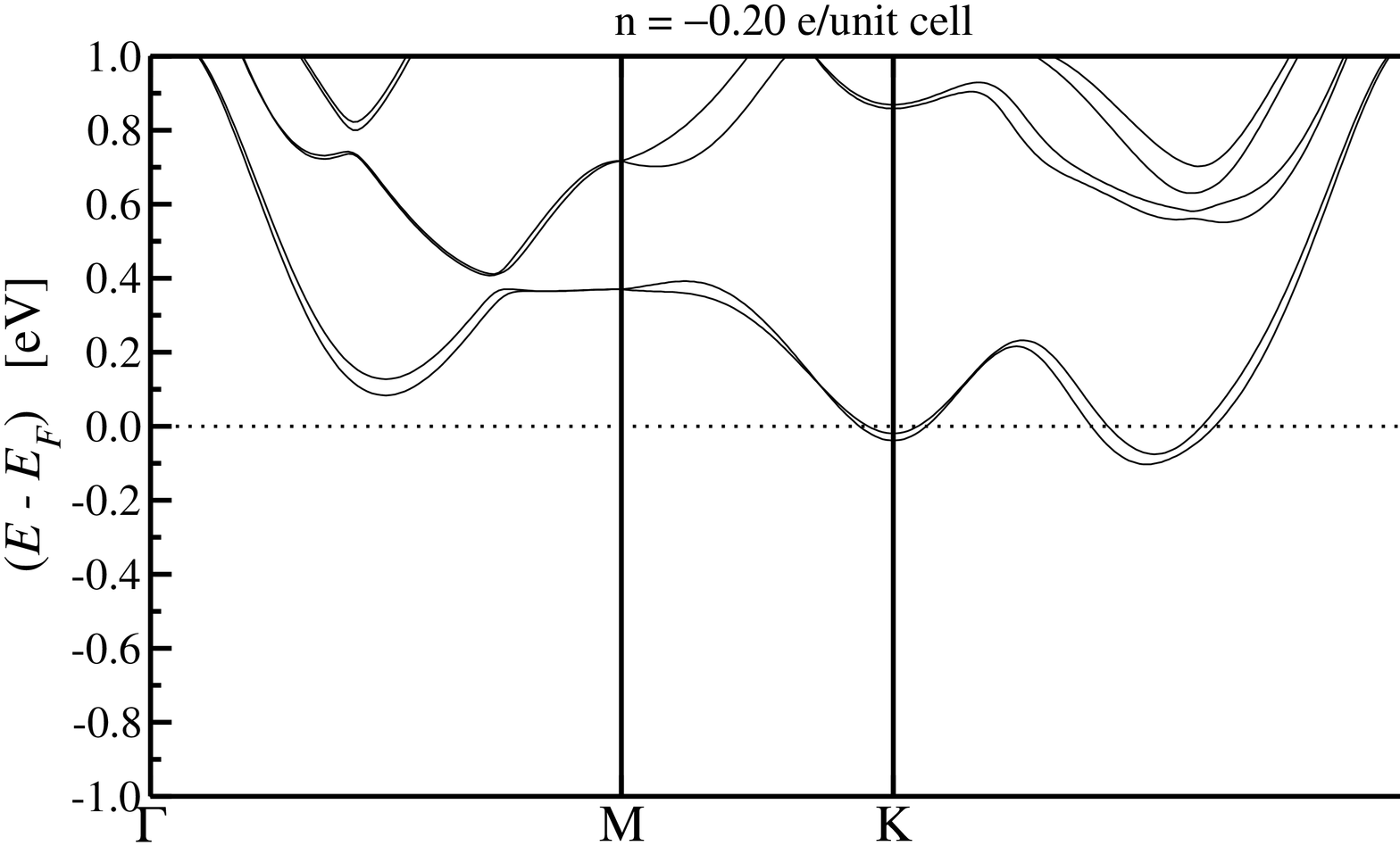}
 \includegraphics[width=0.31\textwidth,clip=,angle=-90]{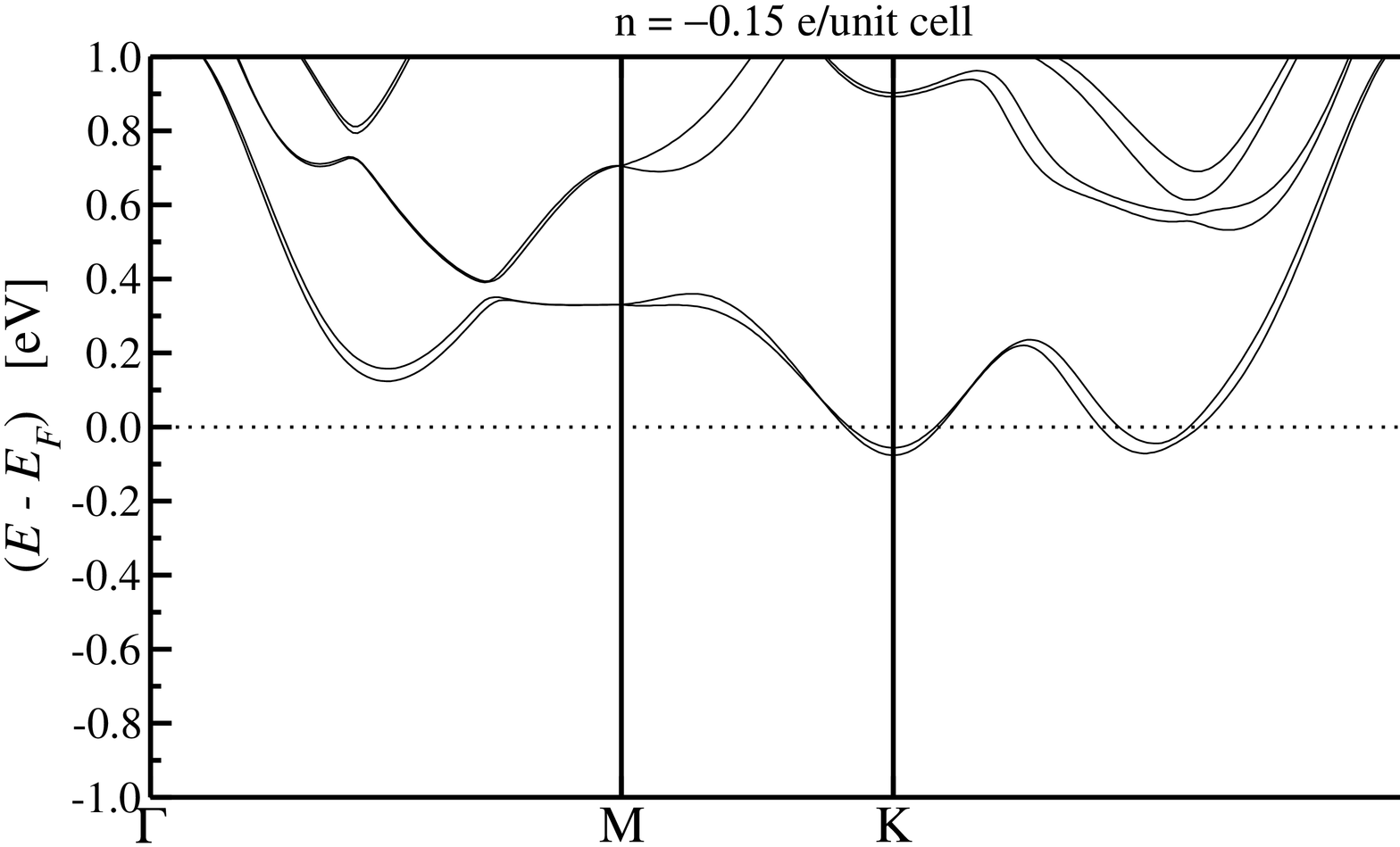}
 \includegraphics[width=0.31\textwidth,clip=,angle=-90]{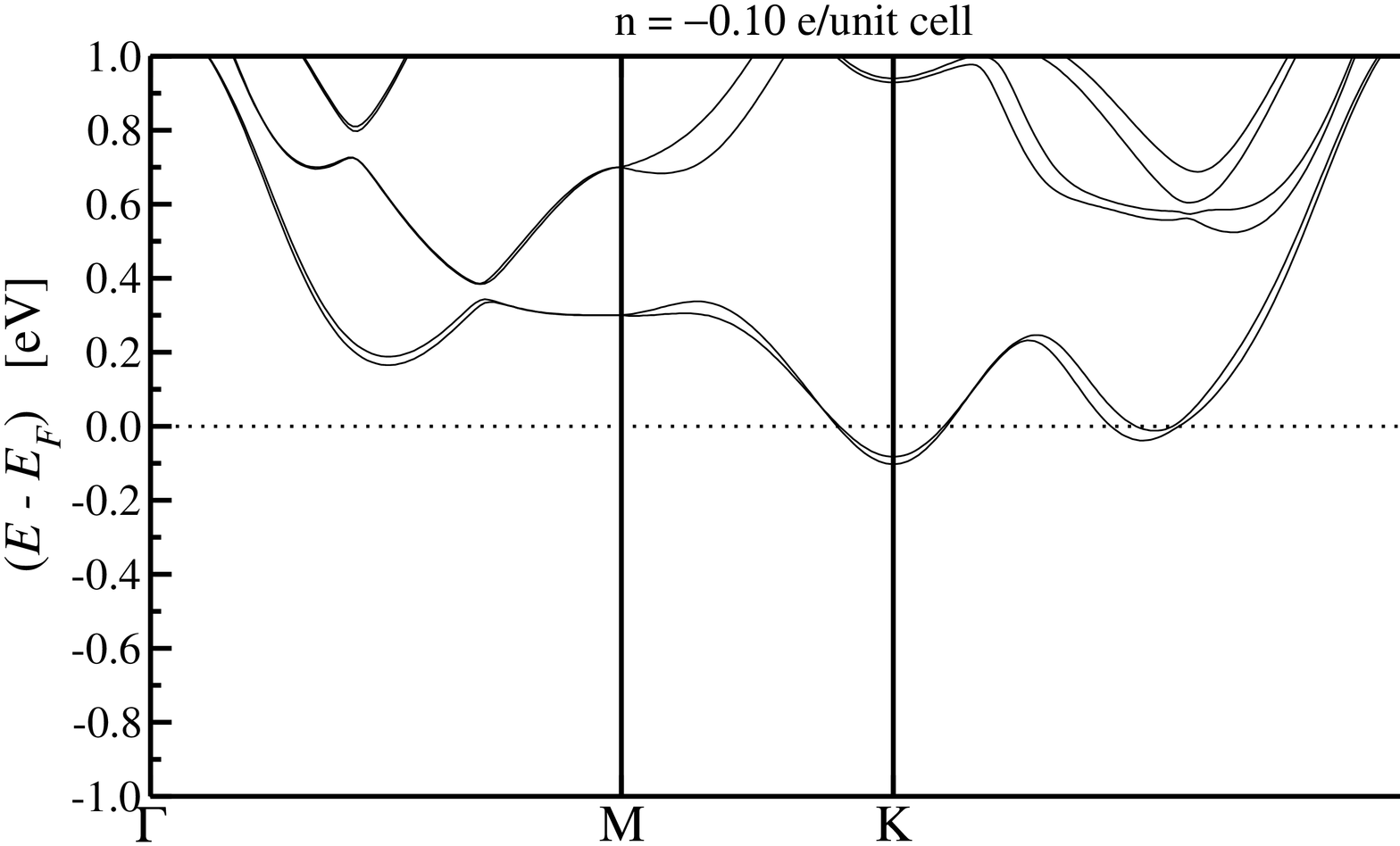}
 \includegraphics[width=0.31\textwidth,clip=,angle=-90]{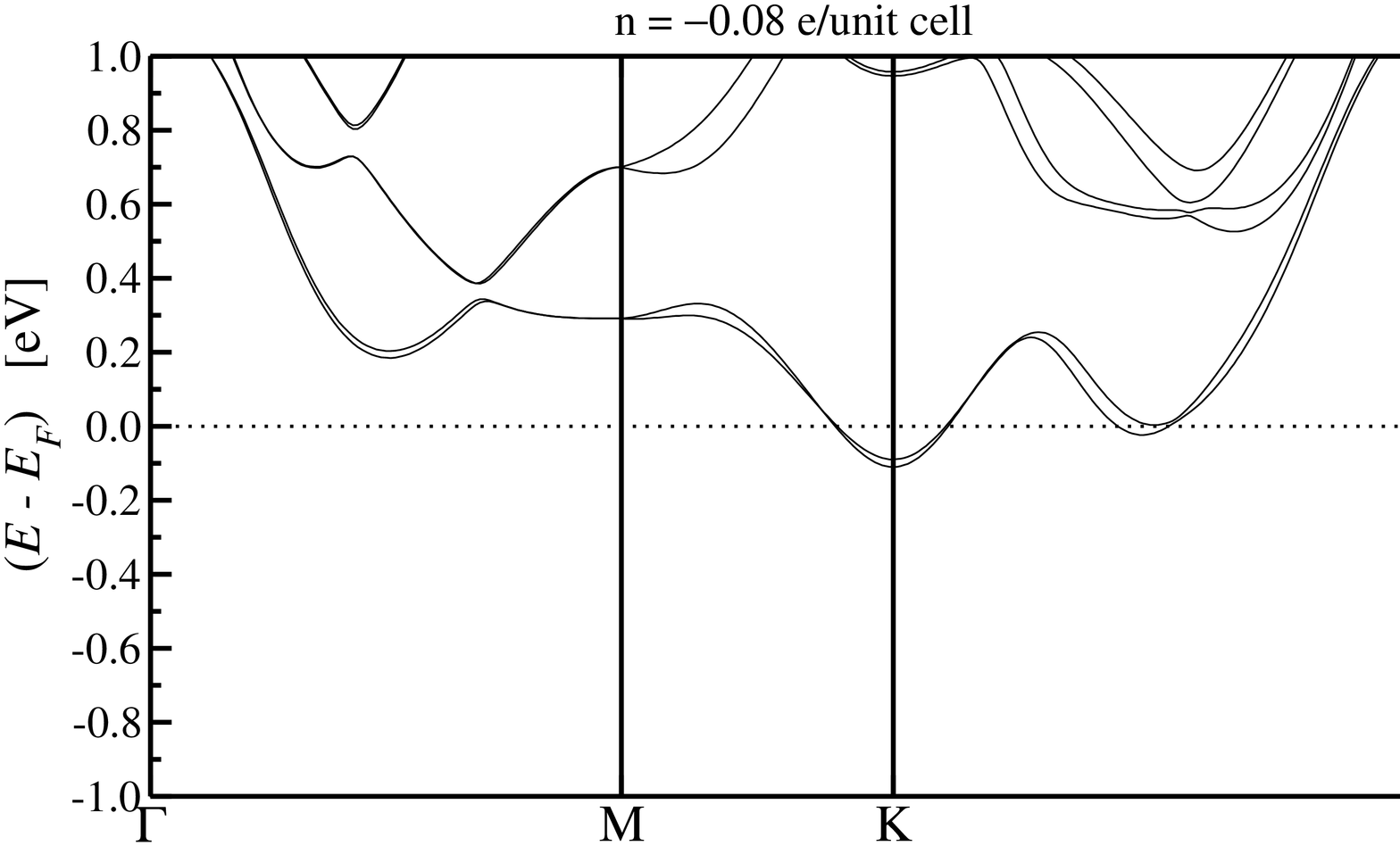}
 \includegraphics[width=0.31\textwidth,clip=,angle=-90]{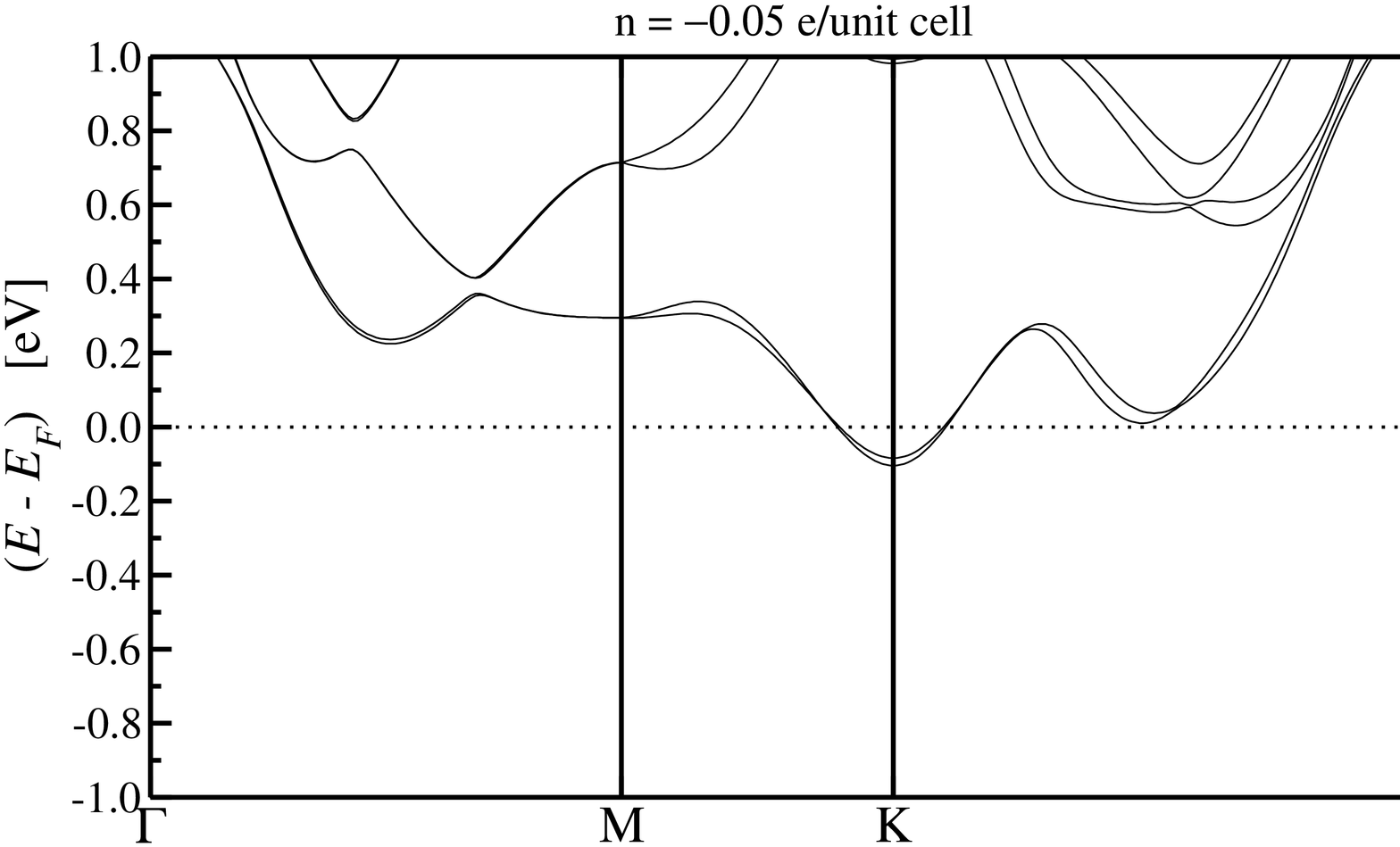}
 \includegraphics[width=0.31\textwidth,clip=,angle=-90]{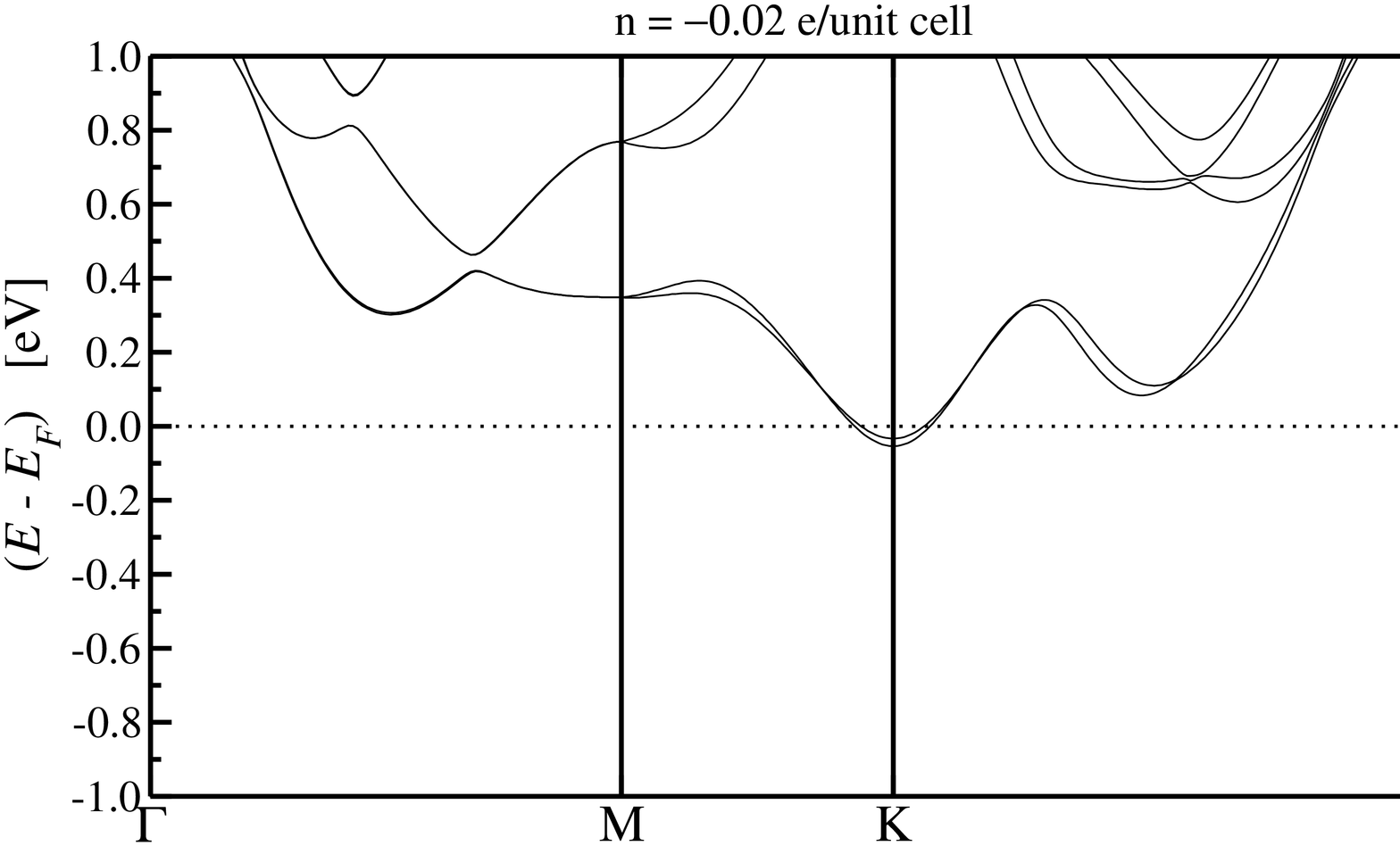}
 \includegraphics[width=0.31\textwidth,clip=,angle=-90]{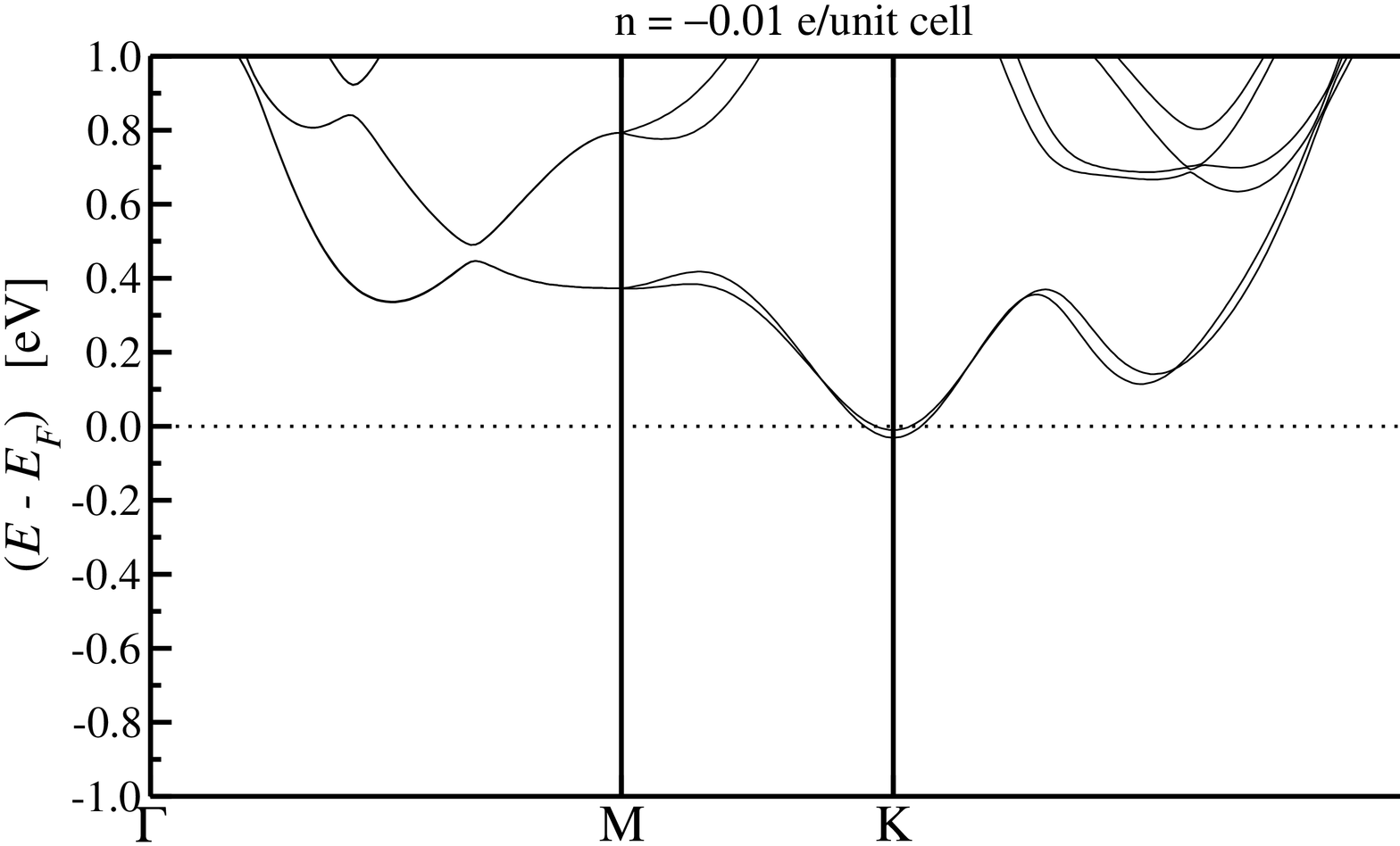}
 \caption{Band structure of monolayer MoSe$_2$ for different doping as indicated in the labels.}
\end{figure*}
\begin{figure*}[hbp]
 \centering
 \includegraphics[width=0.31\textwidth,clip=,angle=-90]{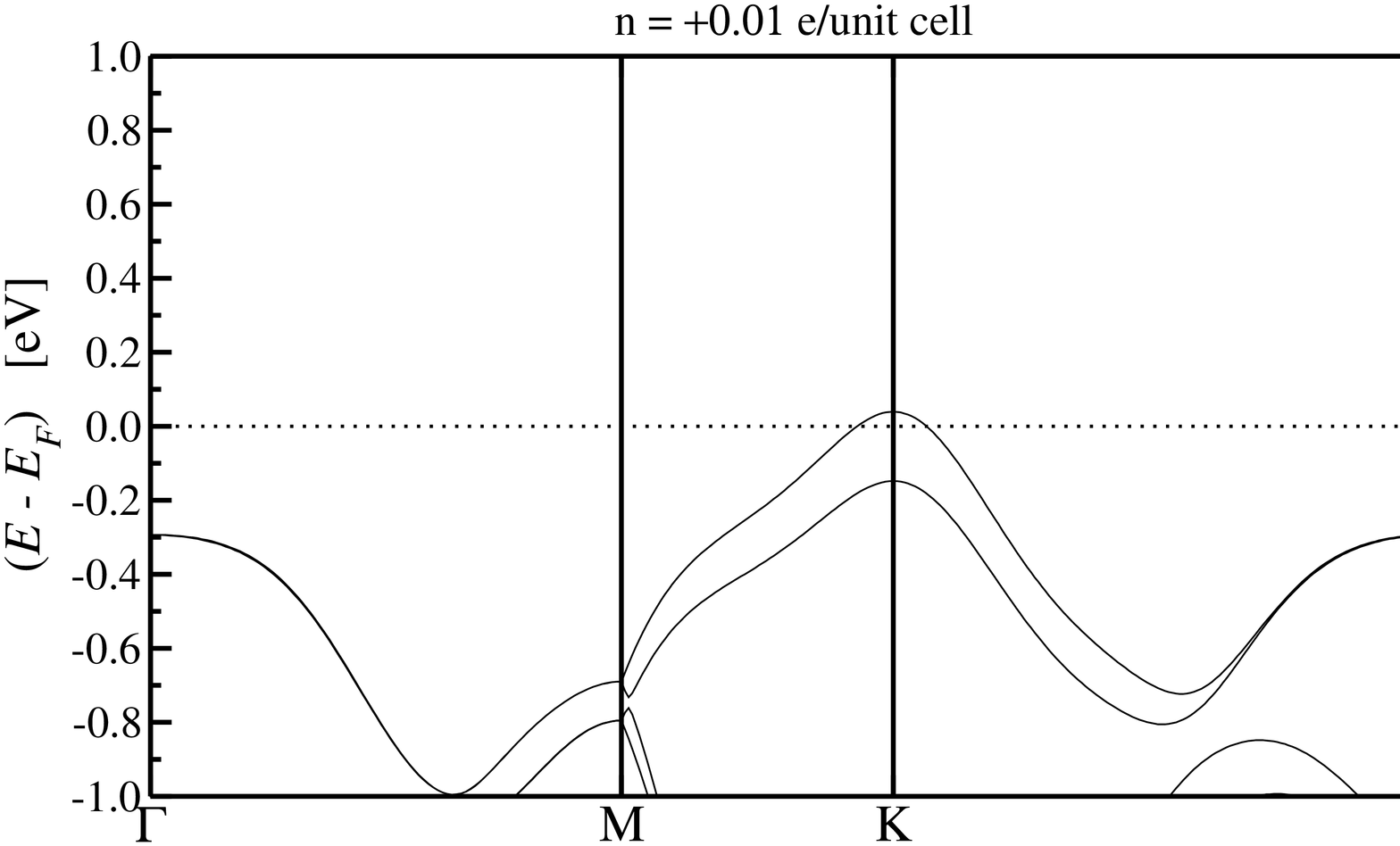}
 \includegraphics[width=0.31\textwidth,clip=,angle=-90]{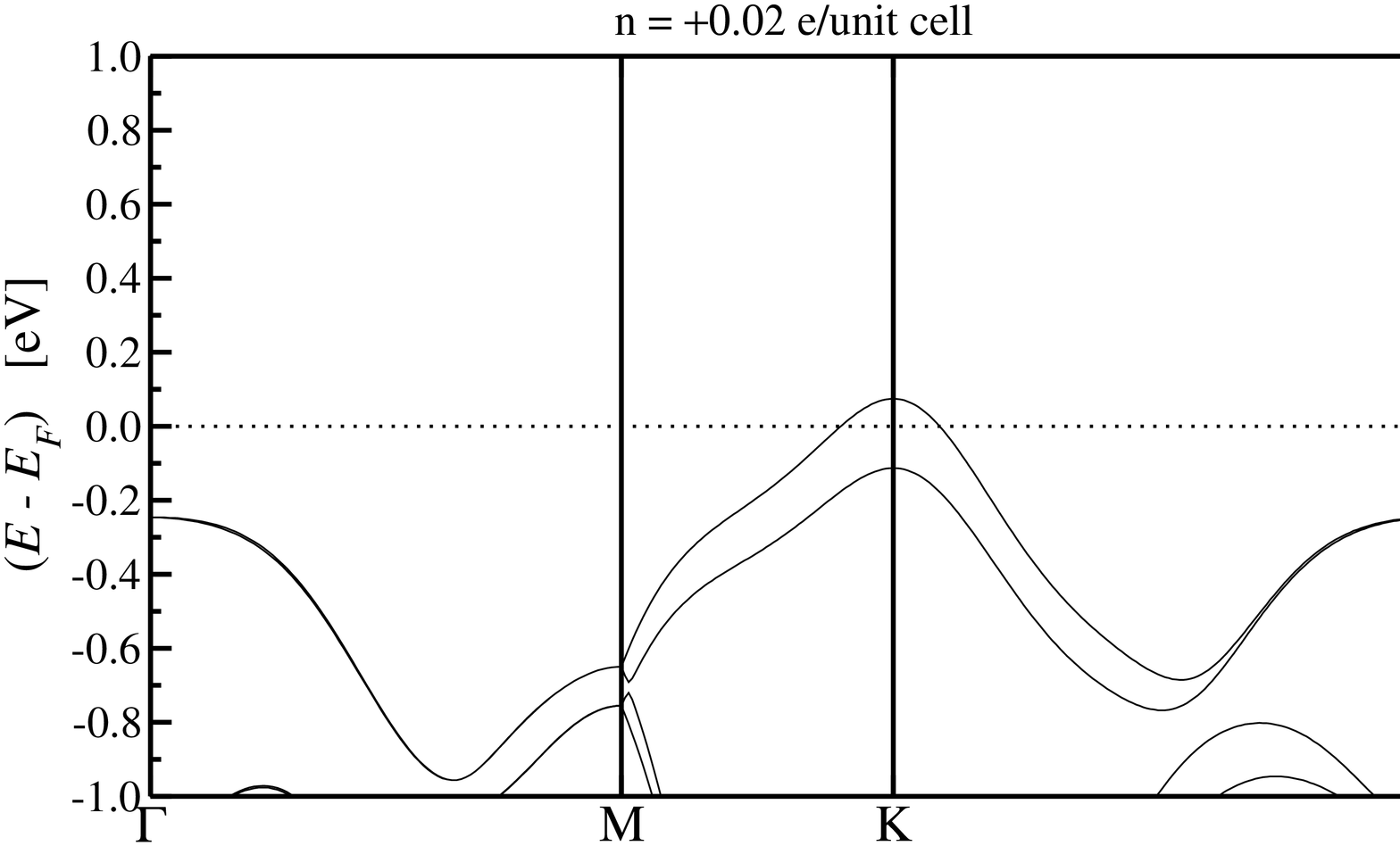}
 \includegraphics[width=0.31\textwidth,clip=,angle=-90]{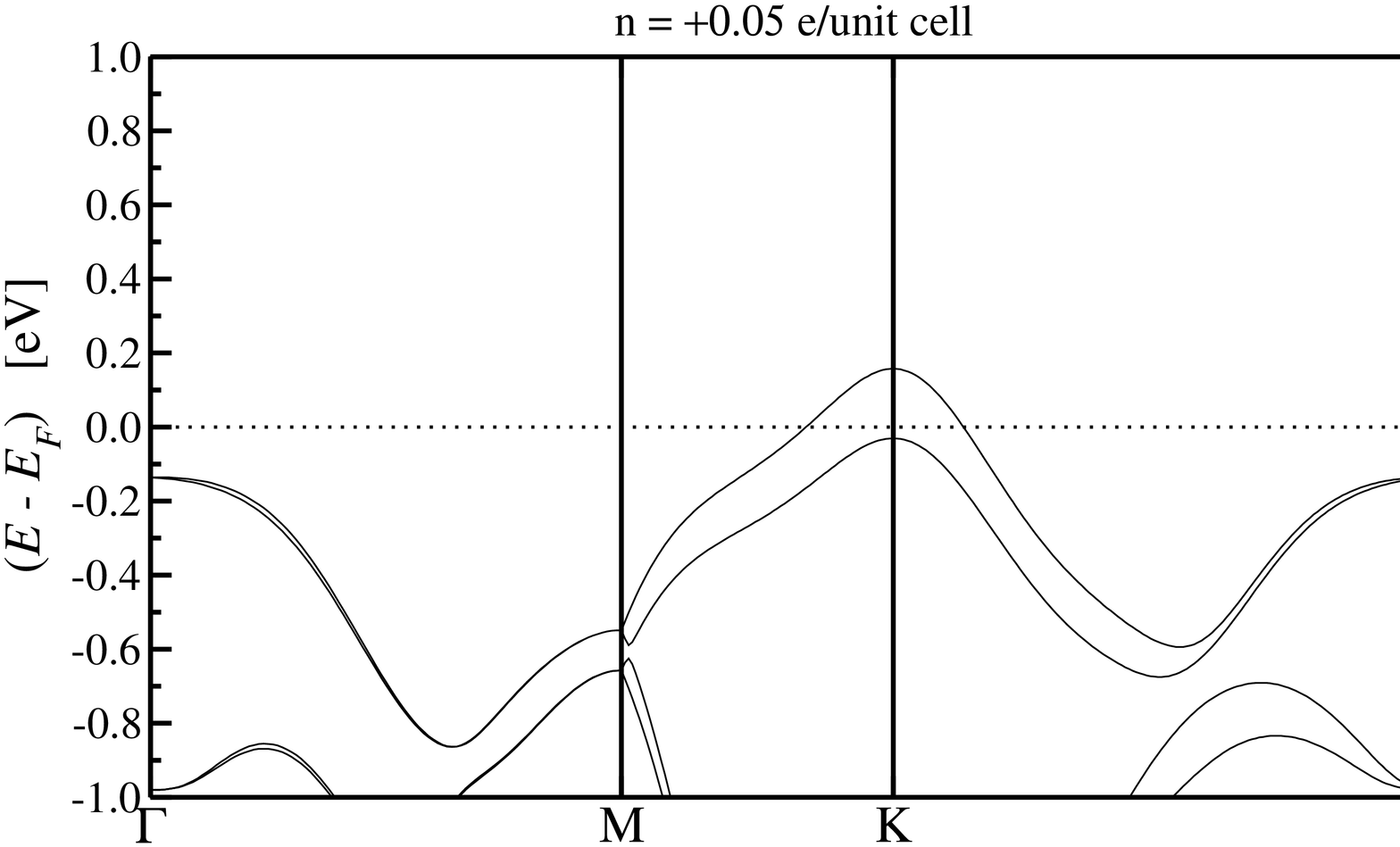}
 \includegraphics[width=0.31\textwidth,clip=,angle=-90]{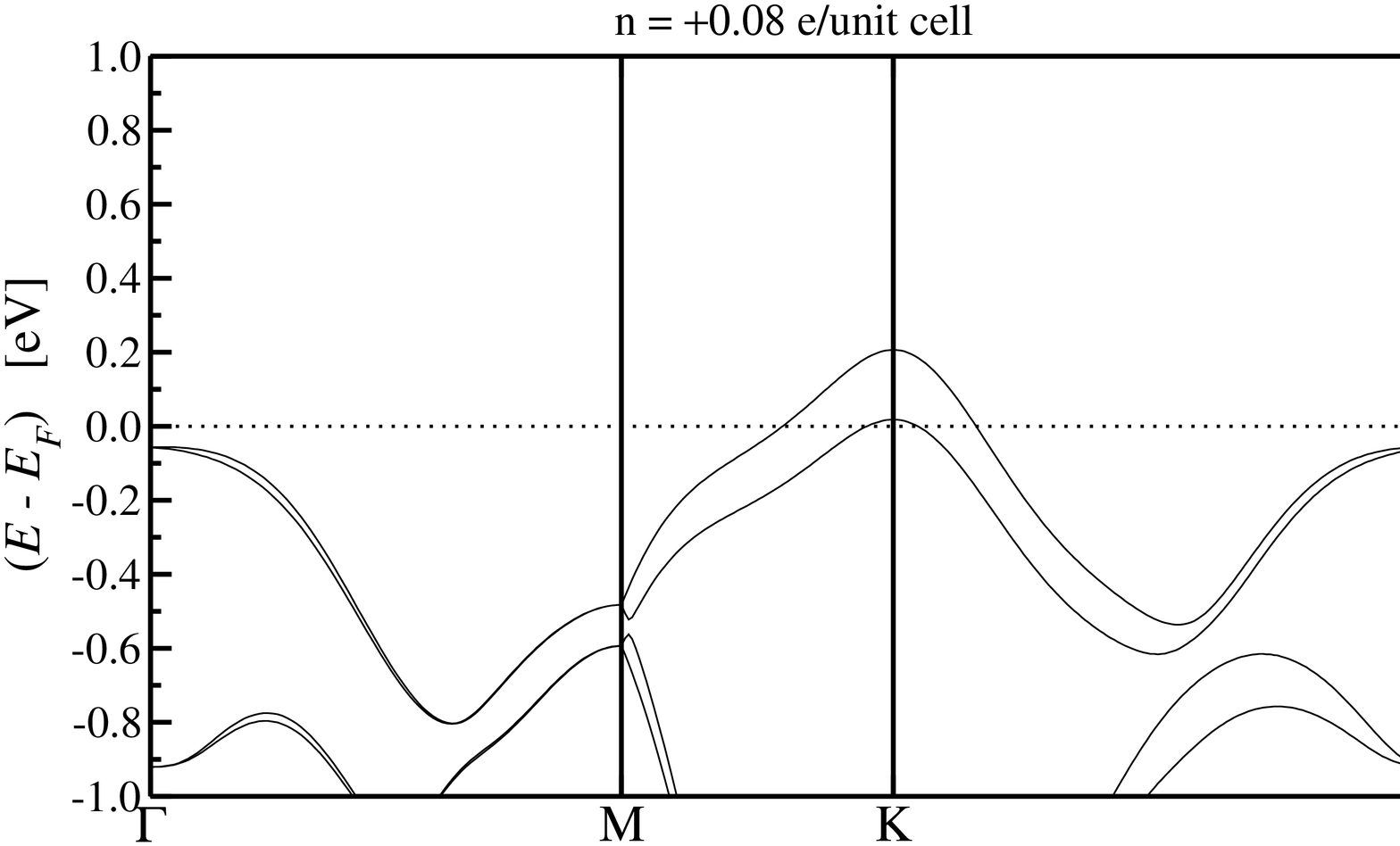}
 \includegraphics[width=0.31\textwidth,clip=,angle=-90]{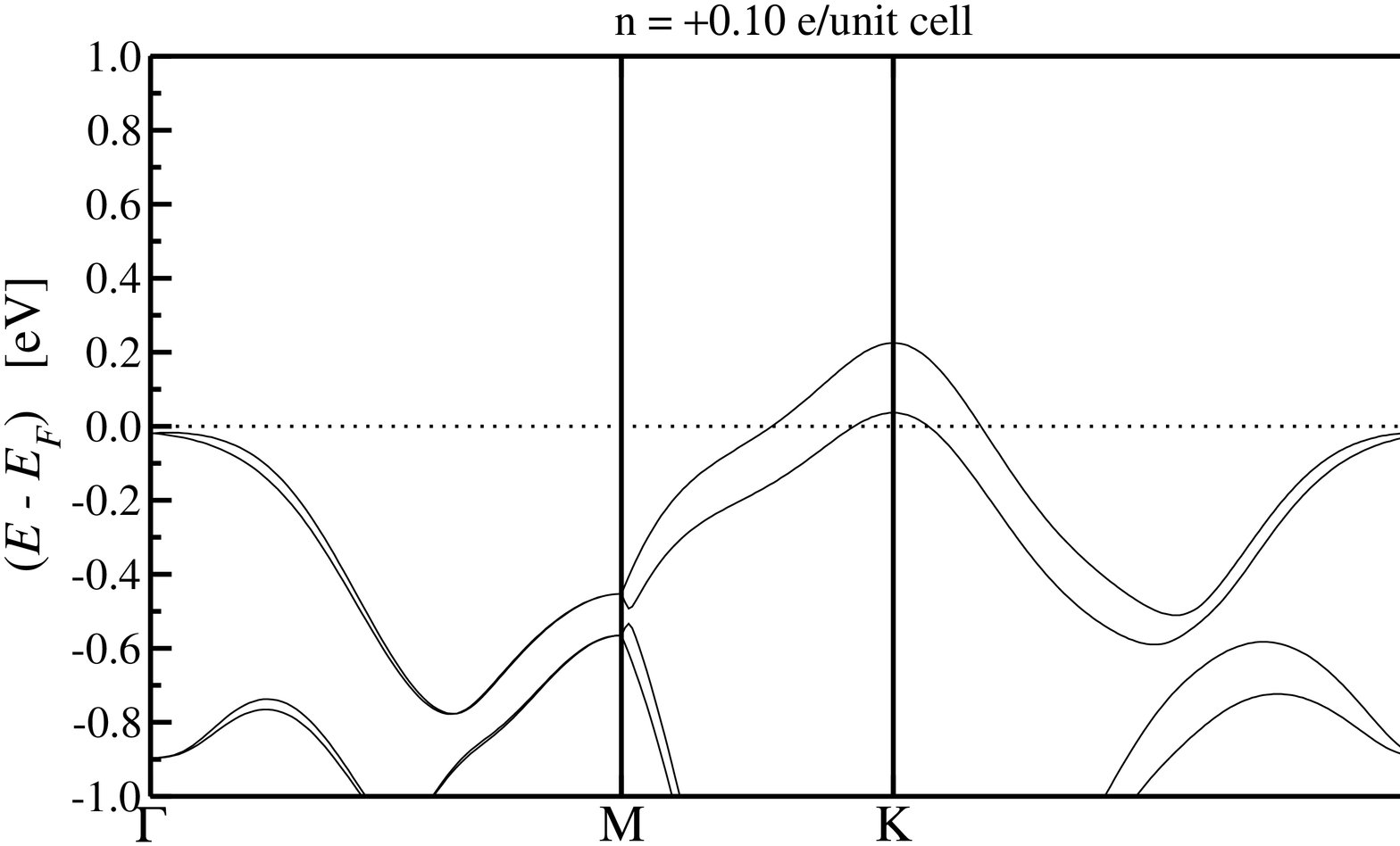}
 \includegraphics[width=0.31\textwidth,clip=,angle=-90]{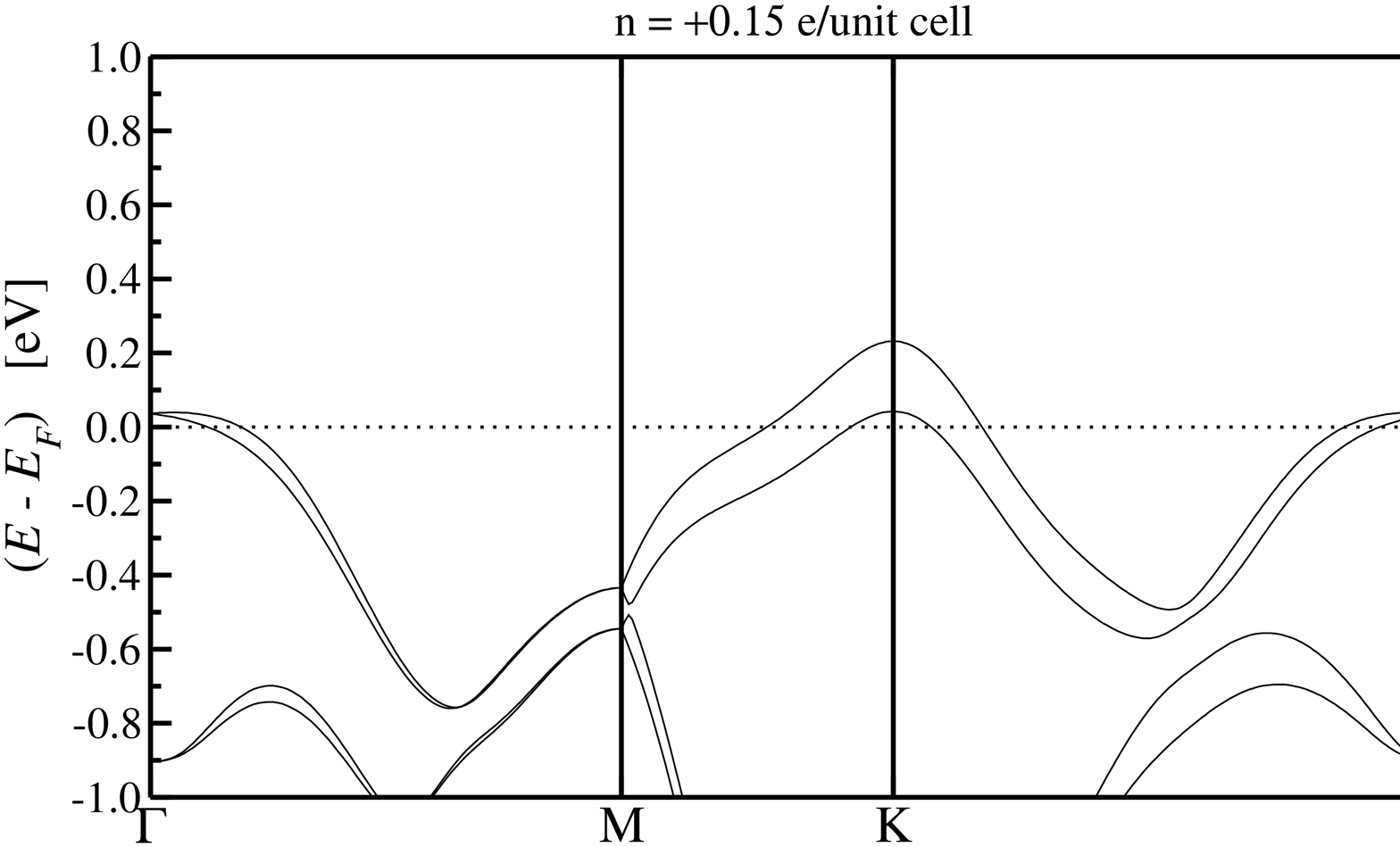}
 \includegraphics[width=0.31\textwidth,clip=,angle=-90]{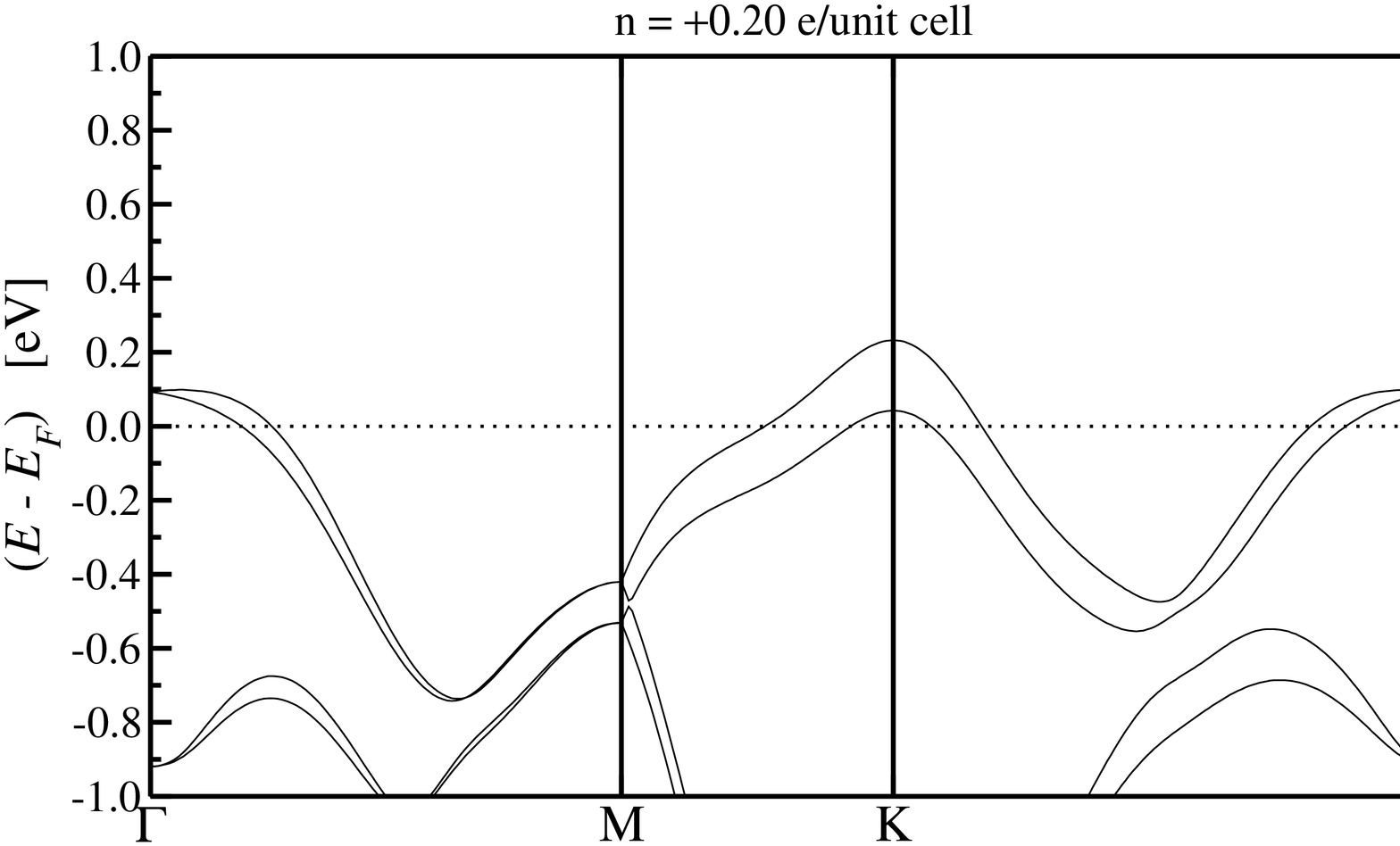}
 \includegraphics[width=0.31\textwidth,clip=,angle=-90]{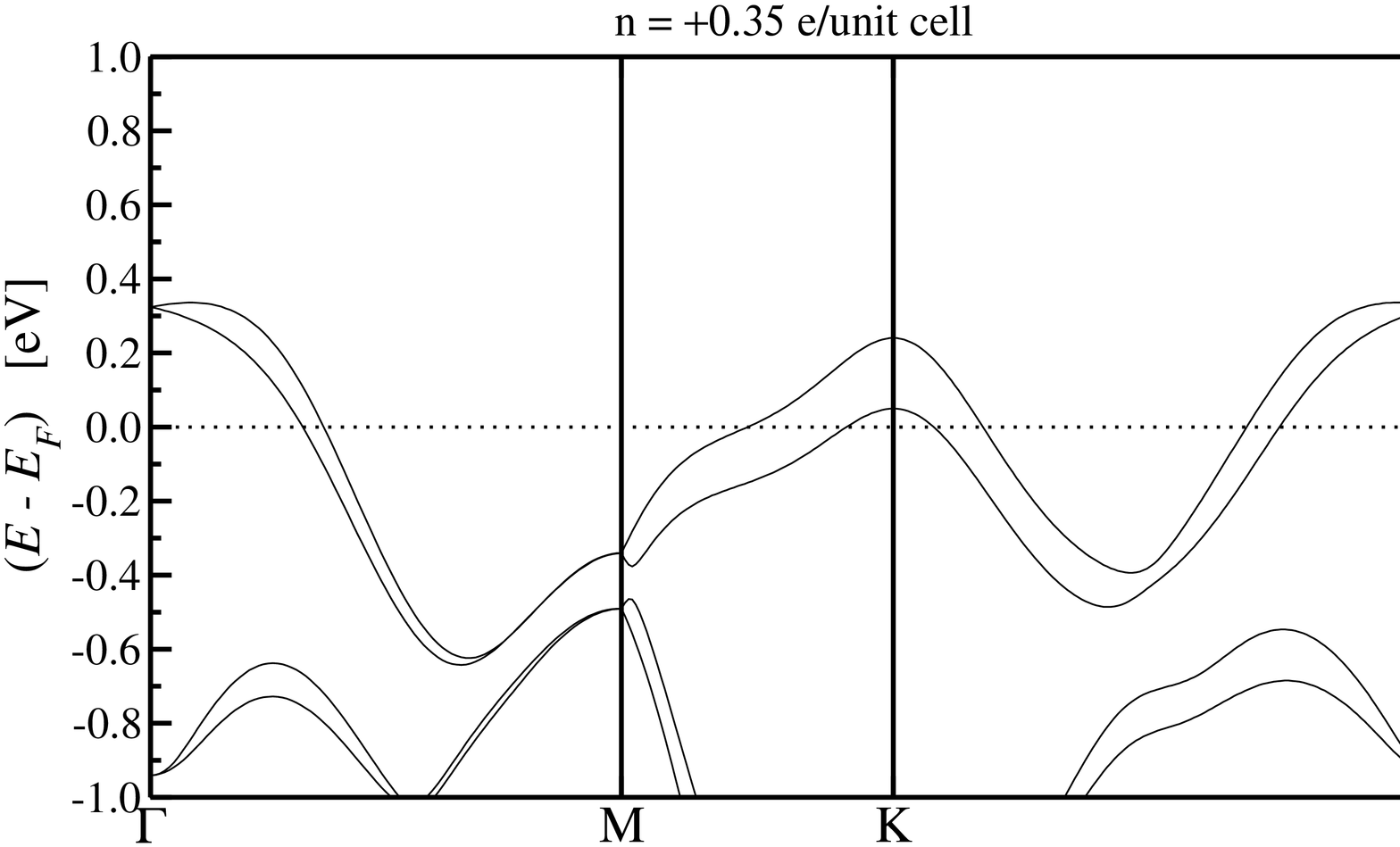}
 \caption{Band structure of monolayer MoSe$_2$ for different doping as indicated in the labels.}
\end{figure*}
\begin{figure*}[hbp]
 \centering
 \includegraphics[width=0.31\textwidth,clip=,angle=-90]{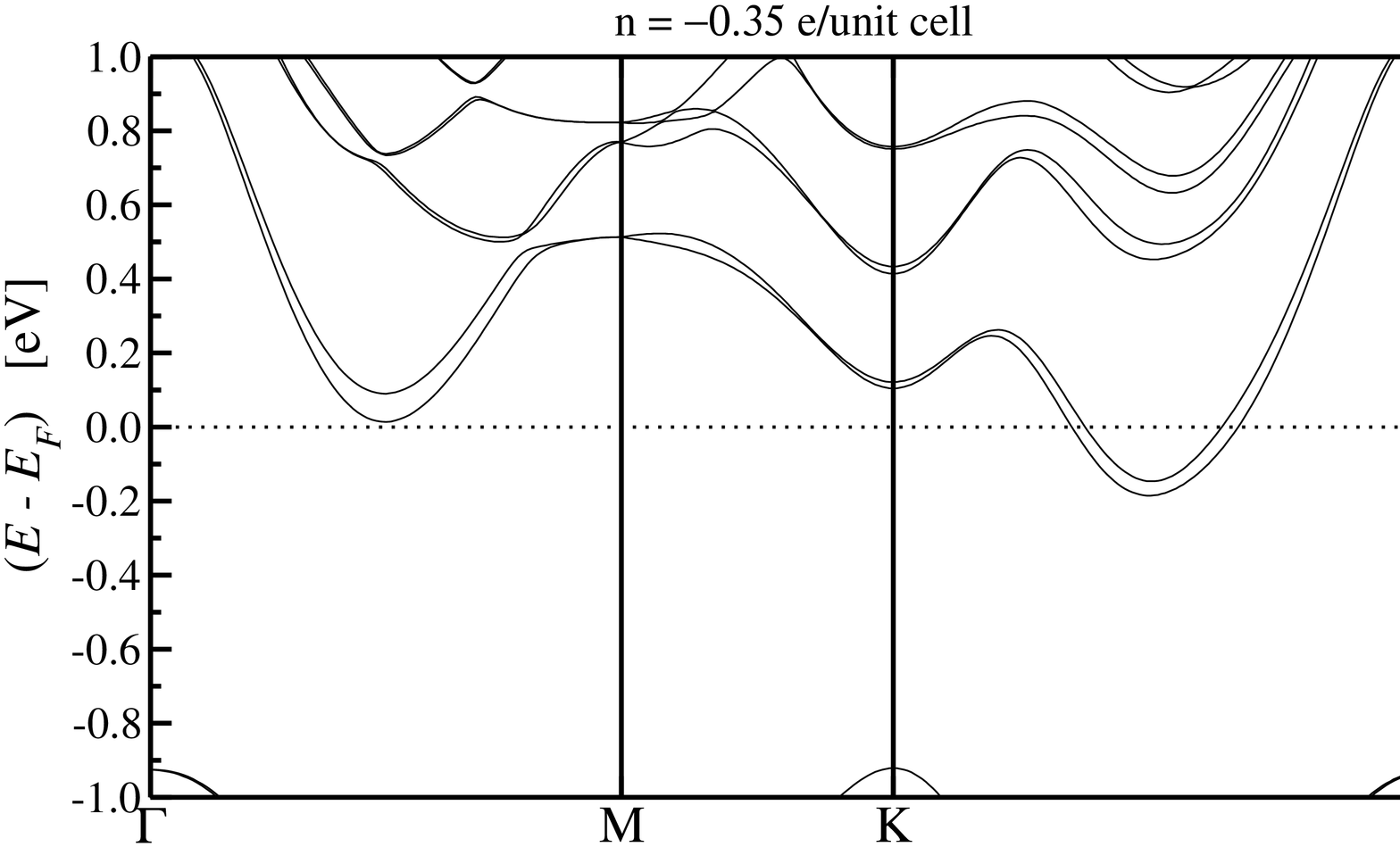}
 \includegraphics[width=0.31\textwidth,clip=,angle=-90]{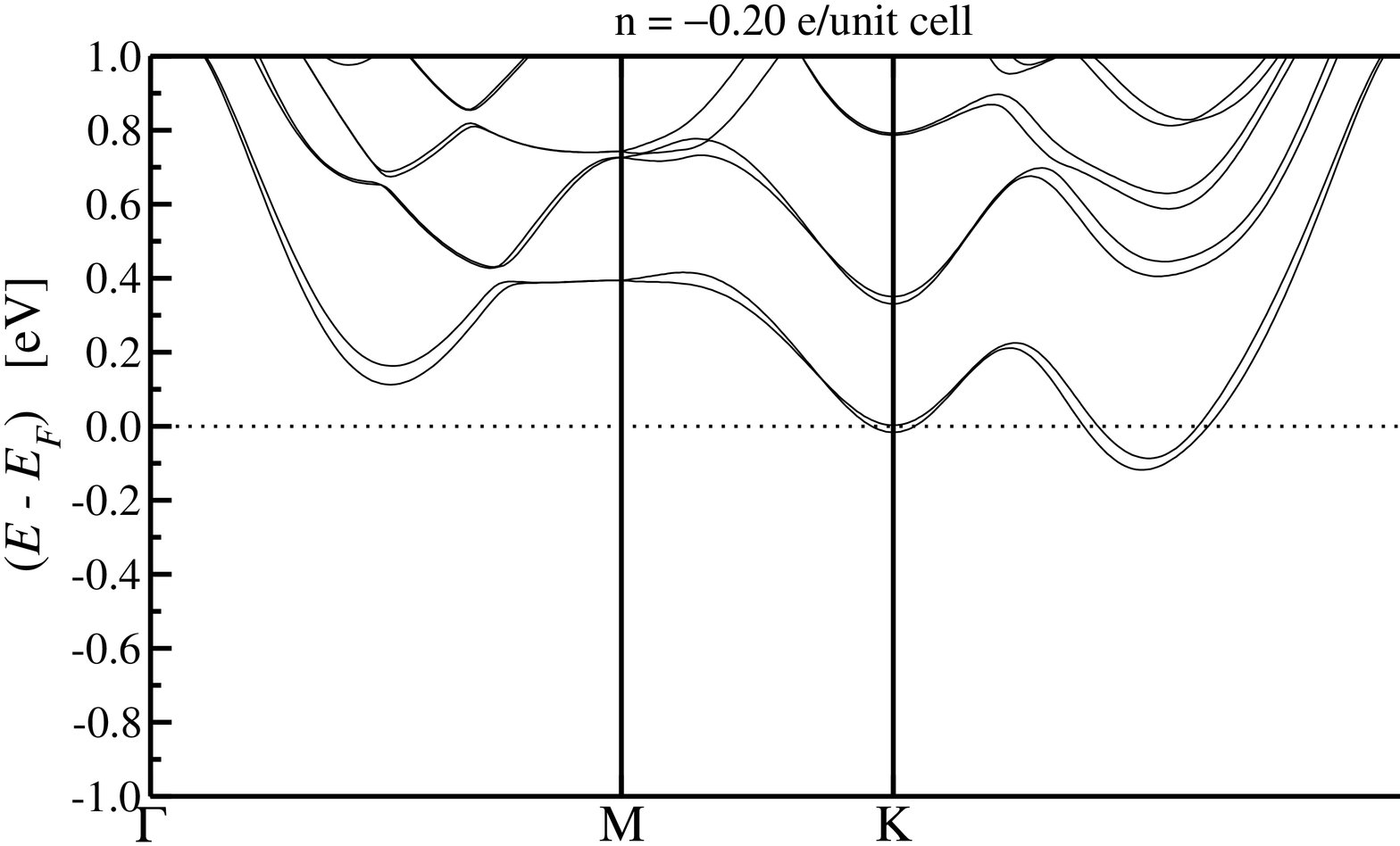}
 \includegraphics[width=0.31\textwidth,clip=,angle=-90]{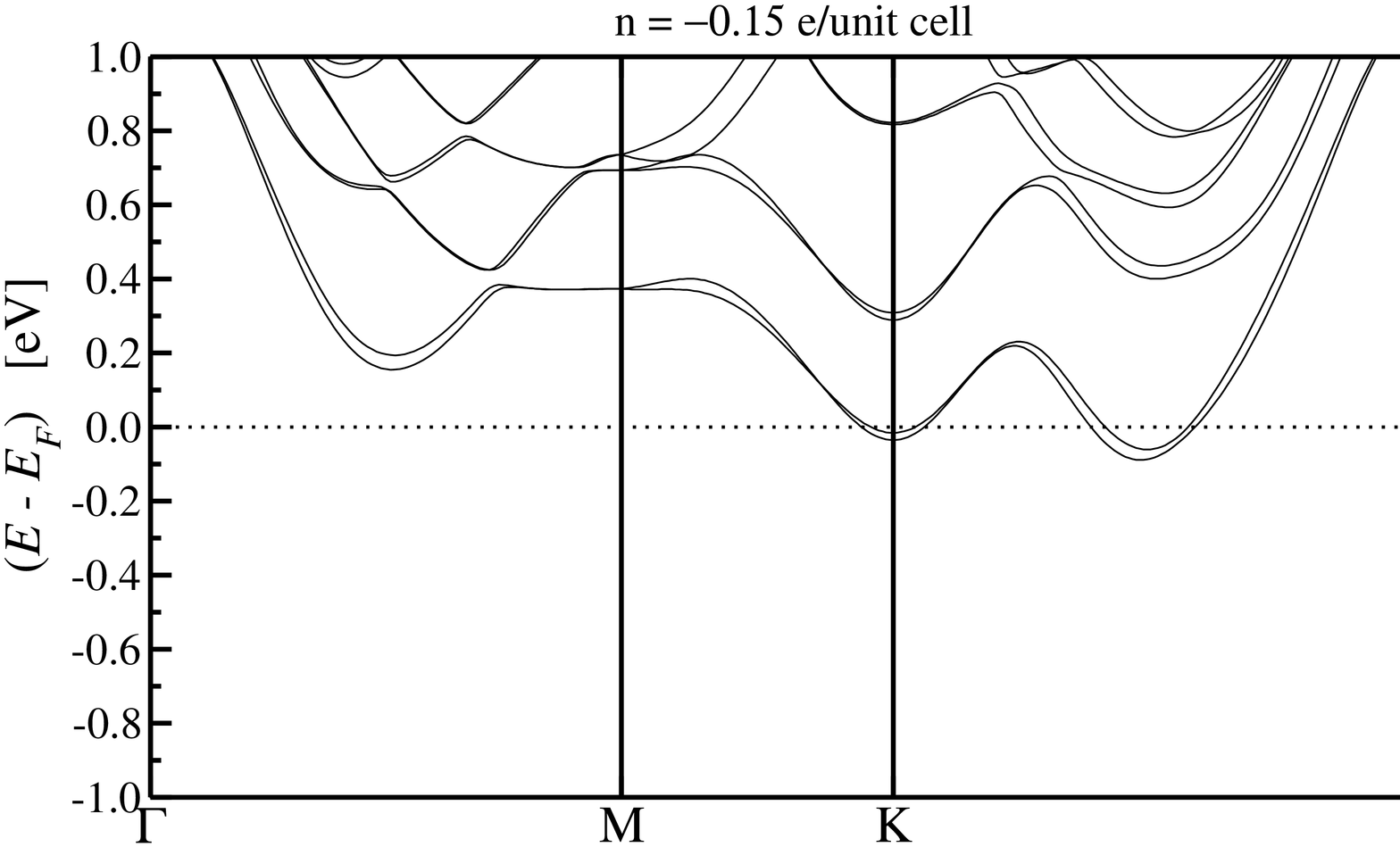}
 \includegraphics[width=0.31\textwidth,clip=,angle=-90]{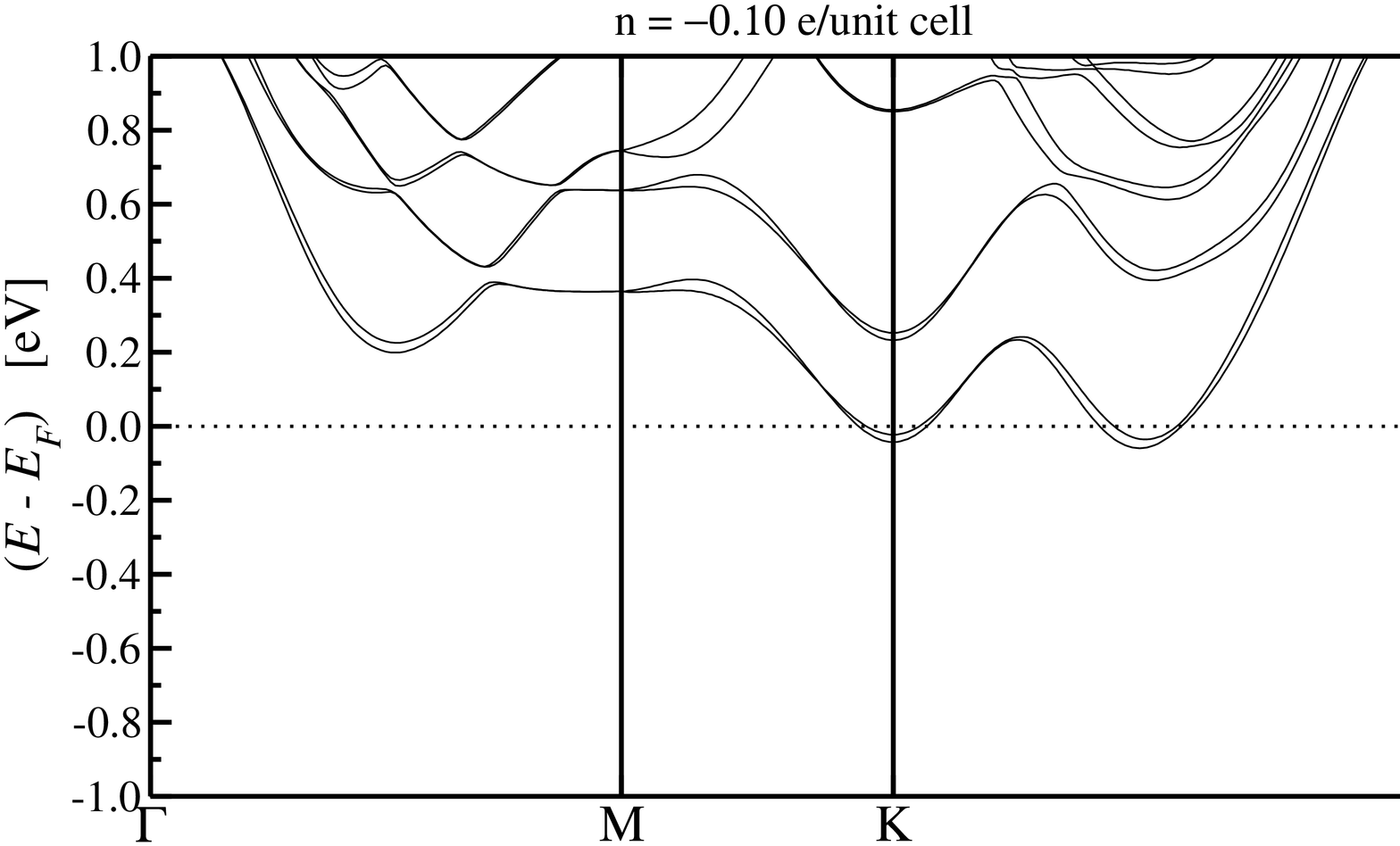}
 \includegraphics[width=0.31\textwidth,clip=,angle=-90]{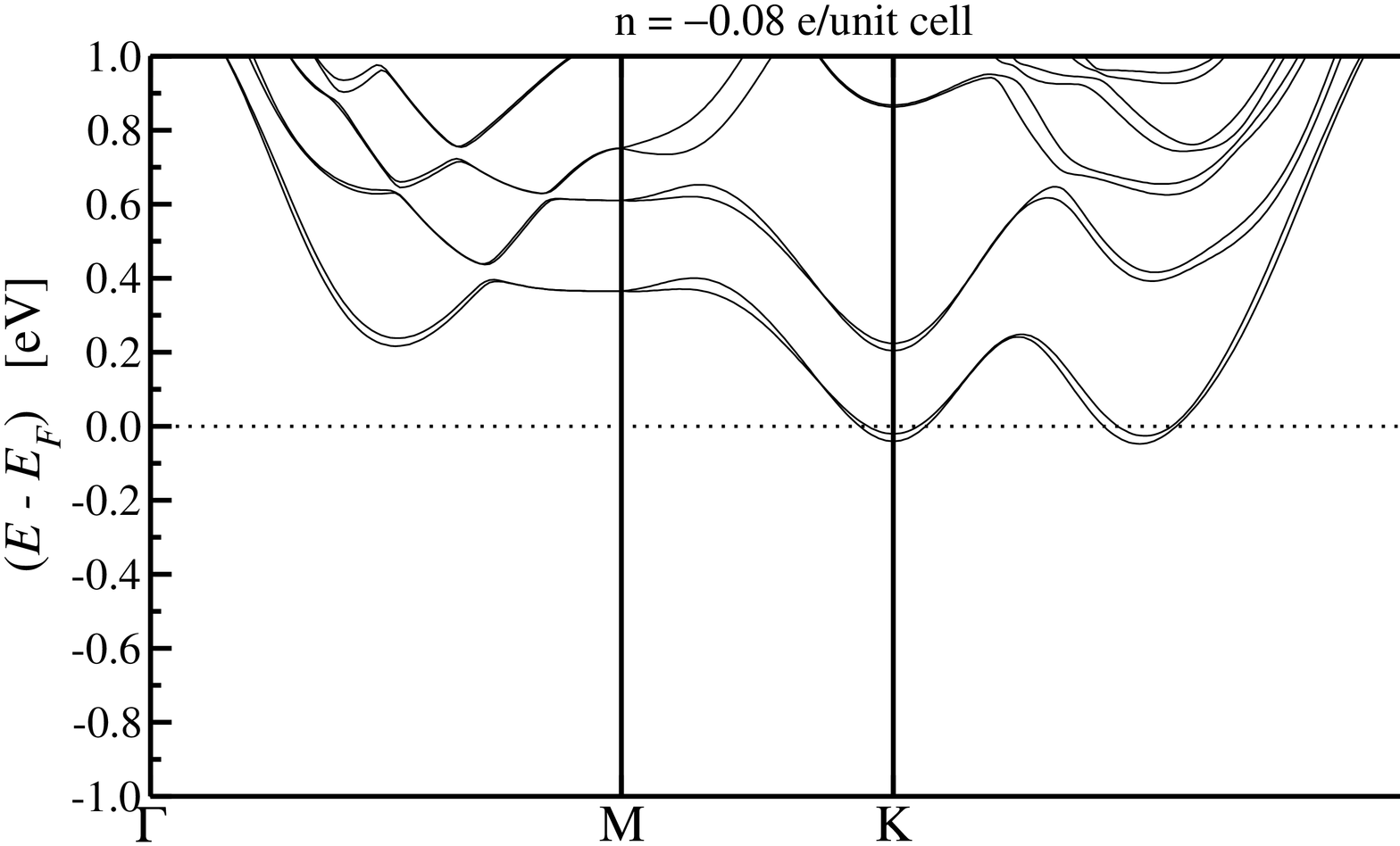}
 \includegraphics[width=0.31\textwidth,clip=,angle=-90]{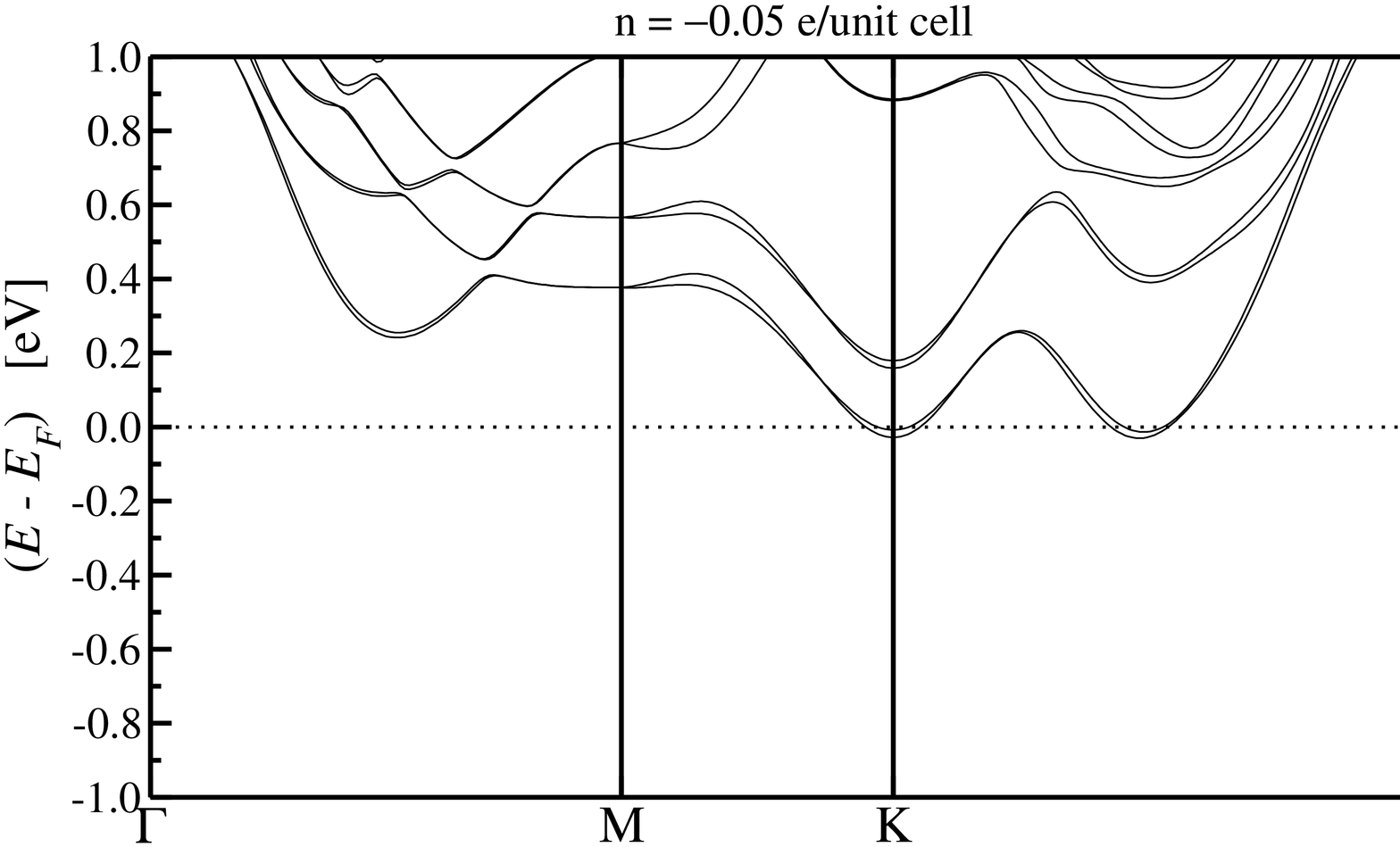}
 \includegraphics[width=0.31\textwidth,clip=,angle=-90]{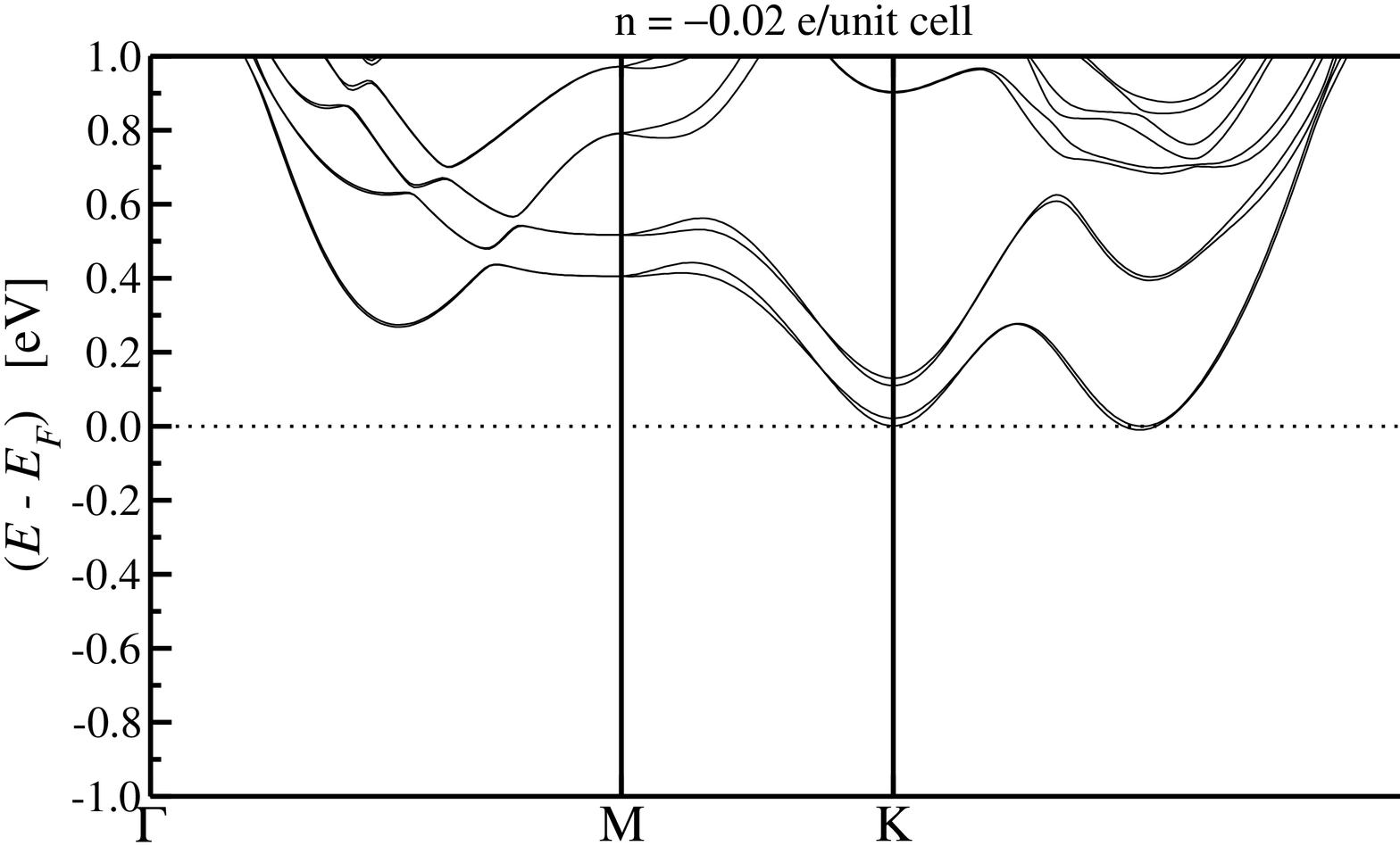}
 \includegraphics[width=0.31\textwidth,clip=,angle=-90]{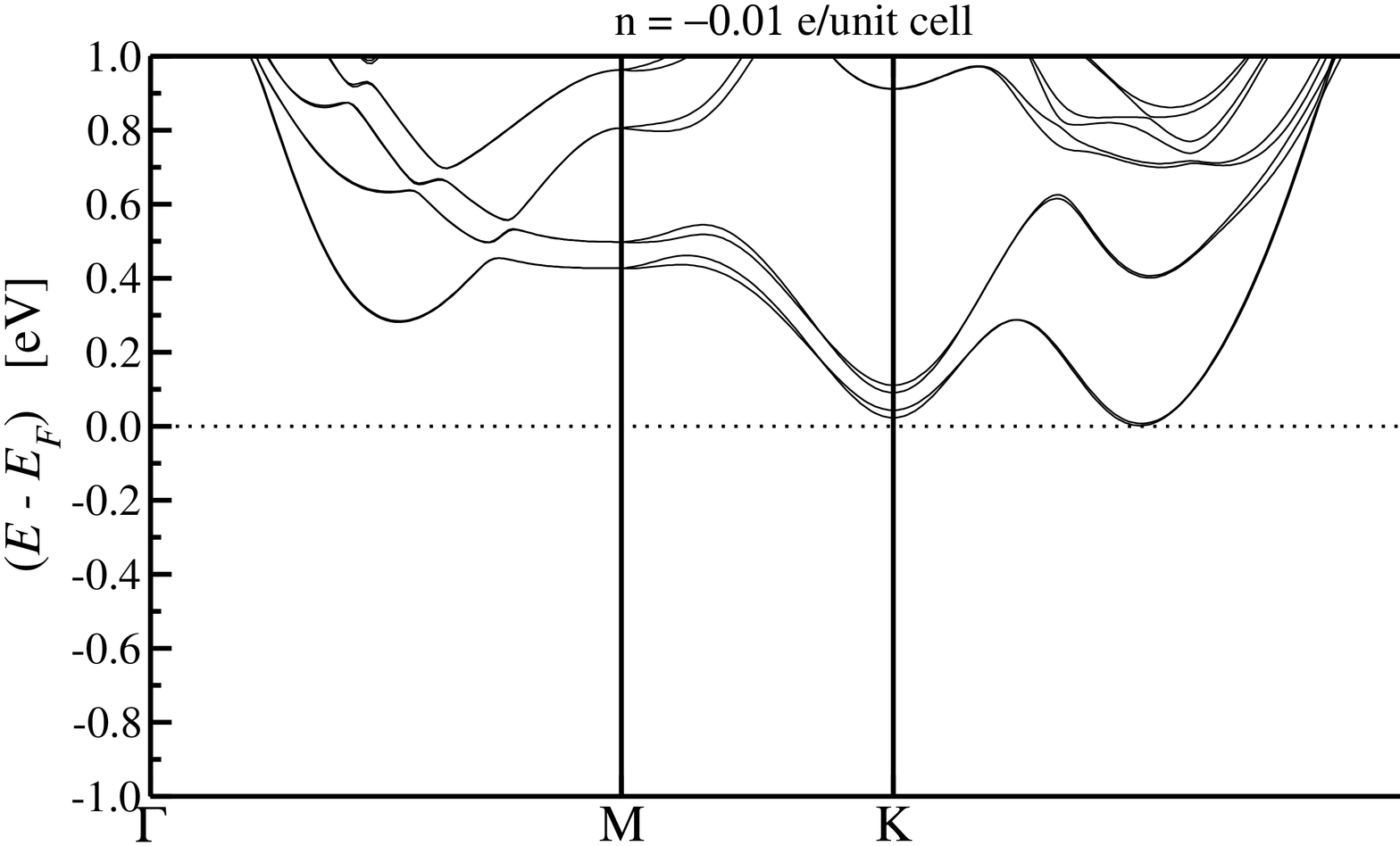}
 \caption{Band structure of bilayer MoSe$_2$ for different doping as indicated in the labels.}
\end{figure*}
\begin{figure*}[hbp]
 \centering
 \includegraphics[width=0.31\textwidth,clip=,angle=-90]{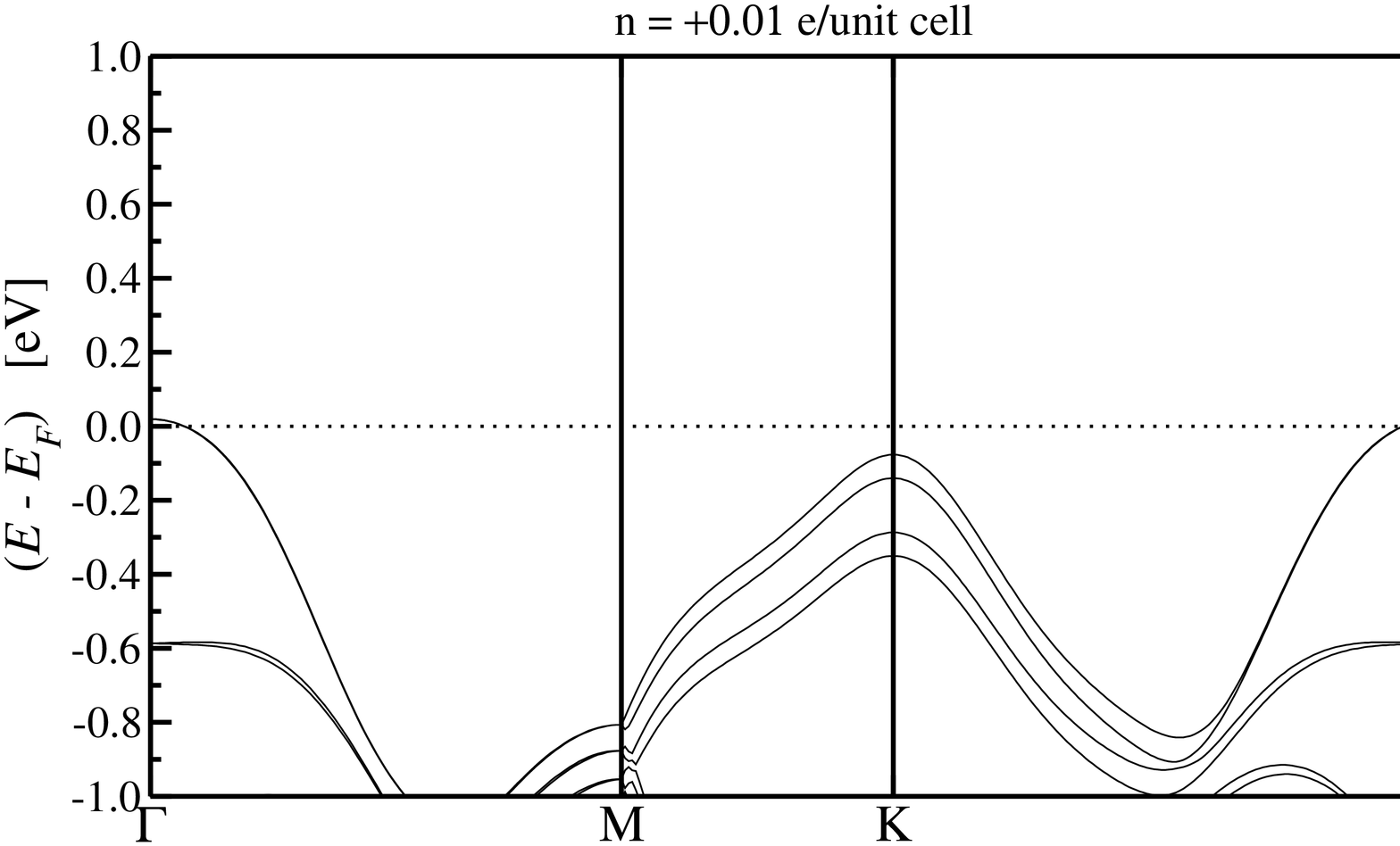}
 \includegraphics[width=0.31\textwidth,clip=,angle=-90]{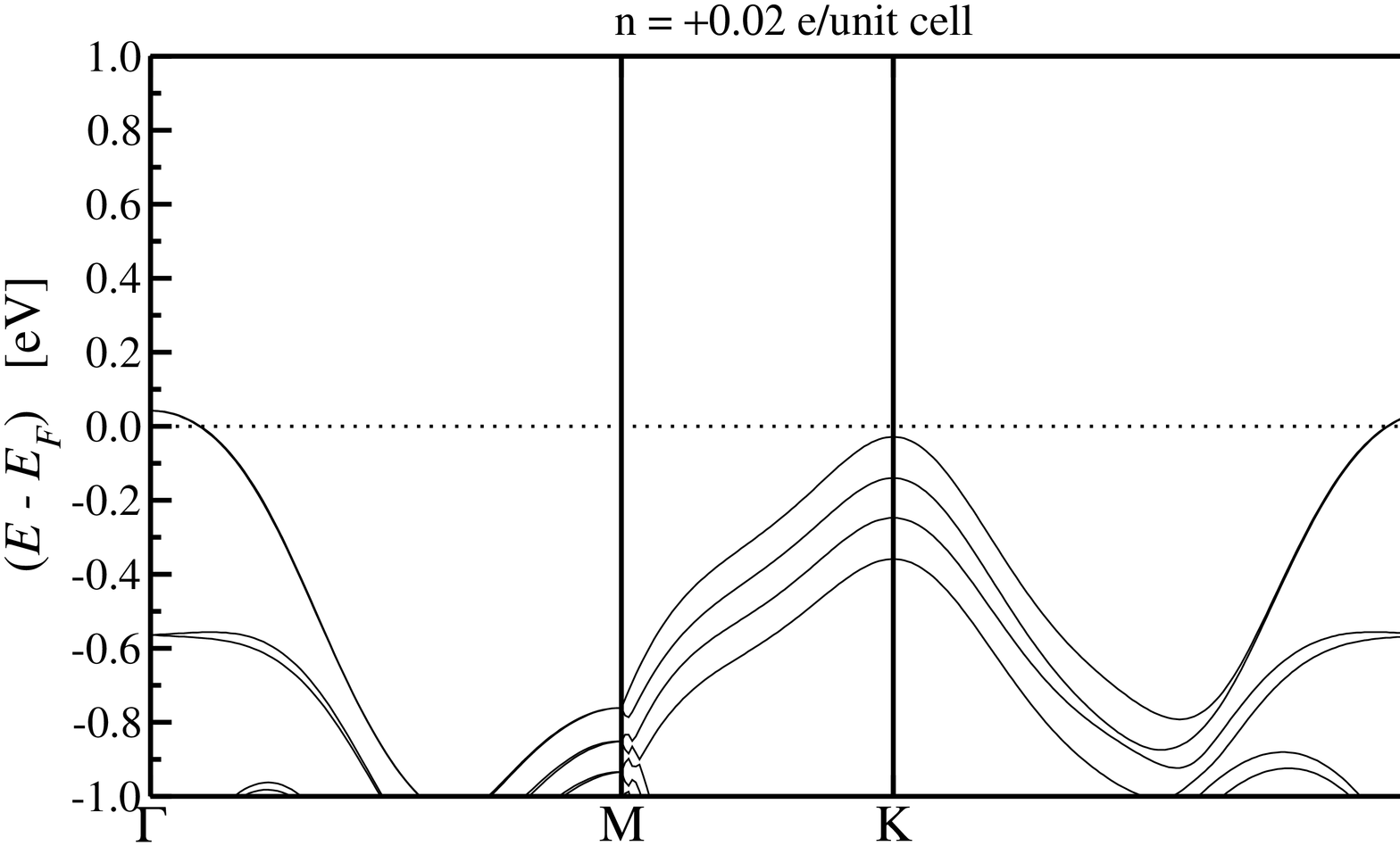}
 \includegraphics[width=0.31\textwidth,clip=,angle=-90]{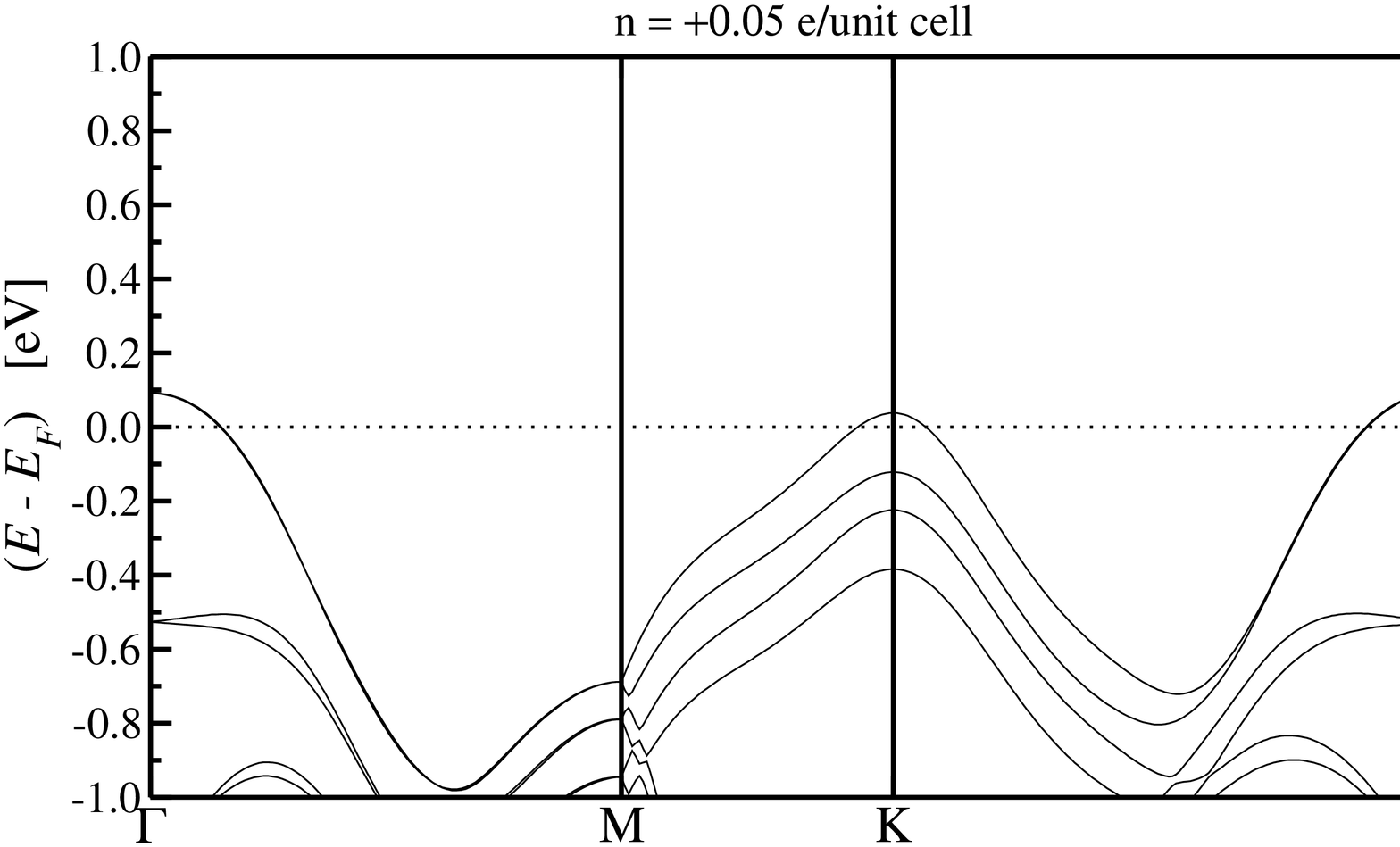}
 \includegraphics[width=0.31\textwidth,clip=,angle=-90]{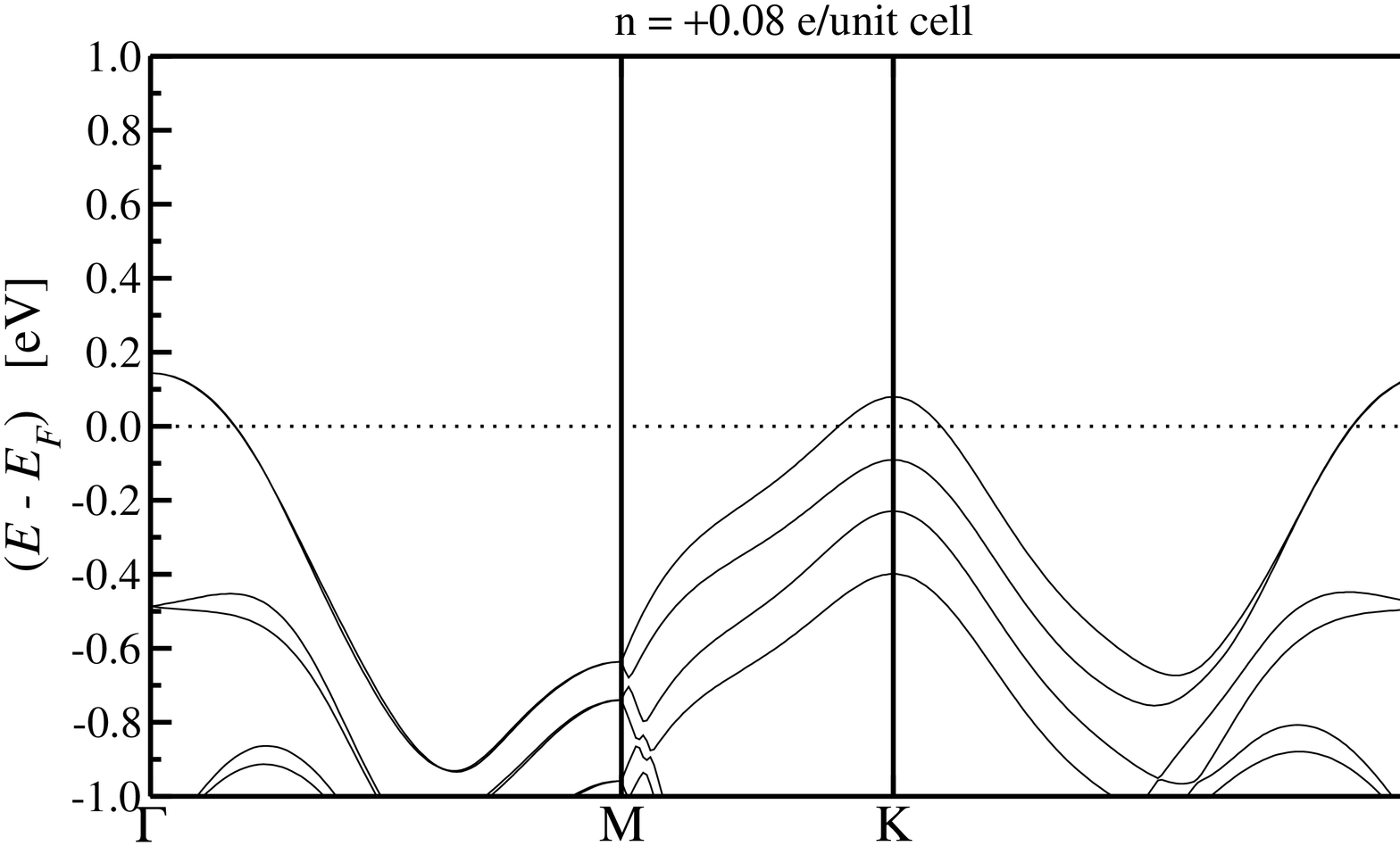}
 \includegraphics[width=0.31\textwidth,clip=,angle=-90]{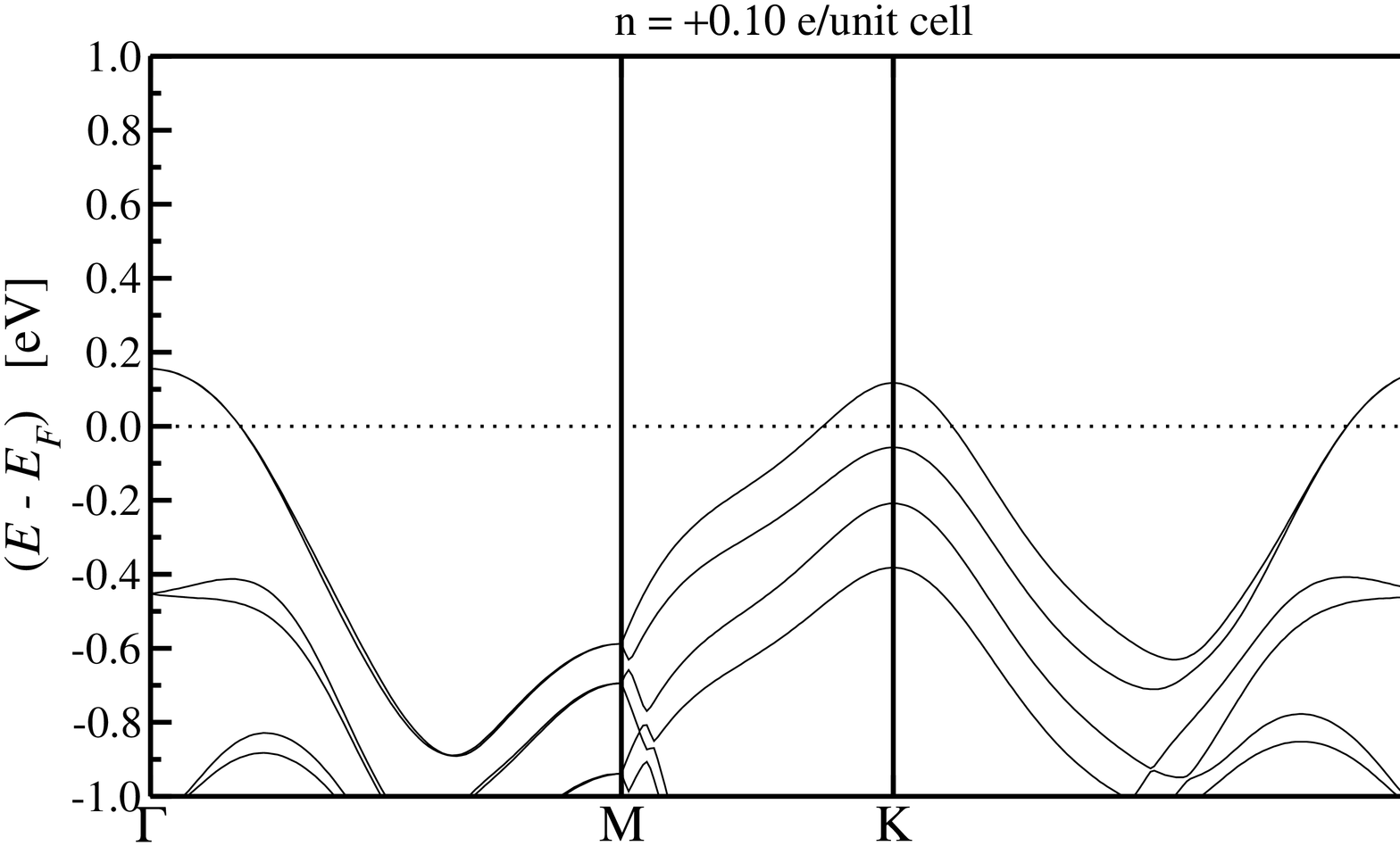}
 \includegraphics[width=0.31\textwidth,clip=,angle=-90]{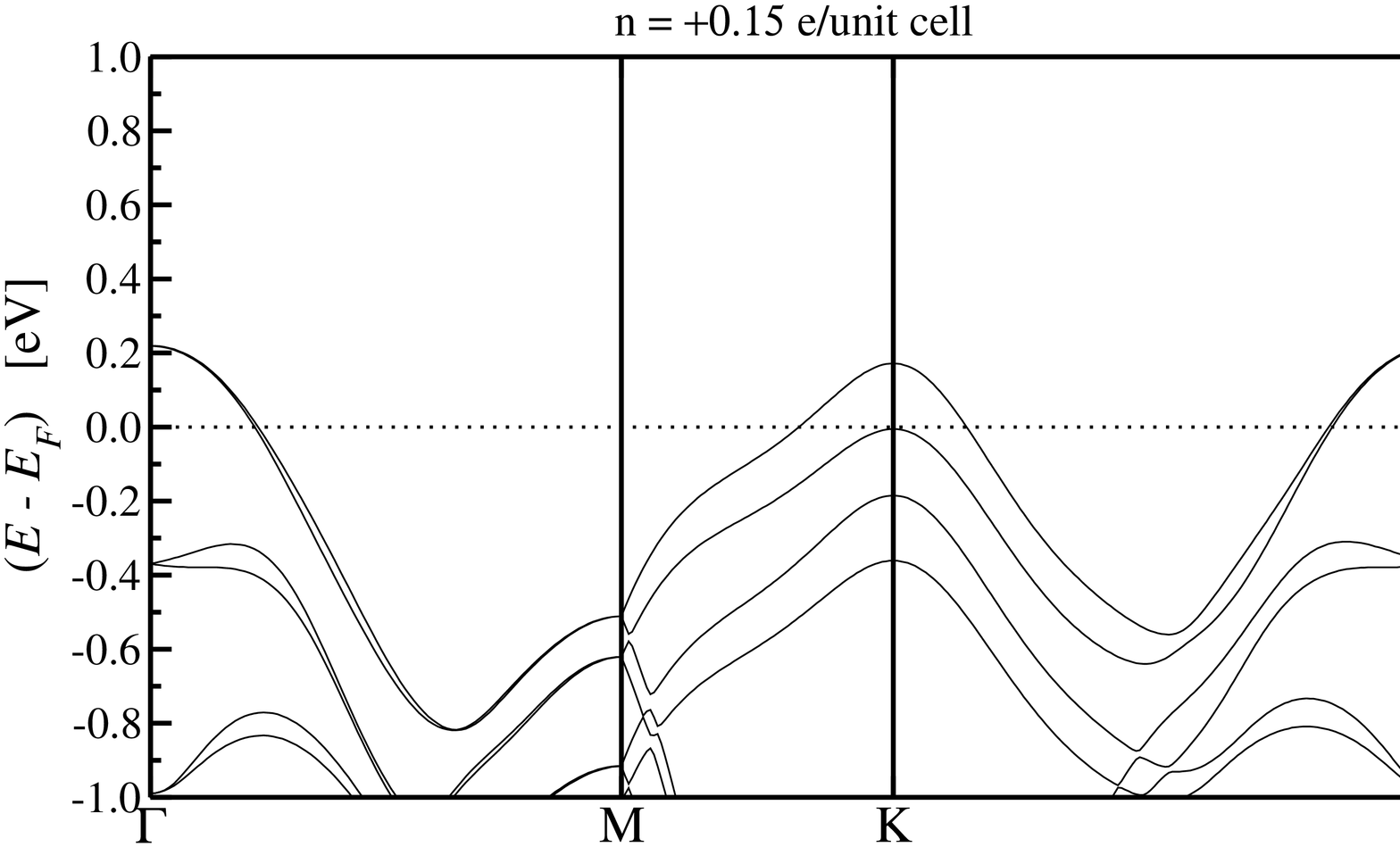}
 \includegraphics[width=0.31\textwidth,clip=,angle=-90]{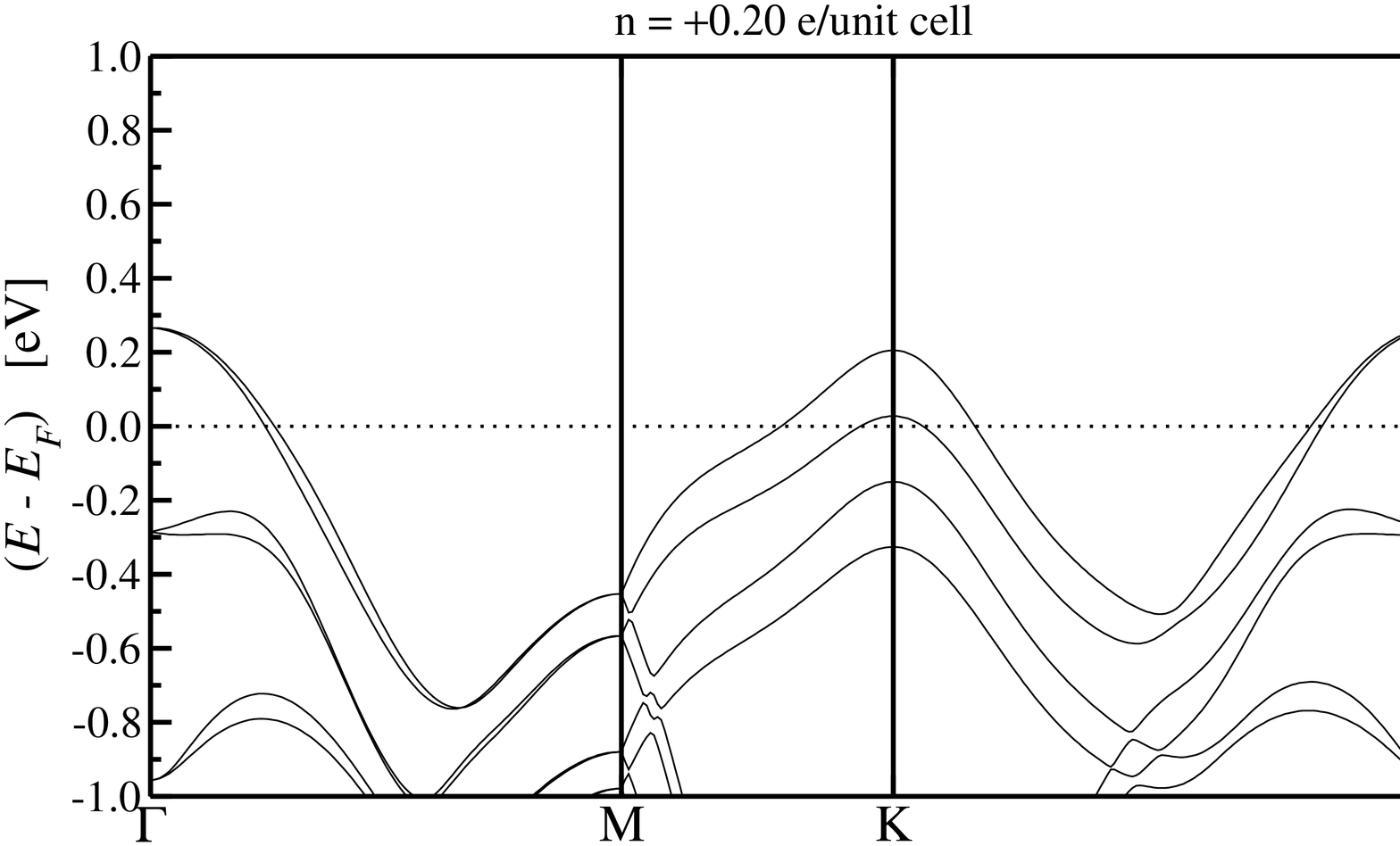}
 \includegraphics[width=0.31\textwidth,clip=,angle=-90]{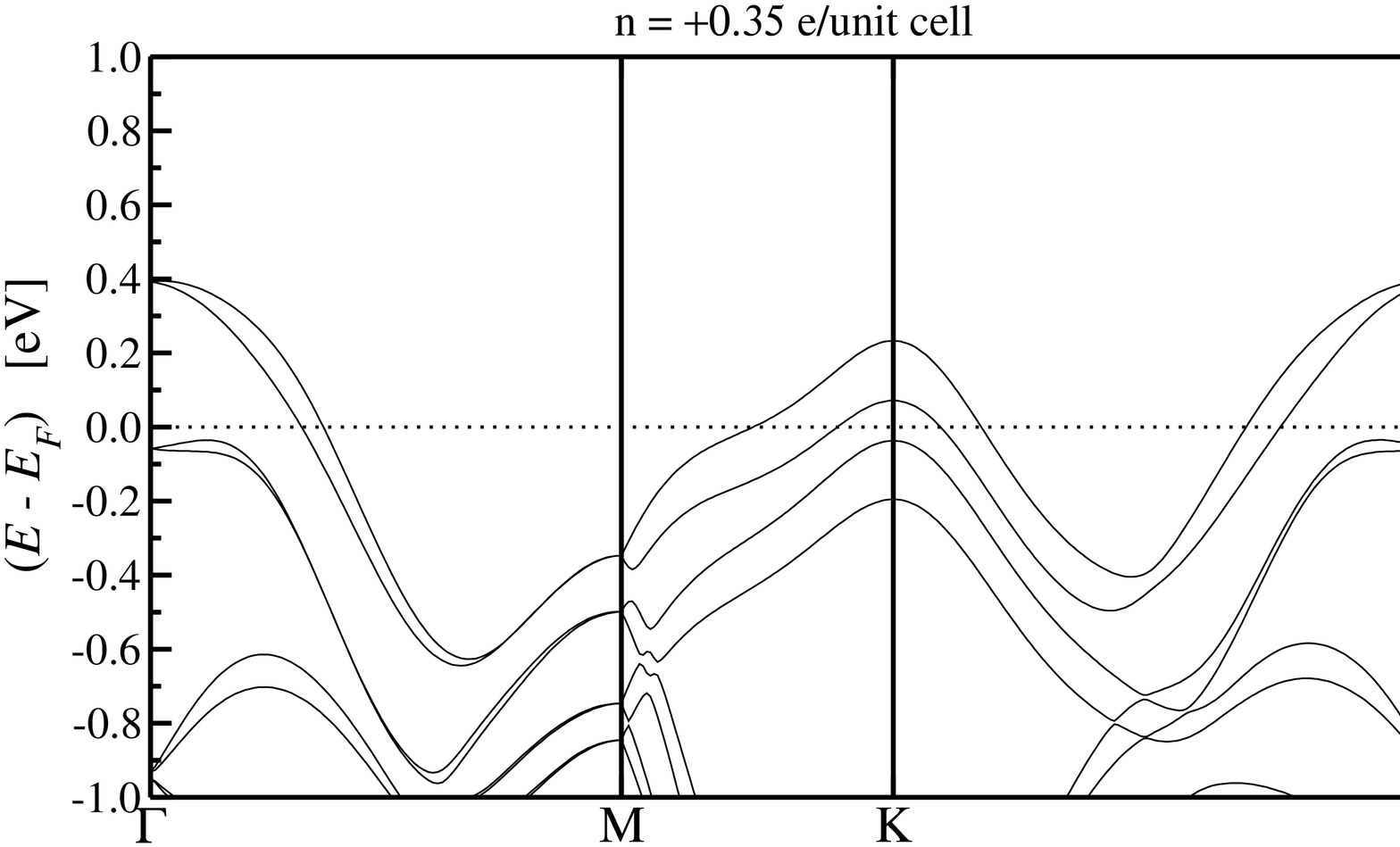}
 \caption{Band structure of bilayer MoSe$_2$ for different doping as indicated in the labels.}
\end{figure*}
\begin{figure*}[hbp]
 \centering
 \includegraphics[width=0.31\textwidth,clip=,angle=-90]{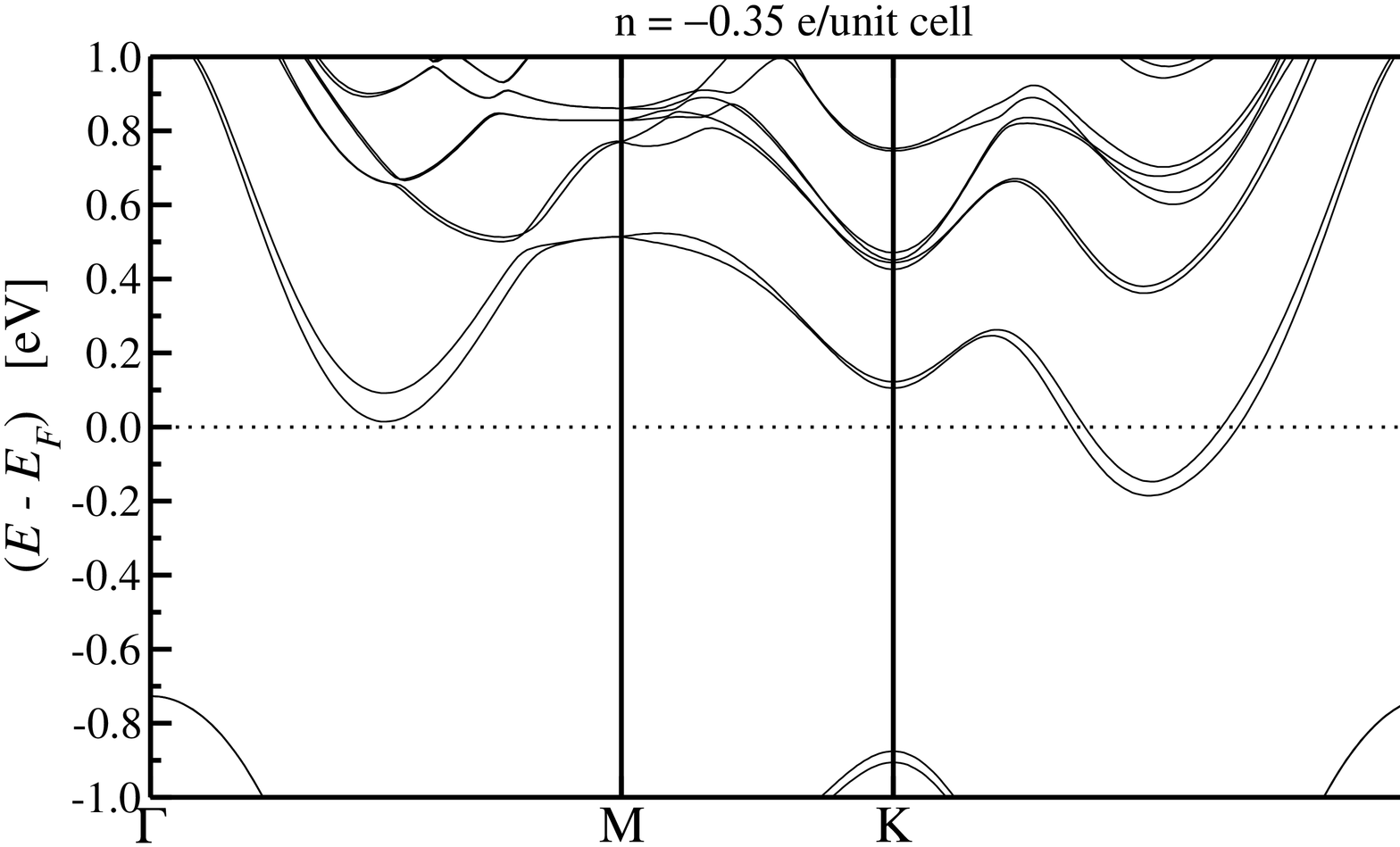}
 \includegraphics[width=0.31\textwidth,clip=,angle=-90]{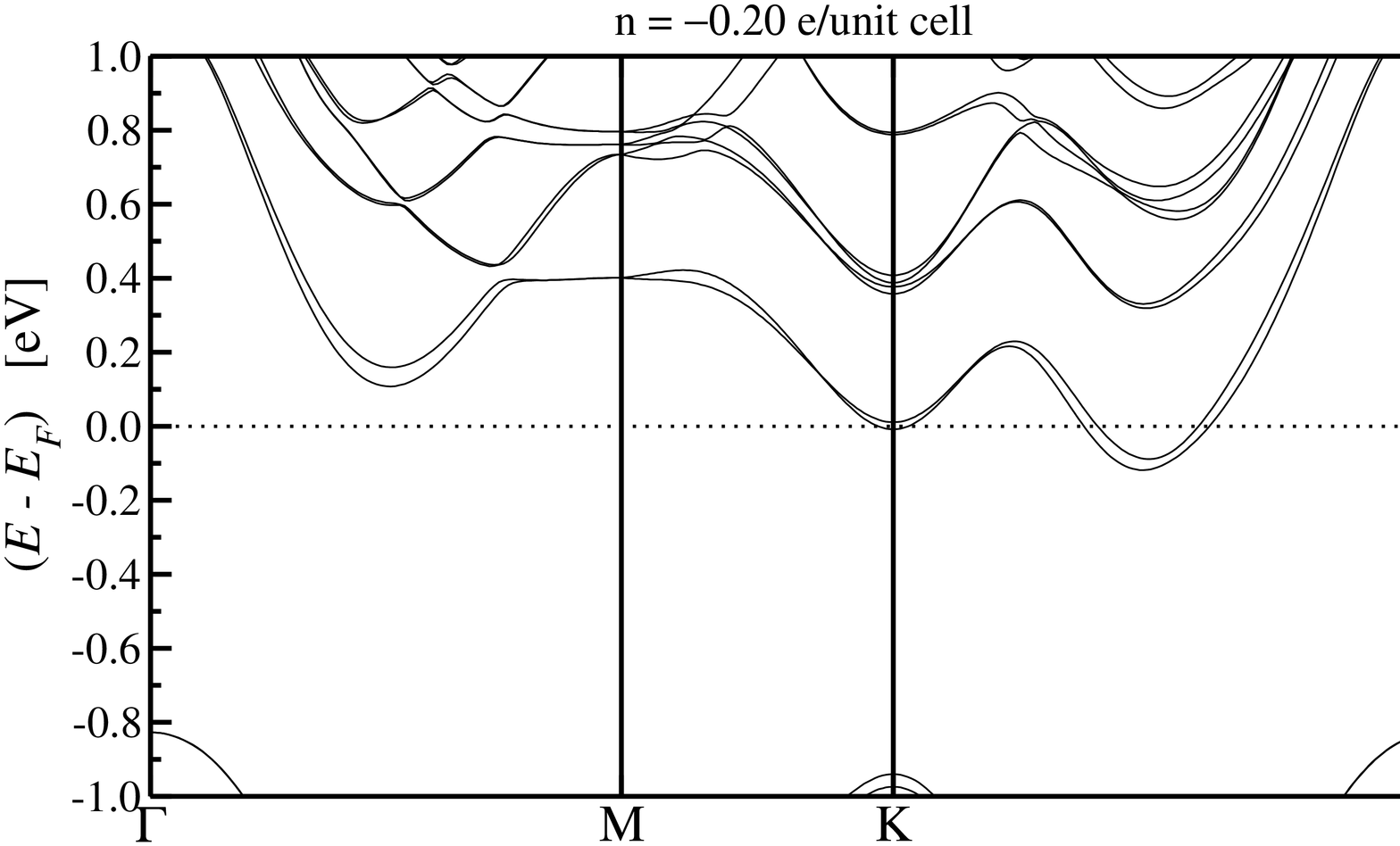}
 \includegraphics[width=0.31\textwidth,clip=,angle=-90]{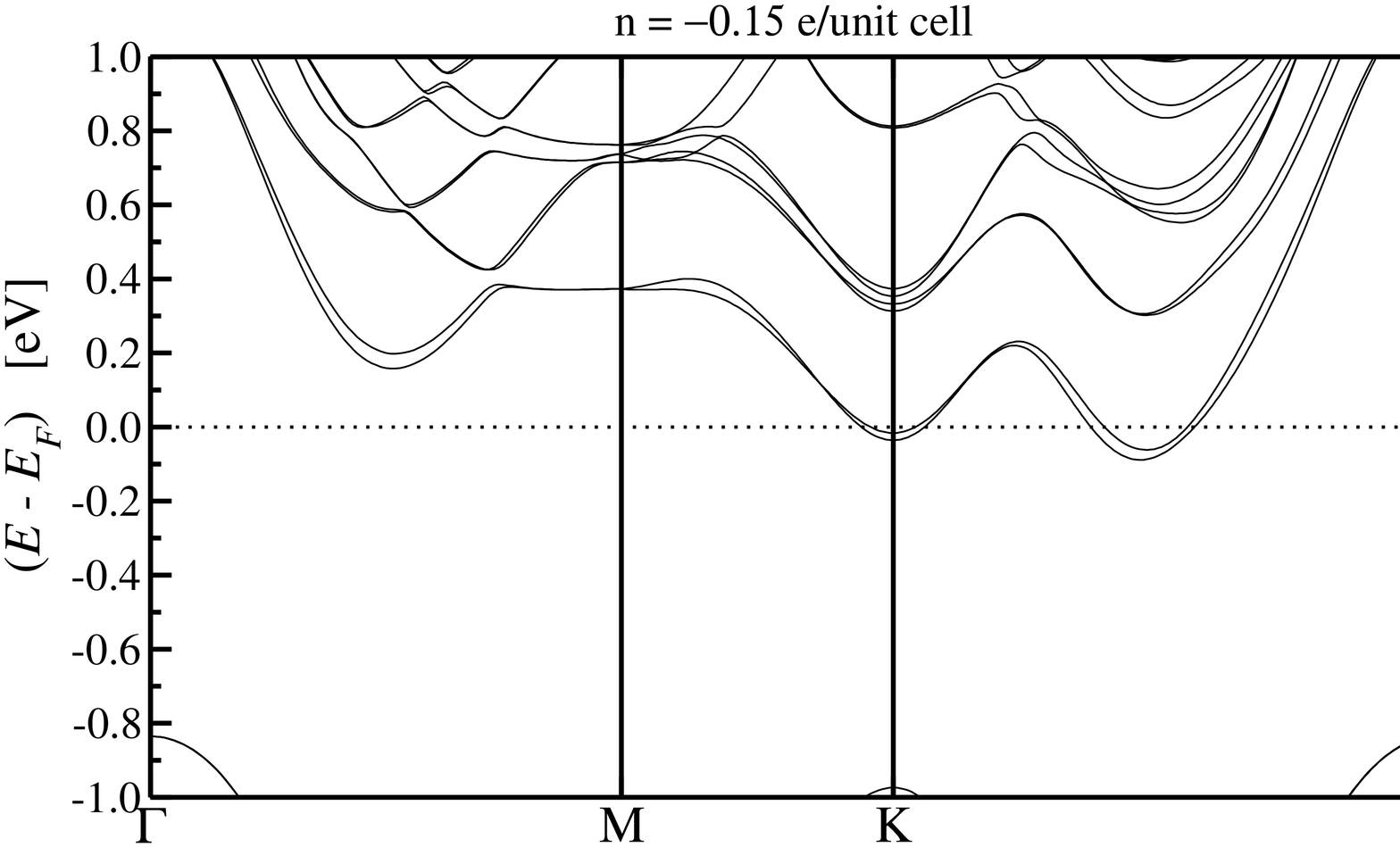}
 \includegraphics[width=0.31\textwidth,clip=,angle=-90]{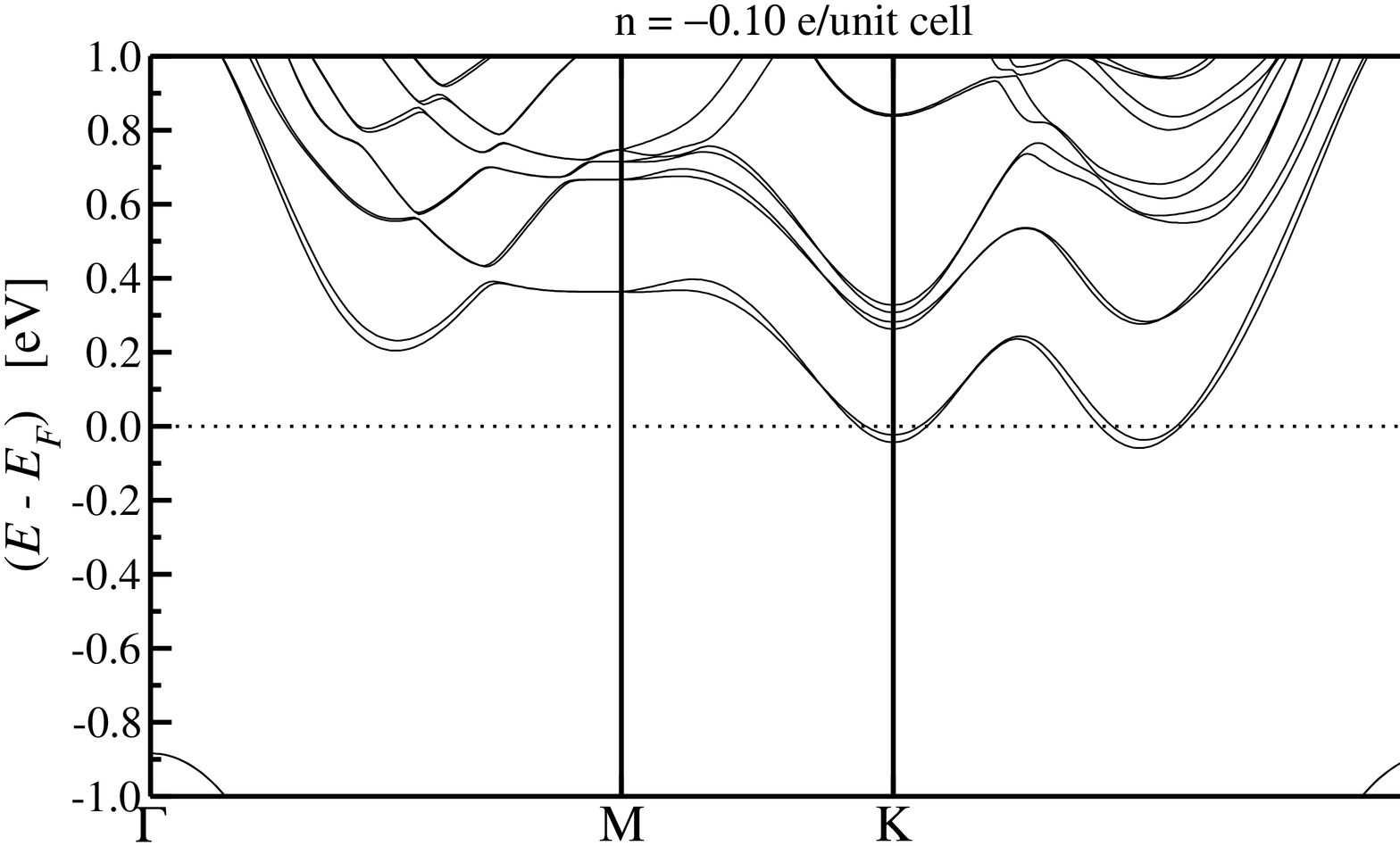}
 \includegraphics[width=0.31\textwidth,clip=,angle=-90]{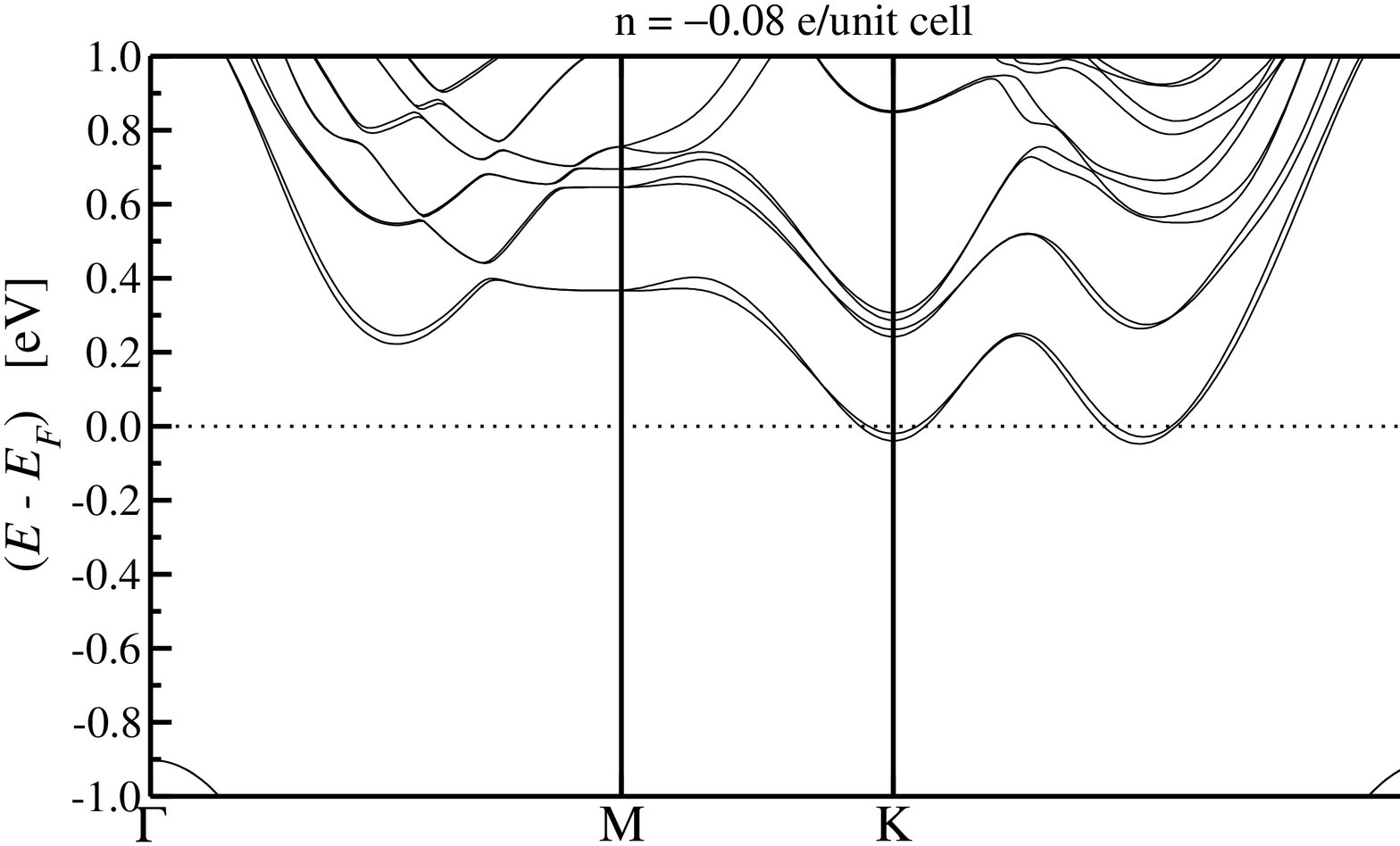}
 \includegraphics[width=0.31\textwidth,clip=,angle=-90]{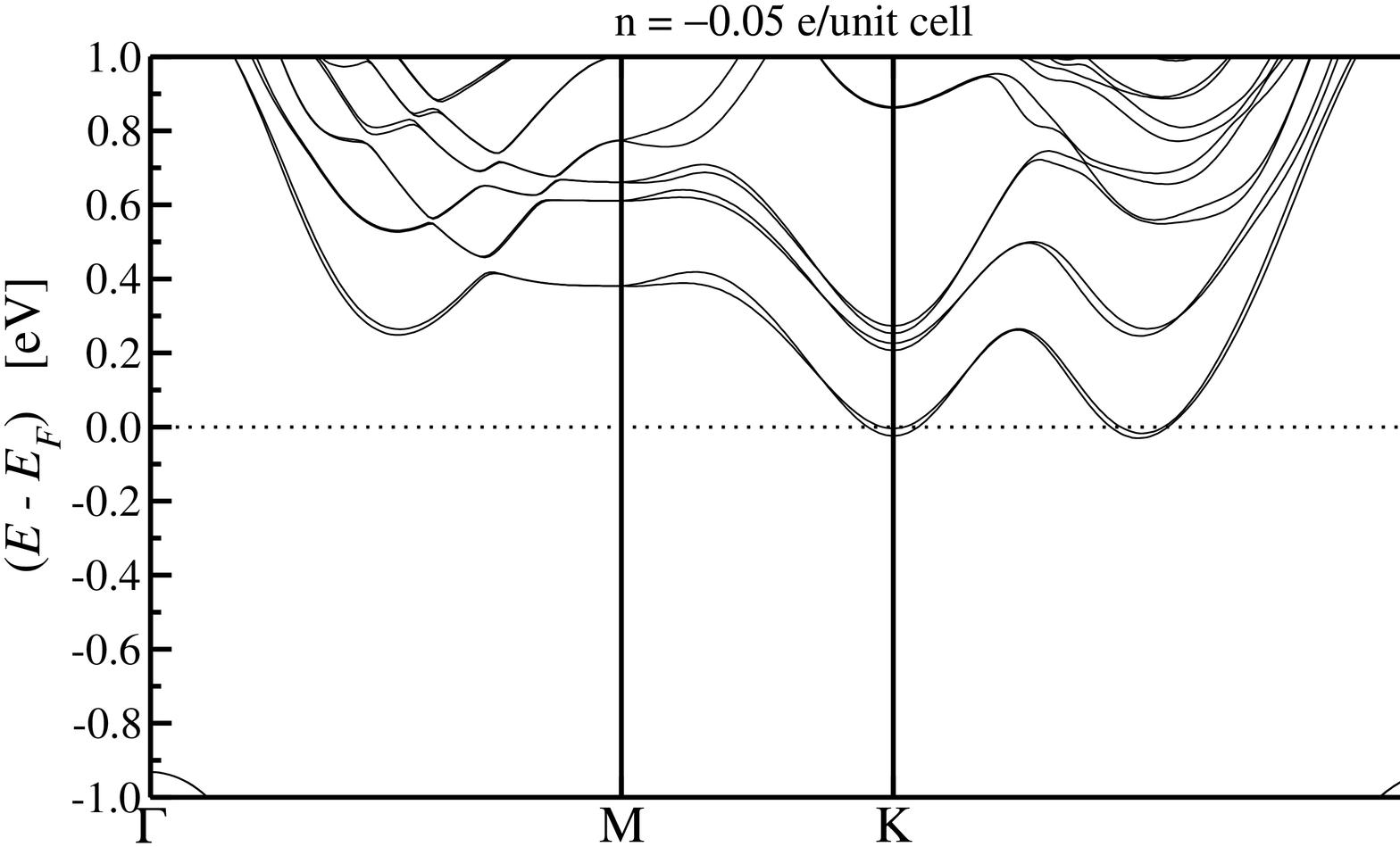}
 \includegraphics[width=0.31\textwidth,clip=,angle=-90]{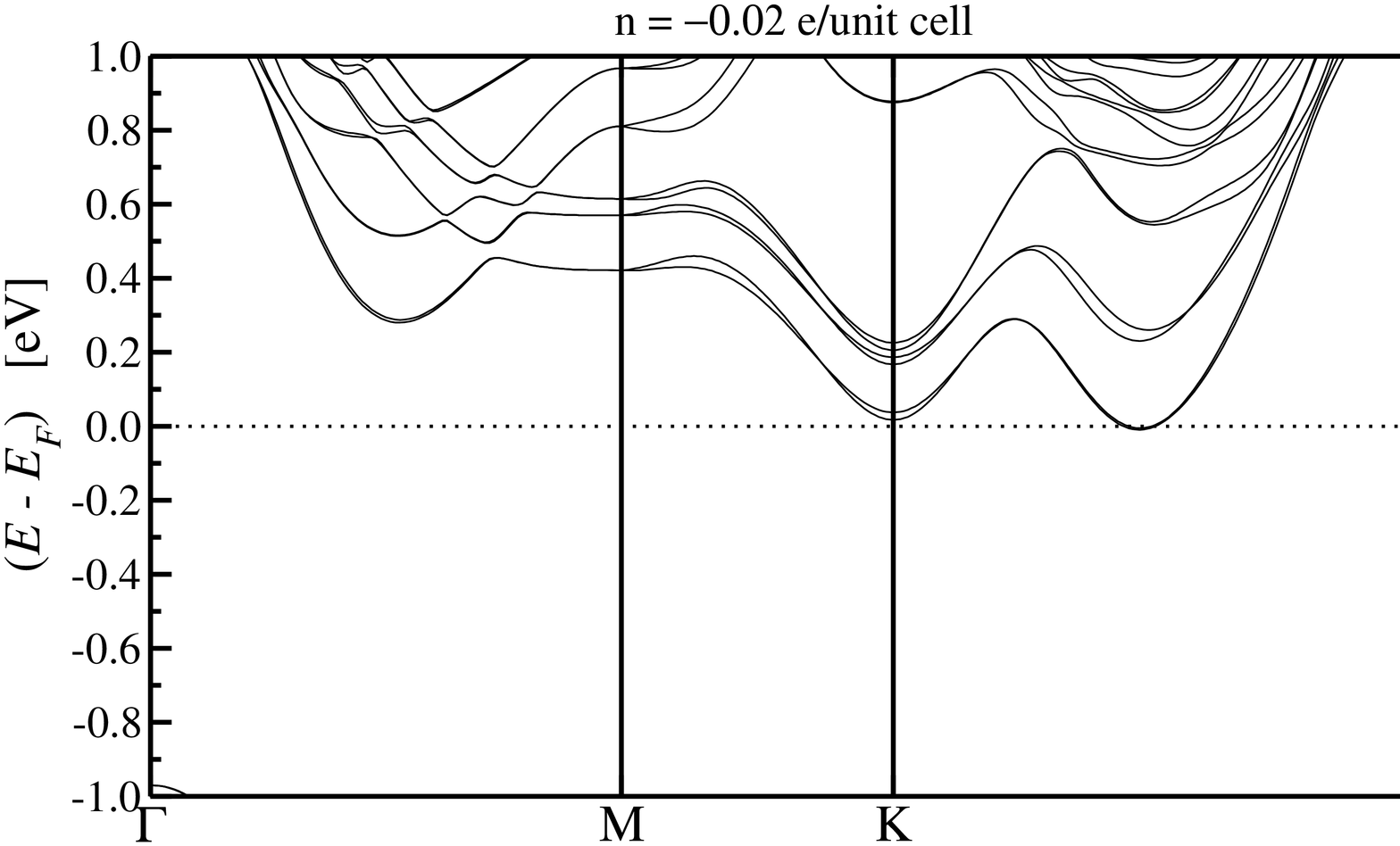}
 \includegraphics[width=0.31\textwidth,clip=,angle=-90]{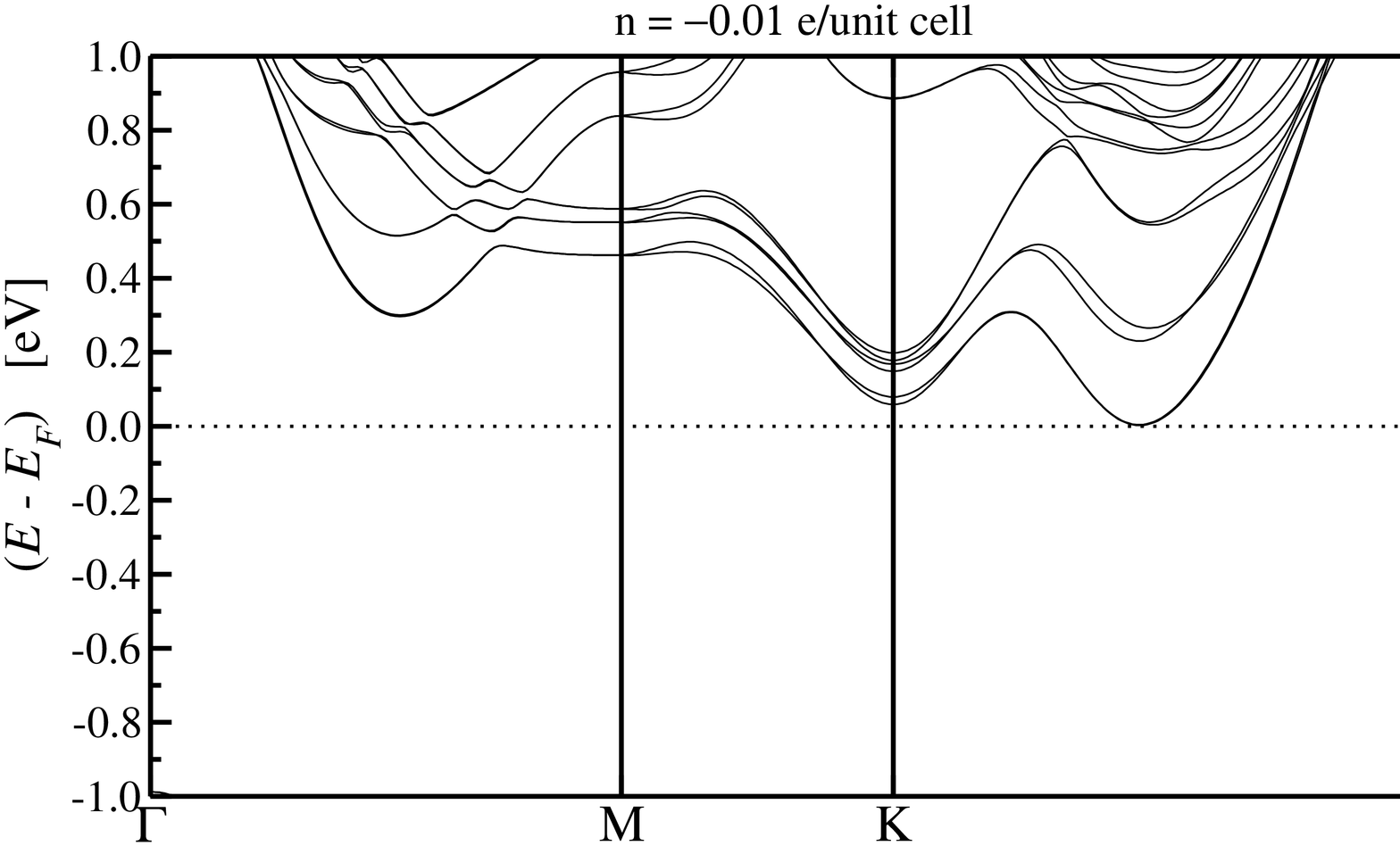}
 \caption{Band structure of trilayer MoSe$_2$ for different doping as indicated in the labels.}
\end{figure*}
\begin{figure*}[hbp]
 \centering
 \includegraphics[width=0.31\textwidth,clip=,angle=-90]{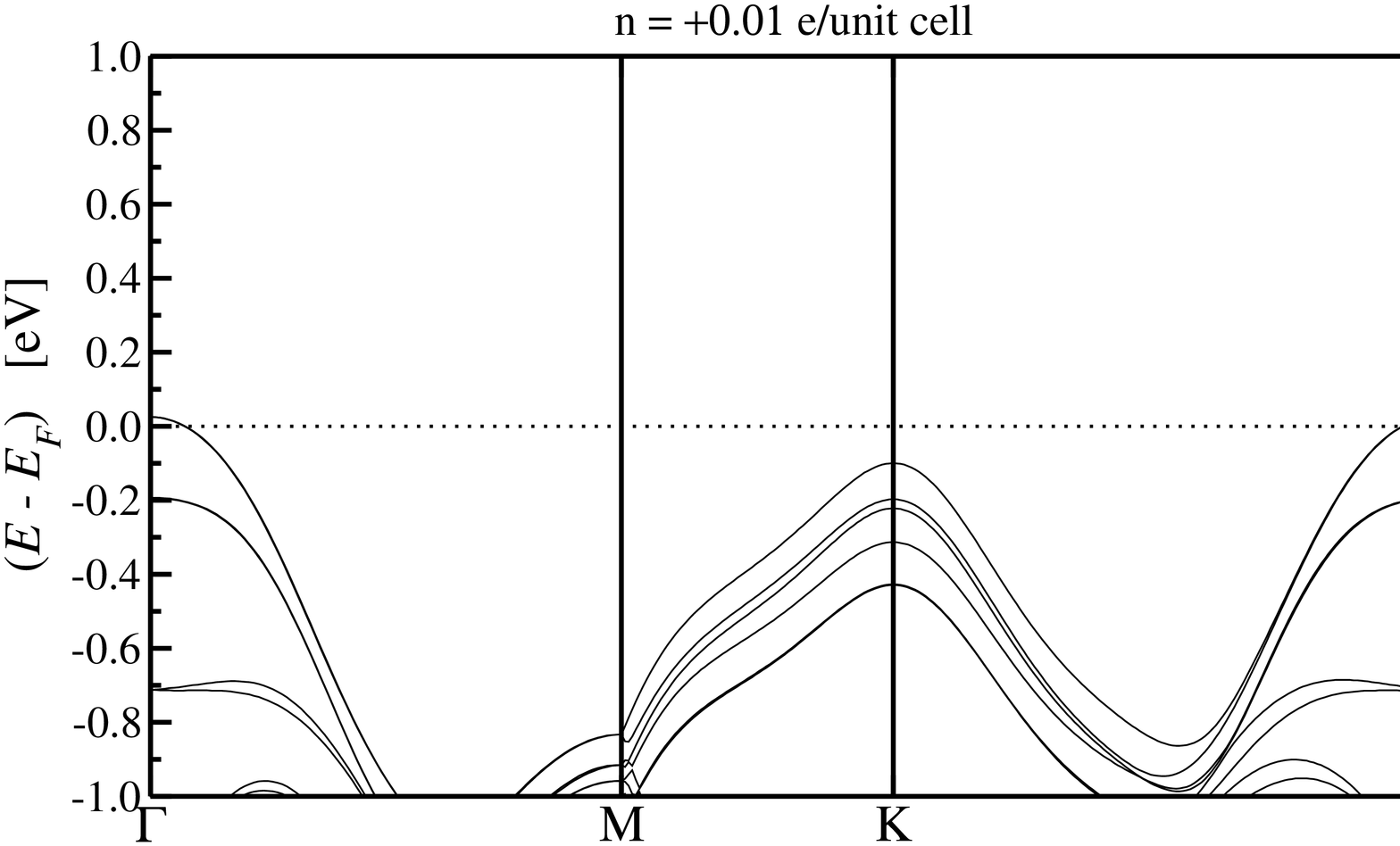}
 \includegraphics[width=0.31\textwidth,clip=,angle=-90]{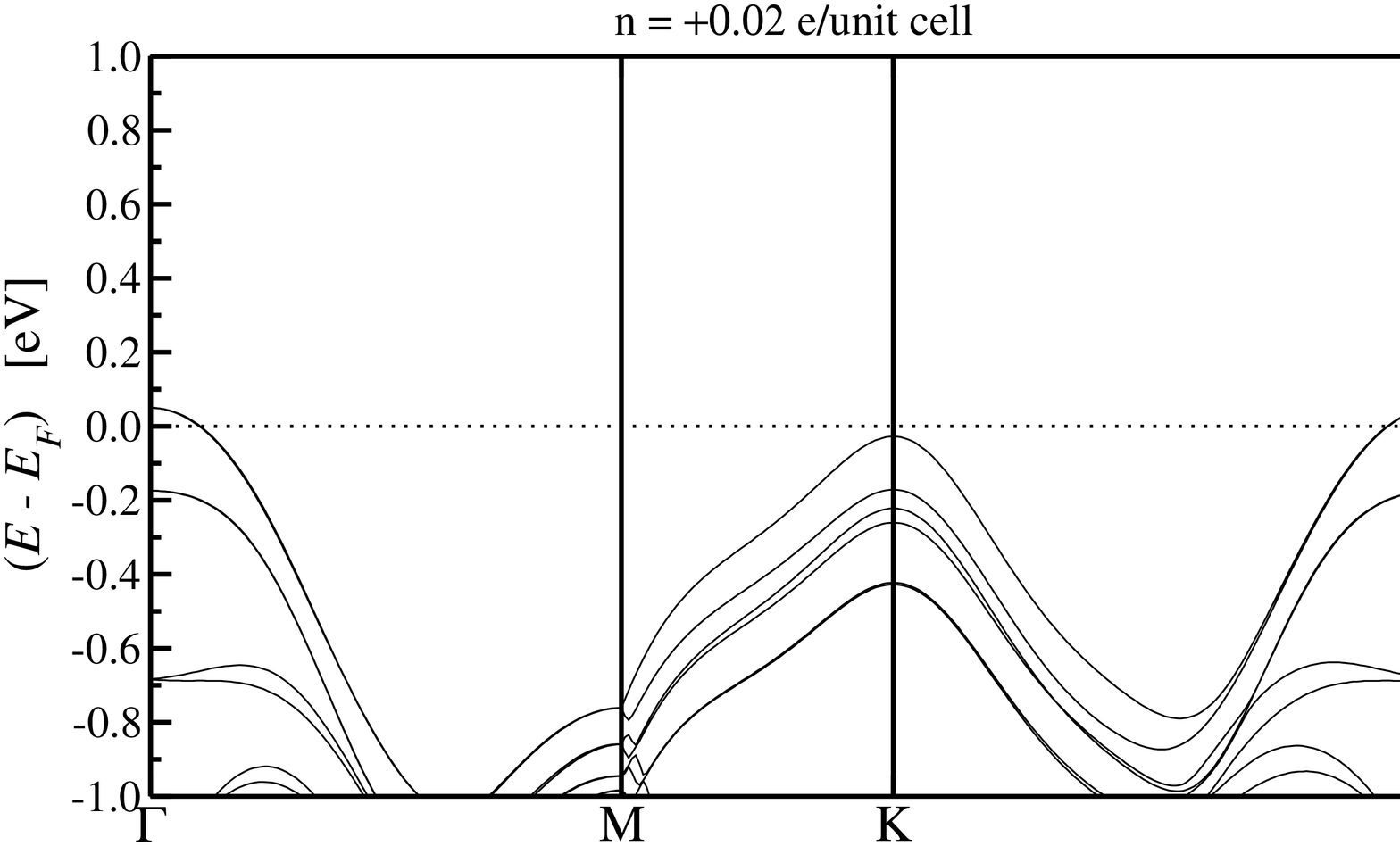}
 \includegraphics[width=0.31\textwidth,clip=,angle=-90]{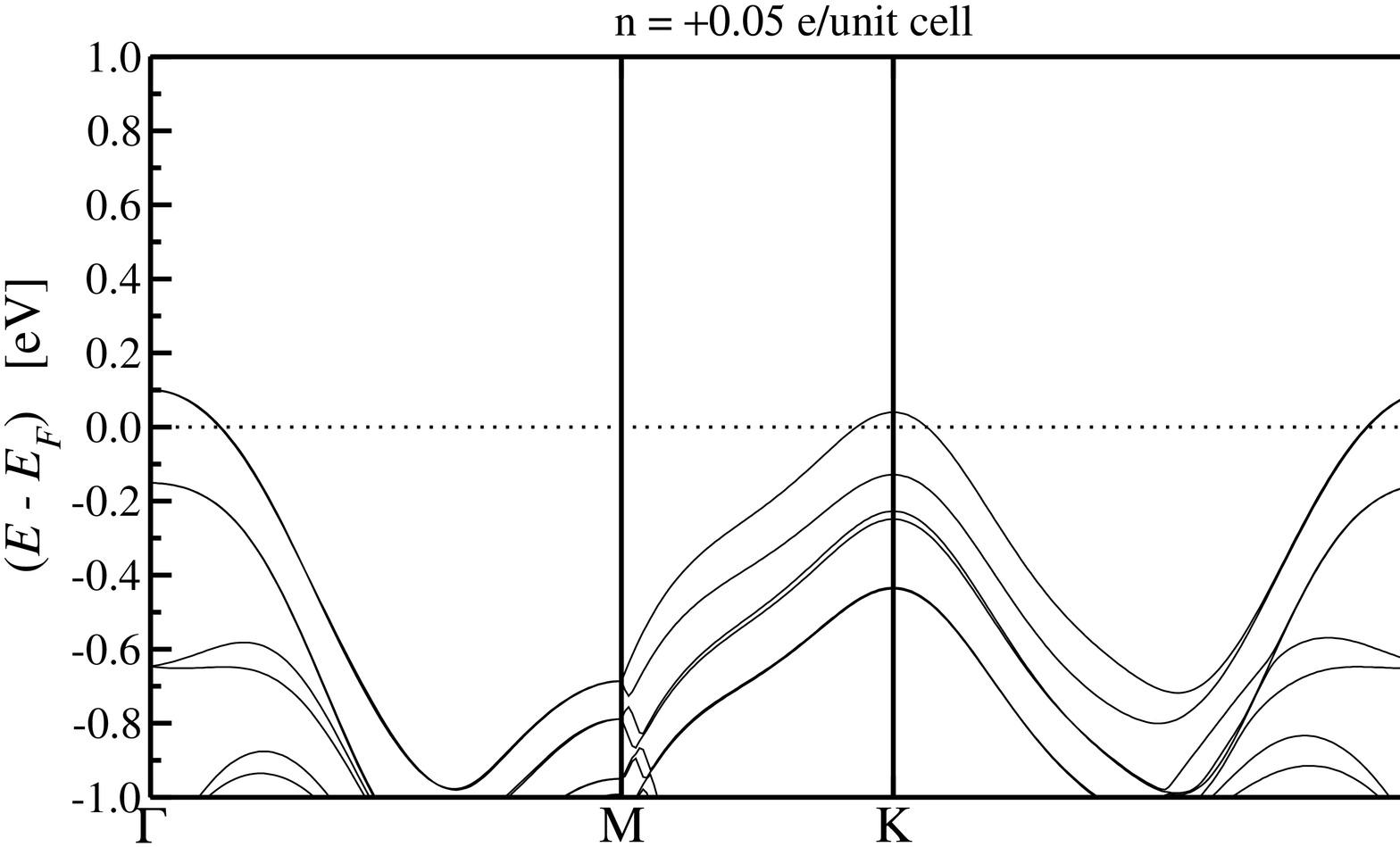}
 \includegraphics[width=0.31\textwidth,clip=,angle=-90]{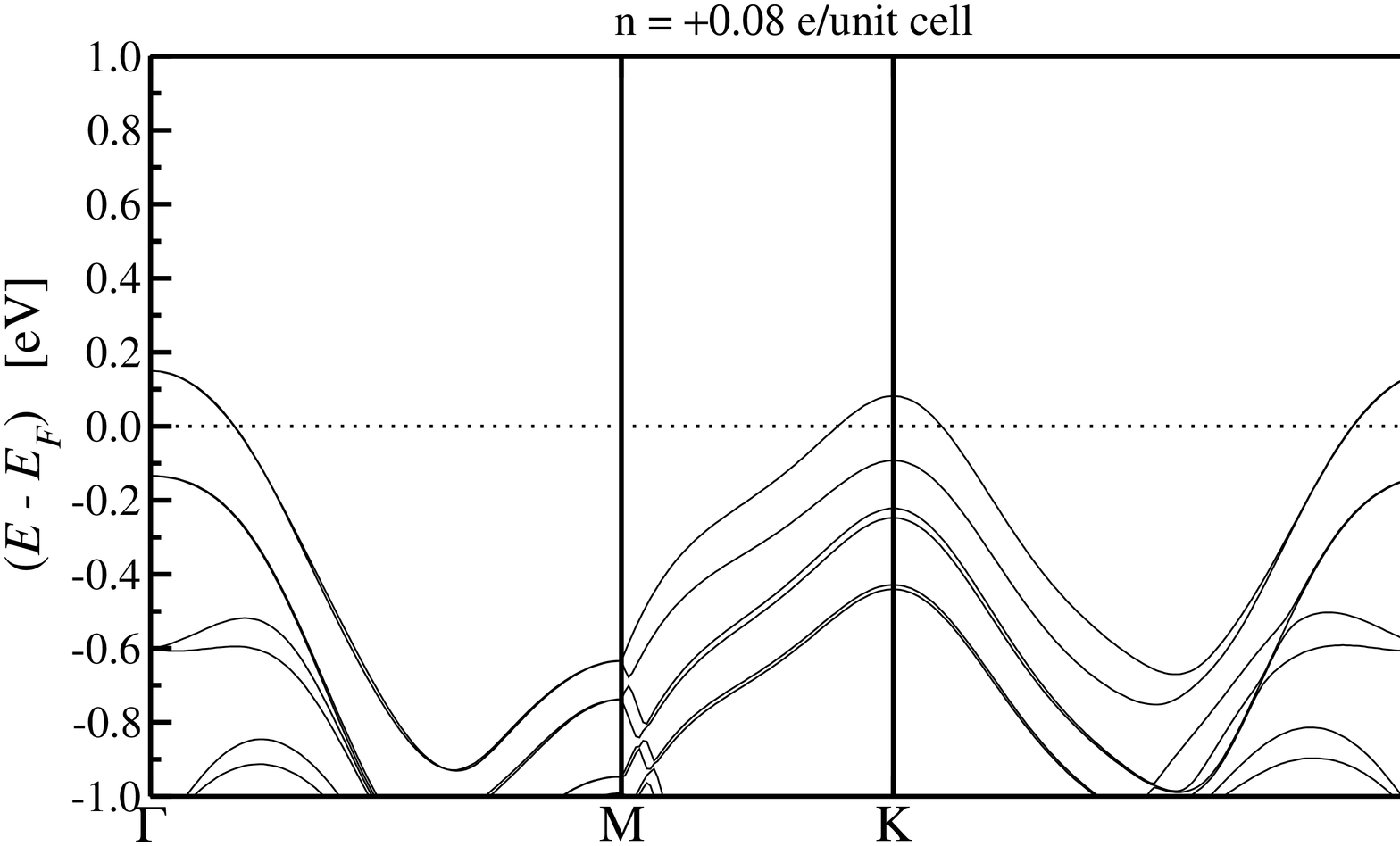}
 \includegraphics[width=0.31\textwidth,clip=,angle=-90]{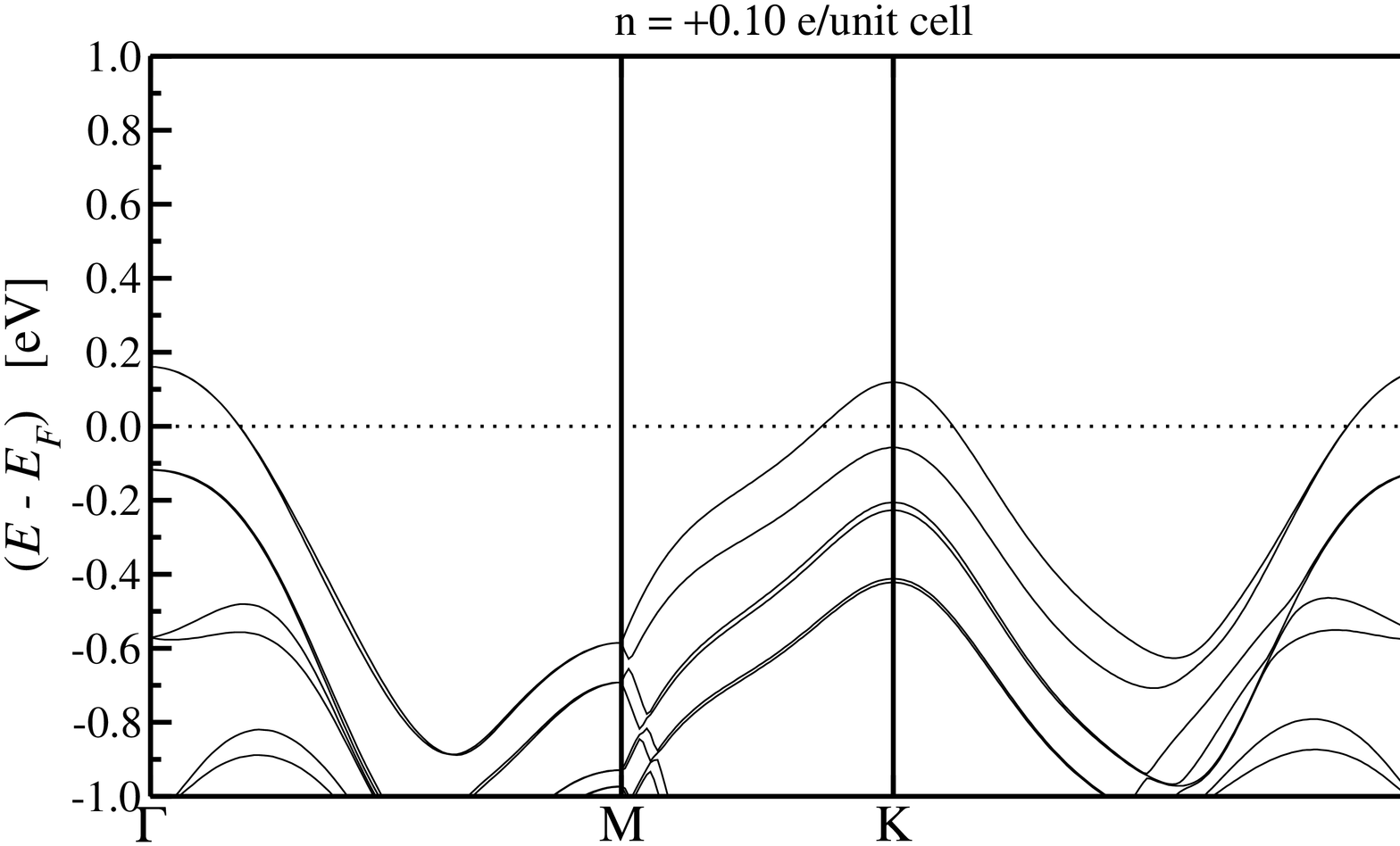}
 \includegraphics[width=0.31\textwidth,clip=,angle=-90]{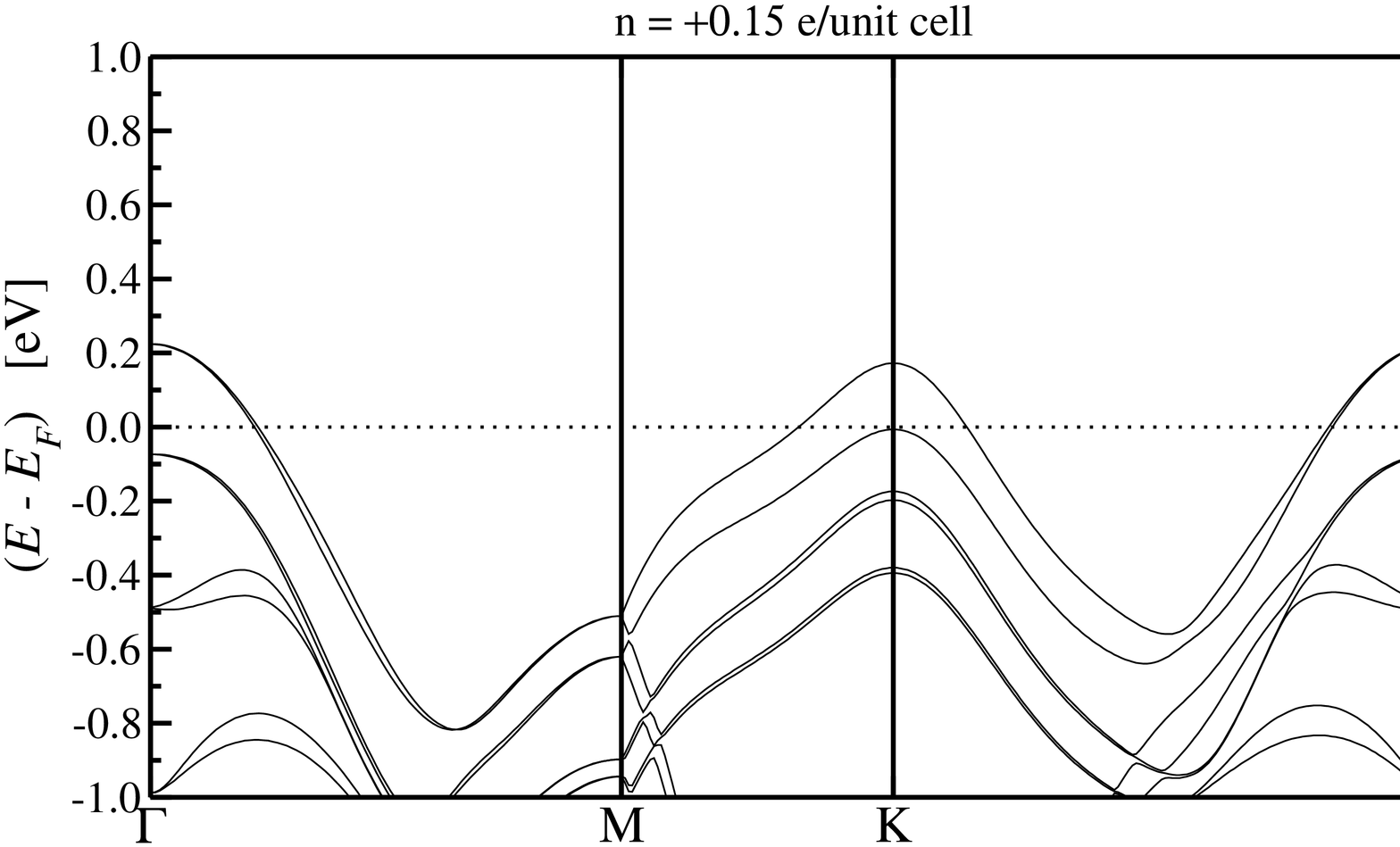}
 \includegraphics[width=0.31\textwidth,clip=,angle=-90]{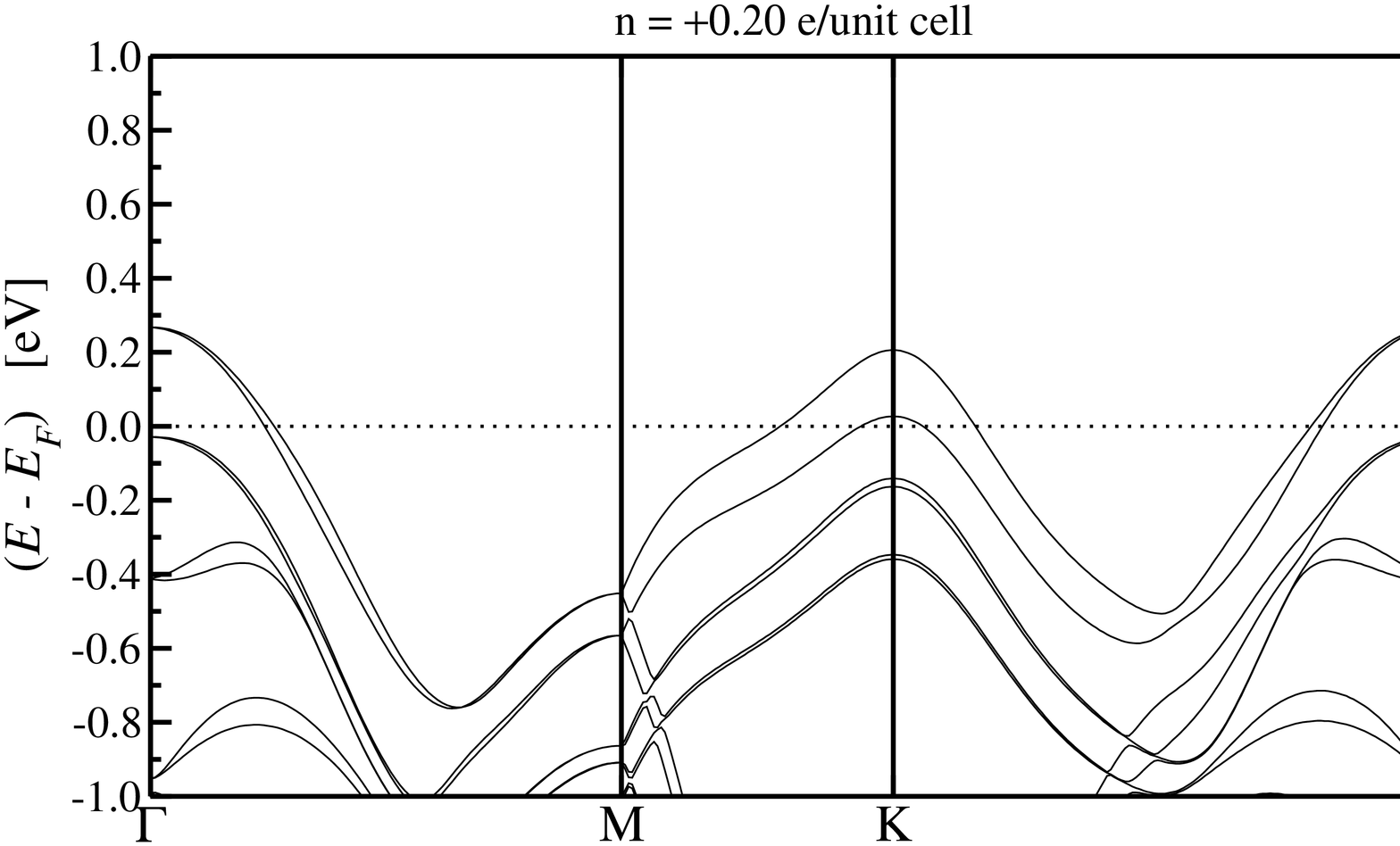}
 \includegraphics[width=0.31\textwidth,clip=,angle=-90]{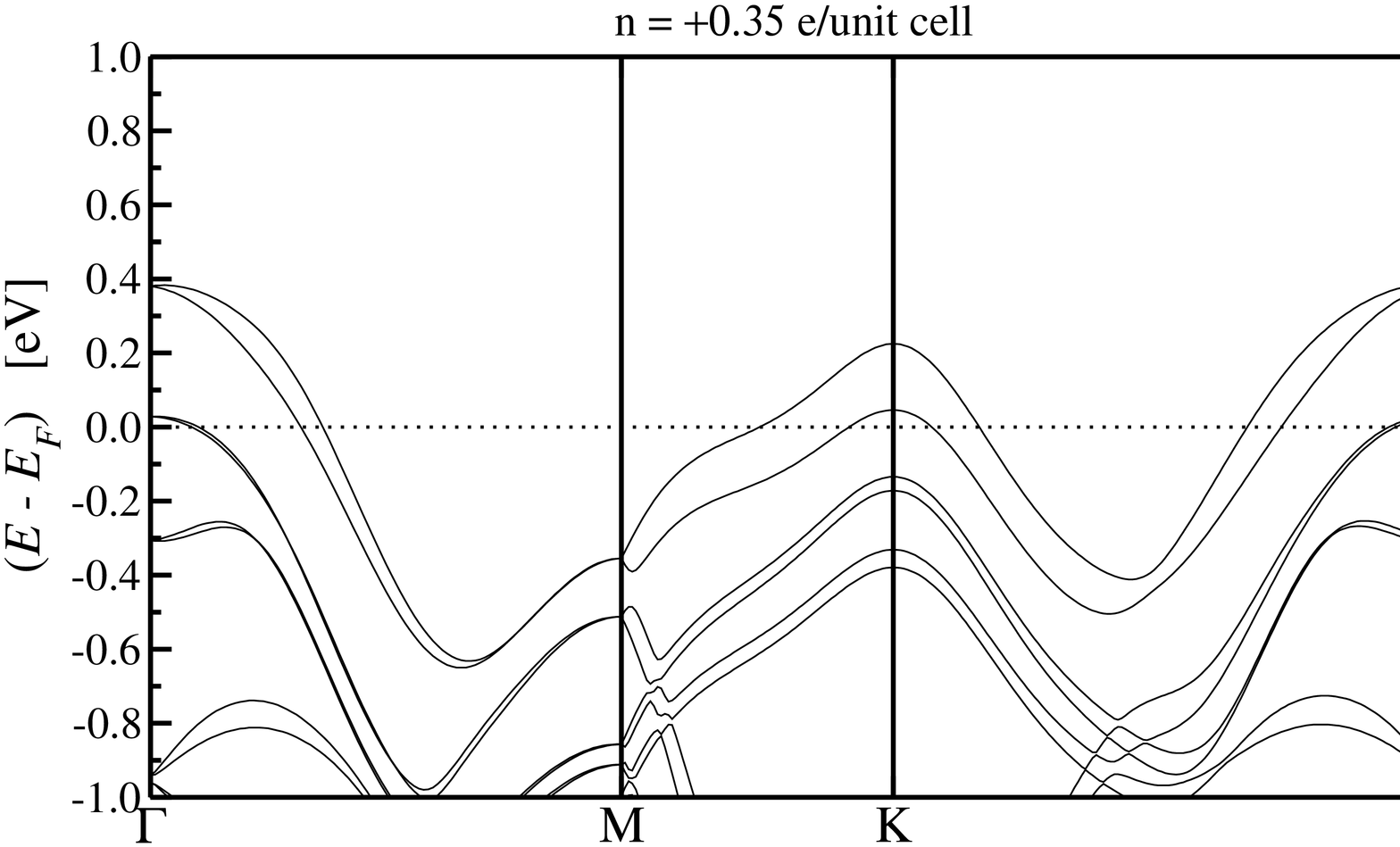}
 \caption{Band structure of trilayer MoSe$_2$ for different doping as indicated in the labels.}
\end{figure*}

\clearpage
\subsection{Molybdenum ditelluride}
\begin{figure*}[hbp]
 \centering
 \includegraphics[width=0.31\textwidth,clip=,angle=-90]{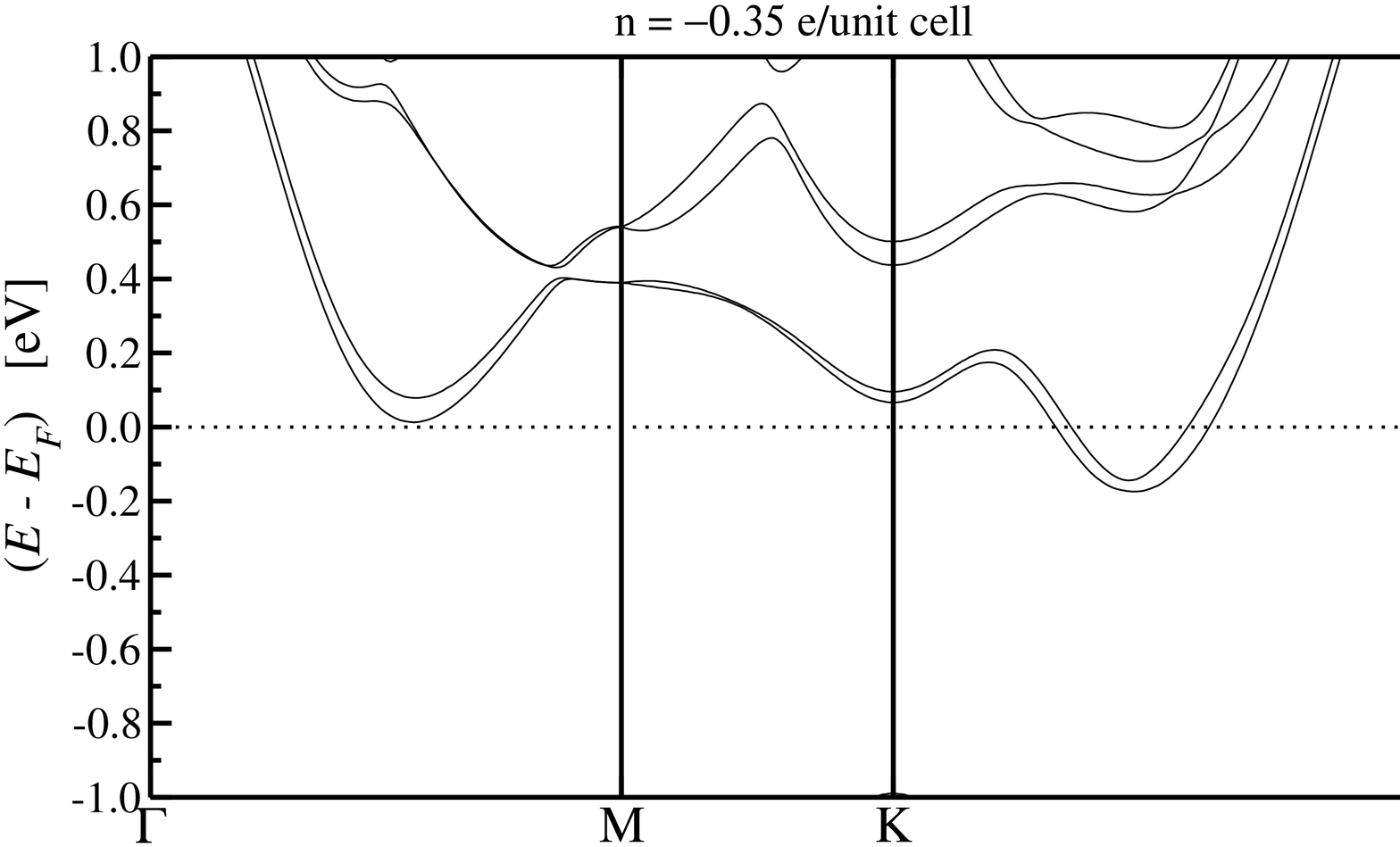}
 \includegraphics[width=0.31\textwidth,clip=,angle=-90]{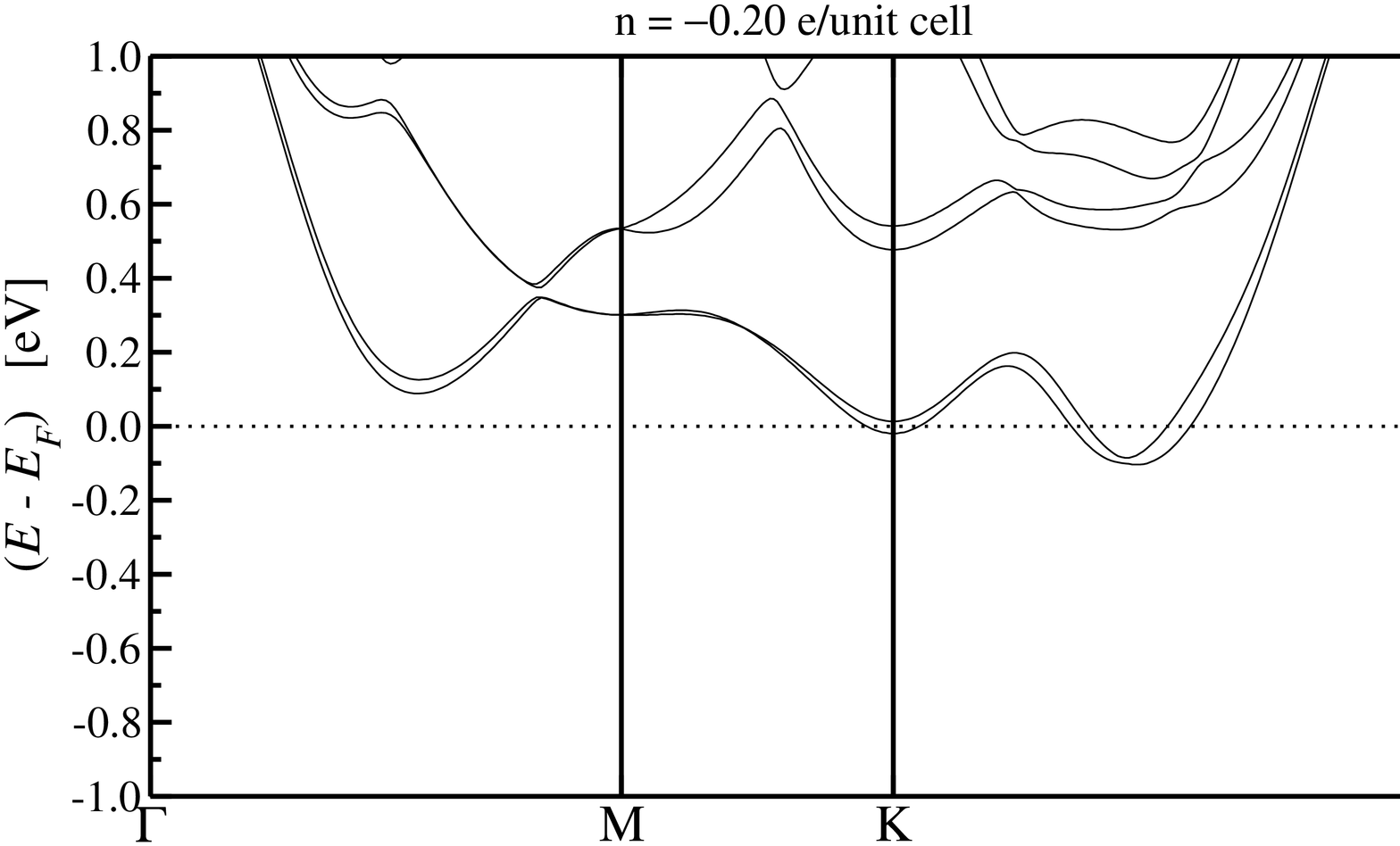}
 \includegraphics[width=0.31\textwidth,clip=,angle=-90]{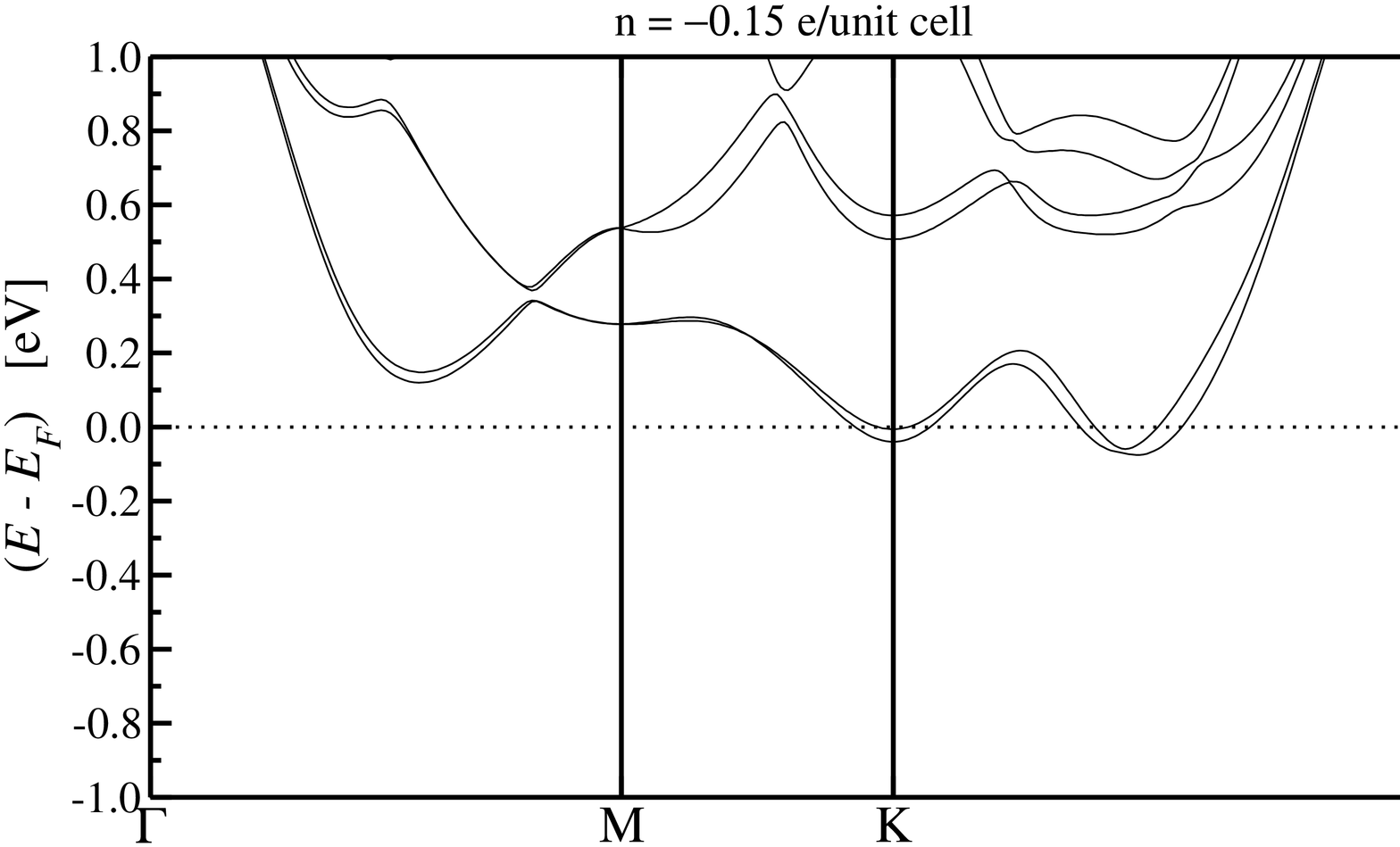}
 \includegraphics[width=0.31\textwidth,clip=,angle=-90]{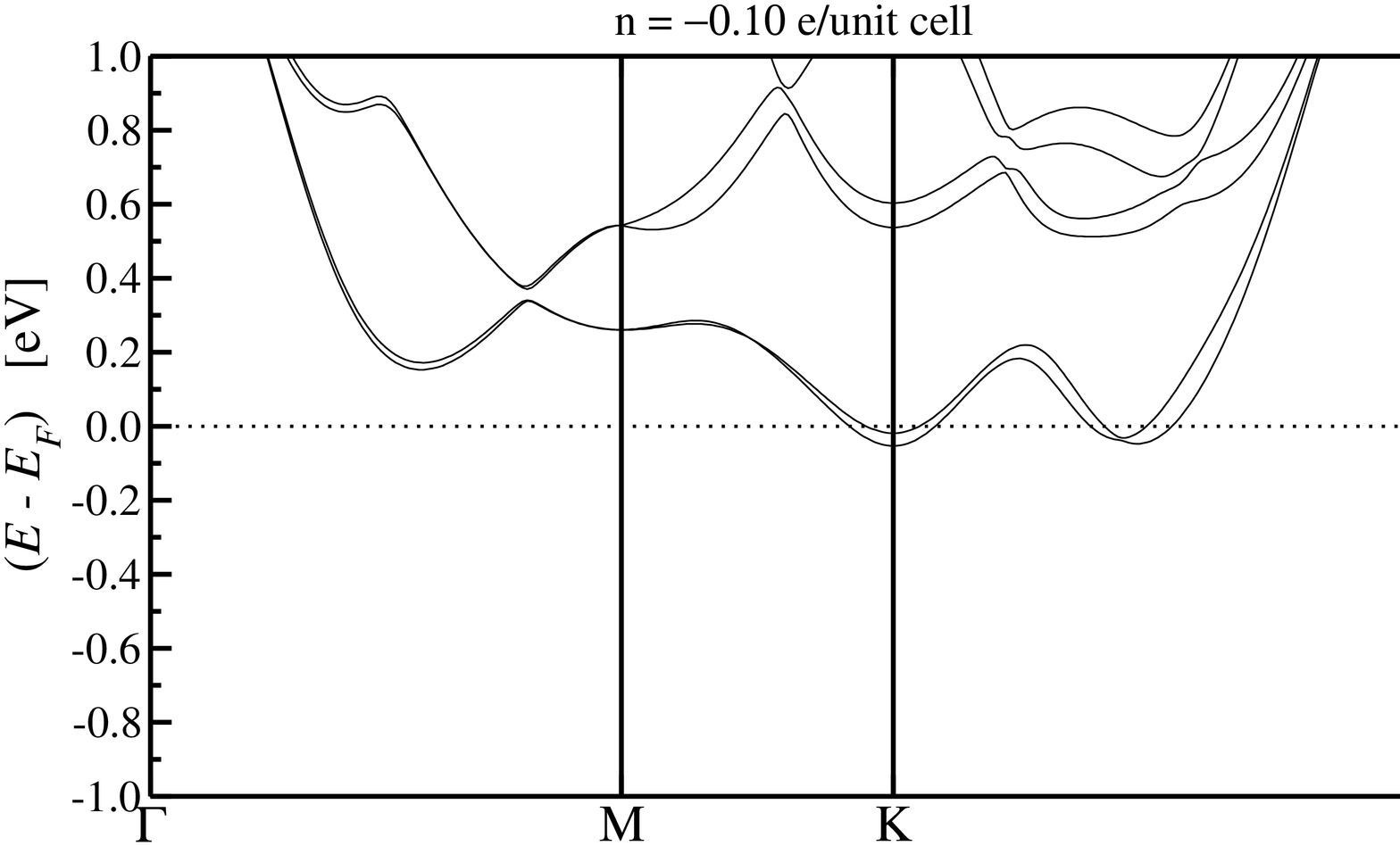}
 \includegraphics[width=0.31\textwidth,clip=,angle=-90]{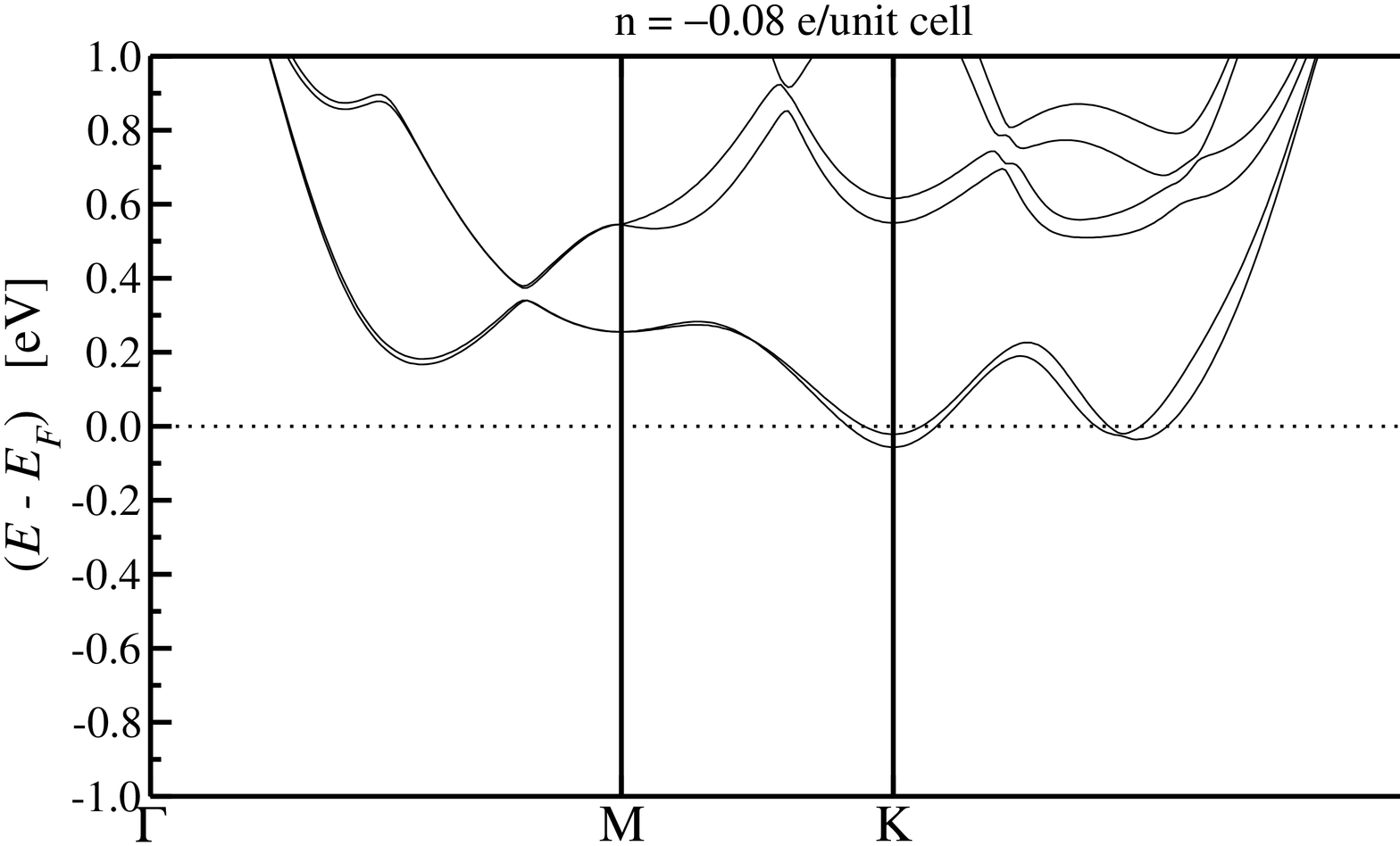}
 \includegraphics[width=0.31\textwidth,clip=,angle=-90]{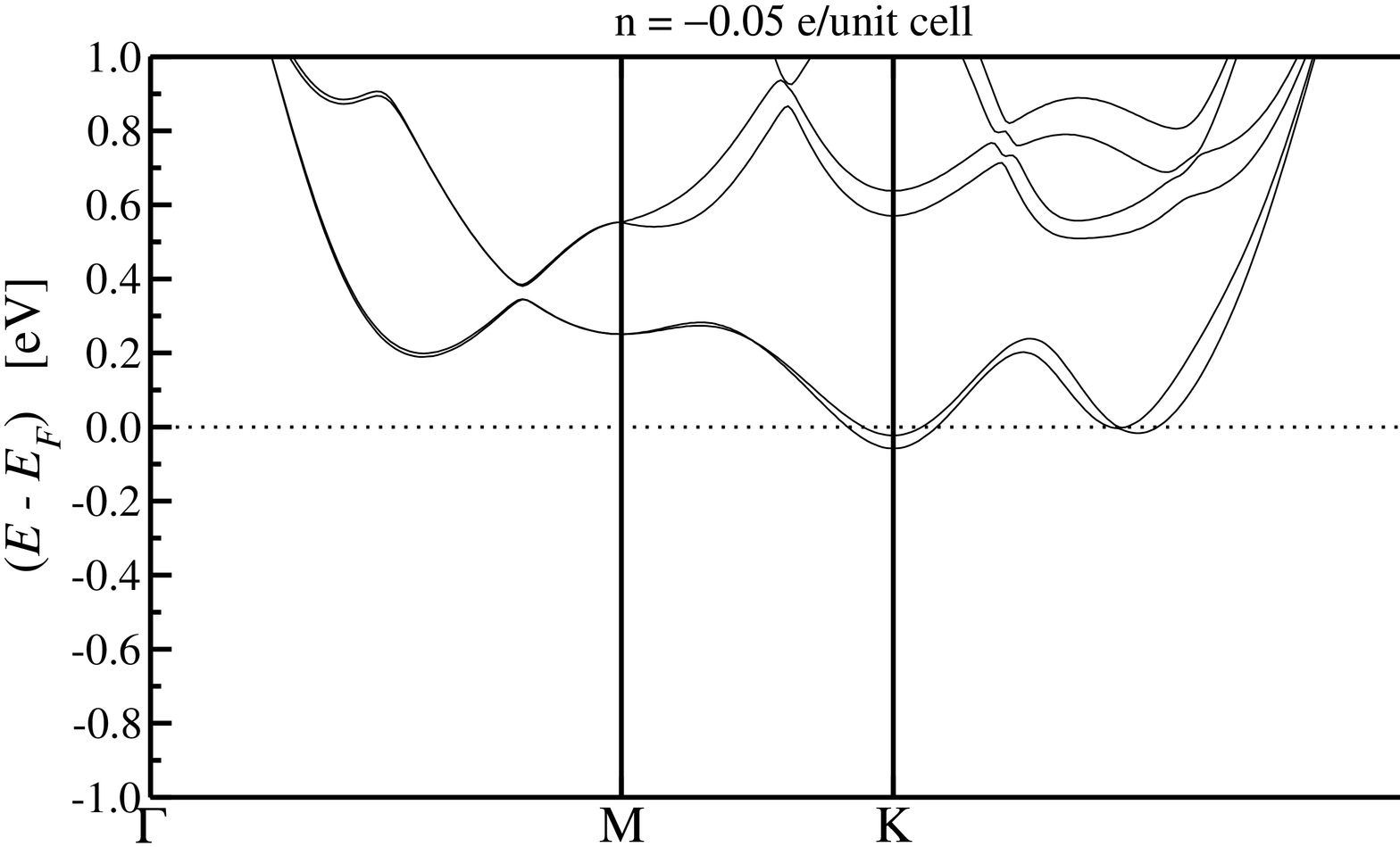}
 \includegraphics[width=0.31\textwidth,clip=,angle=-90]{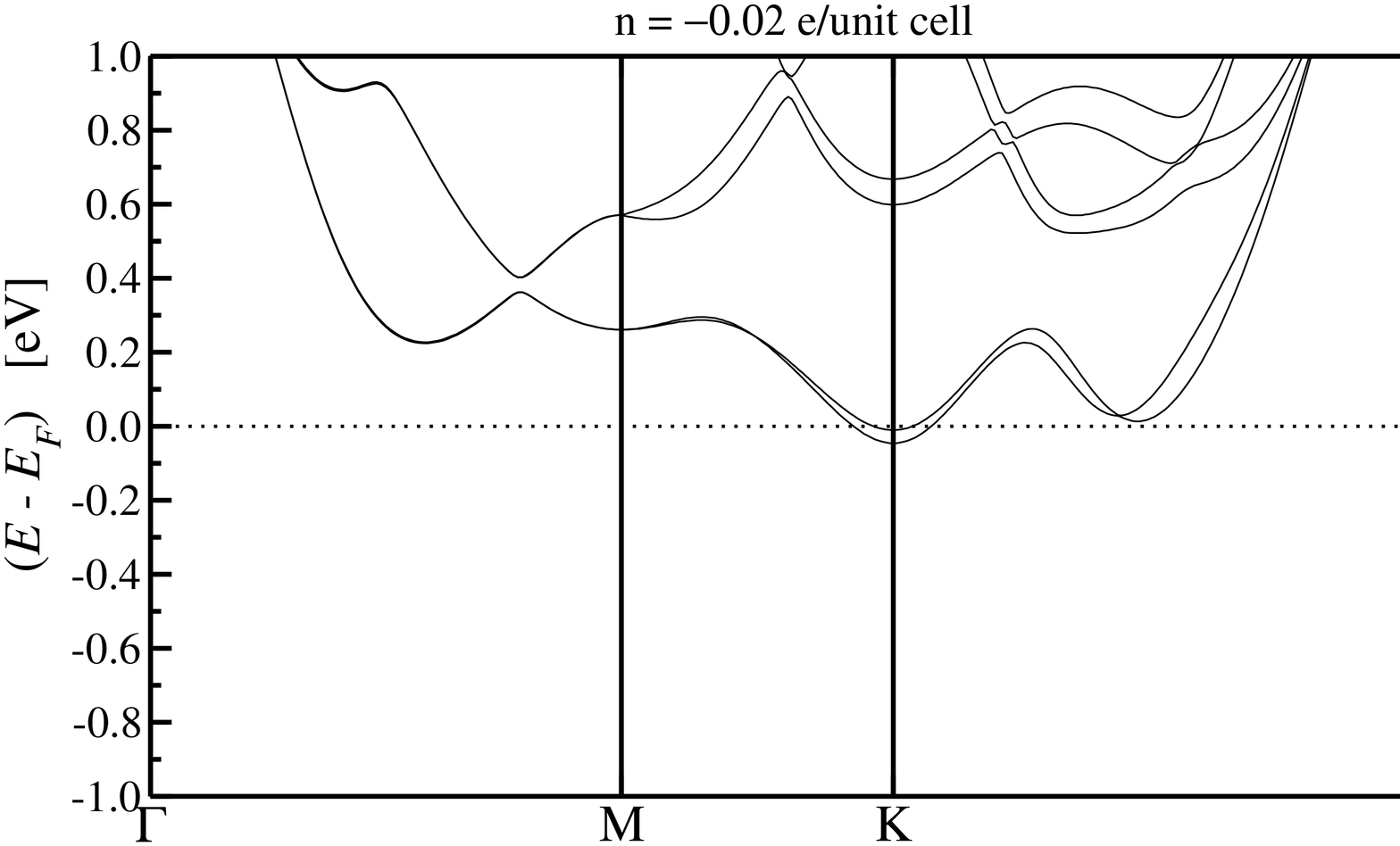}
 \includegraphics[width=0.31\textwidth,clip=,angle=-90]{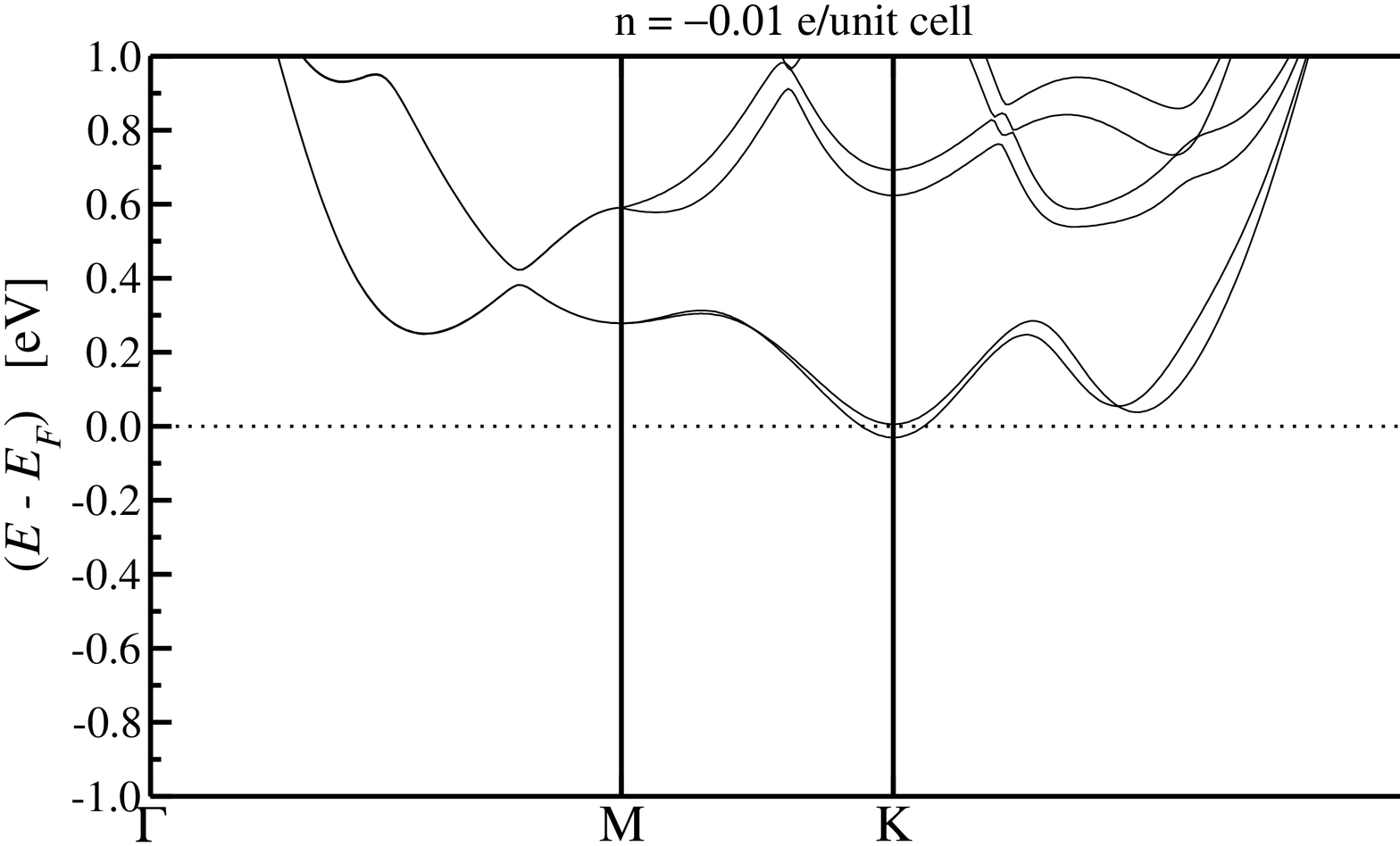}
 \caption{Band structure of monolayer MoTe$_2$ for different doping as indicated in the labels.}
\end{figure*}
\begin{figure*}[hbp]
 \centering
 \includegraphics[width=0.31\textwidth,clip=,angle=-90]{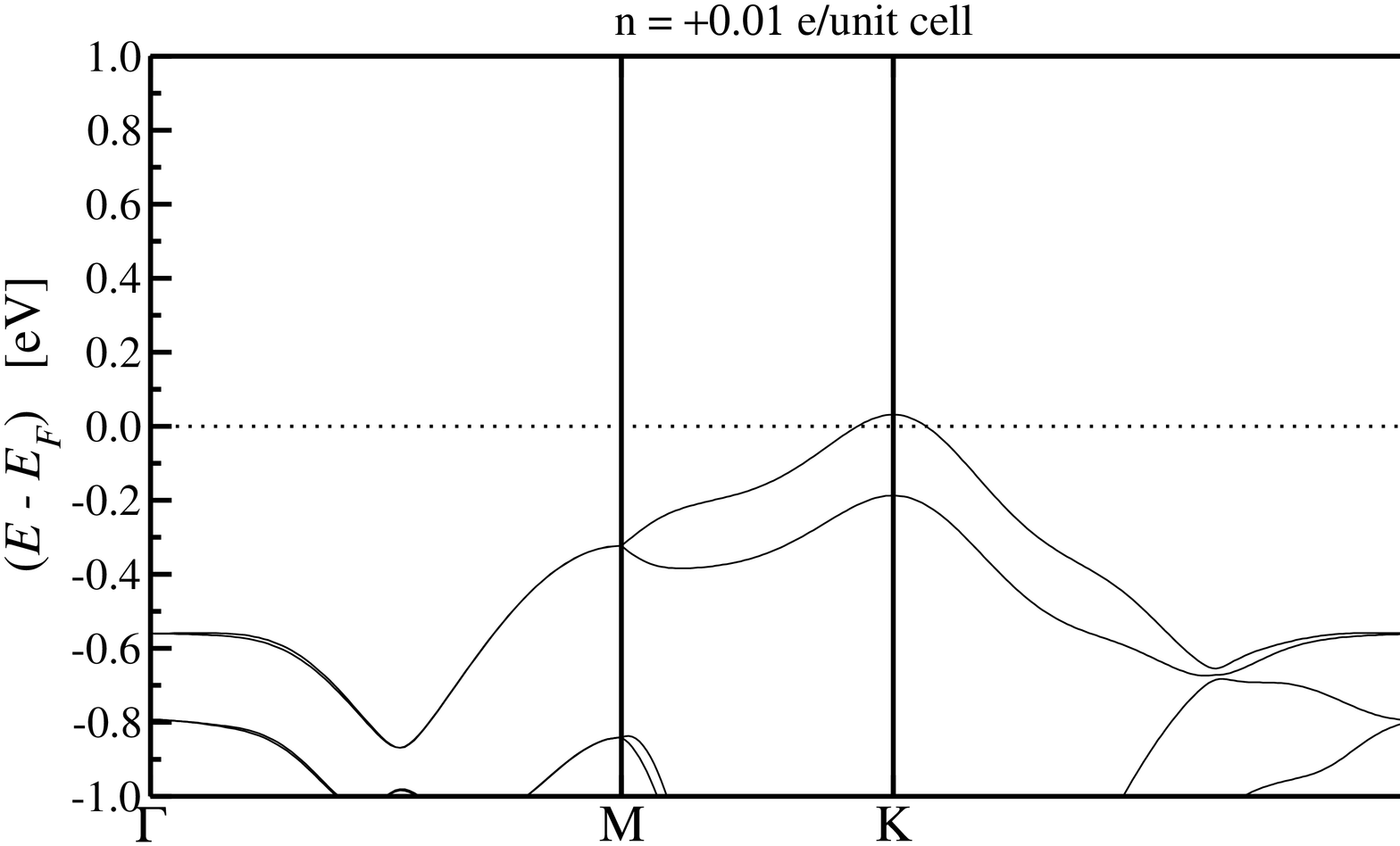}
 \includegraphics[width=0.31\textwidth,clip=,angle=-90]{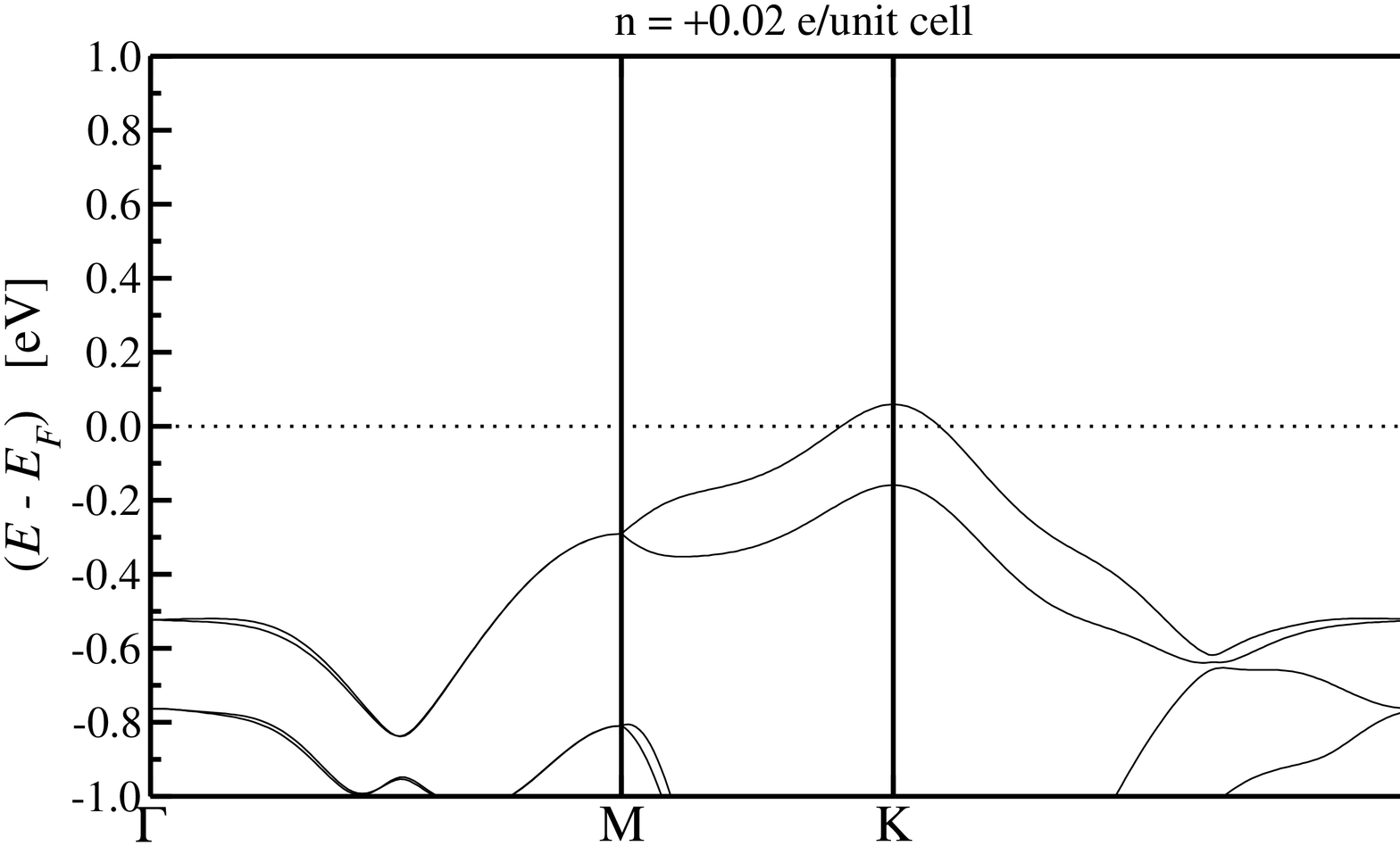}
 \includegraphics[width=0.31\textwidth,clip=,angle=-90]{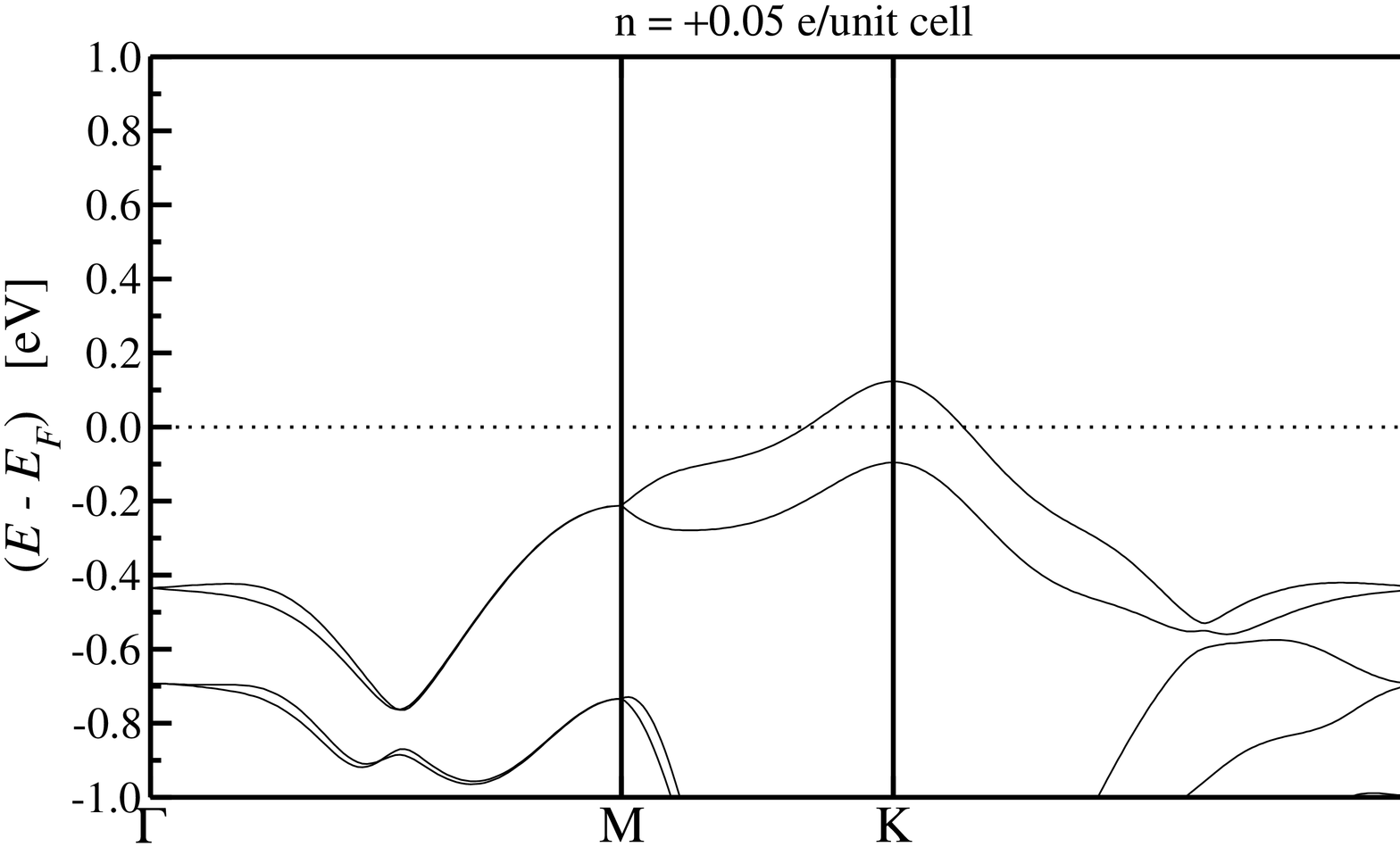}
 \includegraphics[width=0.31\textwidth,clip=,angle=-90]{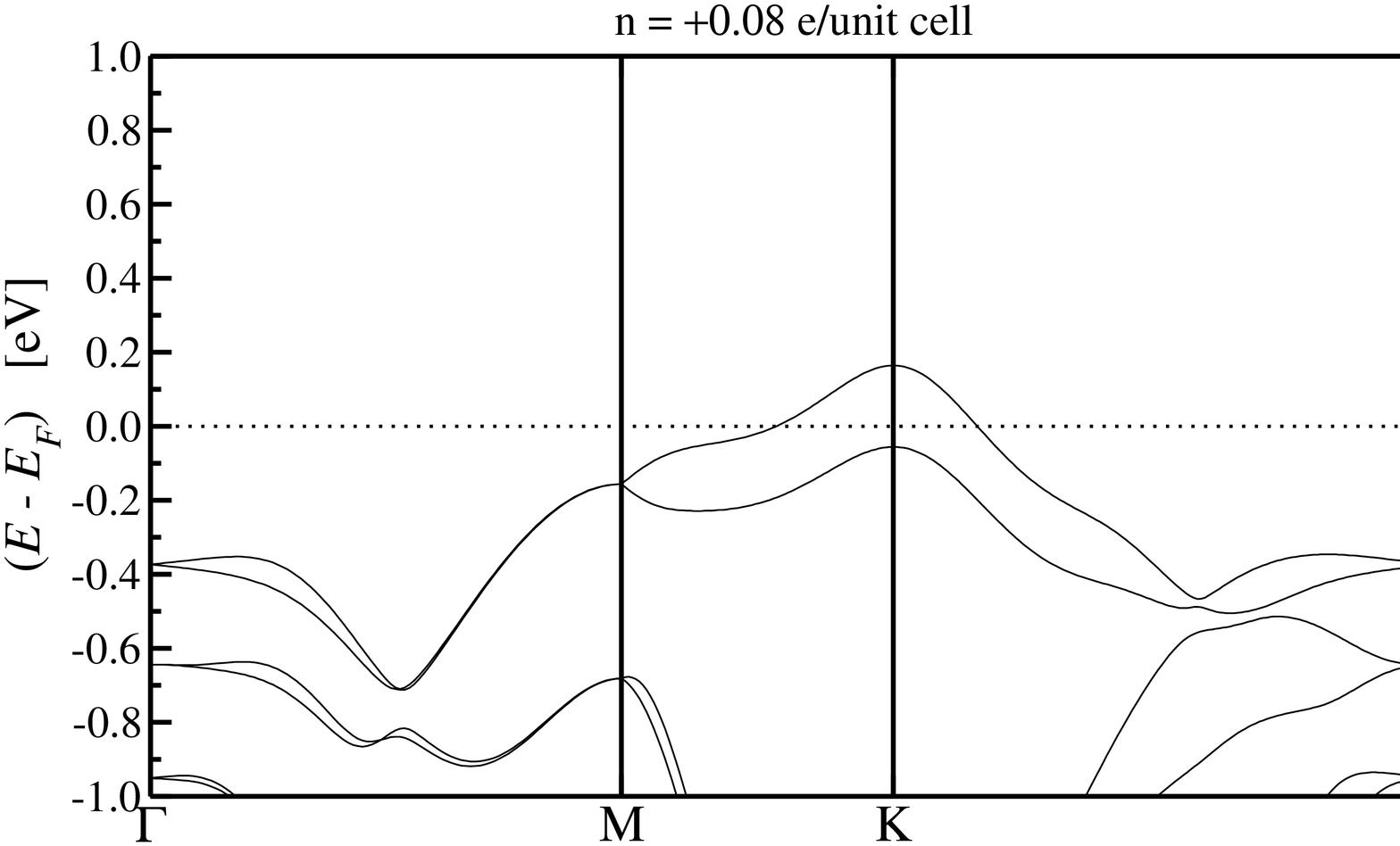}
 \includegraphics[width=0.31\textwidth,clip=,angle=-90]{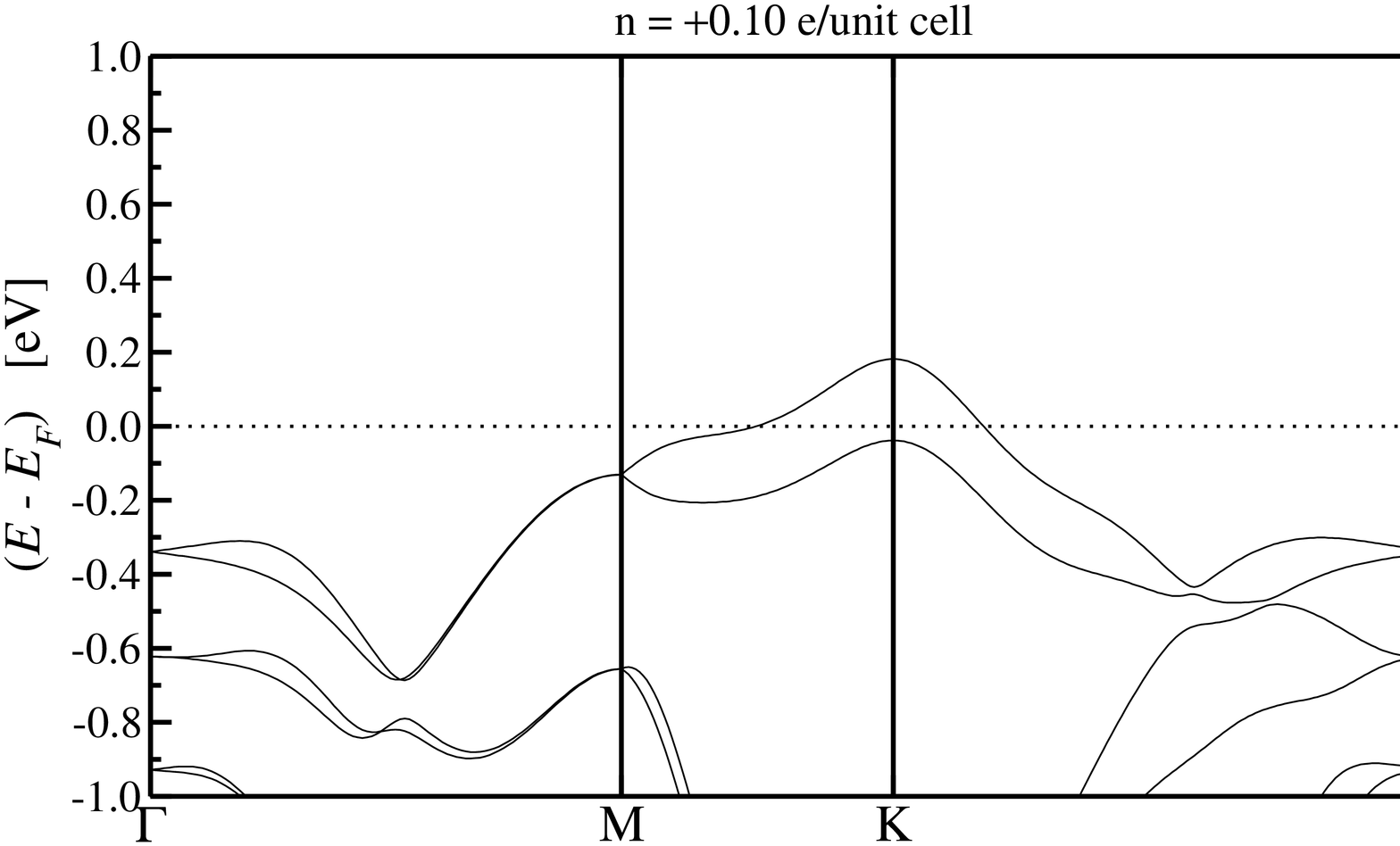}
 \includegraphics[width=0.31\textwidth,clip=,angle=-90]{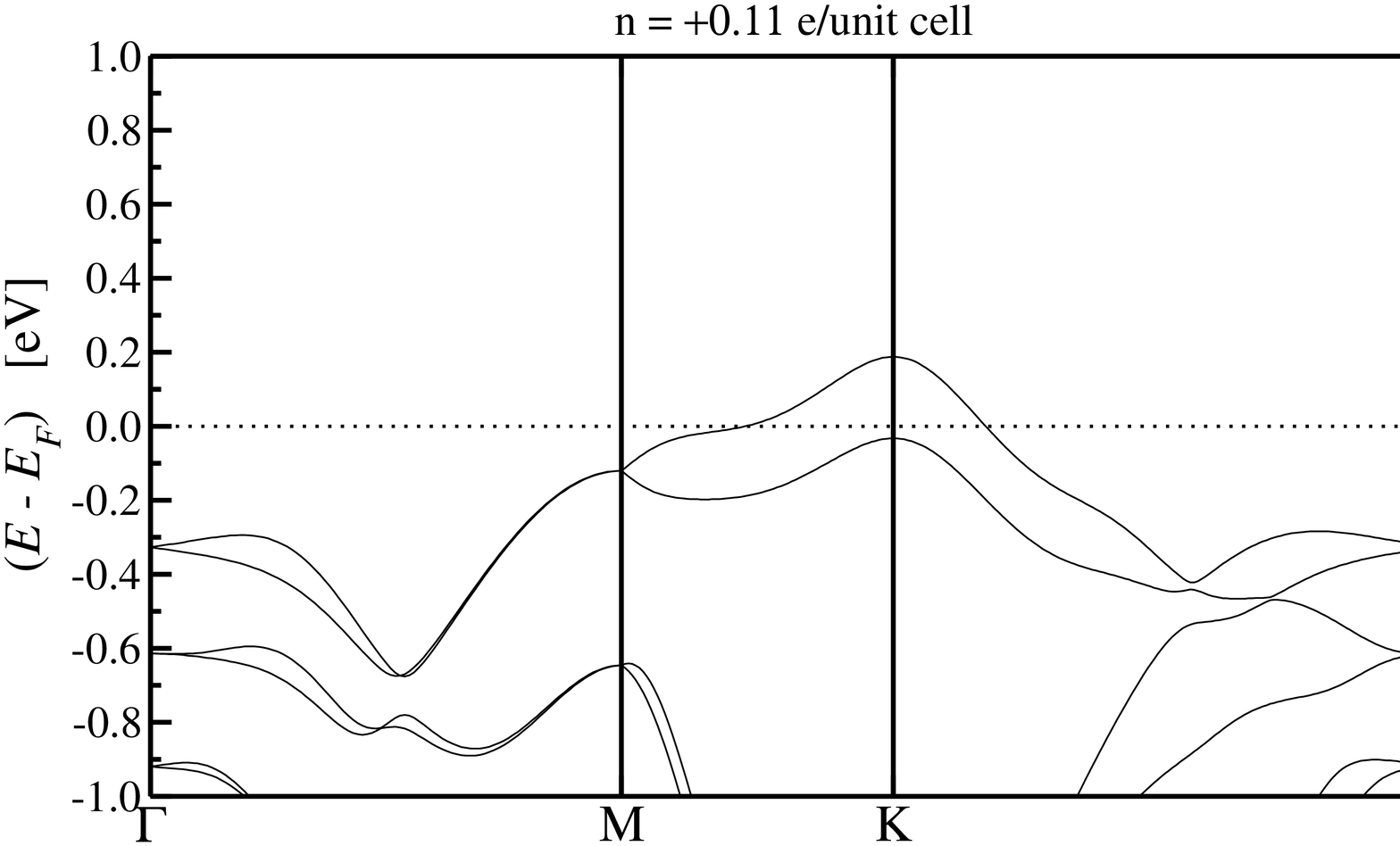}
 \includegraphics[width=0.31\textwidth,clip=,angle=-90]{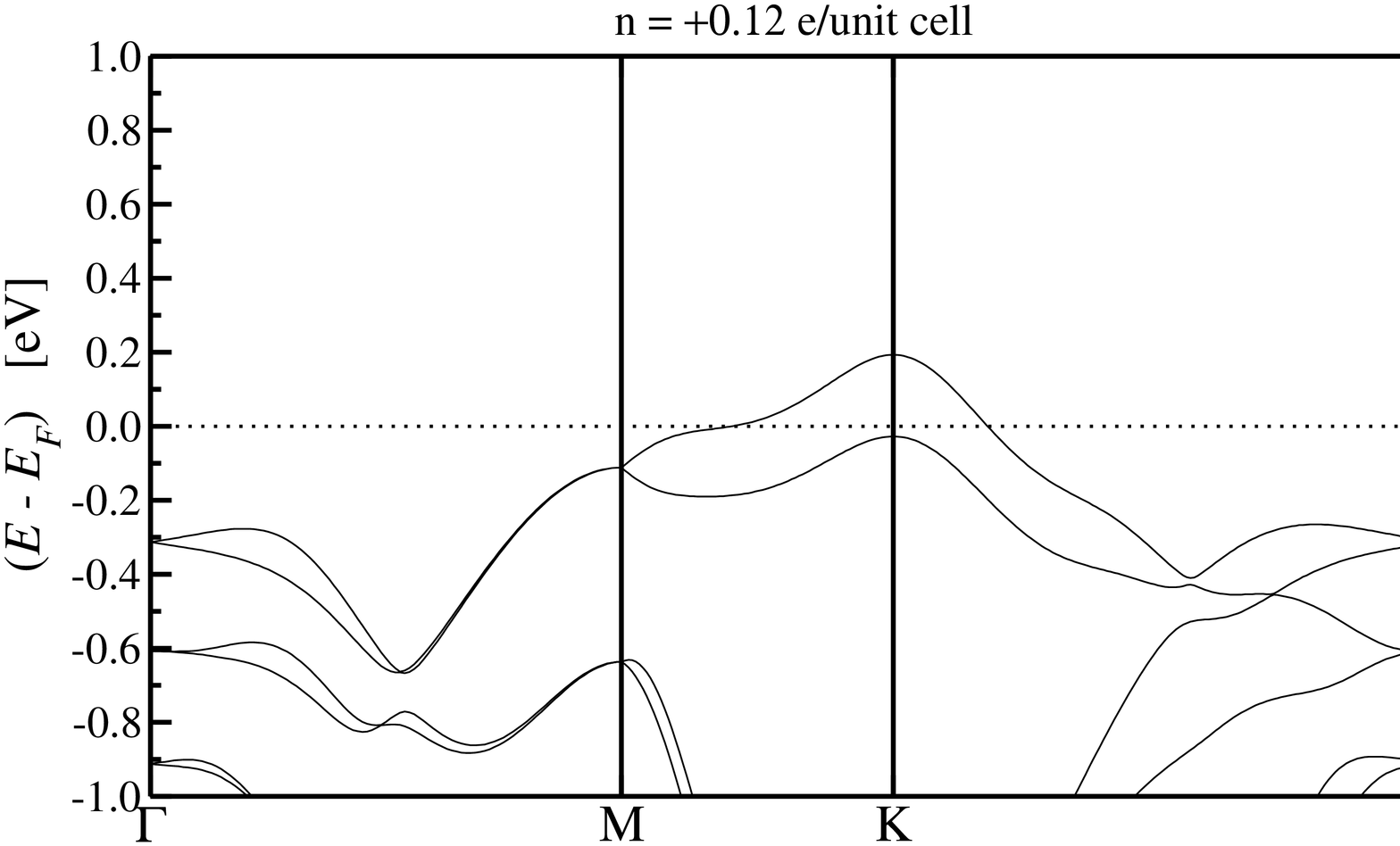}
 \includegraphics[width=0.31\textwidth,clip=,angle=-90]{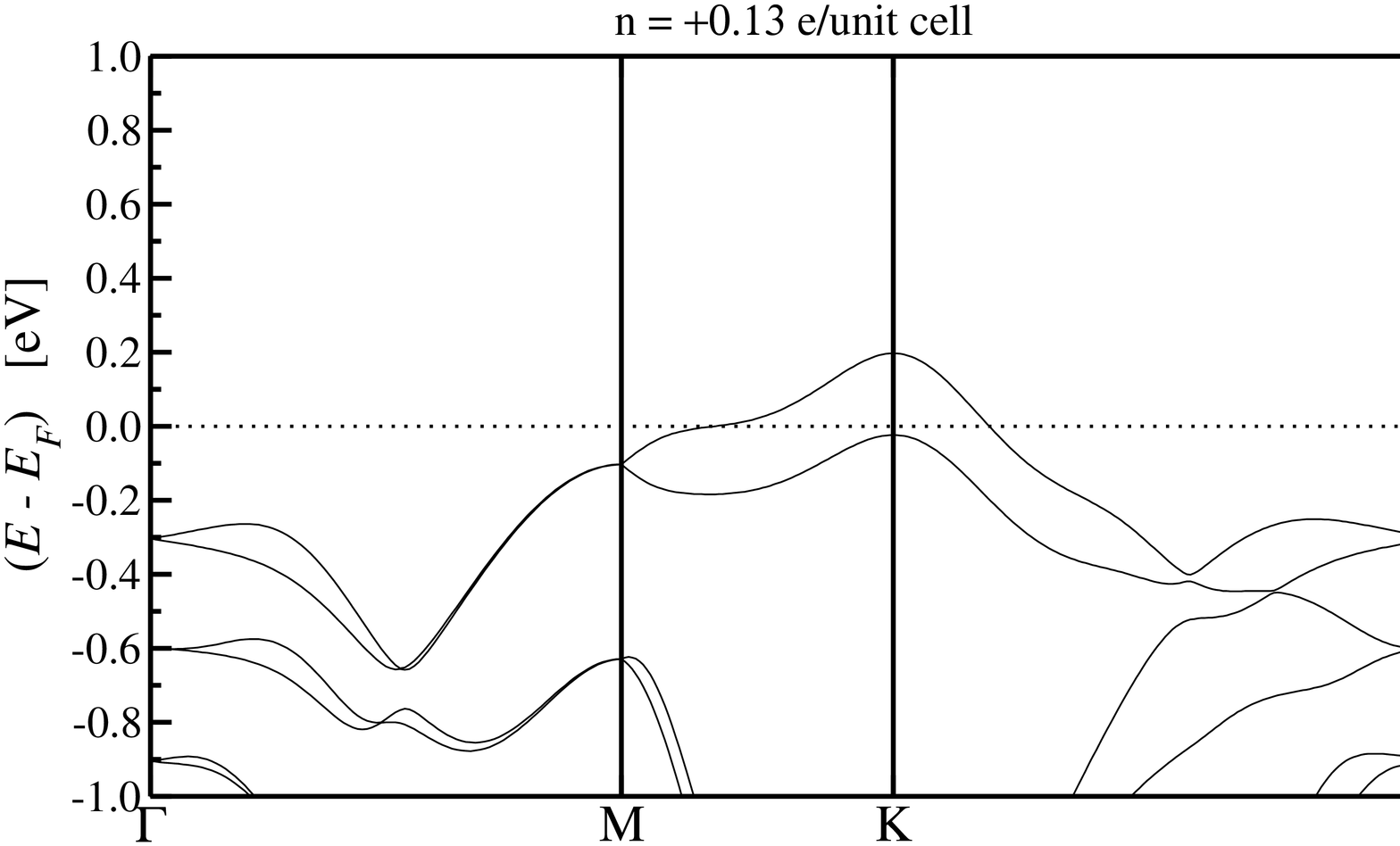}
 \caption{Band structure of monolayer MoTe$_2$ for different doping as indicated in the labels.}
\end{figure*}
\begin{figure*}[hbp]
 \centering
 \includegraphics[width=0.31\textwidth,clip=,angle=-90]{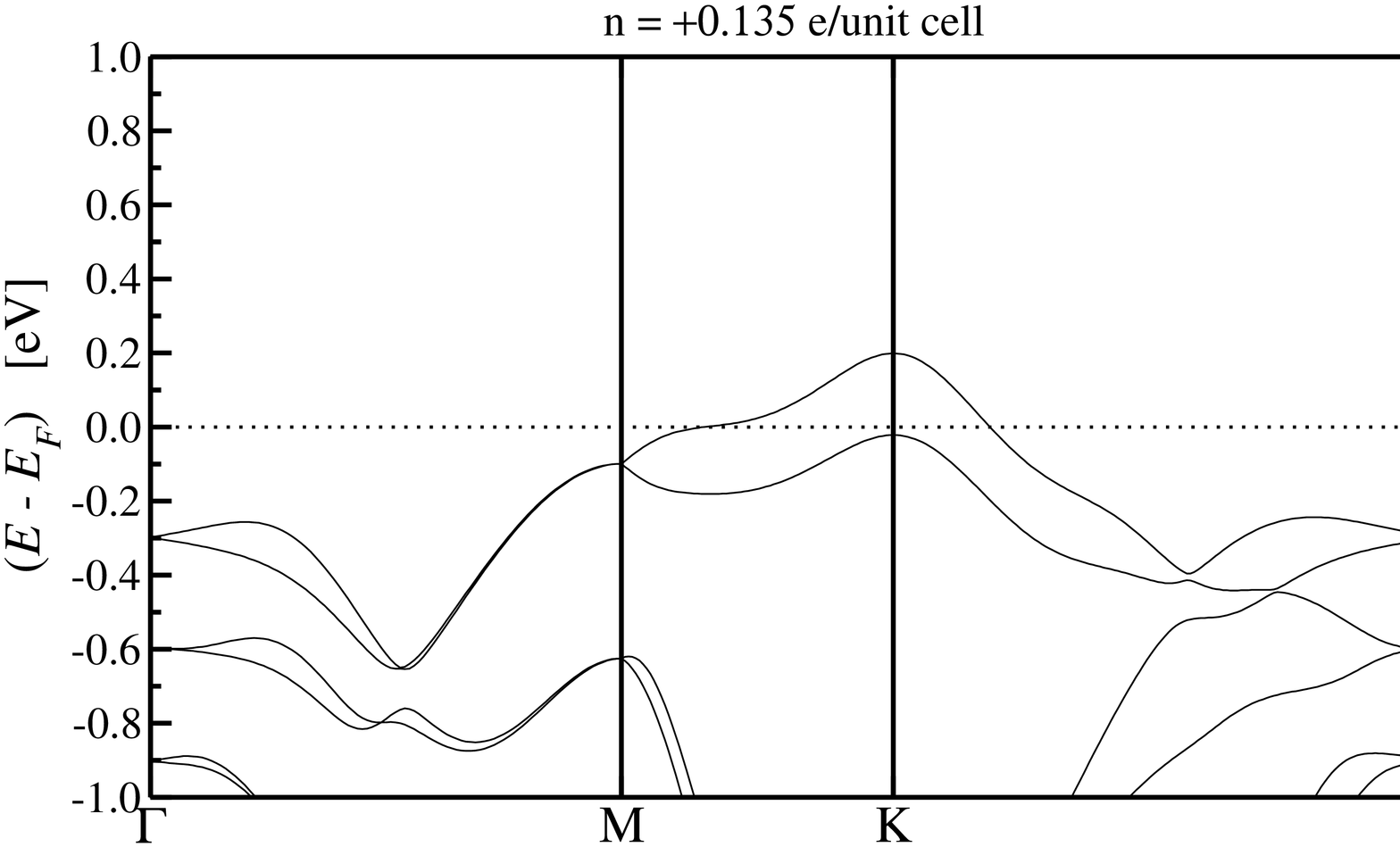}
 \includegraphics[width=0.31\textwidth,clip=,angle=-90]{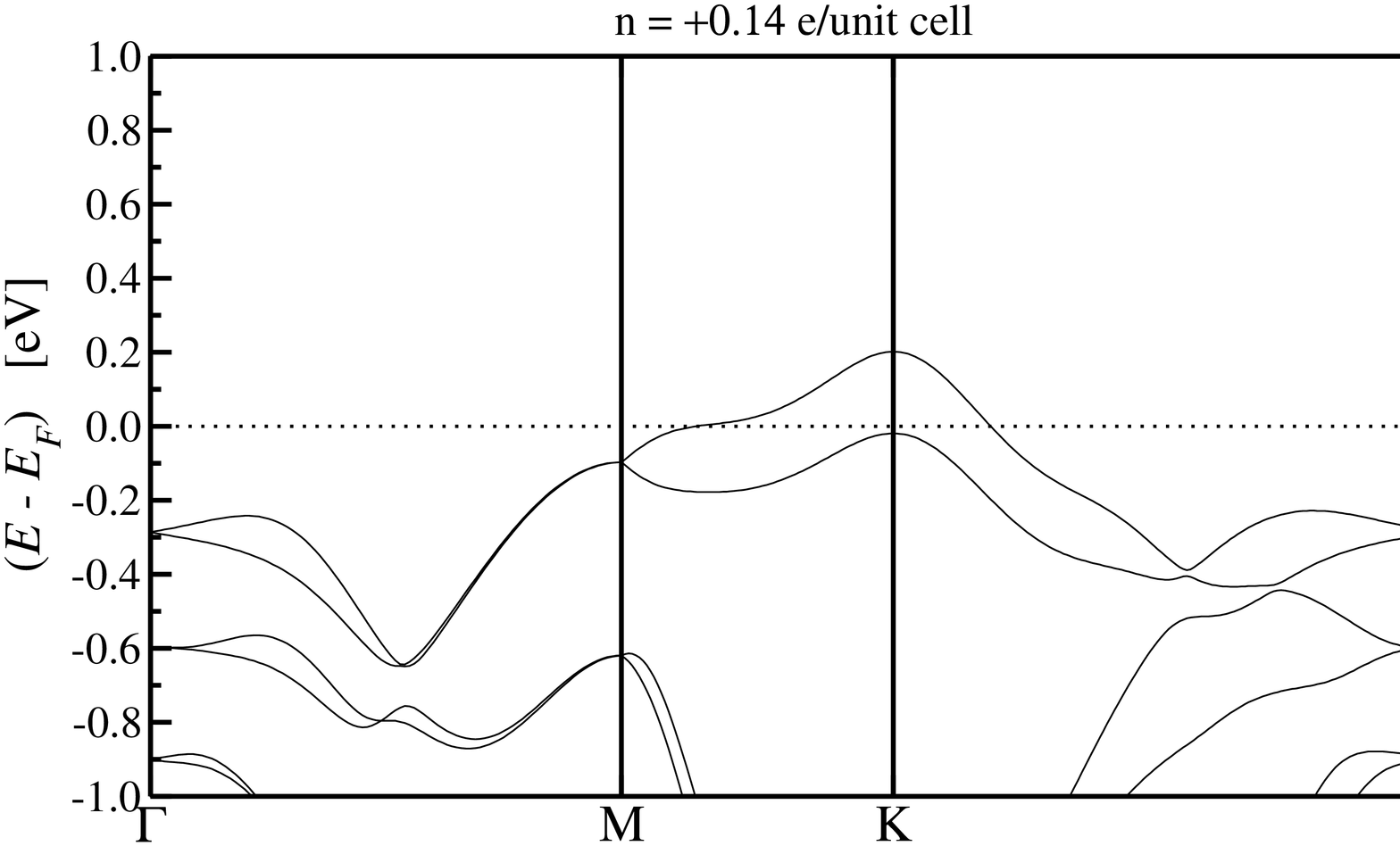}
 \includegraphics[width=0.31\textwidth,clip=,angle=-90]{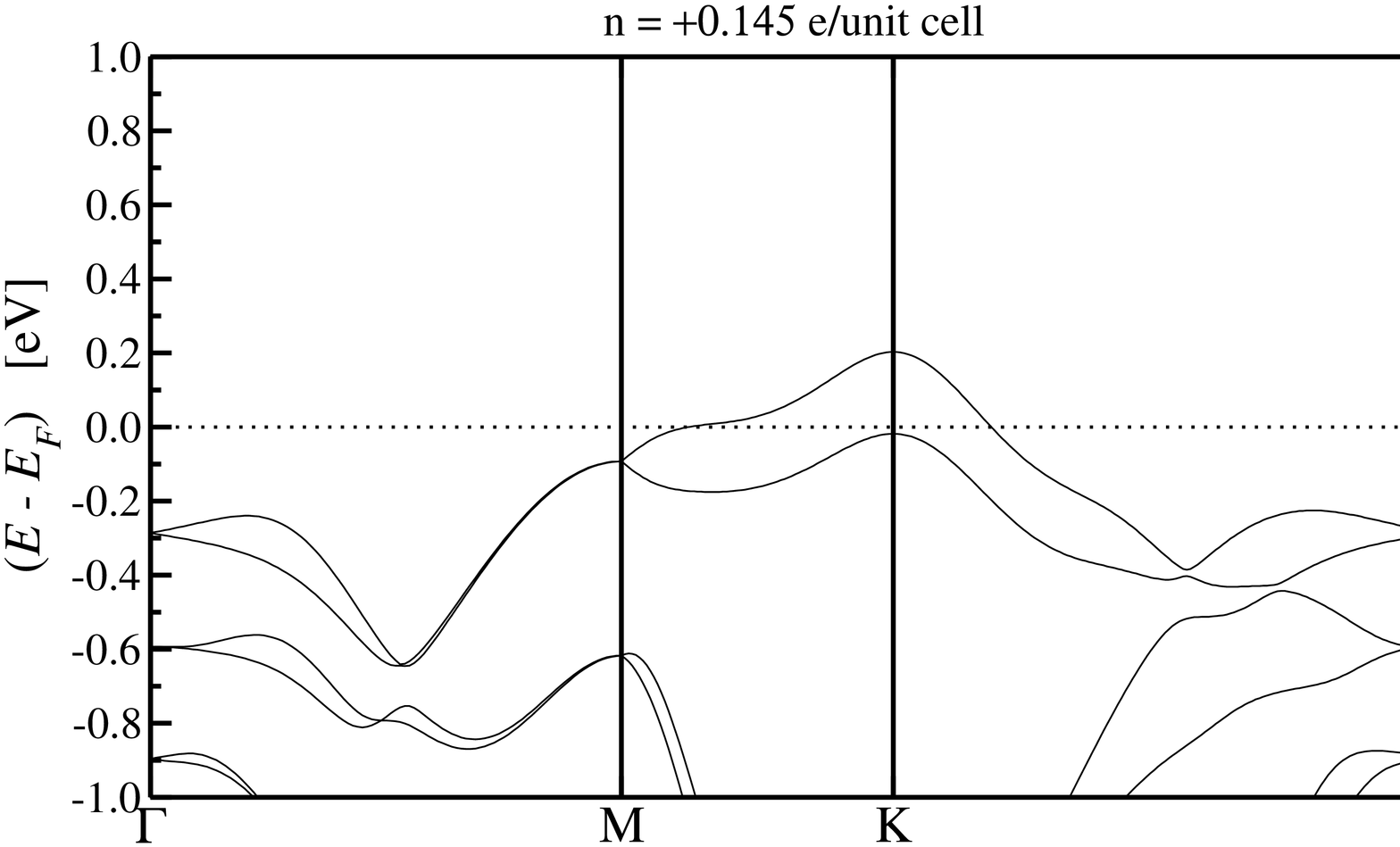}
 \includegraphics[width=0.31\textwidth,clip=,angle=-90]{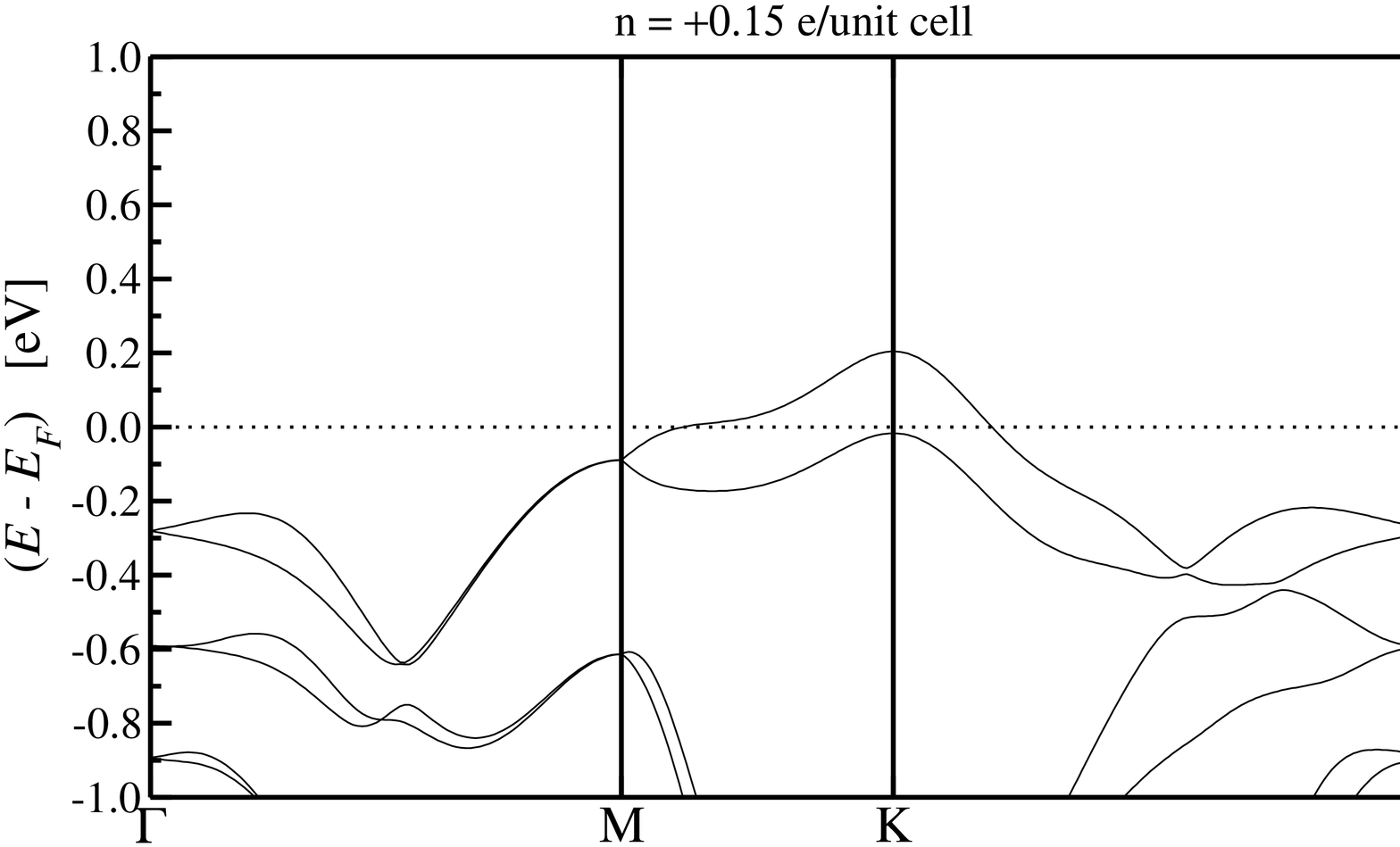}
 \includegraphics[width=0.31\textwidth,clip=,angle=-90]{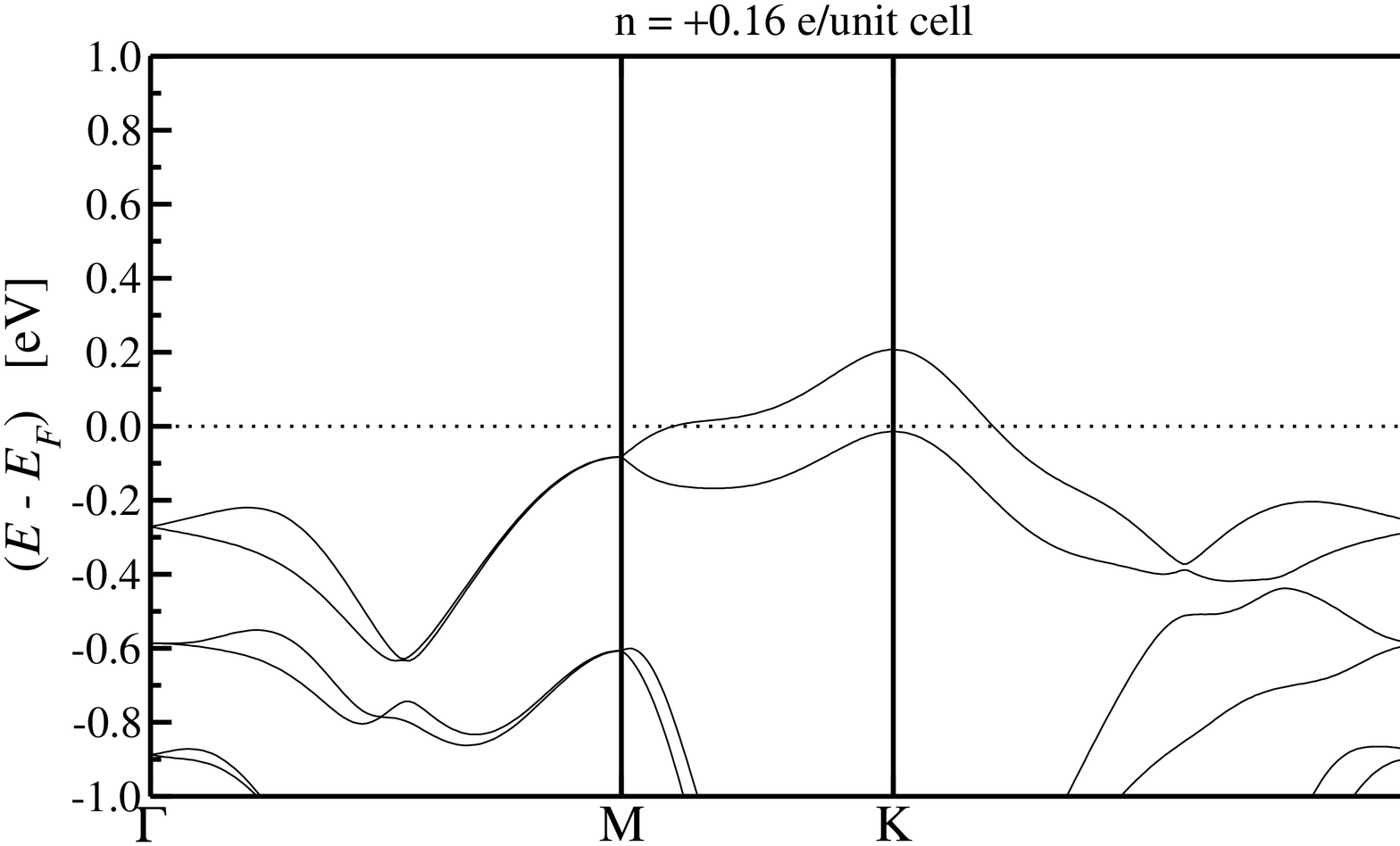}
 \includegraphics[width=0.31\textwidth,clip=,angle=-90]{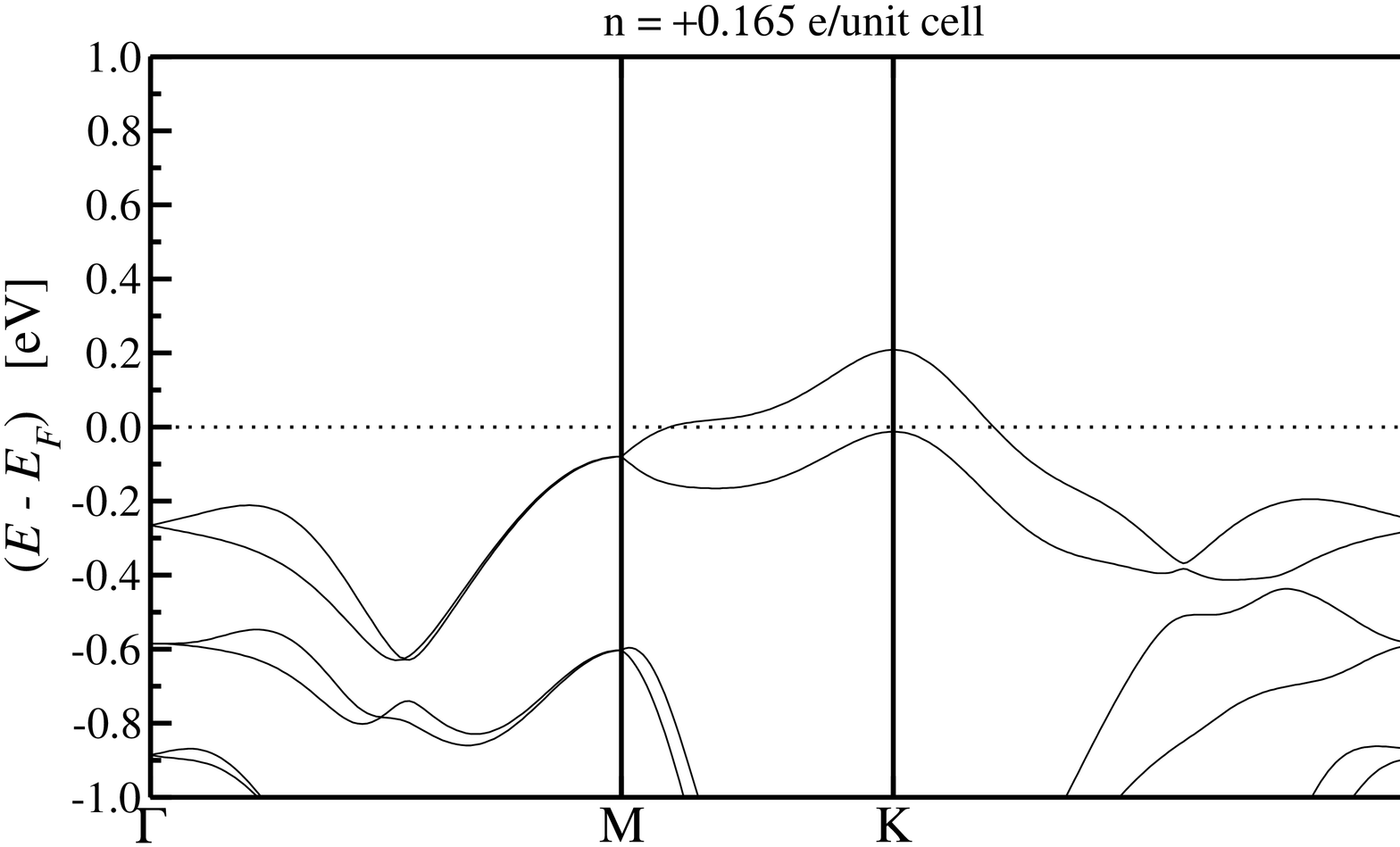}
 \includegraphics[width=0.31\textwidth,clip=,angle=-90]{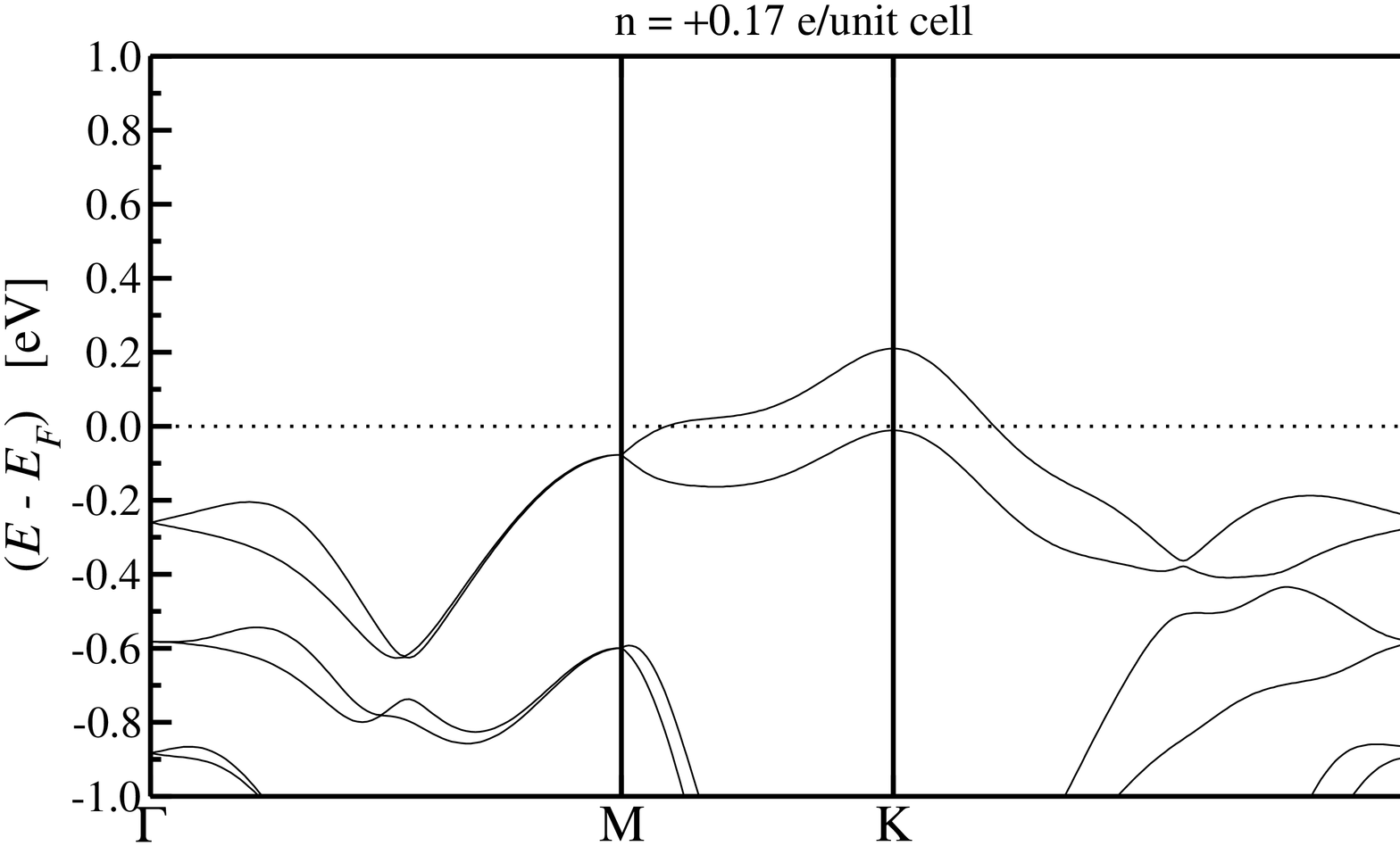}
 \includegraphics[width=0.31\textwidth,clip=,angle=-90]{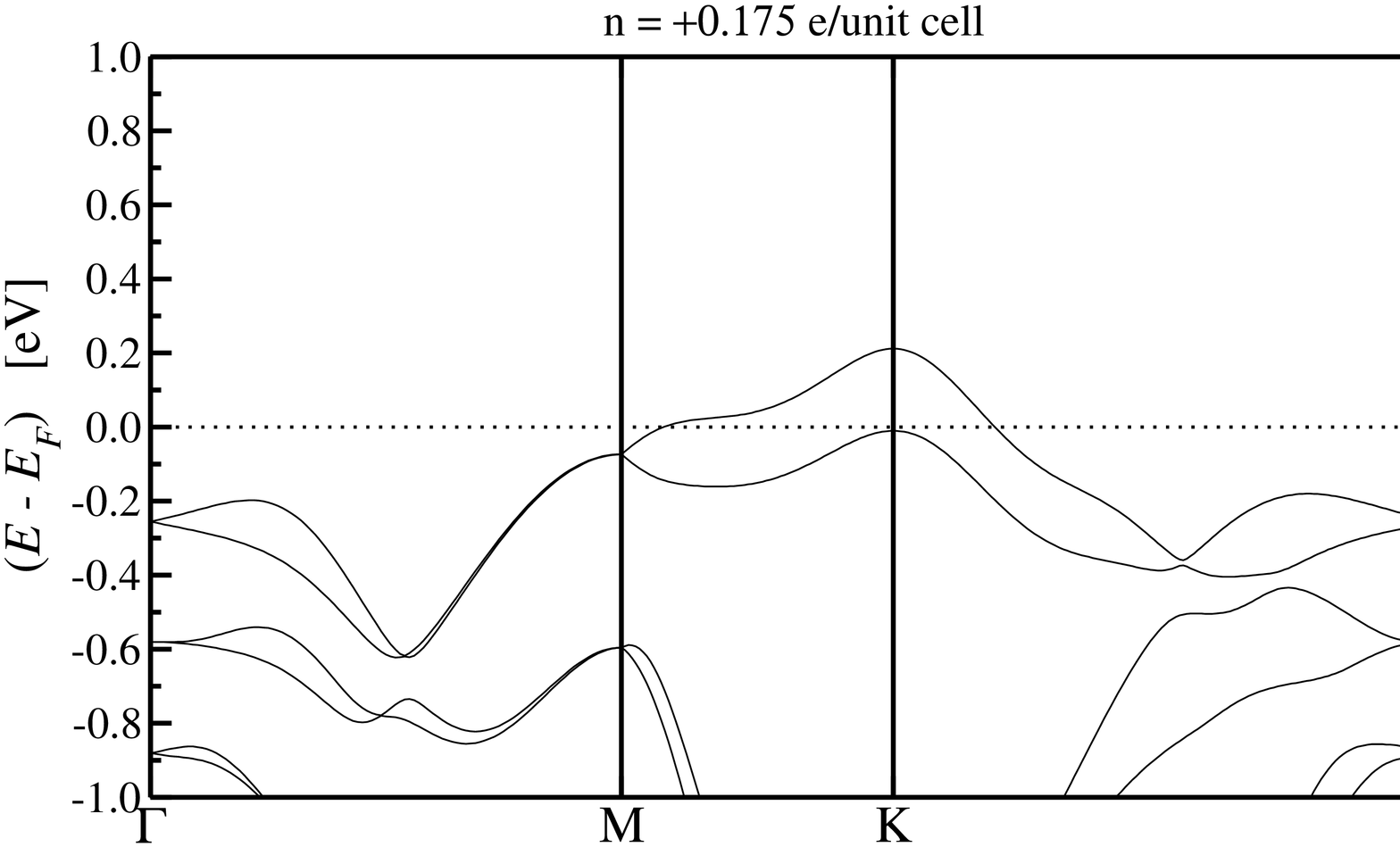}
 \caption{Band structure of monolayer MoTe$_2$ for different doping as indicated in the labels.}
\end{figure*}
\begin{figure*}[hbp]
 \centering
 \includegraphics[width=0.31\textwidth,clip=,angle=-90]{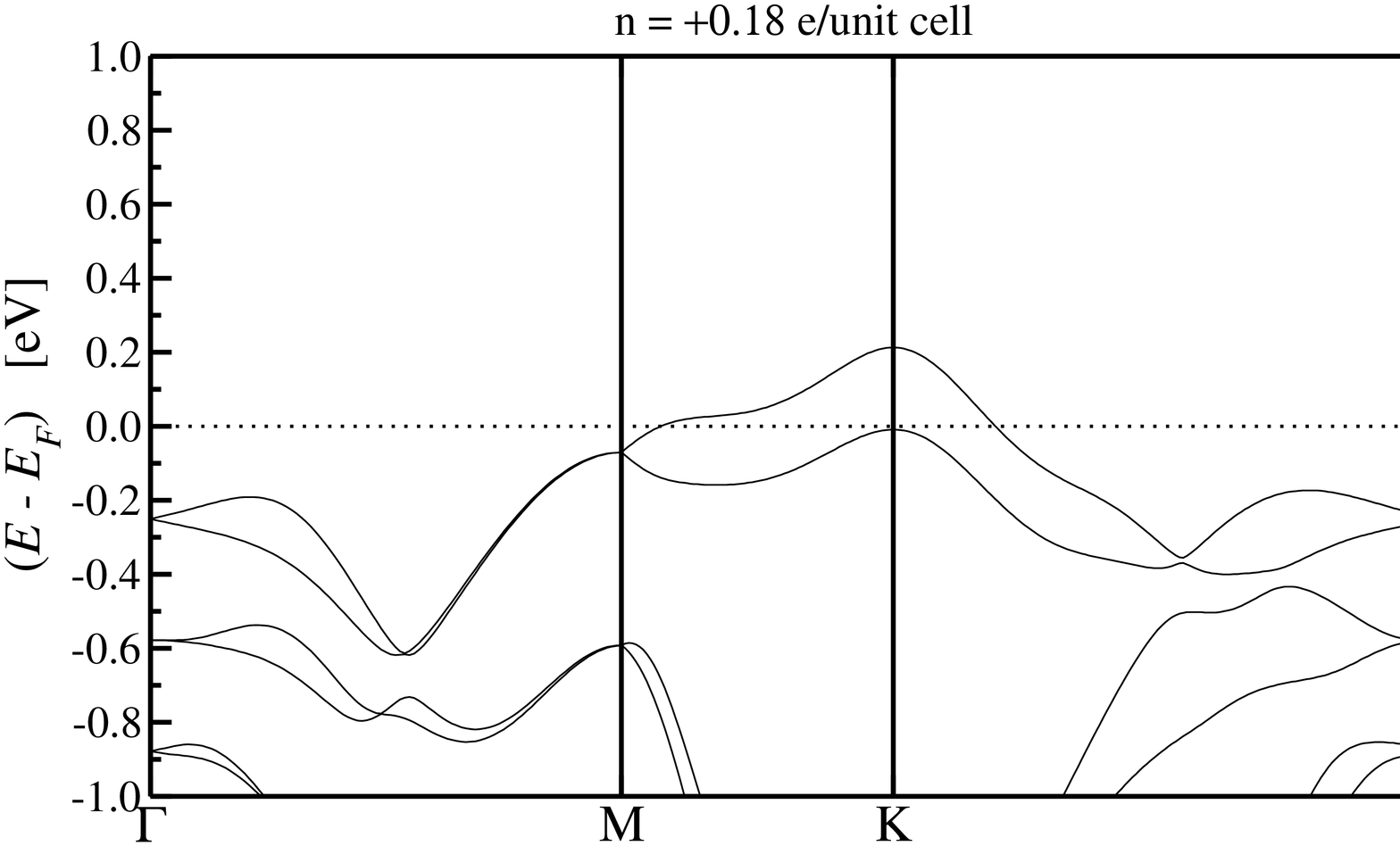}
 \includegraphics[width=0.31\textwidth,clip=,angle=-90]{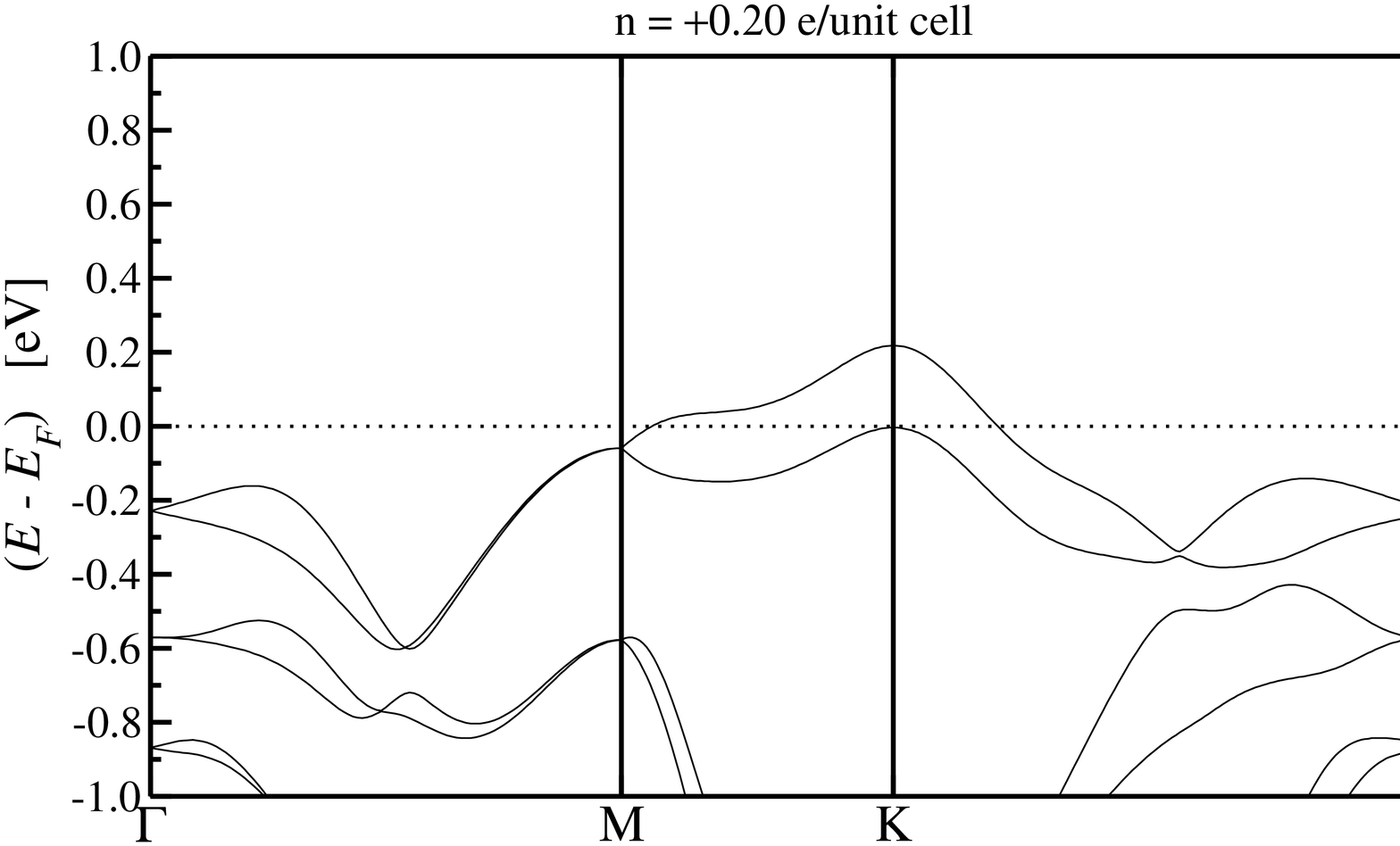}
 \includegraphics[width=0.31\textwidth,clip=,angle=-90]{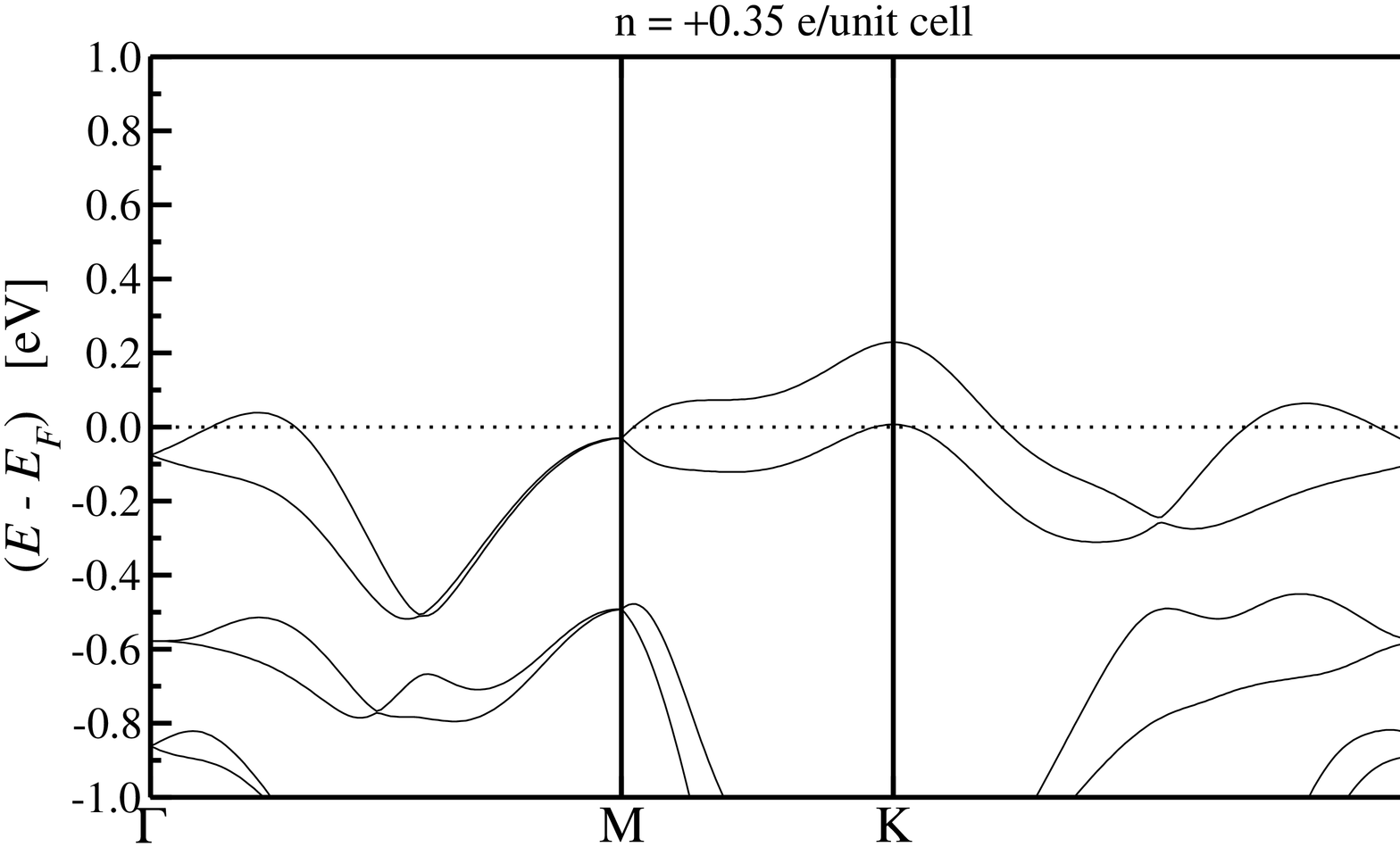}
 \caption{Band structure of monolayer MoTe$_2$ for different doping as indicated in the labels.}
\end{figure*}
\begin{figure*}[hbp]
 \centering
 \includegraphics[width=0.31\textwidth,clip=,angle=-90]{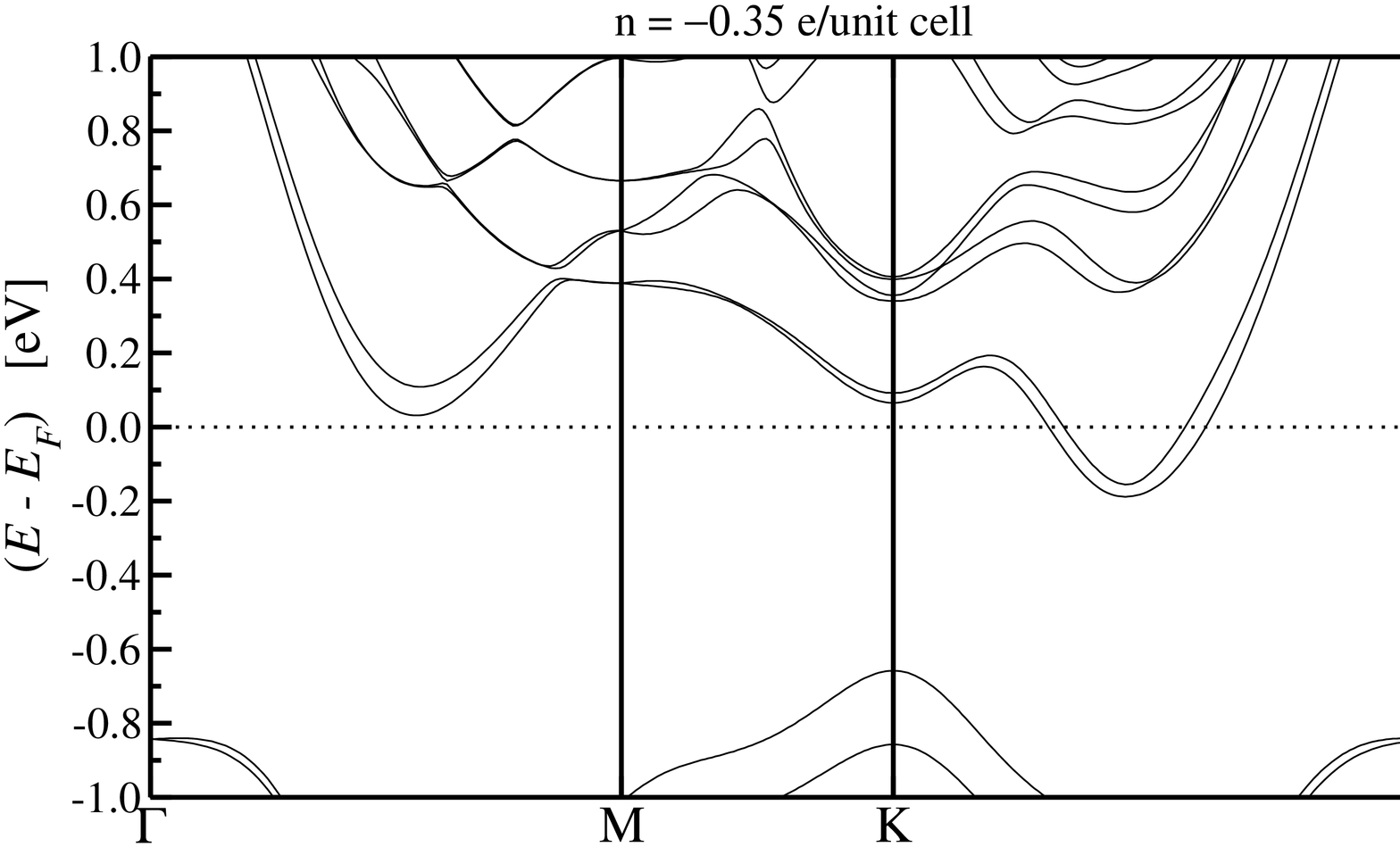}
 \includegraphics[width=0.31\textwidth,clip=,angle=-90]{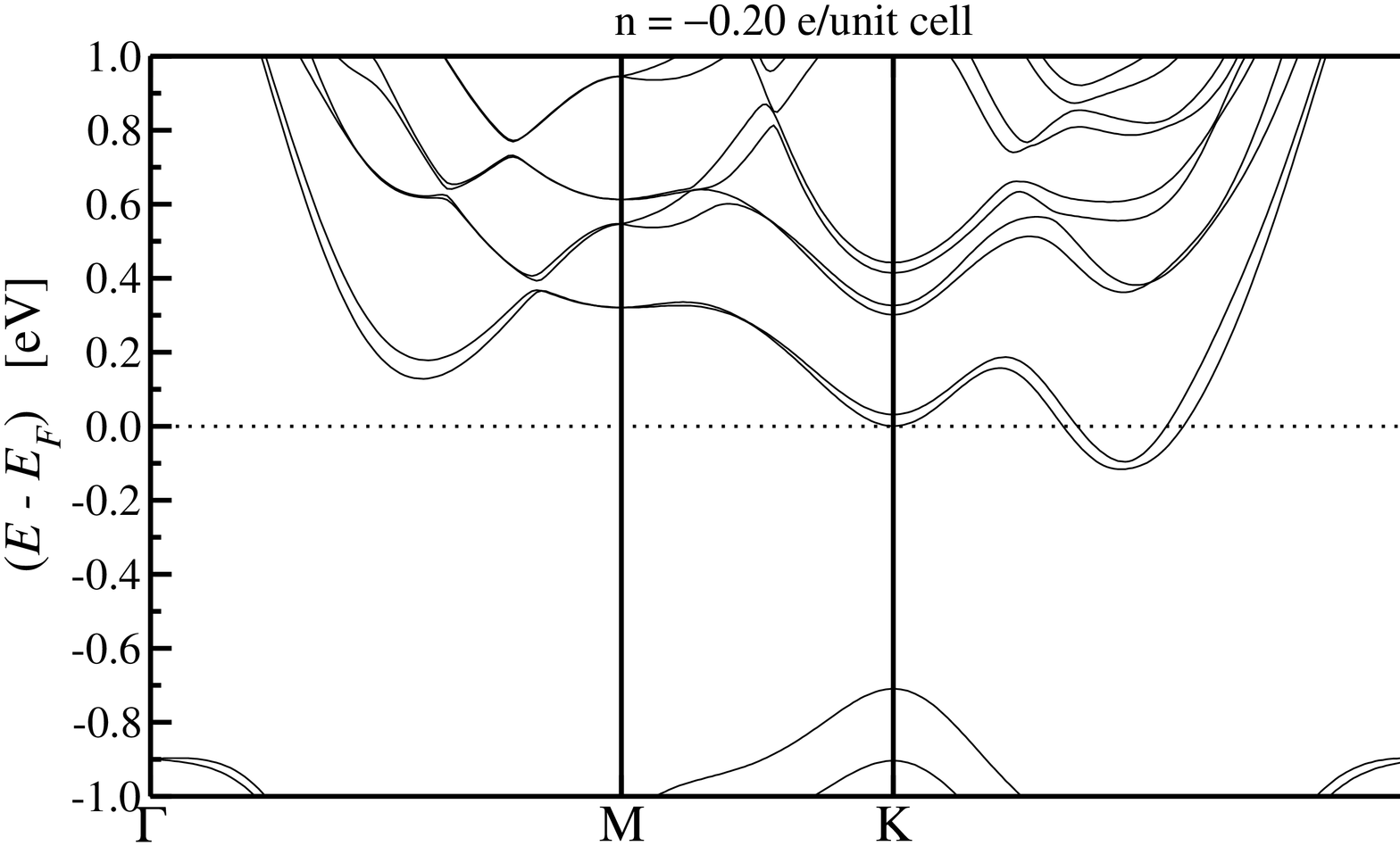}
 \includegraphics[width=0.31\textwidth,clip=,angle=-90]{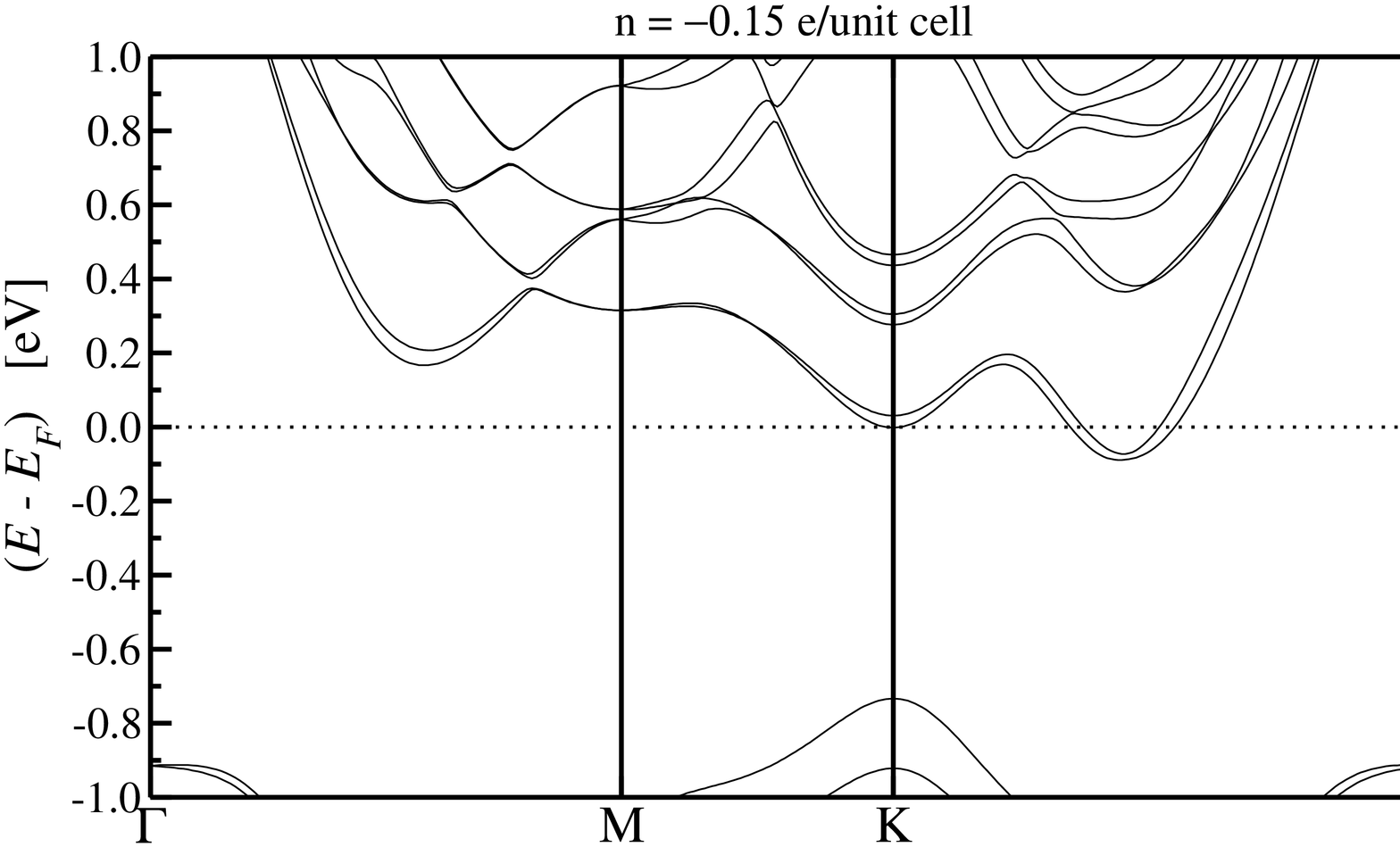}
 \includegraphics[width=0.31\textwidth,clip=,angle=-90]{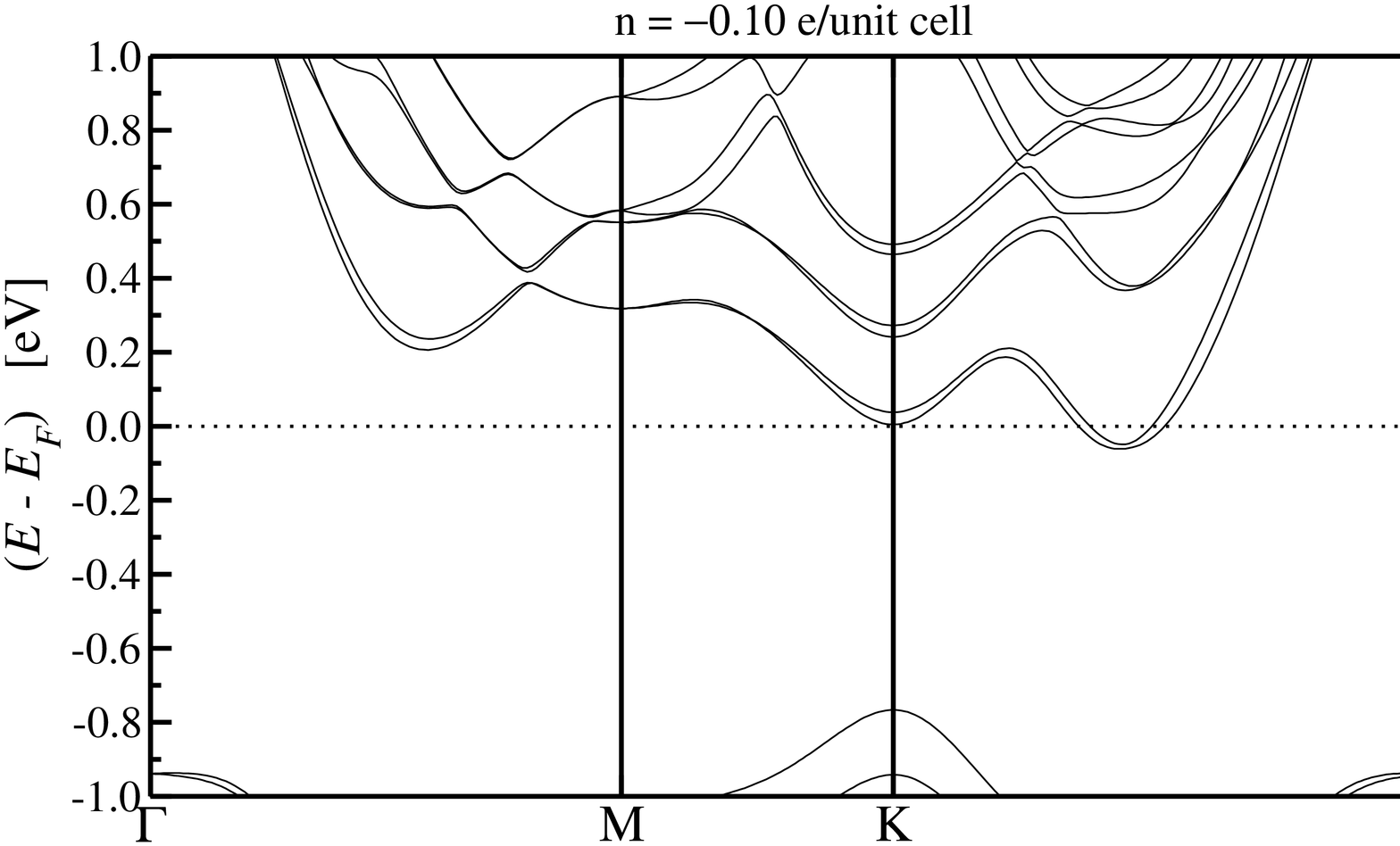}
 \includegraphics[width=0.31\textwidth,clip=,angle=-90]{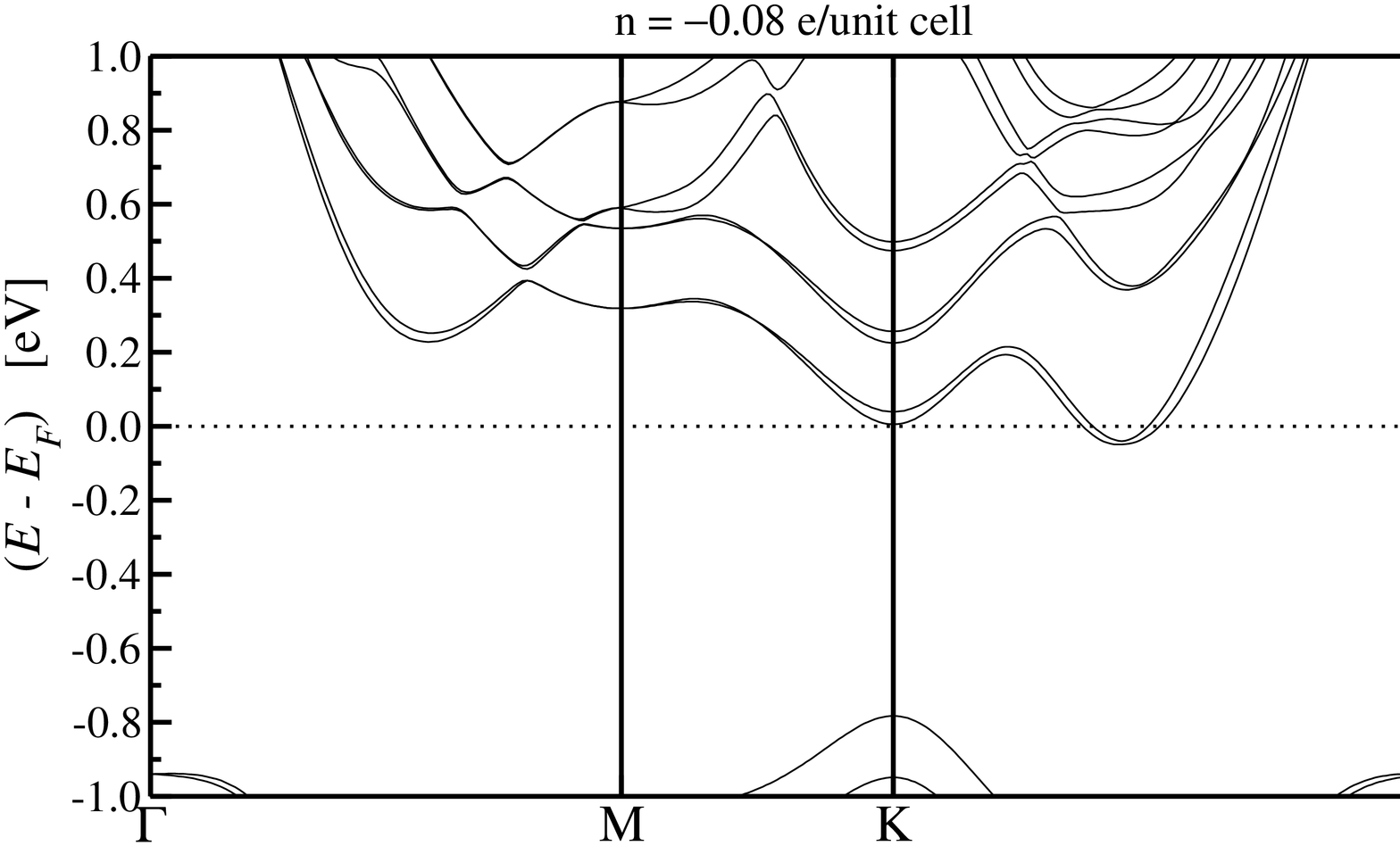}
 \includegraphics[width=0.31\textwidth,clip=,angle=-90]{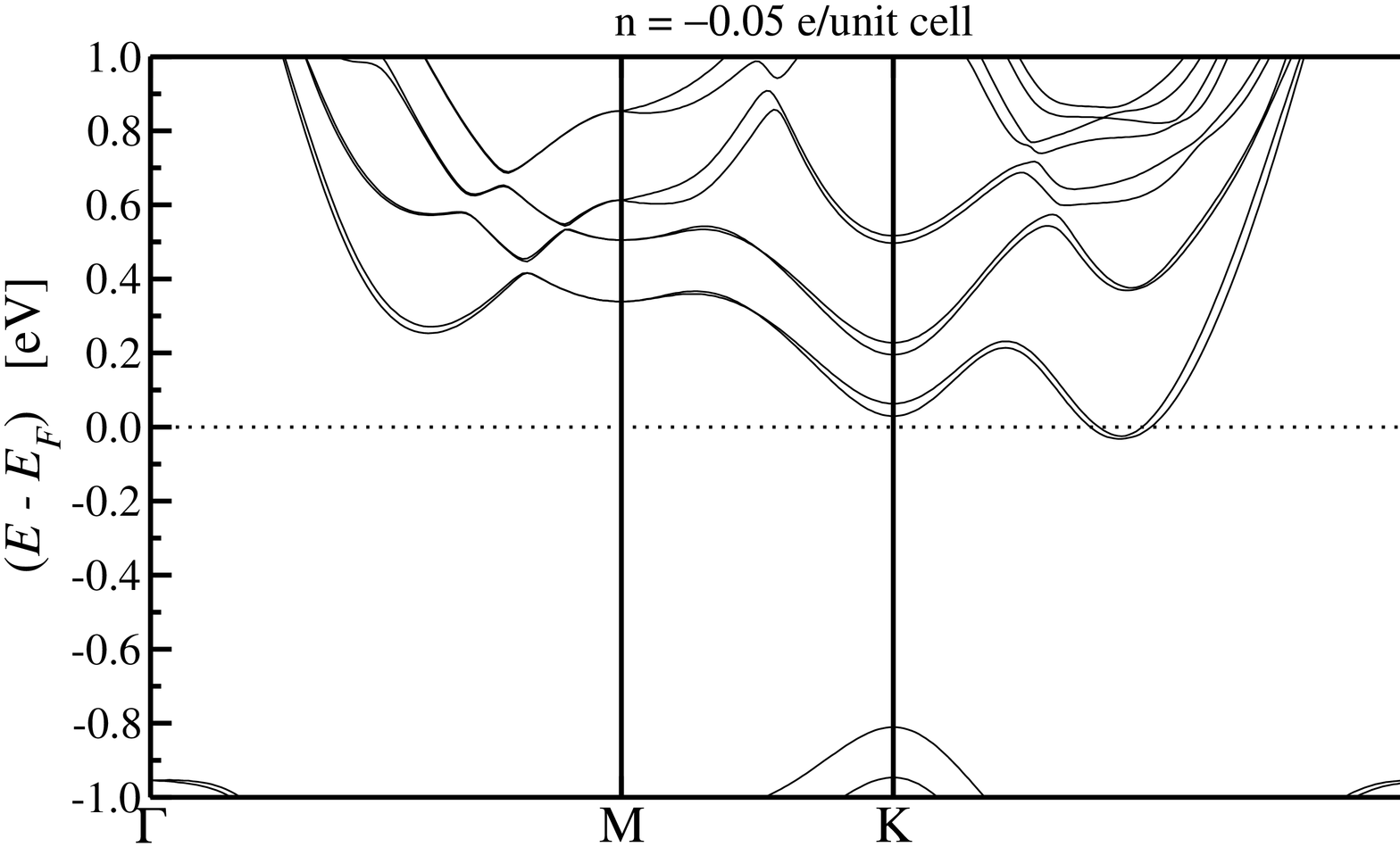}
 \includegraphics[width=0.31\textwidth,clip=,angle=-90]{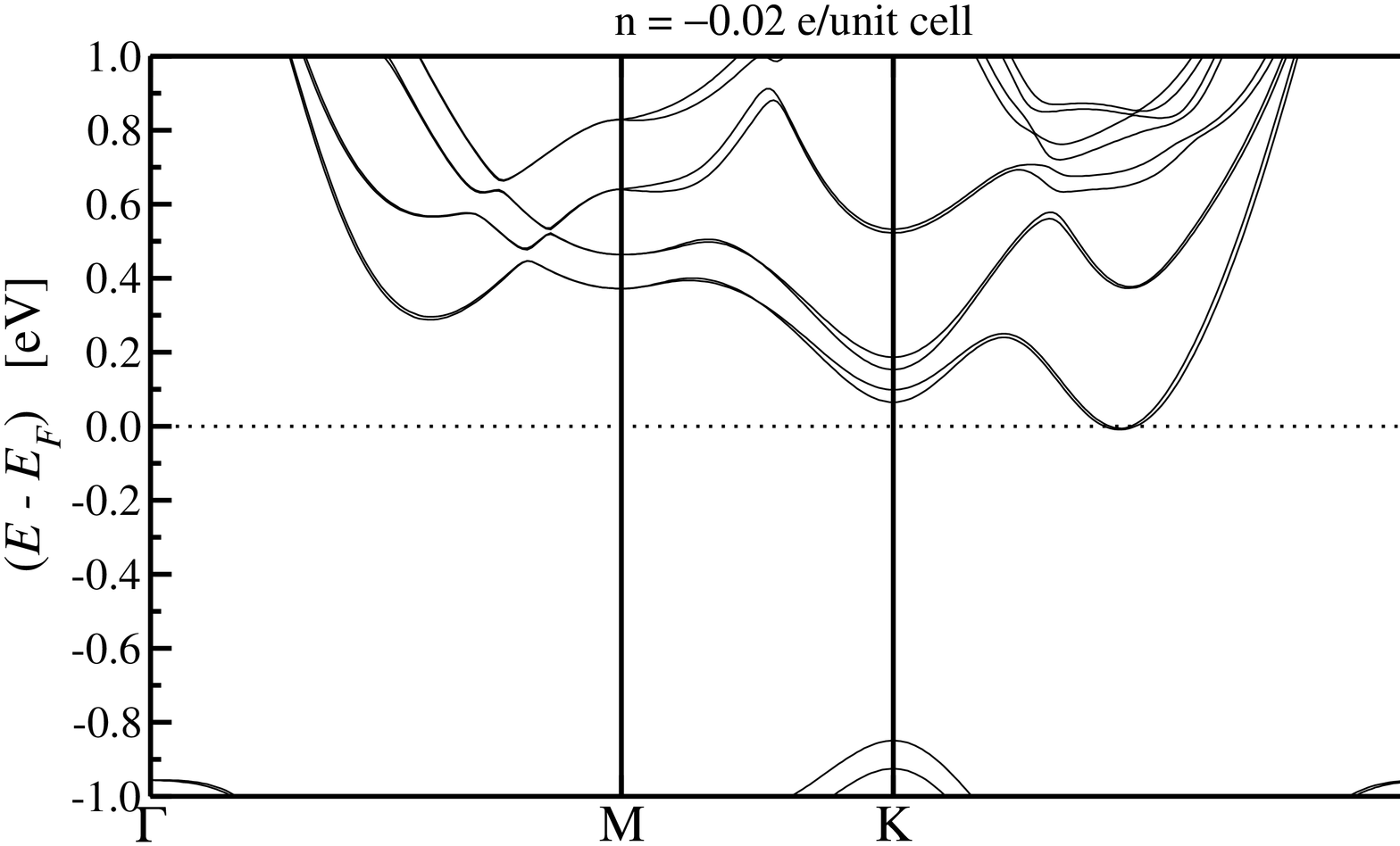}
 \includegraphics[width=0.31\textwidth,clip=,angle=-90]{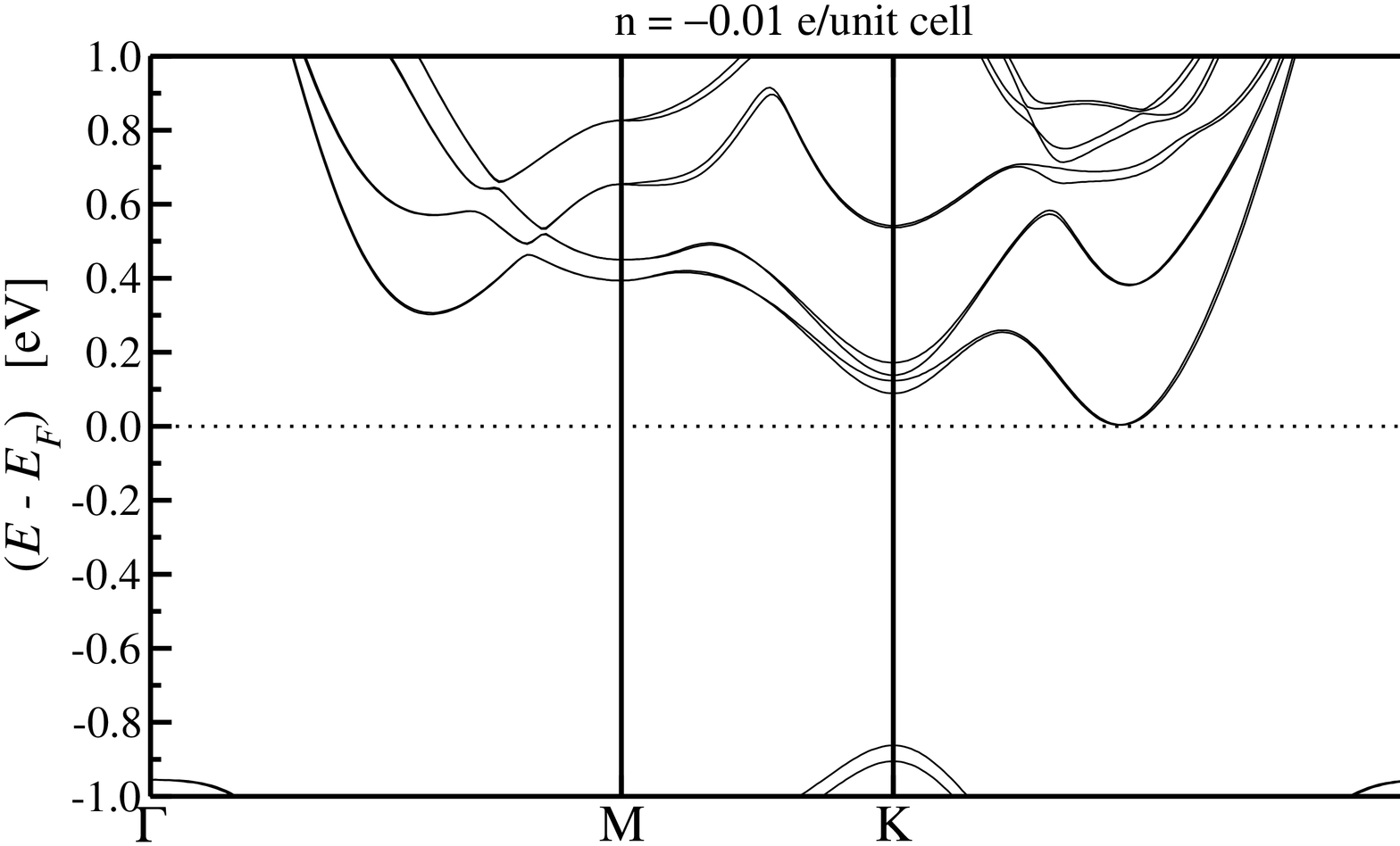}
 \caption{Band structure of bilayer MoTe$_2$ for different doping as indicated in the labels.}
\end{figure*}
\begin{figure*}[hbp]
 \centering
 \includegraphics[width=0.31\textwidth,clip=,angle=-90]{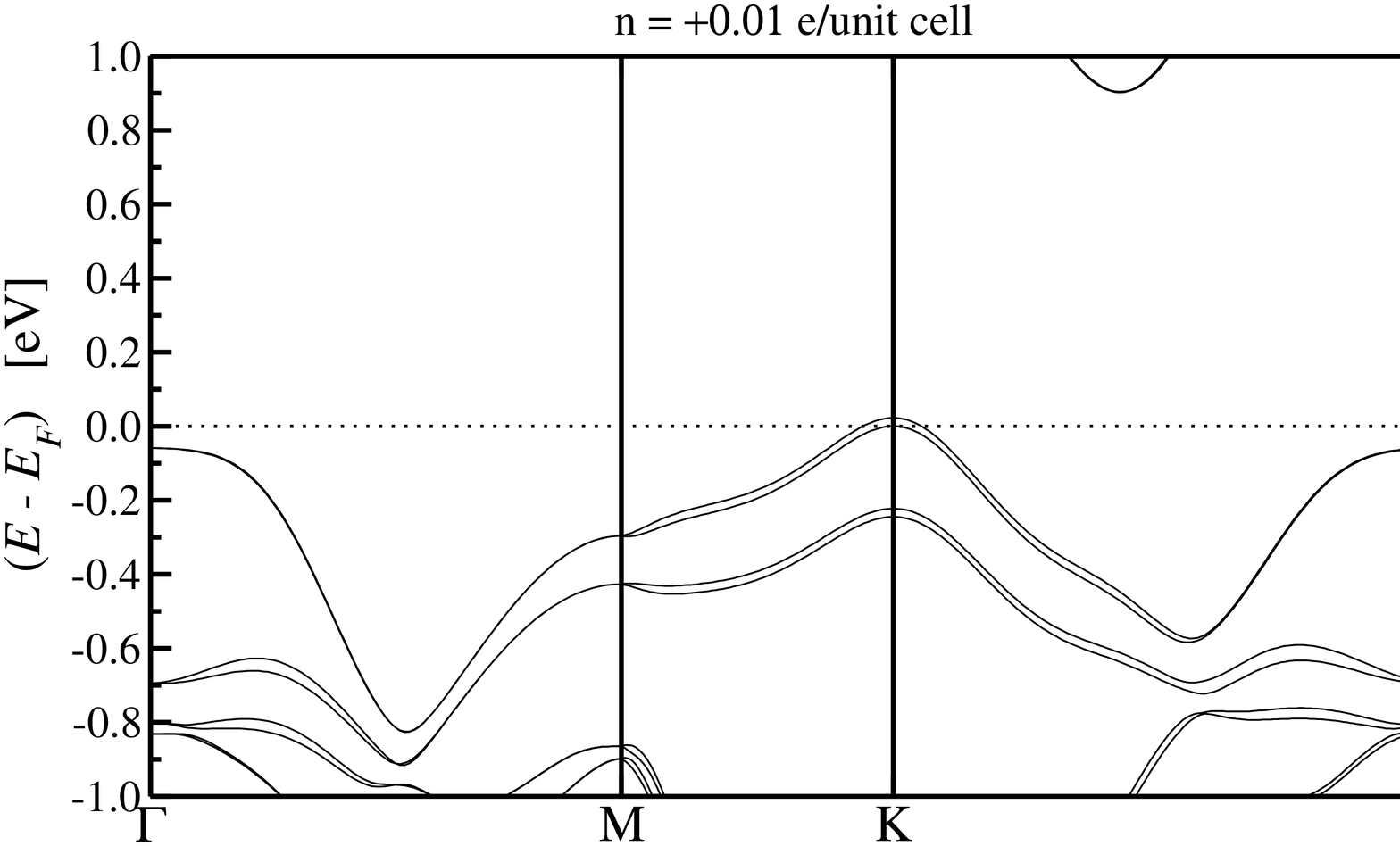}
 \includegraphics[width=0.31\textwidth,clip=,angle=-90]{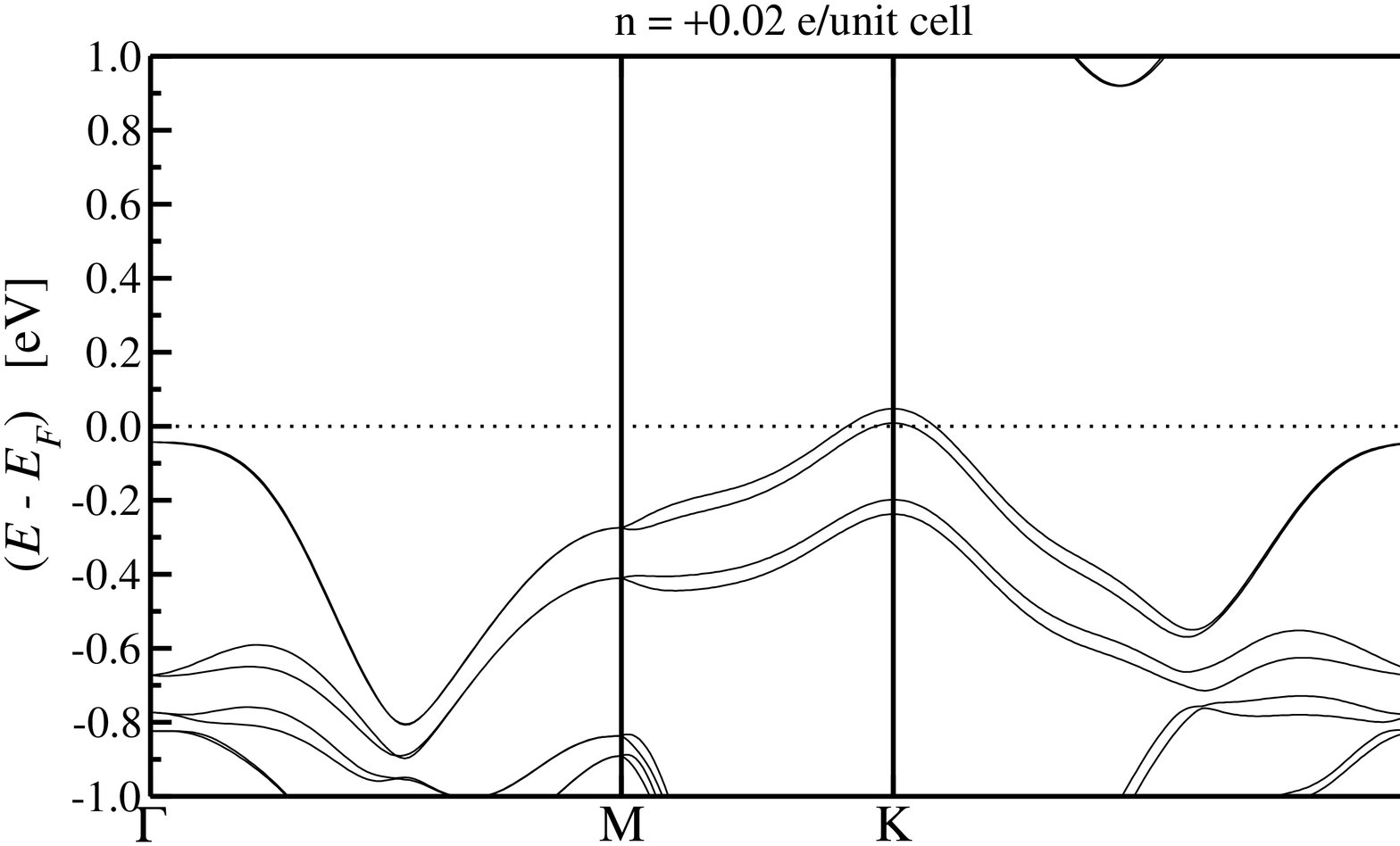}
 \includegraphics[width=0.31\textwidth,clip=,angle=-90]{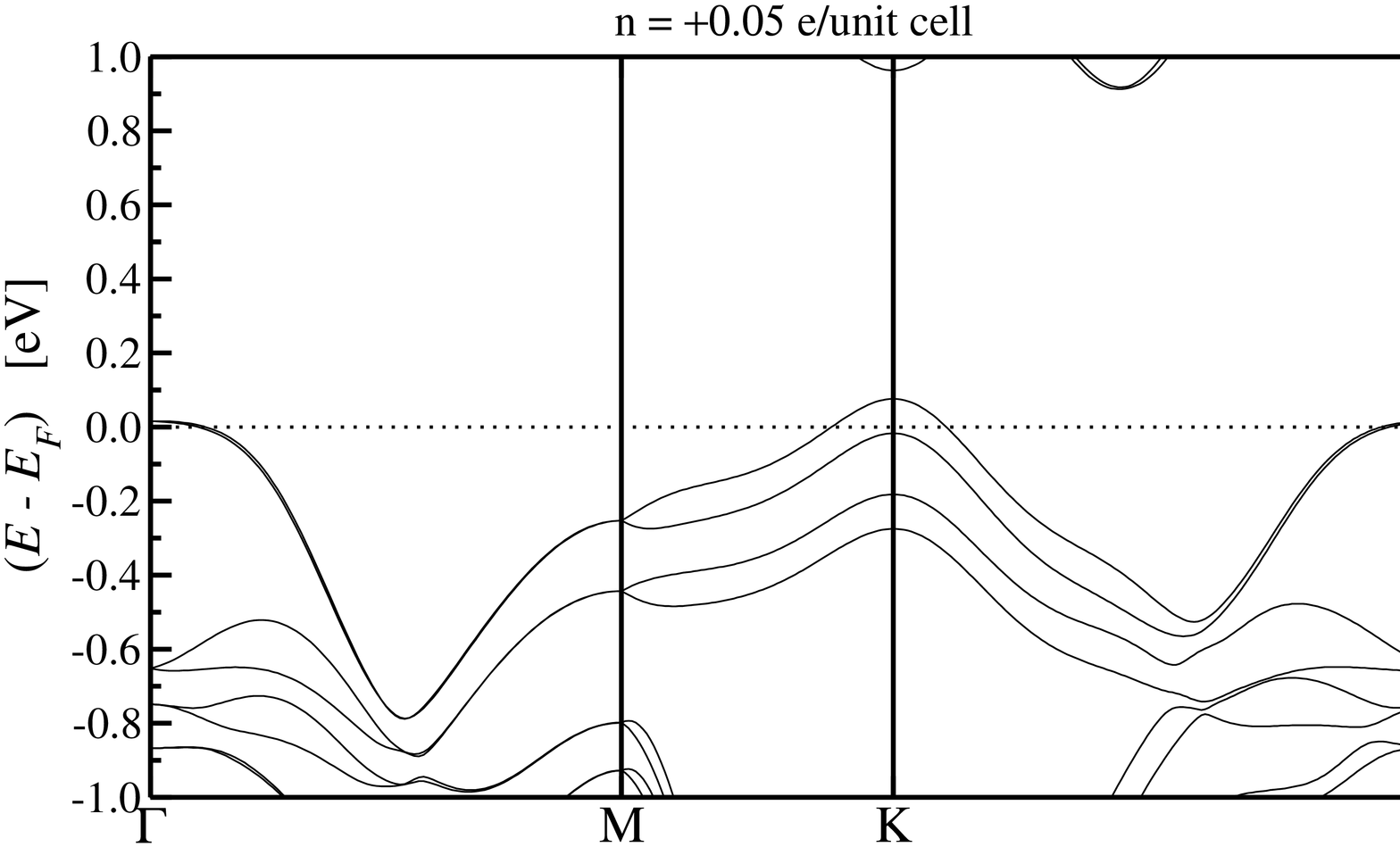}
 \includegraphics[width=0.31\textwidth,clip=,angle=-90]{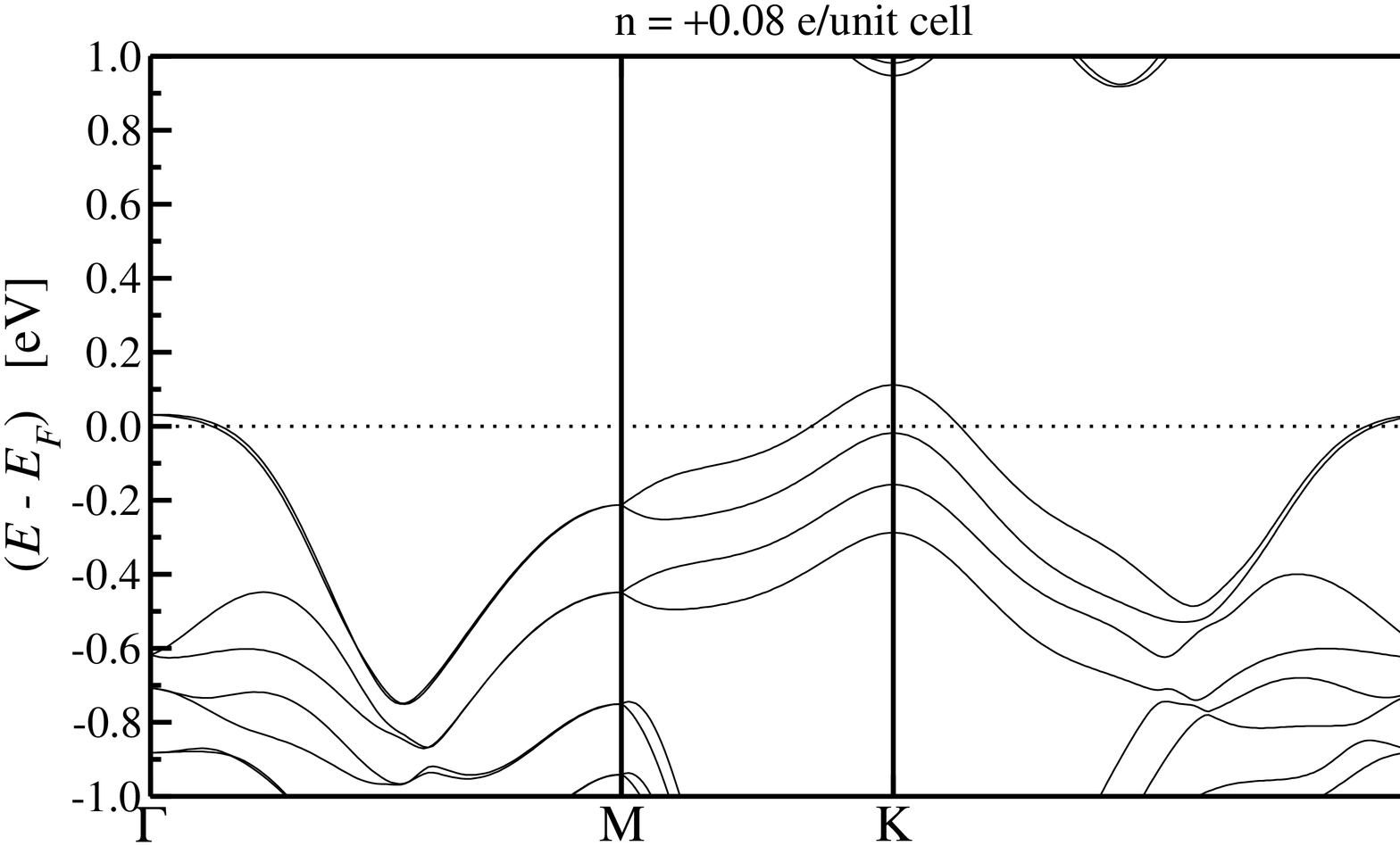}
 \includegraphics[width=0.31\textwidth,clip=,angle=-90]{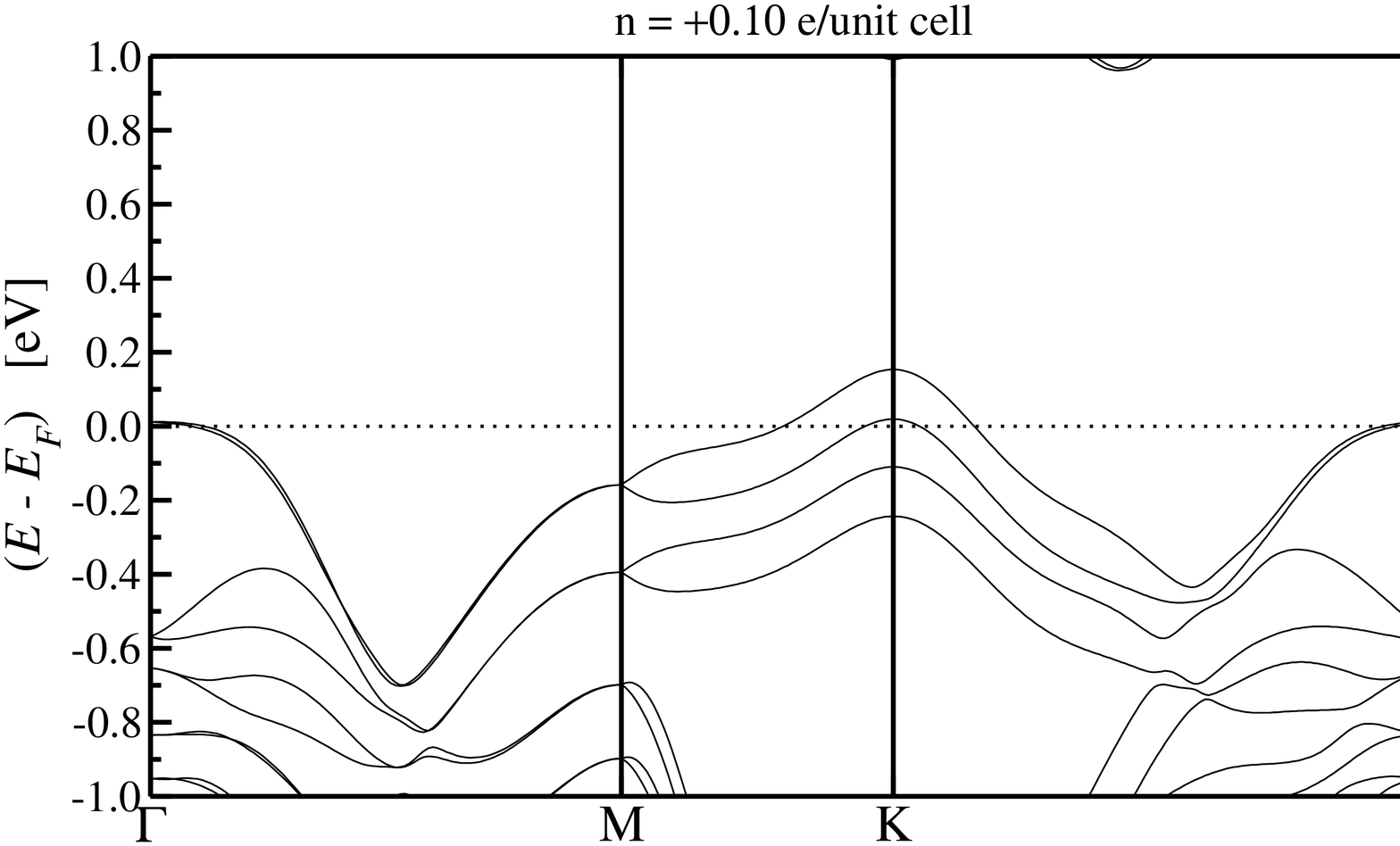}
 \includegraphics[width=0.31\textwidth,clip=,angle=-90]{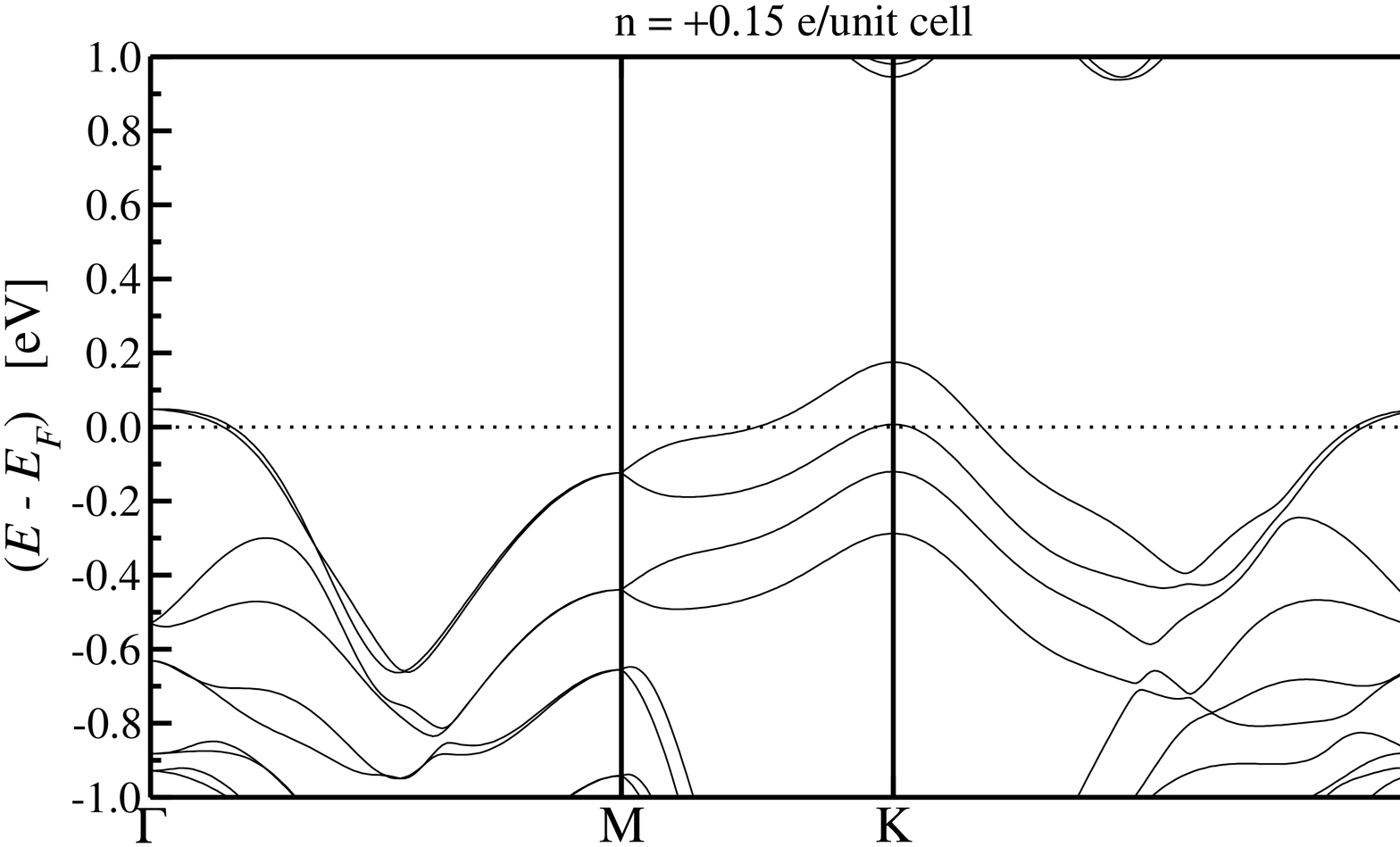}
 \includegraphics[width=0.31\textwidth,clip=,angle=-90]{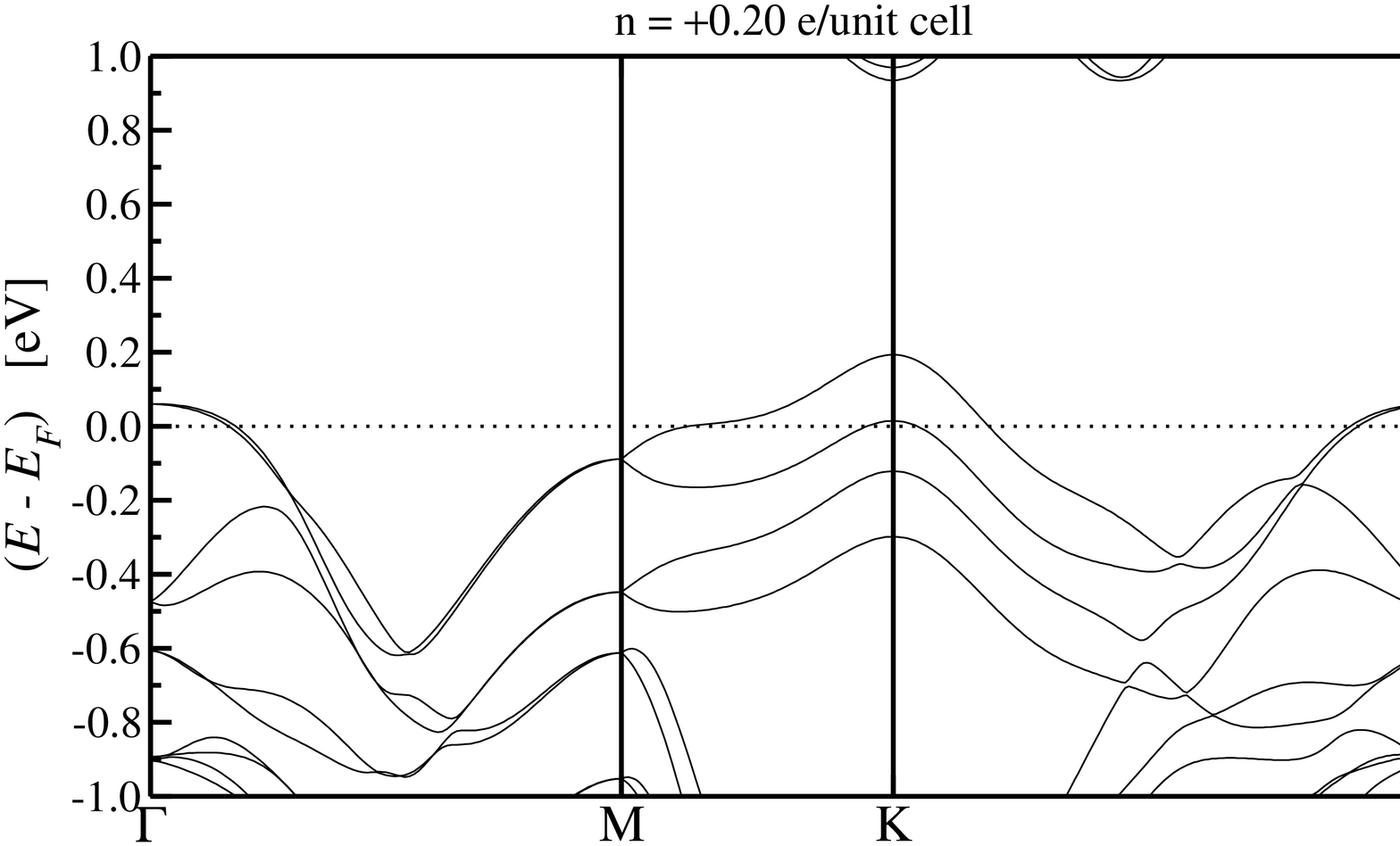}
 \includegraphics[width=0.31\textwidth,clip=,angle=-90]{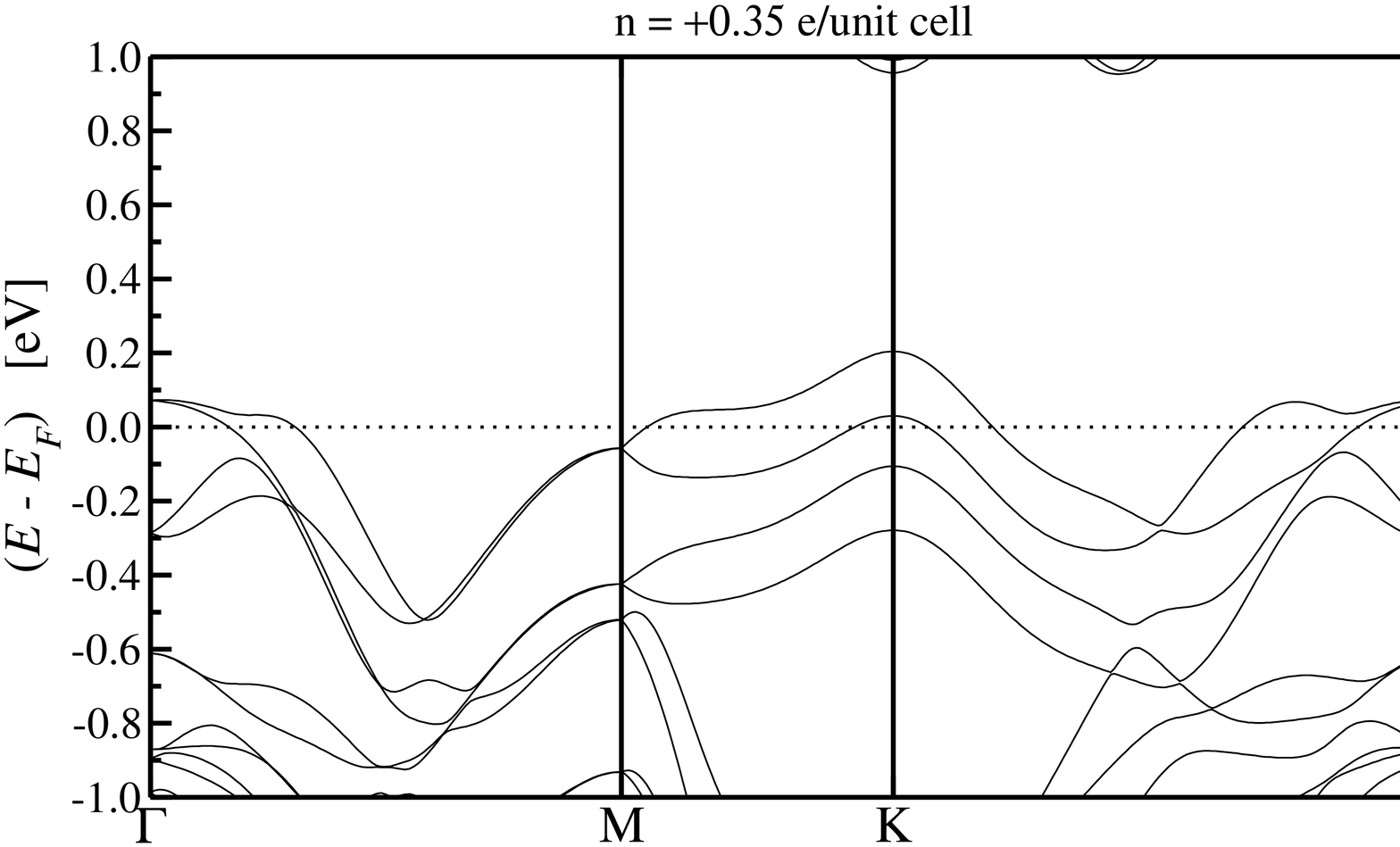}
 \caption{Band structure of bilayer MoTe$_2$ for different doping as indicated in the labels.}
\end{figure*}
\begin{figure*}[hbp]
 \centering
 \includegraphics[width=0.31\textwidth,clip=,angle=-90]{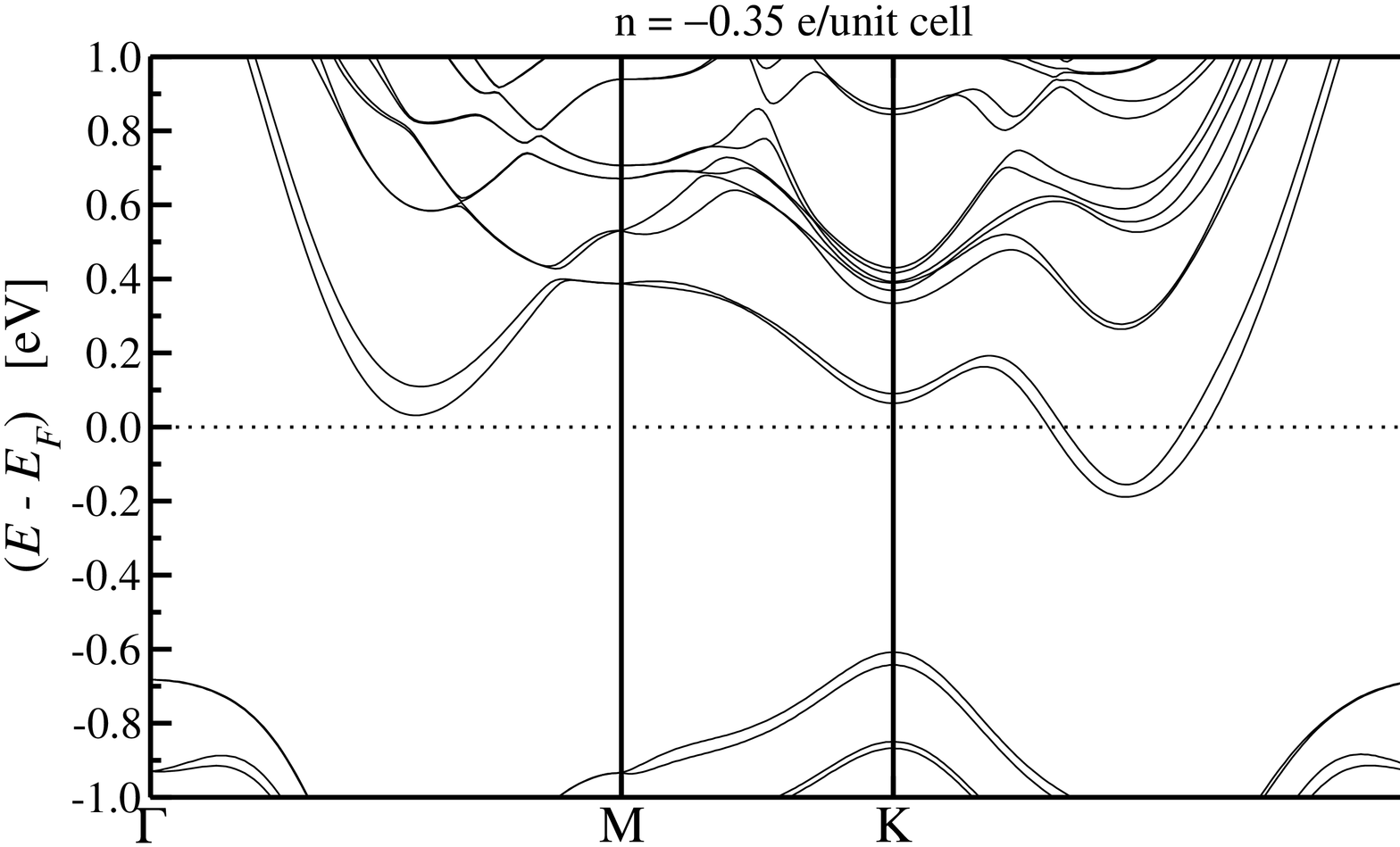}
 \includegraphics[width=0.31\textwidth,clip=,angle=-90]{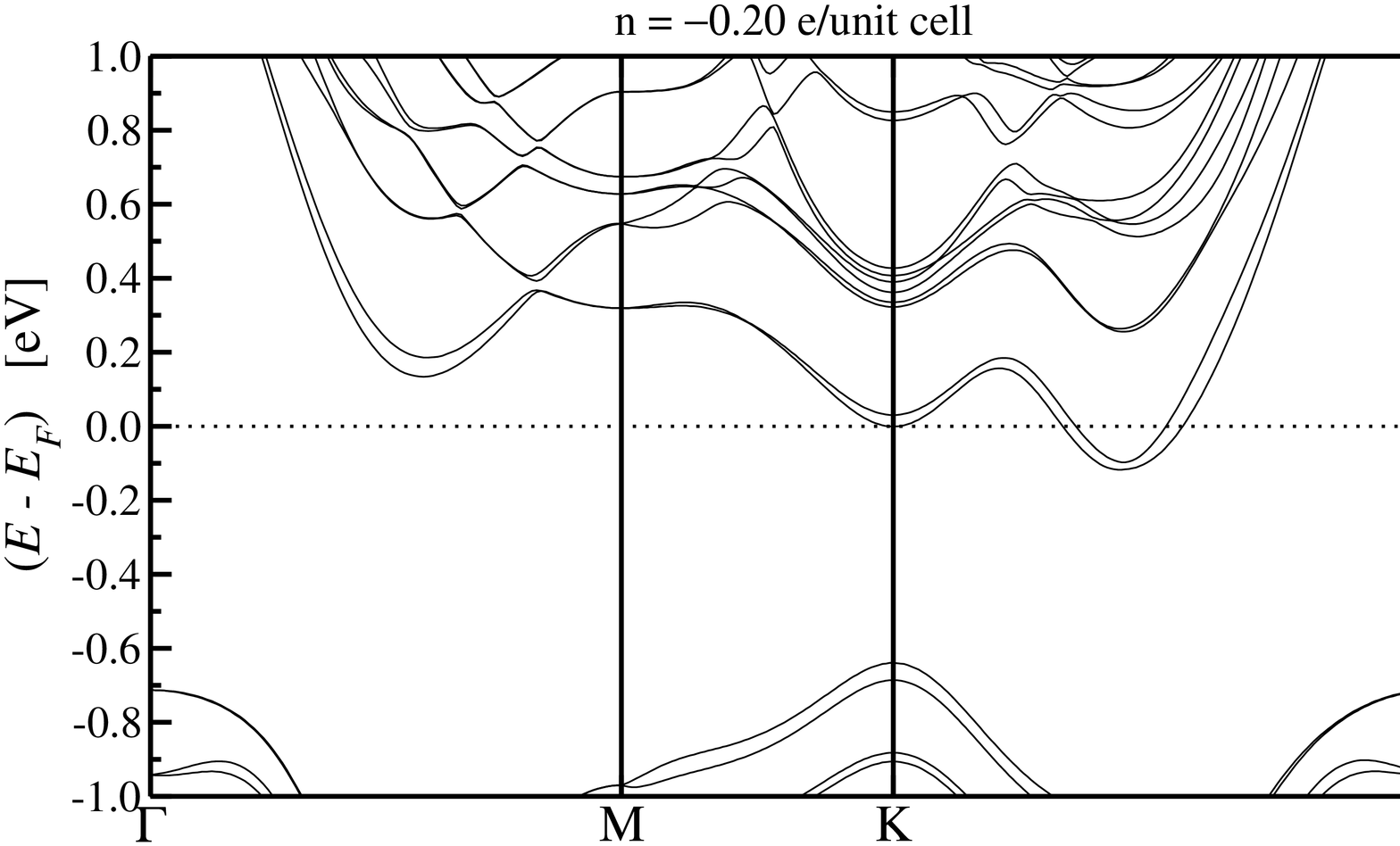}
 \includegraphics[width=0.31\textwidth,clip=,angle=-90]{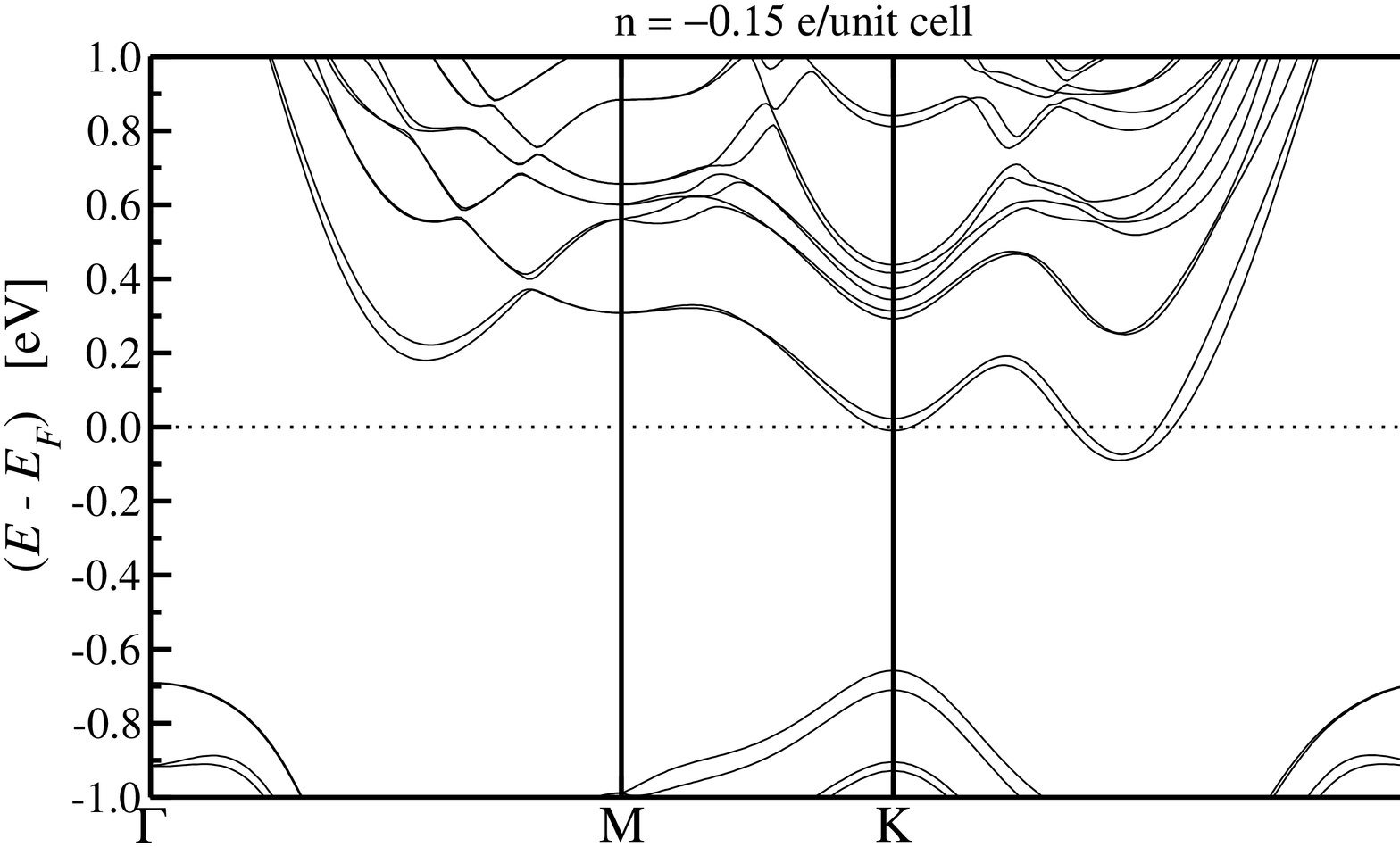}
 \includegraphics[width=0.31\textwidth,clip=,angle=-90]{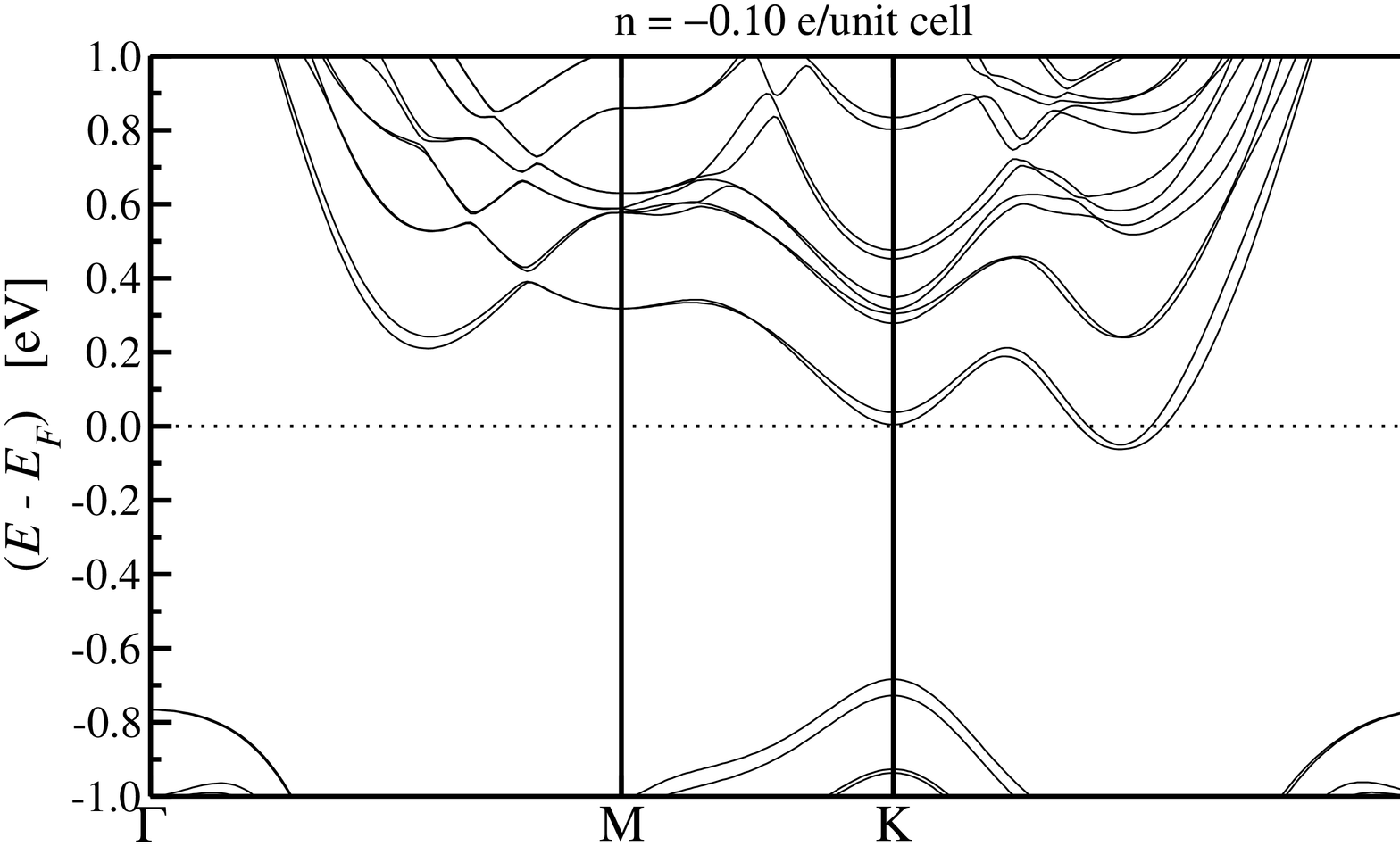}
 \includegraphics[width=0.31\textwidth,clip=,angle=-90]{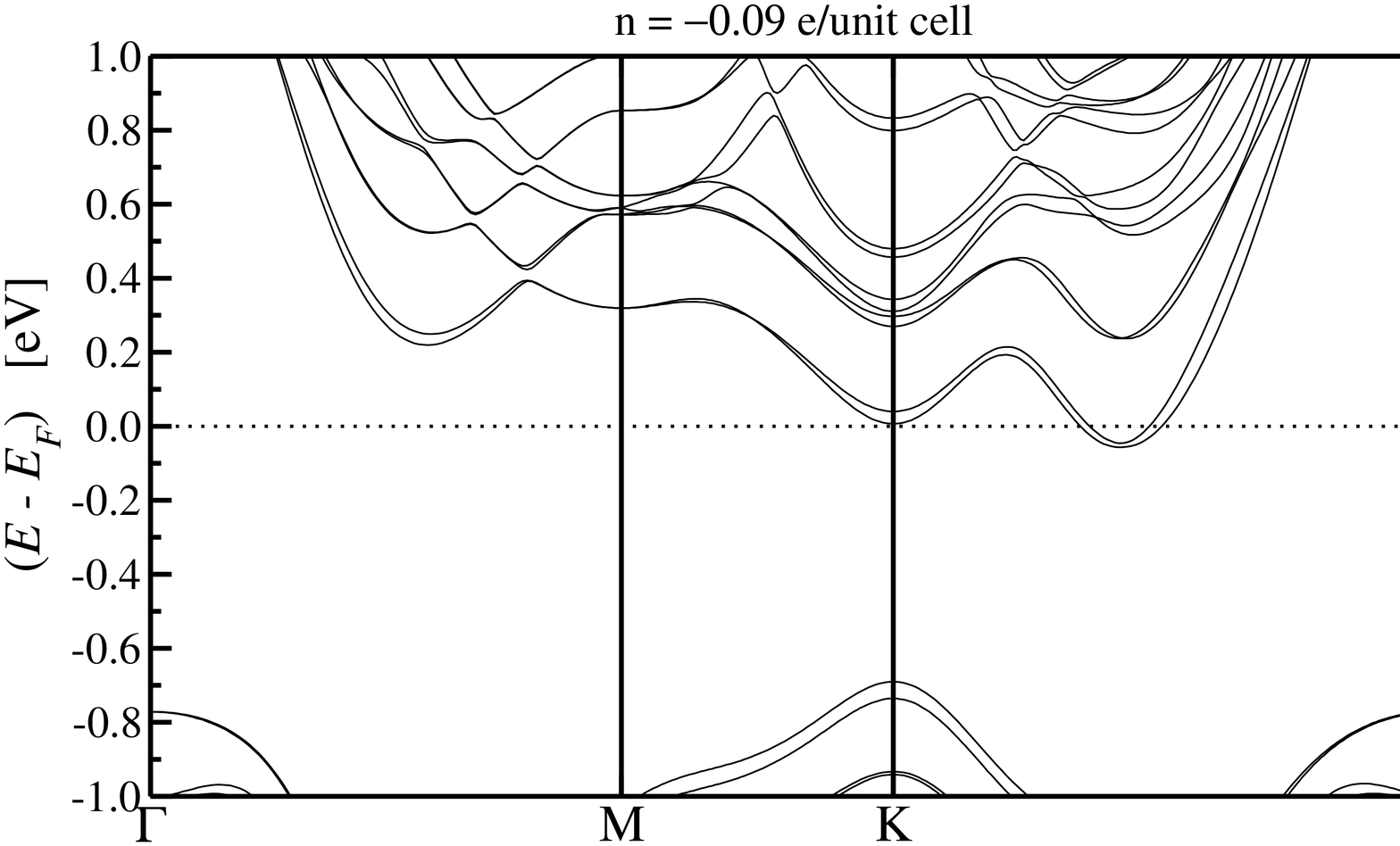}
 \includegraphics[width=0.31\textwidth,clip=,angle=-90]{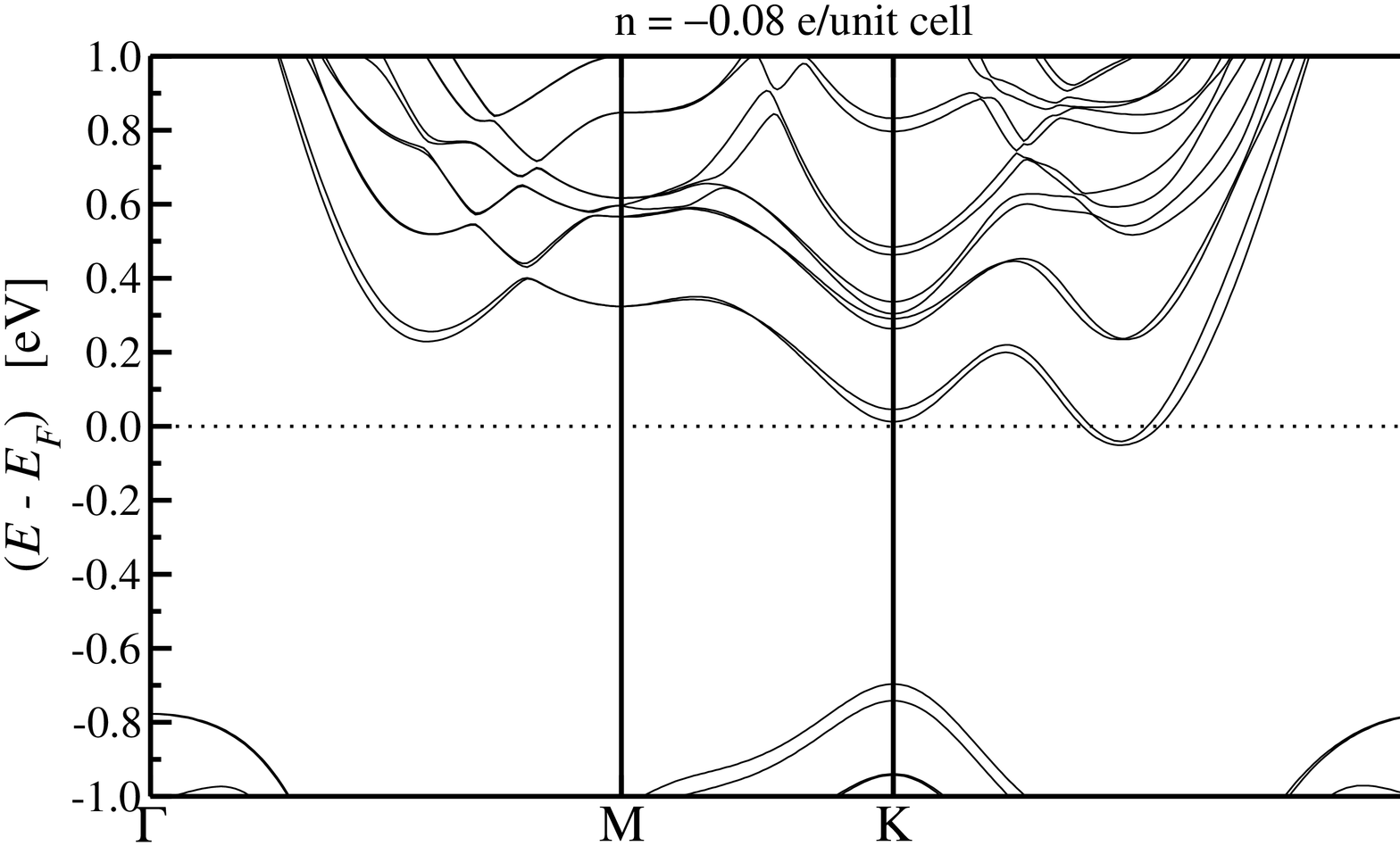}
 \includegraphics[width=0.31\textwidth,clip=,angle=-90]{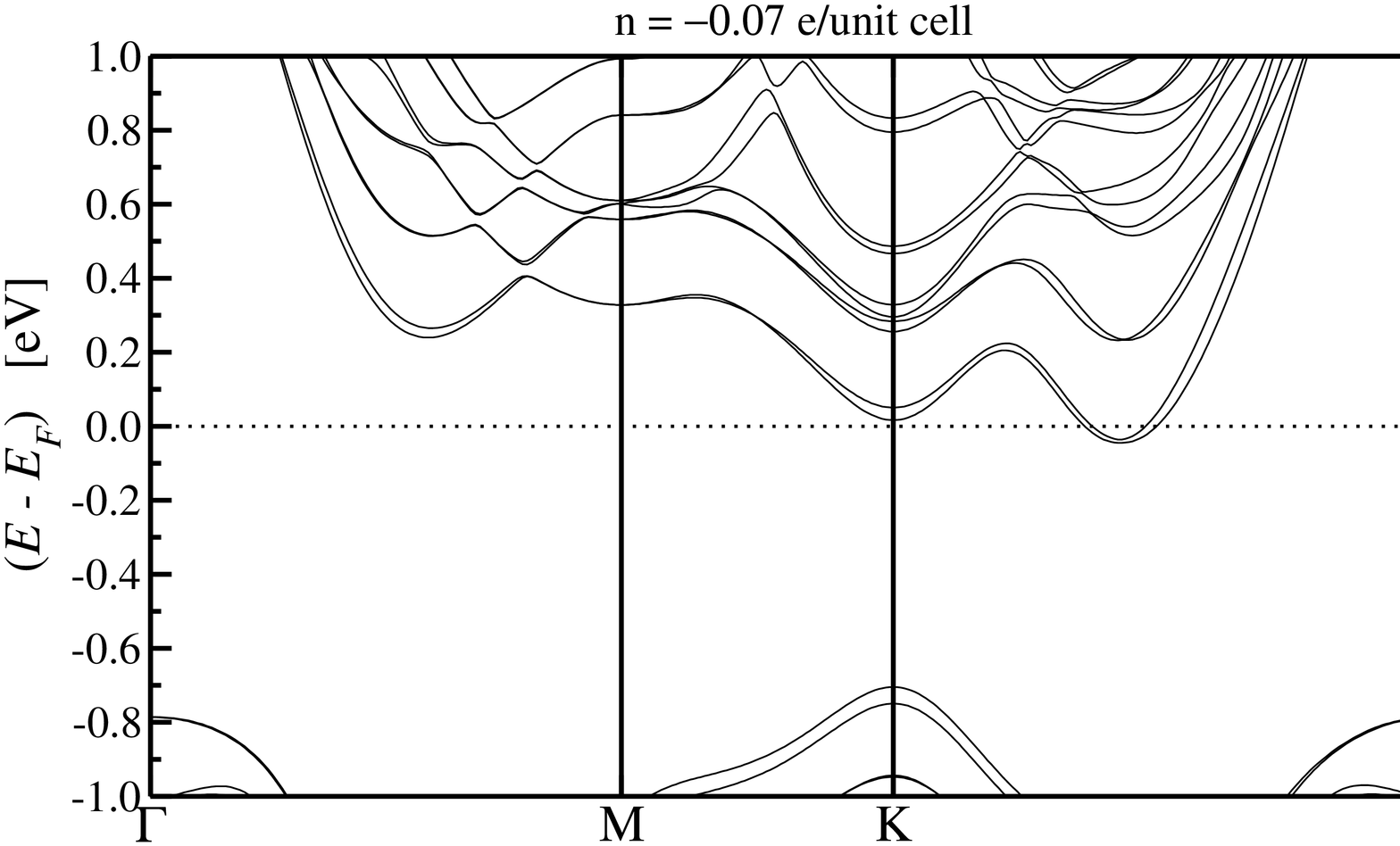}
 \includegraphics[width=0.31\textwidth,clip=,angle=-90]{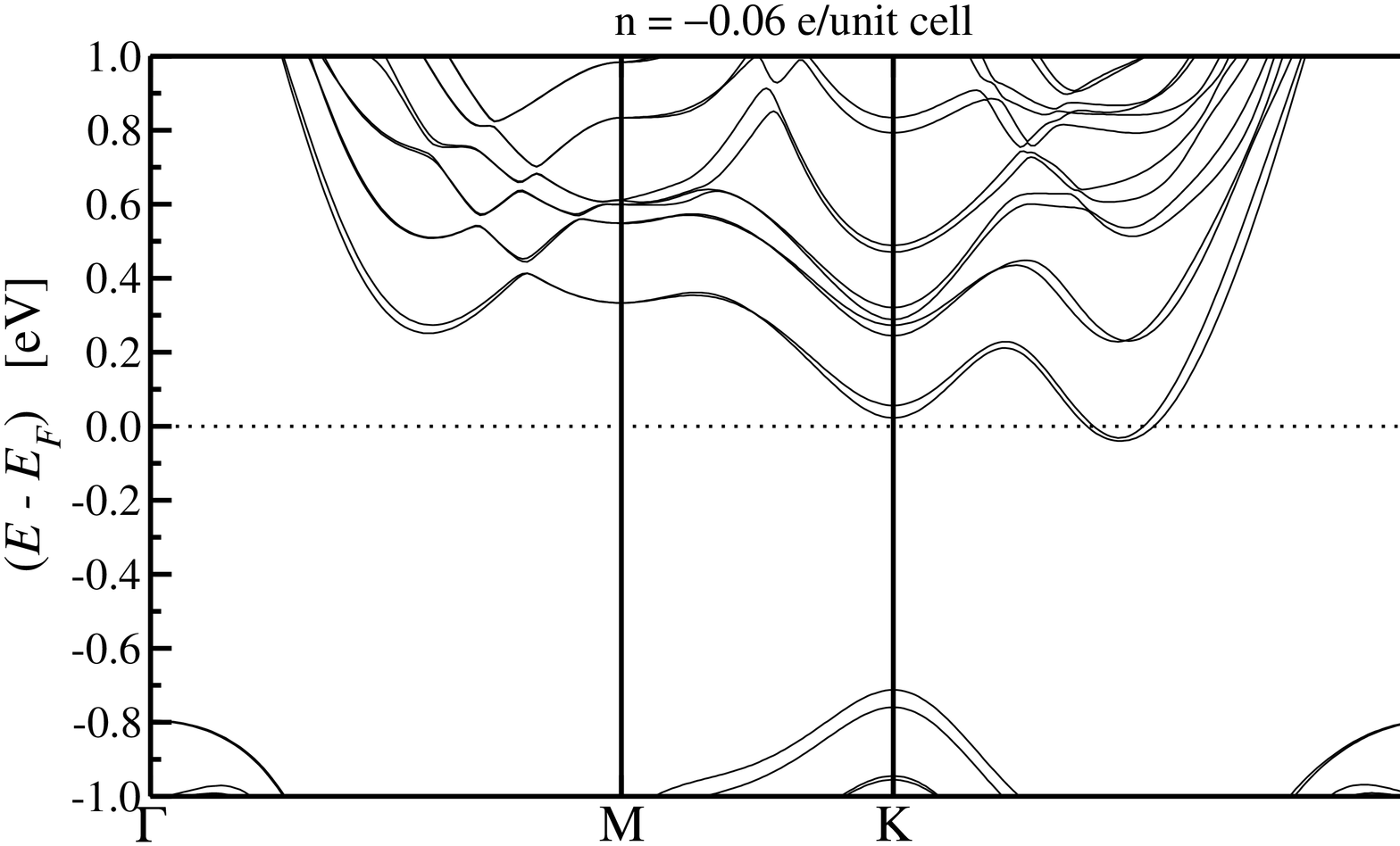}
 \caption{Band structure of trilayer MoTe$_2$ for different doping as indicated in the labels.}
\end{figure*}
\begin{figure*}[hbp]
 \centering
 \includegraphics[width=0.31\textwidth,clip=,angle=-90]{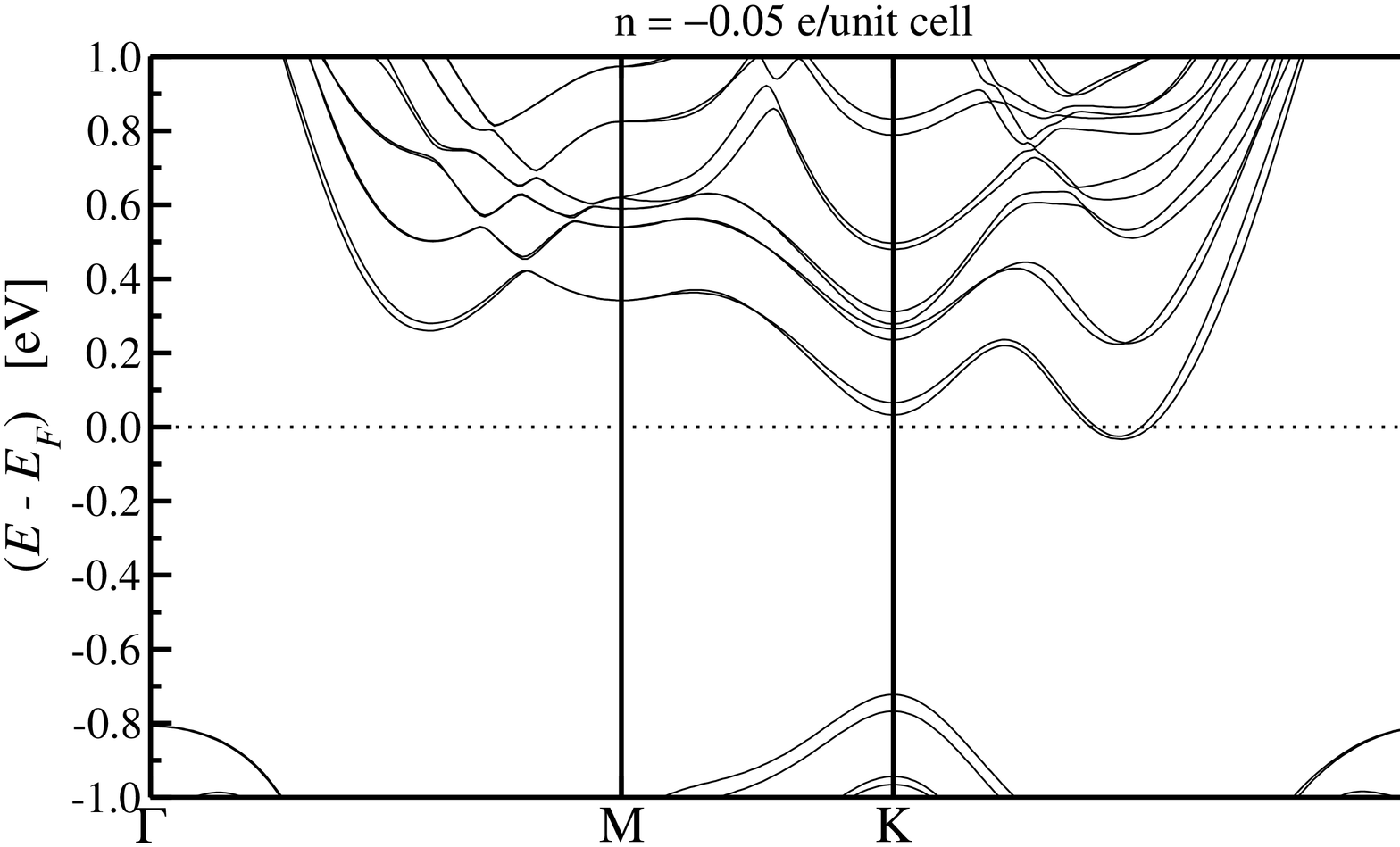}
 \includegraphics[width=0.31\textwidth,clip=,angle=-90]{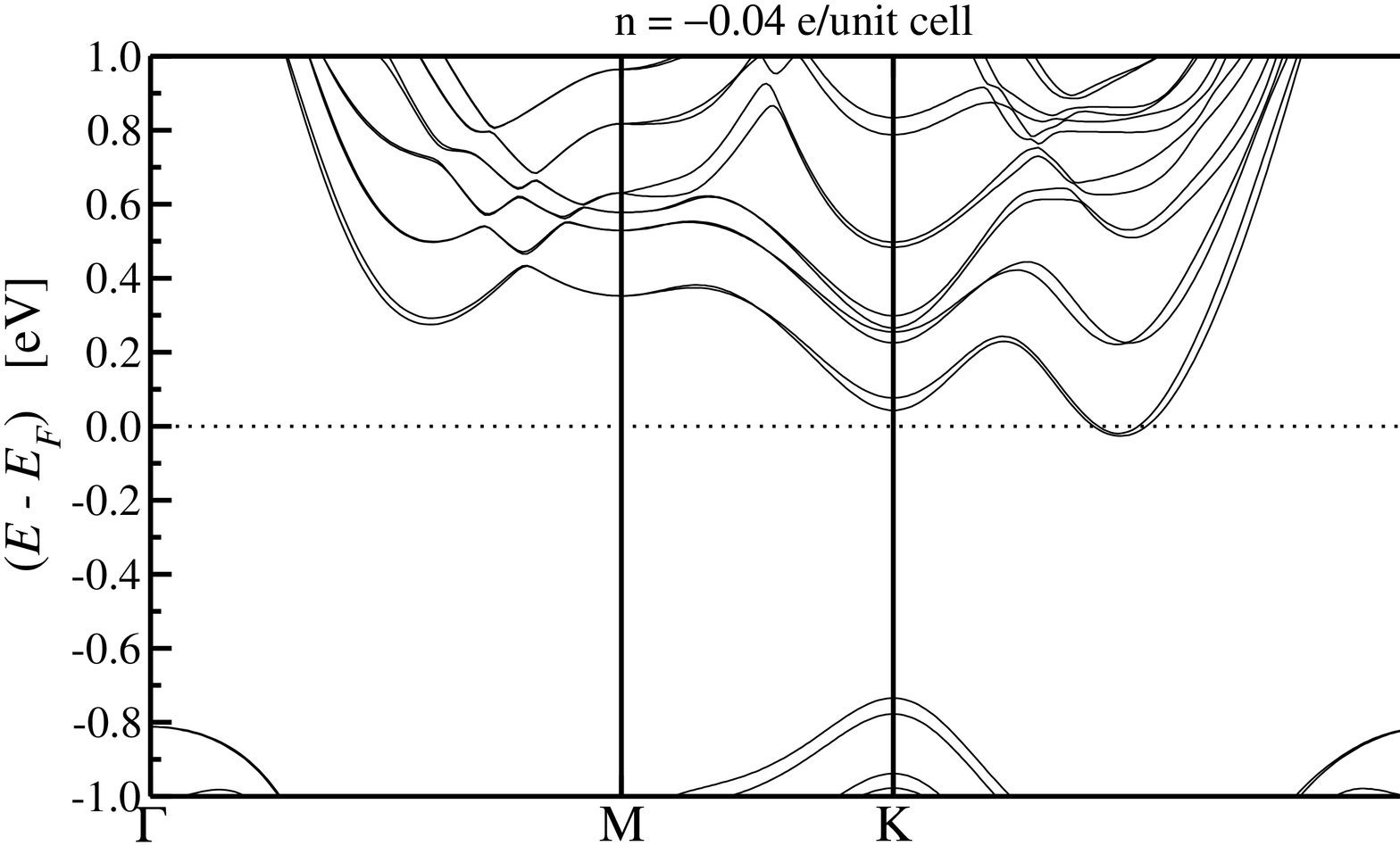}
 \includegraphics[width=0.31\textwidth,clip=,angle=-90]{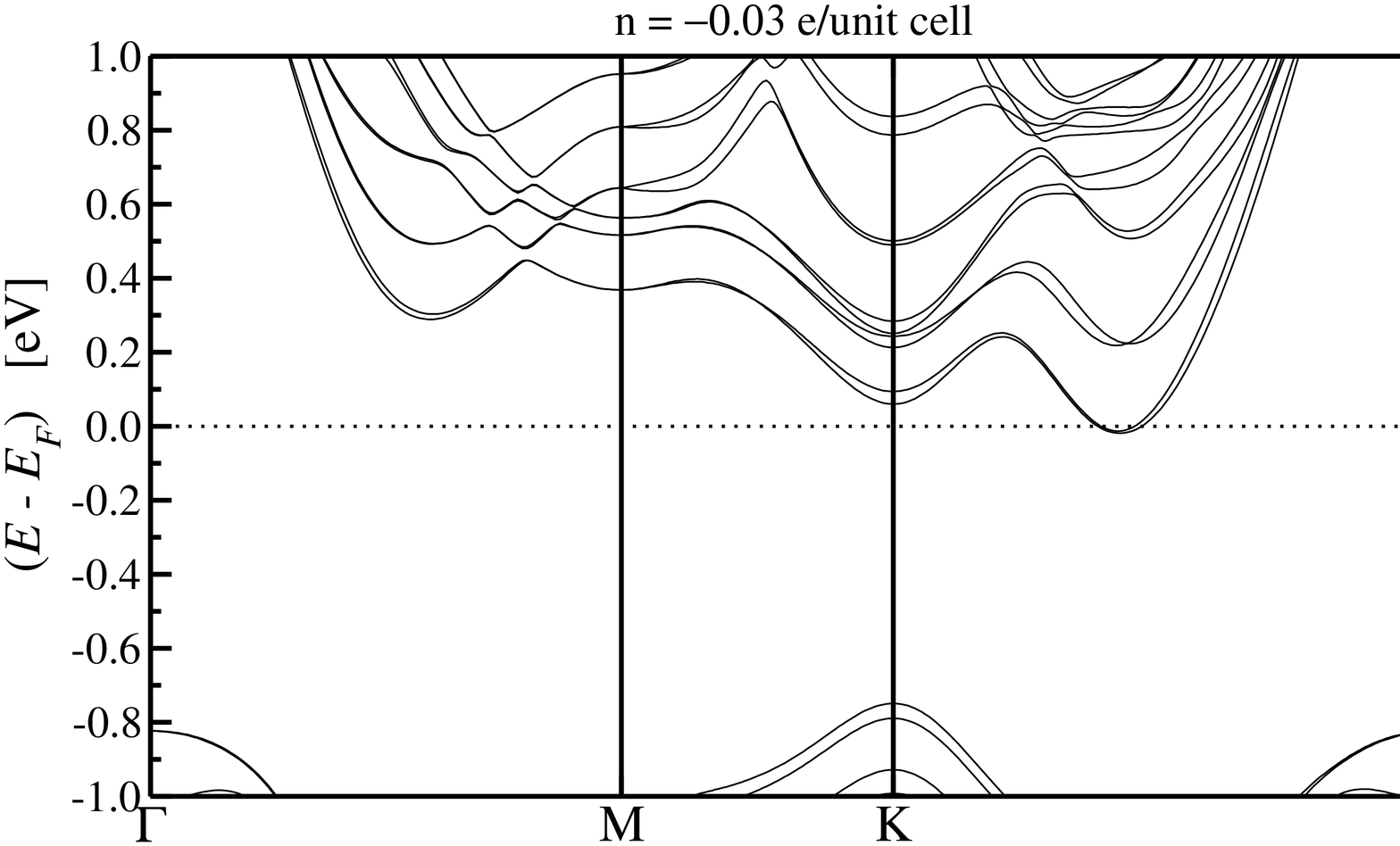}
 \includegraphics[width=0.31\textwidth,clip=,angle=-90]{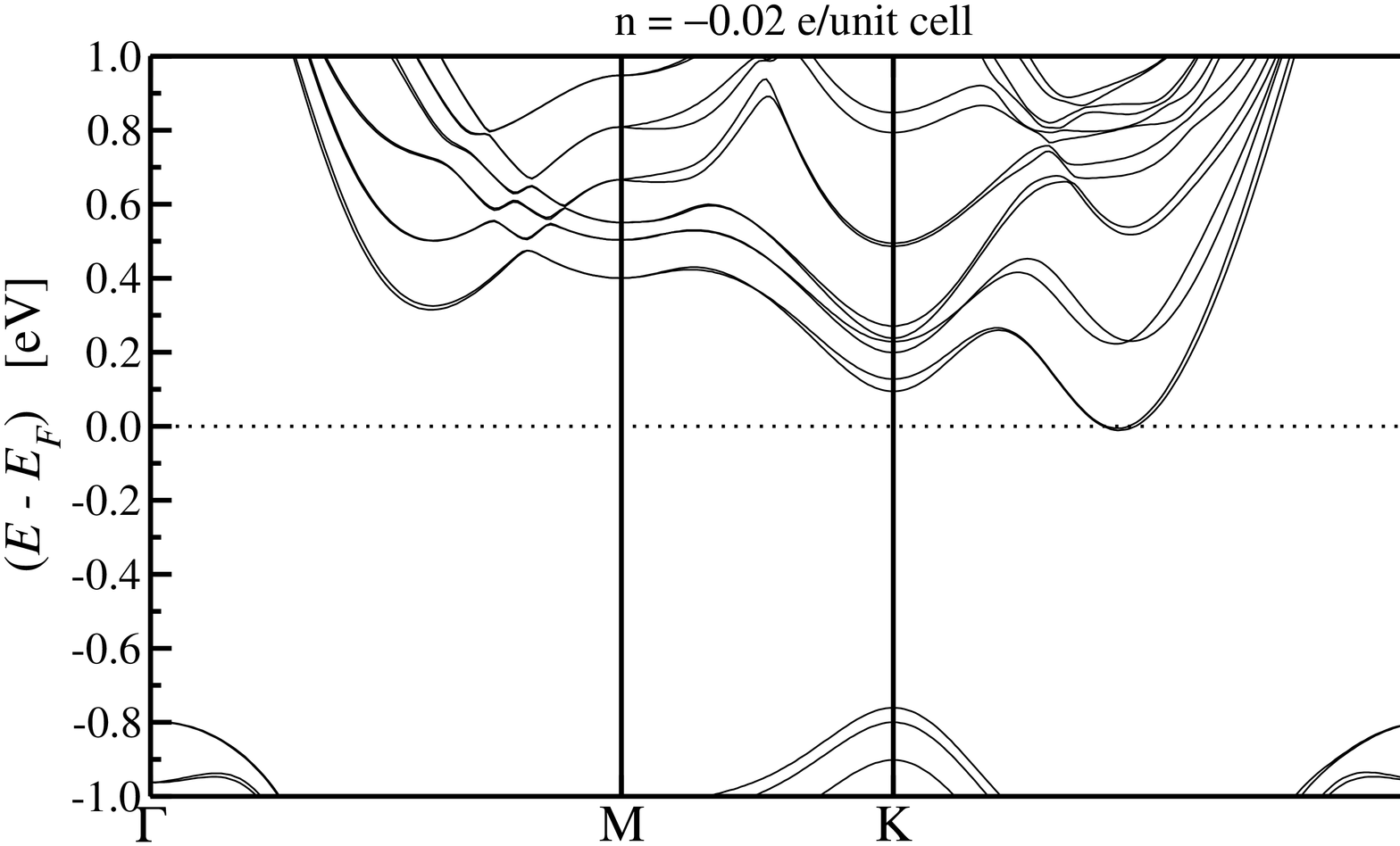}
 \includegraphics[width=0.31\textwidth,clip=,angle=-90]{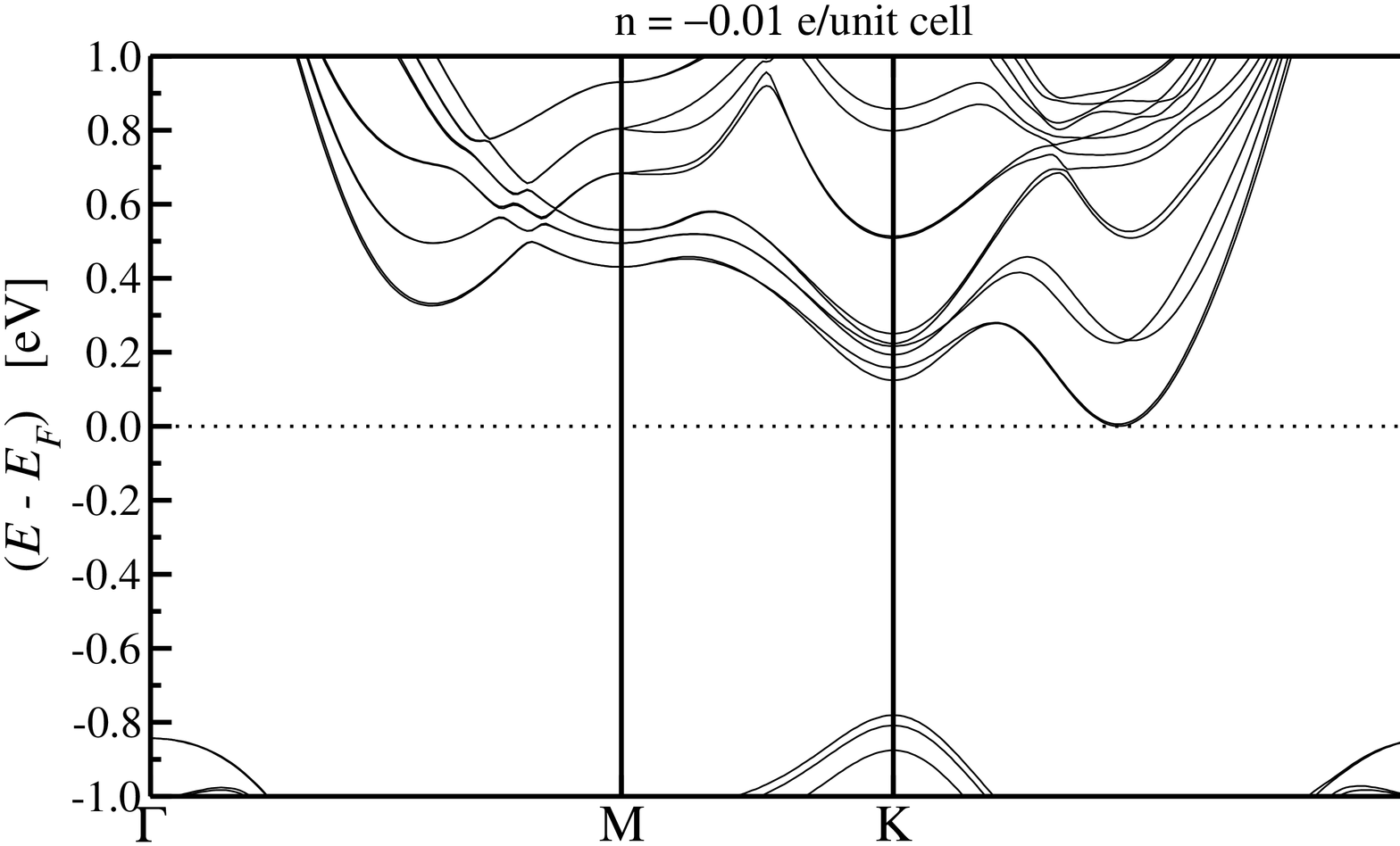}
 \includegraphics[width=0.31\textwidth,clip=,angle=-90]{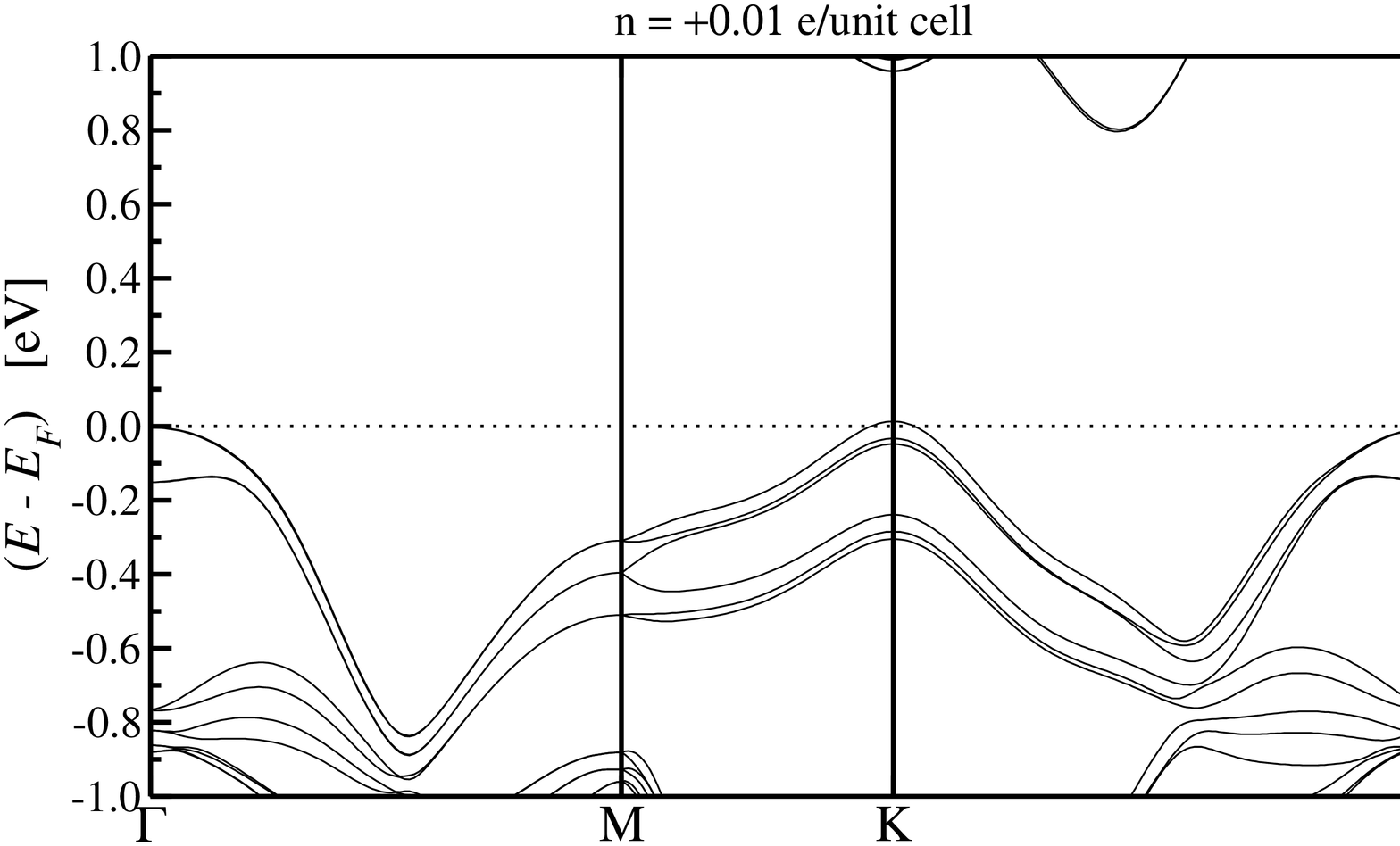}
 \includegraphics[width=0.31\textwidth,clip=,angle=-90]{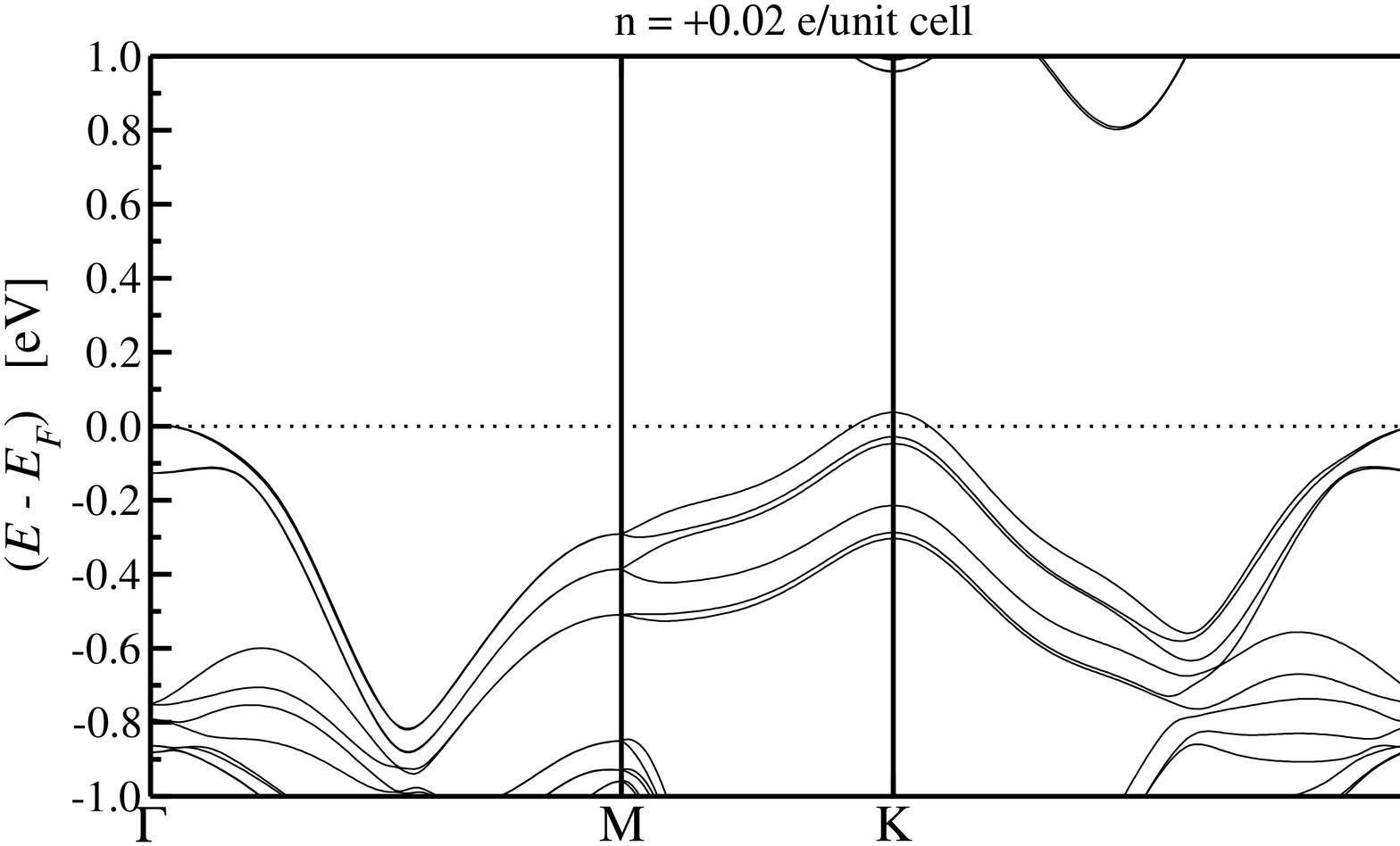}
 \includegraphics[width=0.31\textwidth,clip=,angle=-90]{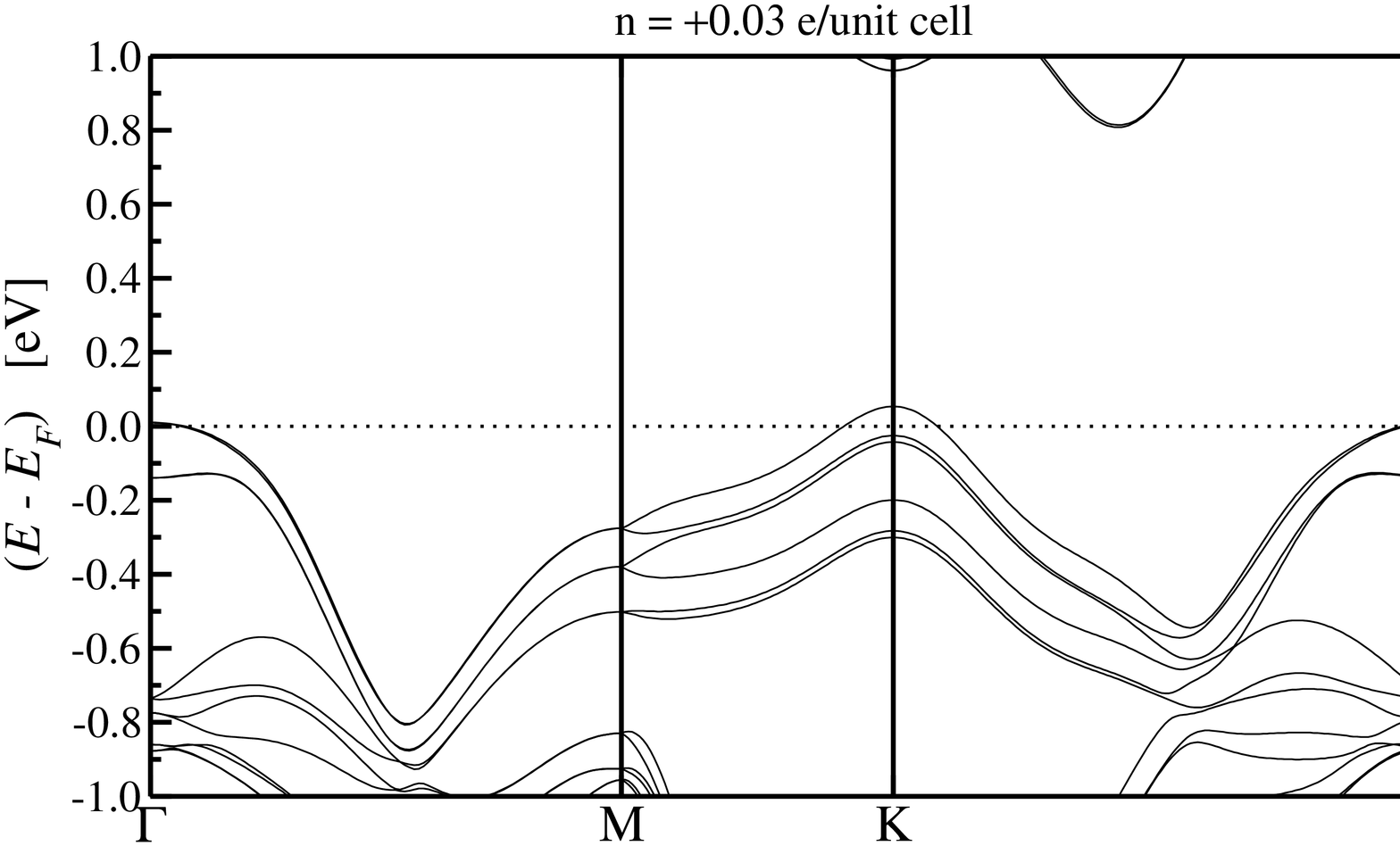}
 \caption{Band structure of trilayer MoTe$_2$ for different doping as indicated in the labels.}
\end{figure*}
\begin{figure*}[hbp]
 \centering
 \includegraphics[width=0.31\textwidth,clip=,angle=-90]{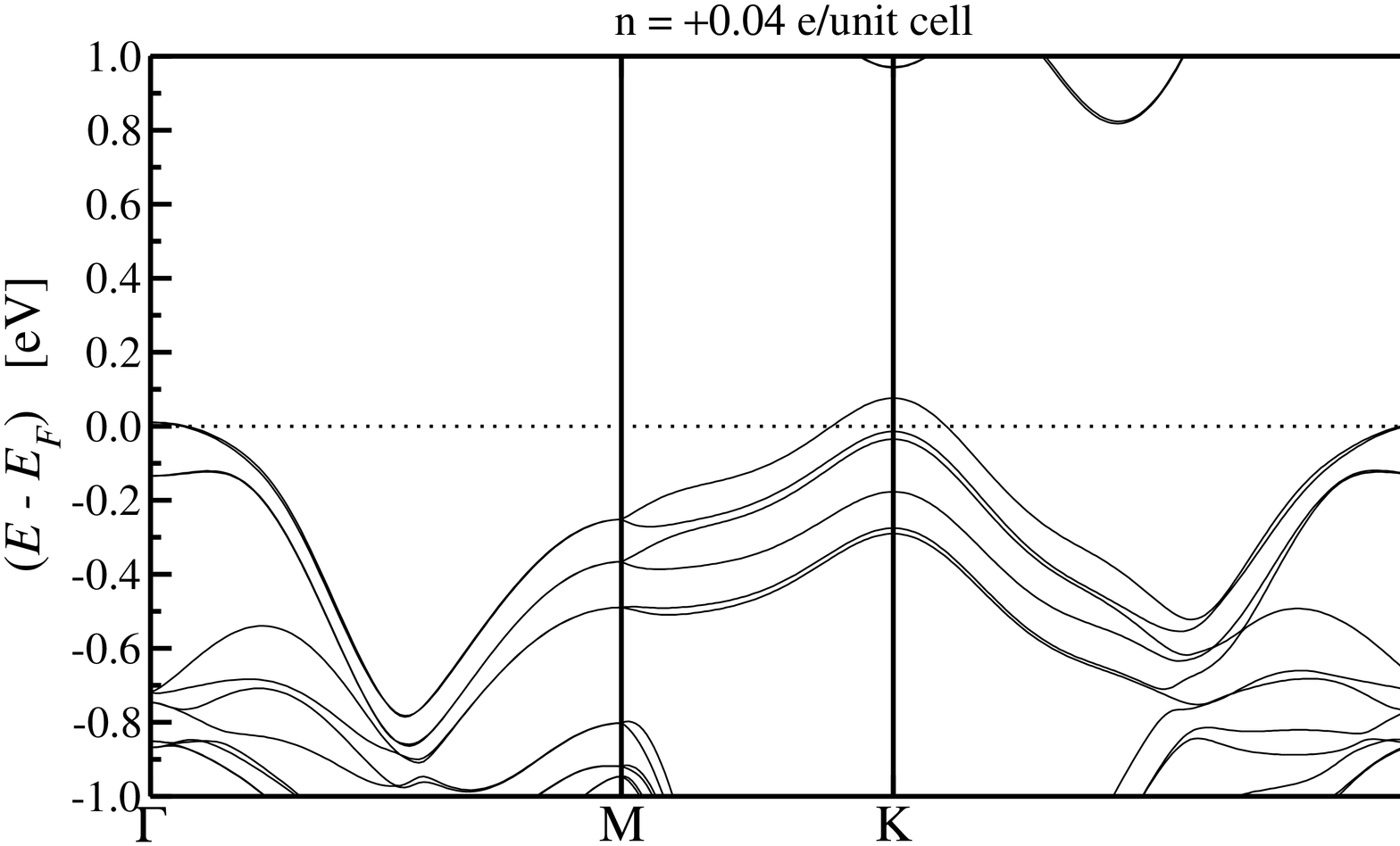}
 \includegraphics[width=0.31\textwidth,clip=,angle=-90]{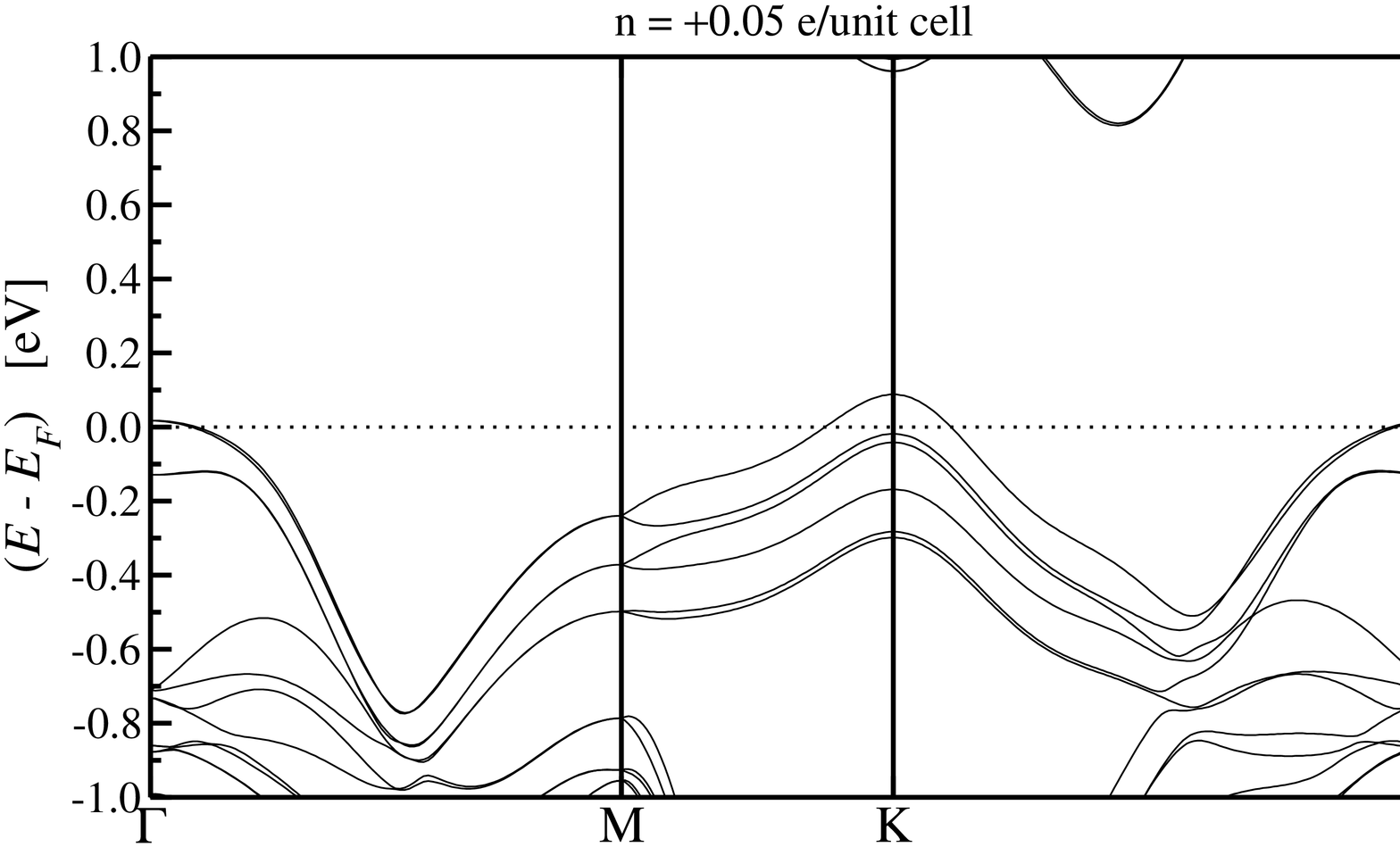}
 \includegraphics[width=0.31\textwidth,clip=,angle=-90]{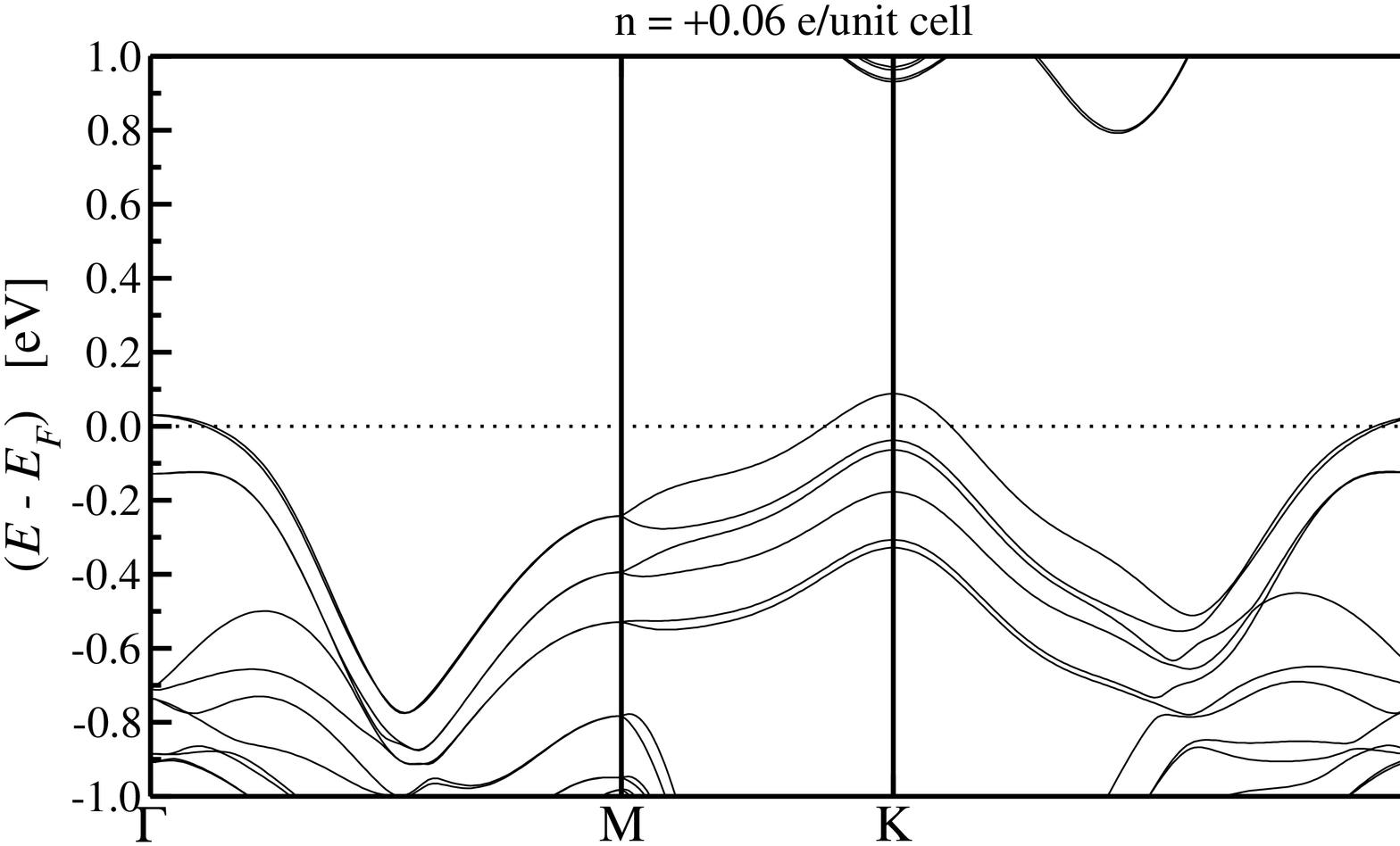}
 \includegraphics[width=0.31\textwidth,clip=,angle=-90]{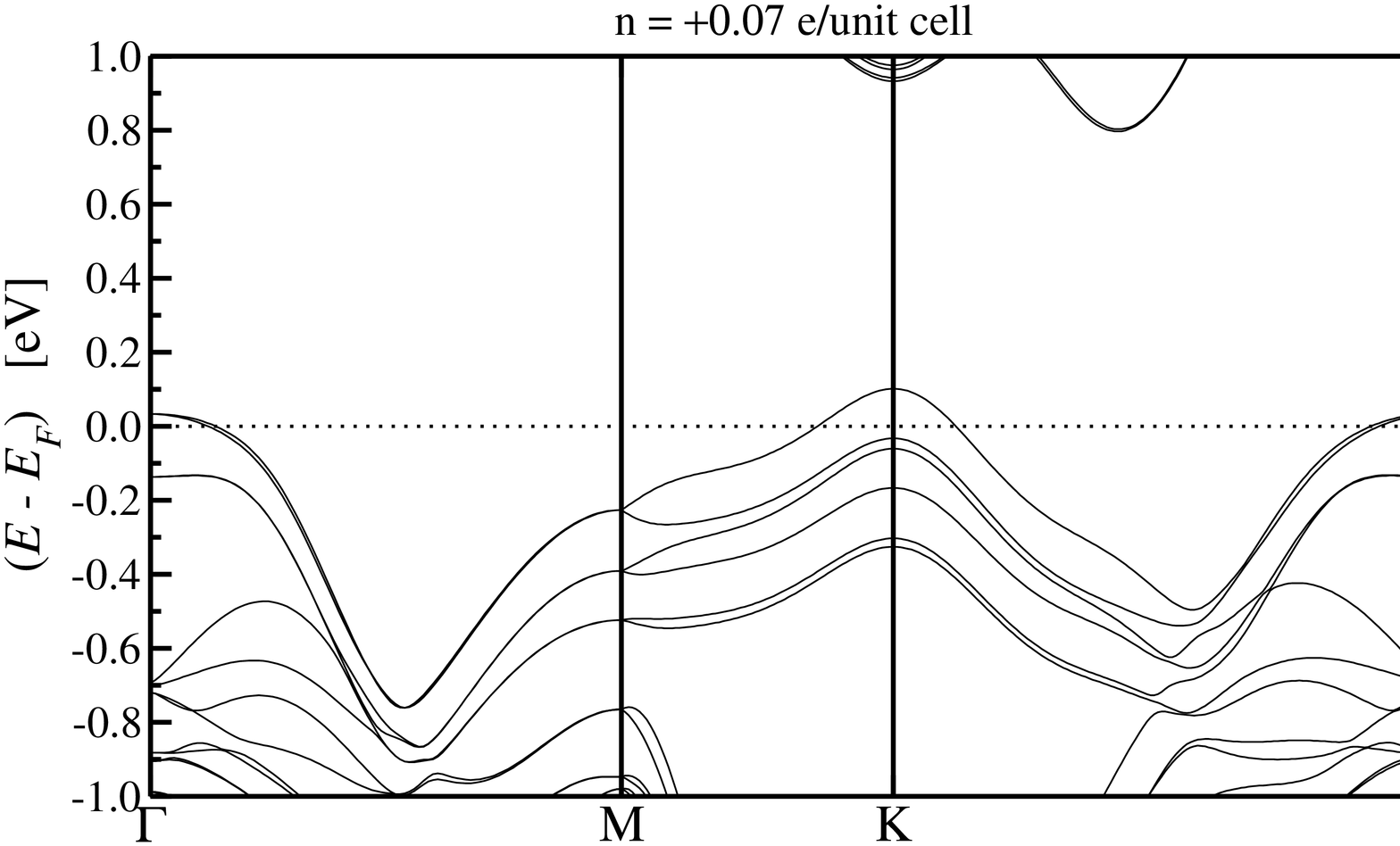}
 \includegraphics[width=0.31\textwidth,clip=,angle=-90]{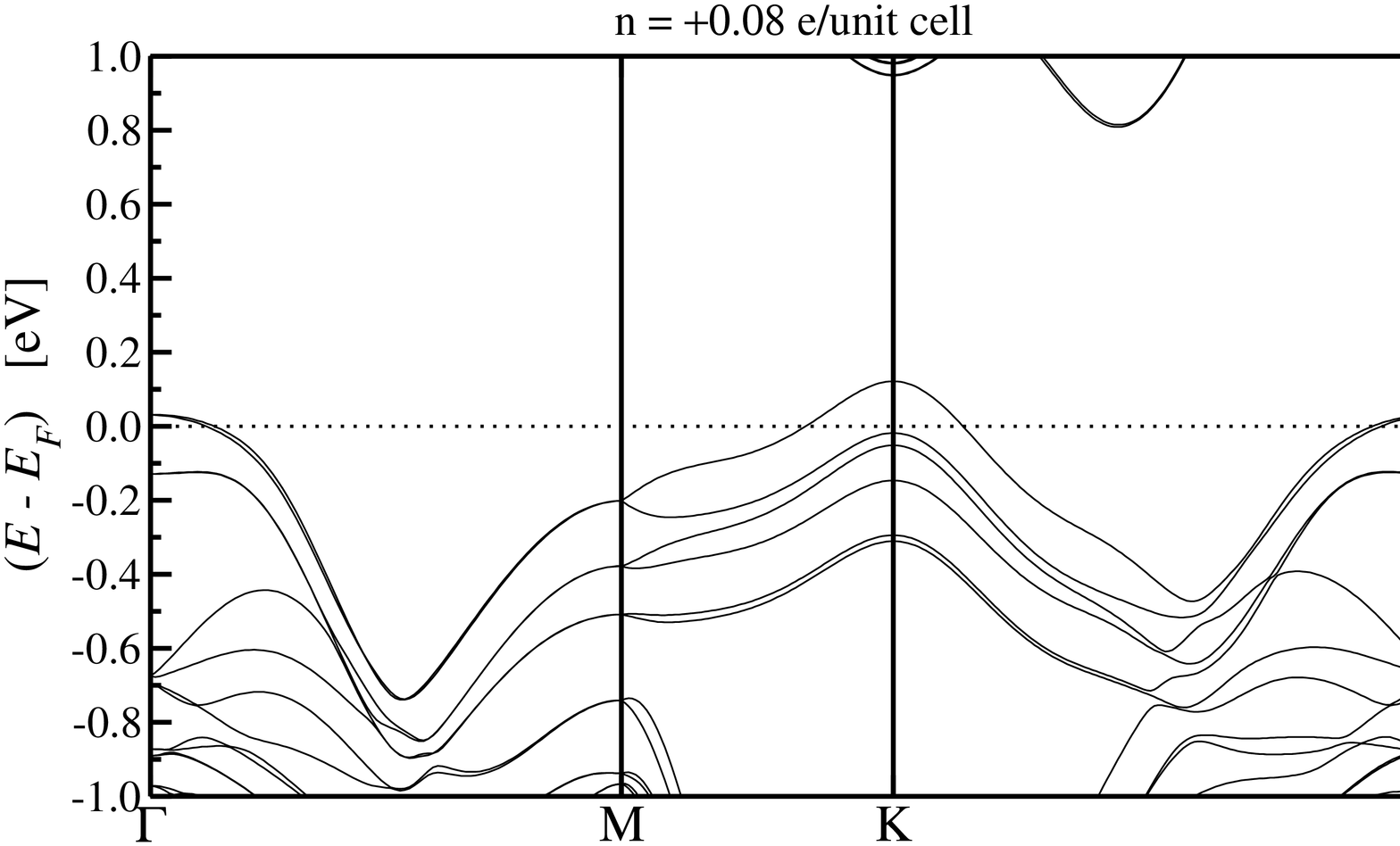}
 \includegraphics[width=0.31\textwidth,clip=,angle=-90]{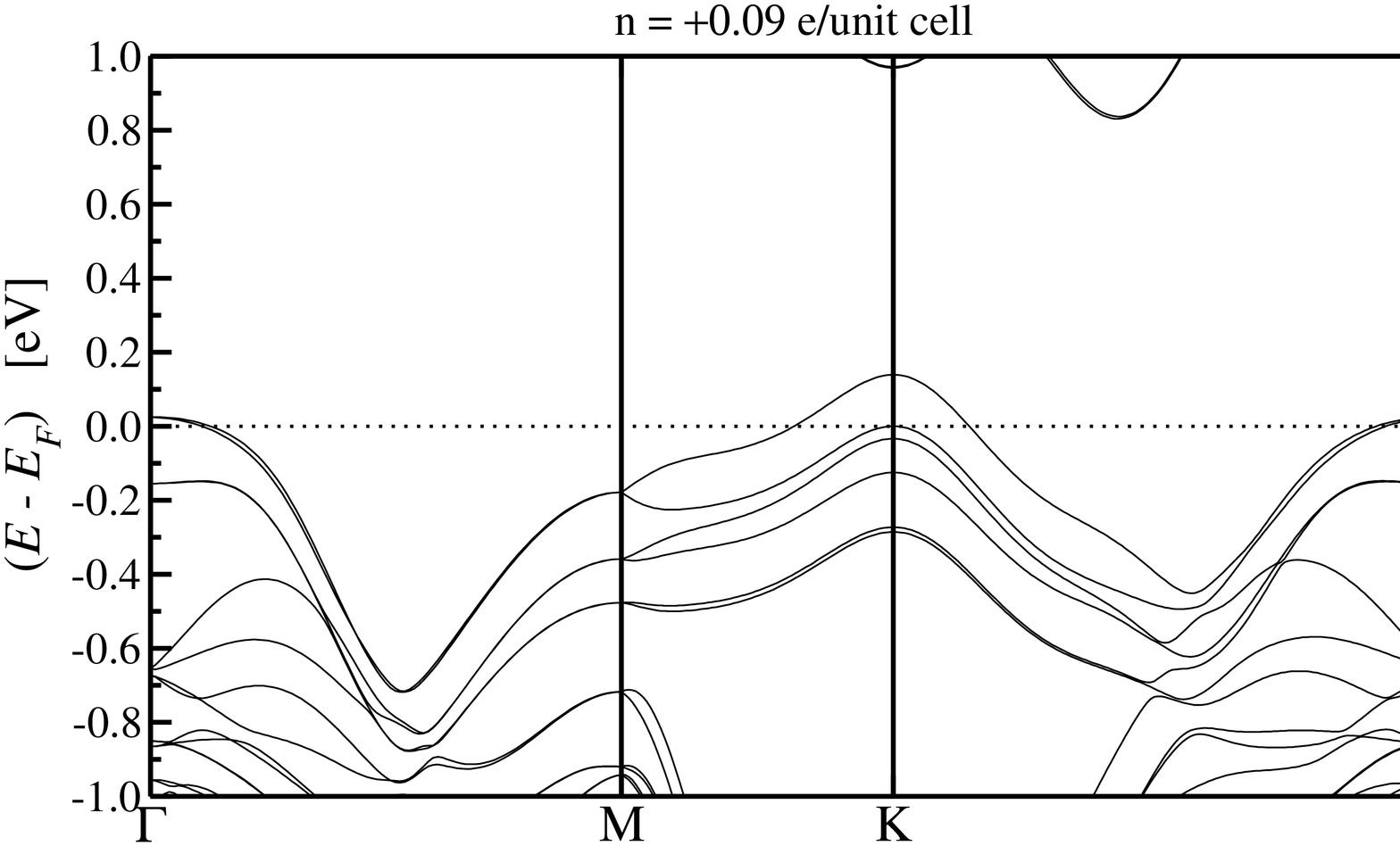}
 \includegraphics[width=0.31\textwidth,clip=,angle=-90]{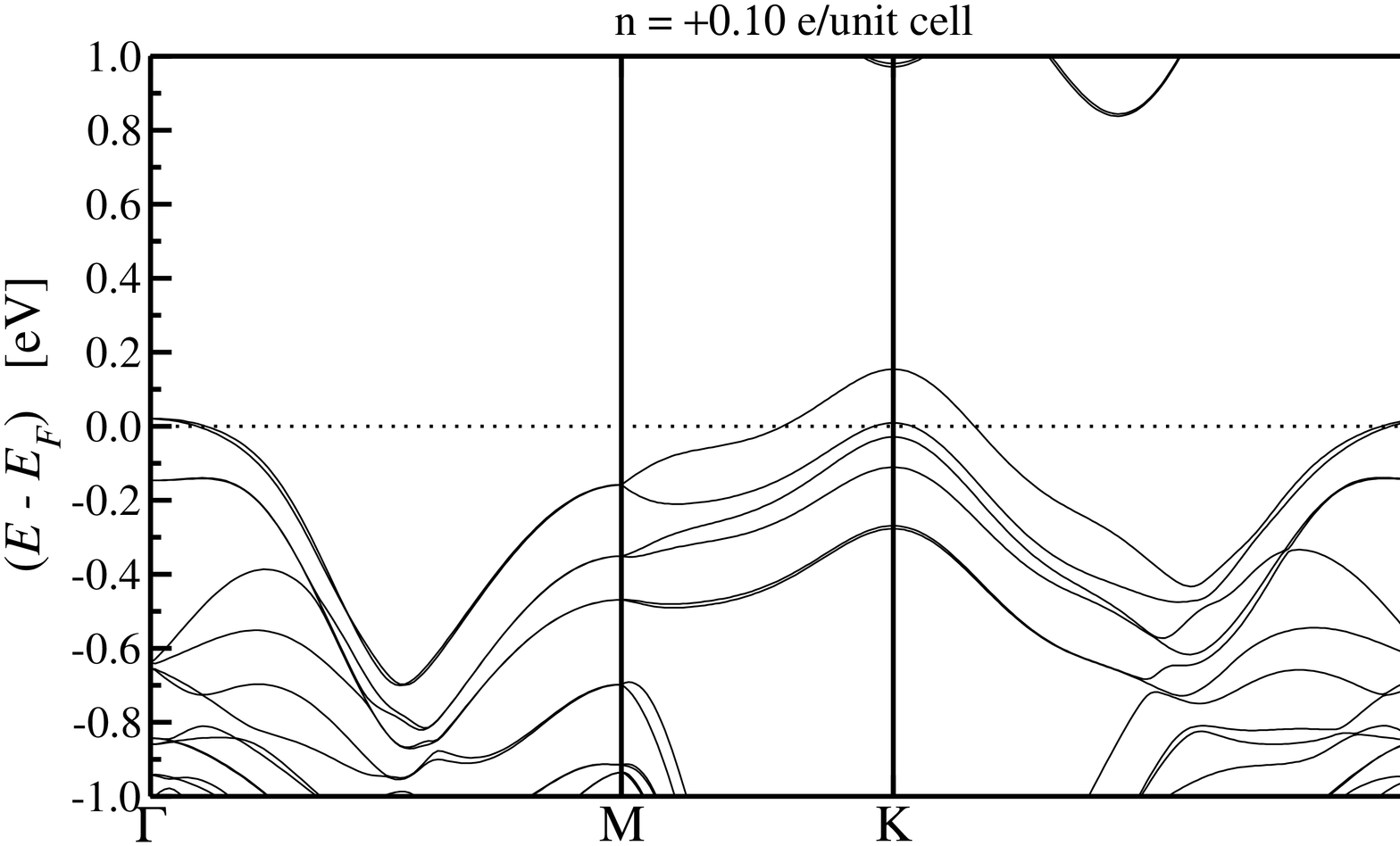}
 \includegraphics[width=0.31\textwidth,clip=,angle=-90]{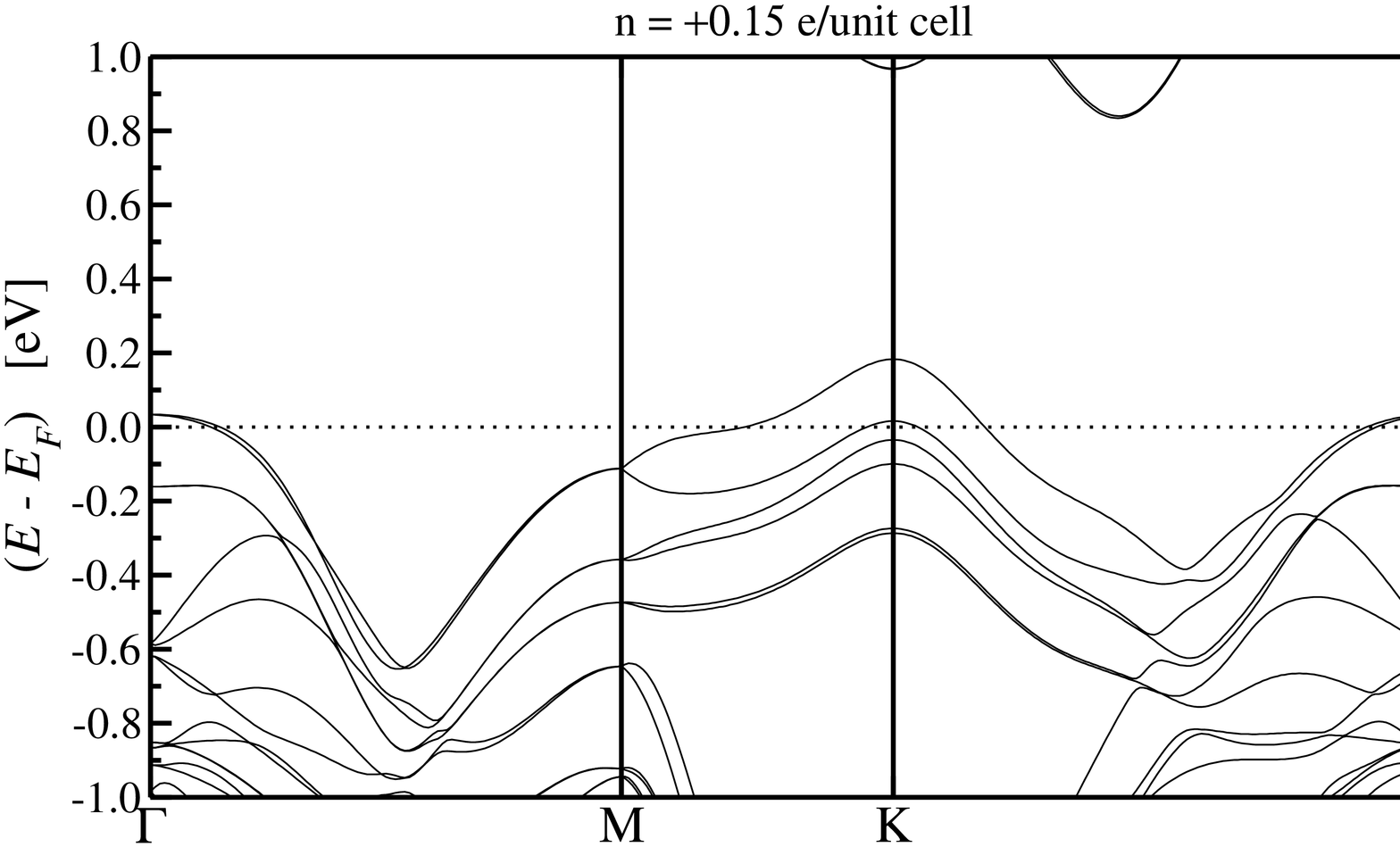}
 \caption{Band structure of trilayer MoTe$_2$ for different doping as indicated in the labels.}
\end{figure*}
\begin{figure*}[hbp]
 \centering
 \includegraphics[width=0.31\textwidth,clip=,angle=-90]{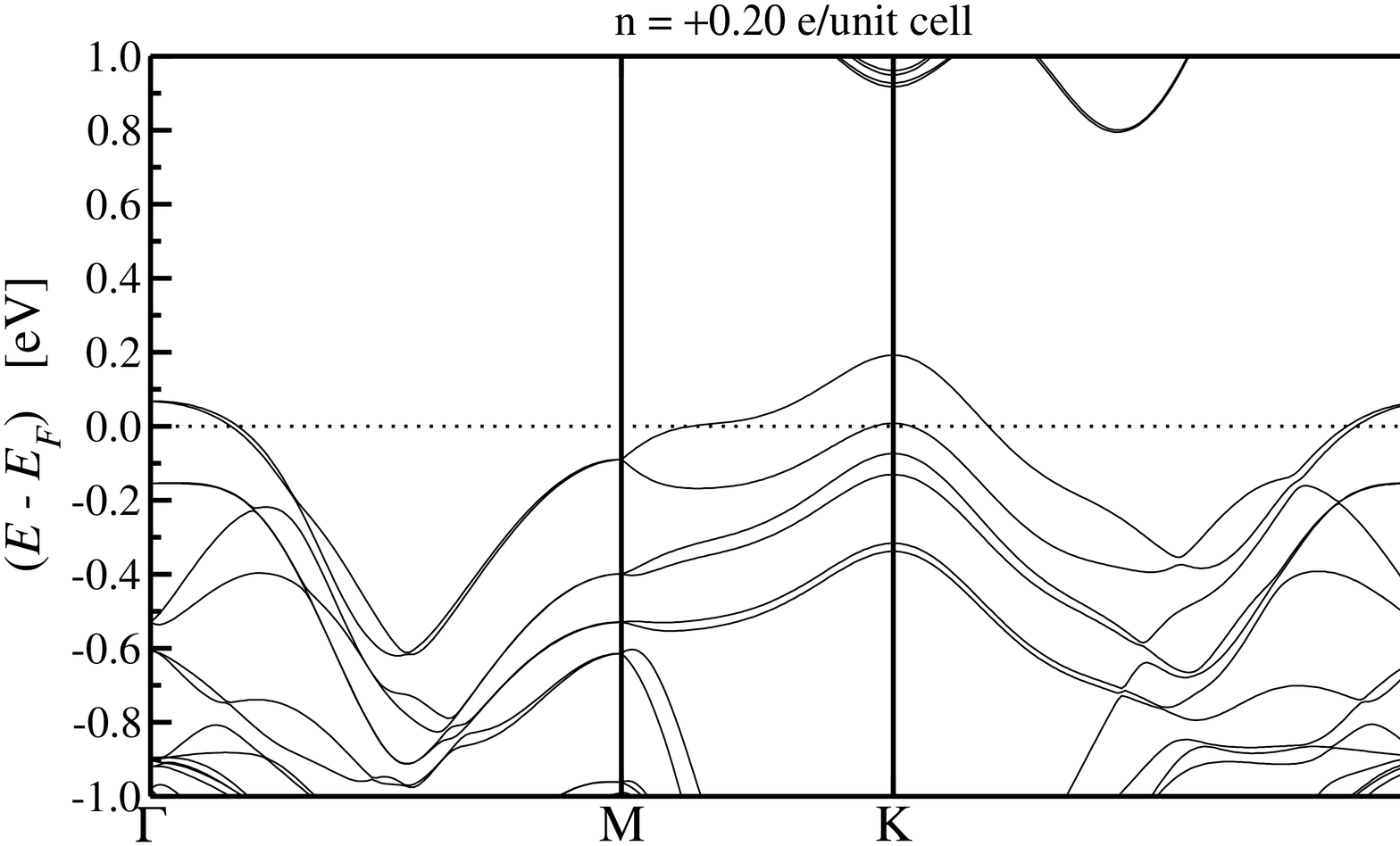}
 \includegraphics[width=0.31\textwidth,clip=,angle=-90]{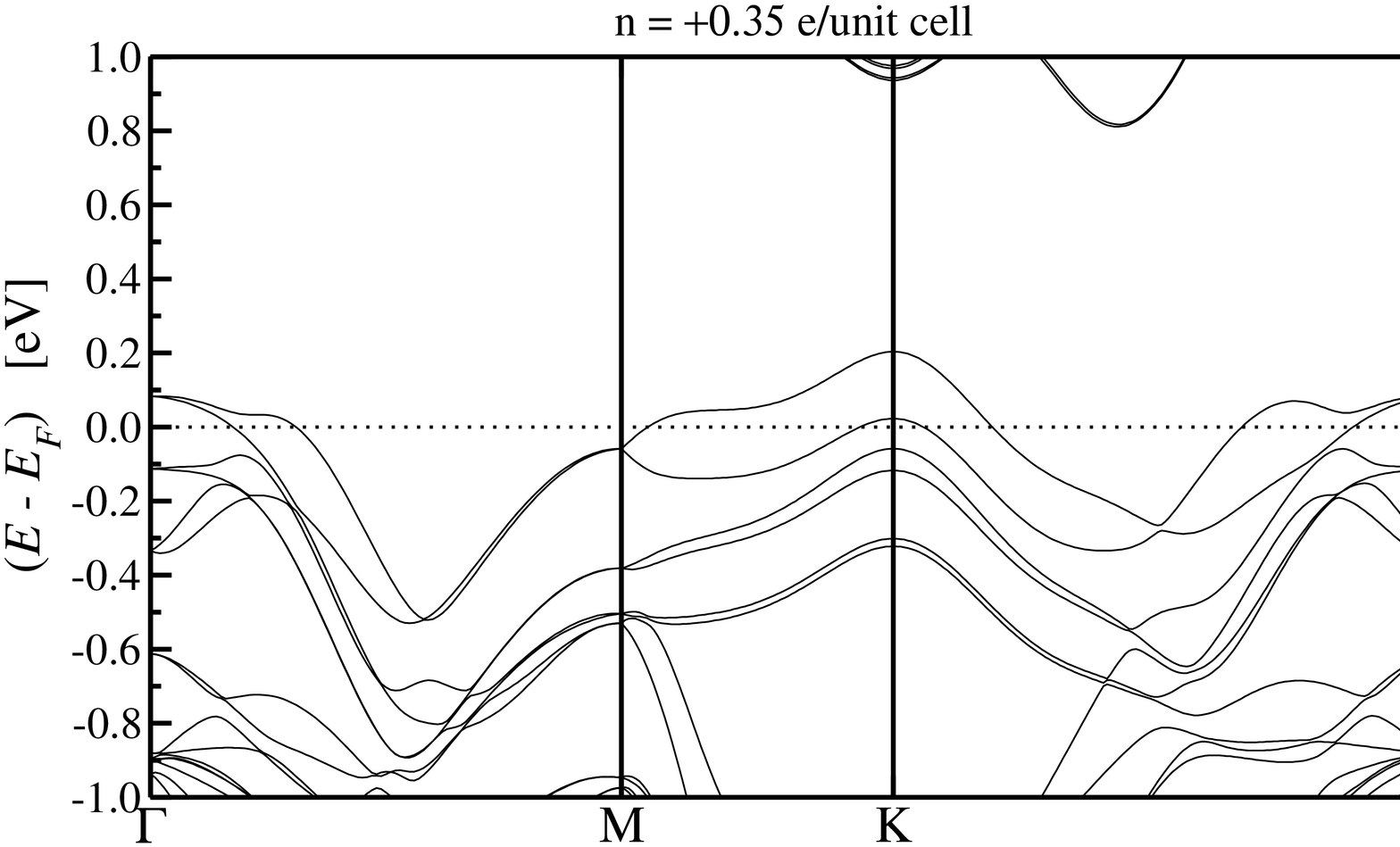}
 \caption{Band structure of trilayer MoTe$_2$ for different doping as indicated in the labels.}
\end{figure*}

\clearpage
\subsection{Tungsten disulfide}
\begin{figure*}[hbp]
 \centering
 \includegraphics[width=0.31\textwidth,clip=,angle=-90]{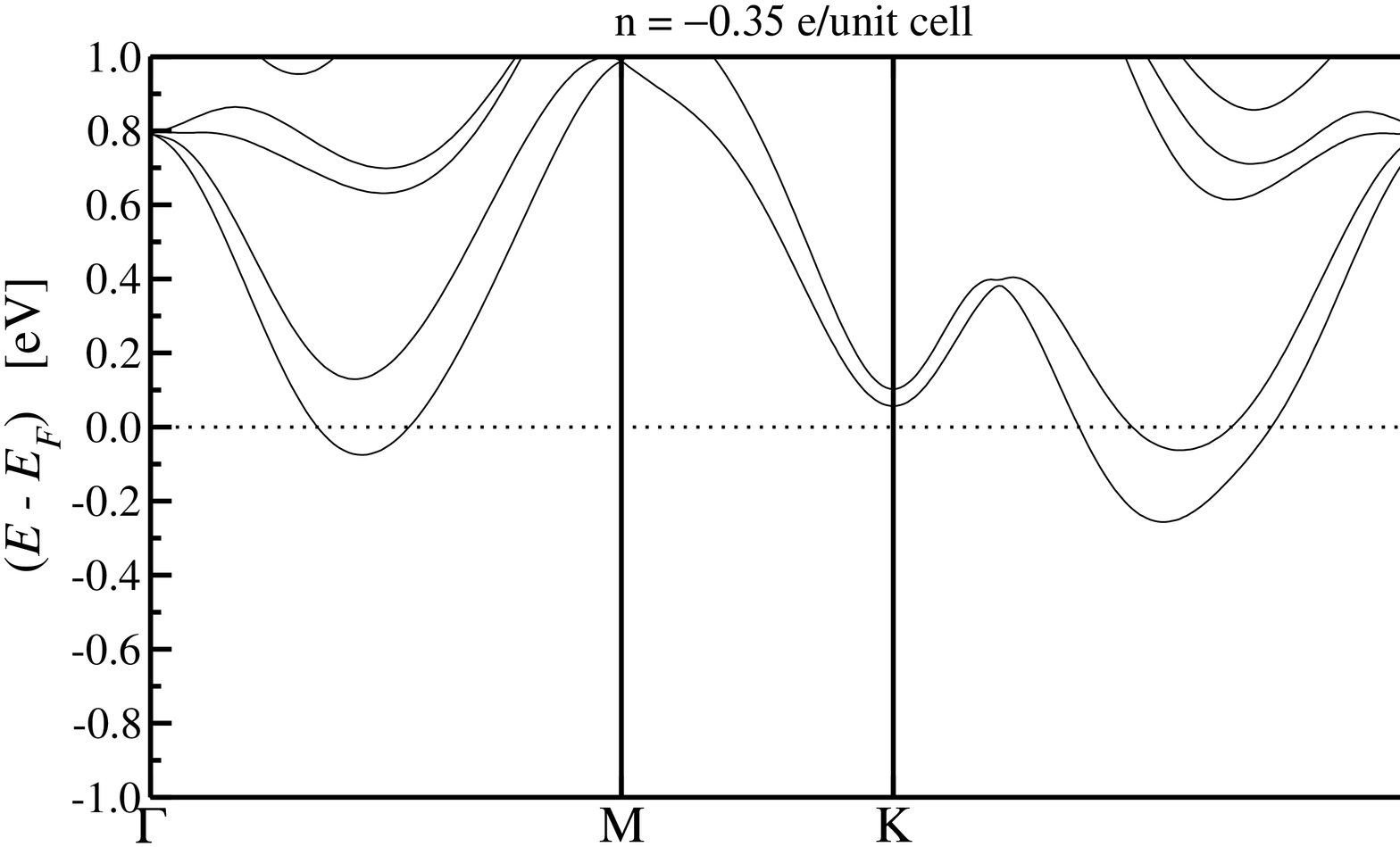}
 \includegraphics[width=0.31\textwidth,clip=,angle=-90]{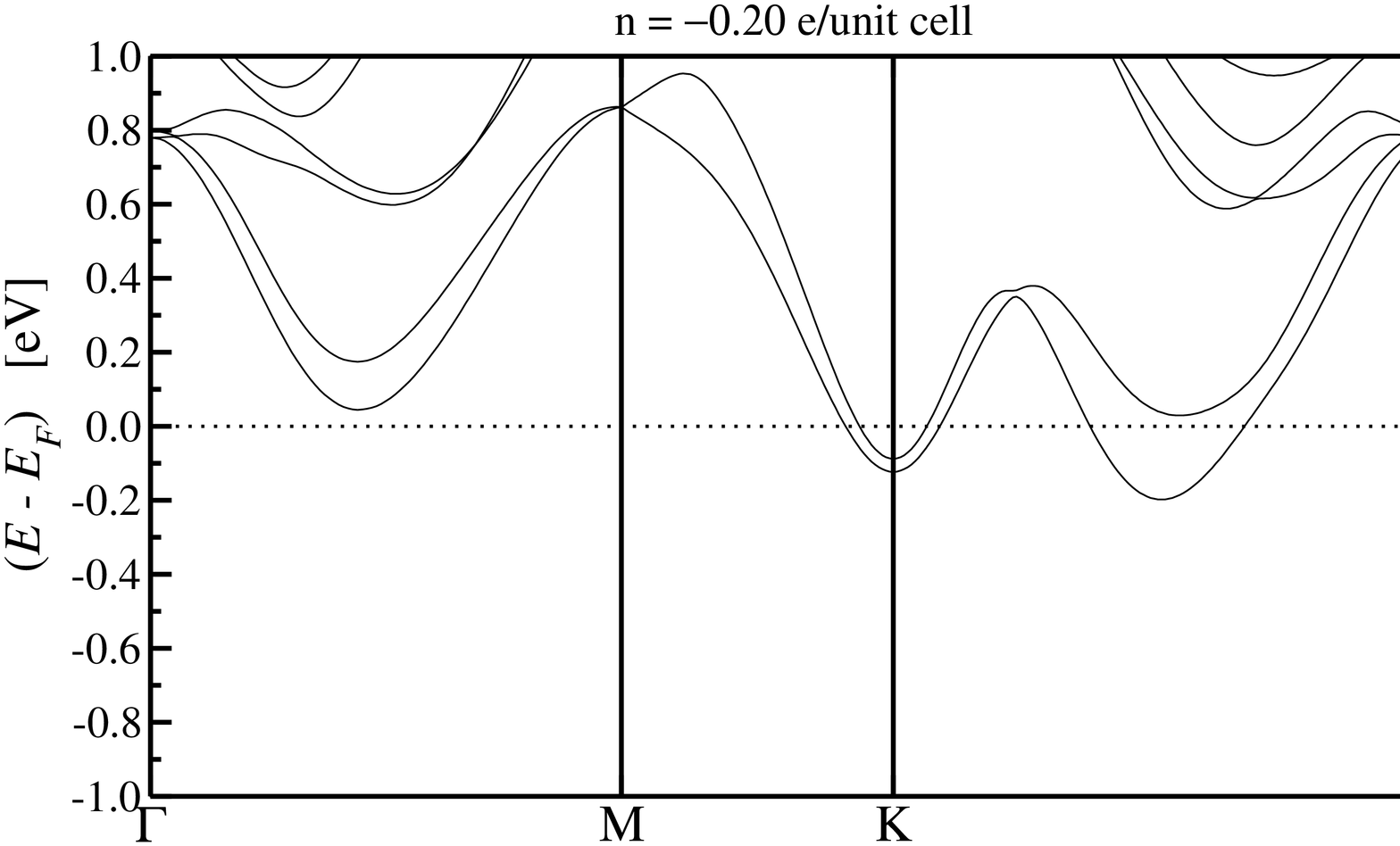}
 \includegraphics[width=0.31\textwidth,clip=,angle=-90]{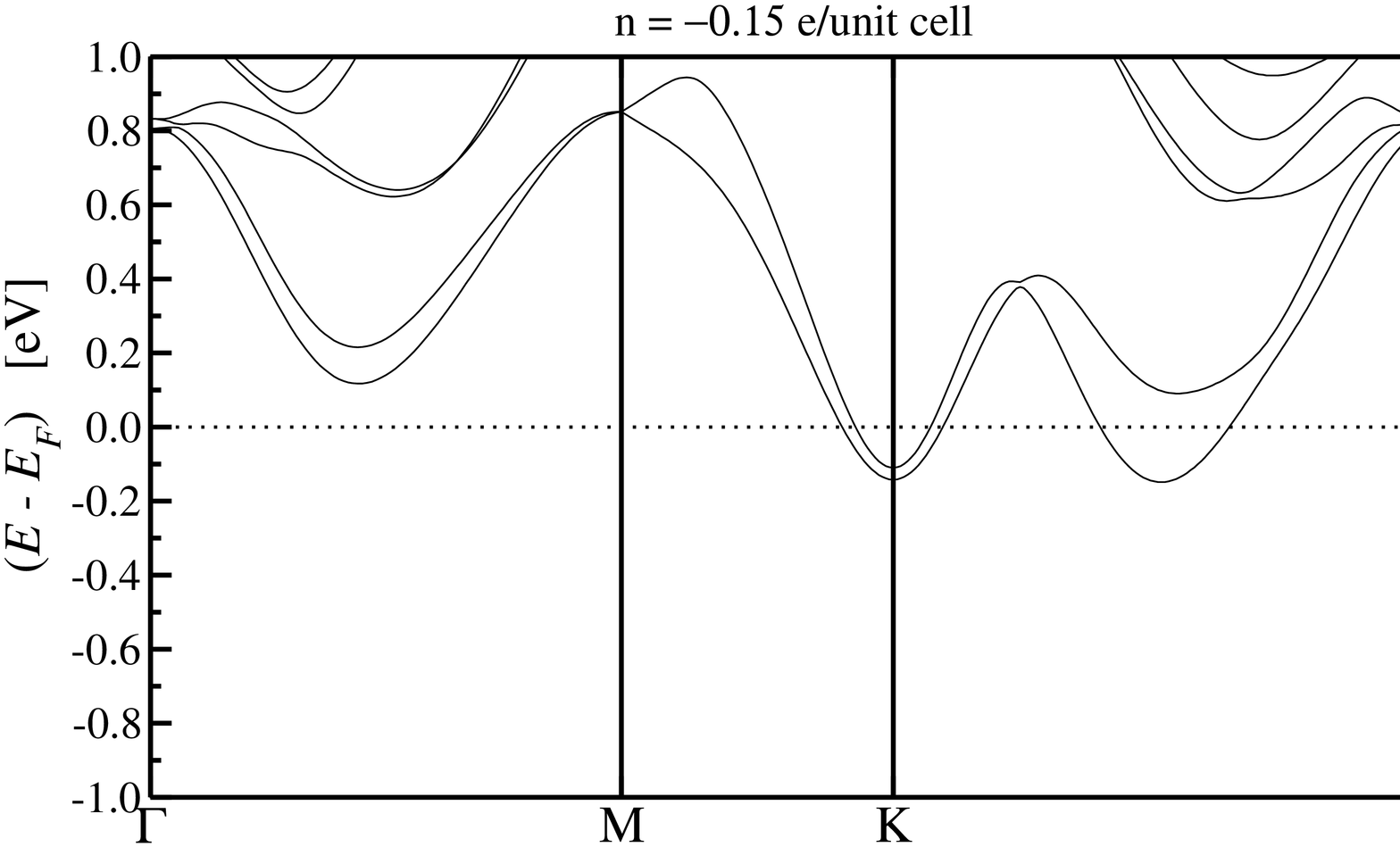}
 \includegraphics[width=0.31\textwidth,clip=,angle=-90]{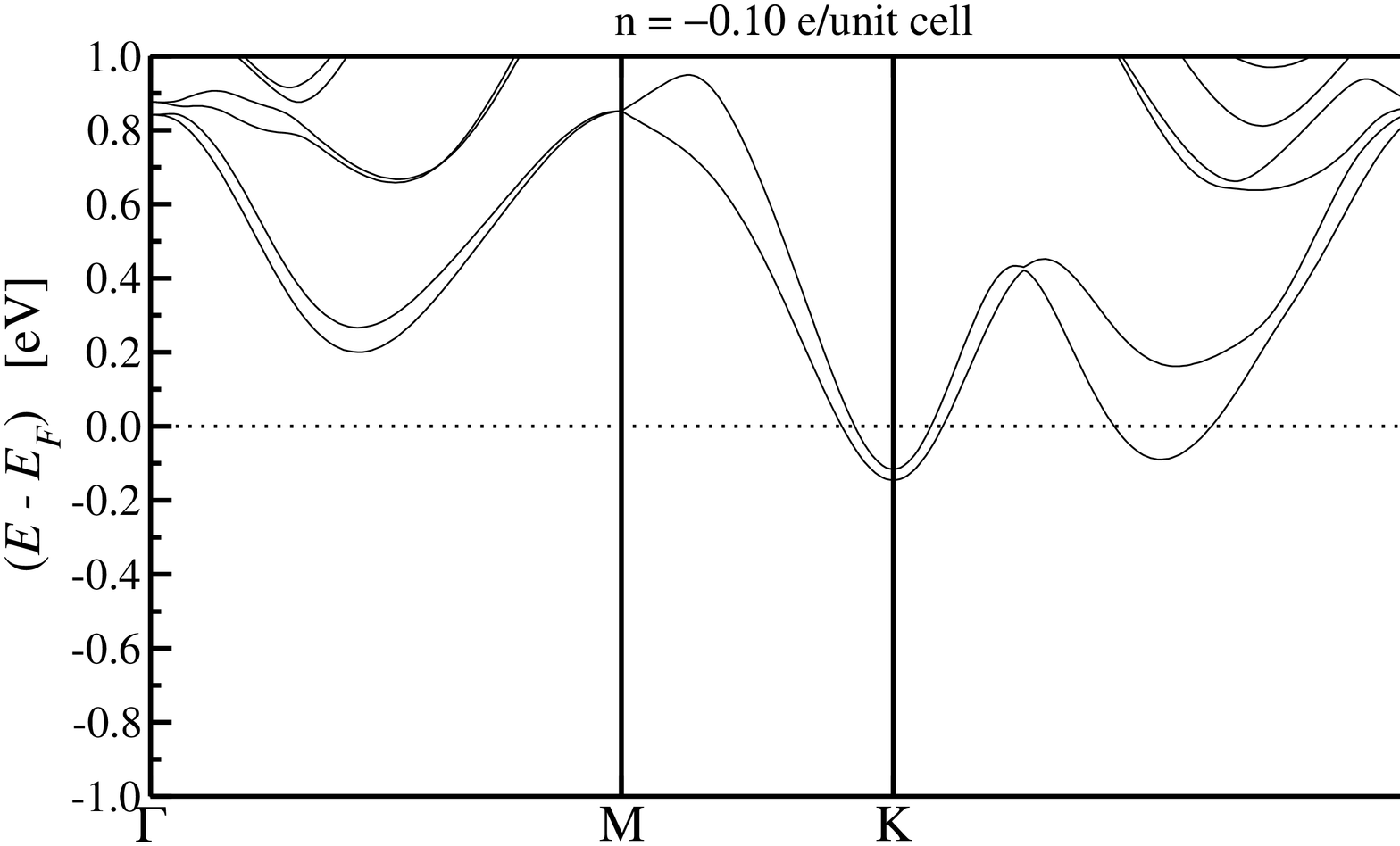}
 \includegraphics[width=0.31\textwidth,clip=,angle=-90]{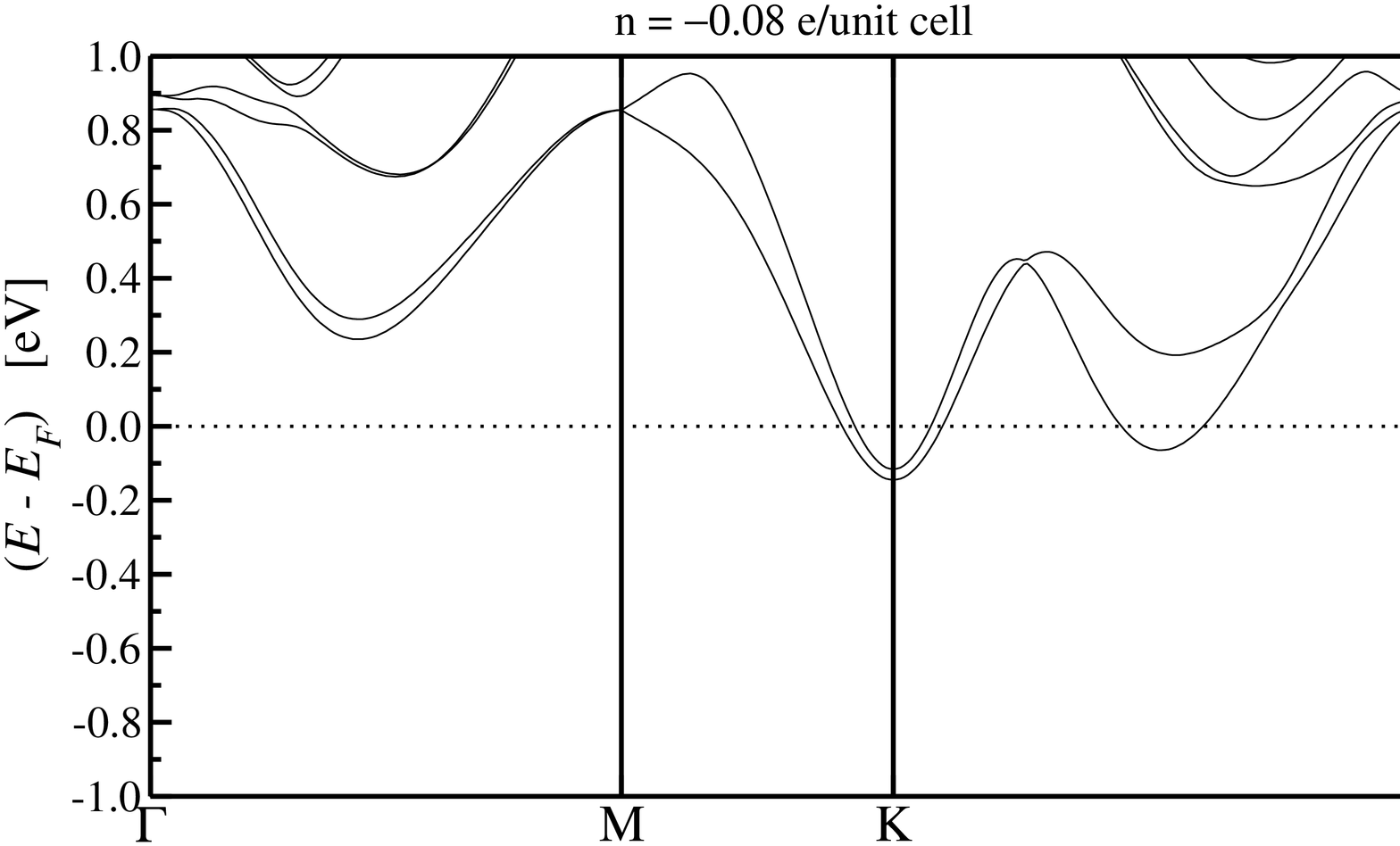}
 \includegraphics[width=0.31\textwidth,clip=,angle=-90]{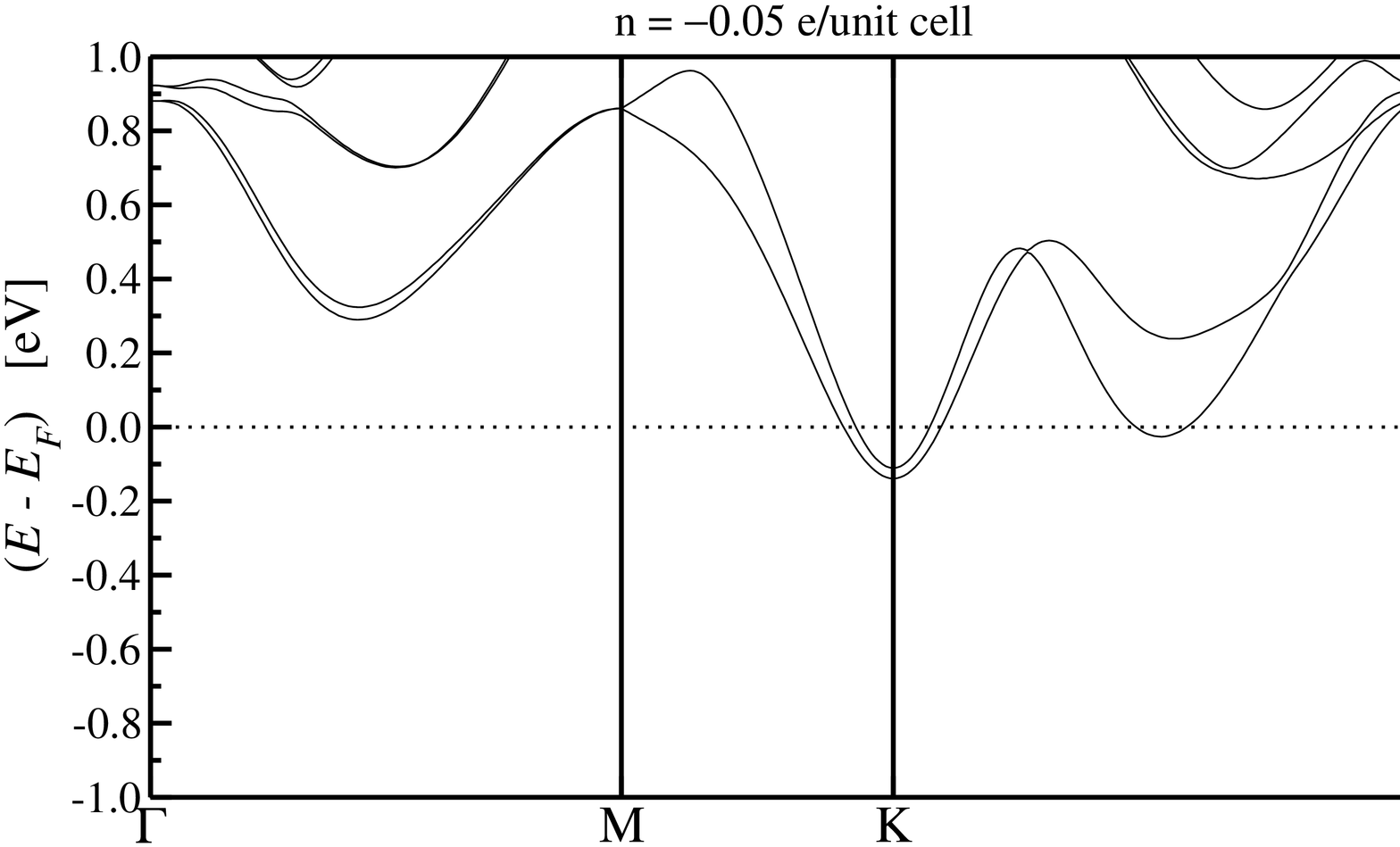}
 \includegraphics[width=0.31\textwidth,clip=,angle=-90]{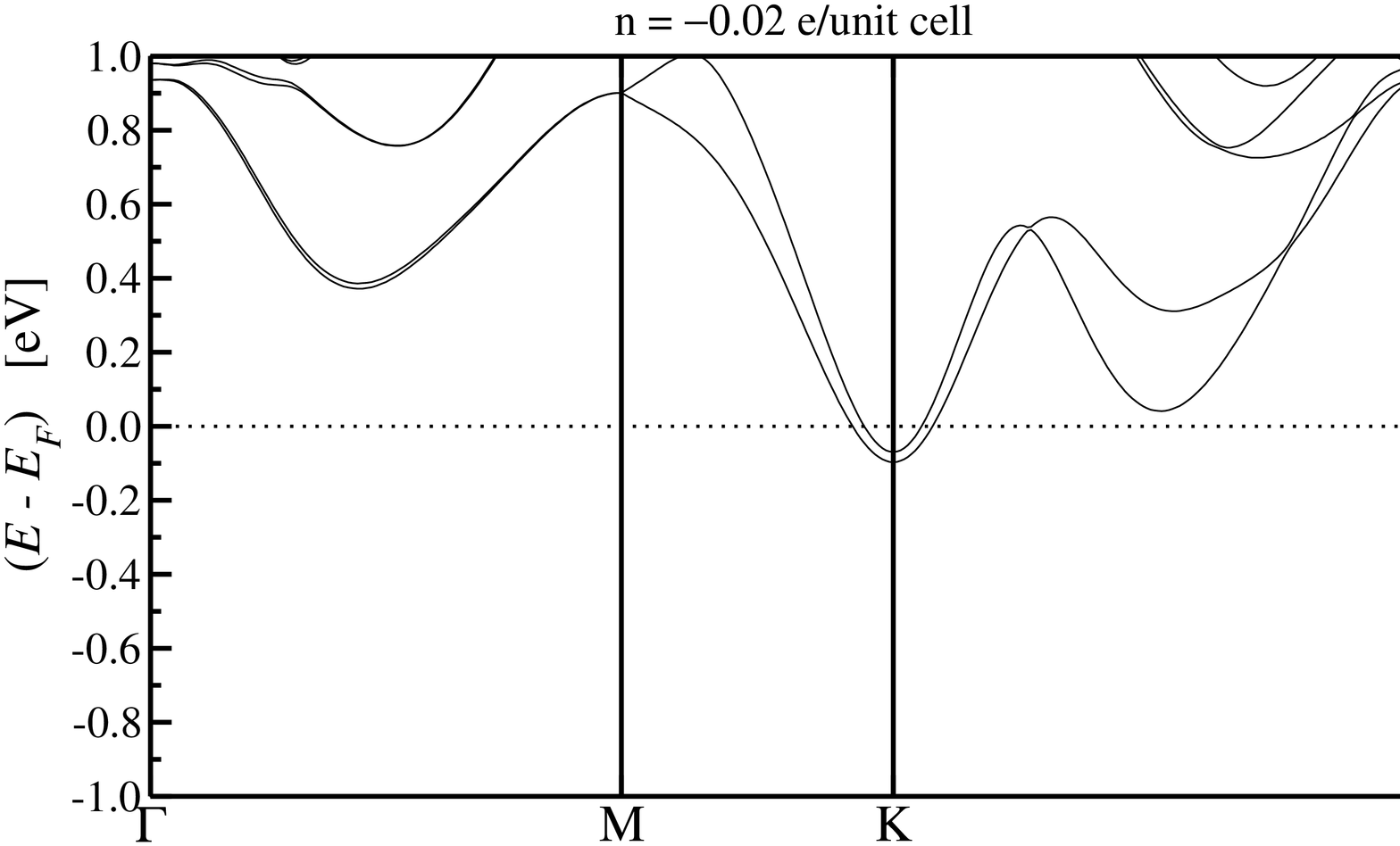}
 \includegraphics[width=0.31\textwidth,clip=,angle=-90]{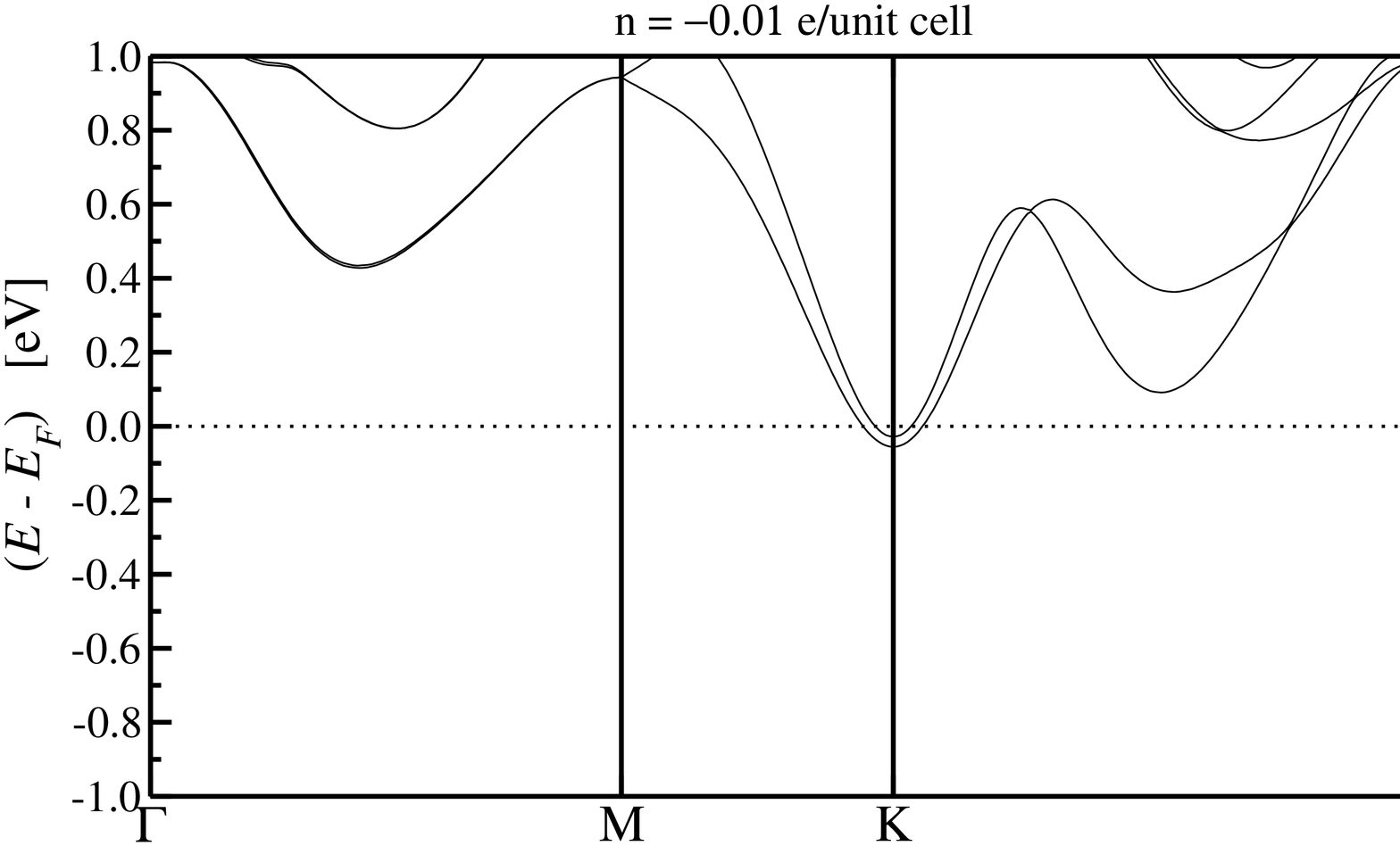}
 \caption{Band structure of monolayer WS$_2$ for different doping as indicated in the labels.}
\end{figure*}
\begin{figure*}[hbp]
 \centering
 \includegraphics[width=0.31\textwidth,clip=,angle=-90]{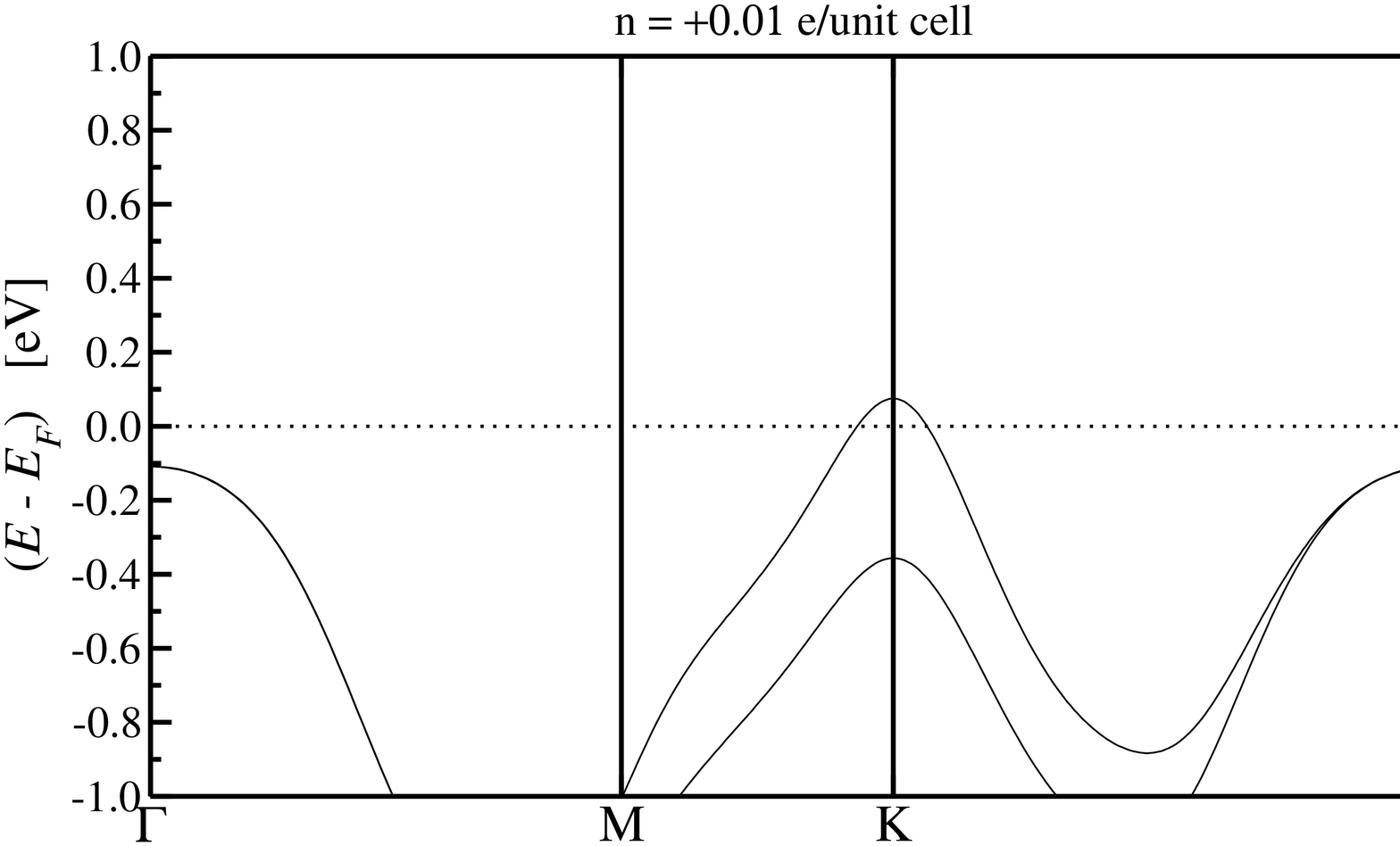}
 \includegraphics[width=0.31\textwidth,clip=,angle=-90]{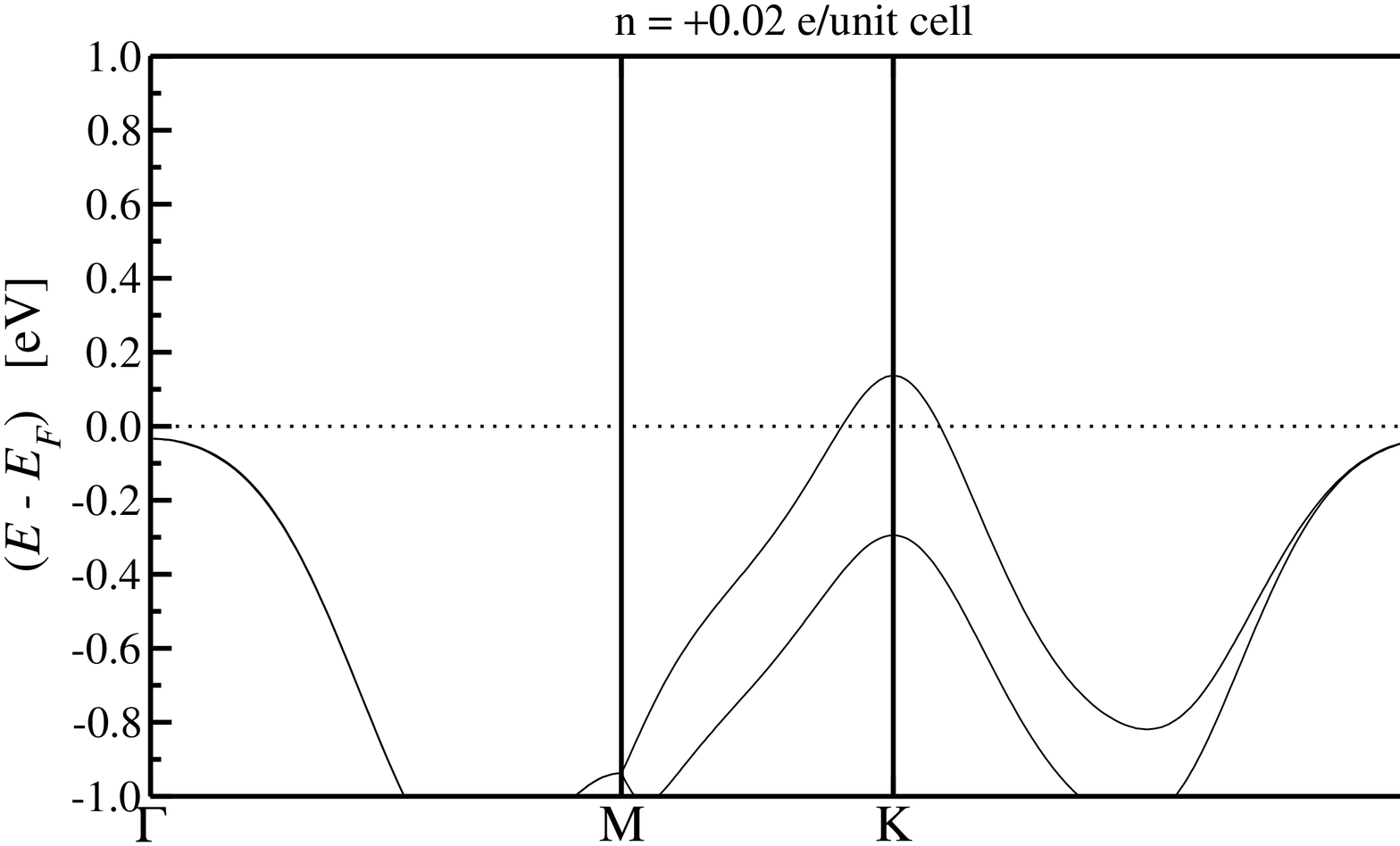}
 \includegraphics[width=0.31\textwidth,clip=,angle=-90]{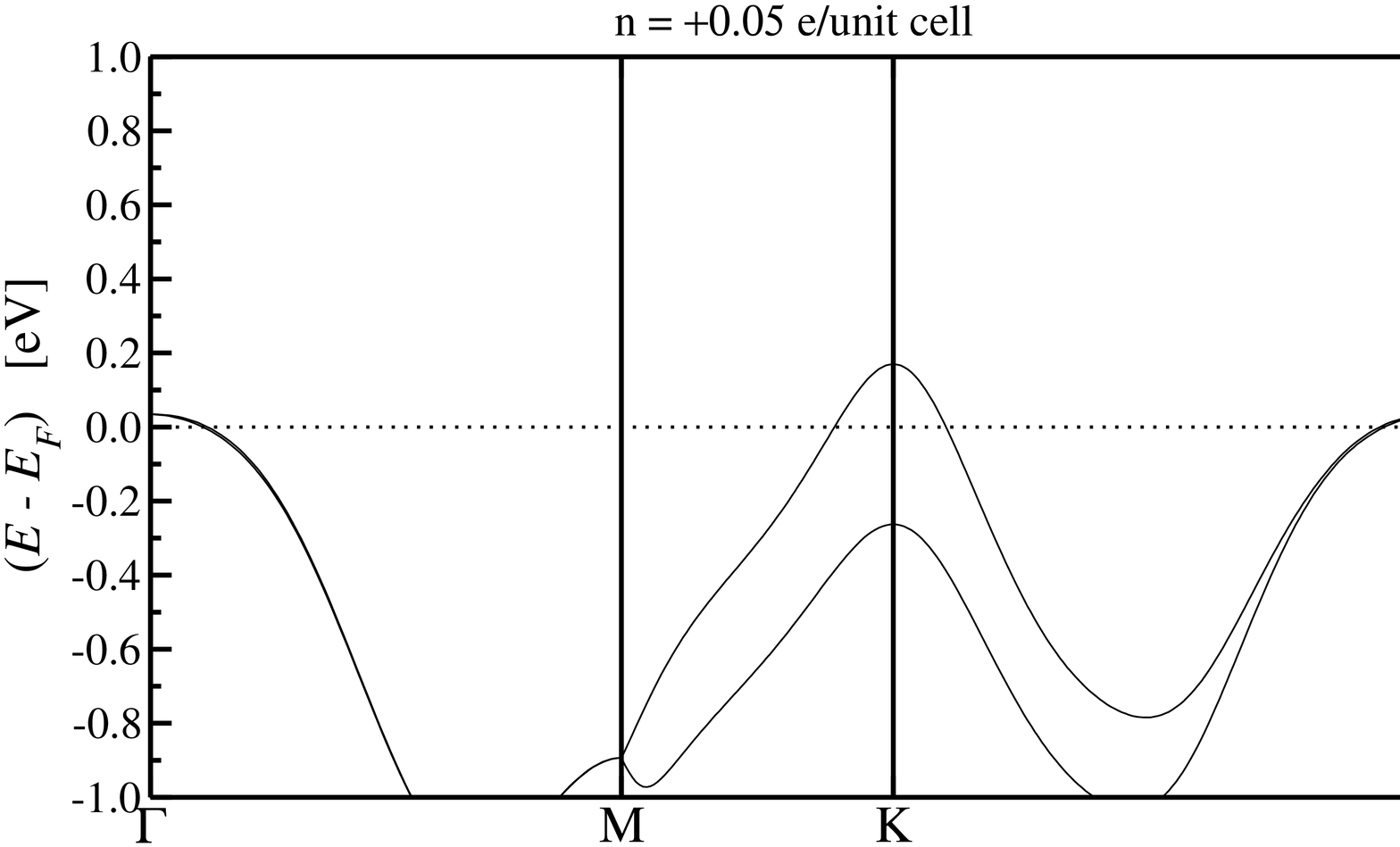}
 \includegraphics[width=0.31\textwidth,clip=,angle=-90]{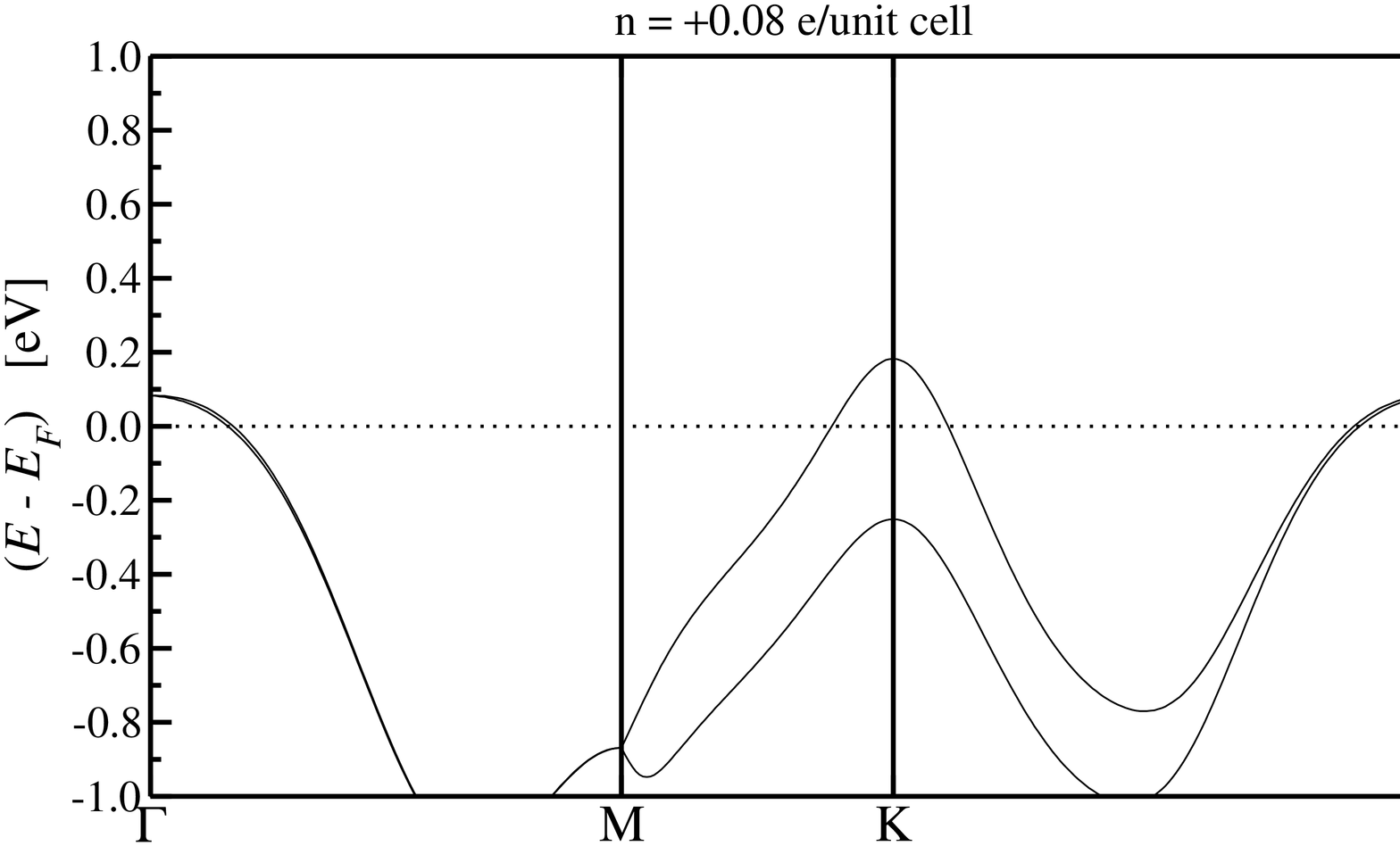}
 \includegraphics[width=0.31\textwidth,clip=,angle=-90]{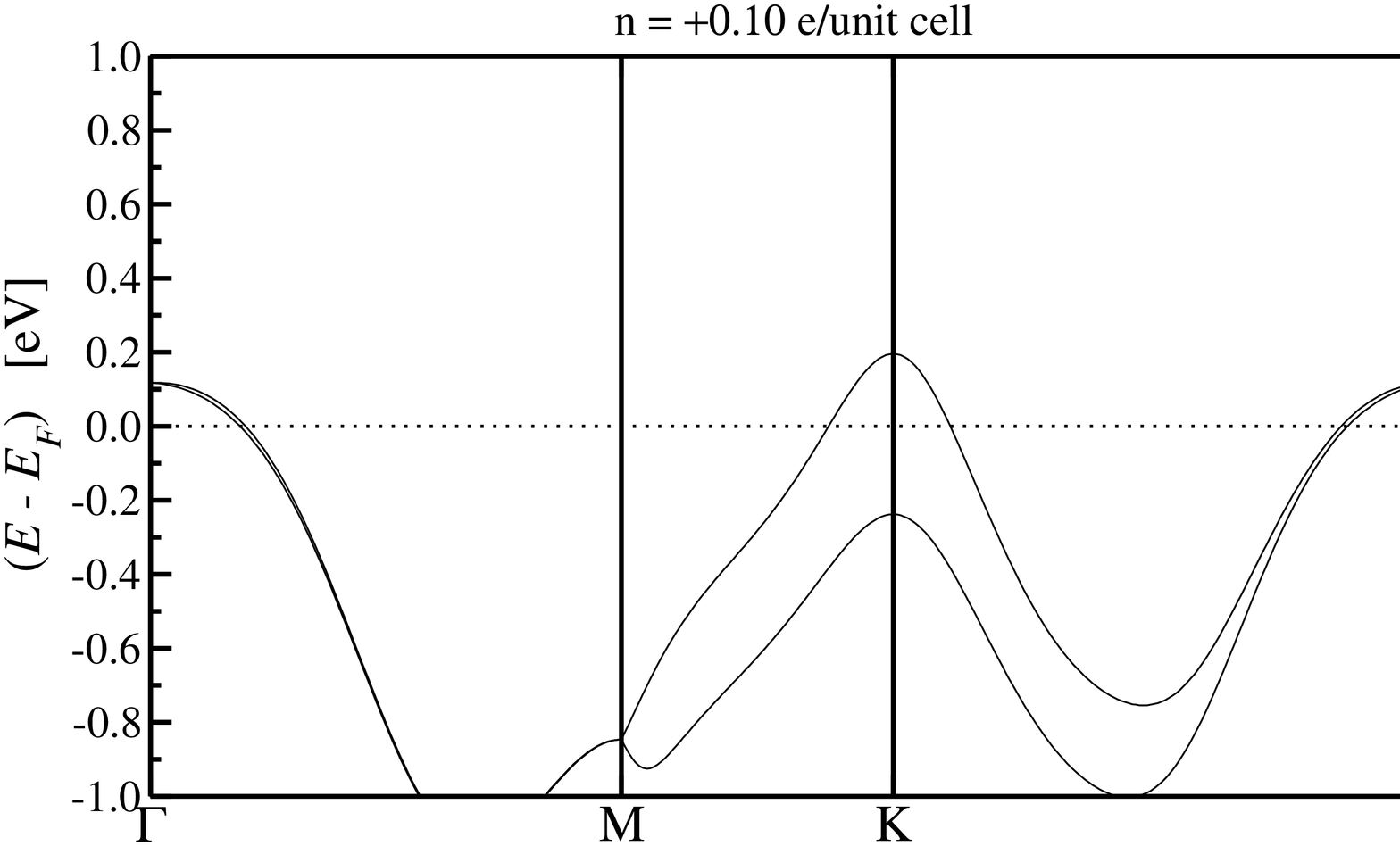}
 \includegraphics[width=0.31\textwidth,clip=,angle=-90]{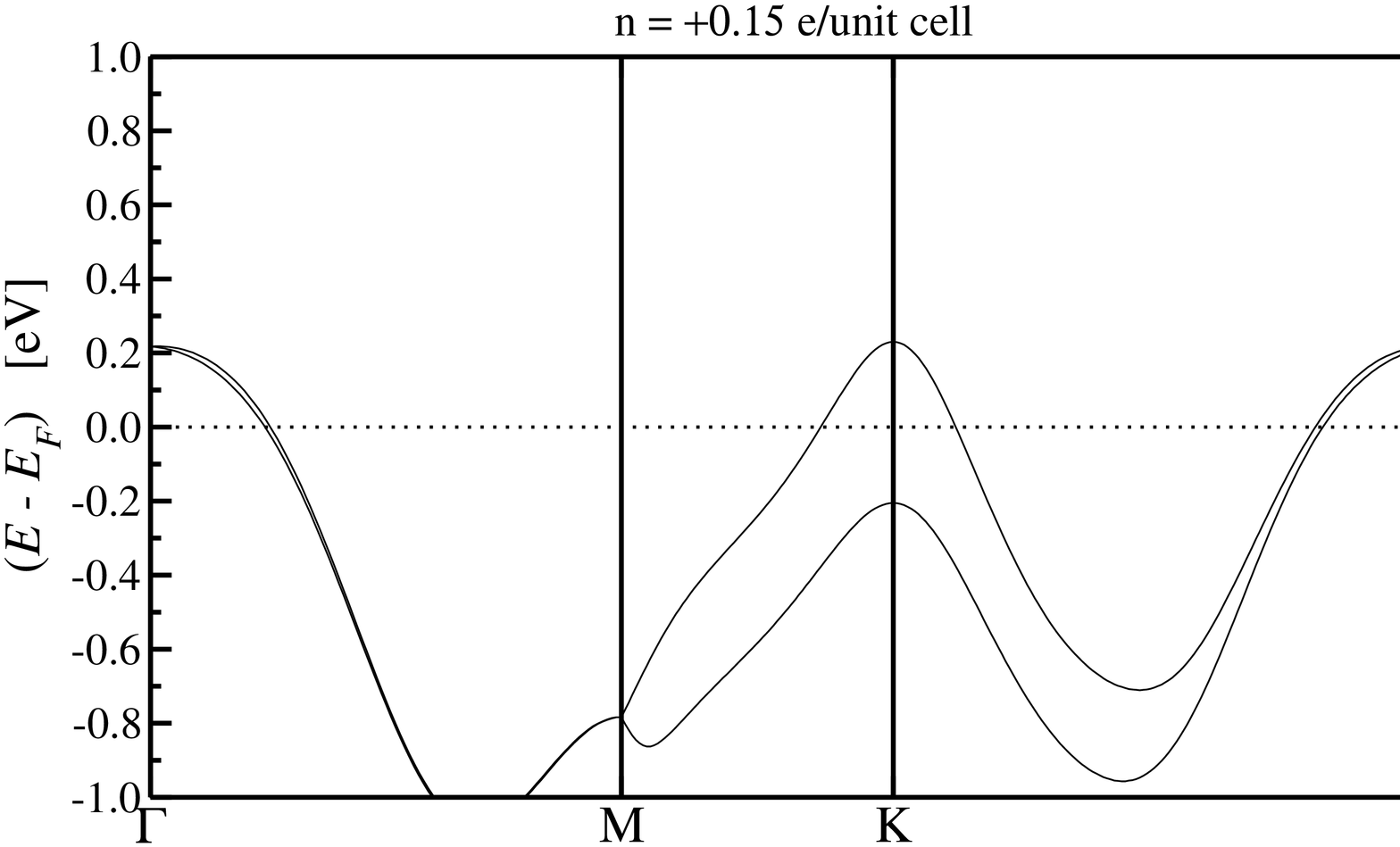}
 \includegraphics[width=0.31\textwidth,clip=,angle=-90]{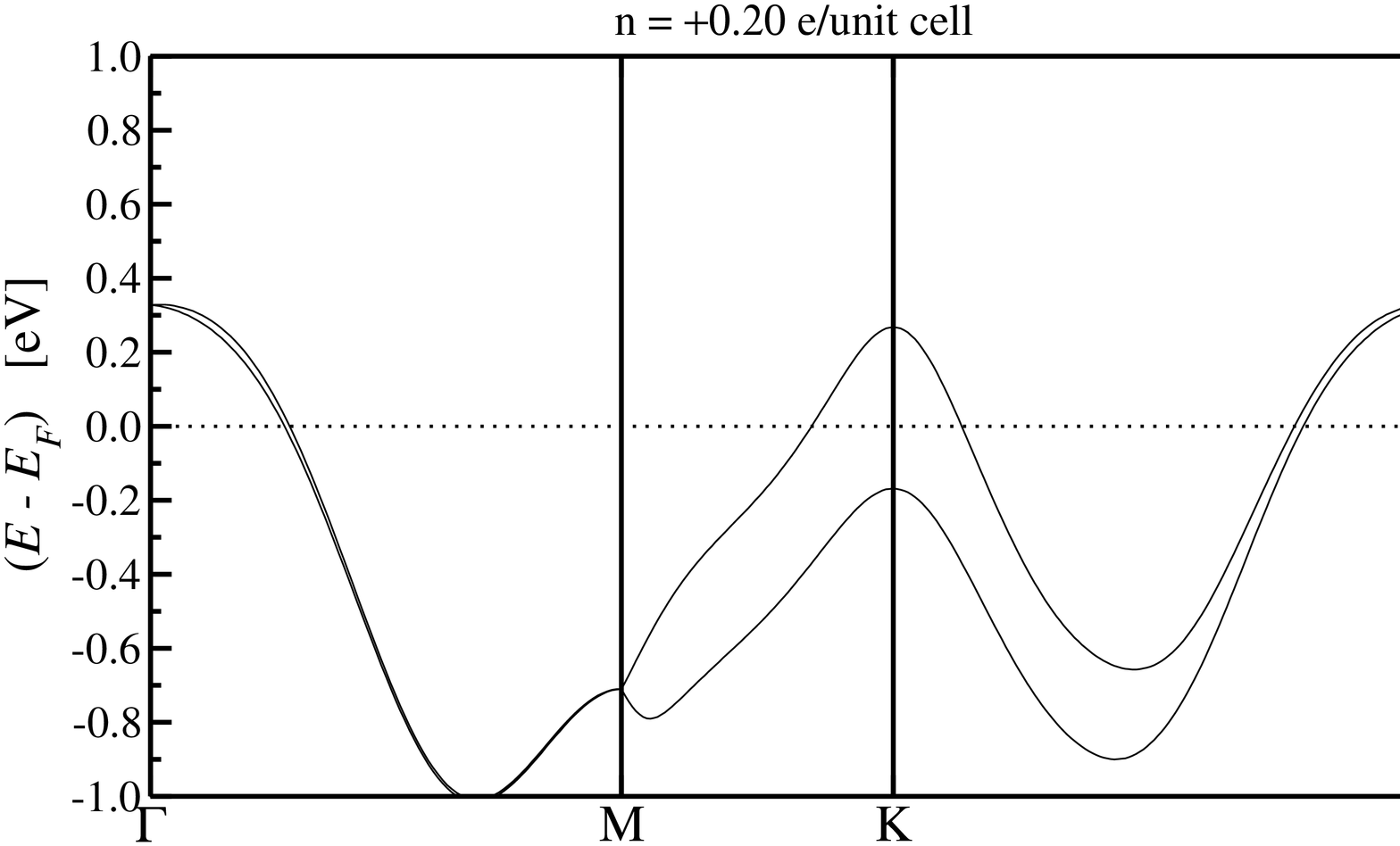}
 \includegraphics[width=0.31\textwidth,clip=,angle=-90]{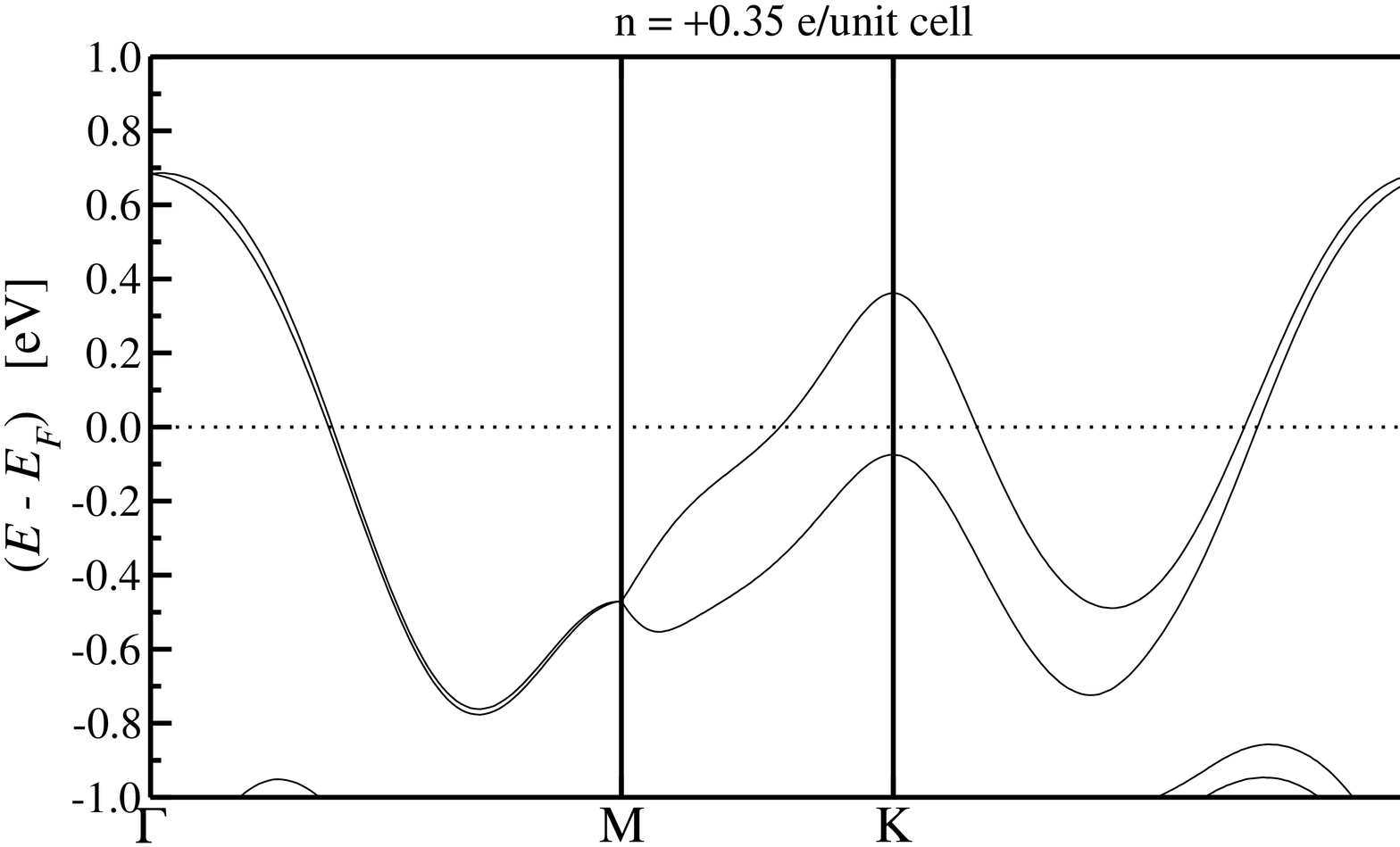}
 \caption{Band structure of monolayer WS$_2$ for different doping as indicated in the labels.}
\end{figure*}
\begin{figure*}[hbp]
 \centering
 \includegraphics[width=0.31\textwidth,clip=,angle=-90]{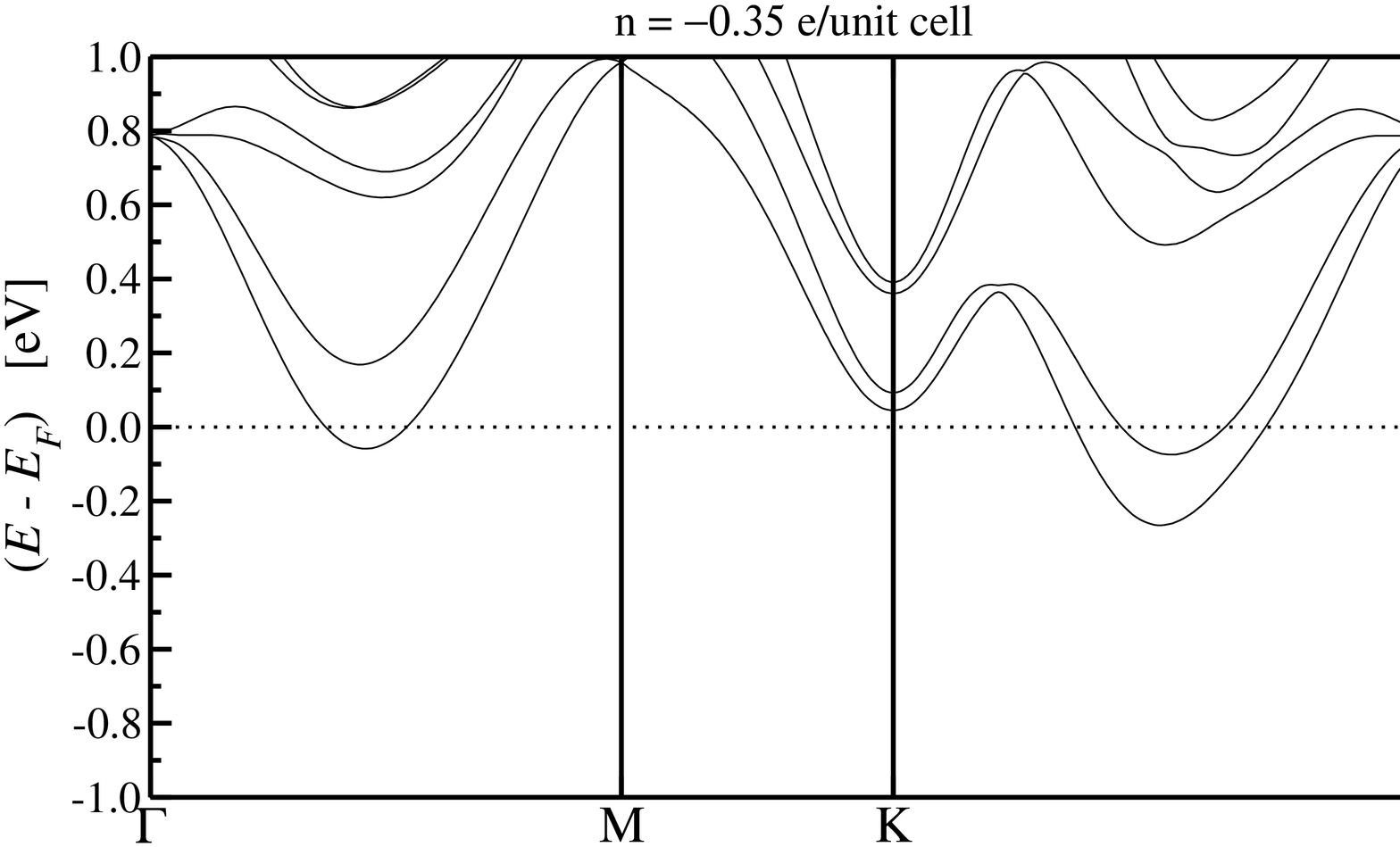}
 \includegraphics[width=0.31\textwidth,clip=,angle=-90]{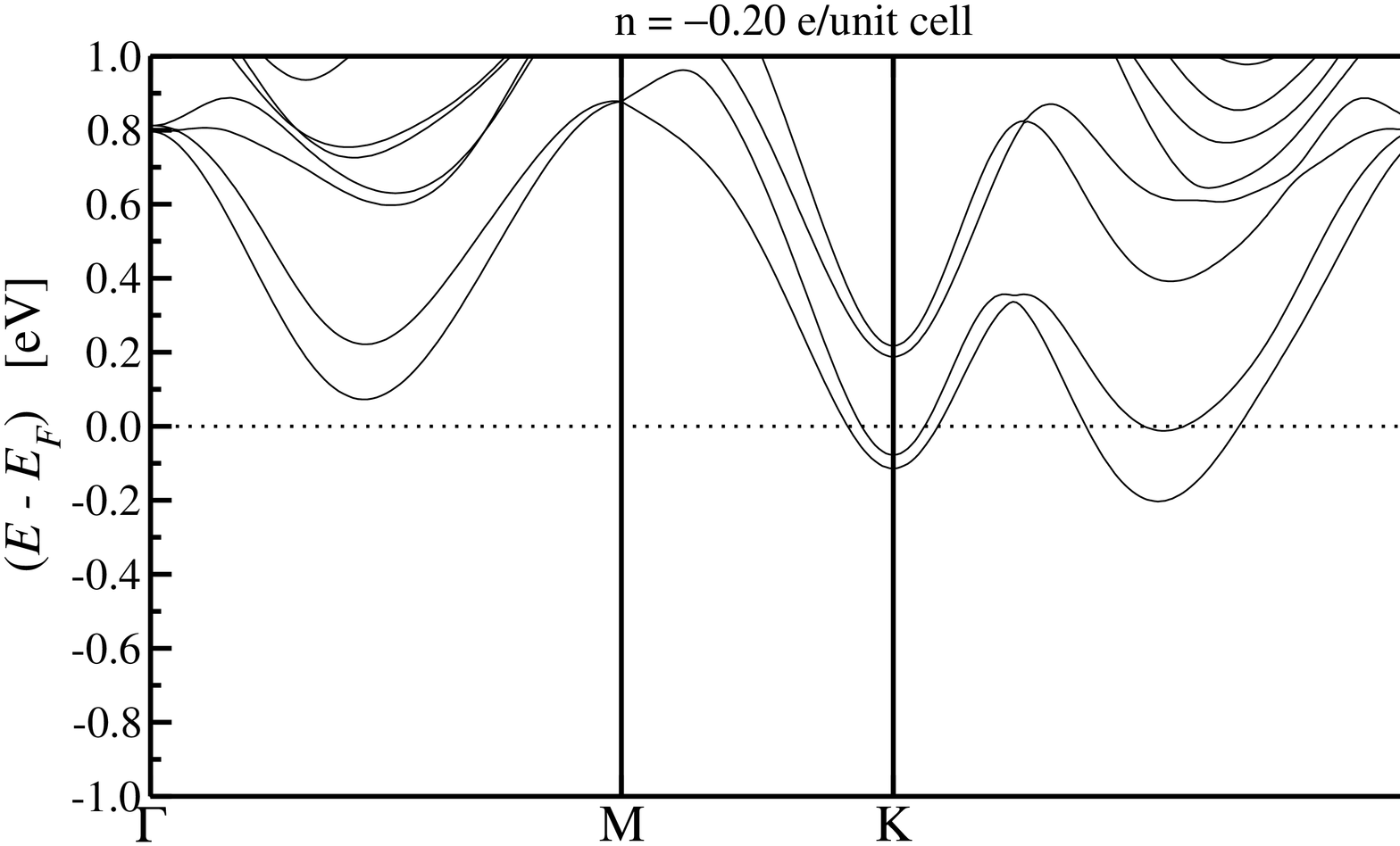}
 \includegraphics[width=0.31\textwidth,clip=,angle=-90]{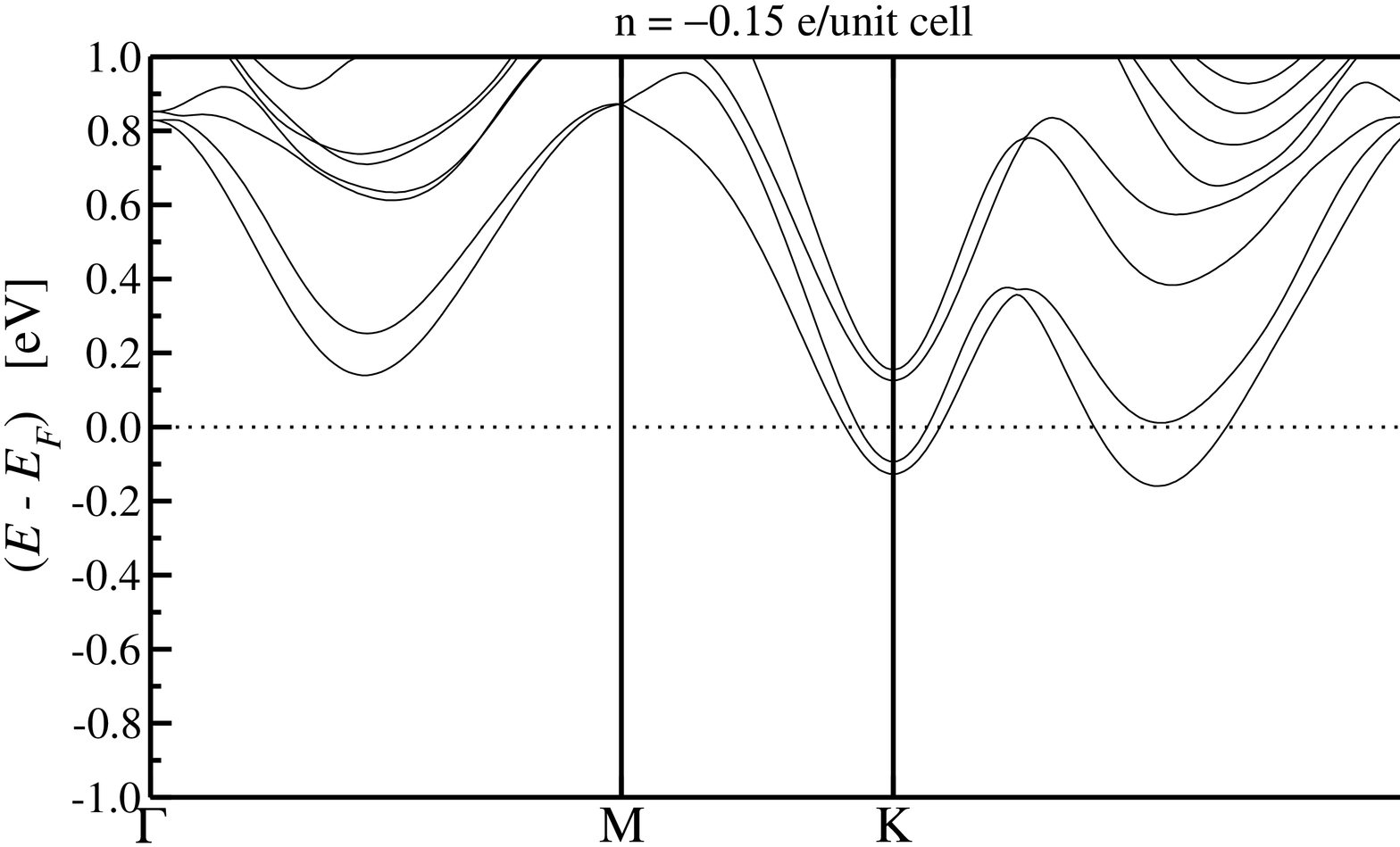}
 \includegraphics[width=0.31\textwidth,clip=,angle=-90]{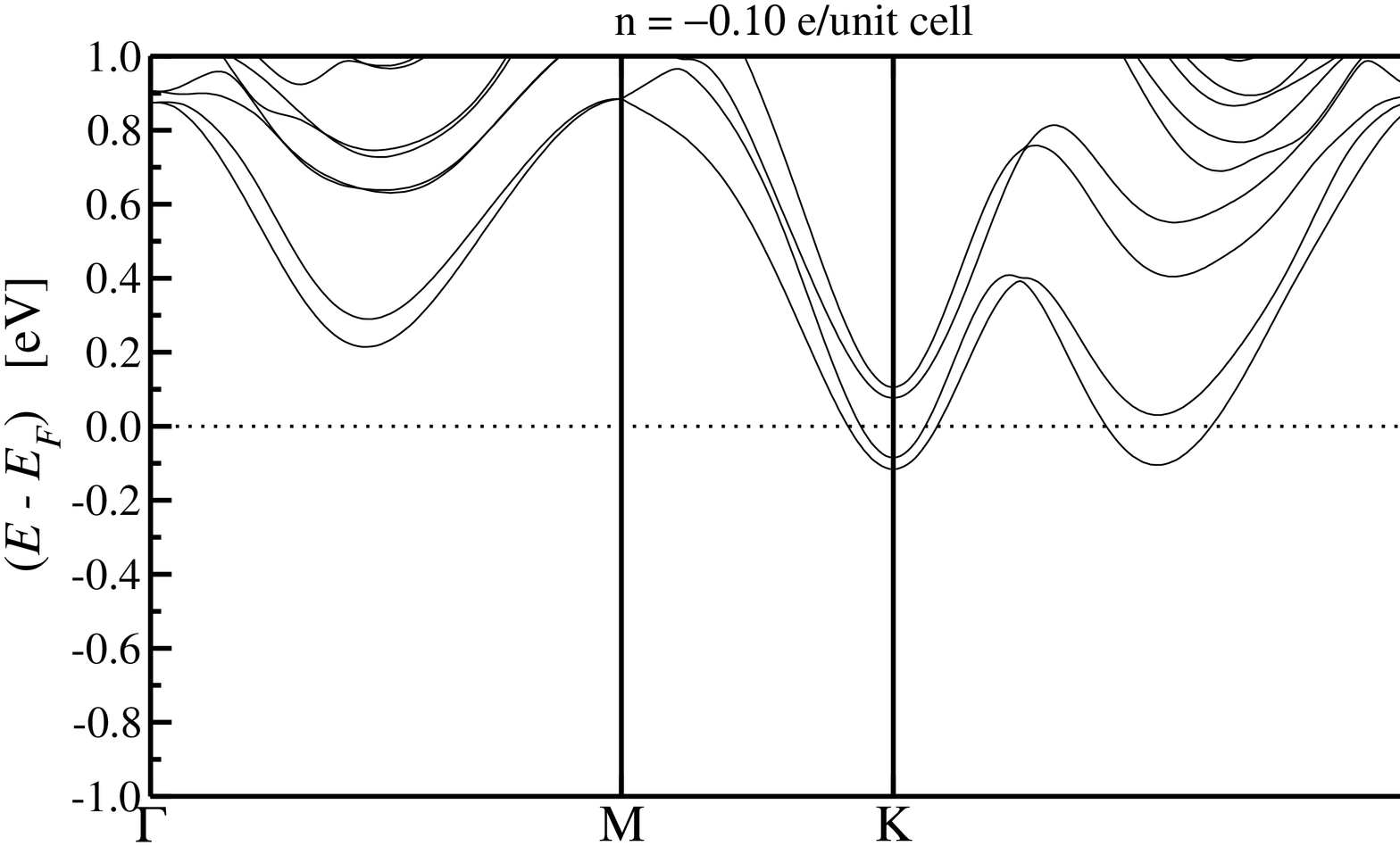}
 \includegraphics[width=0.31\textwidth,clip=,angle=-90]{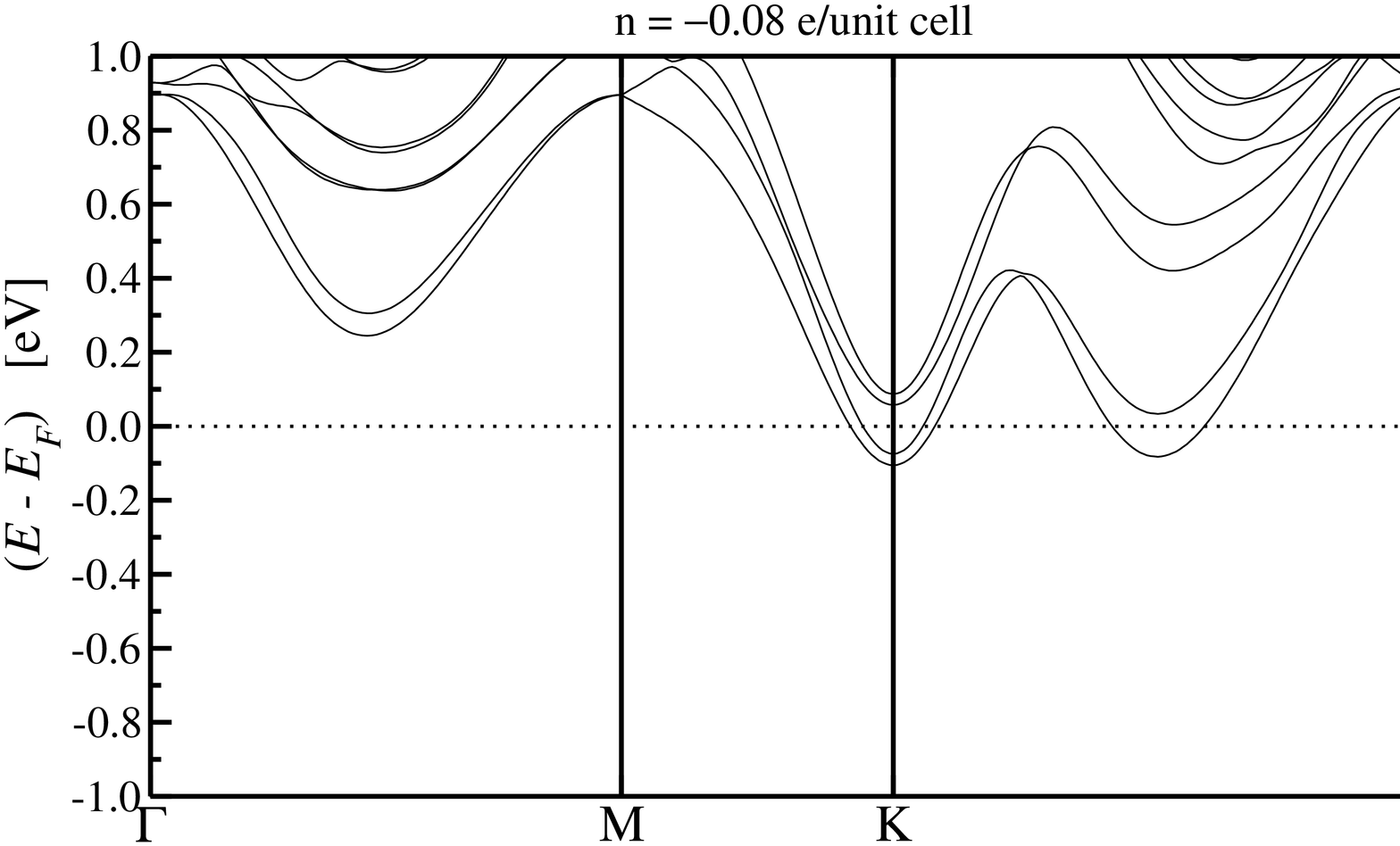}
 \includegraphics[width=0.31\textwidth,clip=,angle=-90]{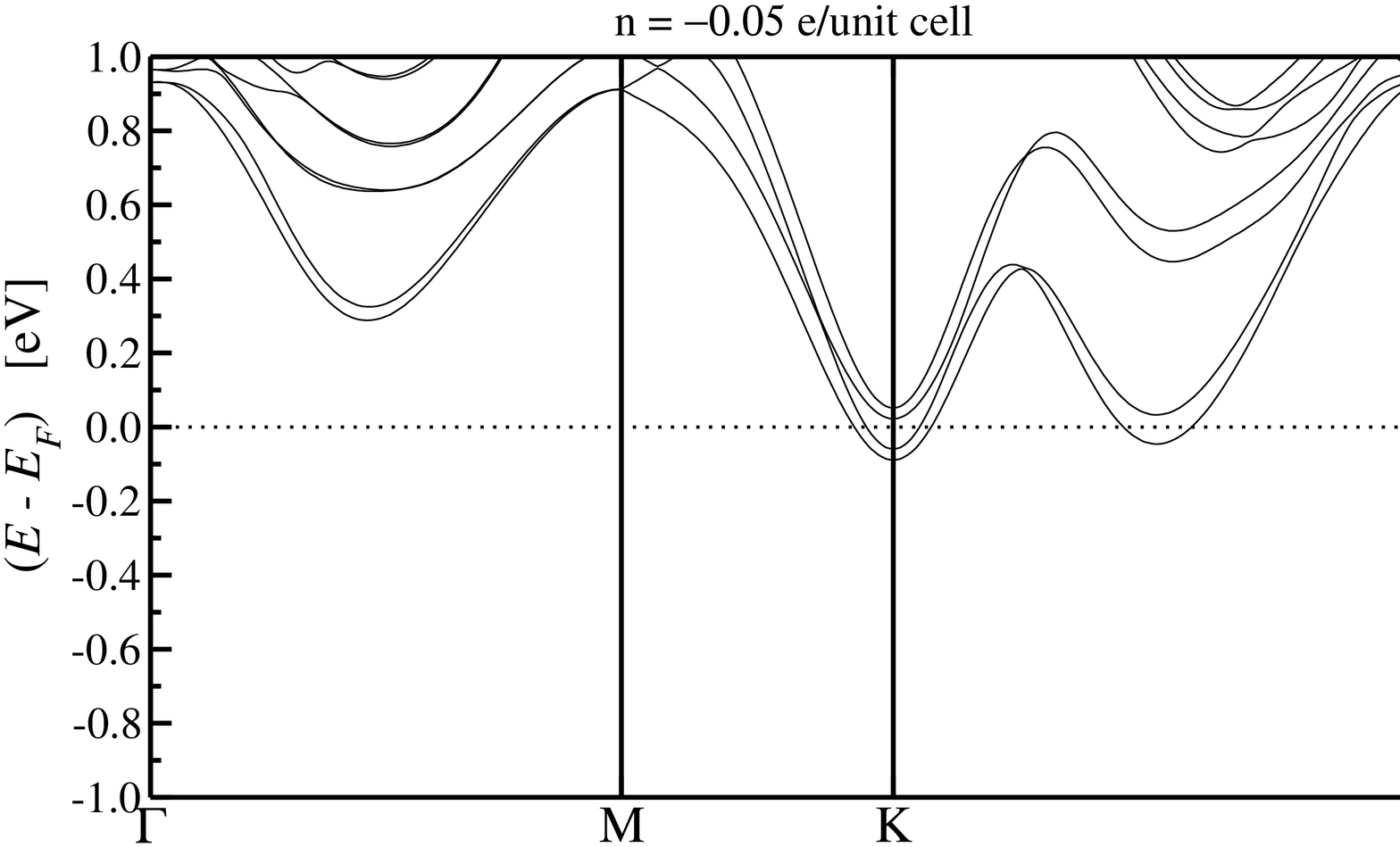}
 \includegraphics[width=0.31\textwidth,clip=,angle=-90]{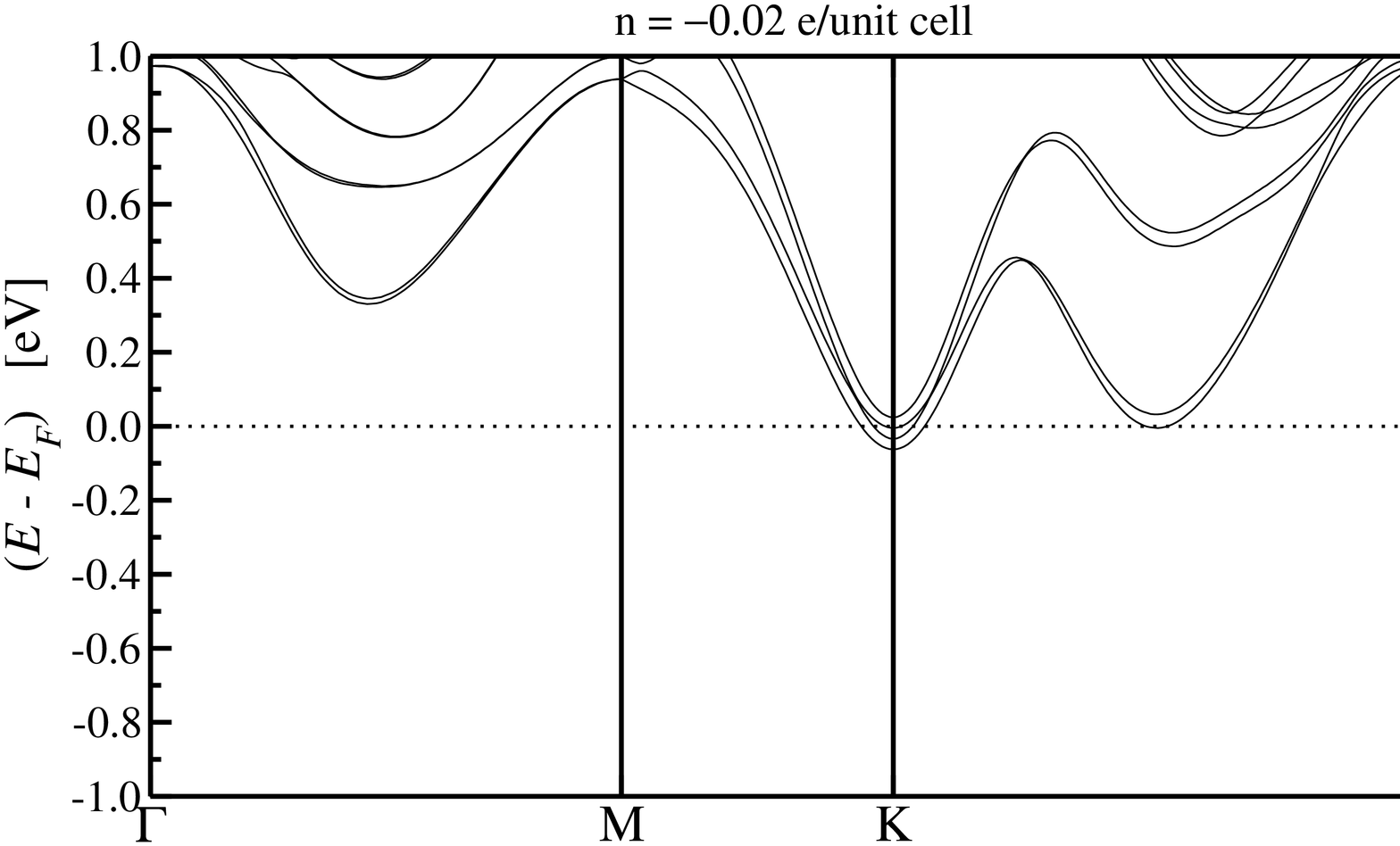}
 \includegraphics[width=0.31\textwidth,clip=,angle=-90]{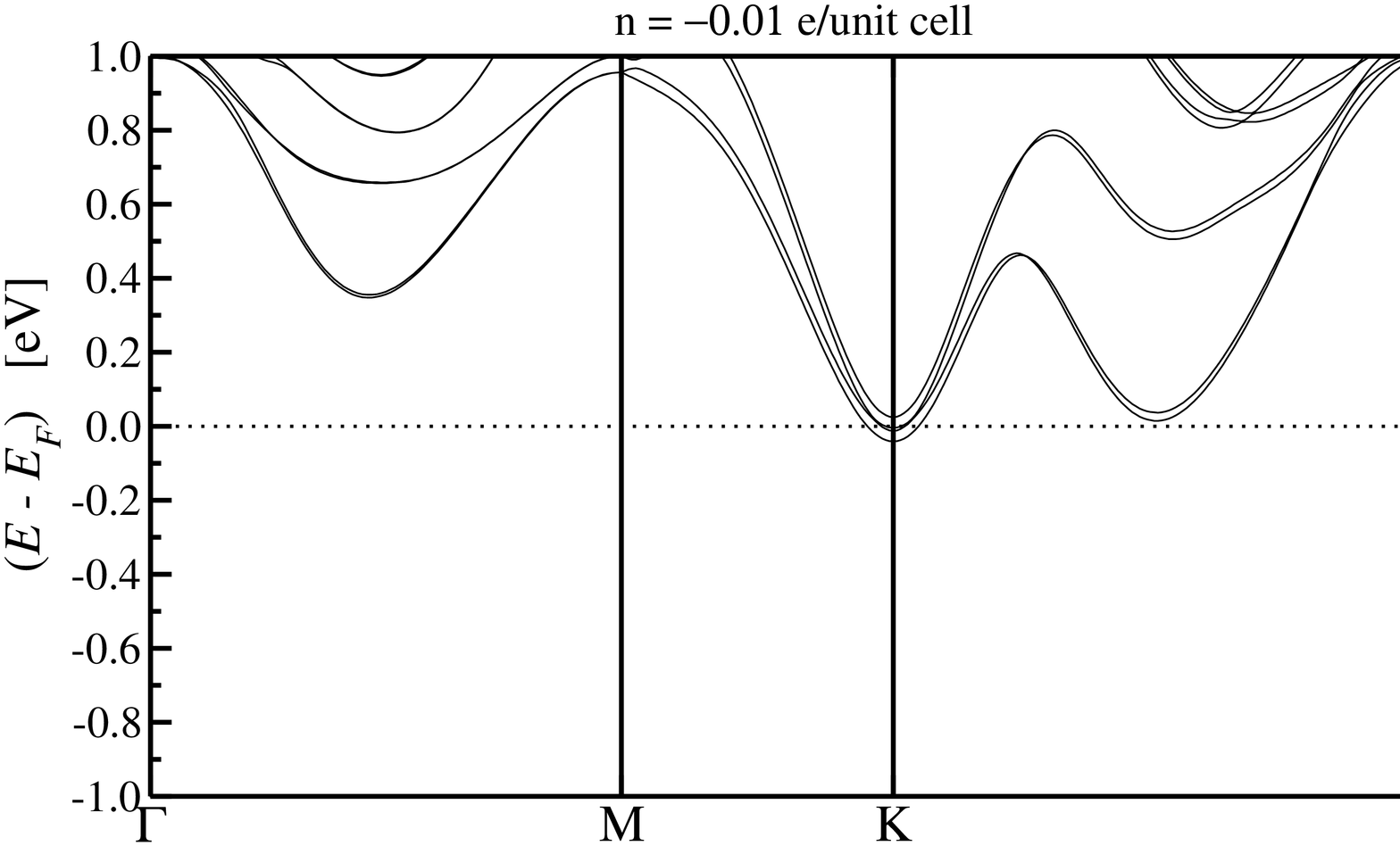}
 \caption{Band structure of bilayer WS$_2$ for different doping as indicated in the labels.}
\end{figure*}
\begin{figure*}[hbp]
 \centering
 \includegraphics[width=0.31\textwidth,clip=,angle=-90]{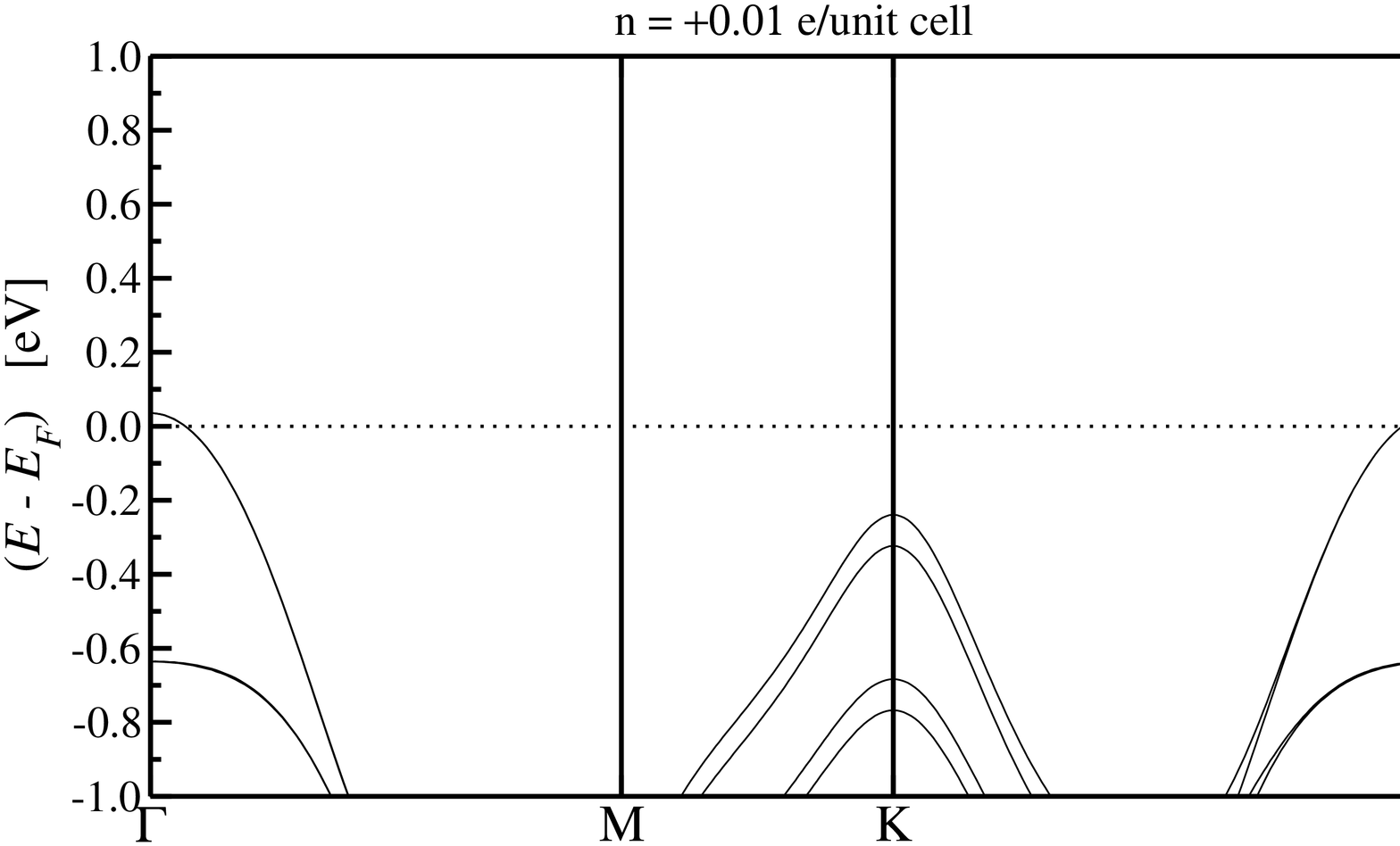}
 \includegraphics[width=0.31\textwidth,clip=,angle=-90]{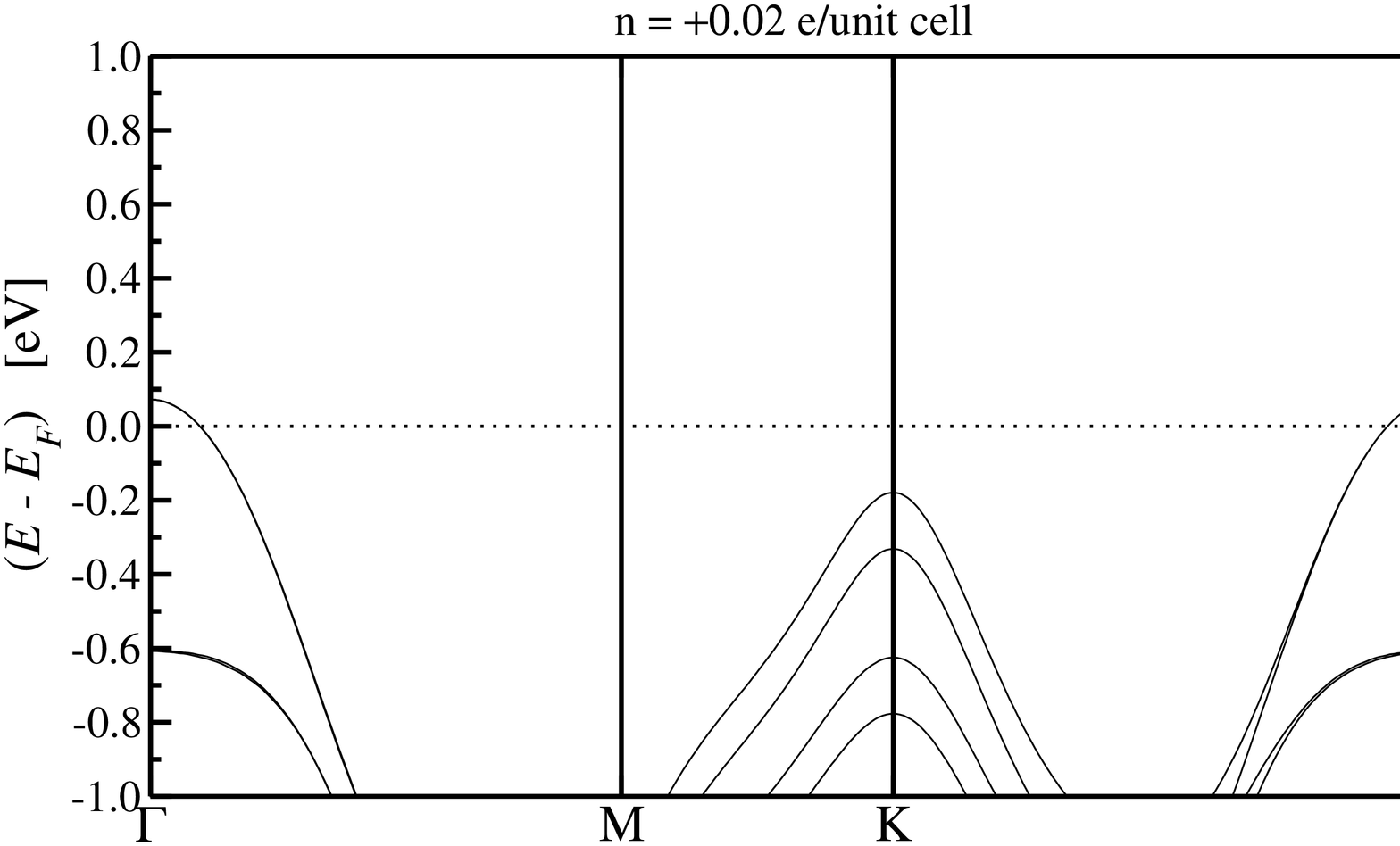}
 \includegraphics[width=0.31\textwidth,clip=,angle=-90]{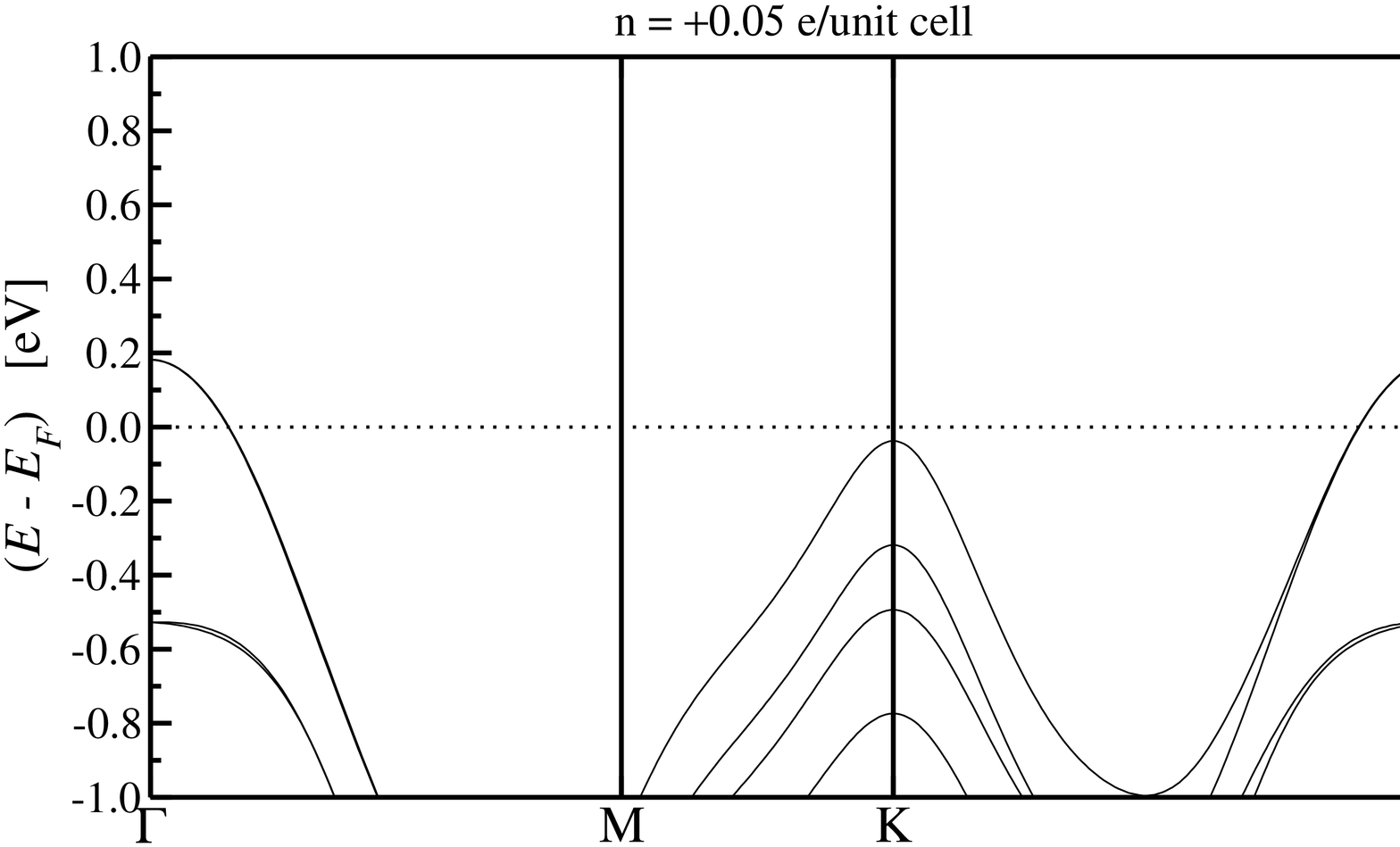}
 \includegraphics[width=0.31\textwidth,clip=,angle=-90]{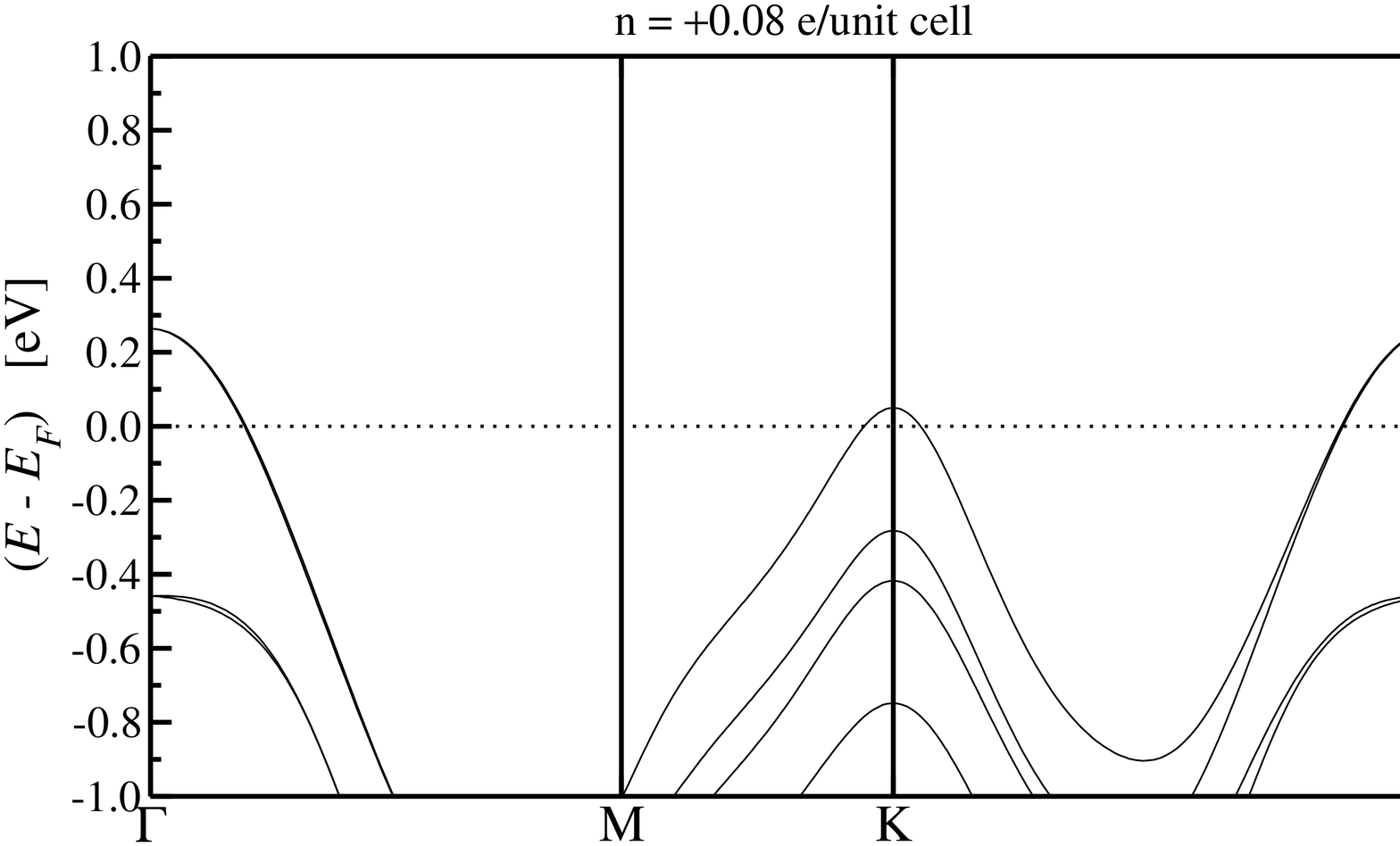}
 \includegraphics[width=0.31\textwidth,clip=,angle=-90]{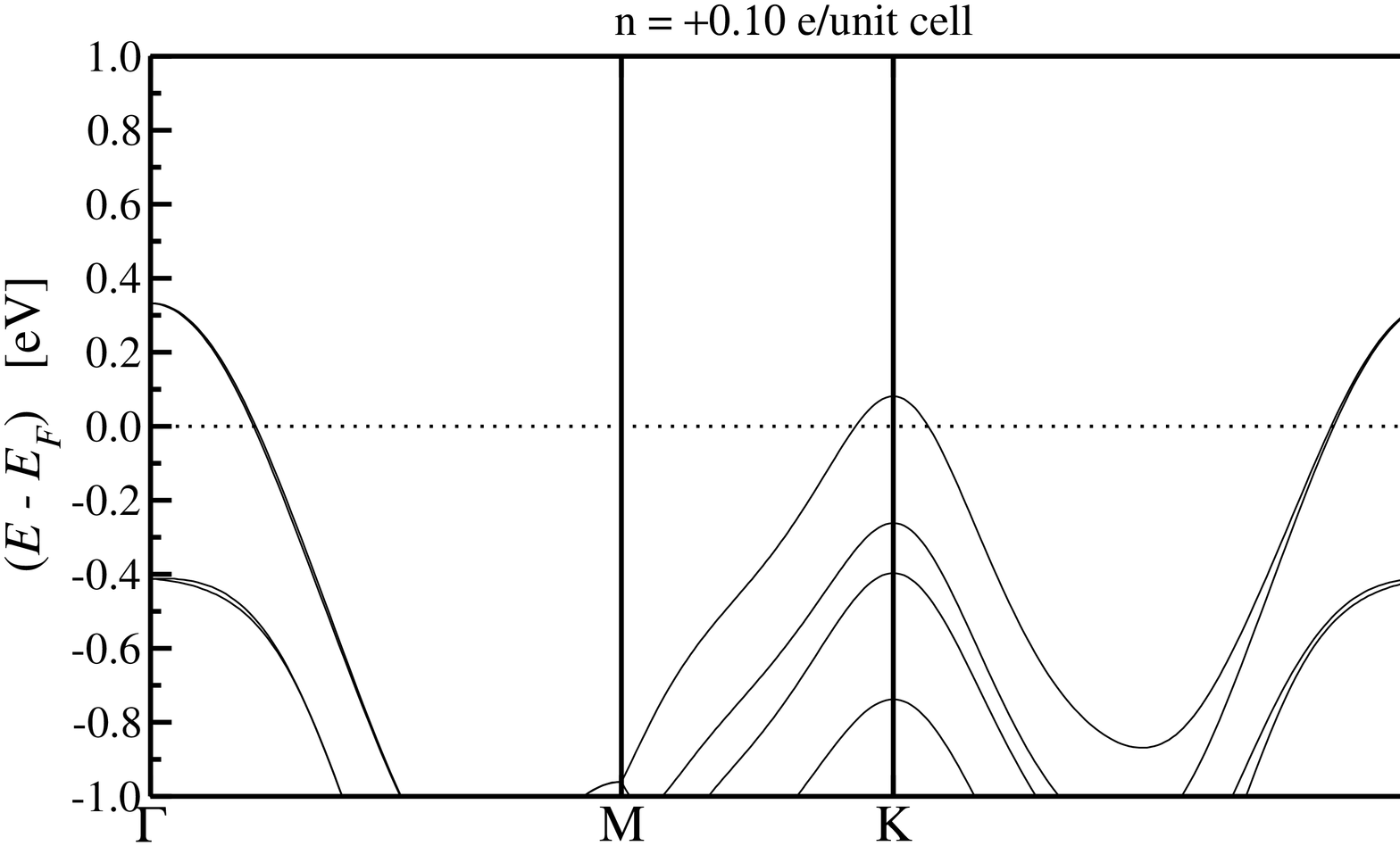}
 \includegraphics[width=0.31\textwidth,clip=,angle=-90]{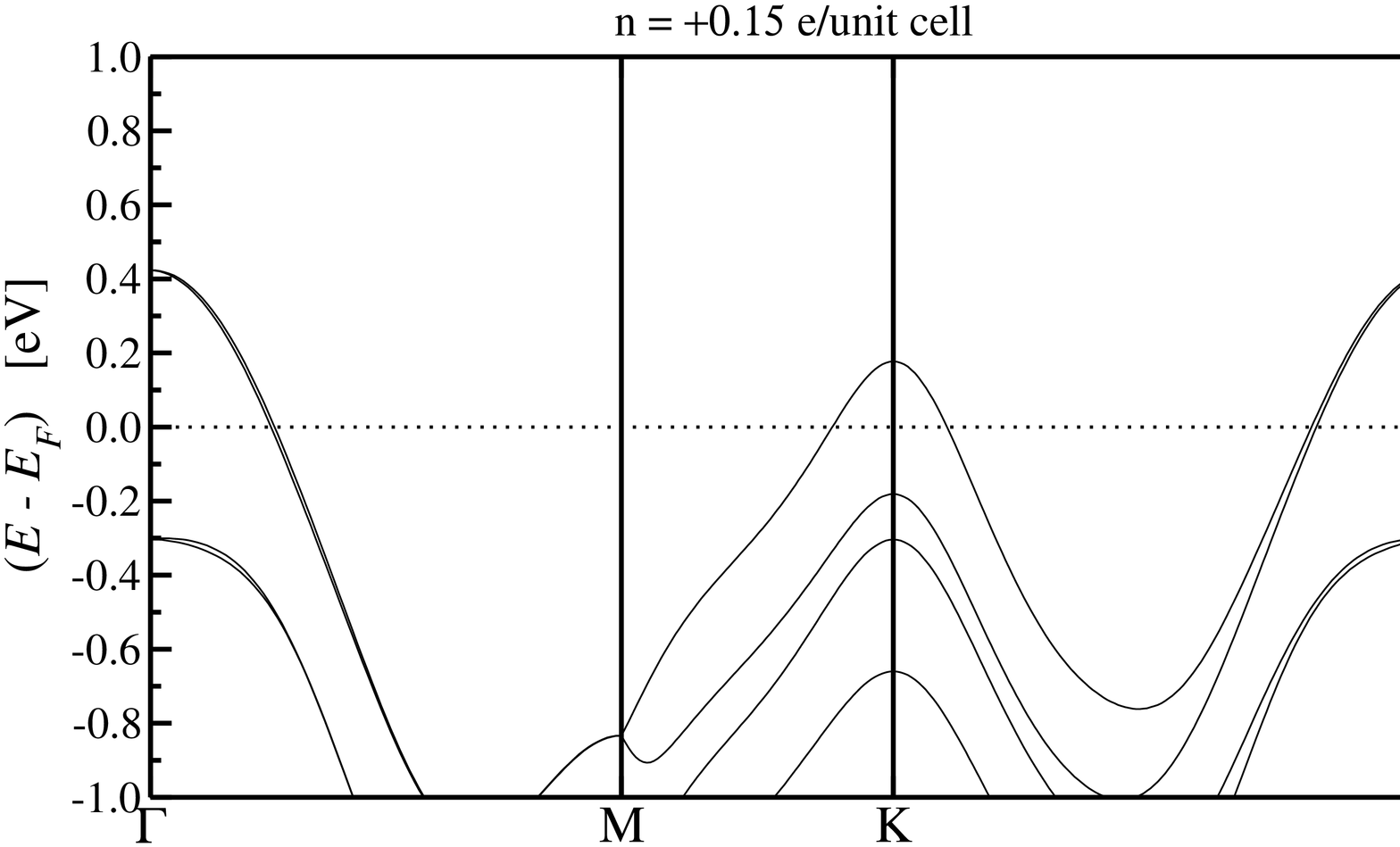}
 \includegraphics[width=0.31\textwidth,clip=,angle=-90]{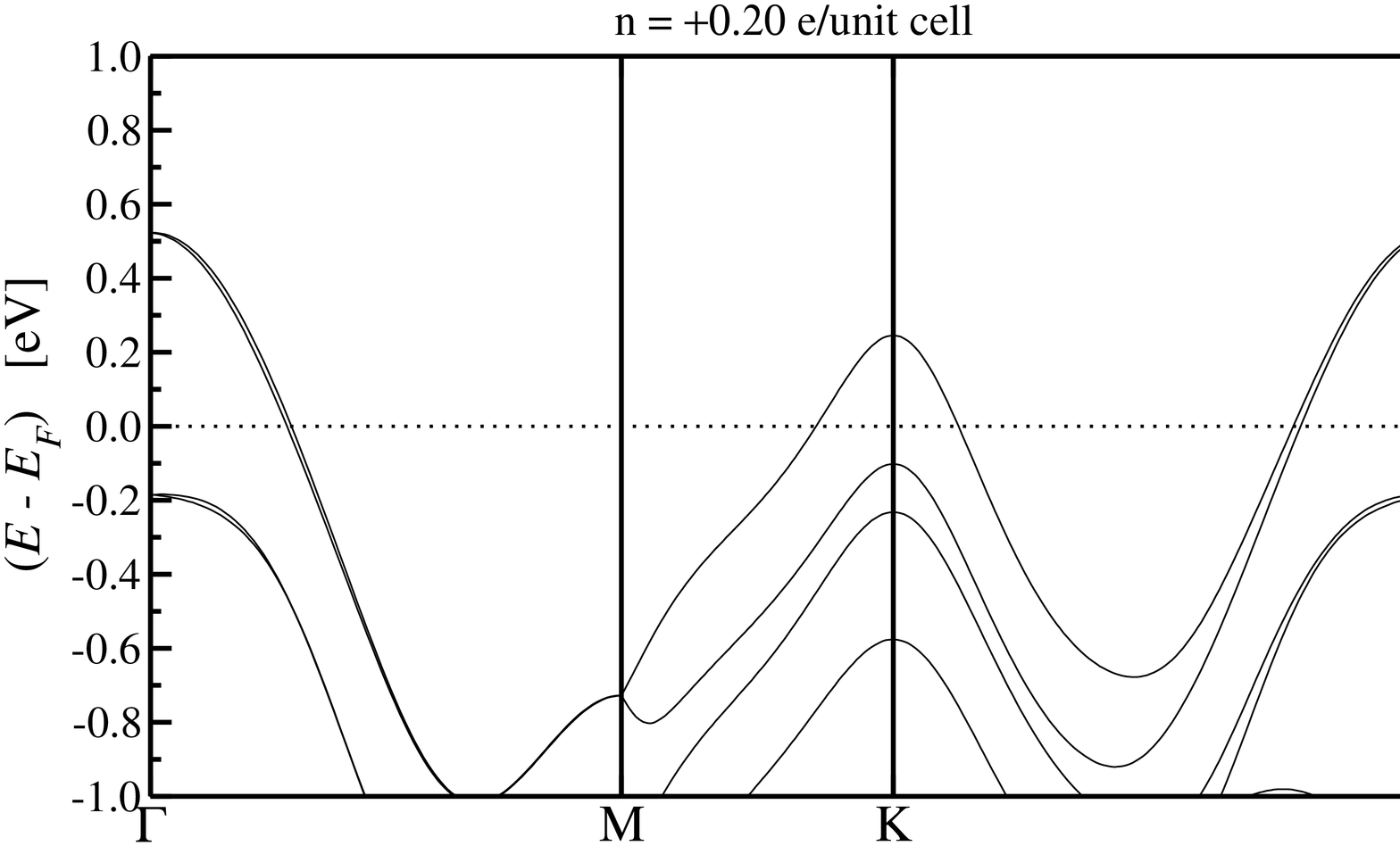}
 \includegraphics[width=0.31\textwidth,clip=,angle=-90]{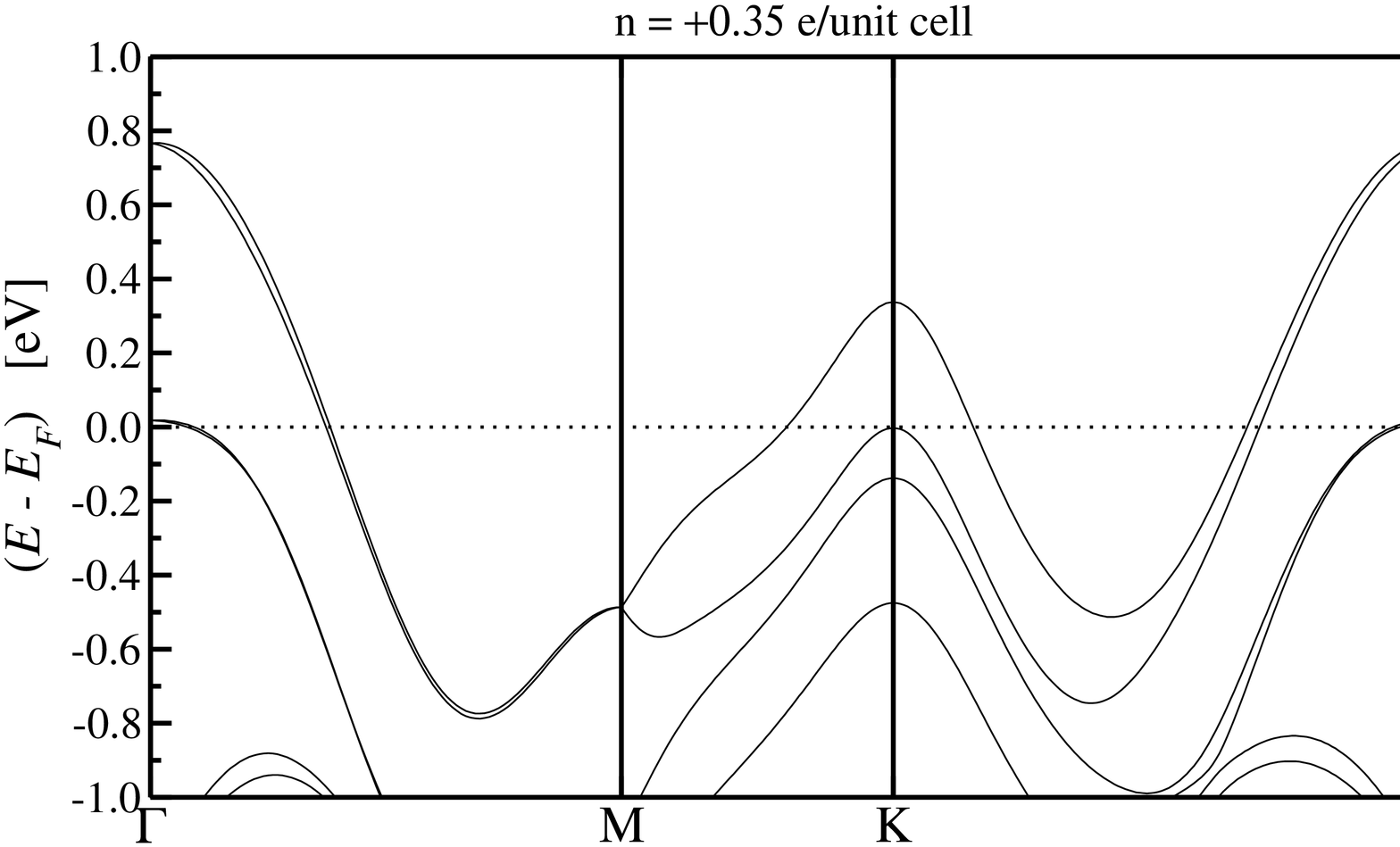}
 \caption{Band structure of bilayer WS$_2$ for different doping as indicated in the labels.}
\end{figure*}
\begin{figure*}[hbp]
 \centering
 \includegraphics[width=0.31\textwidth,clip=,angle=-90]{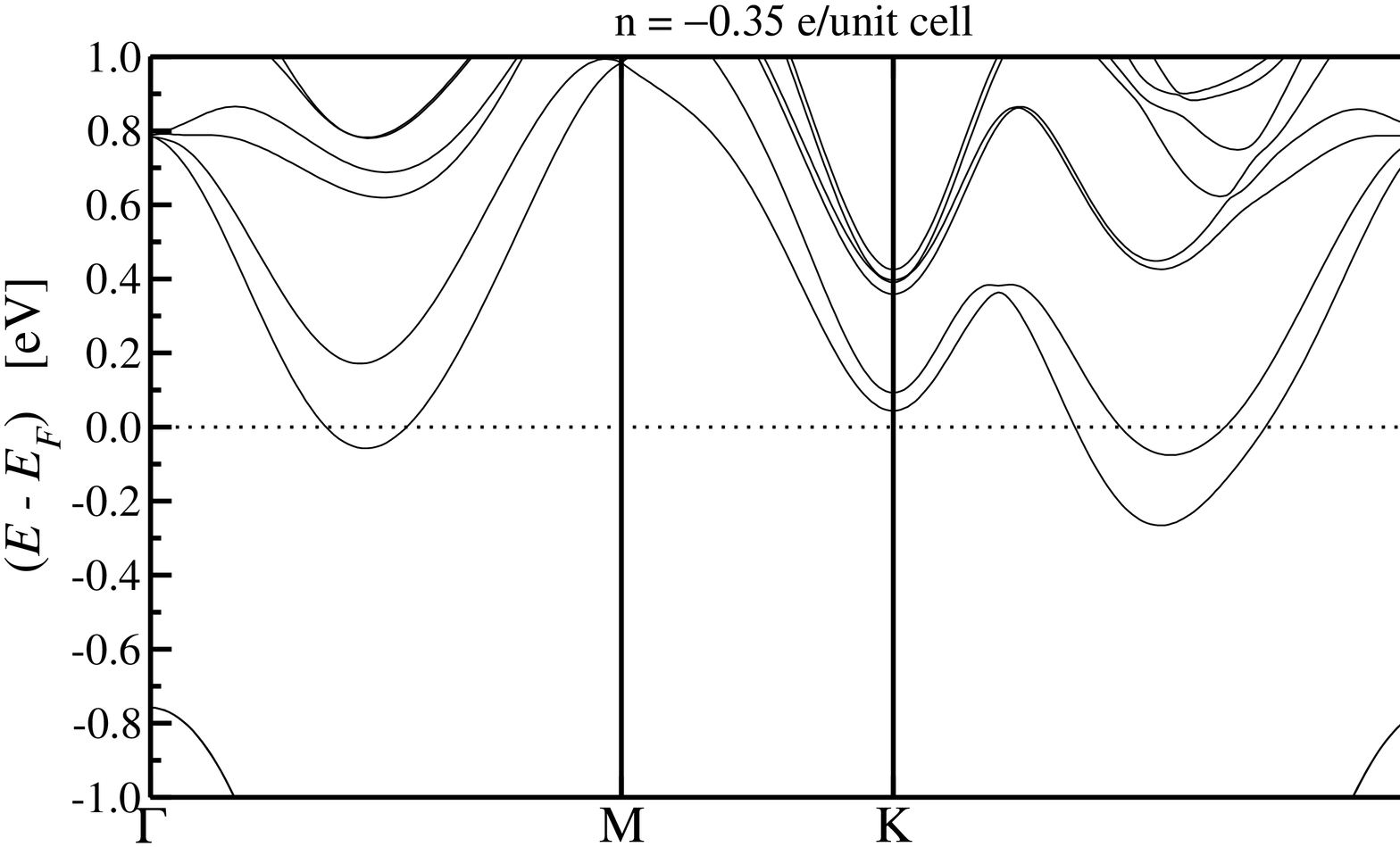}
 \includegraphics[width=0.31\textwidth,clip=,angle=-90]{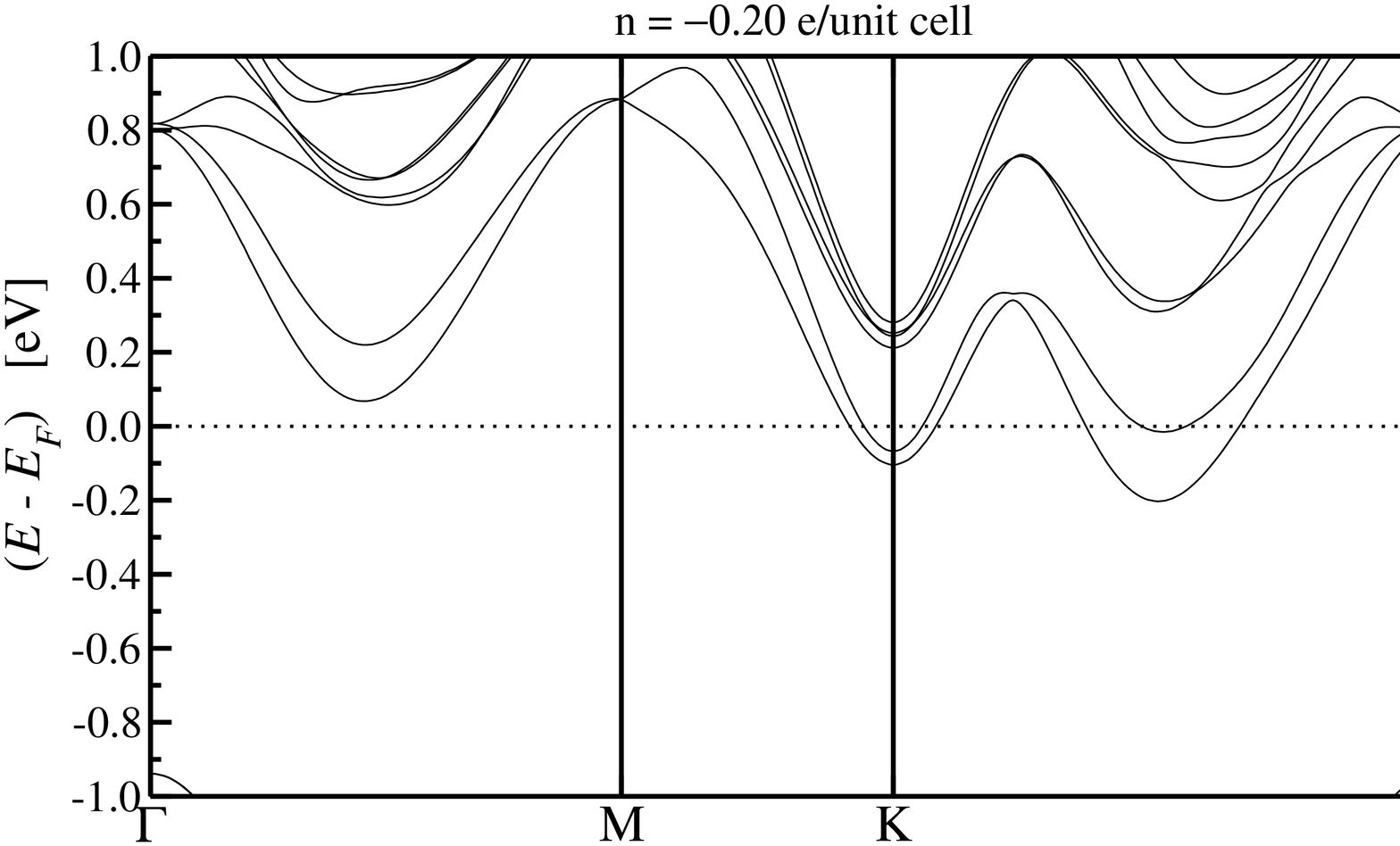}
 \includegraphics[width=0.31\textwidth,clip=,angle=-90]{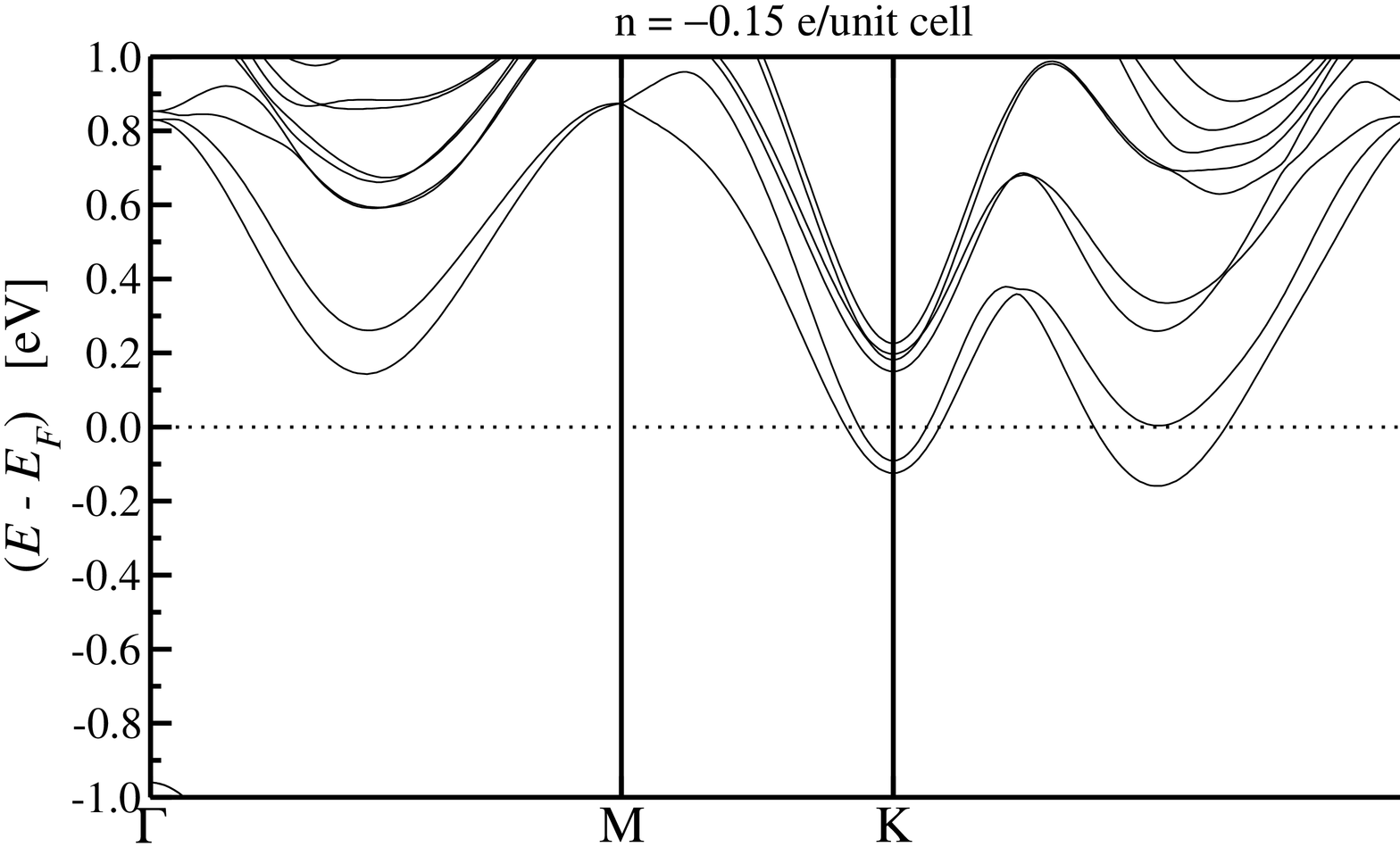}
 \includegraphics[width=0.31\textwidth,clip=,angle=-90]{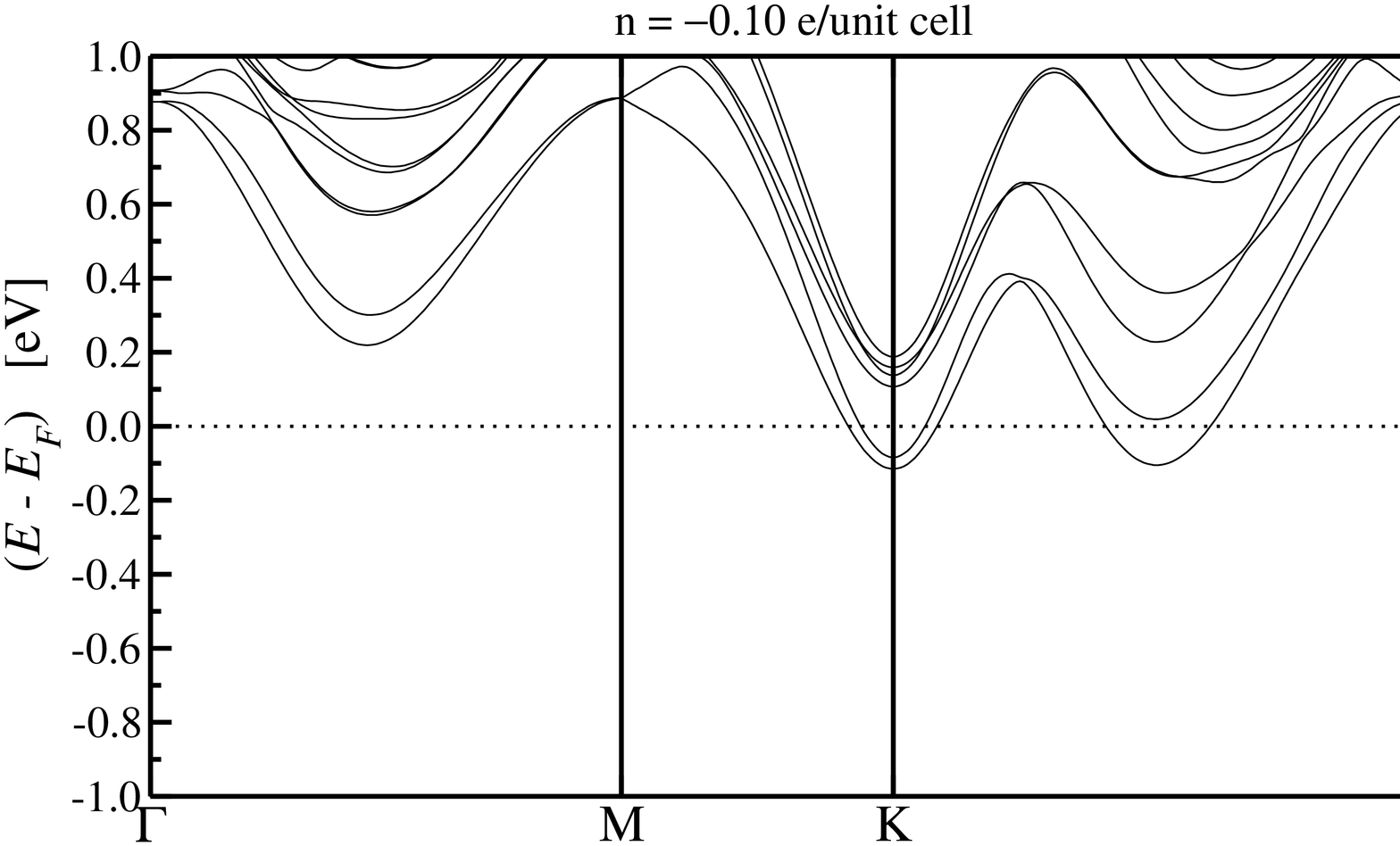}
 \includegraphics[width=0.31\textwidth,clip=,angle=-90]{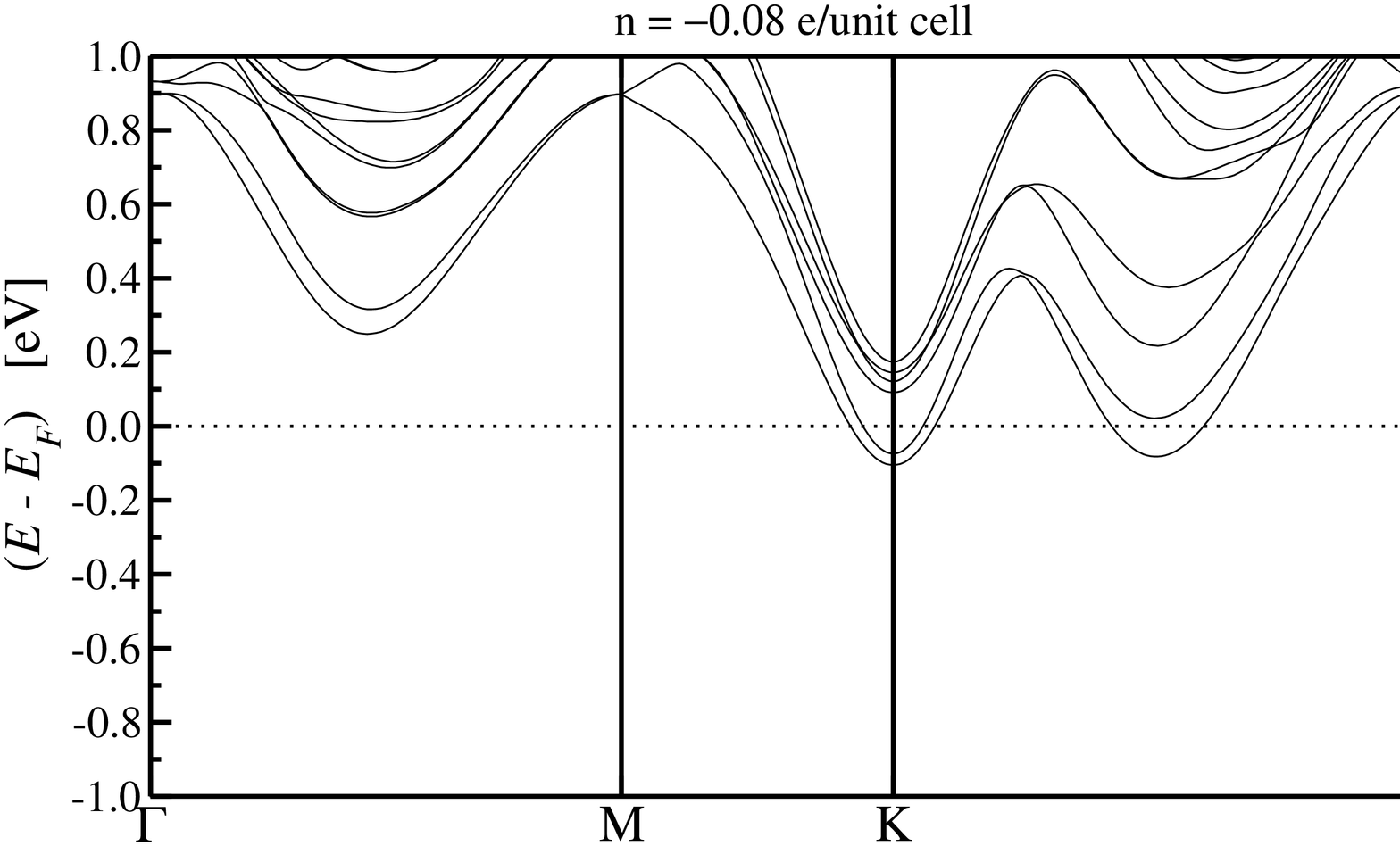}
 \includegraphics[width=0.31\textwidth,clip=,angle=-90]{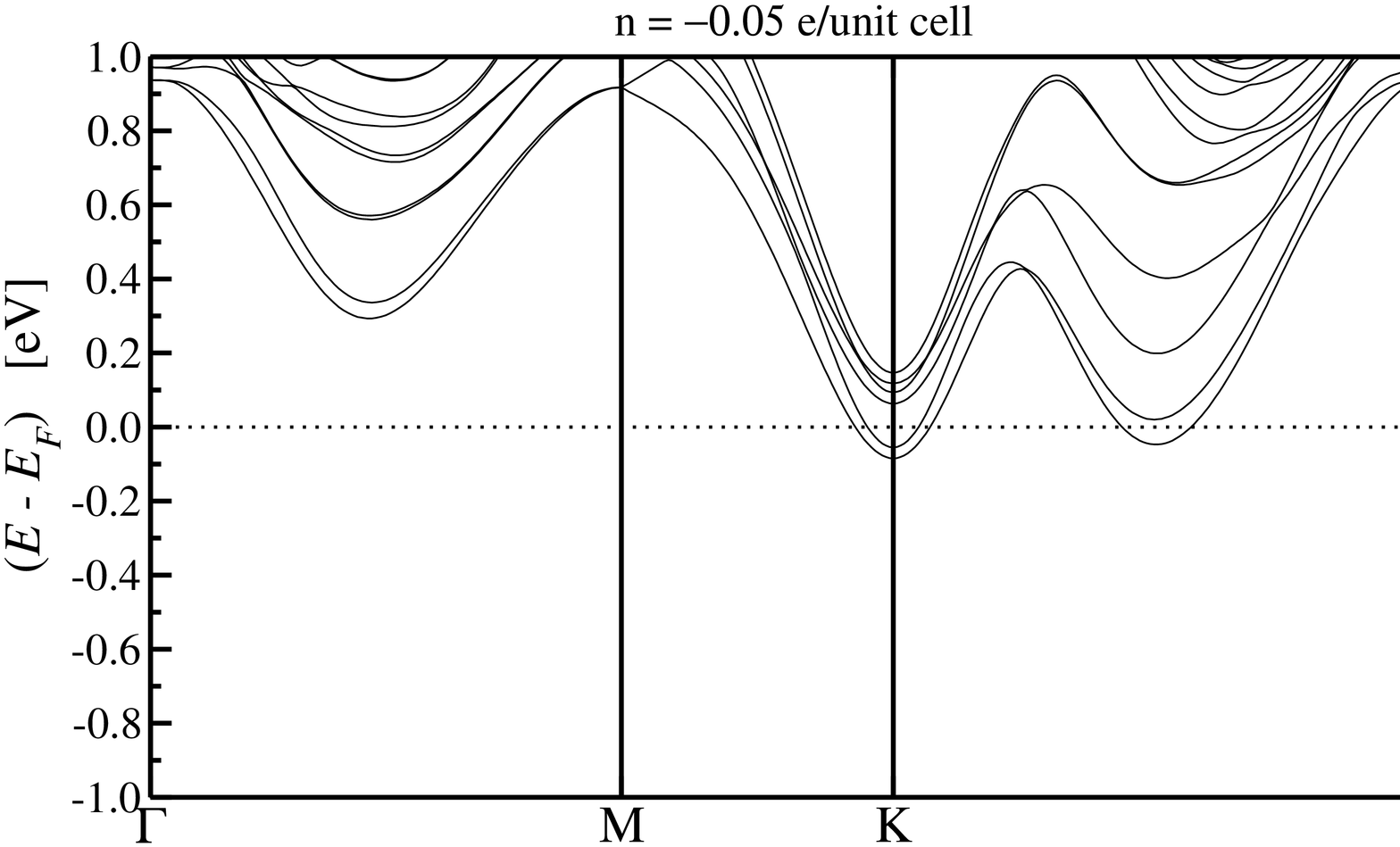}
 \includegraphics[width=0.31\textwidth,clip=,angle=-90]{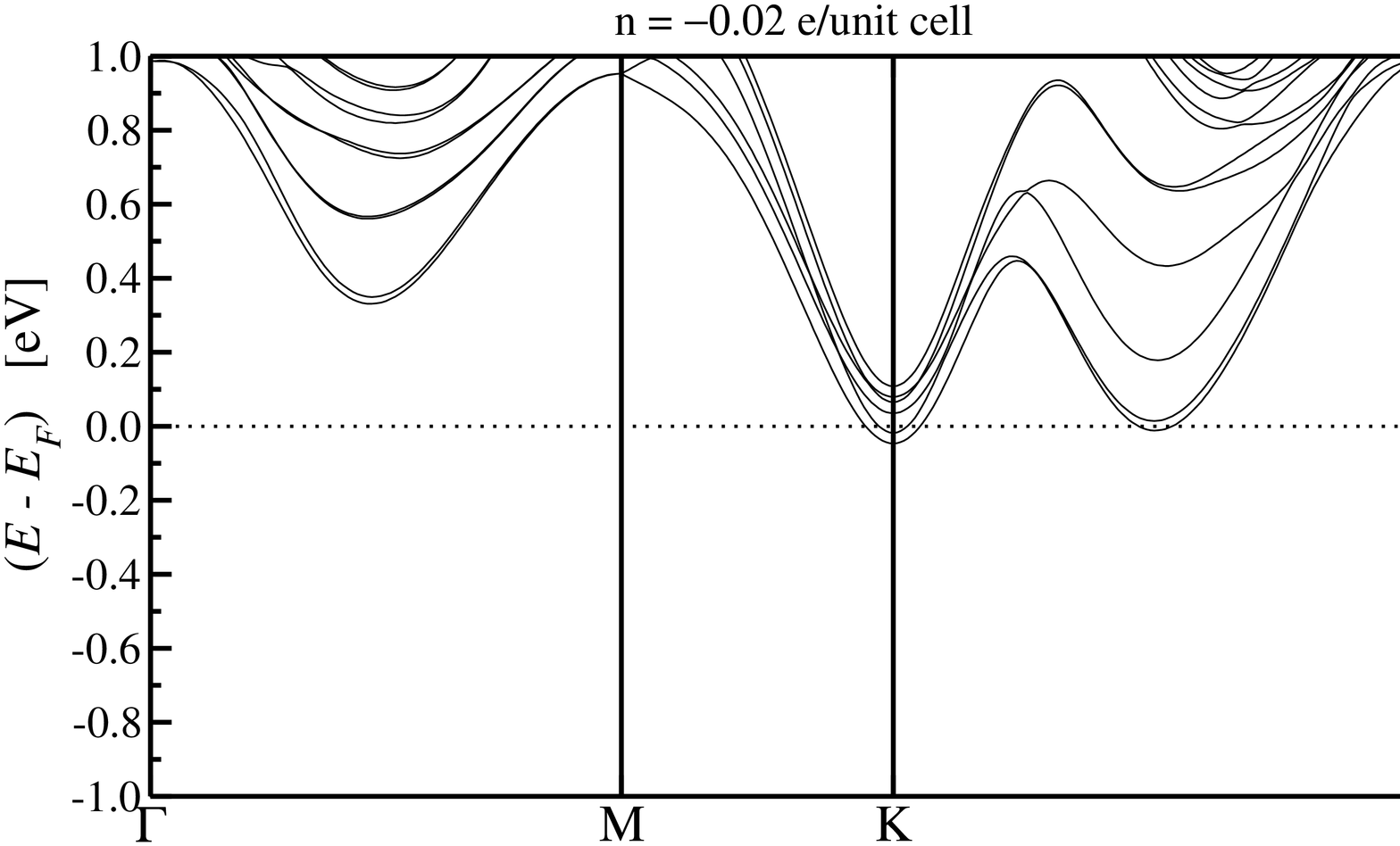}
 \includegraphics[width=0.31\textwidth,clip=,angle=-90]{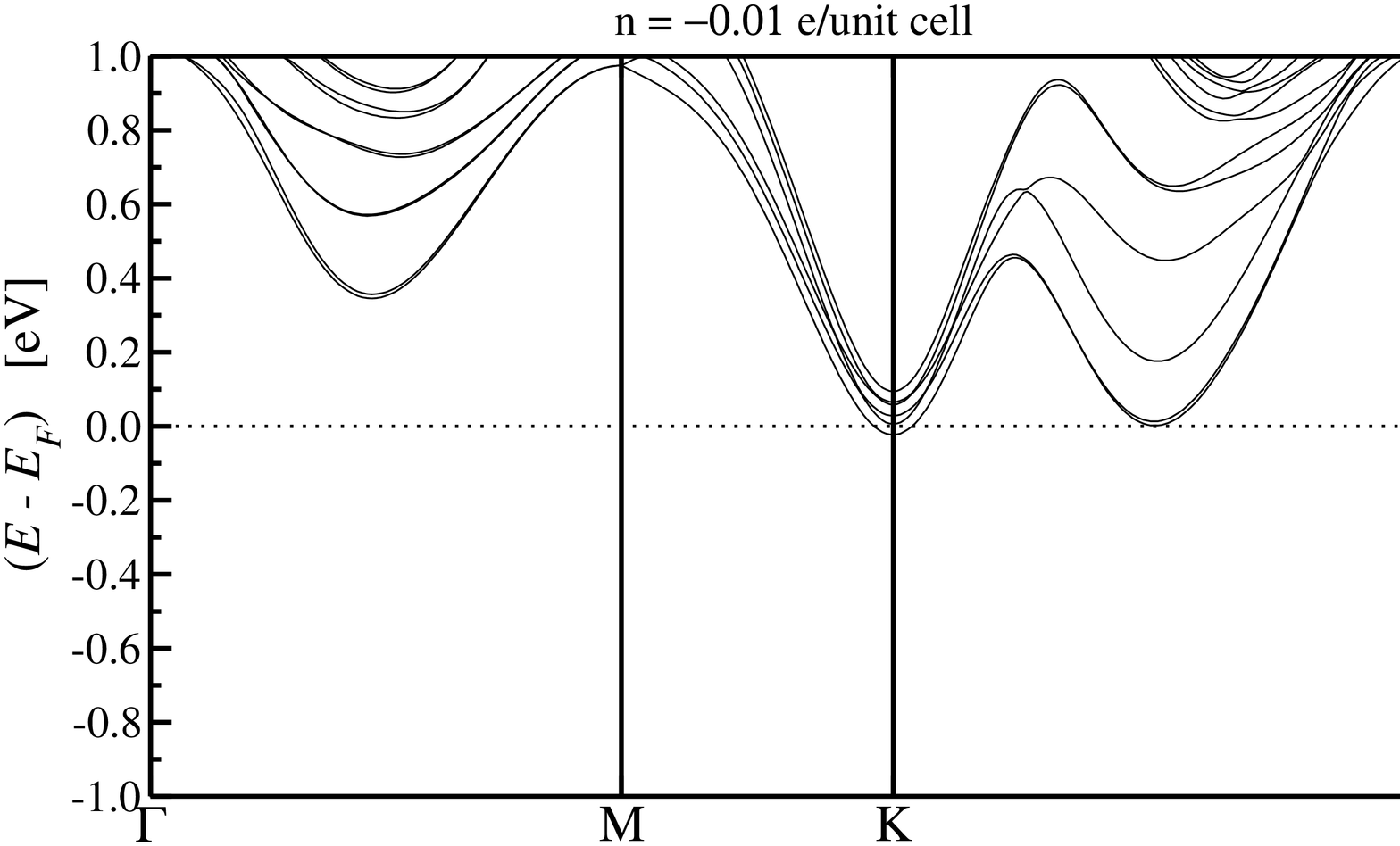}
 \caption{Band structure of trilayer WS$_2$ for different doping as indicated in the labels.}
\end{figure*}
\begin{figure*}[hbp]
 \centering
 \includegraphics[width=0.31\textwidth,clip=,angle=-90]{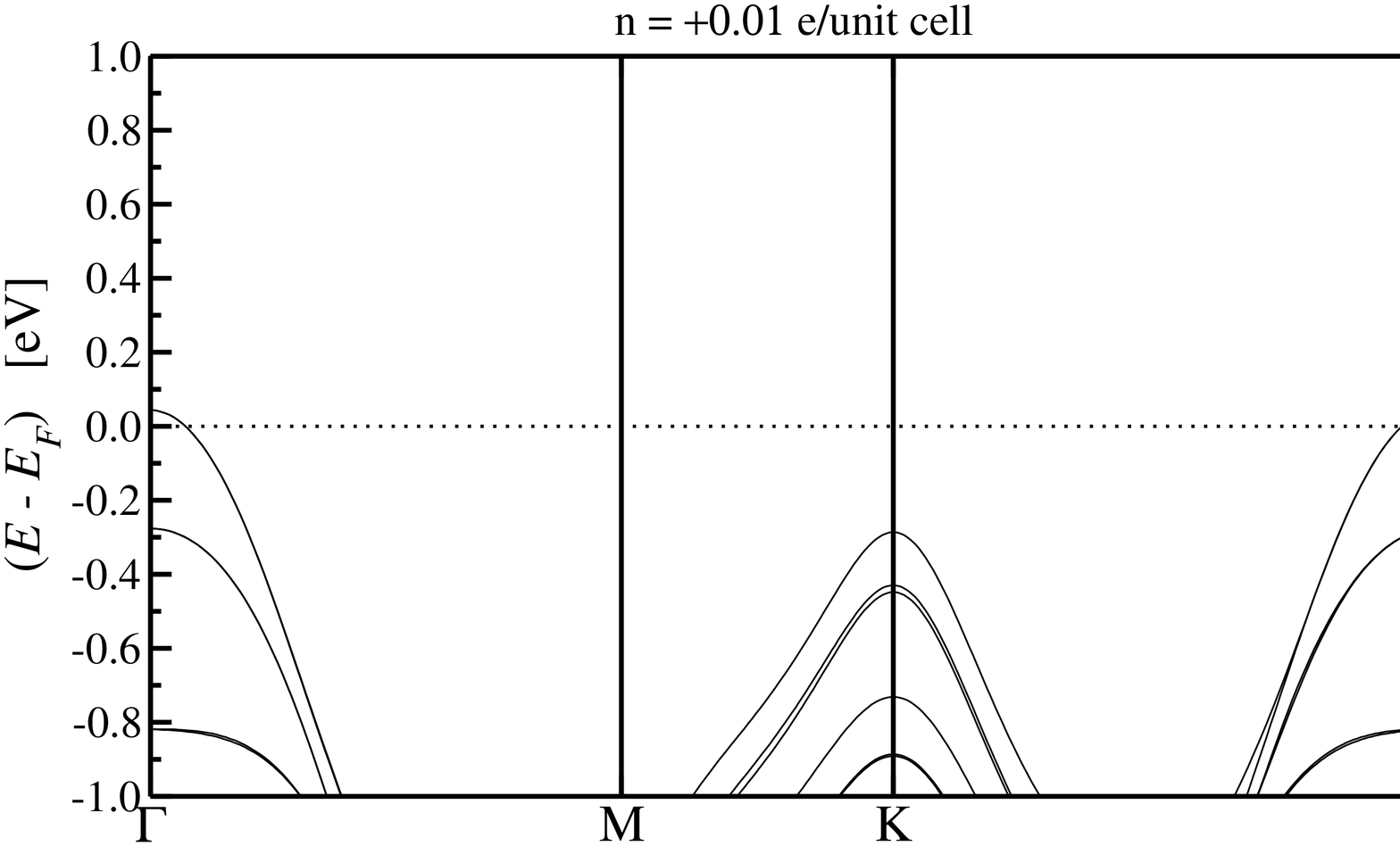}
 \includegraphics[width=0.31\textwidth,clip=,angle=-90]{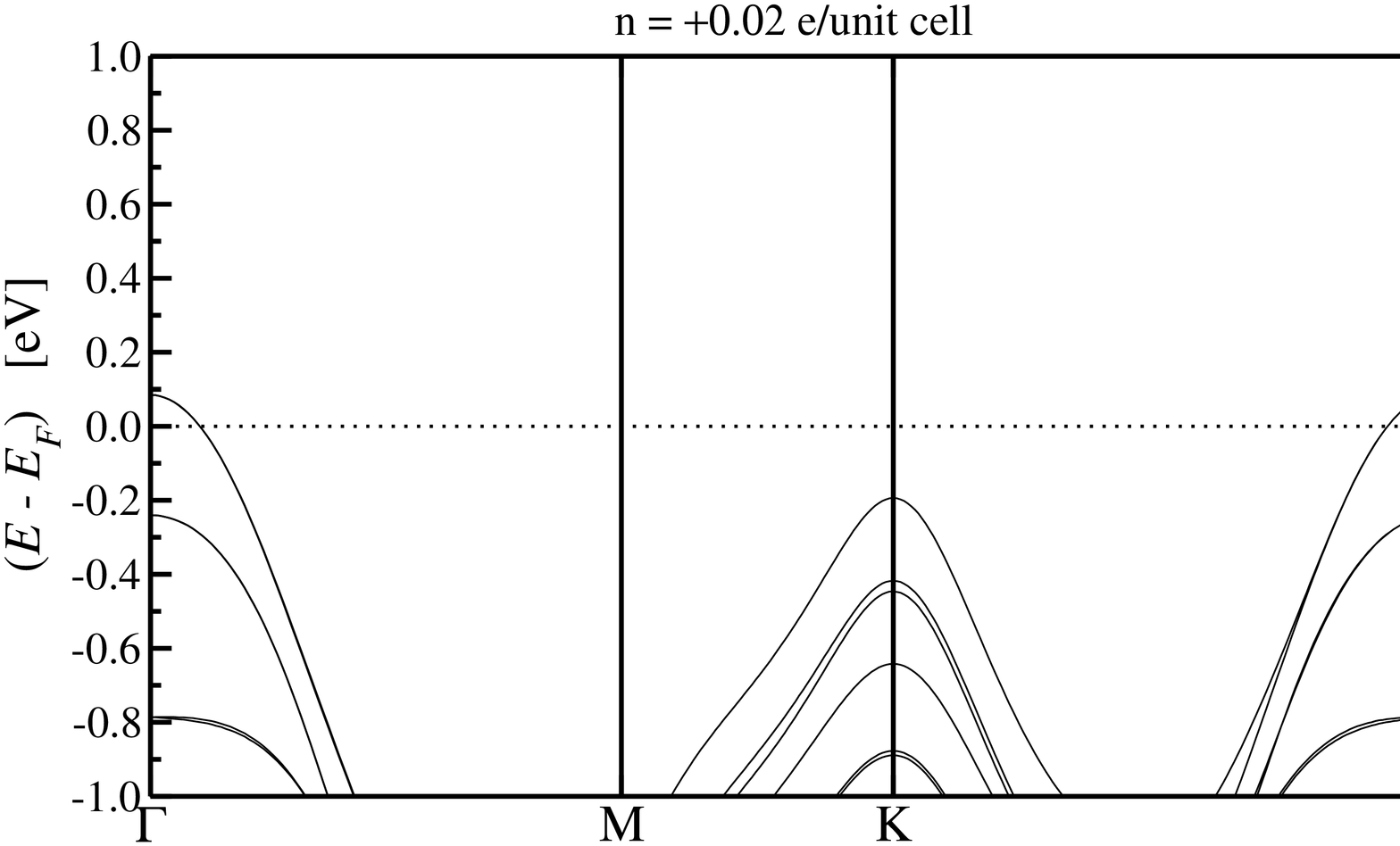}
 \includegraphics[width=0.31\textwidth,clip=,angle=-90]{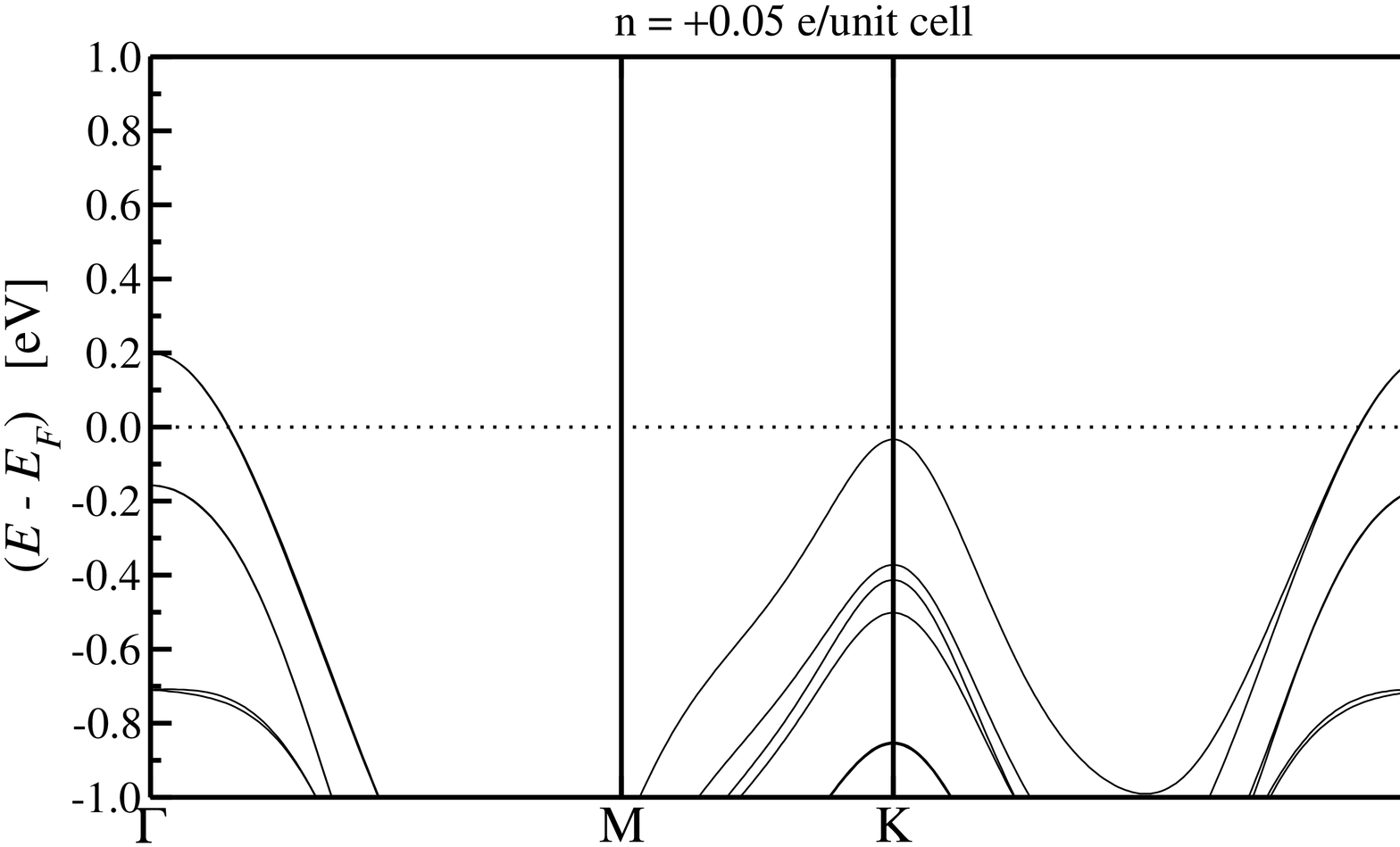}
 \includegraphics[width=0.31\textwidth,clip=,angle=-90]{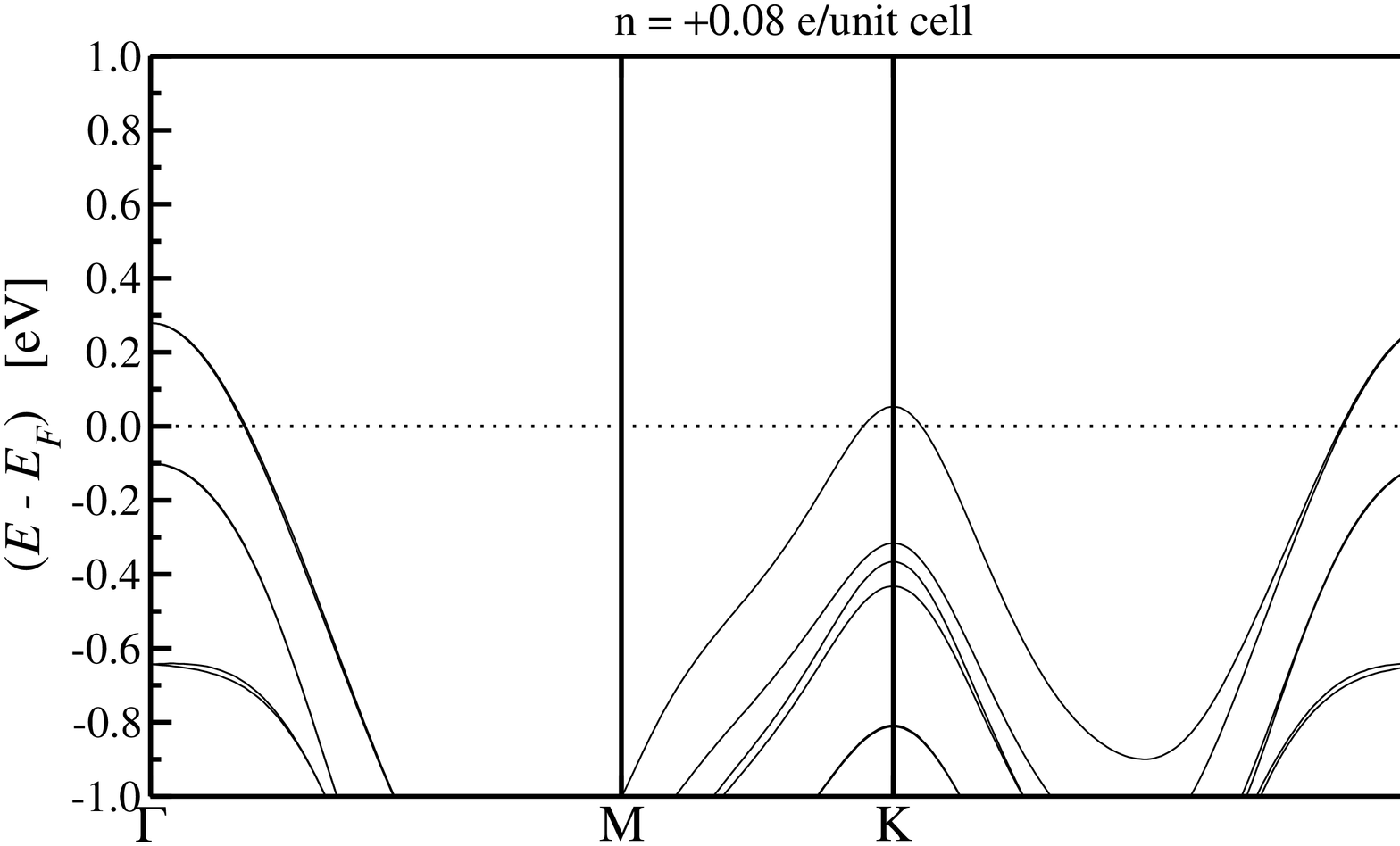}
 \includegraphics[width=0.31\textwidth,clip=,angle=-90]{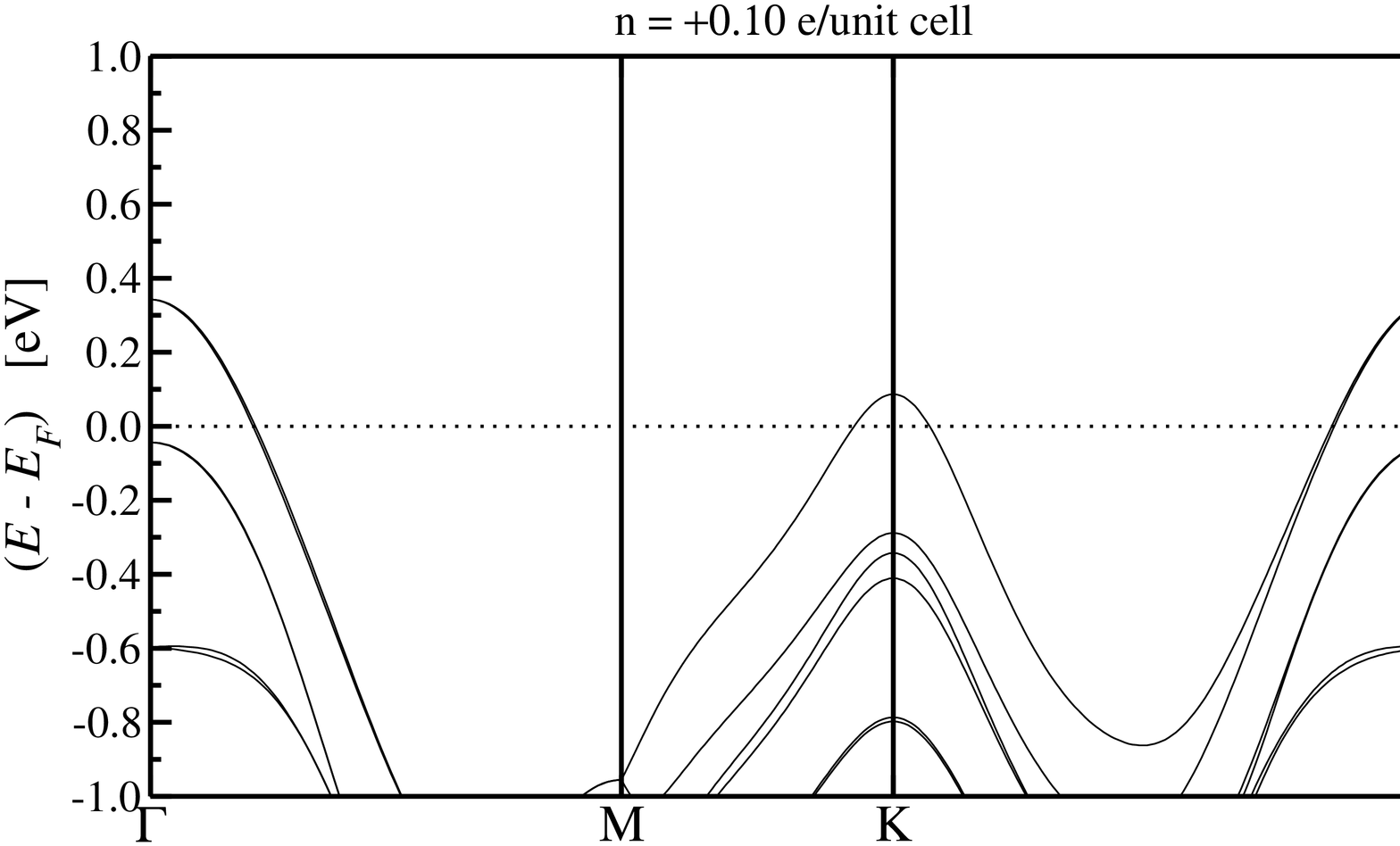}
 \includegraphics[width=0.31\textwidth,clip=,angle=-90]{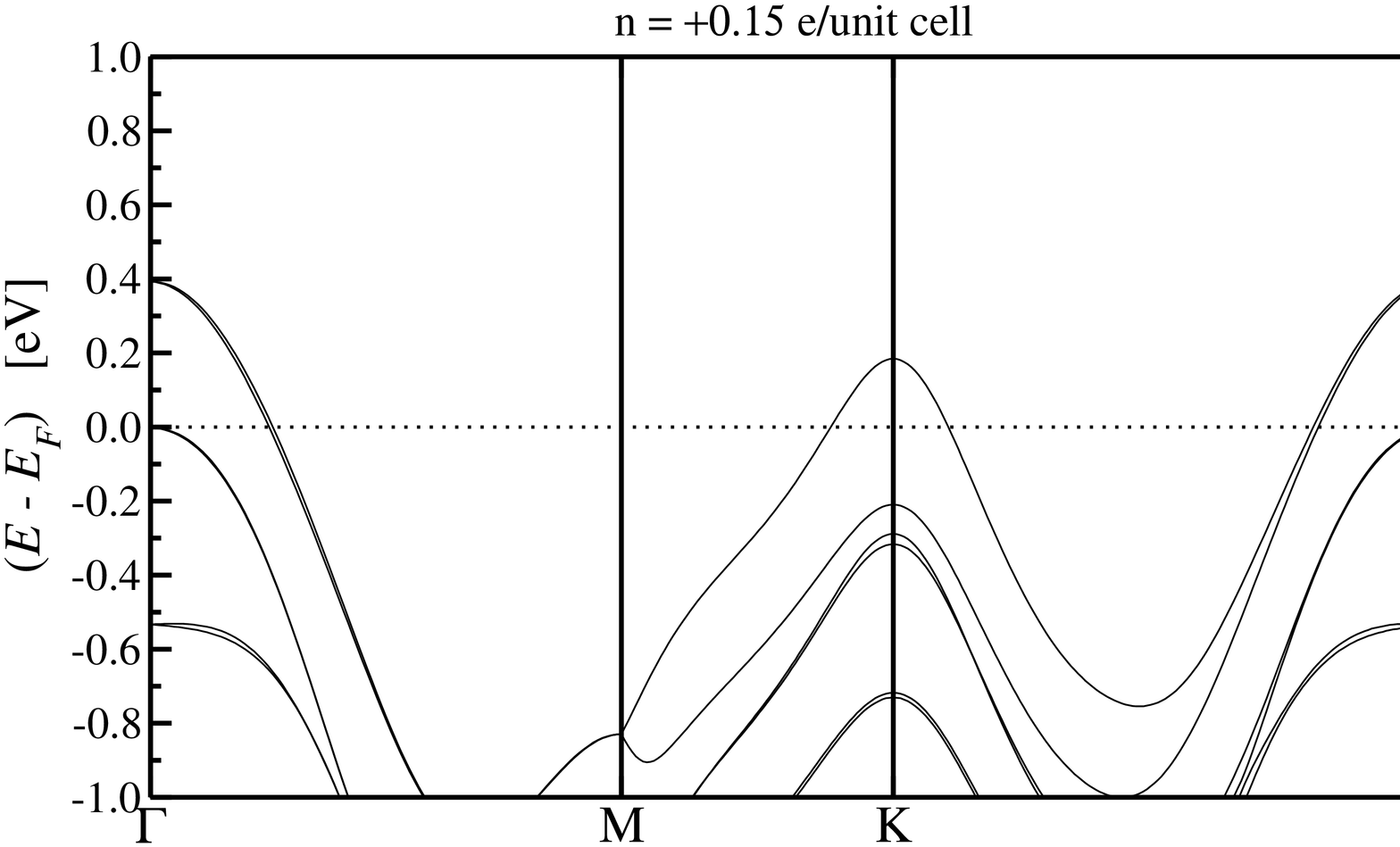}
 \includegraphics[width=0.31\textwidth,clip=,angle=-90]{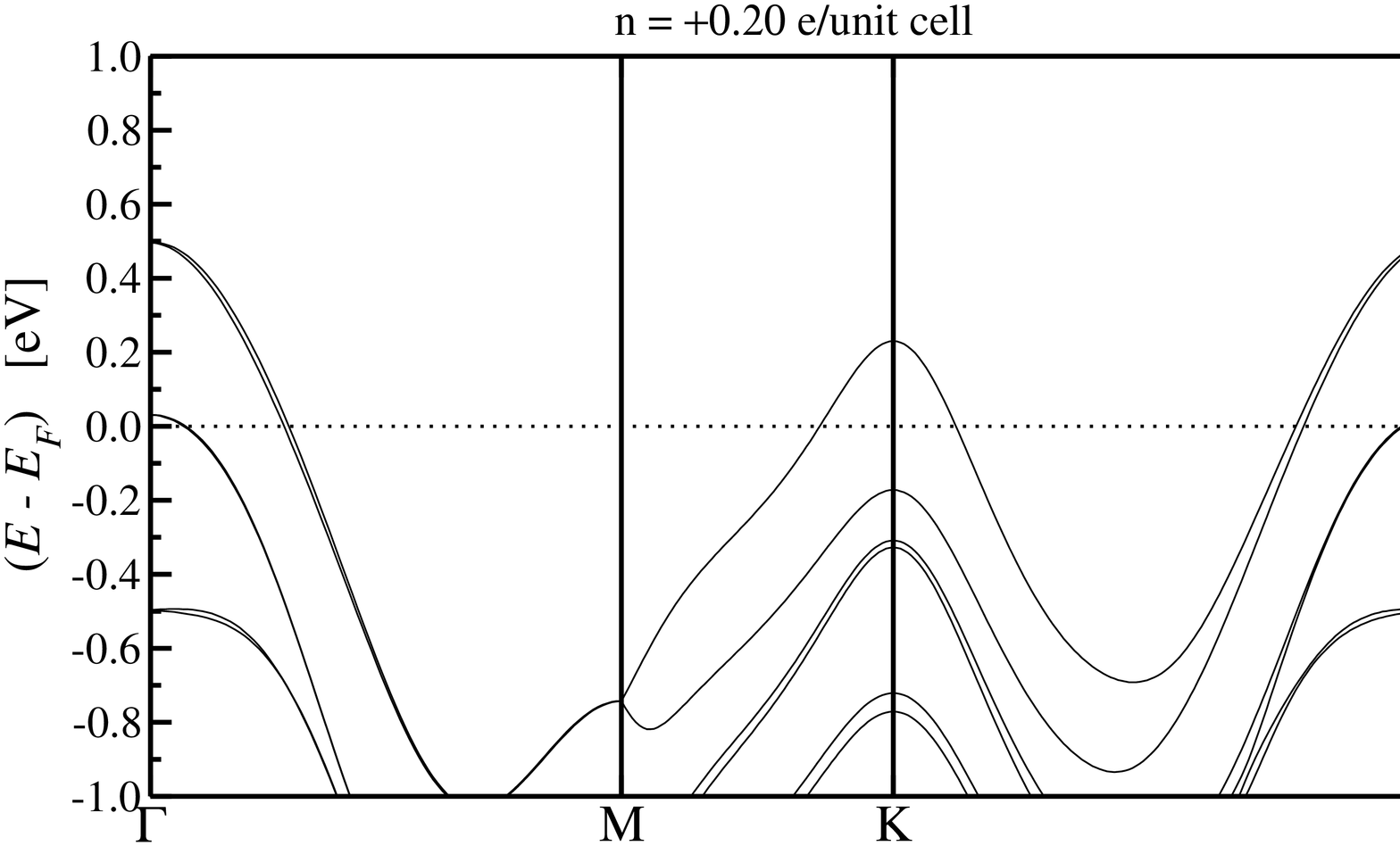}
 \includegraphics[width=0.31\textwidth,clip=,angle=-90]{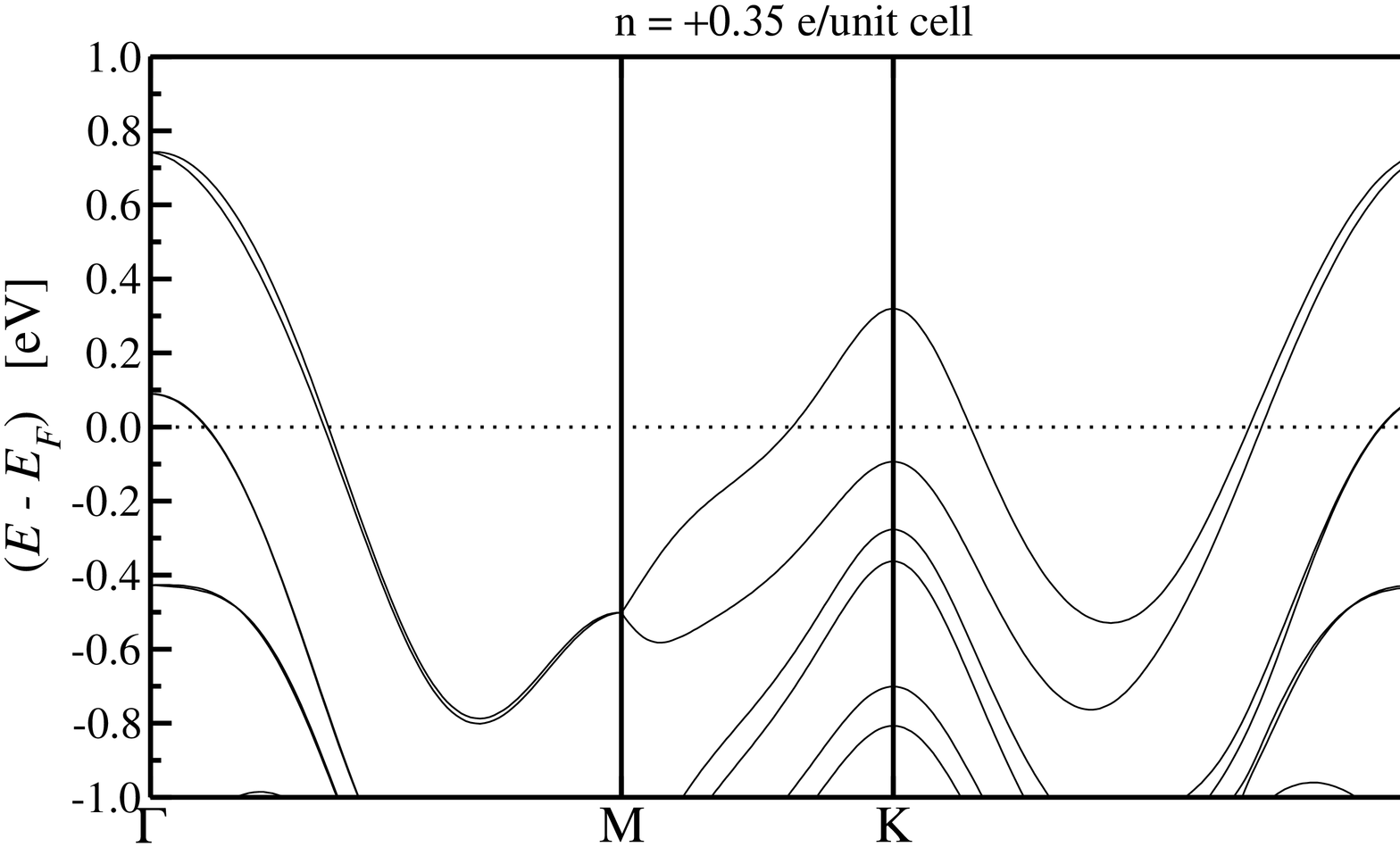}
 \caption{Band structure of trilayer WS$_2$ for different doping as indicated in the labels.}
\end{figure*}

\clearpage
\subsection{Tungsten diselenide}
\begin{figure*}[hbp]
 \centering
 \includegraphics[width=0.31\textwidth,clip=,angle=-90]{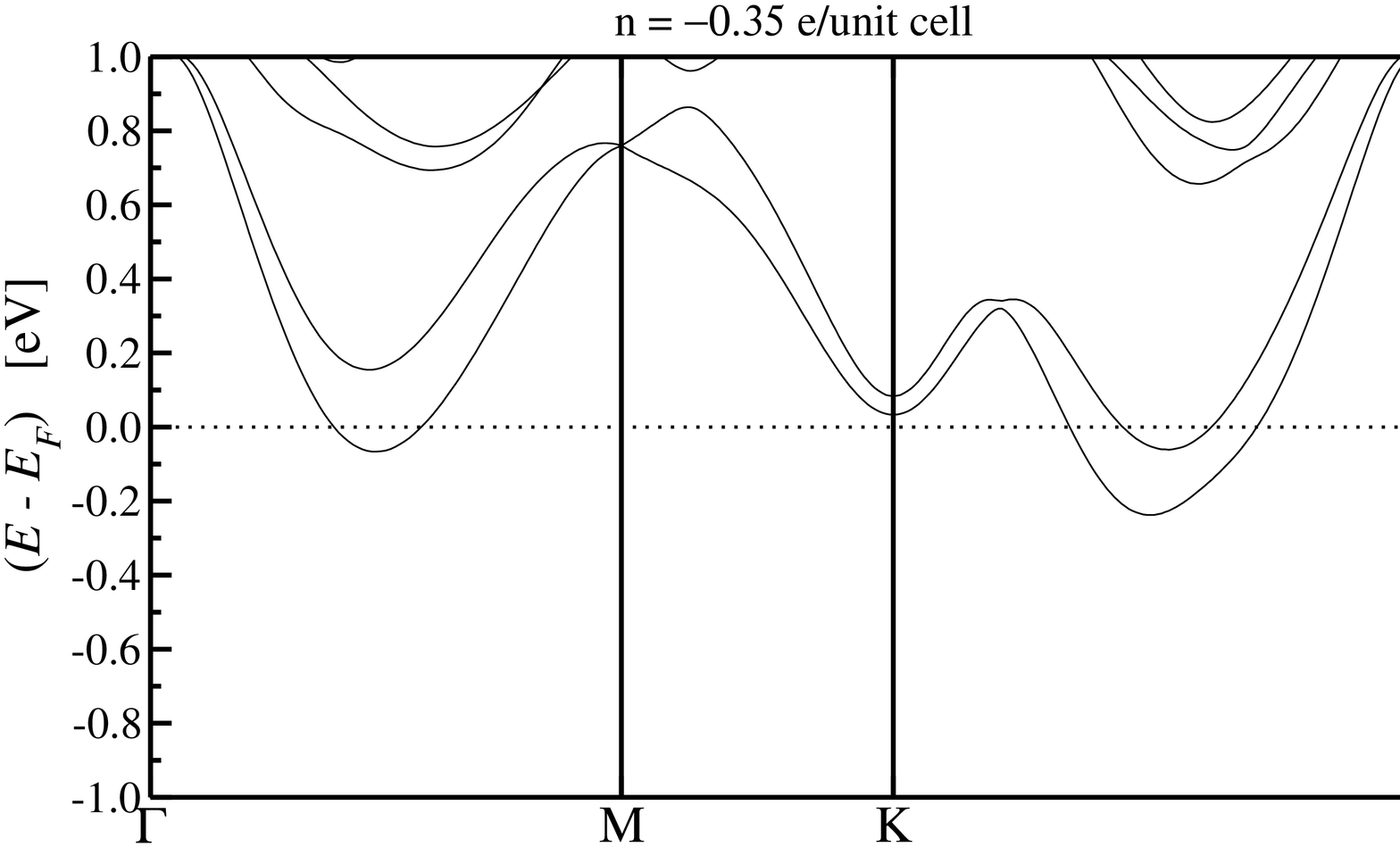}
 \includegraphics[width=0.31\textwidth,clip=,angle=-90]{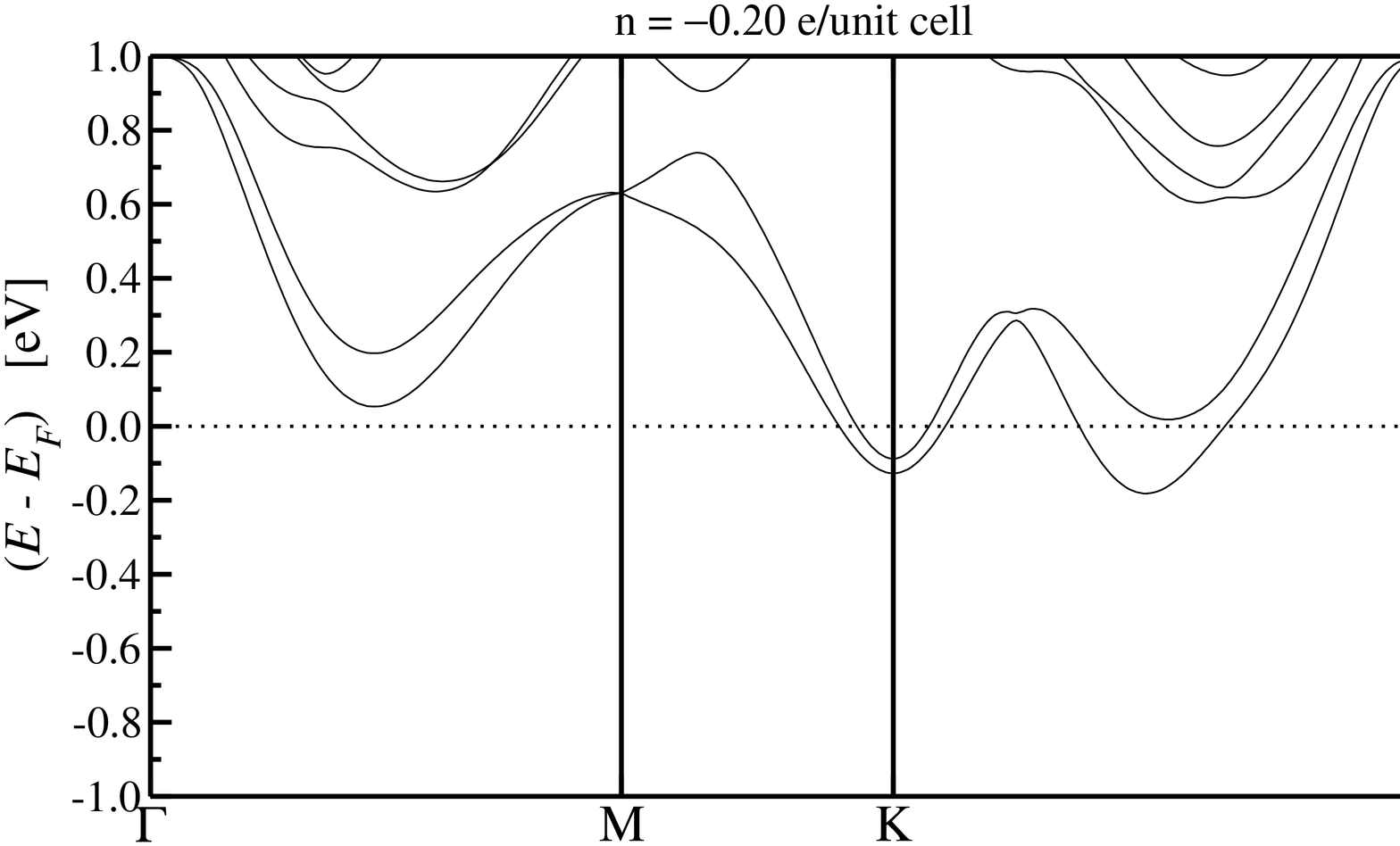}
 \includegraphics[width=0.31\textwidth,clip=,angle=-90]{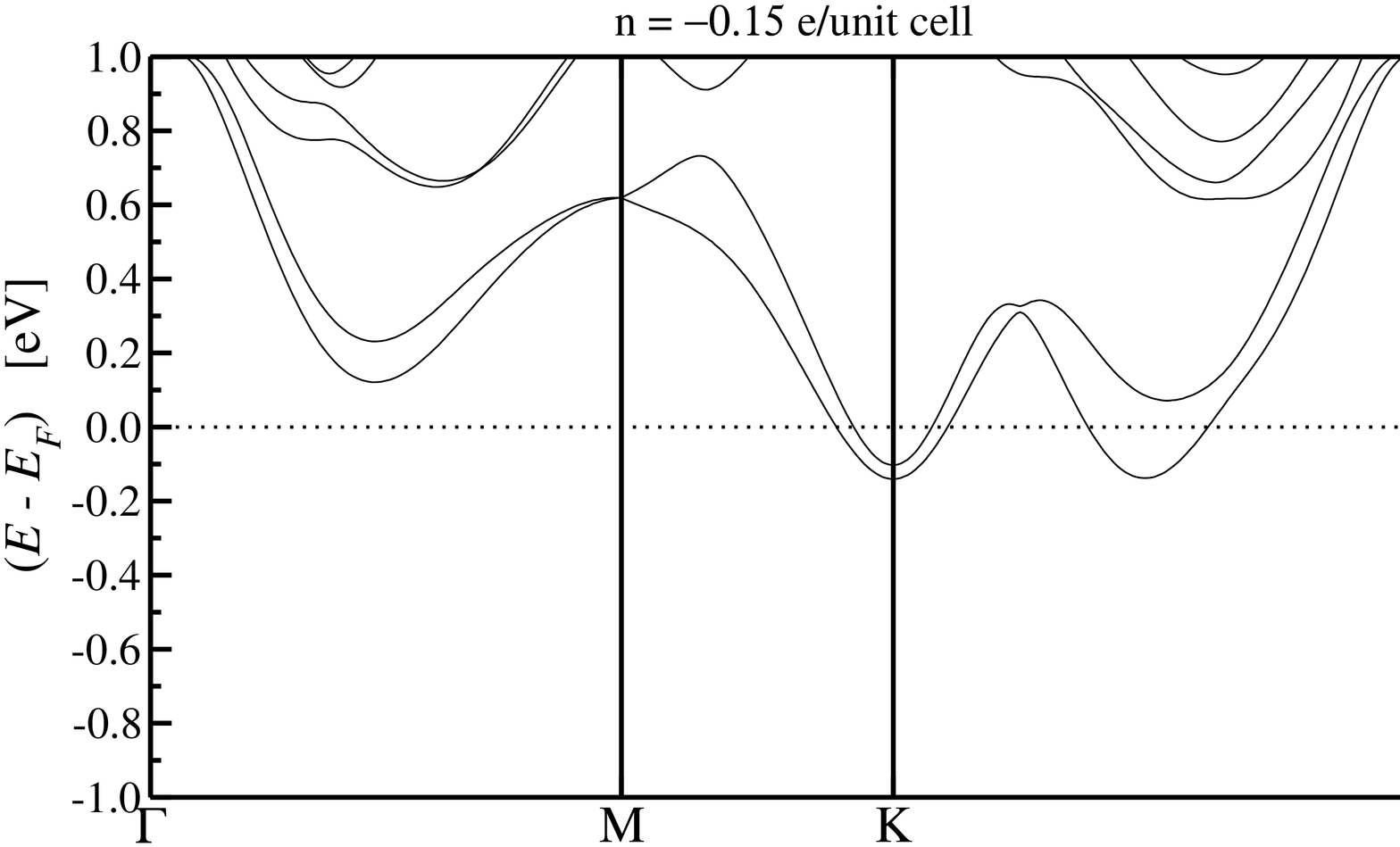}
 \includegraphics[width=0.31\textwidth,clip=,angle=-90]{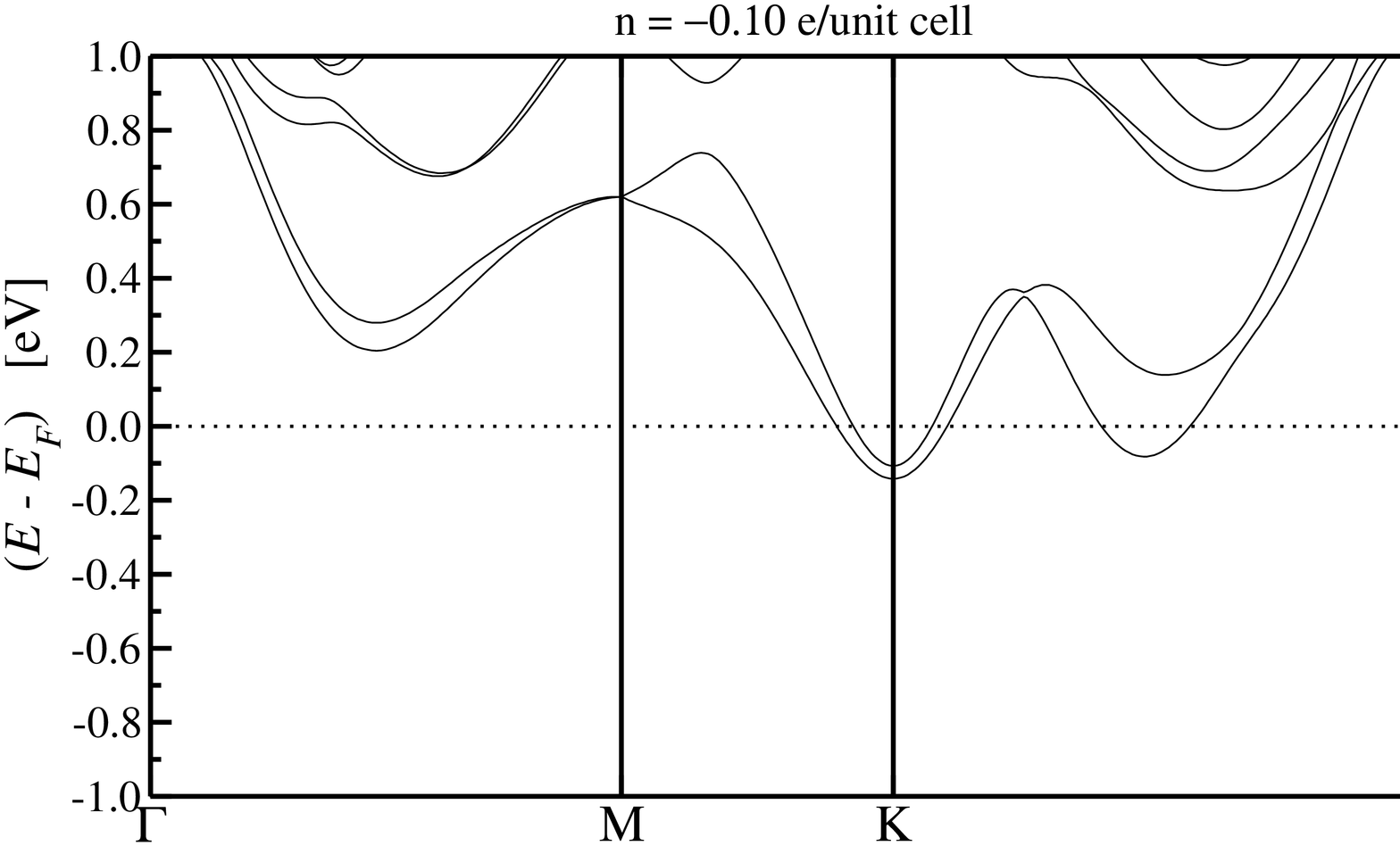}
 \includegraphics[width=0.31\textwidth,clip=,angle=-90]{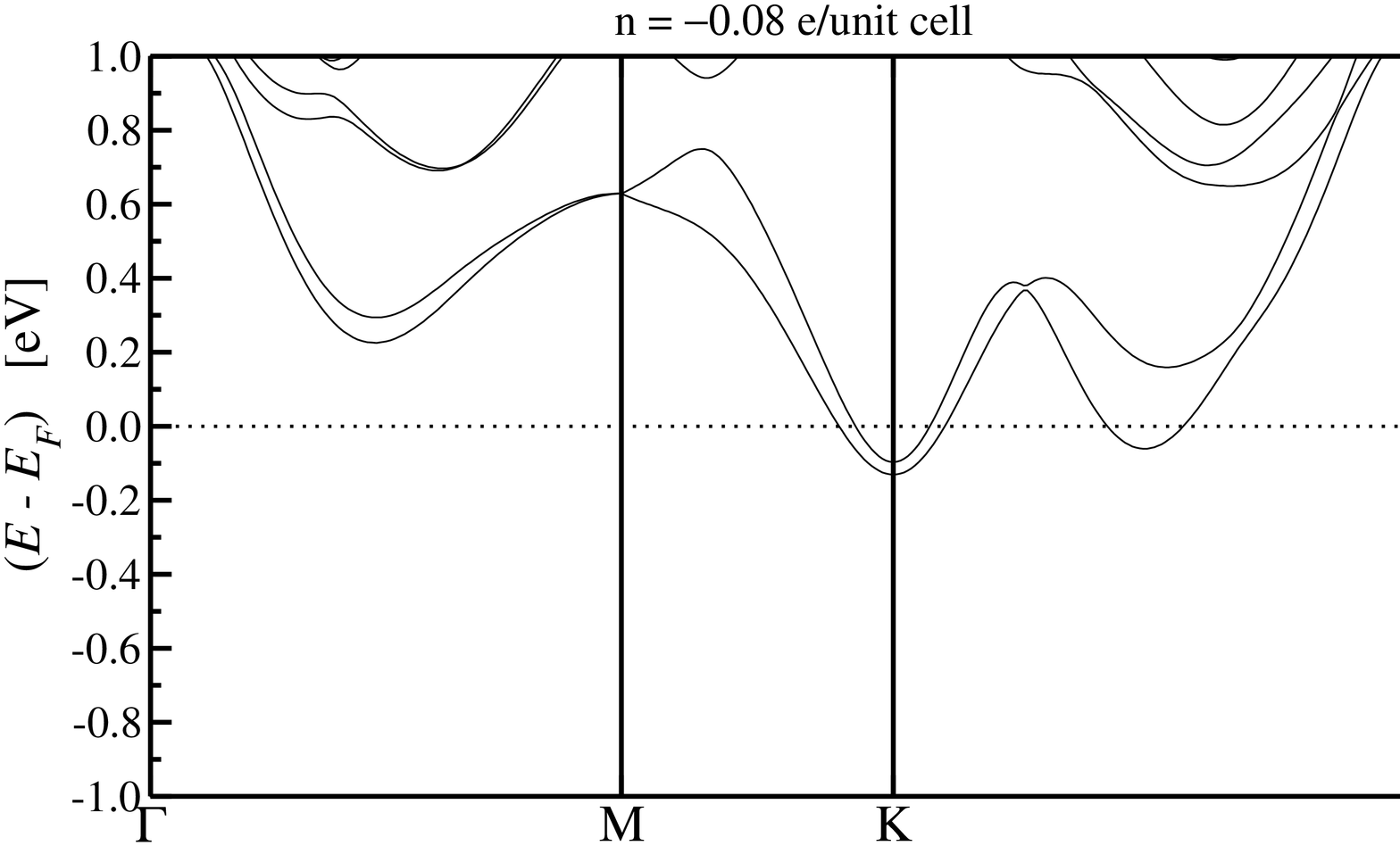}
 \includegraphics[width=0.31\textwidth,clip=,angle=-90]{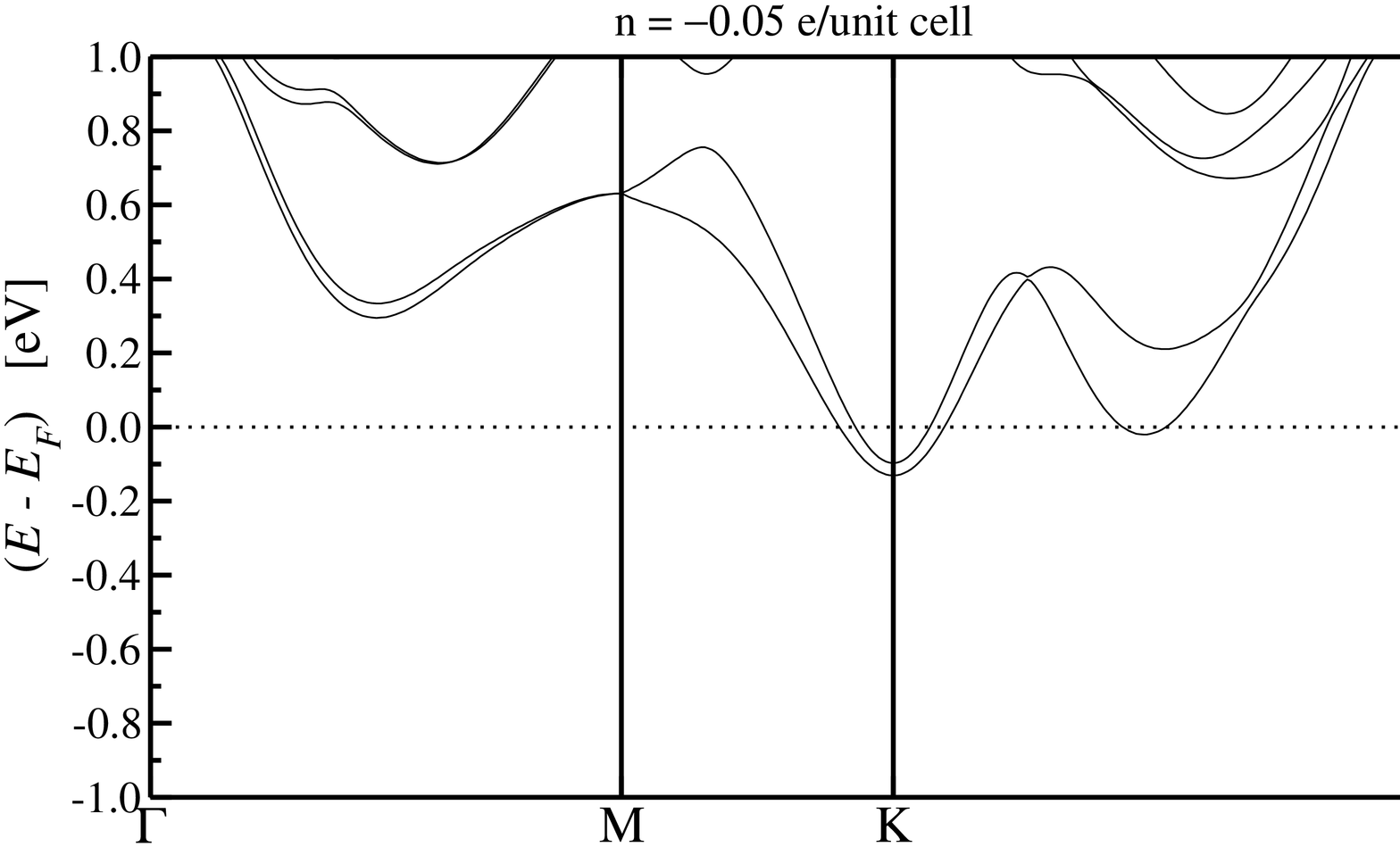}
 \includegraphics[width=0.31\textwidth,clip=,angle=-90]{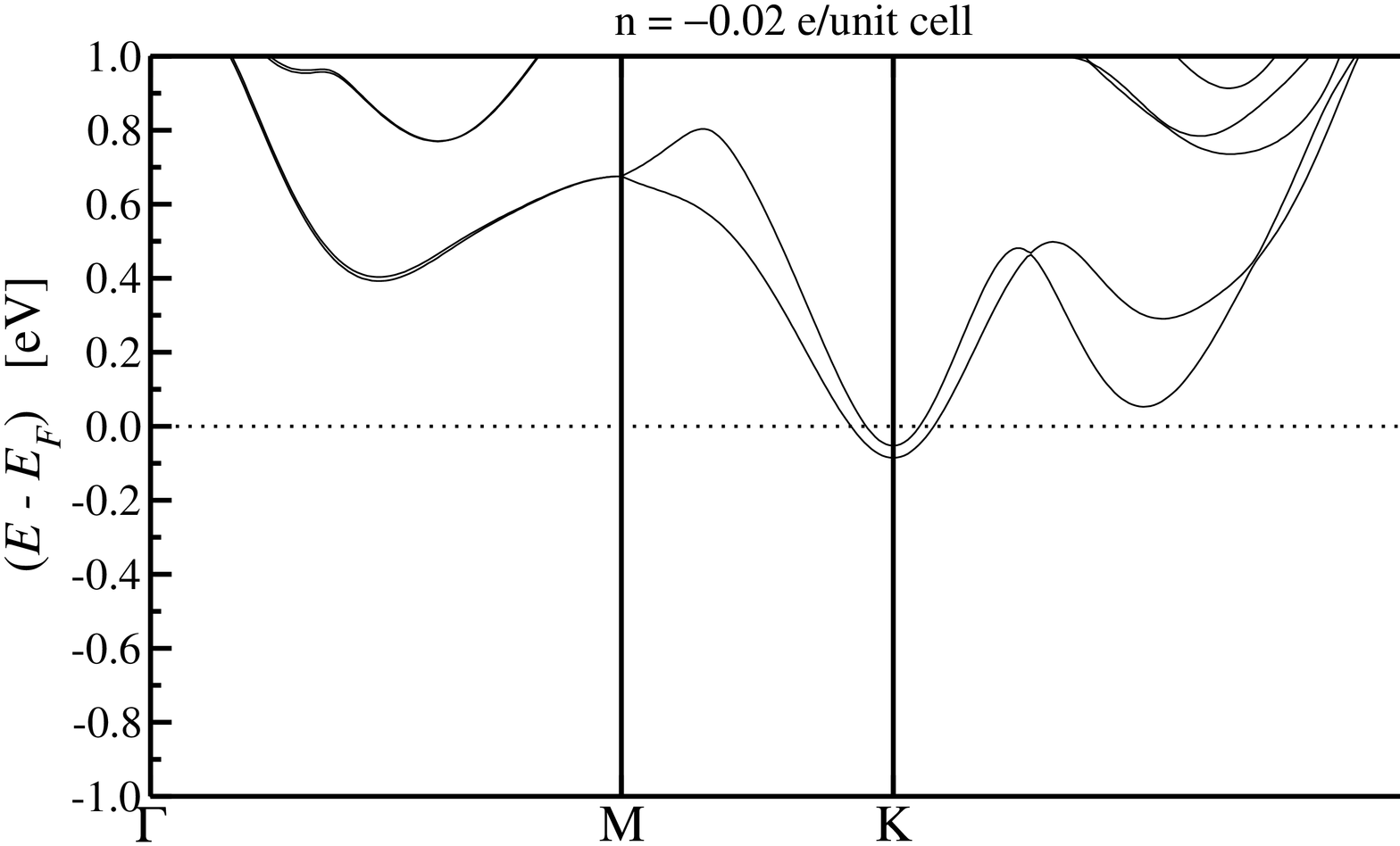}
 \includegraphics[width=0.31\textwidth,clip=,angle=-90]{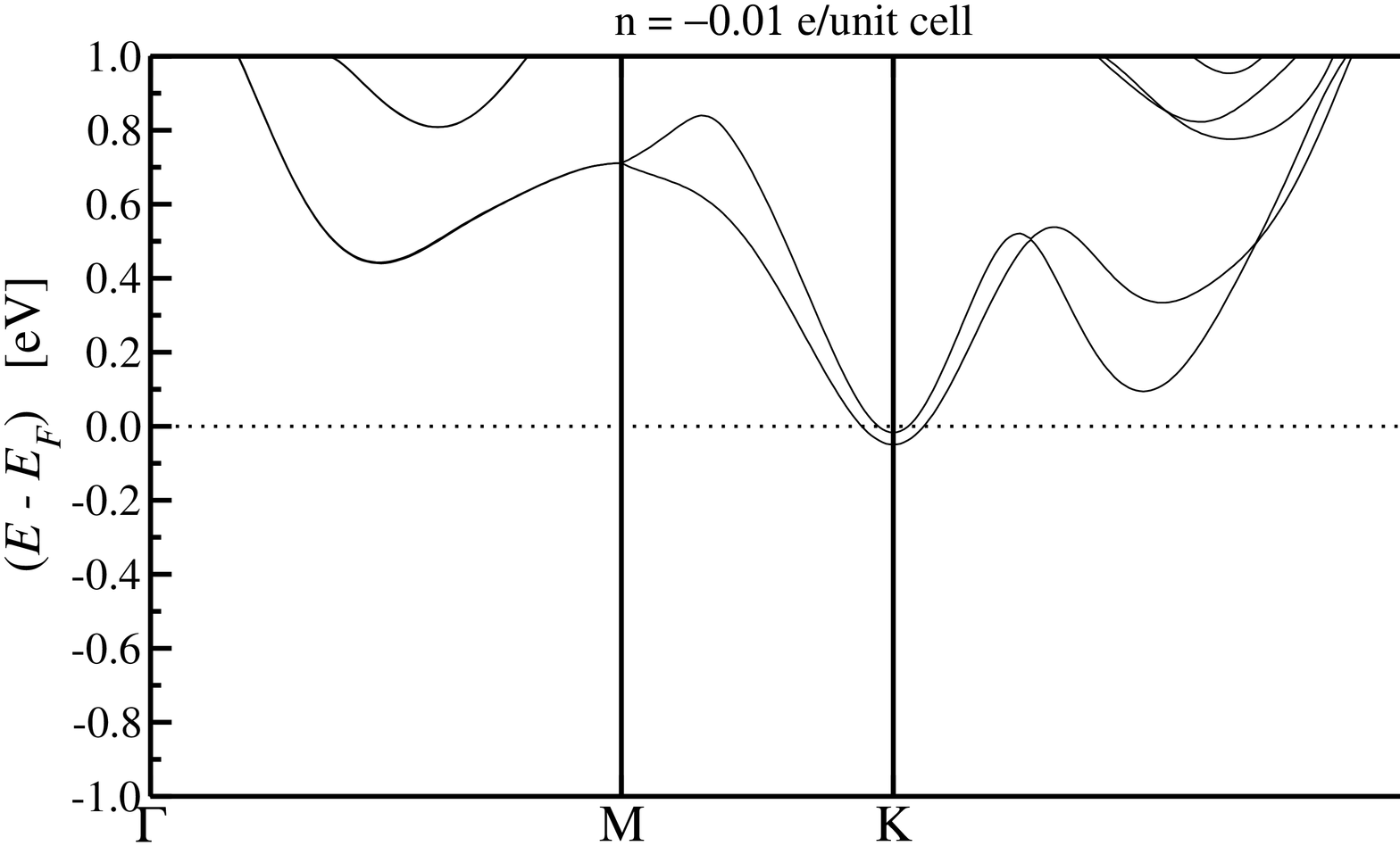}
 \caption{Band structure of monolayer WSe$_2$ for different doping as indicated in the labels.}
\end{figure*}
\begin{figure*}[hbp]
 \centering
 \includegraphics[width=0.31\textwidth,clip=,angle=-90]{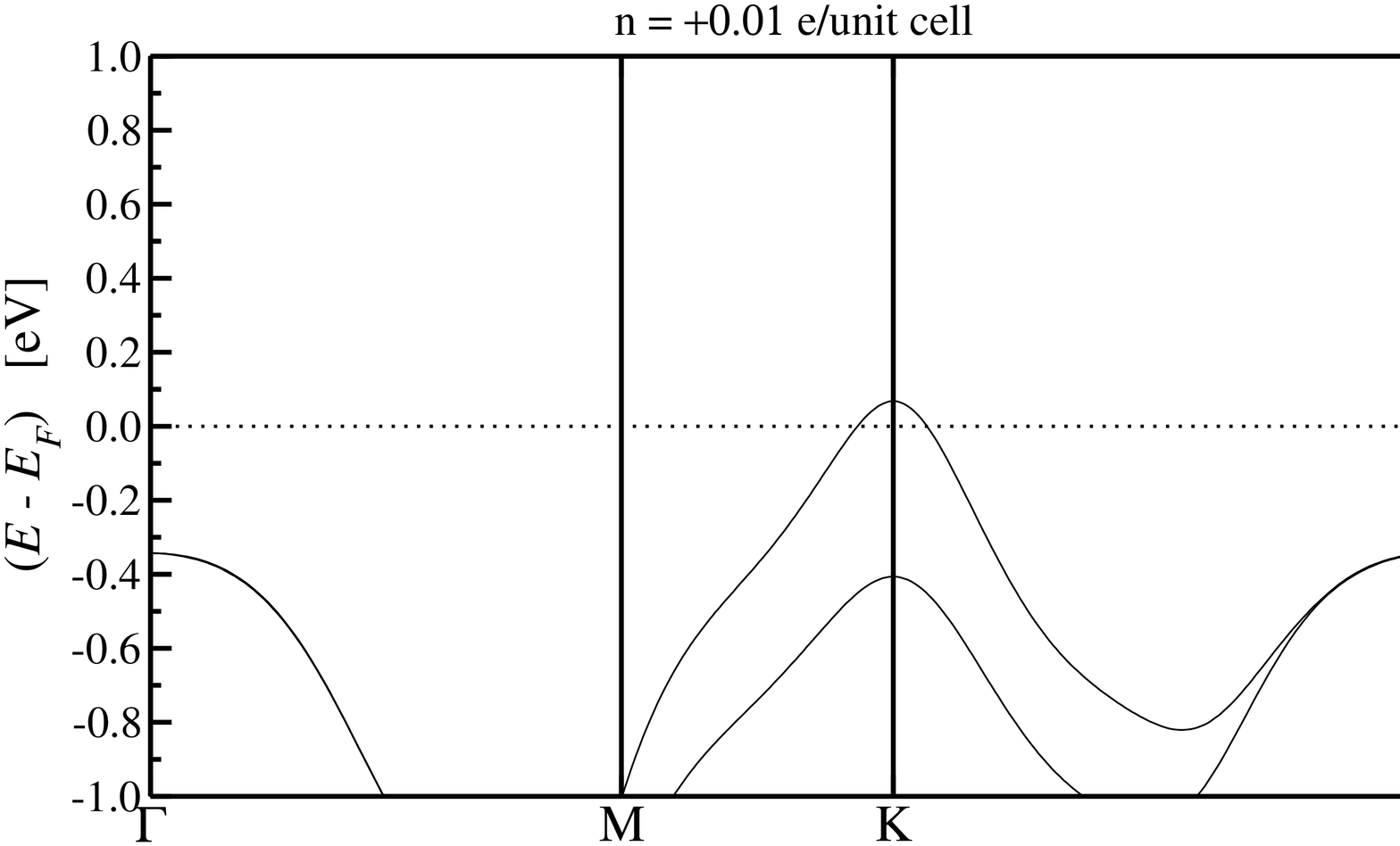}
 \includegraphics[width=0.31\textwidth,clip=,angle=-90]{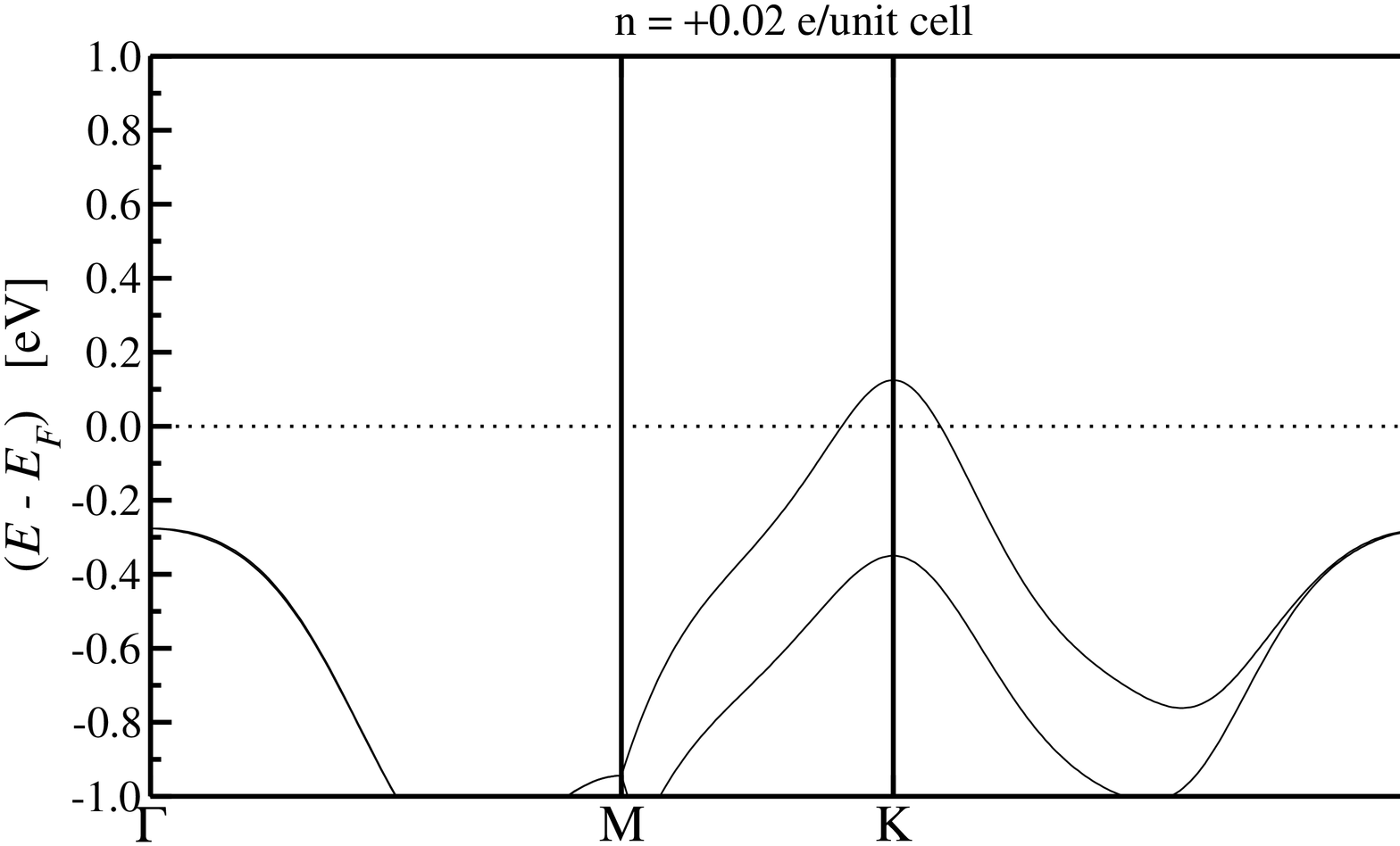}
 \includegraphics[width=0.31\textwidth,clip=,angle=-90]{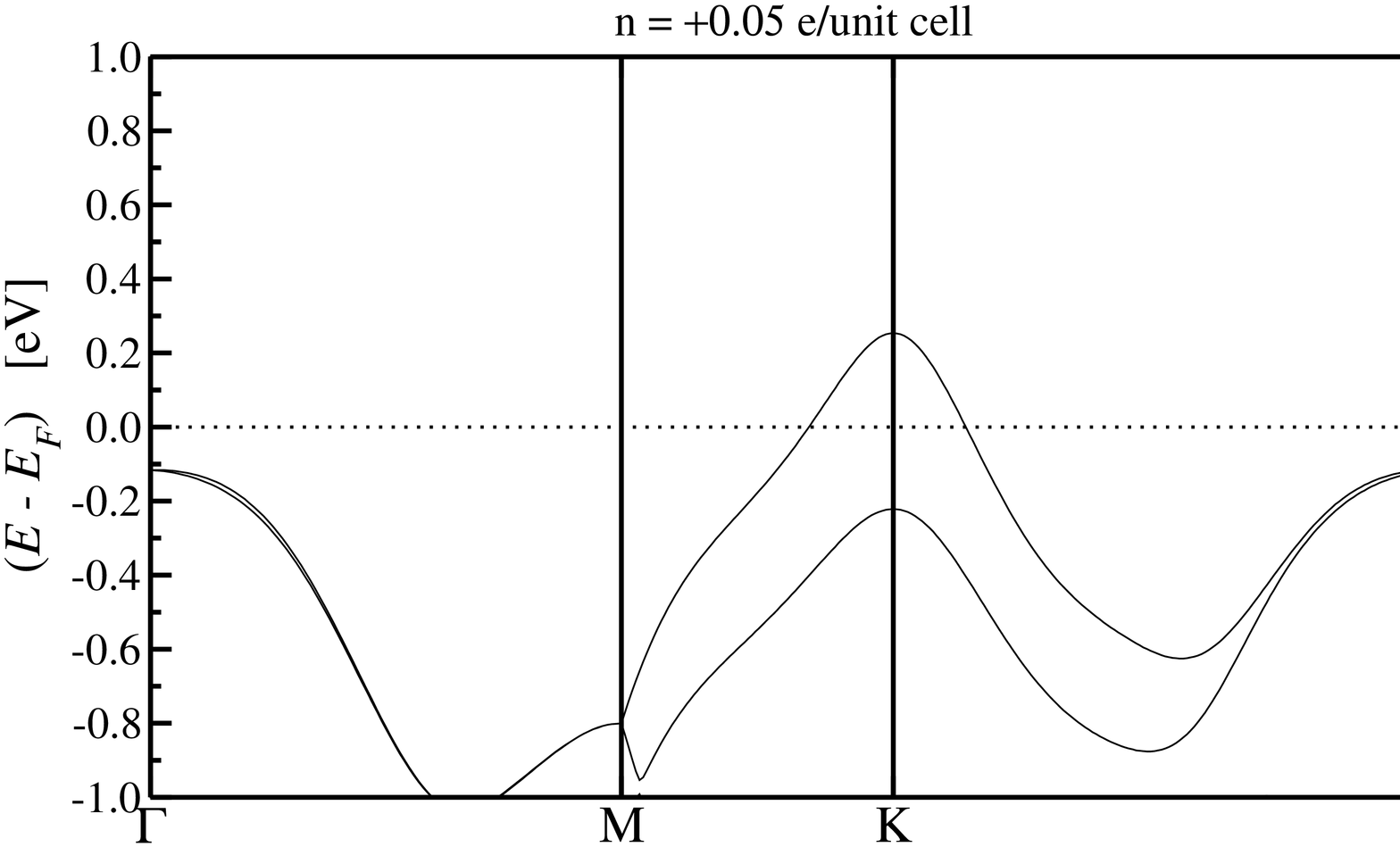}
 \includegraphics[width=0.31\textwidth,clip=,angle=-90]{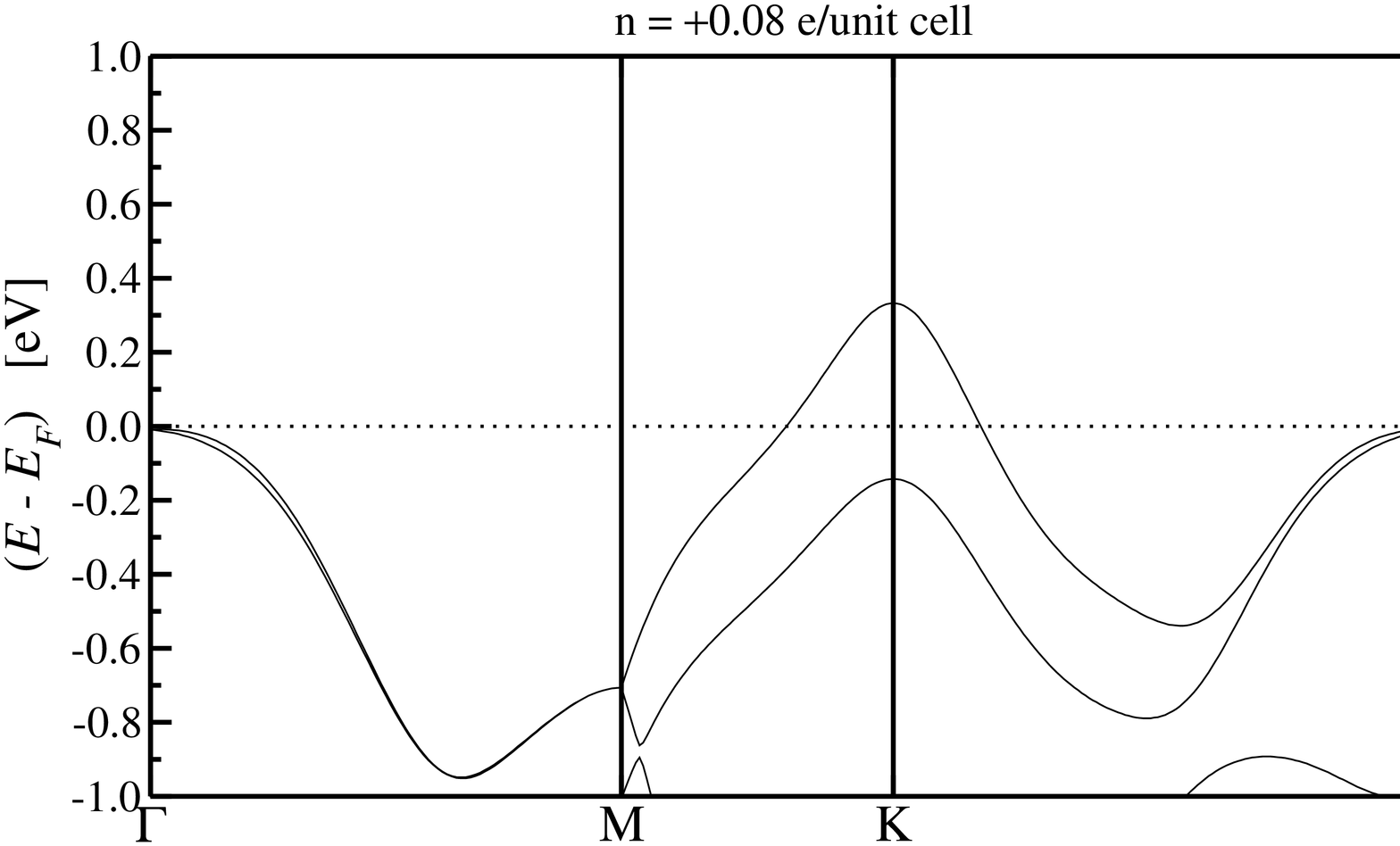}
 \includegraphics[width=0.31\textwidth,clip=,angle=-90]{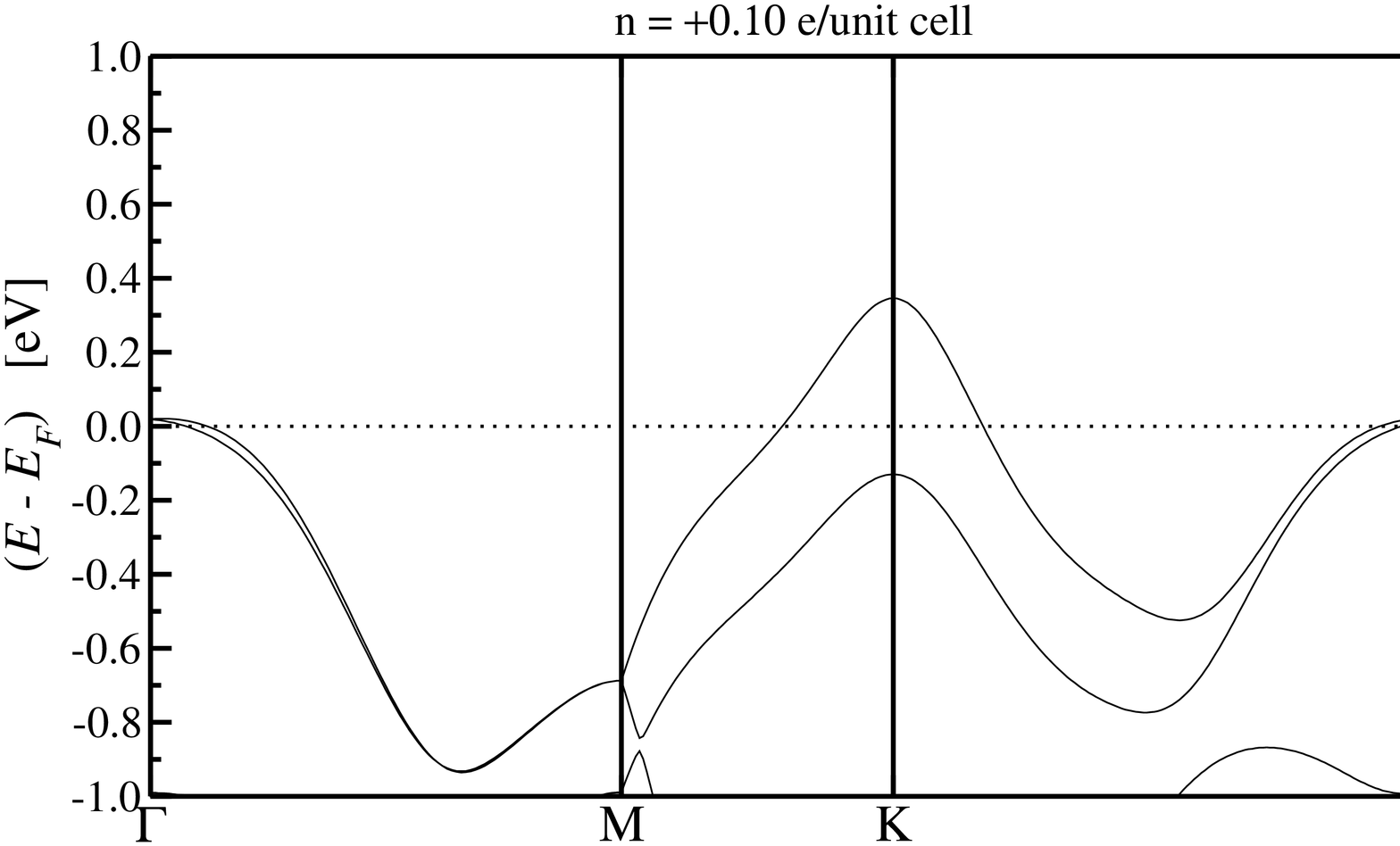}
 \includegraphics[width=0.31\textwidth,clip=,angle=-90]{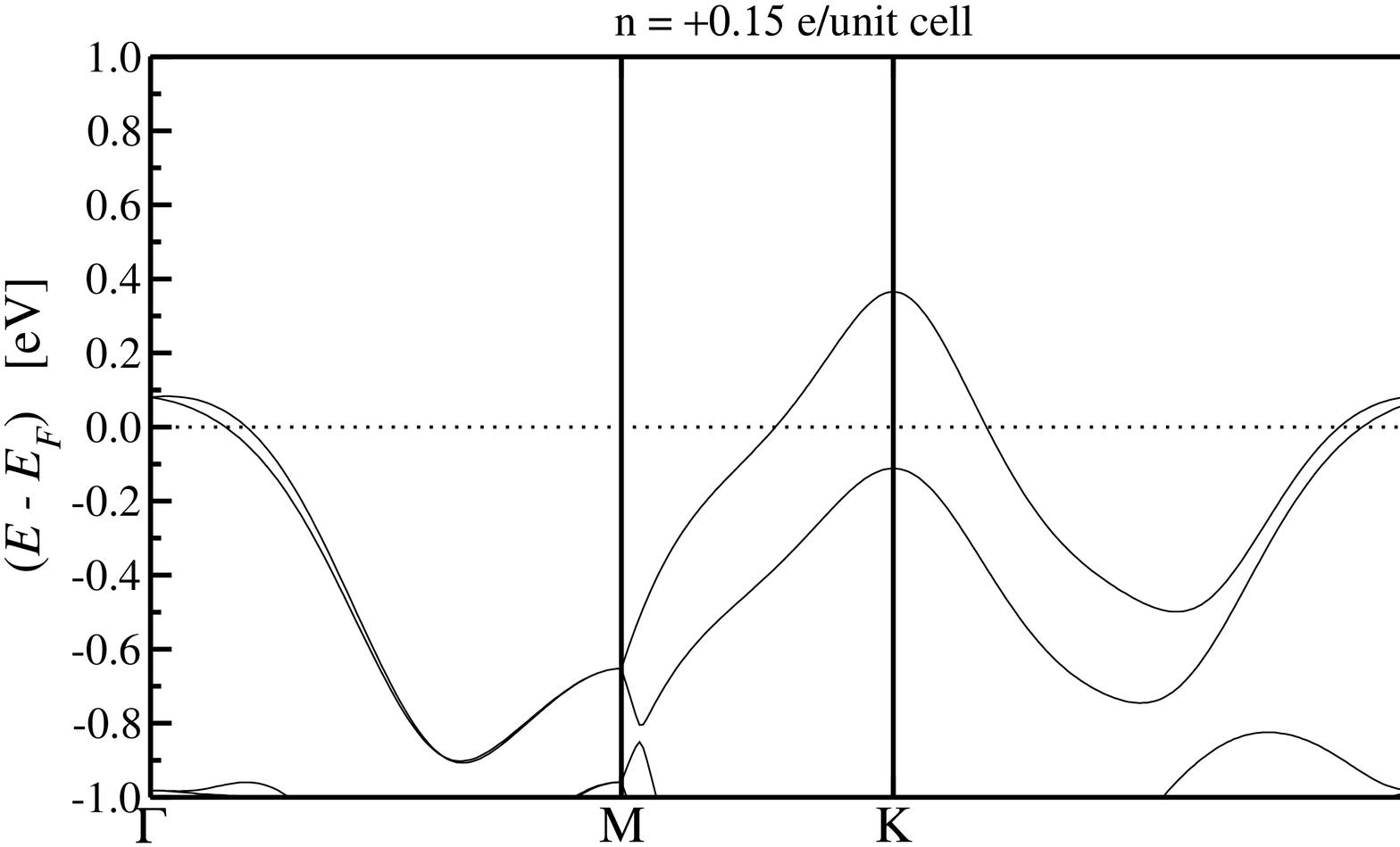}
 \includegraphics[width=0.31\textwidth,clip=,angle=-90]{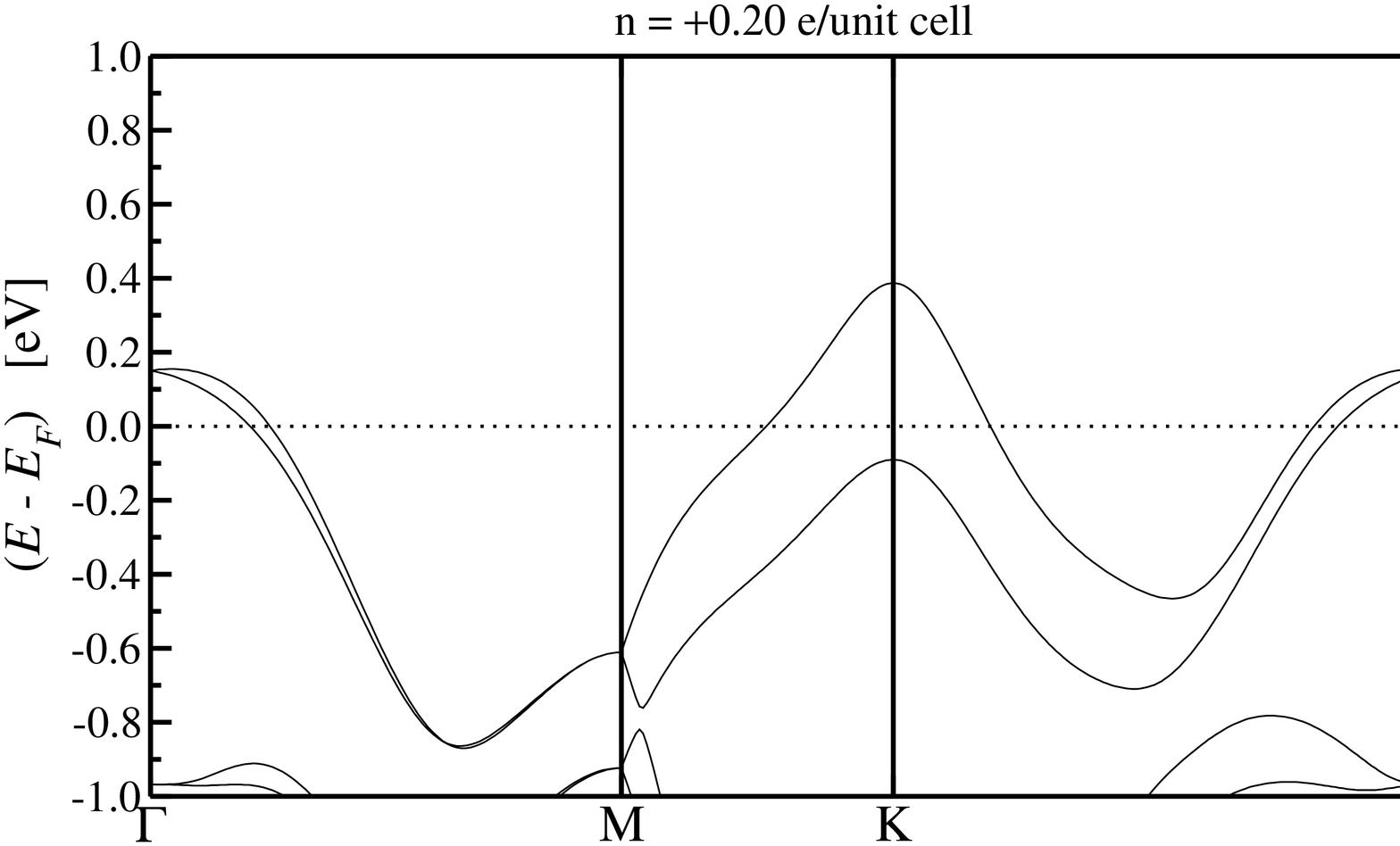}
 \includegraphics[width=0.31\textwidth,clip=,angle=-90]{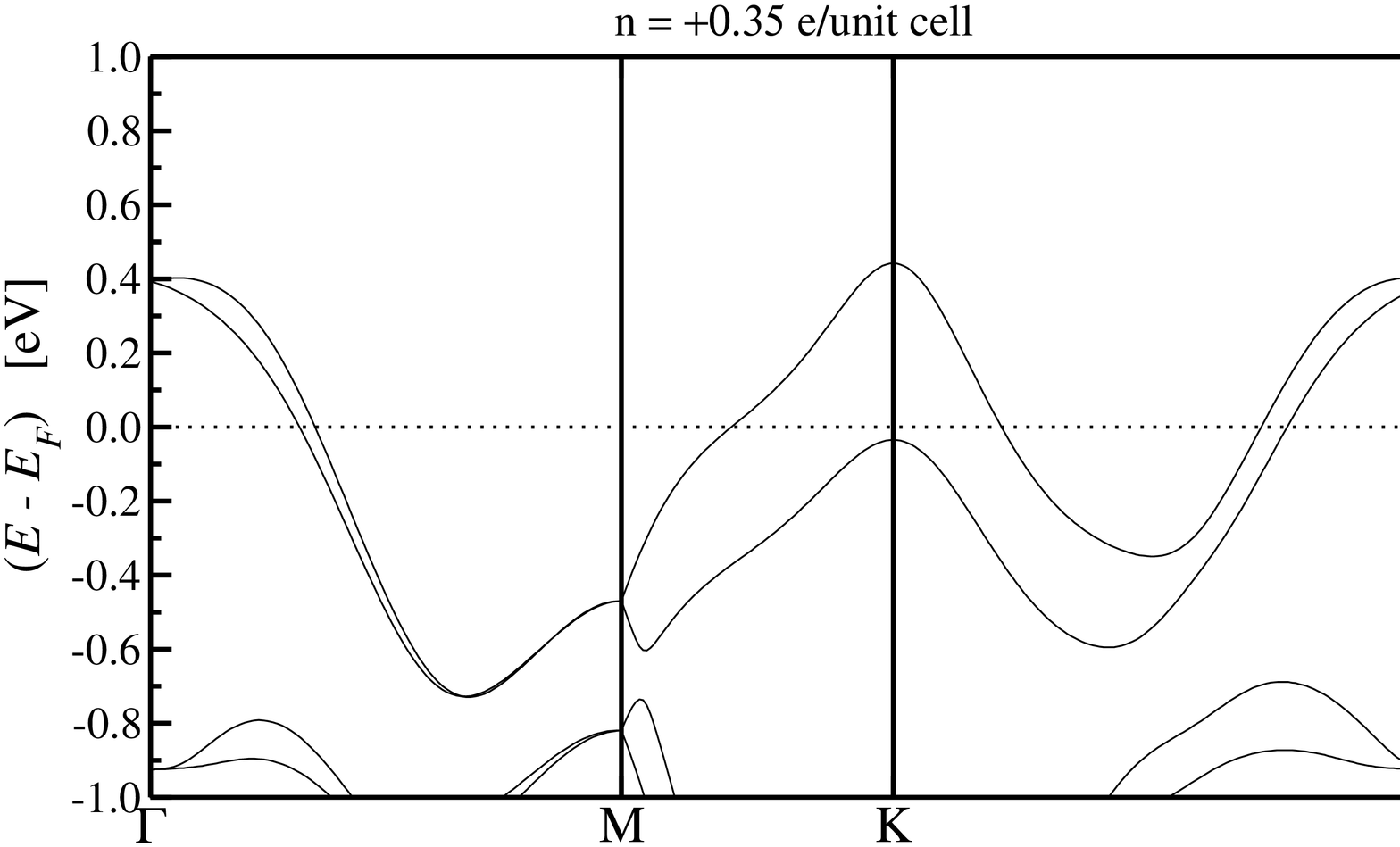}
 \caption{Band structure of monolayer WSe$_2$ for different doping as indicated in the labels.}
\end{figure*}
\begin{figure*}[hbp]
 \centering
 \includegraphics[width=0.31\textwidth,clip=,angle=-90]{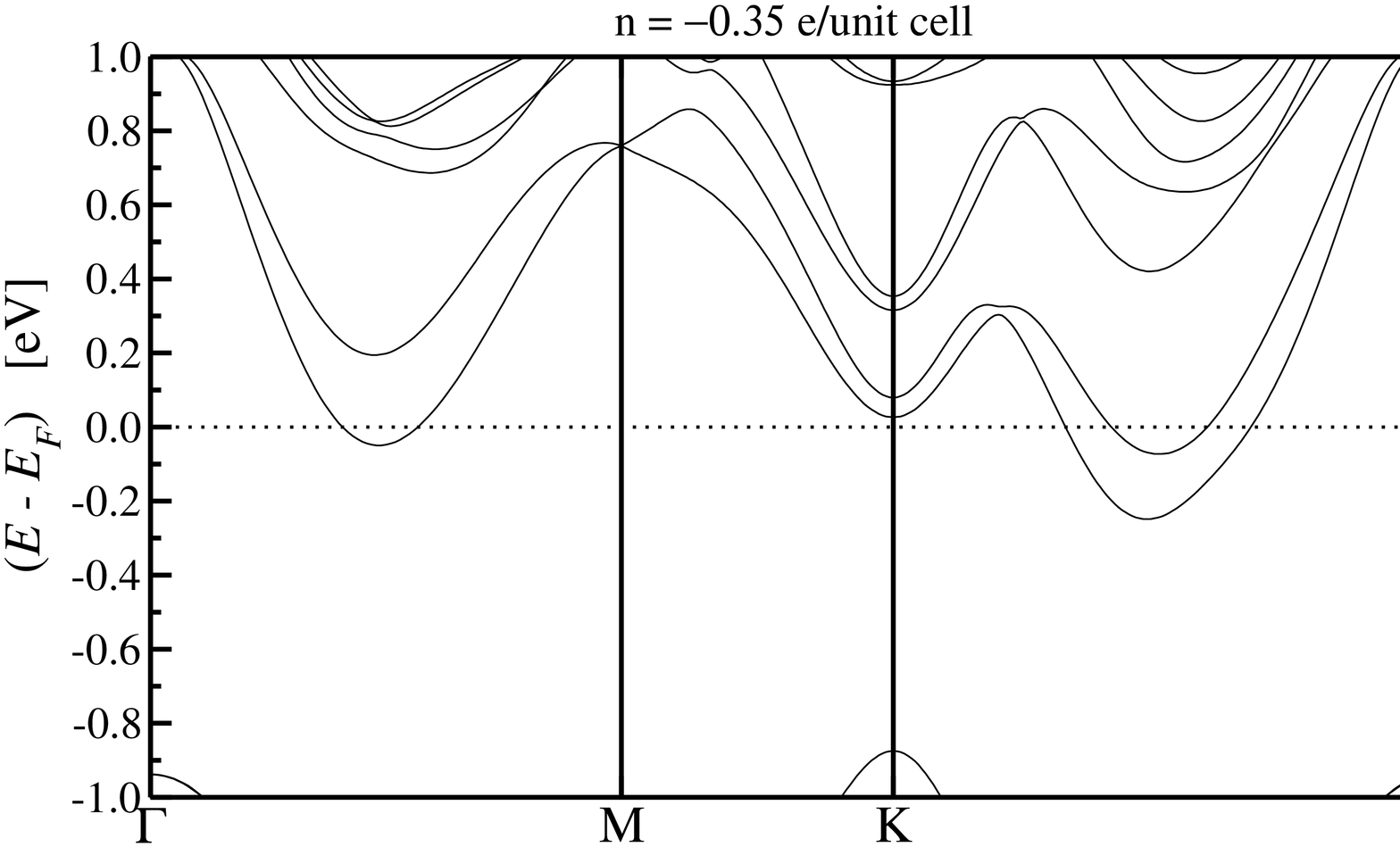}
 \includegraphics[width=0.31\textwidth,clip=,angle=-90]{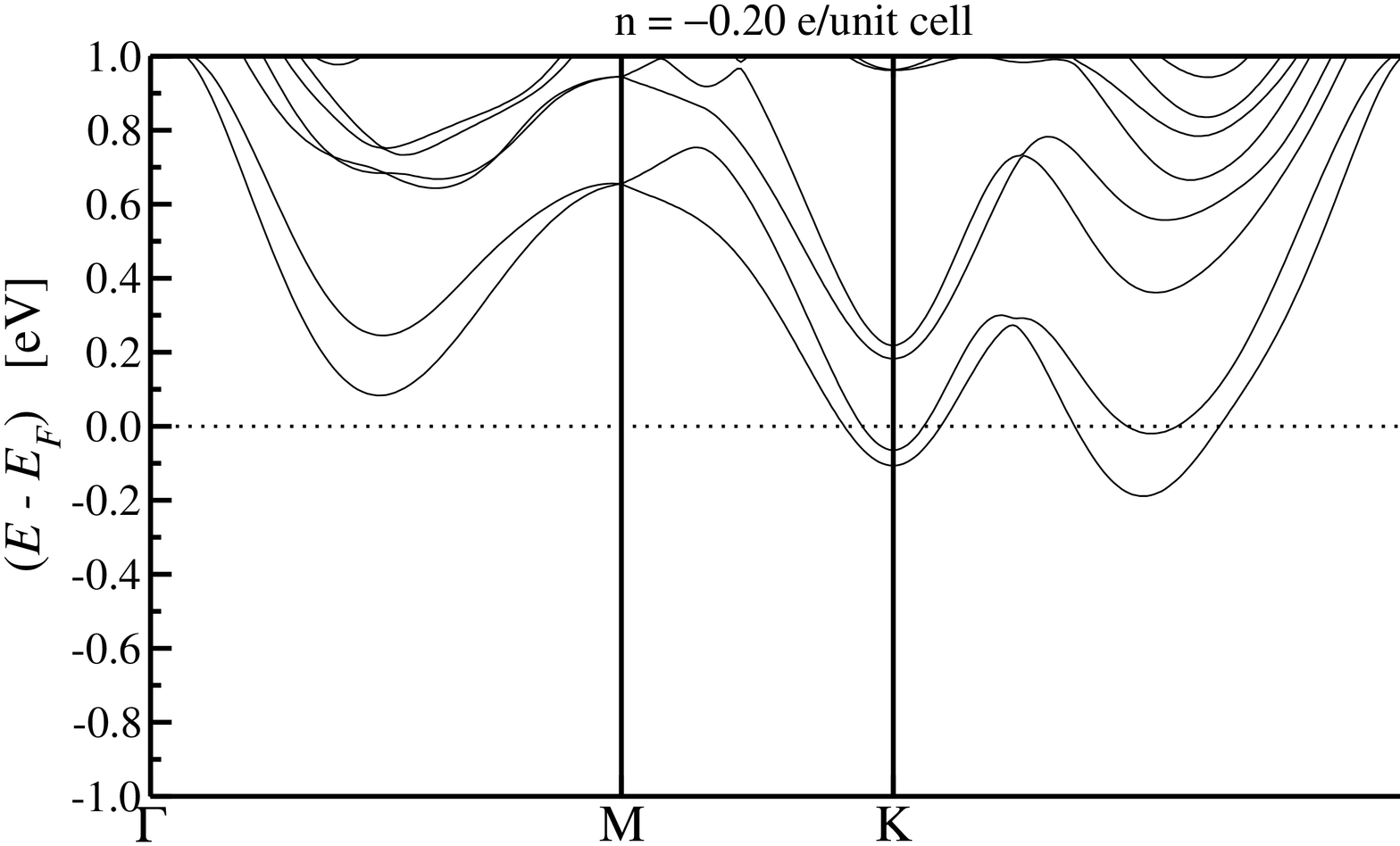}
 \includegraphics[width=0.31\textwidth,clip=,angle=-90]{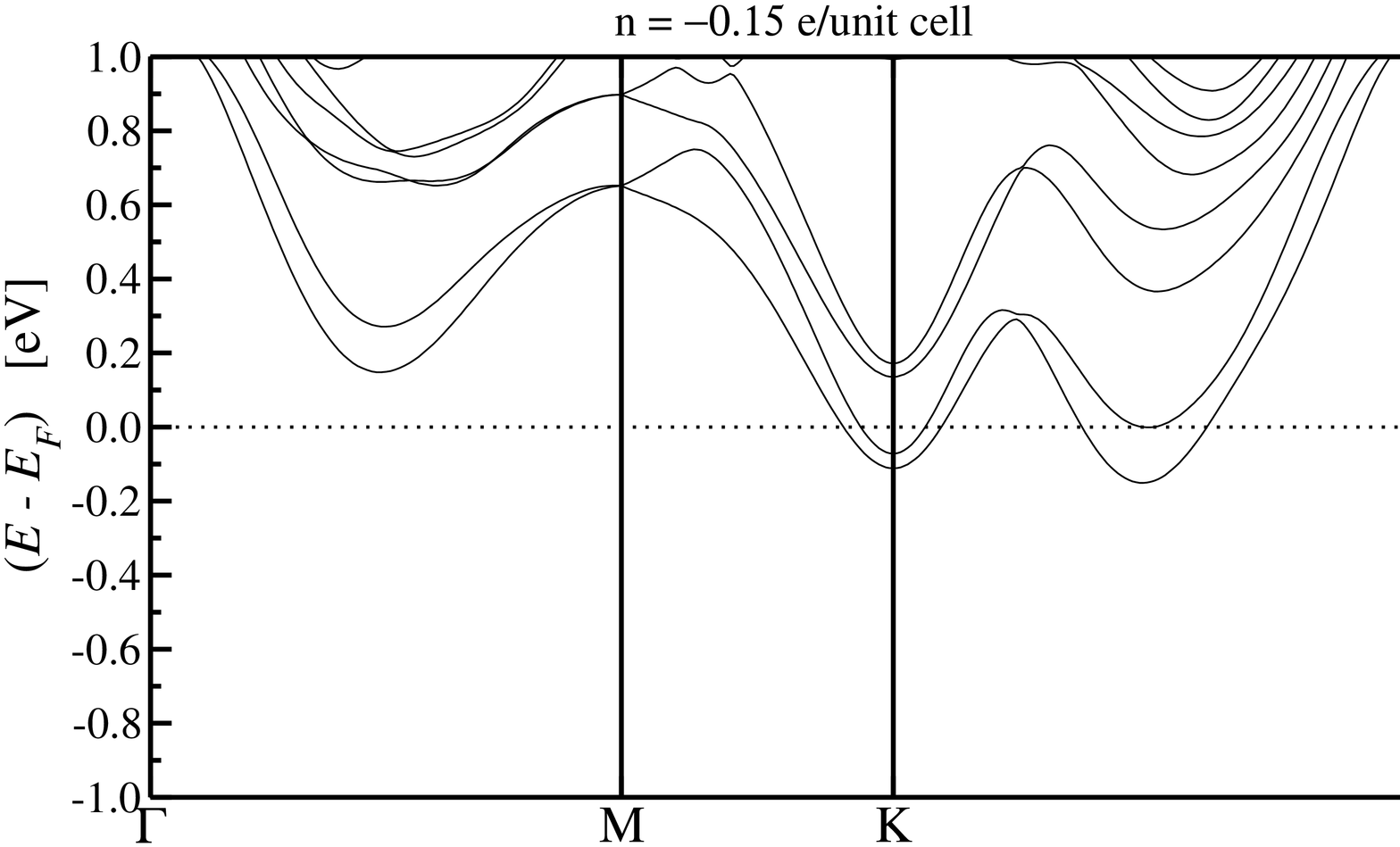}
 \includegraphics[width=0.31\textwidth,clip=,angle=-90]{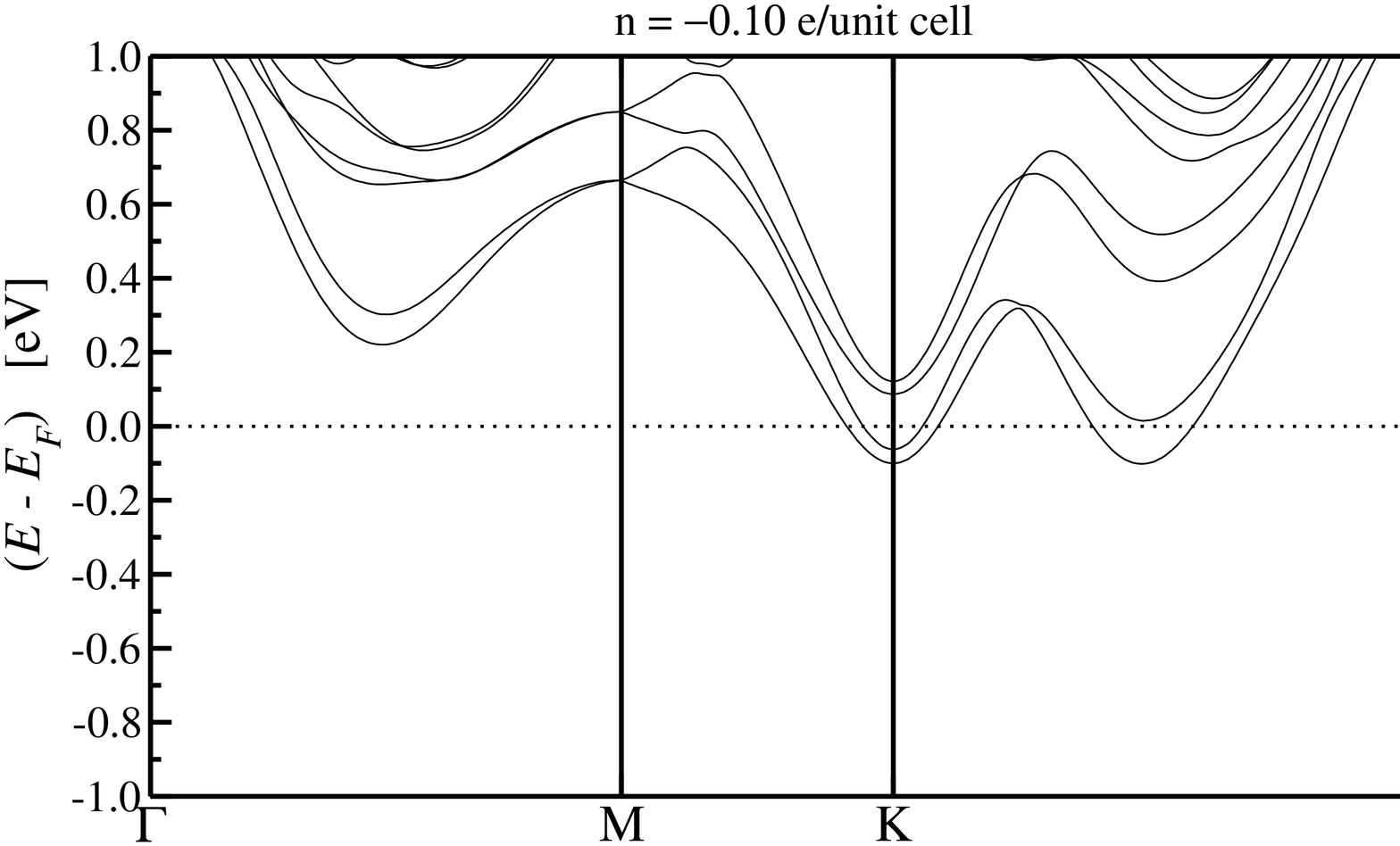}
 \includegraphics[width=0.31\textwidth,clip=,angle=-90]{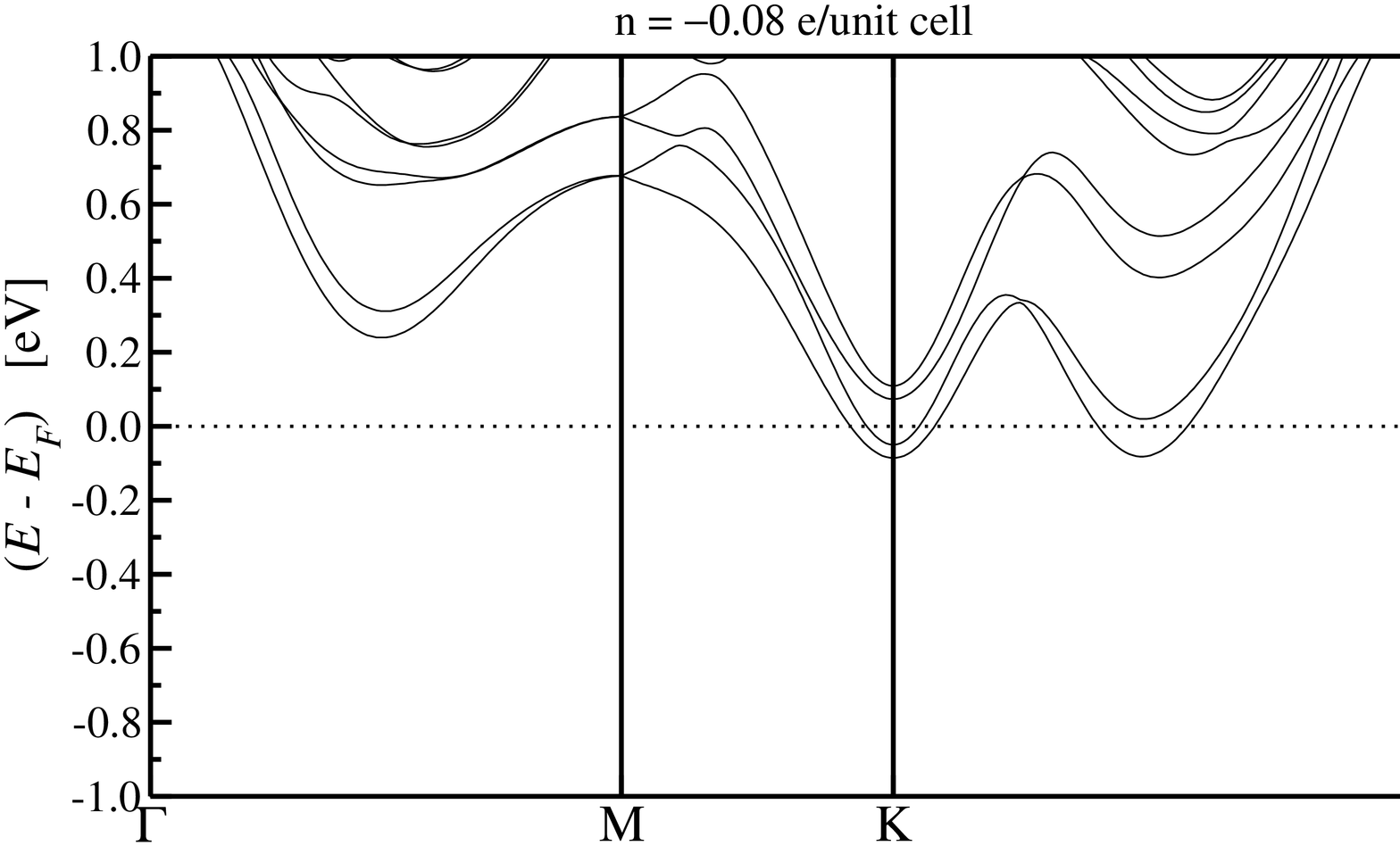}
 \includegraphics[width=0.31\textwidth,clip=,angle=-90]{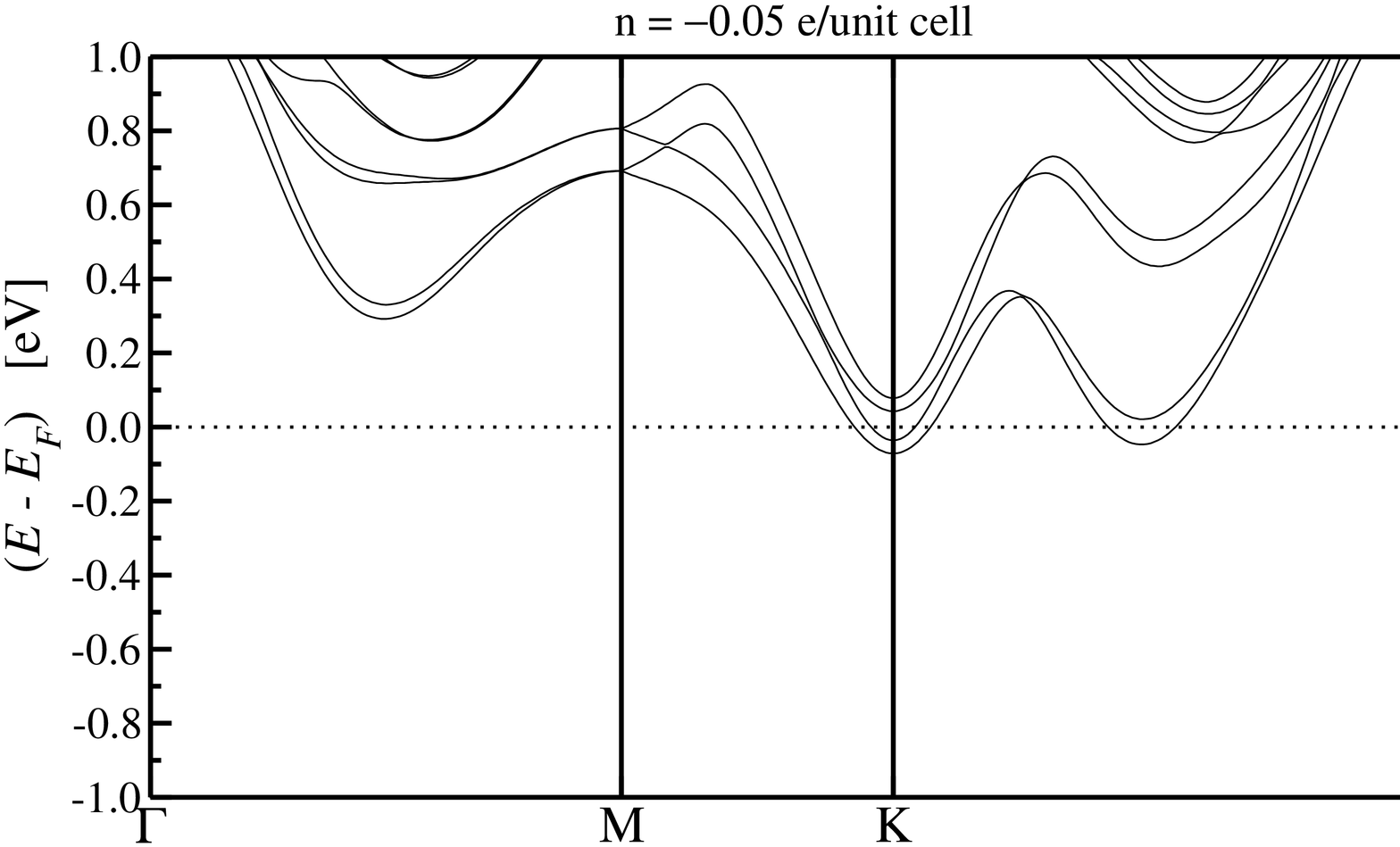}
 \includegraphics[width=0.31\textwidth,clip=,angle=-90]{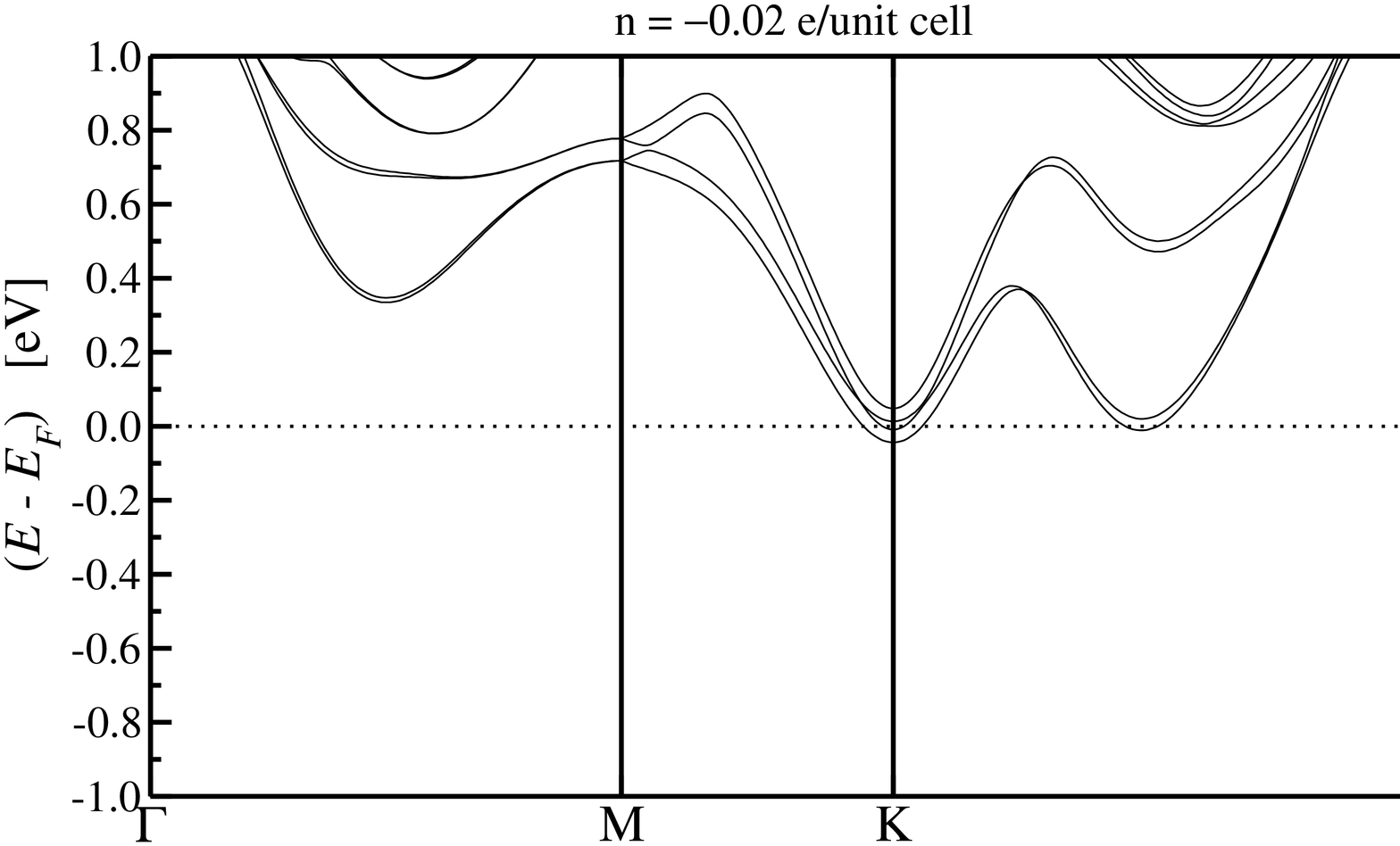}
 \includegraphics[width=0.31\textwidth,clip=,angle=-90]{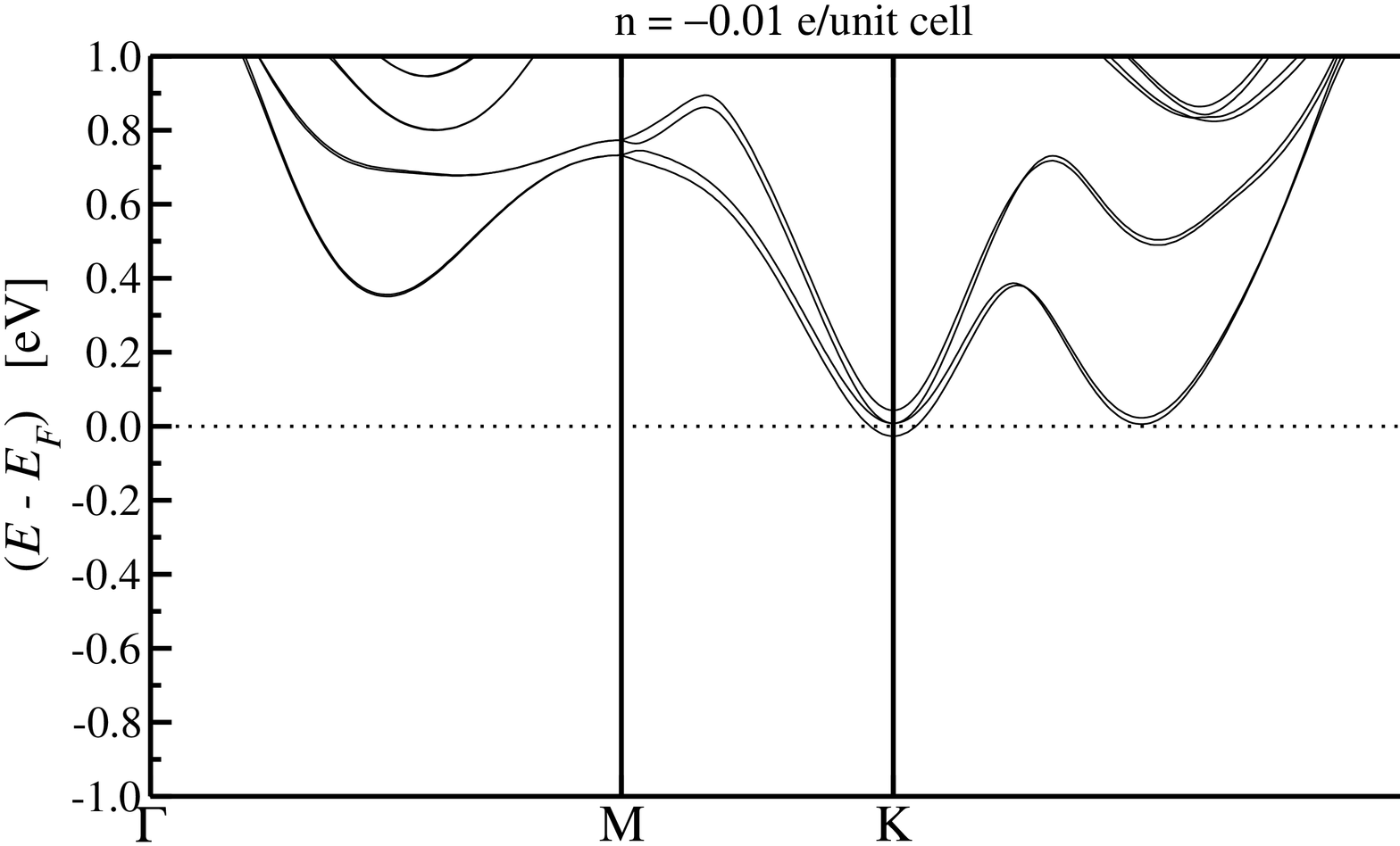}
 \caption{Band structure of bilayer WSe$_2$ for different doping as indicated in the labels.}
\end{figure*}
\begin{figure*}[hbp]
 \centering
 \includegraphics[width=0.31\textwidth,clip=,angle=-90]{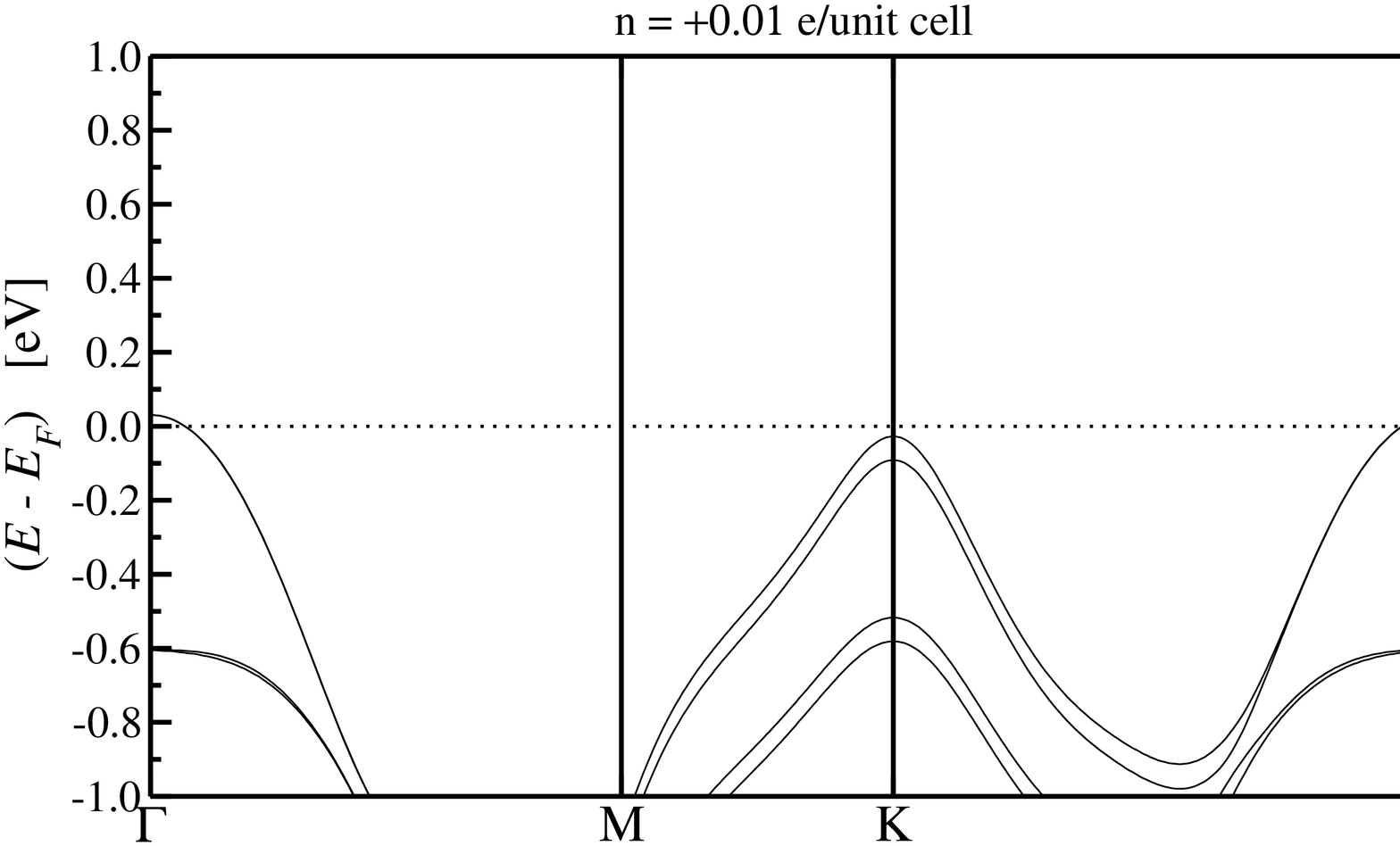}
 \includegraphics[width=0.31\textwidth,clip=,angle=-90]{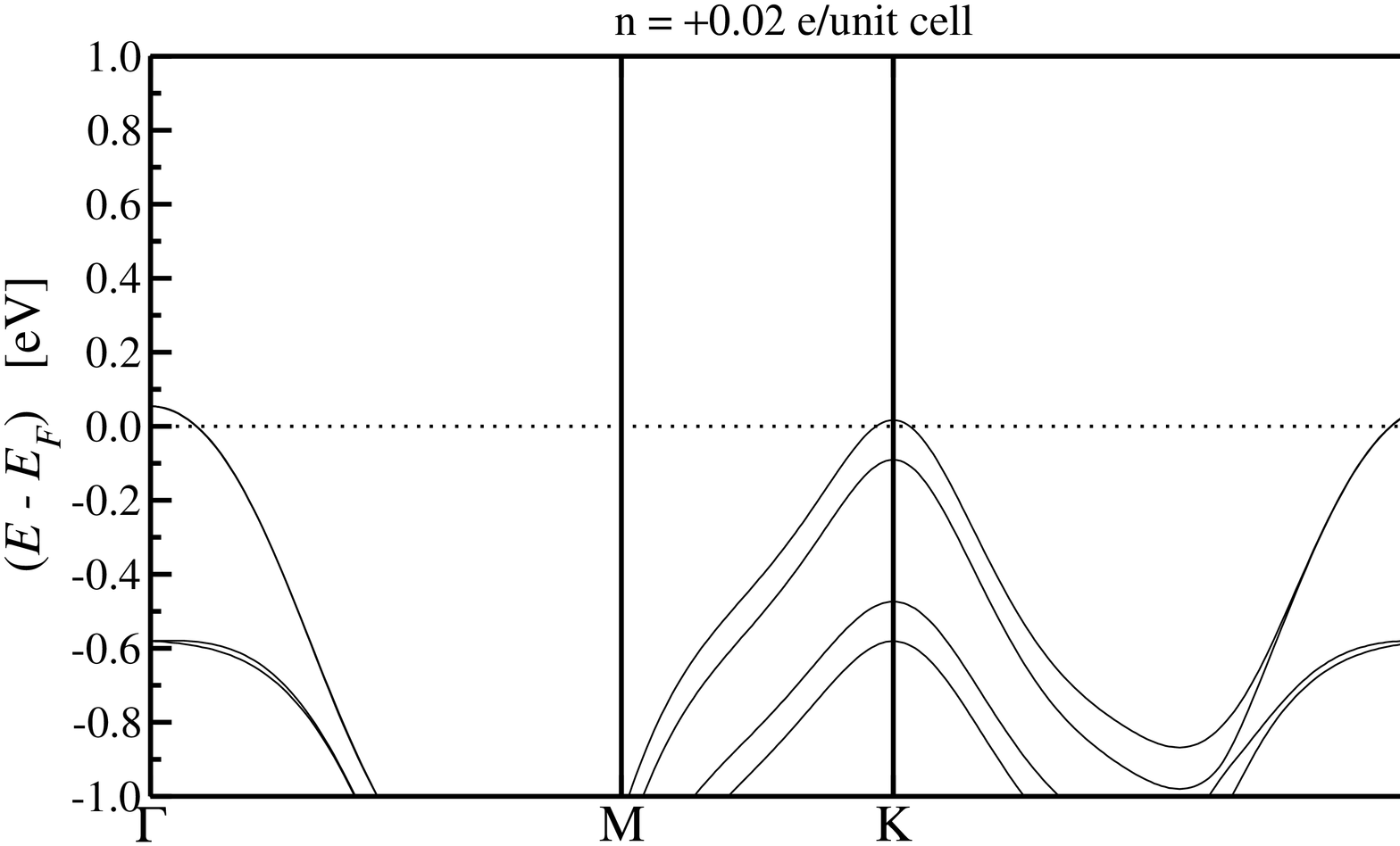}
 \includegraphics[width=0.31\textwidth,clip=,angle=-90]{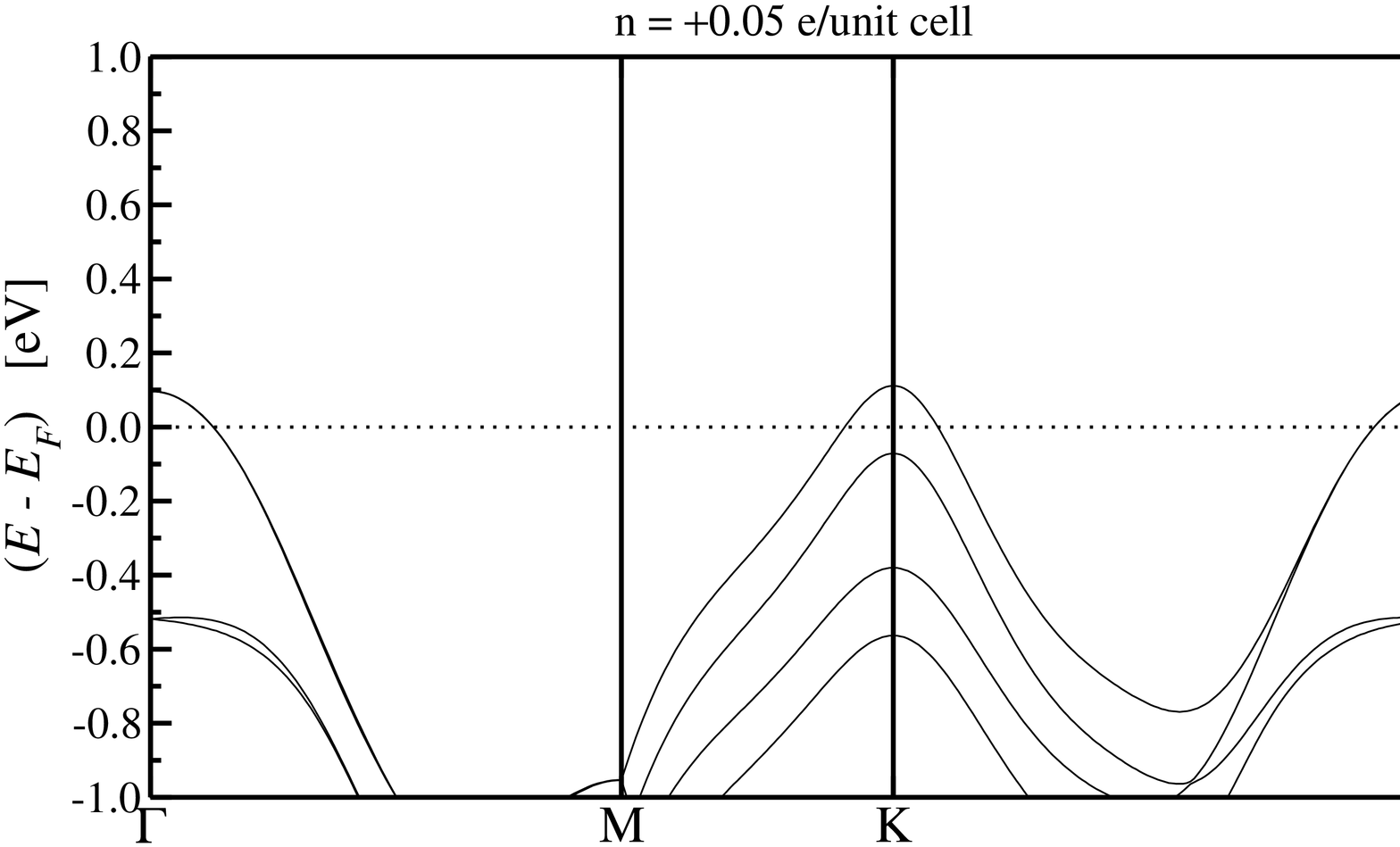}
 \includegraphics[width=0.31\textwidth,clip=,angle=-90]{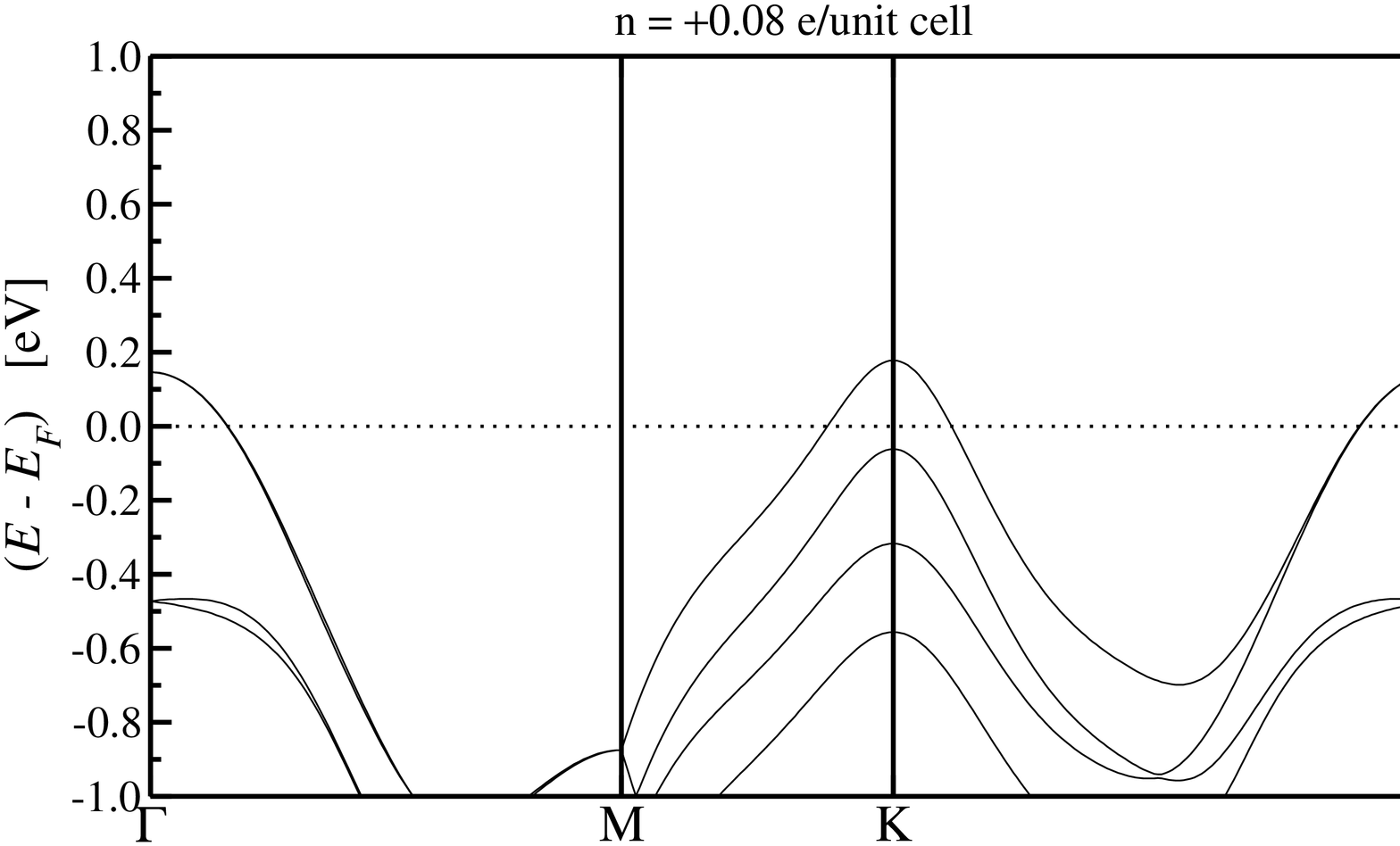}
 \includegraphics[width=0.31\textwidth,clip=,angle=-90]{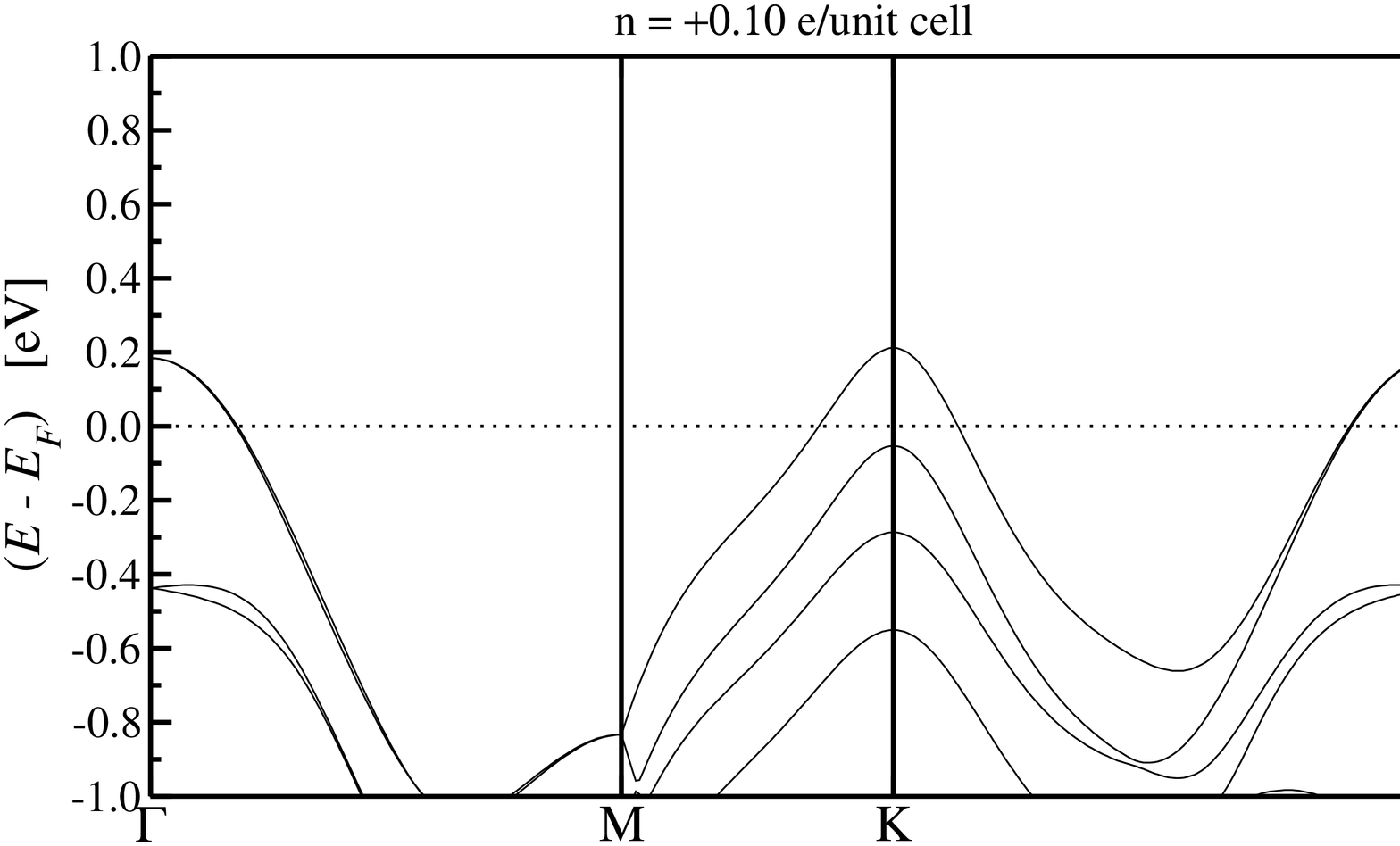}
 \includegraphics[width=0.31\textwidth,clip=,angle=-90]{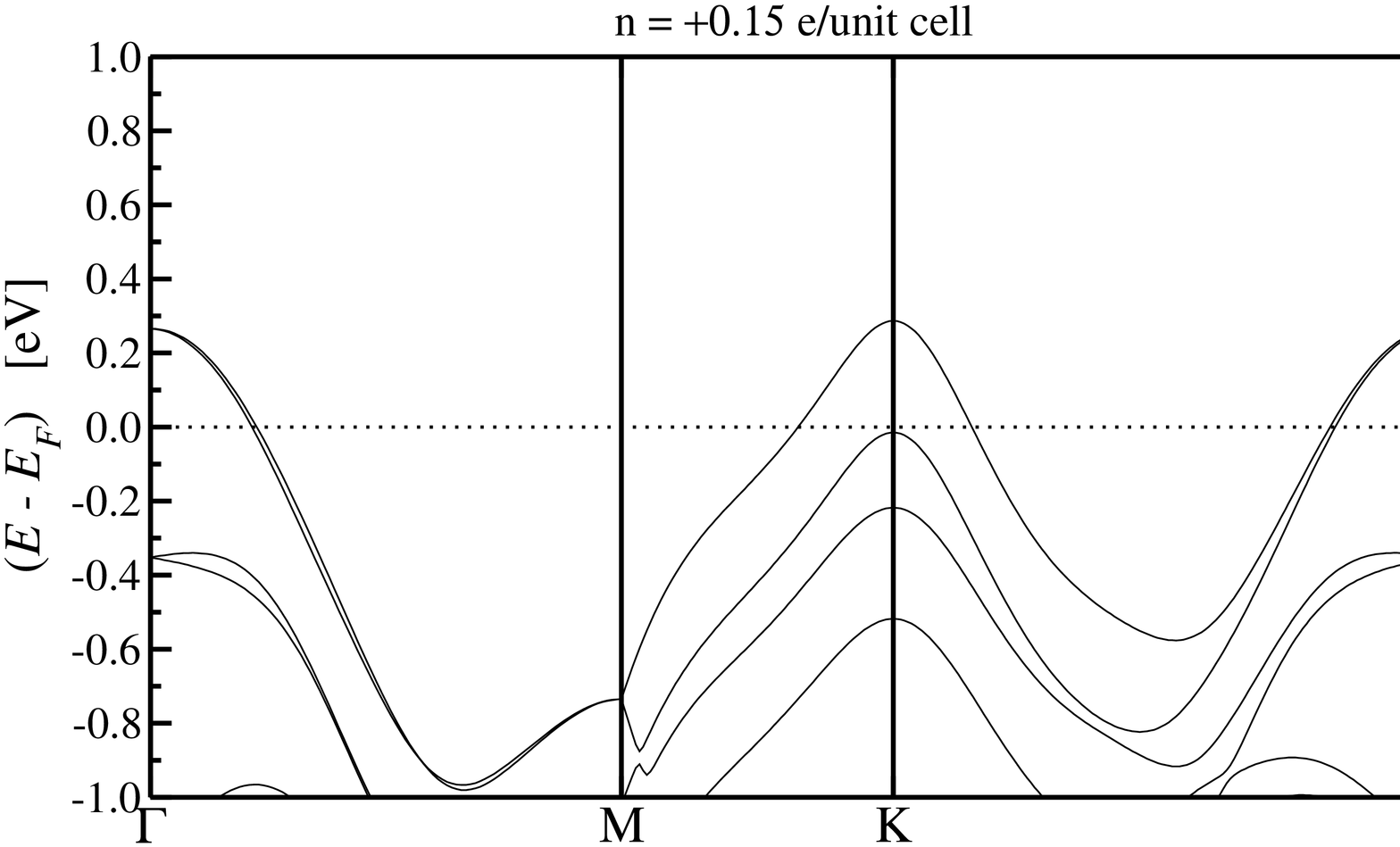}
 \includegraphics[width=0.31\textwidth,clip=,angle=-90]{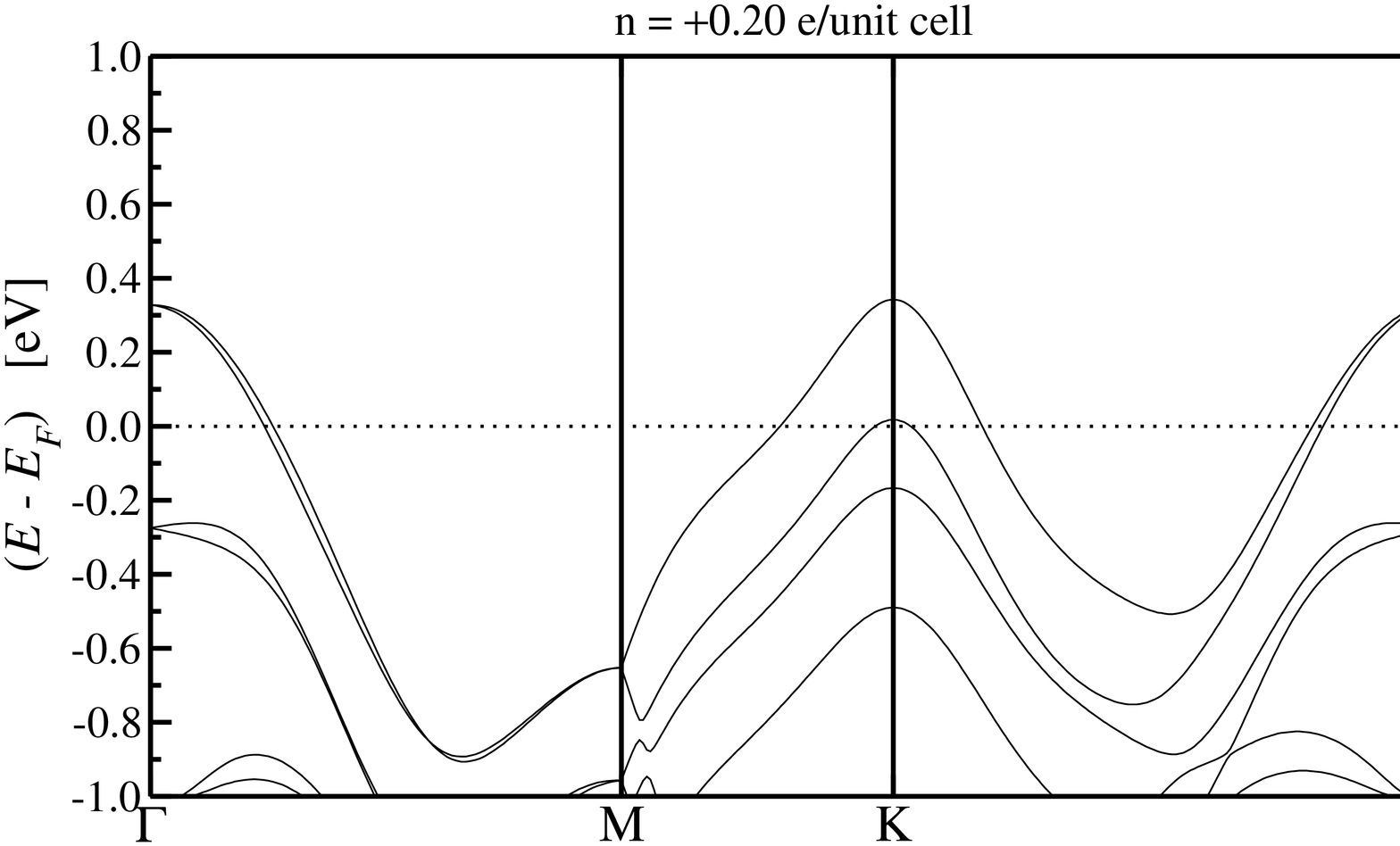}
 \includegraphics[width=0.31\textwidth,clip=,angle=-90]{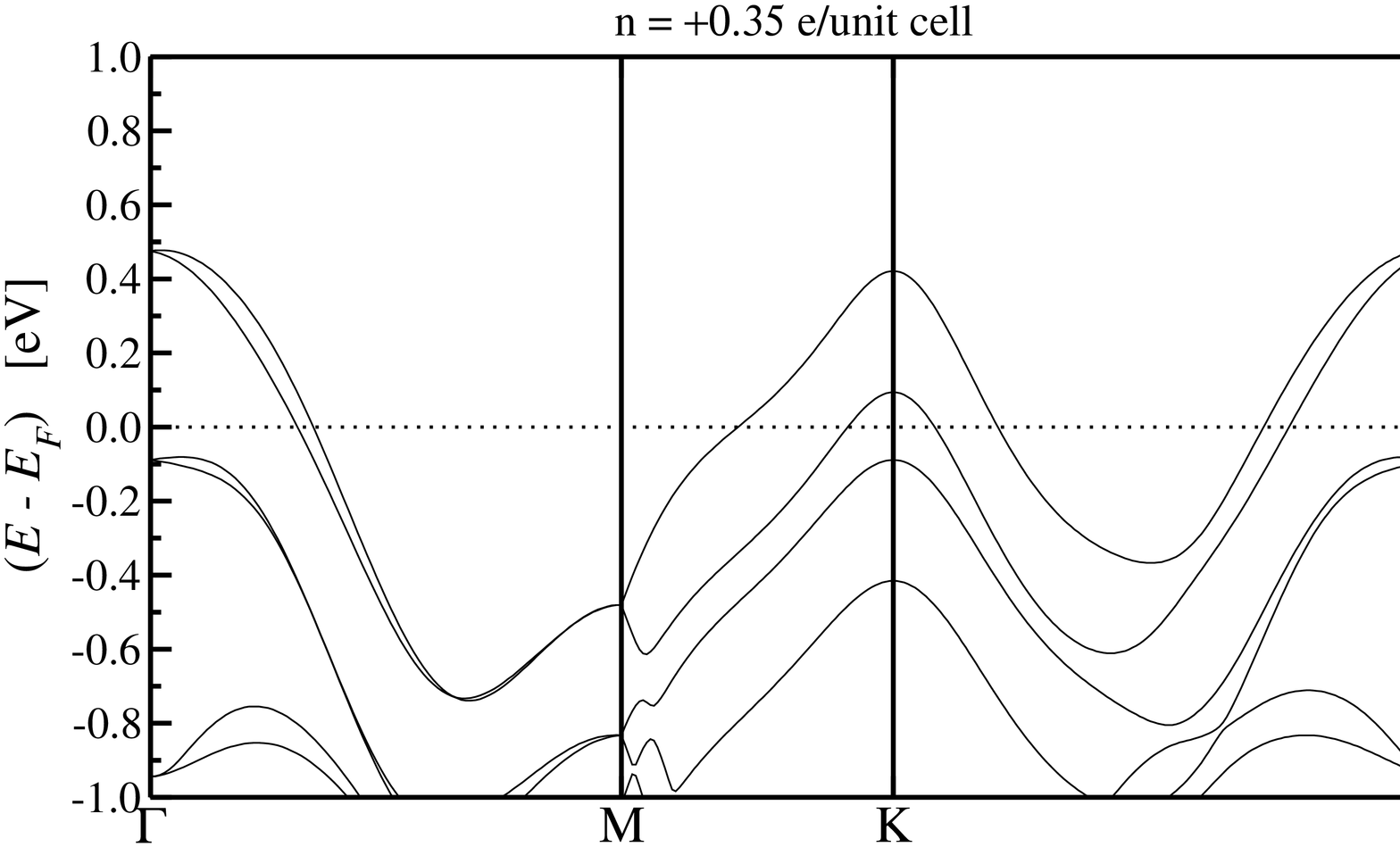}
 \caption{Band structure of bilayer WSe$_2$ for different doping as indicated in the labels.}
\end{figure*}
\begin{figure*}[hbp]
 \centering
 \includegraphics[width=0.31\textwidth,clip=,angle=-90]{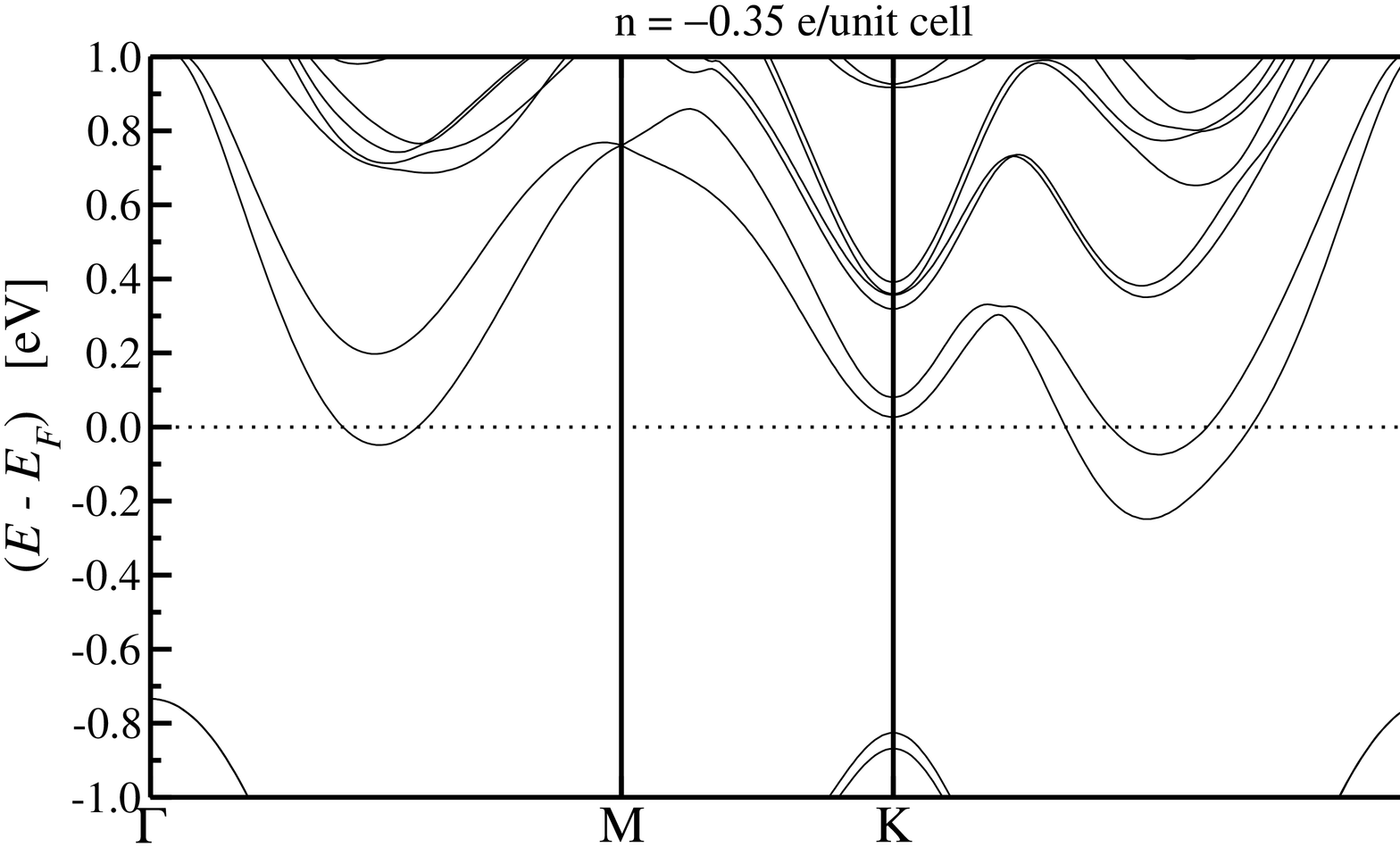}
 \includegraphics[width=0.31\textwidth,clip=,angle=-90]{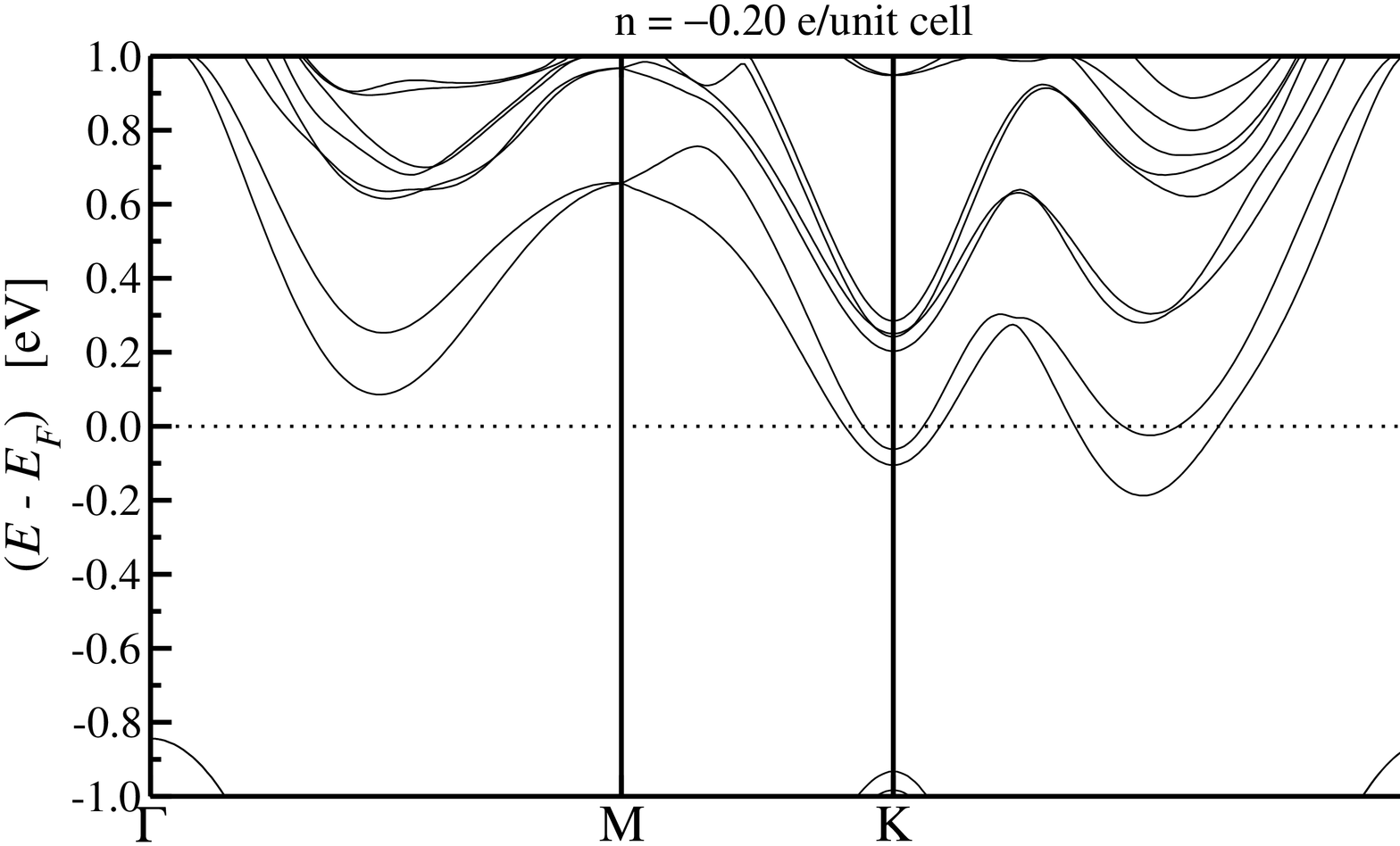}
 \includegraphics[width=0.31\textwidth,clip=,angle=-90]{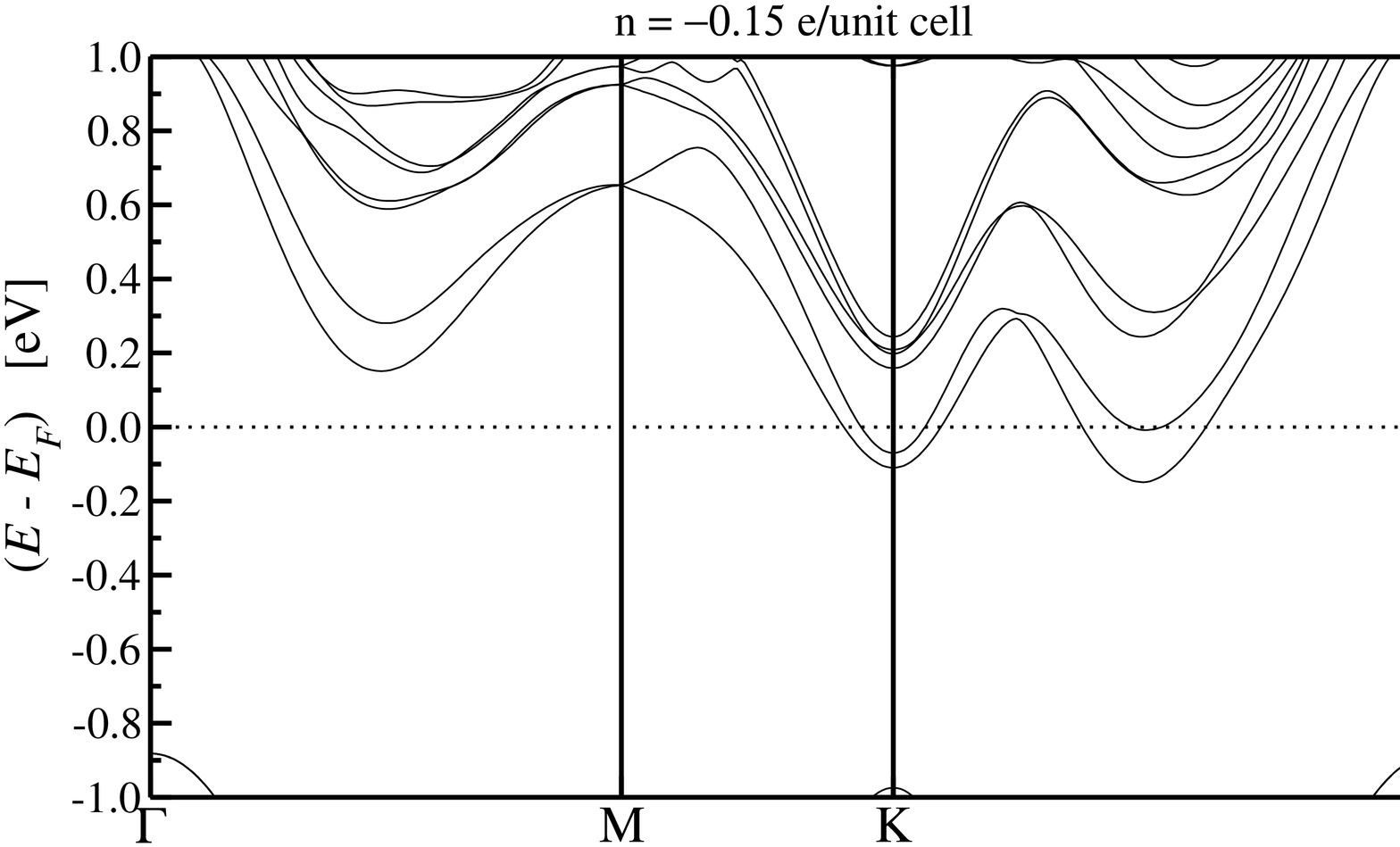}
 \includegraphics[width=0.31\textwidth,clip=,angle=-90]{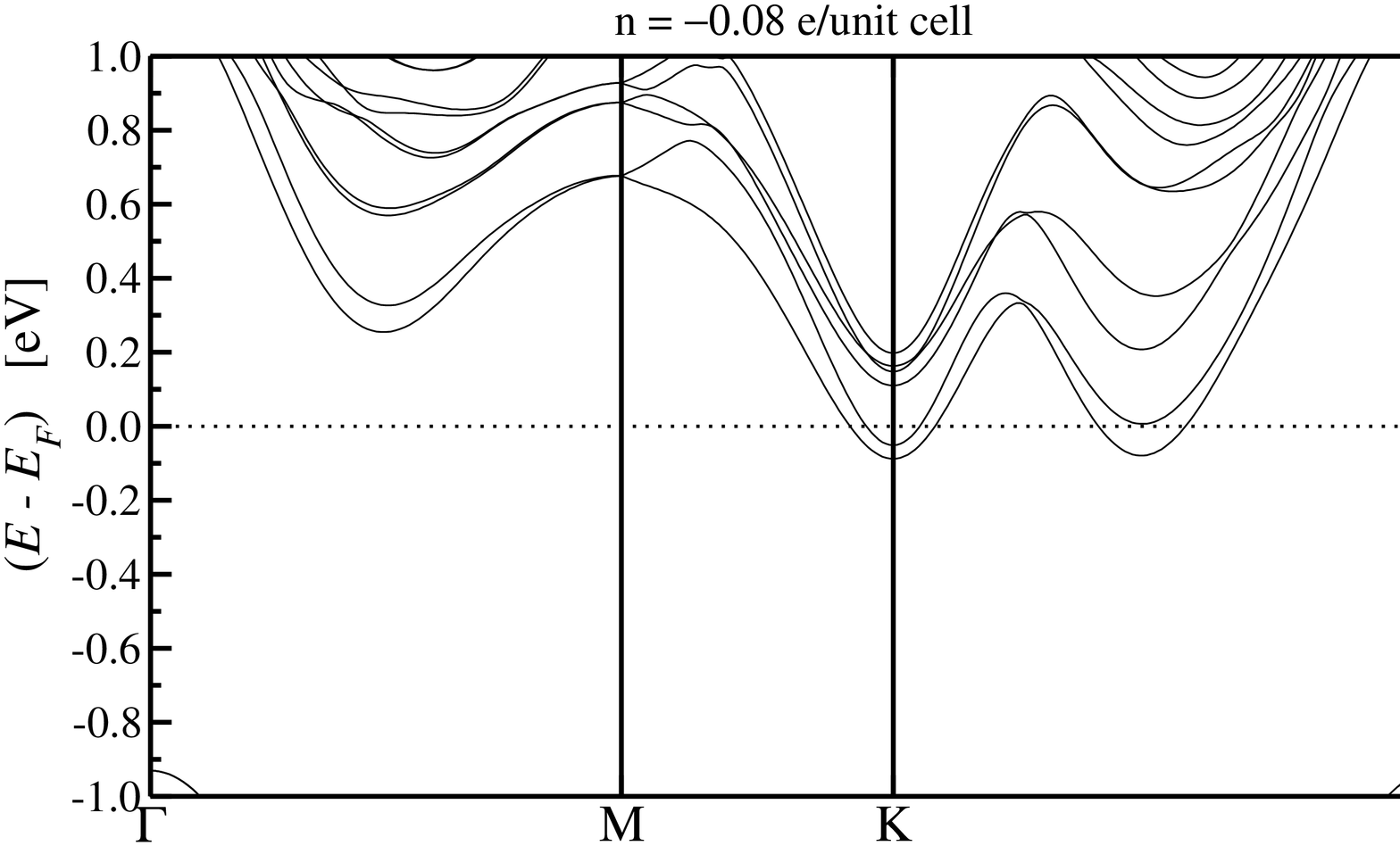}
 \includegraphics[width=0.31\textwidth,clip=,angle=-90]{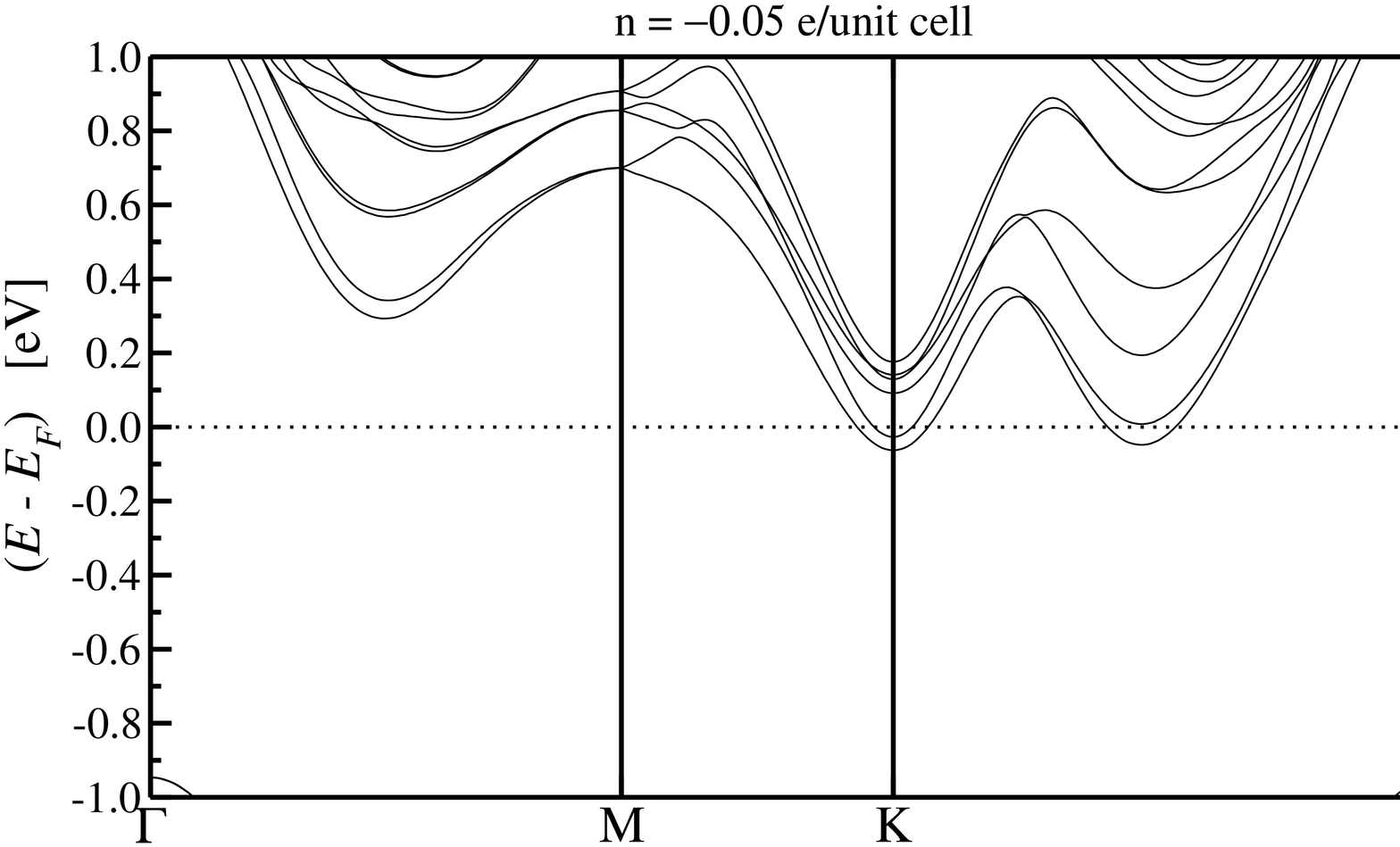}
 \includegraphics[width=0.31\textwidth,clip=,angle=-90]{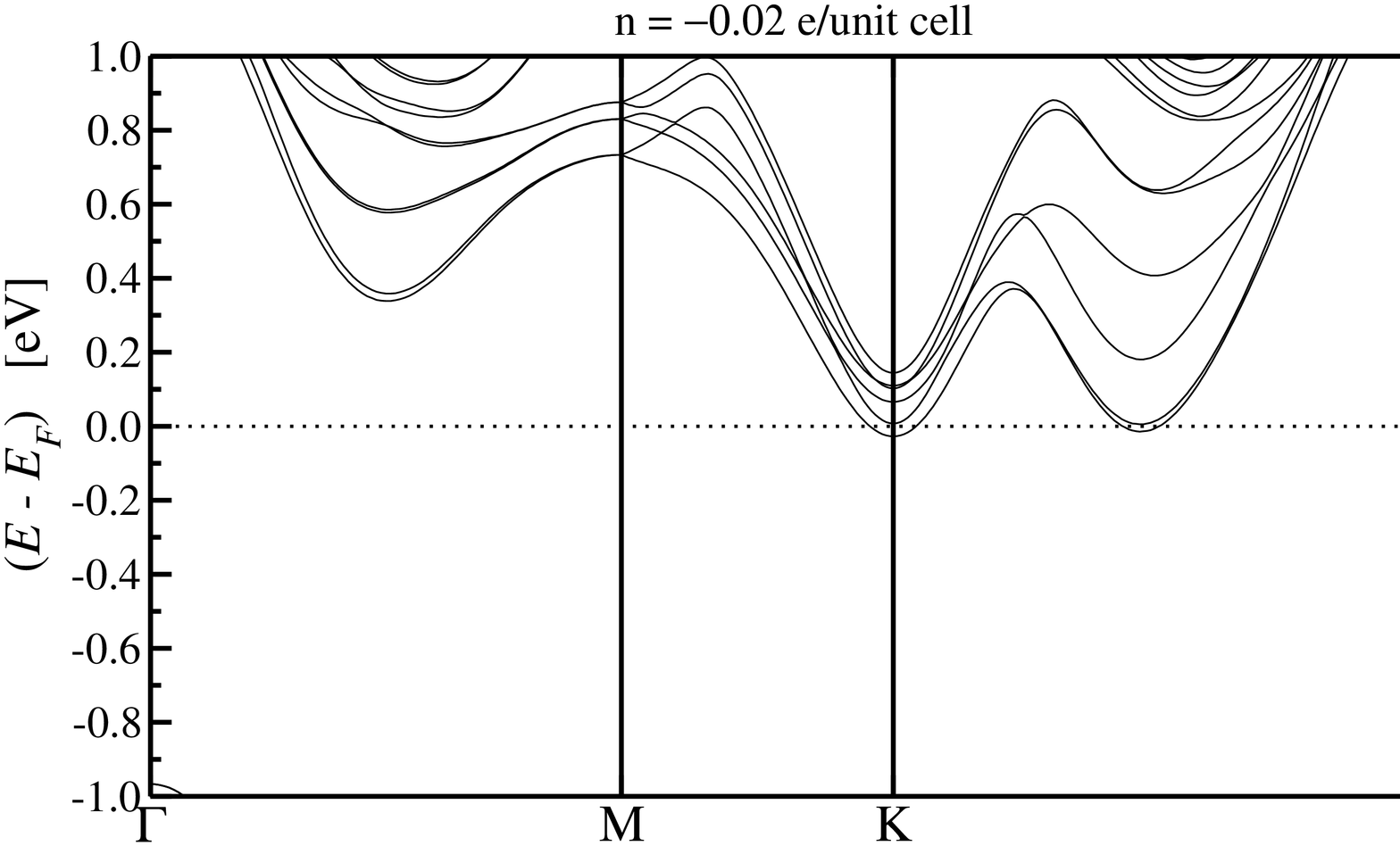}
 \includegraphics[width=0.31\textwidth,clip=,angle=-90]{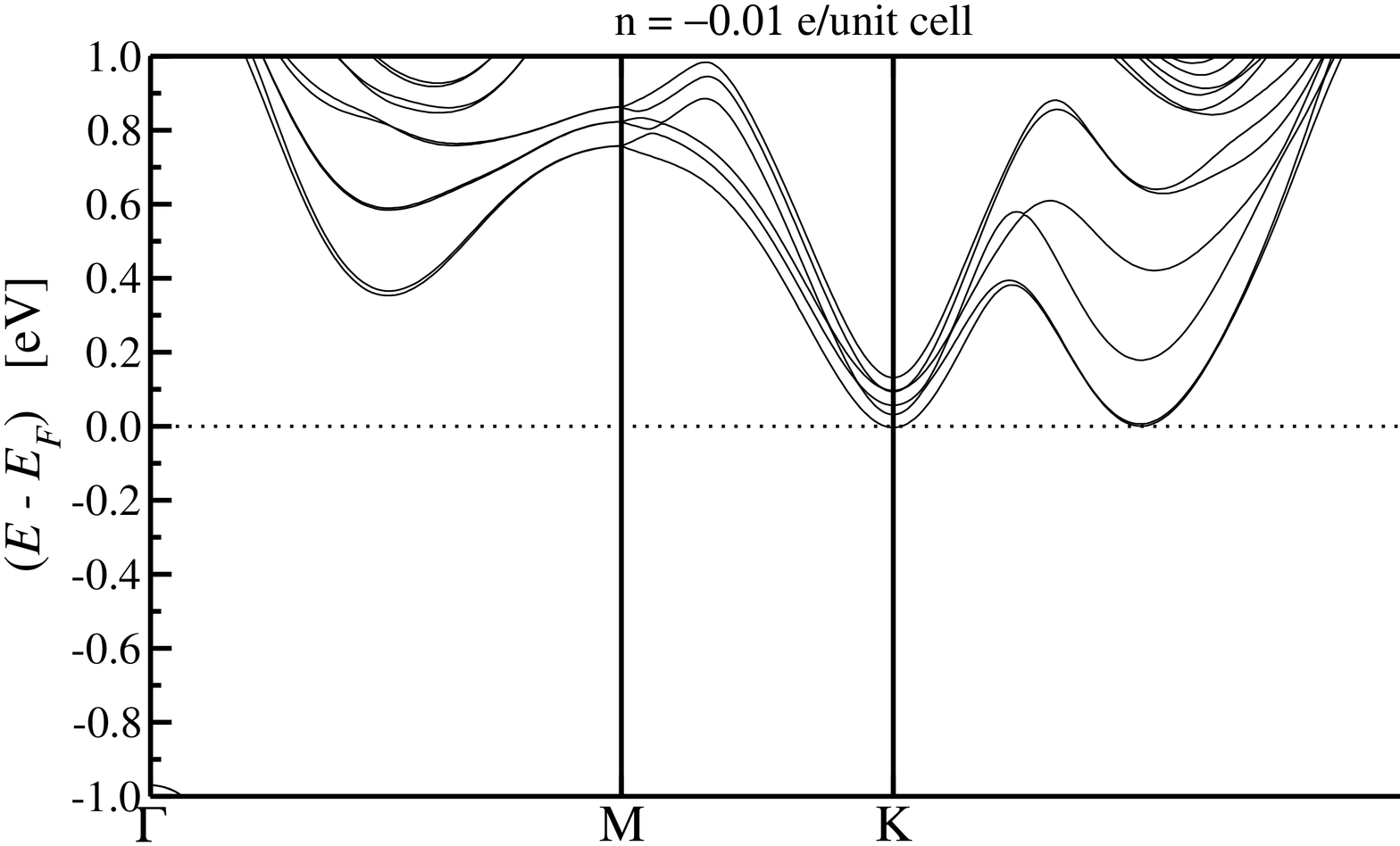}
 \includegraphics[width=0.31\textwidth,clip=,angle=-90]{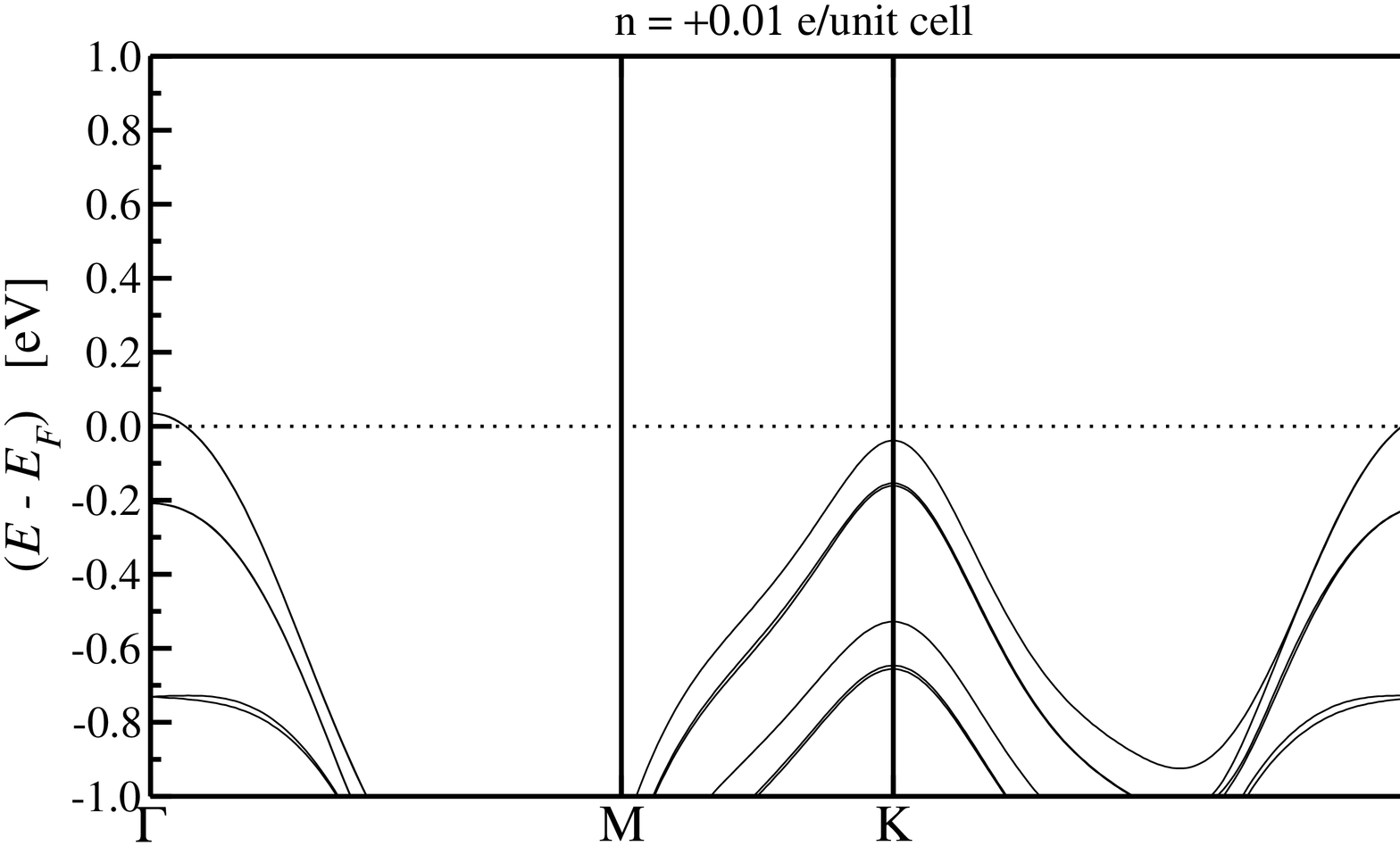}
 \caption{Band structure of trilayer WSe$_2$ for different doping as indicated in the labels.}
\end{figure*}
\begin{figure*}[hbp]
 \centering
 \includegraphics[width=0.31\textwidth,clip=,angle=-90]{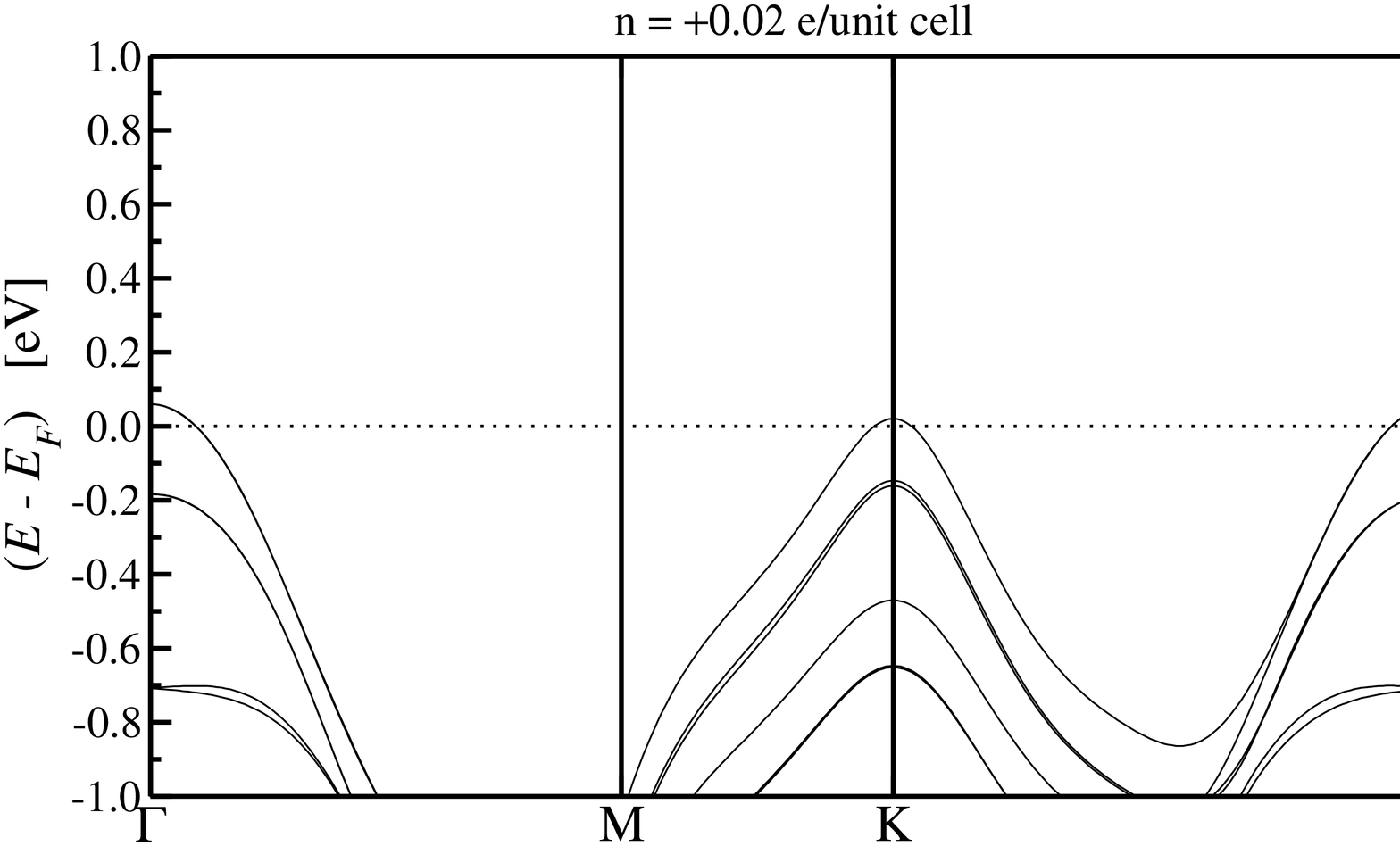}
 \includegraphics[width=0.31\textwidth,clip=,angle=-90]{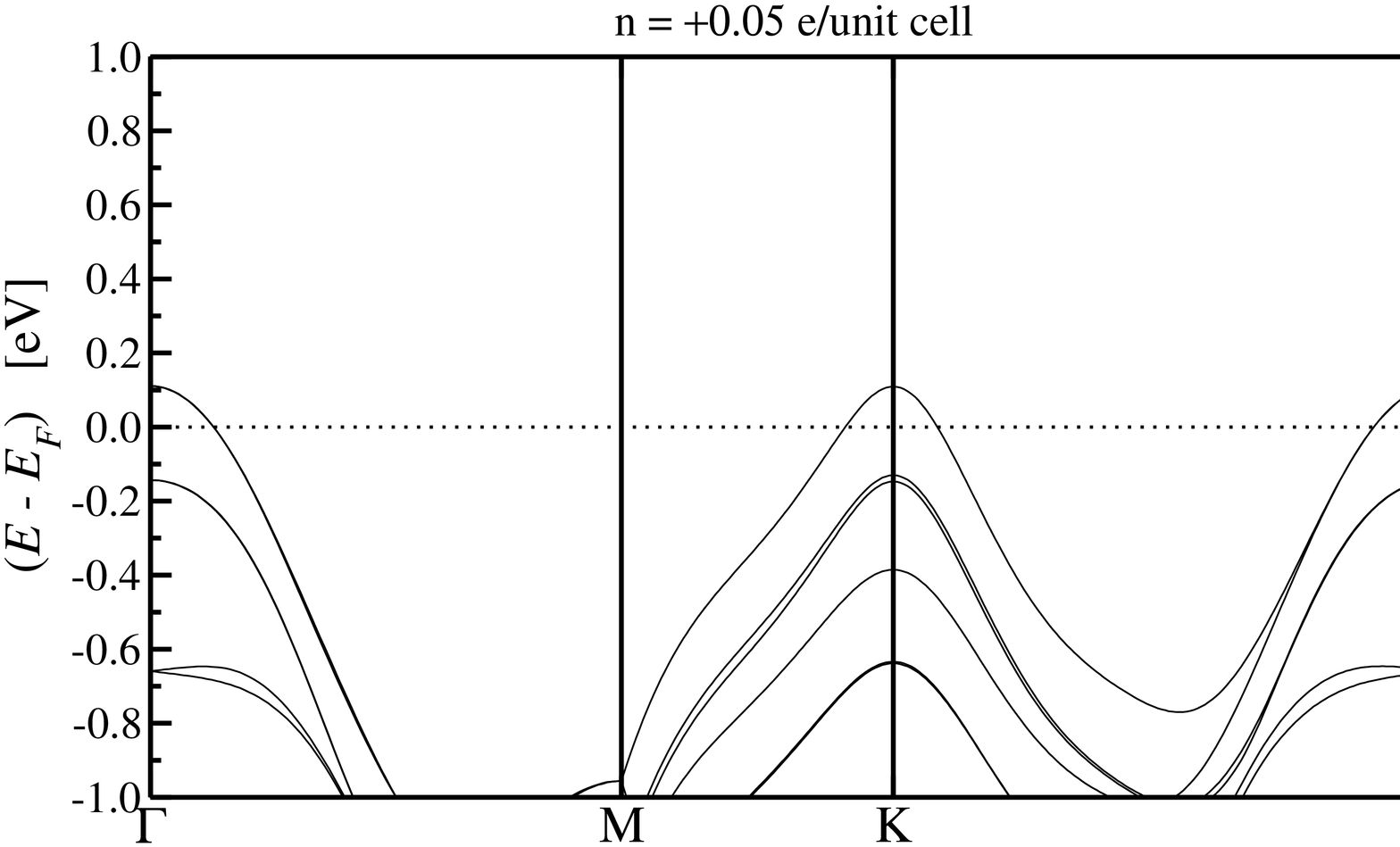}
 \includegraphics[width=0.31\textwidth,clip=,angle=-90]{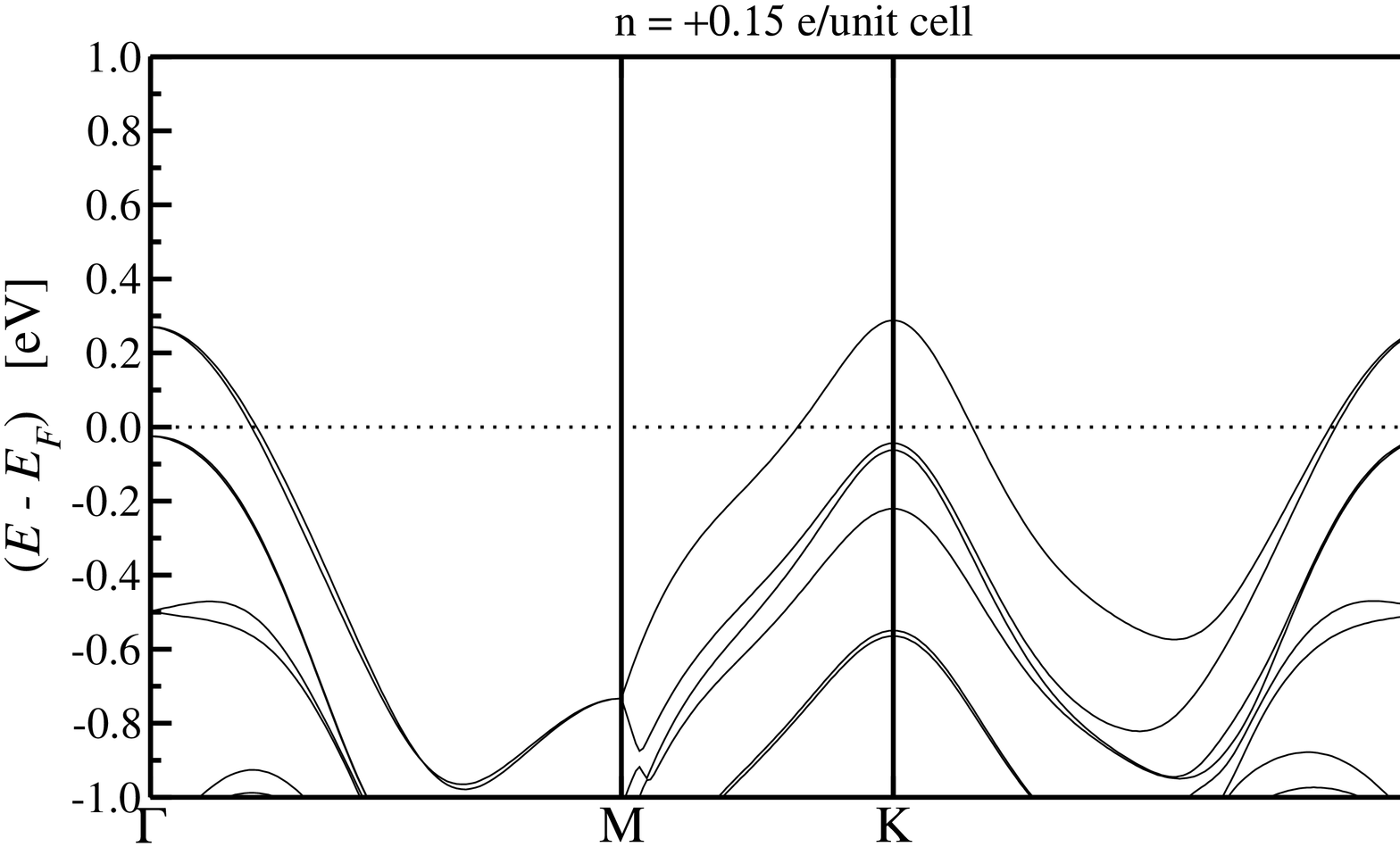}
 \includegraphics[width=0.31\textwidth,clip=,angle=-90]{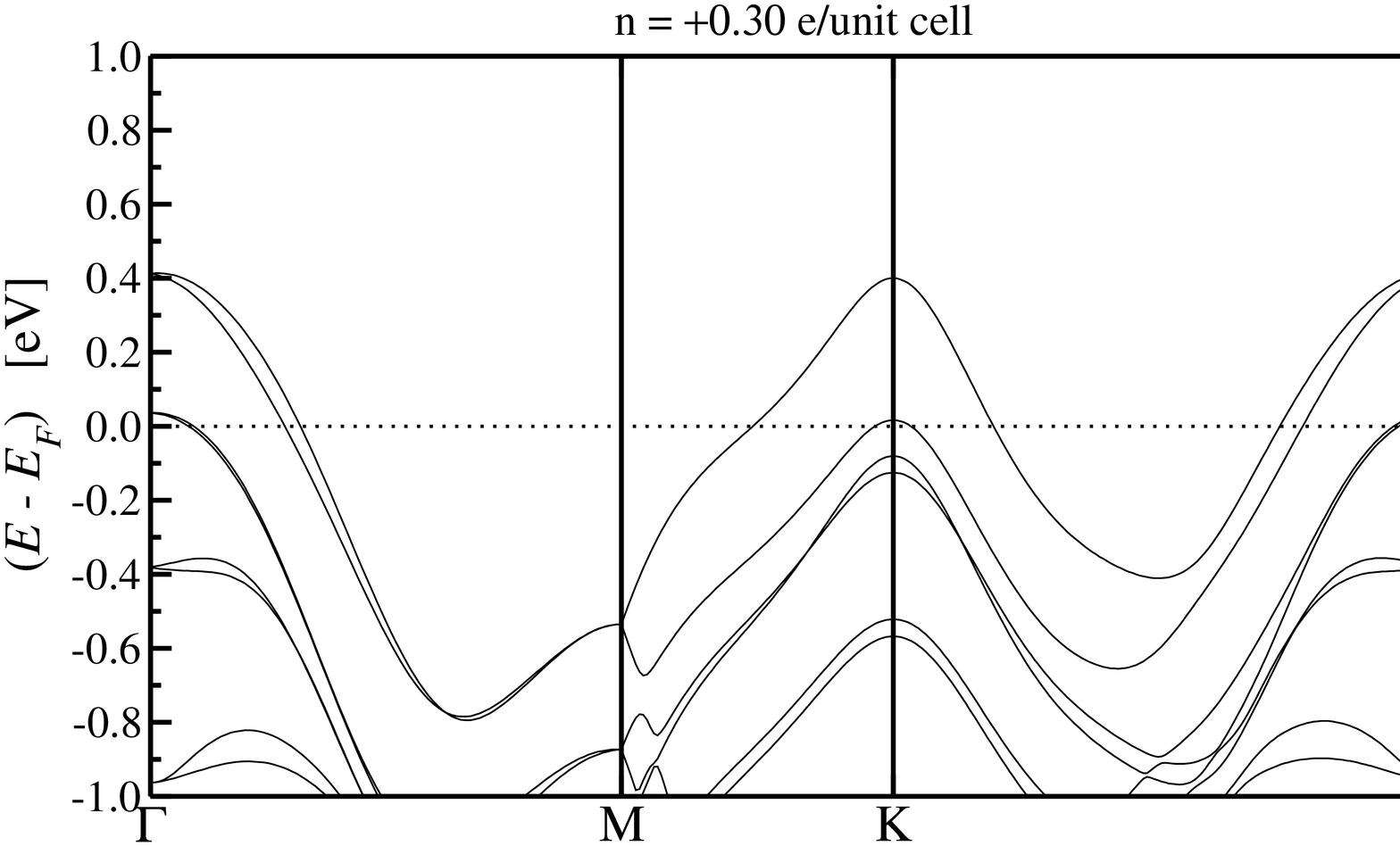}
 \includegraphics[width=0.31\textwidth,clip=,angle=-90]{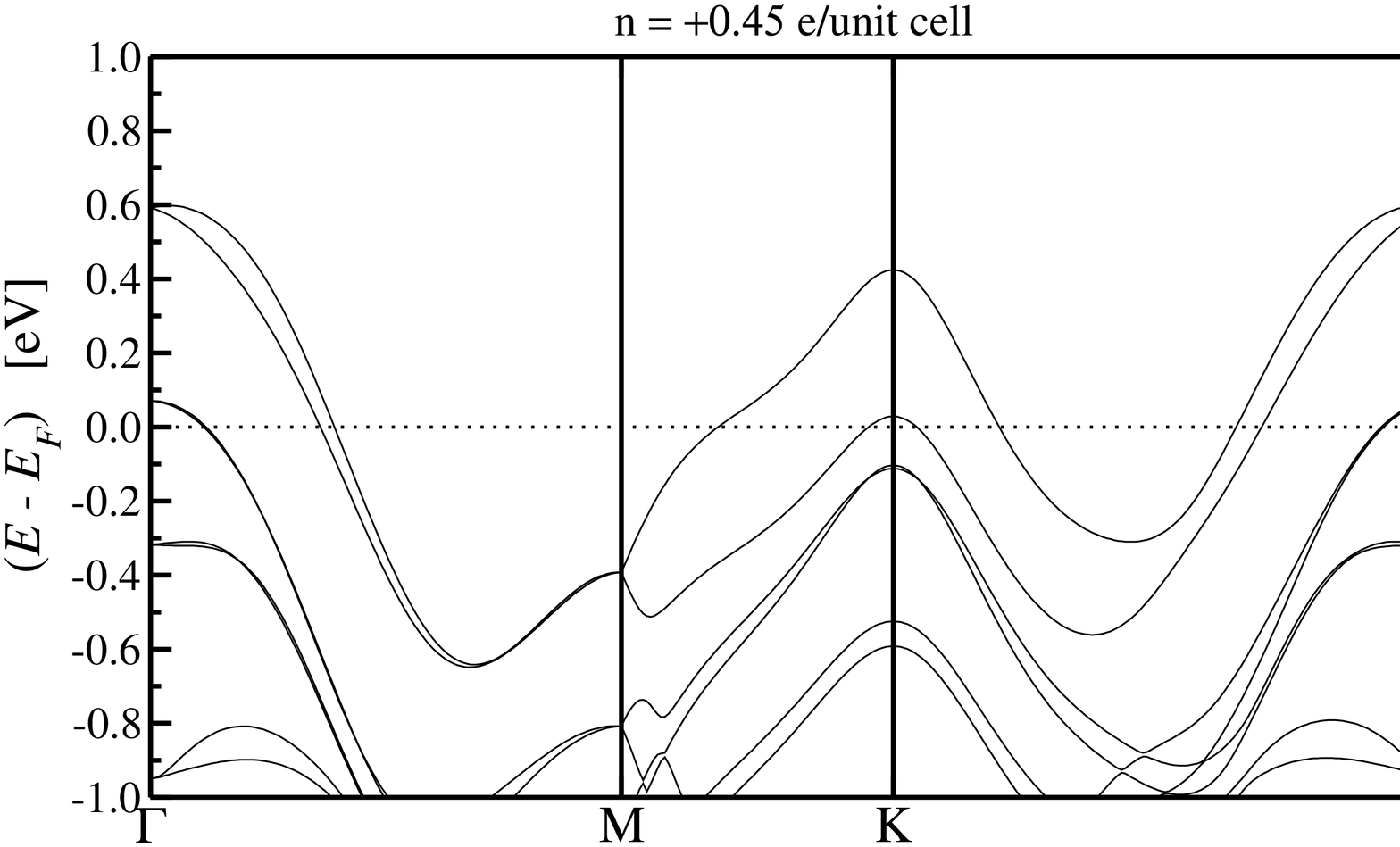}
 \includegraphics[width=0.31\textwidth,clip=,angle=-90]{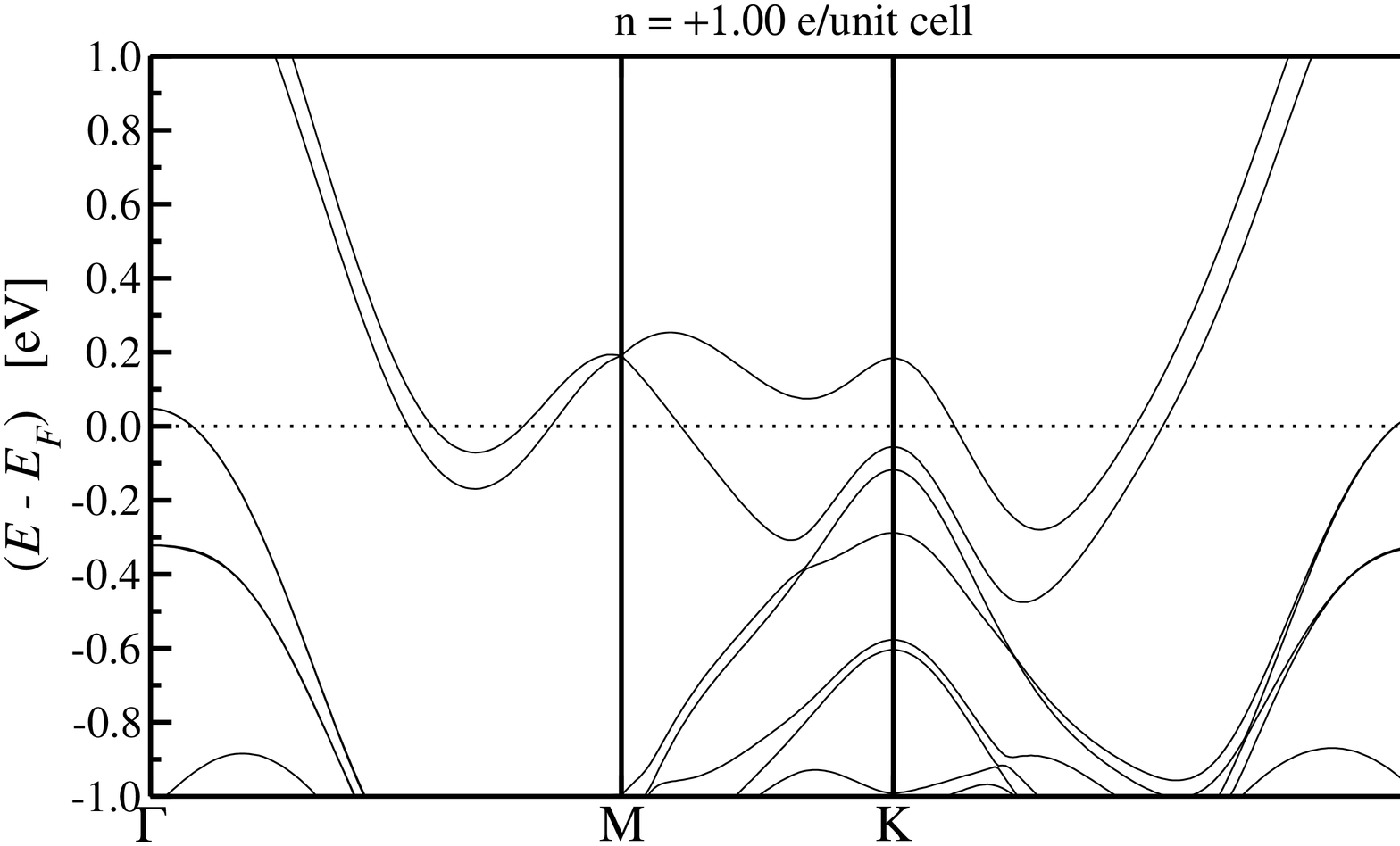}
 \caption{Band structure of trilayer WSe$_2$ for different doping as indicated in the labels.}
\end{figure*}

\clearpage
\section{Hall measurements}
\begin{figure*}[hbp]
 \centering
 \includegraphics[width=0.51\textwidth,clip=]{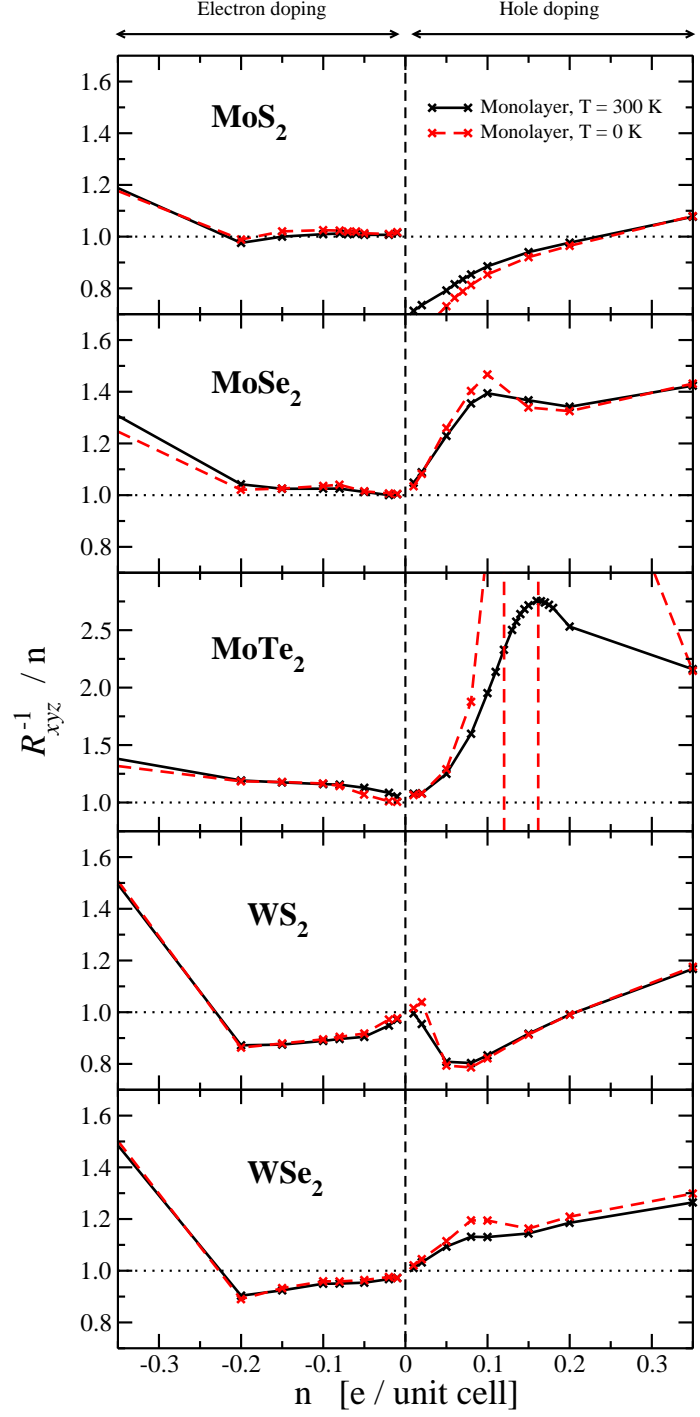}
 \caption{Ratio of the inverse Hall coefficient $R^{-1}_{xyz}$ to the doping charge $\mathrm{n}$ as function of doping
          for all monolayer TMDs and temperatures $T=0\:\mathrm{K}$ and $T=300\:\mathrm{K}$.}
\end{figure*}
\begin{figure*}[hbp]
 \centering
 \includegraphics[width=0.51\textwidth,clip=]{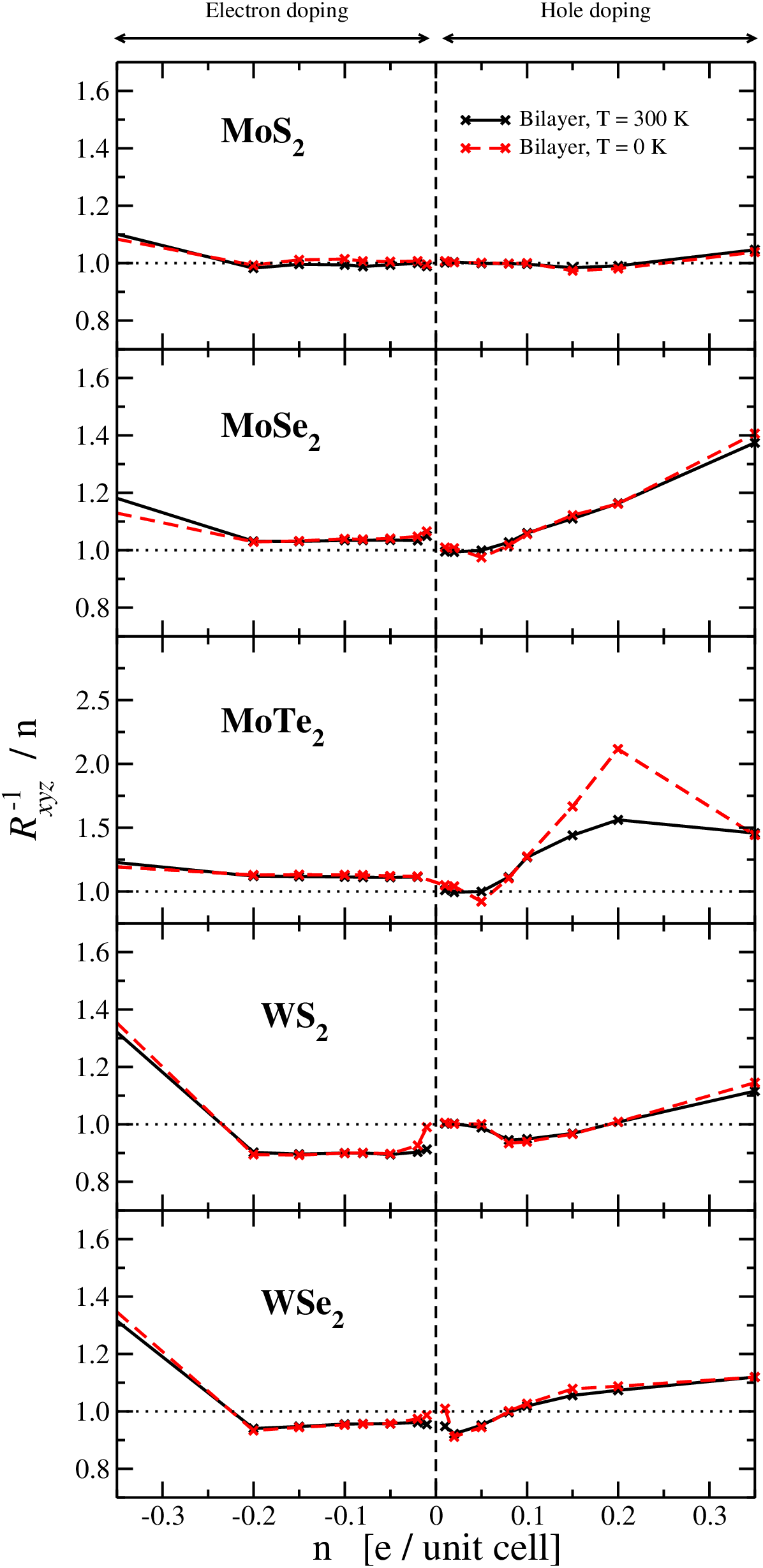}
 \caption{Ratio of the inverse Hall coefficient $R^{-1}_{xyz}$ to the doping charge $\mathrm{n}$ as function of doping
          for all bilayer TMDs and temperatures $T=0\:\mathrm{K}$ and $T=300\:\mathrm{K}$.}
\end{figure*}
\begin{figure*}[hbp]
 \centering
 \includegraphics[width=0.51\textwidth,clip=]{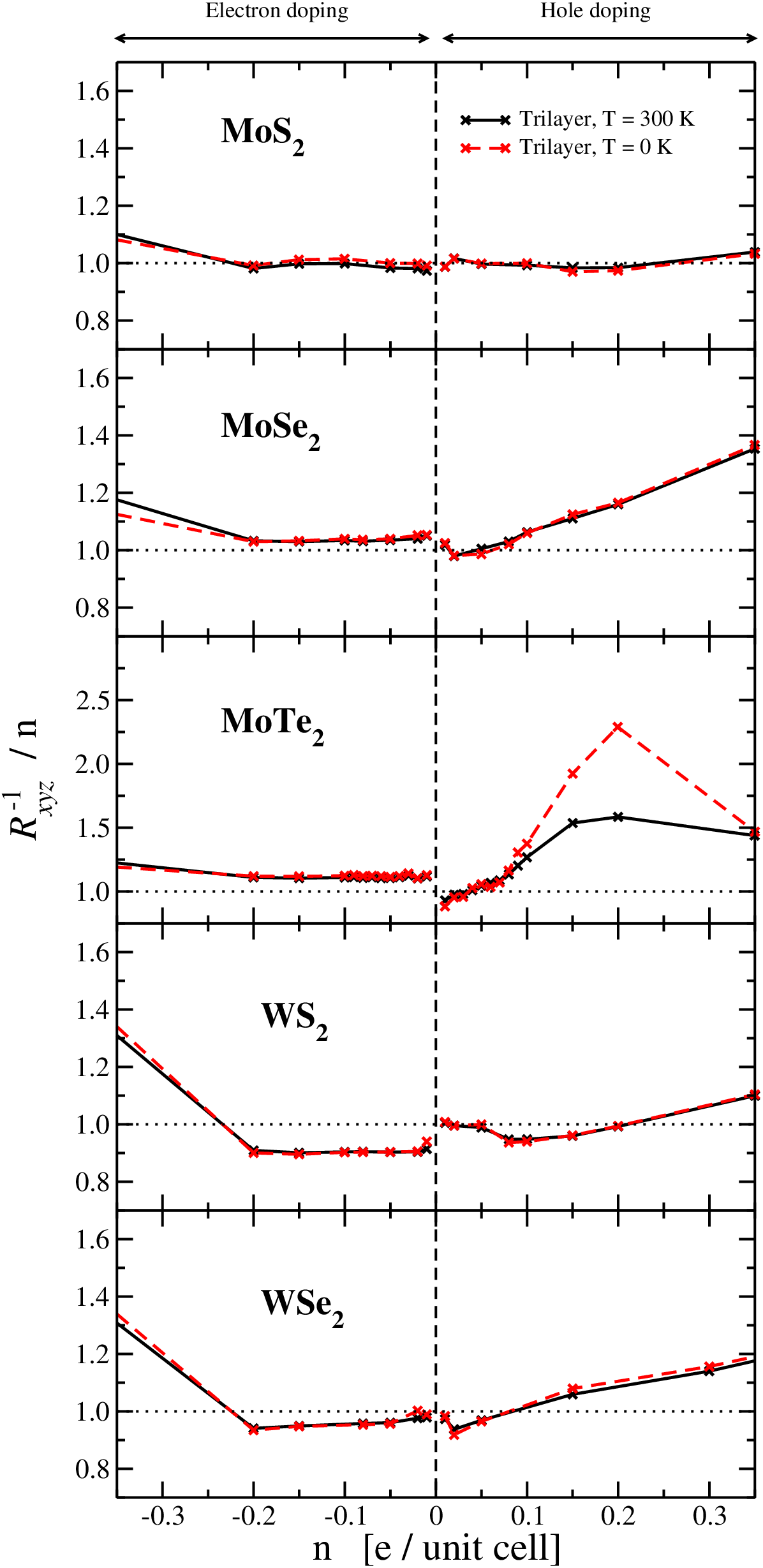}
 \caption{Ratio of the inverse Hall coefficient $R^{-1}_{xyz}$ to the doping charge $\mathrm{n}$ as function of doping
          for all trilayer TMDs and temperatures $T=0\:\mathrm{K}$ and $T=300\:\mathrm{K}$.}
\end{figure*}

\end{document}